\documentclass[11pt,vi,twoside,leftblank]{thesis}
\usepackage{lgrind}
\usepackage{cmap}
\usepackage[english]{babel}
\usepackage[T1]{fontenc}      
\usepackage[latin1]{inputenc}
\usepackage{relsize}
\usepackage{soul}
\usepackage{xspace}
\usepackage[final]{graphicx}  

\usepackage{amsmath,amssymb}
\usepackage{ctable}

\usepackage{pdflscape}
\usepackage[font=normal,format=plain,labelfont=bf,up,textfont=normal,up,justification=justified,singlelinecheck=false]{caption}
\usepackage{chngcntr}

\usepackage{bbm}

\newcommand{\ii}{\textrm {i}}
\newcommand{\ee}{\textrm {e}}
\newcommand{\ga}[1]{\gamma^{#1}}

\newcommand{\unit}{I}
\newcommand{\de}[1]{\partial_{#1}}

\newcommand{\br}[1]{\left( #1 \right)}

\newcommand{\brt}{\br{t}}

\newcommand{\com}[1]{\left[ #1 \right]}
\newcommand{\acom}[1]{\left \{ #1 \right \}}

\newcommand{\ma}[1]{\begin{pmatrix} #1 \end{pmatrix}}

\newcommand{\Ct}{\cos \br{\theta}}
\newcommand{\St}{\sin \br{\theta}}
\newcommand{\qt}{\br{q_x,p_{\rho},t}}
\newcommand{\brw}{\br{\mathbf{x},\mathbf{p},t}}
\newcommand{\brx}{\br{\mathbf{x},t}}

\usepackage{fancyhdr} 
\usepackage{float} 

\usepackage{tikz}
\usetikzlibrary{arrows,decorations.pathmorphing,decorations.markings,backgrounds,positioning,fit,petri,mindmap,spy}

\usepackage{mcode}

\usepackage{url}
\usepackage{cite}
\usepackage{epigraph}

\usepackage{tocloft}

\usepackage[titletoc]{appendix}
\usepackage{xpatch}
\makeatletter
\newcommand\listofappendixname{\appendixname}

\newcommand\listofappendices{%
    \if@twocolumn
      \@restonecoltrue\onecolumn
    \else
      \@restonecolfalse
    \fi
    \chapter*{\listofappendixname
        \@mkboth{%
           \MakeUppercase\listofappendixname}{\MakeUppercase\listofappendixname}}%
    \@starttoc{toa}%
    \if@restonecol\twocolumn\fi
    }
\g@addto@macro\appendix{%
  \addcontentsline{toc}{chapter}{\appendixname}%
  \xpatchcmd{\@part}{\addcontentsline{toc}}{\addcontentsline{toa}}{}{}%
  \xpatchcmd{\@part}{\addcontentsline{toc}}{\addcontentsline{toa}}{}{}%
  \xpatchcmd{\@chapter}{\addcontentsline{toc}}{\addcontentsline{toa}}{}{}%
  \xpatchcmd{\@chapter}{\addcontentsline{toc}}{\addcontentsline{toa}}{}{}%
  \xpatchcmd{\@sect}{\addcontentsline{toc}}{\addcontentsline{toa}}{}{}%
  \xpatchcmd{\@sect}{\addcontentsline{toc}}{\addcontentsline{toa}}{}{}%
}
\makeatother

\DefineNamedColor{named}{RR}  {rgb}{1,0.0,0.0}
\DefineNamedColor{named}{ausgrauen}         {rgb}{0.92,0.93,0.93}
\newcommand{\hlight}{\color{RR}}

\begin{document}

\author{Christian Kohlf\"urst}
\title{Electron-positron pair production \\ in inhomogeneous electromagnetic fields}
\date{2015} 

\newpage
  \thispagestyle{empty}
  \null
  \normalbaselines
  \begin{center}%
    {\openup 0.5cm \Large Christian Kohlf\"urst \par}%
    \vskip 2.5cm%
    {\LARGE \textbf{Electron-positron pair production \\ in inhomogeneous electromagnetic fields} \par }%
    \vskip 3.5cm%
	 \large \textbf{Dissertation} \par
    \vskip 1.0cm%
	 \large zur Erlangung des Doktorgrades der Naturwissenschaften \par
	 an der Naturwissenschaftlichen Fakult\"at der \par
	 Karl-Franzens-Universit\"at Graz \par
    \vskip 4cm \par

\begin{minipage}{0.35\textwidth}
 \begin{flushleft}
  Advisor: \\
 \end{flushleft}
\end{minipage}
\begin{minipage}{0.55\textwidth}
 \begin{flushright}
  Univ-Prof. Dr. rer. nat. Reinhard Alkofer \\
\end{flushright}
\end{minipage}
\vskip 1cm \par
       Institut f\"ur Physik \par
       Fachbereich Theoretische Physik
	\vfill 
	\par
	\vspace{1em plus 2em minus 1em}
   \par \normalbaselines \noindent
  \number\year
  \end{center}%
\normalsize


\cleardoublepage
\setcounter{savepage}{\thepage}

\begin{abstractpage}
 The process of electron-positron pair production is investigated within the phase-space Wigner formalism. 
The similarities between atomic ionization and pair production for homogeneous, but time-dependent linearly polarized electric fields are examined mainly in the regime of multiphoton absorption (field-dependent threshold, above-threshold pair production). Characteristic signatures in the particle spectra are identified (effective mass, channel closing). The non-monotonic dependence of the particle yield on the carrier frequency is discussed as well. The investigations are then extended to spatially inhomogeneous electric fields. New effects arising due to the spatial dependence of the effective mass are discussed in terms of a semi-classical interpretation. An increase in the normalized particle yield is found for various field configurations.
Pair production in inhomogeneous electric and magnetic fields is also studied. The influence of a time-dependent spatially inhomogeneous magnetic field on the momentum spectrum and the particle yield is investigated. The Lorentz invariants are identified to be crucial in order to understand pair production by strong electric fields in the presence of strong magnetic fields.  

\subsection*{Zusammenfassung}

Elektron-Positron Paarerzeugung wird mittels des Wigner-Formalismus im Phasenraum untersucht. Die Gemeinsamkeiten von Atomionisation und Paarerzeugung werden dabei f\"ur homogene, zeitabh\"angige, linear polarisierte elektrische Felder im Bereich der Mehr\-photonen\-absorption  untersucht (feldabh\"angige Produktionsschwelle etc.). Dabei wurden charakteristische Signaturen im Teilchenspektrum gefunden (effektive Masse, Channel Closing). Au\ss erdem wird der nicht-monotone Zusammenhang zwischen Produktionsrate und Feldfrequenz behandelt. Die Untersuchungen werden dann auf r\"aumlich inhomogene Felder erweitert. Neue Effekte, die im Zusammenhang mit einer raumabh\"angigen effektiven Masse stehen, werden mithilfe semi-klassischer Methoden diskutiert. Ein Anstieg der r\"aumlich normierten Produktionsrate f\"ur spezielle Feldkonfigurationen wurde gefunden.
Die Untersuchungen werden um r\"aumlich und zeitlich inhomogene magnetische Felder erweitert, und deren Auswirkungen auf das Teilchenspektrum wird untersucht. Die Lorentzinvarianten werden als ausschlaggebend f\"ur die Produktionsrate durch starke elektrische Felder in Anwesenheit von starken magnetischen Feldern identifiziert.

\end{abstractpage}



\cleardoublepage

\section*{Acknowledgments}

I would like to share my deepest gratitude for the help I received over the last years. This thesis would not have been possible without the permanent support of my colleagues, my friends and my parents. Nevertheless, I want to mention some of the ``contributors'' by name. \\

At first I want to thank my advisor Reinhard Alkofer for ongoing support and giving me free rein to follow my own ideas. I would also like to thank Holger Gies from the FSU Jena. He had a significant impact on my research and consulting him was always valuable. Additionally, I want to thank Tom Heinzl and colleagues at Plymouth University for their generous hospitality. \\

I take this opportunity to express gratitude to all of the Department faculty members for their help and support.
Furthermore, I want to thank the principal investigators of the Doktoratskolleg ``Hadrons in Vacuum, Nuclei and Stars'' for giving me the opportunity to participate in the graduate program. 
I also want to acknowledge the Austrian Science Fund, FWF, for financial support (FWF DK W1203-N16). Additionally, I want to mention the
University of Graz' research core are ``Models and Simulation''. \\

My office mates in Graz and Jena played a big role in the last three years, because we shared a lot of time together. Working on my PhD was often stressful and demanding, but with your help it was an enjoyable part of my life. Hence, my deepest gratitude to Richard Haider, Valentina Verduci, Milan Vujinovic, Mario Giuliani, Alexander Blinne and Nico Seegert. 
Besides my office mates, I want to thank Julia Borchardt, Ana Juricic, Ydalia Delgado Mercado, Matthias Blatnik, Hans-Peter Schadler, Pascal T\"orek, Markus Pak and Alexander Goritschnig. \\

A special thank goes to our football team. It was a pleasure to participate. 
A note to Anita Ulz, conversations with you are and have always been refreshing and entertaining.
A special thanks to Matthias Blatnik for careful proofreading my drafts. \\

Ganz besonderen Dank geb\"uhrt meinen Eltern, Hans und Roswitha, f\"ur fast auf den Tag genau 26 Jahre vollste Unterst\"utzung. Ein Dankesch\"on auch an meinen Bruder David, es macht immer wieder Freude mit dir etwas zu unternehmen. Ganz zum Schluss noch einen besonderen Dank an meine Oma Maria, die diese Arbeit leider nicht mehr miterleben konnte. 


\tableofcontents
\newpage
\listofappendices
\newpage
\listoffigures
\addcontentsline{toc}{chapter}{List of Figures}
\newpage
\listoftables
\addcontentsline{toc}{chapter}{List of Tables}

\pagestyle{plain}

\chapter*{Notation}
\addcontentsline{toc}{chapter}{Notation}

If not stated otherwise, Einstein summation convention is used. Additionally, when Greek letters e.g. $\mu$ are used as indices the following holds: $\mu = 0,~ 1,~ 2,~ 3$. We write the unit matrix as $\unit$ and for the unit vectors we introduce the notation $\mathbf{e}_i$, where $i$ denotes the placement of the $1$.

Throughout this thesis the metric tensor $\eta^{\mu \nu} = \text{diag} \br{1,-1,-1,-1}$ is used. In addition we denote $x = x^{\mu} = \br{t, \mathbf{x}}$. We introduce the notation
\begin{align}
 \de{t} = \frac{\partial}{\partial t} ,\qquad  \de{p_x} = \frac{\partial}{\partial p_x} ,\qquad \partial_{\mu}^r = \frac{\partial}{\partial r^{\mu}} ,\qquad \partial_{\mu}^p = \frac{\partial}{\partial p^{\mu}} ,
\end{align}
in order to distinguish derivatives with respect to spatial quantities from derivatives with respect to momentum quantities. This system is also used for the nabla operator, where we write
\begin{align}
 \boldsymbol{\nabla}_x \quad \text{and} \quad \boldsymbol{\nabla}_p.
\end{align}

\ctable[pos=h,
caption = {Recurrent constants. For the sake of readability, if not stated otherwise we use natural units by setting $c = \hbar = 1$. In order to not distract the reader, we omit writing $m_e$ in chapter three and the first part of chapter four.},
cap={Important constants.},
label = {},
mincapwidth = \textwidth,
]{l c l c}{}{
    \toprule
    $c$ & speed of light & \hspace{1cm} $e$ & coupling constant \\
    \midrule
    $\hbar$ & reduced planck constant & \hspace{1cm}  $\lambda_C$ & Compton wavelength \\ & & & of the electron \\ 
    \midrule
    $m,~ m_e$ & electron mass \\ 
    \bottomrule
}    

\ctable[pos=h,
caption = {Throughout this thesis we will introduce various abbreviations for the most common terms.},
cap = {Important abbreviations.},
label = {},
mincapwidth = \textwidth,
]{l l l l}{}{
    \toprule
    QFT & Quantum Field Theory & \hspace{1cm} & \\
    \midrule
    QED & Quantum Electrodynamics & \hspace{1cm} & \\
    \midrule    
    DE & Differential equation & \hspace{1cm} & \\
    \midrule
    ODE & Ordinary differential equation & \hspace{1cm} & \\
    \midrule
    PDE & Partial differential equation & \hspace{1cm} & \\    
    \midrule
    DHW & Dirac-Heisenberg-Wigner formalism & \hspace{1cm} & \\
     & a.k.a. VGE(Vasak-Gyulassy-Elze) equations & & \\
    \midrule
    QKT & Quantum Kinetic Theory & \hspace{1cm}  &  \\  
    \midrule
    FDM & Finite Difference Method & \hspace{1cm}  &  \\  
    \midrule
    FEM & Finite Element Method & \hspace{1cm}  &  \\  
    \midrule
    FFT & Fast-Fourier Transform & \hspace{1cm}  &  \\      
    \midrule
    WKB & Wentzel-Kramers-Brillouin approximation & \hspace{1cm} & \\  
    \midrule
    CEP & Carrier Envelope Phase & \hspace{1cm}  &  \\  
    \midrule
    FEL & Free Electron Laser & \hspace{1cm}  &  \\  
    \midrule
    CPA & Chirped Pulse Amplification & \hspace{1cm}  &  \\      
    \midrule
    SASE & Self-amplified spontaneous emission  & \hspace{1cm}  &  \\    
    \midrule
    TEM & Transversal Electromagnetic Mode & \hspace{1cm}  &  \\   
    \midrule
    RAM & Random-Access-Memory & \hspace{1cm} & \\
     & (working memory) & & \\    
    \bottomrule
}    
\clearpage
\newpage
\ctable[pos=h,
caption = {Occurring styles for the quantities. A single number shares the style with a four-vector. The various Wigner functions/operators are excluded from this notation.},
cap = {Styles for occurring quantities.},
label = {},
mincapwidth = \textwidth,
]{l c}{}{
    \toprule
    $x$ & \hspace{1cm} single number, \\ & \hspace{1cm} Lorentz four-vector \\
    \midrule    
    $\mathbf{x}$ & \hspace{1cm} vector \\
    \midrule
    $\overline{\mathbf{M}}$ & \hspace{1cm} matrix  \\ 
    \midrule
    $\hat {\mathcal{C}}$ & \hspace{1cm} operator \\ 
    \bottomrule
}    

\ctable[pos=h,
caption = {Style sheet for the Wigner components.},
cap = {Styles for the Wigner components.},
label = {},
mincapwidth = \textwidth,
]{l c}{}{
    \toprule
    $W$ & \hspace{1cm} Wigner function \\ & \hspace{1cm} in quantum mechanics \\
    \midrule        
    $\hat {\mathcal W}$ & \hspace{1cm} covariant Wigner operator \\    
    \midrule        
    $\mathbbm{W}$ & \hspace{1cm} covariant Wigner function \\
    \midrule    
    $\mathbbm{w}$ & \hspace{1cm} equal-time(single-time) Wigner function \\
    \midrule    
    $\hat {\mathbbm{w}}$ & \hspace{1cm} Fourier transformed \\ & \hspace{1cm} equal-time Wigner function \\    
    \midrule
    $\mathbbm{w}_k$ & \hspace{1cm} equal-time Wigner vector \\ & \hspace{1cm} ($k$ holds as a general index)  \\ 
    \midrule
    $\mathbbm{w}_i$ & \hspace{1cm} initial condition for \\ & \hspace{1cm} specific Wigner component  \\     
    \bottomrule
}   

\newpage
\thispagestyle{empty}
\mbox{}

\part{Considerations}
\pagestyle{plain}
\chapter{Introduction}
\pagestyle{fancy}
\epigraph{``It is all a matter of time scale. An event that would be unthinkable in a hundred years may be inevitable in a hundred million.''}{Carl Sagan, \textit{Cosmos}\cite{SaganCosmos}}

The rapid development of Lasers brings us closer and closer to a new field of experimental science: non-linear QED. Although first theoretical studies on pair production and vacuum polarization have been undertaken in the 1930's\cite{Sauter,PhysRev.46.1087,Bethe,HeisenbergEuler}, the intensities needed in order to measure them were out of reach for a long time. However, within the last decade many new facilities have been planned aiming to probe strong-field QED or investigating Laser-matter interactions at high intensities\cite{LaserSys}. 

Using advanced Laser technology we could not only probe time-resolved processes in molecules enabling us to study real-time chemical reactions in great detail \cite{LaserChem}. 
We could also progress in developing new sources of energy being sustainable and ecologically clean. The key technologies towards green energy are possibly fast ignition and laser-driven fusion\cite{RevModPhys.46.325,LaserFusion}.
The progress in X-ray Laser physics is equally interesting. Due to the improvements in creating intense light at very short wavelengths a promising new tool for probing electromagnetic interactions is emerging\cite{LaserXFEL,Ringwald2001107}. Performing experiments with this new light source could lead to a better understanding of high-intensity Laser interactions.
Last but not least, modern Laser systems could probe strong-field QED. At intensities of this magnitude the vacuum becomes polarized and according to various calculations this should lead to remarkable new phenomena\cite{itzykson2012quantum,MarlundQED,HeinzlQED}. One of these anticipated effects is particle pair production, the conversion of light into matter\cite{HeisenbergEuler,Schwinger,PhysRevD.2.1191}.

As it is of essential importance to understand what could happen when switching-on ultra-high intensity Lasers, our goal for this thesis is to broaden the knowledge regarding electron-positron pair production. This topic is deeply connected with general particle creation. Baryogenesis and the asymmetric distribution of matter and anti-matter in the universe is probably the most prominent example.  
Besides matter creation in the early universe, particle creation holds also as one key process in order to interpret astrophysical measurements, for example Hawking radiation\cite{Hawking} or pair instability supernovas\cite{Fraley}.
Nonetheless, particle production is not only restricted to astrophysics. The flux tube model, for example, was introduced in order to describe heavy ion collisions at hadron colliders (see RHIC or the LHC). To interpret such a heavy ion collision the formation of a chromoelectric flux tube\cite{PhysRevD.20.179} has been considered. If the energy stored in these flux tubes is sufficiently high they break creating an additional quark and antiquark, respectively. 

Returning to the particular case of electron-positron pair production the advances in Laser technology provides us novel possibilities. If particle production were feasible in a laboratory under controlled conditions it would provide us a powerful new tool in order to study high-energy processes with unprecedented precision. We would be able to test quantum electrodynamics(QED) at a completely new scale and open the door to new physics (non-linear QED)\cite{doi:10.1142/S0217751X1260010X,RevModPhys.84.1177}. 

This doctoral thesis is basically divided into three parts. Chapters one to four provide the necessary background in order to understand the results obtained.
In section \ref{Sec_His} we give a brief overview on the history having a direct impact on the following discussion on pair production. Certain key inventions in theoretical as well as applied physics had to be made, thus we will mention the most important milestones. 
On the theoretical side, the introduction of a profound theory of the electromagnetic force holds as one of the cornerstones of modern theoretical physics. Moreover, the invention and the subsequent development of Lasers opened up new possibilities in research and revolutionized our everyday world. To respect these achievements, we will discuss Maxwell equations as well as the Euler-Heisenberg Lagrangian and QED. Additionally, we sketch the history of the Laser starting with early concept studies and ending with an overview of current high-power Laser facilities. Furthermore, an examination of the current status of the field is in order. By this means, we motivate the formalism used throughout this thesis by showing its success in describing $N$-particle systems. In section \ref{Mechanism} we introduce the various mechanisms, that could lead to pair production, on the basis of atom physics. Following the concepts developed in ionization physics we are able to derive certain key elements of particle creation leading to the interpretations given in the second part of the thesis.

In the third chapter we analyze the Dirac-Heisenberg-Wigner(DHW) formalism. Derivation of the transport equations describing pair production in up to $3+1$ dimensions is done in great detail. Moreover, we draw connections to pair production in lower dimensions. The findings are supported by an analysis of the various symmetries contained in the transport equations. Additionally, we discuss relevant observables including the charge density and the particle distribution. In the end, we take the classical limit and demonstrate how to obtain the relativistic Vlasov equation.  

The fourth chapter covers all relevant information in order to solve the equations obtained via applying the DHW formalism to the pair production problem. We give an introduction in pseudo spectral methods and consider different solution strategies. We discuss various possibilities to numerically stabilize the computation.
Besides, we discuss various models for the electromagnetic background field regarding time-dependence as well as spatial inhomogeneity. Additionally, we introduce a semi-classical picture in order to interpret the final particle distribution. 

In the second part of the thesis, chapters five to eight, we identify three different regimes of pair production. Based upon this differentiation we present our results accompanied by detailed discussions. As we are trying to cover many aspects of pair production, we split our findings into three different chapters. In chapter five and six we investigate the regime of multiphoton pair production. More precisely, in chapter five we introduce the concept of an ``effective'' particle mass. We discuss the consequences of a field-dependent threshold and examine the significance of the characteristic Laser pulse parameters on the particle phase-space signature. In chapter six we extend the effective mass concept to spatially inhomogeneous problems. Furthermore, we present calculations for $3+1$ dimensions including also the transversal momenta of the created particles. 

In chapter seven, we discuss our findings regarding pair production in an inhomogeneous electric and magnetic background field. Due to the increased complexity introduced by a time-dependent, spatially inhomogeneous magnetic field we perform a first feasibility study. The results are illustrated for a wide range of parameter sets demonstrating how a magnetic field influences the momentum spectrum of the created particles. At last, we draw a conclusion which relates the Lorentz invariants to the total particle production rate.

In the end we summarize our findings and propose a short list of interesting topics about future projects regarding pair production. For the interested reader an appendix with detailed calculations as well as supplementary figures is prepared. Tables with the specific parameter values for all calculations can be found. Additionally, we provide code snippets in MATLAB in appendix \ref{App_Matlab}, which are sufficient in order to perform calculations regarding pair production within the DHW approach.
It should be mentioned, that we followed the ideas in Light et al. \cite{Light} for defining the colormaps used.
 
\newpage 
\thispagestyle{empty}
\begin{tikzpicture}[mindmap]
\begin{scope}[every node/.style={concept, execute at begin node=\hskip0pt,scale=1.1},
pairprod/.append style={
concept color=black, fill=orange!20, line width=1.0ex, text=black, font=\Huge, text width=10em},
text=white,
QKT/.style={concept color=green!50!black,faded/.style={concept color=green!20!black!25,text=black}},
DHW/.style={concept color=blue!75,faded/.style={concept color=blue!20,text=black}},
grow cyclic, 
level 1/.append style={level distance=5.5cm,sibling angle=180},
level 2/.append style={level distance=3.5cm,sibling angle=60},
level 3/.append style={level distance=3cm,sibling angle=60},
level 4/.append style={level distance=2.5cm,sibling angle=30}]
  \node(a) [pairprod] {Pair Production} 
   child[QKT] { node[text width=6.5em] {Phenomenology}
    child[sibling angle = 105] { node[text width=5.5em] {Spatial Inhomogeneity} 
     child { node[text width=3.5em] {Self-Bunching} }
     child { node[text width=3.5em] {Ponderomotive Force} }
     child { node[text width=3.5em] {B-Spin Interaction} }
     }
    child[sibling angle = 20] { node {$M_{\textrm{eff}}$}
     child { node[text width=3.5em] {ATPP} }
     child { node[text width=3.5em] {Channel Closing} }
    }
    child[sibling angle = 100] { node {Model for the field}
     child[faded] { node[text width=3.5em] {Rotating Fields} }
     child { node[text width=3.5em] {Linear Polarization} }
     child { node[text width=3.5em] {Cylindrical Symmetry} }
     child[faded] { node {Chirped Pulses} }
    }    
   }
   child[DHW] { node {Theory}
    child[sibling angle = 140]  { node[text width=6.0em] {Computational Techniques} 
     child { node {Spectral Methods} }
     child[faded] { node {FD} }
     child[faded] { node {FEM} }
     child { node {Symmetries} } 
    }
    child { node[text width=4.5em] {DHW} 
    child { node {$QED_{3+1}$} }
     child { node {$QED_{1+1}$} }
     child { node {$QED_{2+1}$} 
      child[faded] { node {E $\parallel$ B} }
      child { node {E$\perp$B} }
     }
    }
   };
   \end{scope}
\end{tikzpicture}

\pagestyle{plain}
\chapter{Overview}
\pagestyle{fancy}

As pair production is a vast subject of study we can only give a broad overview in this chapter. To do this, we will introduce important theoretical concepts and discuss their applicability. Moreover, we will motivate our studies and relate them to experimental prospects.

\section{Historical remarks}
\label{Sec_His}

\subsection{Electrodynamics}

There have been basically two milestones contributing significantly towards the understanding of electromagnetism.
The first major developments culminated in the formulation of the Maxwell equations describing classical electrodynamics\cite{Maxwell,jackson91}. 
Due to historical reasons we refrain from a fully covariant formalism of the Maxwell equations at this point. Rather, we write
\begin{alignat}{6}
 &\boldsymbol{\nabla} &&\cdot \mathbf{E} &&= \frac{\rho}{\varepsilon_0} ,\\
 &\boldsymbol{\nabla} &&\cdot \mathbf{B} &&= 0 ,\\
 &\boldsymbol{\nabla} &&\times \mathbf{E} &&= -\frac{\partial \mathbf{B}}{\partial t} ,\\
 &\boldsymbol{\nabla} &&\times \mathbf{B} &&= \mu_0 \mathbf{j} + \mu_0 \varepsilon_0 \frac{\partial \mathbf{E}}{\partial t}. 
\end{alignat}
The electric as well as the magnetic field are connected due to the scalar potential $\phi$ and the vector potential $\mathbf{A}$:
\begin{align}
 \mathbf{E} &= -\boldsymbol{\nabla} \phi - \de{t} \mathbf{A} ,\\
 \mathbf{B} &= \boldsymbol{\nabla} \times \mathbf{A}.
\end{align}
The second key element was the development of quantum electrodynamics originating from the formulation of the Euler-Heisenberg effective Lagrangian\cite{HeisenbergEuler}. In two recently published papers\cite{doi:10.1142/S0217751X12600044,DunneHEul}, the impact of the Euler-Heisenberg Lagrangian on theoretical physics is reviewed. Basically, the Euler-Heisenberg Lagrangian expands the Maxwell Lagrangian by nonlinear terms covering all quantum effects arising in a background electromagnetic field
\begin{align}
 \mathcal L &= \frac{e^2}{h c} \int_0^{\infty} \frac{d \eta}{\eta^3} \ee^{-\eta} \times \\
 &\br{\ii \eta^2 \br{\mathbf{E} \cdot \mathbf{B}} 
 \frac{\cos \br{\frac{\eta}{E_0} \sqrt{\mathbf{E}^2 - \mathbf{B}^2 + 2 \ii \br{\mathbf{E} \cdot \mathbf{B}} }} +c.c}{\cos \br{\frac{\eta}{E_0} \sqrt{\mathbf{E}^2 - \mathbf{B}^2 + 2 \ii \br{\mathbf{E} \cdot \mathbf{B}} }} -c.c}
 + E_0^2 + \frac{\eta^2}{3} \br{\mathbf{B}^2 - \mathbf{E}^2} },
\end{align}
where
\begin{equation}
 E_0 = \frac{m^2 c^3}{e \hbar} \approx 1.3 \ 10^{18} \ V/m.
\end{equation}

The term $E_0$, first identified in reference \cite{Sauter}, has been attributed to the critical field strength setting the scale for matter creation in constant electric fields. 
Expanding the Lagrangian for perturbative weak-fields one obtains the leading quantum corrections to the Maxwell Lagrangian
\begin{equation}
 \mathcal{L}_{pert} = \frac{\mathbf{E}^2 - \mathbf{B}^2}{2} + \frac{1}{90 \pi} \frac{\hbar c}{e^2} \frac{1}{E_0^2} \br{ \br{\mathbf E^2 - \mathbf B^2}^2 + 7 \br{\mathbf{E} \cdot \mathbf B}^2}.
\end{equation}
It should be noted that the Lagrangian is entirely formulated in terms of the Lorentz invariants
\begin{equation}
 \mathcal F = -\frac{1}{4} F^{\mu \nu} F_{\mu \nu} = \frac{1}{2} \br{\mathbf{E}^2 - \mathbf{B}^2} ,\qquad \mathcal G = -\frac{1}{4} F^{\mu \nu} \tilde F_{\mu \nu} = \mathbf{E} \cdot \mathbf{B}
\end{equation}
and additionally that the second term is suppressed by the critical field strength $E_0$.

While developing the renormalization scheme and incorporating it into QED forming a fully covariant formulation of the electromagnetic force, the problem of pair production was further investigated with the new techniques. Viewing the pair production process as an evolutionary process in proper-time\cite{Schwinger}, J. Schwinger was able to link matter creation with the constant-field effective Lagrangian
\begin{equation}
 \mathcal L = \frac{1}{2} E^2 - \frac{1}{8 \pi^2} \int_0^{\infty} \frac{ds}{s^3} \br{e E \ s \ \text{cot} \br{e E \ s} - 1 + \frac{1}{3} \br{e E \ s}^2}.
\end{equation}
He identified the imaginary part being of special importance as
\begin{equation}
\text{Im} \ \mathcal L = \frac{\alpha^2}{2 \pi^2} E^2 \sum_{n=1}^{\infty} n^{-2} \exp \br{\frac{-n \pi m^2}{e E}}.
\end{equation}
The term above provides a description of the decay rate of the vacuum in a constant electric field. 

\vspace{5cm}

Moreover, the first term of this sum yields the production rate of a single electron-positron pair:
\begin {equation}
 \dot N = \frac{\alpha^2}{2 \pi^2} E^2 \exp \br{\frac{- \pi m^2}{e E}}.
\end {equation} 

\subsection{Advances in Laser technology}

In 1960 T. H. Maiman\cite{Maiman} created the first fully operating Laser, thus being able to produce light within a small bandwidth. However, the groundwork for the theoretical work has been done by A. Einstein by introducing the concept of absorption, stimulated and spontaneous emission based upon probabilities\cite{Einstein}.
In the subsequent years the advancement in Laser physics was possible due to ongoing progression in reducing the Laser pulse time and the Laser intensity. Nowadays, the world record for the shortest Laser pulse ever created in laboratory is set at $67$ attoseconds\cite{Zhao:12}. Connected with a decrease of the pulse duration is also an increase in the peak intensity. To put it simply, shortening the Laser pulse time for a fixed energy leads inevitably to a higher peak intensity.
One major breakthrough in developing high-intensity Lasers has been the implementation of the so-called chirped pulse amplification(CPA) technique\cite{RevModPhys.78.309,Strickland1985219}. The advantage of the CPA technique is the ability to increase the peak intensity of a Laser pulse without damaging the gain medium. 
Basically, the Laser pulse is stretched effectively lowering the intensity in the gain medium. Amplification of this low-intensity pulse and subsequent assembling allows to build Laser systems that can create the pulses needed in order to study non-linear QED.
At the moment, high-intensity Laser facilities are planned and build at various places all over the world\cite{LaserSys}. For example, high-performance optical Laser systems are operating up to the $PW$-scale\cite{LaserSysOpt} and even more powerful facilities are about to come in the next years. Hence, the creation of Laser pulses reaching intensities of $10^{26}$ $W/cm^2$ ($0.01 \ E_0$) is probably possible in the near future. One should add at this point, that $E_0$ does not draw the line between pair production and no pair production. Rather, the time-dependency of the electric field and the chance for multiple Laser shots make probing particle creation via light-light scattering also feasible at lower field strengths. 

Another interesting candidate for pair production experiments are X-Ray Laser systems\cite{LaserSysX}. The basic principle of a X-Ray Laser has been described in \cite{Bonifacio1984373}. At first, one has to create electrons with tiny emittance. Then this bunch of electrons is accelerated in an undulator. Due to the applied magnetic field and due to the so-called micro-bunching, the electrons emit coherent light which adds up while the electrons are accelerated in the undulator. When the electrons are deflected at the end of the accelerator only the photons remain. This remarkable feature of self-amplified spontaneous emission(SASE) opened up the possibility to develop free-electron Lasers(FEL) in order to generate light with a very short wavelength and thus high photon energy. Facilities operating with FELs are LCLS in Stanford\cite{LaserSysX2} and DESY in Hamburg\cite{LaserSysX}. 

\vspace{5cm}

Special focus is on the SLAC E-144 experiment\cite{PhysRevLett.79.1626,PhysRevD.60.092004}. This has been the first earth-based experiment, where inelastic light-by-light scattering with real photons was involved\cite{Reiss}. The experiment was performed recording positrons stemming from collisions of high-energy electrons from optical terawatt pulses. The interpretation of the observed positron momentum agreed within experimental uncertainties with the theoretical calculations of a two-step scattering process. In the first step high-energy photons are created due to non-linear Compton scattering. In the experiment, a high-energy electron could absorb $n$ Laser photons with energy $\omega_0$ and subsequently emit a single high-energy photon
\begin{equation}
 e^- + n \omega_0 \to \acute e^- + \omega.
\end{equation}
In the second step the photons are then transformed into matter due to multiphoton Breit-Wheeler reaction\cite{PhysRev.46.1087,Pike}. In a Breit-Wheeler process multiple interacting photons produce a particle-antiparticle pair
\begin{equation}
 \omega + n \omega_0 \to e^+ e^-.
\end{equation}

\section{Theoretical considerations}

Over the last 80 years various theoretical methods have been used in order to study strong-field QED effects. In a recent review\cite{DunneELI}, many theoretical tools are presented focusing on their applicability on describing pair production. Nevertheless, we shall give a brief overview on nowadays most important techniques. At first, there are semi-classical methods\cite{PhysRevD.2.1191,PhysRevD.79.065027,PhysRevLett.104.250402,PhysRevD.82.045007} capable of describing many aspects of pair production including the dynamically assisted Schwinger effect\cite{PhysRevLett.101.130404,PhysRevD.85.025004,Schneider,Strobel20141153,Kleinert2013104}. Besides WKB approximations one also has to mention instanton techniques\cite{PhysRevD.65.105002,PhysRevD.72.065001,PhysRevD.72.105004,PhysRevD.74.065015}.
Then, there are numerically more sophisticated methods, which can be combined under the term quantum kinetic approaches. Results for the particle yield are obtained by performing computations in a phase-space approach\cite{Vasak1987462,PhysRevD.44.1825,PhysRevA.48.1869,Best1993169,PhysRevD.47.4639,Zhuang1996311,Ochs1998351,PhysRevD.82.105026,PhysRevLett.107.180403,Hebenstreit,Berenyi} or by solving the quantum Vlasov equation\cite{PhysRevLett.67.2427,PhysRevD.45.4659,PhysRevD.58.125015,Schmidt,PhysRevD.60.116011,PhysRevLett.87.193902,PhysRevLett.89.153901}. In this way, pair production in homogeneous, time-dependent electric fields has been investigated profoundly\cite{PhysRevLett.102.150404,Orthaber201180,Kohlfurst,Otto2015335,PhysRevD.90.125033,PhysRevD.90.025021}.
A completely different possibility is provided by an analysis of the imaginary part of the QED effective action\cite{PhysRevD.78.036008}. Last but not least, also Monte-Carlo techniques have been applied in order to understand the matter creation process\cite{PhysRevD.87.105006,Kasper,PhysRevD.90.025016}. For a detailed review regarding pair production, the interested reader may have a look at \cite{DunneHEul,Ruffini20101,RevModPhys.84.1177}.

At the beginning, pair production was investigated in terms of constant electric fields only\cite{Schwinger}. Due to the arising of new tools applicable for studying matter creation, the focus shifted to time-dependent electric fields still neglecting magnetic fields entirely\cite{PhysRevD.2.1191,PROP:PROP19770250111}. Research of pair production for arbitrarily complicated time-dependent electric fields\cite{Vinnik,Kim,PhysRevD.83.065028,He,PhysRevD.89.085001,PhysRevD.90.113004,PhysRevB.92.035401} as well as first calculations for space-dependent background fields shed further light on our understanding of the formation of matter\cite{PhysRevD.74.065015,PhysRevD.72.065001,PhysRevD.78.025011,PhysRevD.82.105026,PhysRevD.75.045013,Han201099}. Additionally, parallel as well as collinear electric and magnetic fields have been considered\cite{Tarakanov,PhysRevLett.102.080402,Tanji20091691}. In order to account for the difficulties in creating a substantial amount of particles in an experiment, new setups have been proposed and new tools for optimizing the pulse shape have been developed\cite{PhysRevD.88.045028,Hebenstreit2014189}. \\

Due to breakthroughs in atom physics there has been tremendous progress in the understanding of ionization processes.
When investigating atomic ionization one distinguishes between electron tunneling and photon absorption \cite{Keldysh,Ammosov-1986-Tunnel,PhysRevLett.71.1994,Popov}. Introducing the Keldysh parameter $\gamma$ we obtain an indicator which of the two effects is dominating. In case of a linearly polarized many-cycle field parameterized by 
\begin{equation}
 E \brt = \varepsilon \cos \br{\omega t},
\end{equation}
with the field strength $\varepsilon$ and the field frequency $\omega$, the Keldysh parameter takes the form
\begin{equation}
 \gamma = \sqrt{\frac{E_I}{2 U_p}}, \qquad \text{where} \qquad U_p = \frac{1}{2} \frac{e^2 \varepsilon^2}{m \omega^2}.
\end{equation}
In the equation above we have introduced the ionization energy $E_I$ as well as the ponderomotive potential $U_p$.
Generally, the Keldysh parameter is written 
\begin{equation}
 \gamma = \frac{\omega}{\omega_T}, 
\end{equation}
where $\omega_T$ gives the inverse of a hypothetical tunneling time.
In other words, the Keldysh parameter compares the oscillation period of the Laser (crucial for photon absorption) with the time it takes for a tunneling process to happen. The ratio of these two quantities provides an estimate which process is more likely and therefore dominating. In case of $\gamma \ll 1$, the applied electric field is only slowly varying in time increasing the chances for electron tunneling, because the Coulomb barrier is strongly suppressed on a longer time scale.
On the other hand, a Keldysh parameter of $\gamma \gg 1$ indicates, that photon absorption dominates the ionization process. A possible interpretation is, that the time-dependent electric field is close to the peak strength only at short intervals in time.

At this point we have to place emphasis on the fact, that the Keldysh parameter, if defined as above, is only meaningful for many-cycle, linearly polarized fields. Imagine a circularly polarized field\cite{PhysRevD.89.085001}, then the intensity at every instant in time is equal to $I/\sqrt{2}$, where $I$ is the peak intensity of the Laser pulse. Hence, the time for a tunneling process to happen is dramatically increased compared to the linearly polarized case leading to tunneling dominance even at higher field frequencies $\omega$.

Another way of studying the effects of strong fields on particles is to probe atoms and ions via few-cycle pulses \cite{RevModPhys.72.545}. In a few-cycle pulse the number of oscillations of the electric field is limited leading to the emergence of further implications. In this case one has to introduce a modified Keldysh parameter taking into account the short lifespan of the Laser pulse, too. 

\vspace{5cm}

In complete analogy to atomic physics the Keldysh parameter was also taken up for pair production. Here, the parameter is defined as
\begin{equation}
 \gamma = \frac{m \omega}{e \varepsilon},
\end{equation}
where $\omega$ is again associated with the photon energy and $\varepsilon$ with the electric field strength.
Identical to the previous discussion on the Keldysh parameter one can in principle distinguish the different regimes by the value of $\gamma$. In case of field strengths lower than the critical field strength $E_0$ the production rate of a single electron-positron pair is close to the vacuum decay rate. The Heisenberg-Euler exponential to the leading order \cite{PhysRevD.2.1191,PROP:PROP19770250111} then becomes
\begin{equation}
 \dot N \approx \begin{cases}  \exp \br{-\frac{\pi m^2 c^3}{e \hbar \varepsilon}}, & \gamma \ll 1 \\ \br{\frac{e \varepsilon}{m \omega}}^{4mc^2/ \br{\hbar \omega}}, & \gamma \gg 1\end{cases}
\end{equation}
As mentioned earlier there has been much progress in understanding the tunneling process also known as Schwinger effect. Moreover, a detailed analysis of multiphoton pair production has led to the introduction of the effective mass concept and subsequently to the discovery of a field dependent threshold \cite{Heinzl}. 
However, also the intermediate region has been investigated. By superimposing a long and strong electric field ($\gamma \ll 1$) with a short, weak field the so-called dynamically assisted Schwinger effect was found \cite{PhysRevLett.101.130404}. The quintessence of these findings are that an interplay of fields, which would be associated with completely different regimes when on their own, could greatly boost the pair production rate. Unsurprisingly, sophisticated setups, like the ones suggested in Sch\"utzhold et al. \cite{PhysRevLett.101.130404}, hold as promising candidates for an at least indirect measurement of the Schwinger effect.

\section{Laser beams}
\label{Kap_Exp}
In this thesis we focus on pair production via light-light interactions. This means, that particle creation is achieved either via scattering of a high-energy photon in a strong electric field or by formation of a standing light-wave. For theoretical investigations regarding particle creation in Laser-nucleus collisions, see Di Piazza et al. \cite{RevModPhys.84.1177}. 

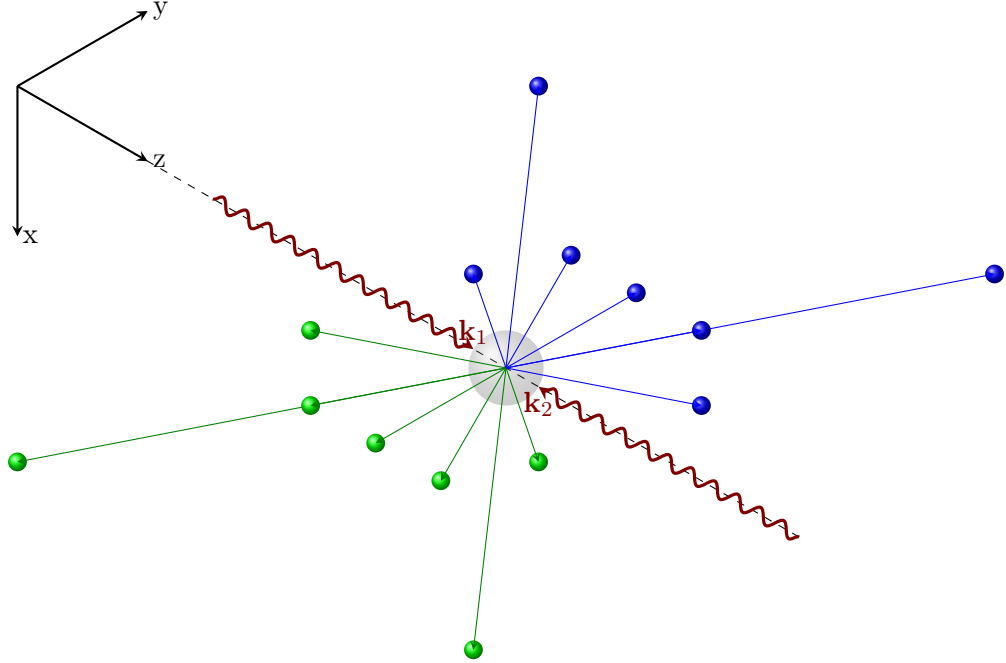
\begin{figure}[htb]
\begin{center}
\begin{tikzpicture}[x={(0.866cm,-0.5cm)}, y={(0.866cm,0.5cm)}, z={(0cm,1cm)}, scale=1.0,
    >=stealth, %
    inner sep=0pt, outer sep=2pt,%
    axis/.style={thick,->},
    wave/.style={thick,color=#1,smooth},
]

    \colorlet{darkgreen}{green!50!black}
    \colorlet{darkred}{red!50!black}

    \newcommand{\inter}[3]{
      \node[circle,shading=ball,minimum width=#3cm,opacity=0.25,ball color=gray] (ball) at (#1,#2) {};
    }
    \newcommand{\electron}[5]{
      \node[circle,shading=ball,minimum width=#3cm,ball color=blue] (ball) at (#1,#2) {};
      \draw[blue, ->]       (7.5,0) -- (#1,#2);
    }    
    \newcommand{\positron}[5]{
      \node[circle,shading=ball,minimum width=#3cm,ball color=green] (ball) at (#1,#2) {};
      \draw[darkgreen, ->]       (7.5,0) -- (#1,#2);
    }      

    \coordinate (O) at (0, 0, 0);
    \draw[axis] (O) -- +(2, 0,   0) node [right] {z};
    \draw[axis] (O) -- +(0,  2.0, 0) node [right] {y};
    \draw[axis] (O) -- +(0,  0,   -2) node [right] {x}; 
        
    \draw[dashed] (12,0,0) -- (O);

    \inter{7.5}{0}{1} 
    \electron{7.5}{2}{0.25}{0.1}{0.44};
    \positron{11}{-4}{0.25}{0.8}{1};
    \positron{5}{-5}{0.25}{-0.6}{-1.1};
    \electron{6}{1}{0.25}{0.2}{1.3};
    \positron{6.5}{-2}{0.25}{0.5}{-1};
    \positron{8.5}{-2}{0.25}{1.4}{-1};
    \positron{5.5}{-1}{0.25}{-0.5}{-0.4};
    
    \positron{7.5}{-2}{0.25}{0.1}{0.44};
    \electron{4}{4}{0.25}{0.8}{1};
    \electron{10}{5}{0.25}{-0.6}{-1.1};
    \positron{9}{-1}{0.25}{0.2}{1.3};
    \electron{8.5}{2}{0.25}{0.5}{-1};
    \electron{6.5}{2}{0.25}{1.4}{-1};
    \electron{9.5}{1}{0.25}{-0.5}{-0.4}; 
        
    \draw[darkred, ->, decorate, decoration={snake}, very thick]       (3, 0, 0) -- (7, 0, 0) node [above] {$\textbf{k}_1$};             
    \draw[darkred, ->, decorate, decoration={snake}, very thick]       (12, 0, 0) -- (8, 0, 0) node [below] {$\textbf{k}_2$};     
        
\end{tikzpicture} 
\end{center}
\caption[Schematic picture displaying pair production via light-light interaction.]{Schematic picture displaying pair production via light-light interaction. Two incoming lightwaves $k_1$ and $k_2$ have either a sufficiently high field strength (Schwinger effect) or are energetic enough (multiphoton effect) in order to produce particles. These particles are created in the interaction region(grey sphere) and subsequently accelerated due to the high field strength. In this picture the pair production process is completely symmetric.}
\label{Exp_collision}
\end{figure}

In the following we will give an idealized and highly simplified picture in order to describe a possible experimental setup using two Laser pulses as illustrated in Fig. \ref{Exp_collision}. An introduction on how Lasers operate as well as a summary on different Laser types and details on the applicability of Lasers can be found in Svelto \cite{Svelto}. 

\begin{figure}[htb]
\begin{center}
\begin{tikzpicture}[x={(0.866cm,-0.5cm)}, y={(0.866cm,0.5cm)}, z={(0cm,1cm)}, scale=1.0,
    >=stealth, %
    inner sep=0pt, outer sep=2pt,%
    axis/.style={thick,->},
    wave/.style={thick,color=#1,smooth},
]
    \colorlet{darkgreen}{green!50!black}
    \colorlet{darkred}{red!50!black}

    \coordinate (O) at (0, 0, 0);
    \draw[axis] (O) -- +(8, 0,   0) node [right] {z};
    \draw[axis] (O) -- +(0,  2.5, 0) node [right] {y};
    \draw[axis] (O) -- +(0,  0,   -2) node [right] {x};

    \draw[wave=blue, variable=\z,samples at={0,0.25,...,7}]
        plot (\z,{sin(2*\z r)},0)node[anchor=west]{$\textbf{B}$};
    \foreach \z in{0, 0.25,...,7}
        \draw[color=blue,->] (\z,0,0) -- (\z,{sin(2*\z r)},0);                
        
    \draw[wave=green, variable=\z,samples at={0,0.25,...,7}]
        plot (\z,0,{sin(2*\z r)})node[anchor=south]{$\textbf{E}$};    
    \foreach \z in{0, 0.25,...,7}
        \draw[color=green,->] (\z,0,0) -- (\z,0,{sin(2*\z r)});        

\end{tikzpicture}  
\end{center}
\caption[Schematic view of a transverse electromagnetic mode.]{Schematic view of a TEM wave propagating in z-direction. The electric and magnetic field are perpendicular to each other at every point in spacetime. The electric field points towards $-x$-direction and the magnetic field towards $y$-direction.}
\label{TEM_wave}
\end{figure}
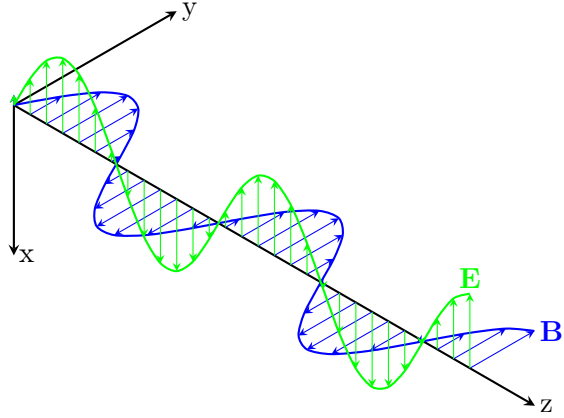

For the sake of simplicity, it is sufficient to introduce a Laser pulse as a decomposition of longitudinal and transversal modes giving the beam profile times a time-dependent factor yielding propagation. Here, we want to focus on the transversal electromagnetic modes(TEM) in particular. The characteristic feature of a TEM is, that the electric and the magnetic parts of the field vanish in direction of propagation. Assuming the pulse propagates in $z$-direction, we obtain an intensity profile in the $xy$-plane, see Fig. \ref{TEM_wave}. In order to classify the various transverse mode patterns a Laser can produce, one introduces labels denoting the different mode orders. The common notation is TEM$_{mn}$, where $m$ and $n$ give the number of modes in $x$ and $y$ direction, respectively. One example for a TEM are the so-called Hermite-Gaussian modes. If one describes a Laser field in this way the electric component reads
\begin{equation}
 E \br{\mathbf{x}} = \varepsilon \ \frac{w_0}{w \br{z}} \ H_m \br{\sqrt{2} \frac{x}{w \br{z}} } H_n \br{\sqrt{2} \frac{y}{w \br{z}} } \ee^{- \br{x^2+y^2}/ w \br{z}^2} \ee^{\ii \Phi \br{\mathbf{x}}},
\end{equation}
where $H_m$, $H_n$ are the Hermite polynomials, $\Phi \br{\mathbf{x}}$ is a phase factor and $w \br{z}$ the beam radius. As we want to focus on the spatial beam profile we have not stated a time-dependent factor of $\ee^{\ii \omega t}$. A thorough derivation of the Hermite-Gaussian modes as well as Laguerre-Gaussian modes (in case of cylindrical symmetry) can be found in various articles\cite{Beam}, where the corresponding electric fields are introduced as solutions of the paraxial wave equation
\begin{equation}
 \br{\de{x}^2 + \de{y}^2 + 2 \ii ~ k ~ \de{z}} E \br{\mathbf{x}} = 0. \label{Helmholtz}
\end{equation}

\vspace{5cm}

As TEM are solutions to \eqref{Helmholtz}, they automatically solve Maxwell equations for specific boundary conditions. This is an important fact from a theoretical point of view, because it allows for a consistent description of the pair production process. In chapter four, see especially section \ref{Kap_Field}, we will demonstrate how to obtain models for the interaction region of two Laser pulses.

\section{Mechanisms for pair production}
\label{Mechanism}
In this section we want to give an overview on the different mechanisms, that can cause the formation
of a particle-antiparticle pair. In order to graphically support the arguments we use an old picture 
illustrating a vacuum state: the Dirac sea picture. Although we are aware of the fact, that it is
outdated and not compatible with modern quantum field theoretical knowledge it still holds in many ways as a simple tool for demonstration purposes. 

Generally speaking, the basic idea is to consider the vacuum state as a material having a clearly separated conduction band (continuum) and valence band (Dirac sea). The band gap in
between these two bands is assumed to be of the rest mass of particle plus corresponding antiparticle; in our case the mass of electron plus positron. In the vacuum state there are by definition no electrons in the continuum. The Dirac sea, however, is completely filled with electrons having negative energy. As all observable particles must have positive energy the problem of particle creation can be reduced to the issue of bringing a particle from the Dirac sea to the continuum.
As every empty slot in one of these bands can be associated with an antiparticle the ideas formulated above hold vice-versa. 

\begin{figure}[ht]
    \centering
    \begin{minipage}[t]{0.45\linewidth}
        \hspace{-0.5cm}
        \includegraphics[width=1.15\linewidth]{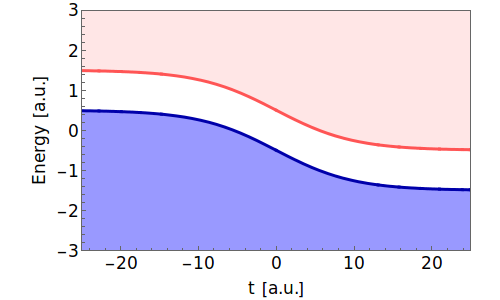}
        \label{Dirac0}
    \end{minipage}
    \hfill
    \begin{minipage}[t]{0.45\linewidth}
        \hspace{-1cm}
        \includegraphics[width=1.15\linewidth]{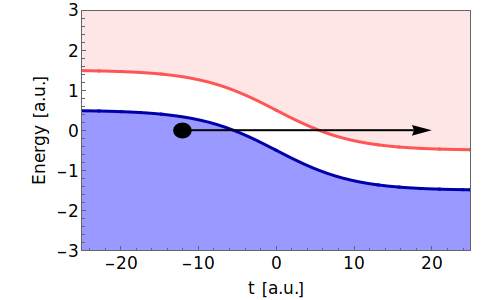}
        \label{Dirac1}
    \end{minipage}
    \caption[Pictorial description of the Schwinger effect.]{Pictorial description of the Schwinger effect (right-hand side). The dark blue band pictures the Dirac sea and the light red band pictures the continuum. The band gap (white) holds as a forbidden region for real particles. When a strong electric field is applied to the system, the energy bands are deformed in such a way, that the electron (black dot) can tunnel to the continuum.}
    \label{Dirac1_Full}    
\end{figure}

In order to illustrate the various mechanisms leading to pair production we have to apply a strong electric field to the vacuum state. The external potential deforms the bands of the state as illustrated in Fig. \ref{Dirac1_Full}. As a consequence, various options for exciting an electron from the Dirac sea to the continuum become possible. In case the field strength of the applied electric field is of the order of $E_0$ the bands are deformed strongly enabling the electron to simply tunnel from one band to the other. This phenomenon is called Schwinger effect. A schematic picture of such a tunneling process is shown in Fig. \ref{Dirac1_Full}.

\begin{figure}[ht]
    \centering
    \begin{minipage}[t]{0.45\linewidth}
        \hspace{-0.5cm}
        \includegraphics[width=1.15\linewidth]{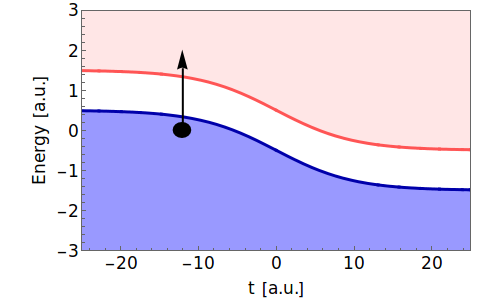}
        \label{Dirac2}
    \end{minipage}
    \hfill
    \begin{minipage}[t]{0.45\linewidth}
        \hspace{-1cm}
        \includegraphics[width=1.15\linewidth]{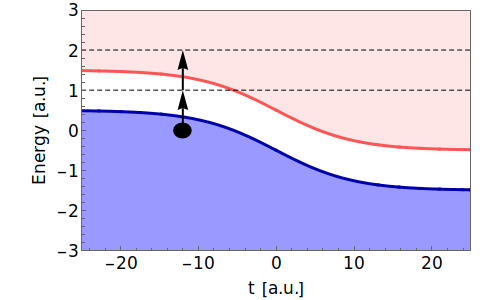}
        \label{Dirac3}
    \end{minipage}
    \caption[Pictorial description of (multi)photon absorption.]{Pictorial description of (multi)photon absorption, where an electron absorbs $n$ photons at the same time. On the left-hand side, one highly energetic photon is absorbed by an electron (black dot) in the Dirac sea (dark blue region). The final energy of the electron is high enough in order to overcome the band gap(white region) and reach the continuum (light red region). However, the energy of a single photon could be not enough in order to produce a particle pair. Hence, a $n=2$ absorption process is shown on the right-hand side.}
    \label{Dirac2_Full}    
\end{figure}

\begin{figure}[htb]
    \centering
    \begin{minipage}[t]{0.45\linewidth}
        \hspace{-0.5cm}
        \includegraphics[width=1.15\linewidth]{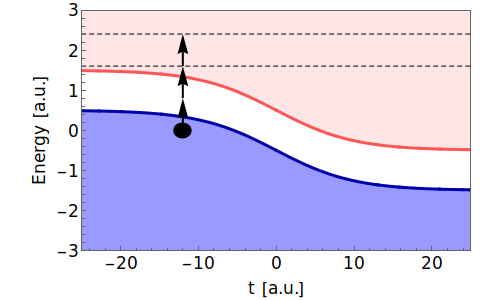}
        \label{Dirac4}
    \end{minipage}
    \hfill
    \begin{minipage}[t]{0.45\linewidth}
        \hspace{-1cm}
        \includegraphics[width=1.15\linewidth]{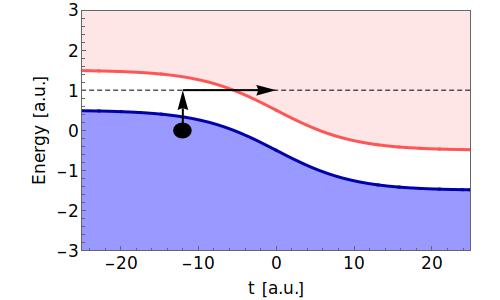}
        \label{Dirac5}
    \end{minipage}
    \caption[Pictorial description of above-threshold pair production and the dynamically assisted Schwinger effect.]{On the left hand side, above-threshold pair production is shown. Although particle creation via absorption of $n$ photons is possible there is a non-vanishing probability of an $n+m$ absorption process resulting in particles produced with higher net momentum. On the right-hand side, the so-called dynamically assisted Schwinger effect, the combination of photon absorption and tunneling process, is illustrated. The electron can absorb a photon and subsequently tunnel to the continuum (light red band) through the now lowered tunneling barrier.}
    \label{Dirac3_Full}
\end{figure}

A second possibility is the excitation of an electron by absorbing high-energy photons. In such a case the electron absorbs the energy of a photon resulting in an electron net energy high enough to overcome the band gap (Fig. \ref{Dirac2_Full}). If a multiple number of photons is absorbed simultaneously one refers to the effect as multiphoton pair production (Fig. \ref{Dirac2_Full}).  
Moreover, assuming a $n$-photon absorption process leads to pair production the chances for a $n+m$-photon process are nonzero. As $n,m$ are arbitrary non-negative integer numbers, this inevitably yields a clear signature in phase space due to the difference in the net momenta of the produced particles. Based on a well-known process in atomic ionization, this effect is called above-threshold pair production (Fig. \ref{Dirac3_Full}). 

As observed in theoretical calculations only recently there is also the possibility of a combined effect. In case the photon energy absorbed by the electron is not sufficient in order to push the electron to the continuum two possibilities remain. Either the electron emits the energy in terms of photons or it tunnels from a virtual state to the continuum. The idea of such a dynamically assisted Schwinger effect is illustrated in Fig. \ref{Dirac3_Full}.

\clearpage
 
\section{Quantum kinetic methods}

In order to study pair production we rely on the Wigner function approach \cite{Vasak1987462,Schleich,Lee1995147}. When working with a phase-space formulation of quantum physics, the problem is formulated mathematically via quasi-probability distributions. Although there are several different options available we will only consider the Wigner distribution.

In order to introduce the Wigner function approach and subsequently discuss characteristic features of kinetic theories we will introduce the phase-space formulation for quantum mechanical problems. Kinetic approaches are established tools in order to study e.g. quantum optics \cite{Schleich} or plasma physics \cite{Nicholson,Chen,Baumjohann,Liboff}. Hence, we present only a brief introduction here as one can find extensive literature on the subject. By this means, an introduction in terms of a $N$-particle quantum system is more appropriate due to the fact that the particles, once created, obey the rules of plasma physics. Nevertheless, a quantum mechanical treatment contains all approximations and characteristics one can also find in a QFT approach.  

In quantum mechanics the Wigner quasi-probability distribution for a pure state is given by
\begin{equation}
 W \br{\mathbf{x}, \mathbf{p}, t} = \int d^3s \ \ee^{-\ii \mathbf{p} \cdot \mathbf{s}} \psi^{*} \br{\mathbf{x} + \frac{1}{2} \mathbf{s}} \psi \br{\mathbf{x} - \frac{1}{2} \mathbf{s}}.
\end{equation}
The vectors $\mathbf{x}$, $\mathbf{p}$ and $\mathbf{s}$ are the center-of-mass, momentum and relative coordinates. Generalizing the definition above to mixed states we obtain 
\begin{equation}
 W \br{\mathbf{x}, \mathbf{p}, t} = \int d^3s \ \ee^{-\ii \mathbf{p} \cdot \mathbf{s}} \langle \mathbf{x} + \frac{1}{2} \mathbf{s} | \hat \rho | \mathbf{x} - \frac{1}{2} \mathbf{s} \rangle, \label{Wigner_nr}
\end{equation}
with the density matrix $\hat \rho$. The Wigner quasi-probability distribution possesses certain interesting mathematical properties. Despite the fact, that $W \br{\mathbf{x}, \mathbf{p}, t}$ is real, it is not non-negative definite. Hence, the first axiom of probability theory is violated resulting in the fact, that $W \br{\mathbf{x}, \mathbf{p}, t}$ cannot hold as an ordinary distribution function. 
Another important aspect is, that unphysical values can be integrated out when it comes to observables and thus real physical quantities. 
Following Case \cite{Case}, we interpret an observable as the average of the corresponding phase-space quantity with probability weight $W \br{\mathbf{x}, \mathbf{p}, t}$
\begin{equation}
 \langle \mathcal O \brt \rangle = \int \int d^3x \ d^3p \ W \br{\mathbf{x}, \mathbf{p}, t} \tilde{\mathcal O} \br{\mathbf{x}, \mathbf{p}, t}.
\end{equation}
If one is integrating over momenta only, one obtains the probability distribution in the spatial variables
\begin{equation}
 \int d^3p \ W \br{\mathbf{x}, \mathbf{p}, t} = \langle \mathbf{x} | \hat \rho | \mathbf{x} \rangle.
\end{equation}
Similarly, integration over the spatial variables yields the momentum distribution
\begin{equation}
 \int d^3x \ W \br{\mathbf{x}, \mathbf{p}, t} = \langle \mathbf{p} | \hat \rho | \mathbf{p} \rangle.
\end{equation}

The Wigner function approach is an interesting tool in order to study quantum mechanical problems. However, in this thesis we have to account for dynamical particle creation. As this is not possible in ordinary non-relativistic quantum mechanics we have to expand the formulation introduced above. The generalization of \eqref{Wigner_nr} to relativistic situations appearing in astrophysics has been done in \cite{Groot,Baym}. Moreover, due to the replacement of quantum mechanics with quantum field theory, another implication is the need for gauge invariant quantities. Hence, we have to introduce an additional term ensuring that the QFT formulation of the Wigner function is indeed covariant \cite{Elze1986706,Elze1986402,Vasak1987462}.

We have already defined the Wigner function for a mixed state in \eqref{Wigner_nr}. In order to account for the dynamics of a physical system, we have to find the corresponding equation of motion. This is done via the von-Neumann equation in phase space\cite{BBGKY,Nicholson,Schleich} 
\begin{equation}
 \de{t} \hat{\rho} = -\frac{\ii}{\hbar} \com{\hat H, \hat \rho}
\end{equation}
and multiplication by $\langle \mathbf{x} + \frac{1}{2} \mathbf{s} |$ and $| \mathbf{x} - \frac{1}{2} \mathbf{s} \rangle$.
As a concrete example, we may introduce the $N$-particle Hamiltonian \cite{BBGKY}:
\begin{equation}
 H_N = \sum_{i=1}^N \frac{\hat{\mathbf{p}}_i^2}{2 m} + \sum_{i=1}^N V \br{\hat{\mathbf{x}}_i} + \sum_{1 \le i \le j \le N} W \br{\hat{\mathbf{x}}_i, \hat{\mathbf{x}}_j},
\end{equation}
where, for the sake of simplicity, we have assumed that only particle-pair interactions are allowed ($W \br{\hat{\mathbf{x}}_i, \hat{\mathbf{x}}_j}$) and all external vector potentials vanish. Hence, there are only external scalar potentials present ($V \br{\hat{\mathbf{x}}_i}$).
In the classical limit the Wigner function $W \br{\mathbf{x}, \mathbf{p}, t}$ basically becomes the $N$-particle distribution function $f_N \br{\mathbf{x}_l, \mathbf{p}_l,t}$, where $l=1,2,\ldots,N$. Altogether, this yields after a lengthy calculation the quantum Liouville equation
\begin{equation}
\br{ \de{t} + \sum_i \frac{\mathbf{p}_i}{m} \cdot \boldsymbol{\nabla}_{x_i} + \sum_i \br{\mathbf{F}_i \cdot \boldsymbol{\nabla}_{{p}_i} + \sum_{j \neq i} \mathbf{K}_{ij} \cdot \boldsymbol{\nabla}_{p_i}}} f_N = 0. \label{BBGKY0}
\end{equation}
In order to simplify the notation we have introduced the abbreviations
\begin{alignat}{6}
 \mathbf{F}_i = -\boldsymbol{\nabla}_{x_i} V \br{\mathbf{x}_i} ,\qquad \mathbf{K}_{ij} = -\boldsymbol{\nabla}_{x_i} W \br{\mathbf{x}_i, \mathbf{x}_j}.
\end{alignat}
As solving the differential equation \eqref{BBGKY0} is virtually impossible due to the fact, that one equally takes into account all $N$ particles an approximation is needed. Hence, we rewrite equation \eqref{BBGKY0} by introducing $k$-particle distribution functions (with $k=1,2,3, \ldots N$). By this way, we are able to determine $N-1$ equations taking the form
\begin{equation}
 \br{\de{t} + \sum_i^k \frac{\mathbf{p}_i}{m} \cdot \boldsymbol{\nabla}_{x_i} + \br{ \sum_i^k \br{\mathbf{F}_i \cdot \boldsymbol{\nabla}_{{p}_i} + \sum_{j \neq i}^k \mathbf{K}_{ij} \cdot \br{\boldsymbol{\nabla}_{p_i} - \boldsymbol{\nabla}_{p_j}}}}} f_k = \hat \Delta f_{k+1},
\end{equation}
where $\hat \Delta$ is an integro-differential operator. Additionally, we are able to determine 
an equation for the $N$-particle distribution function (compare with \eqref{BBGKY0}):
\begin{equation}
 \br{\de{t} + \sum_i^N \frac{\mathbf{p}_i}{m} \cdot \boldsymbol{\nabla}_{x_i} + \br{ \sum_i^N \br{\mathbf{F}_i \cdot \boldsymbol{\nabla}_{{p}_i} + \sum_{j \neq i}^N \mathbf{K}_{ij} \cdot \br{\boldsymbol{\nabla}_{p_i} - \boldsymbol{\nabla}_{p_j}}}}} f_N = 0.
\end{equation}

All in all, we have obtained a system of $N$ coupled equations. This tower of equations makes up the Born-Bogoliubov-Green-Kirkwood-Yvon(BBGKY) hierarchy being equivalent to the Liouville equation. In order to simplify the mathematical formulation a truncation scheme or closure procedure has to be applied. In this example as well as in chapter three, we use a Hartree approximation. Assuming that the particles are non-interacting the tower of equations in the BBGYK hierarchy already truncates with the one-particle distribution function. Within this approximation one obtains
\begin{equation}
 \br{\de{t} + \frac{\mathbf{p}}{m} \cdot \boldsymbol{\nabla}_x + \mathbf{F} \cdot \boldsymbol{\nabla}_{{p}}} f \br{\mathbf{x}, \mathbf{p}, t} = 0.
\end{equation}

If we were about to use a Hamiltonian describing charged particles in an external electromagnetic field we would have to define 
\begin{equation}
 \acute H_N = \sum_{i=1}^N \frac{\br{\hat{\mathbf{q}}_i - e \mathbf{A} \br{\hat{\mathbf{x}}_i}}^2}{2 m} + \sum_{i=1}^N e \phi \br{\hat{\mathbf{x}}_i} + \sum_{1 \le i \le j \le N} W \br{\hat{\mathbf{x}}_i, \hat{\mathbf{x}}_j},
\end{equation}
with the potential $\br{\phi, \mathbf{A}}$. A similar derivation then yields the Vlasov equation 
\begin{equation}
 \br{\de{t} + \mathbf{v} \cdot \boldsymbol{\nabla}_x + \frac{\mathbf{F}}{m} \cdot \boldsymbol{\nabla}_{{v}}} f \br{\mathbf{x}, \mathbf{v}, t} = 0.
\end{equation}

In chapter three, see section \ref{Kap_EoM}, we will introduce an approximation of Hartree type via ensemble averaging over the electromagnetic field thus neglecting all collision terms.   
Replacing the quantum gauge fields by a mean electromagnetic field \cite{Vasak1987462} one obtains
\begin{equation}
 F^{\mu \nu} \br{\mathbf{x}} \approx \langle \hat F^{\mu \nu} \br{\mathbf{x}} \rangle. 
\end{equation}
Due to this averaging the operator valued fields become ordinary c-number fields and similarly the BBGKY hierarchy truncates at the one-body level.

At this point some remarks on the photon interpretation are in order. As we are working with an ensemble averaged background field the expression "photon" is problematic, because we are basically switching to a formalism using ``averaged energy packages'' instead. Throughout this thesis we will nevertheless refer to photon pair production or photon energy when interpreting the results. The validity of using the term "photon" is based upon findings obtained from different approaches, because it fits the physical interpretation of the results. 

Another major point when introducing pair production via the Wigner function approach is the classical limit. For the case under consideration here, this question reduces to how to obtain the Vlasov equation when formulating the Wigner function in terms of QED. A detailed mathematical calculation, already given in Shin et al. \cite{PhysRevA.48.1869}, will be presented in chapter three, section \ref{Kap_Class}. At this point we simply want to place emphasis on the fact, that taking the limit $\hbar \to 0$ does not universally lead to classical physics. As discussed in Case \cite{Case}, quantum physics and classical physics are not separated by a mere limit. Depending on the circumstances, one has to introduce sophisticated methods in order to show how they are related. Luckily, in section \ref{Kap_Class} dimensional analysis is enough to obtain the Vlasov equation for particles and antiparticles, respectively.
 
\subsection*{Pair production within quantum kinetic methods}

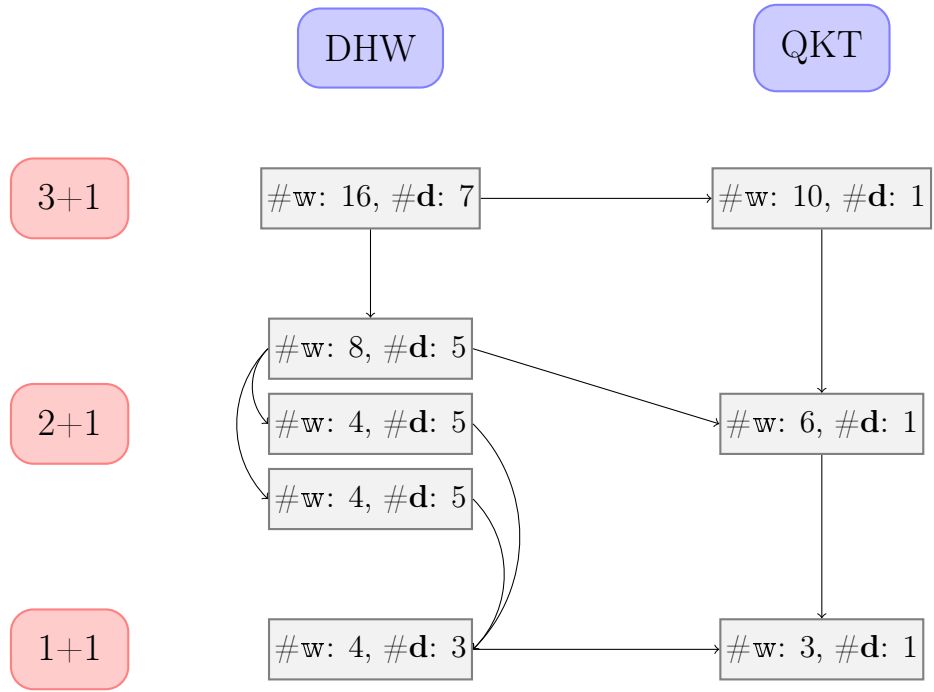
\begin{figure}
\begin{center}
\begin{tikzpicture}
[Form/.style={rectangle,draw=blue!50,fill=blue!20,thick,
inner sep=10pt,minimum size=6mm, rounded corners=3mm,},
Dim/.style={rectangle,draw=red!50,fill=red!20,thick,
inner sep=10pt,minimum size=6mm, rounded corners=3mm,},
Txt/.style={rectangle,draw=black!50,fill=black!5,thick,
inner sep=2pt,minimum size=8mm}]

\Large

\node at (4,9)[Form] {DHW};
\node at (10,9)[Form] {QKT};

\node at (0,1)[Dim] {1+1};
\node at (0,4)[Dim] {2+1};
\node at (0,7)[Dim] {3+1};

\large

\node (n1) at (4,7)[Txt] {\#$\mathbbm{w}$: 16, \#\textbf{d}: 7};
\node (n2) at (10,7)[Txt] {\#$\mathbbm{w}$: 10, \#\textbf{d}: 1};

\node (n3) at (4,3)[Txt] {\#$\mathbbm{w}$: 4, \#\textbf{d}: 5};
\node (n4) at (4,4)[Txt] {\#$\mathbbm{w}$: 4, \#\textbf{d}: 5};
\node (n5) at (4,5)[Txt] {\#$\mathbbm{w}$: 8, \#\textbf{d}: 5};
\node (n6) at (10,4)[Txt] {\#$\mathbbm{w}$: 6, \#\textbf{d}: 1};

\node (n7) at (4,1)[Txt] {\#$\mathbbm{w}$: 4, \#\textbf{d}: 3};
\node (n8) at (10,1)[Txt] {\#$\mathbbm{w}$: 3, \#\textbf{d}: 1};

\normalsize

\draw [->] (n1.east) -- (n2.west);
\draw [->] (n5.east) -- (n6.west);
\draw [->] (n7.east) -- (n8.west);

\draw [->] (n2.south) -- (n6.north);
\draw [->] (n6.south) -- (n8.north);

\draw [->] (n1.south) -- (n5.north);
\draw [->] (n5.west) to [bend right=45] (n3.west);
\draw [->] (n5.west) to [bend right=45] (n4.west);

\draw [->] (n3.east) to [bend left=45] (n7.east);
\draw [->] (n4.east) to [bend left=45] (n7.east);

\end{tikzpicture} 
\end{center}
\caption[Illustration of the number of non-trivial Wigner components depending on the dimension of the problem.]{The number of non-trivial Wigner components $\# \mathbbm{w}$ as well as the number of dimensions $\mathbf{d}$ of the corresponding phase-space domain are depicted for $n+1$ dimensional representations of the problem. A further reduction is possible if the background field exhibits specific symmetries. In $2+1$ dimensions the problem can be formulated in three different ways depending on the chosen basis, while in the homogeneous limit(QKT) only an ODE has to be solved.}
\label{DHW_pic1}
\end{figure}

Treating the process of pair production with phase-space methods the problem is formulated in terms of a system of PDEs. Calculations are only feasible if, besides the time-dependence, the actual phase-space domain exhibits maximally $3$ dimensions (e.g. one spatial and two momentum directions). However, in the following chapters we will demonstrate how a selectively chosen background field configuration together with constraints on the degrees of freedom can nevertheless lead to a viable problem. This gradual decline is illustrated in Fig. \ref{DHW_pic1} in order to support our arguments graphically. Obviously, the dimension of the phase-space reduces drastically in the homogeneous limit, because the problem becomes formulated purely in momentum-space. Another consequence is, that all operators turn into local operators and due to various mappings the problem can be reduced to solving an ODE.
In case of a spatially inhomogeneous problem the problem can often be separated into various subsystems. Instead of solving a PDE formulated on a $6$-dimensional domain, isolating lower dimensional hyperplanes is often advantageous. The result for the full phase-space is then obtained by solving the subsystem of PDEs multiple times with varying external parameters.  

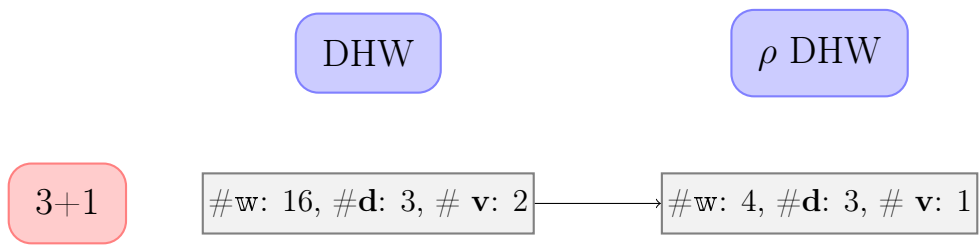
\begin{figure}
\begin{center}
\begin{tikzpicture}
[Form/.style={rectangle,draw=blue!50,fill=blue!20,thick,
inner sep=10pt,minimum size=6mm, rounded corners=3mm,},
Dim/.style={rectangle,draw=red!50,fill=red!20,thick,
inner sep=10pt,minimum size=6mm, rounded corners=3mm,},
Txt/.style={rectangle,draw=black!50,fill=black!5,thick,
inner sep=2pt,minimum size=8mm}]

\Large

\node at (4,9)[Form] {DHW};
\node at (10,9)[Form] {$\rho$ DHW};

\node at (0,7)[Dim] {3+1};

\large

\node (n1) at (4,7)[Txt] {\#$\mathbbm{w}$: 16, \#\textbf{d}: 3, \# \textbf{v}: 2};
\node (n2) at (10,7)[Txt] {\#$\mathbbm{w}$: 4, \#\textbf{d}: 3, \# \textbf{v}: 1};

\normalsize

\draw [->] (n1.east) -- (n2.west);

\end{tikzpicture} 
\end{center}
\caption[Illustration of the difference in the number of non-trivial Wigner components for ordinary and cylindrically symmetric problems.]{Exploiting the symmetries of a cylindrically symmetric electric field and vanishing magnetic field ($\rho DHW$) results in a significantly easier formulation. Although the dimensionality $ \#\textbf{d}$ of the problem does not change, the number of non-trivial Wigner components $ \# \mathbbm{w}$ as well as the number of free parameters $ \# \mathbf{v}$ is reduced.}
\label{DHW_pic2}
\end{figure}

Finding and understanding the symmetries of the underlying physical process can also simplify the equations significantly. In chapter three we provide a detailed analysis of various symmetries found in the different formulations of the problem. A special emphasis is on field configurations, which exhibit cylindrical symmetry. In case the applied magnetic field is zero, the cylindrical symmetry can be exploited resulting in a system of equations barely more complex than in a $1+1$ dimensional problem, as shown in Fig. \ref{DHW_pic2}. 

\clearpage

\clearpage

\section{Particle dynamics}
\label{Kap_Dyn}
Besides investigating the mechanisms leading to pair production, understanding the particle dynamics after particle creation is equally important. Only by controlling particle trajectories, a subsequent complete conversion back to photons can be avoided. Additionally, the control of the particle bunches is of great concern regarding building up an experiment including detectors. In fact, it is essential to have a (anti-)particle beam being easily detectable, because the number of particles produced via light-light scattering will be only slightly higher than the background noise. Hence, reducing the background noise and systematic errors is a crucial point in order to verify/falsify the predictions successfully.

Basically, the dynamics of the produced particles can be described via the Dirac equation \cite{Dirac}:
\begin{equation}
 \br{\ii \ga{\mu} \de{\mu} - e \ga{\mu} A_{\mu} \br{r} - m} \Psi \br{r} = 0,
\end{equation}
where $A_{\mu}$ characterizes the Laser field and $\Psi \br{r}$ is the electron bispinor. Of particular interest in this thesis, is the special case of monochromatic plane-waves. Hence, we introduce a vector potential leading to a monochromatic, linearly polarized electric field. Such a vector potential takes the form\cite{RevModPhys.84.1177}:   
\begin{equation}
 A^{\mu} = -\frac{\varepsilon}{\omega} A_0^{\mu} \cos \br{\omega \phi}.
\end{equation}
At this point, we may introduce the parameter $\xi_0$ giving the root-mean-squared intensity of the pulse (compare with the definition of the Keldysh parameter \cite{Keldysh}):
\begin{equation}
 \xi_0 = \frac{|e| \varepsilon}{m \omega}.
\end{equation}
Furthermore, we find for the electron's effective momentum \cite{PhysRevLett.109.100402}:
\begin{equation}
 q_{\mu} = p_{\mu} + \frac{\xi_0^2 m^2}{4 k \cdot p} k_{\mu}
\end{equation}
and its effective mass
\begin{equation}
 \sqrt{q_0^2} = m_{\ast} = m \sqrt{1+ \frac{\xi_0^2}{2}}.
\end{equation}
Hence, instead of electrons we could switch to a picture working with free quasi-electrons having mass $m_{\ast}$.

Throughout the following chapters we will establish a semi-classical interpretation of our findings. The whole purpose of introducing an approach, based upon evaluating the Lorentz force, is to improve our understanding of the pair production process. In chapter four, section \ref{Kap_SemiClass}, we will discuss the value of being able to easily determine particle trajectories via the Lorentz force
\begin{equation}
 \frac{d}{dt} \br{\gamma \dot x_m} = e \br{\mathbf{E} \br{\mathbf{x},t} + \dot{\mathbf{x}} \times \mathbf{B} \br{\mathbf{x},t}}_m.
\end{equation}

\vspace{5cm}

Evaluating the force on a particle seeded at an intermediate time $t_0$ gives a good impression on what to expect from the DHW calculations. However, such a semi-classical interpretation cannot cover all features of quantum physics. Due to its limitation it is e.g. impossible to determine the production probability quantitatively.

In order to discuss additional quantum effects, like spin-field interactions, we have to rely on the Dirac equation. However, we do not solve the Dirac equation numerically in this thesis. Rather we base our interpretations upon findings in the literature. The Stern-Gerlach experiment\cite{Gerlach}, for example, holds as the most prominent example for magnetic field-spin interactions. This interaction becomes obvious when taking the non-relativistic limit leading to the Pauli equation\cite{Pauli}. The corresponding Hamiltonian governing the particle dynamics yields
\begin{equation}
 H = \frac{\br{\mathbf{q} - e \mathbf{A}}^2}{2m} + e \phi - \frac{e \hbar}{2 m c} \boldsymbol{\sigma} \cdot \mathbf{B}
\end{equation}
First evidence of this effect can be seen in chapter seven, where the particle distribution is split in two parts in similarity to the Stern-Gerlach experiment.

\section{Objectives}

At this point we want to put emphasis on the key features of pair production treated within this thesis. Although the Schwinger effect usually attracts most attention, a major part of the thesis deals with multiphoton pair production. The first part of this thesis (up to chapter four) can be seen as a preparation in order to discuss our findings. This includes a thorough derivation of the transport equations describing the pair production process mathematically as well as a complete review on computational methods and solution strategies. The results are interpreted and discussed in the second part of the thesis. Among the questions we want to answer are the following: \\

Is the concept of an effective mass viable in the regime of multiphoton pair production? Can we link the theoretical concept of quasi-particles acquiring an effective mass with observables, thus can we ``measure'' this effective mass?

We will answer these questions in the context of a spatially homogeneous electric field. This includes an examination of the total particle yield discussing e.g. field-dependent thresholds. Furthermore, the momentum spectra of the particles created is analyzed and connections to atomic ionization processes are made. \\

\vspace{5cm}

What happens when focusing a Laser pulse? How does a spatial focus affect the pair production process?

We discuss the results for a cylindrically symmetric, spatially confined electric field in order to mimic a Laser focus. We expand the concept of an effective mass to spatially inhomogeneous fields. The outcome of our calculations is compared with results obtained through homogeneous field configurations. Concepts, that are well-established in plasma physics, are applied in order to explain the particles momentum spectra. \\

When examining pair production via light-by-light scattering, which role does the magnetic field play?

We perform calculations regarding pair production in the plane for electric and magnetic fields. In particular, we introduce a modified field energy mimicking the Lorentz invariants. Furthermore, we analyze and interpret the results for pair production in inhomogeneous electromagnetic background fields and identify connections to this modified field energy.  

\pagestyle{plain}
\chapter{Dirac-Heisenberg-Wigner formalism}
\pagestyle{fancy}
\label{sec_DHW}
In this chapter we will investigate the transport equations derived within the Dirac-Heisenberg-Wigner (DHW) formalism\cite{PhysRevD.44.1825,Zhuang1996311,Ochs1998351,Vasak1987462}. This means, that we will analyze the equations of motions governing pair production in $3+1$ dimensions in section \ref{Sec_Wigner} and section \ref{Kap_EoM}. Then, we will simplify these transport equations by incorporating various symmetries in order to obtain a formalism applicable also for lower-dimensional problems. In section \ref{Kap_SpatHom} we will take the homogeneous limit and show how the DHW formalism is related to Quantum Kinetic Theory (QKT). Moreover, the DHW formalism can be used to study classical phenomena. In section \ref{Kap_Class} we will demonstrate, that disregarding all quantum corrections the DHW equations yield
the relativistic Vlasov equation\cite{PhysRevA.48.1869}. Moreover, we will show how to take advantage of cylindrically symmetric background fields greatly simplifying the DHW equations in $3+1$ dimensions.

\section{Wigner operator}
\label{Sec_Wigner}
Before writing down the Wigner operator, we have to specify the matter fields as well as the gauge field.
As we want to study electron-positron pair production in electromagnetic fields we state the QED Lagrangian
\begin{equation}
\mathcal L \br{\Psi, \bar{\Psi}, A} = \frac{1}{2} \br{\ii \bar{\Psi} \gamma^{\mu} \mathcal{D}_{\mu} \Psi - \ii \bar{\Psi} \mathcal{D}_{\mu}^{\dag} \gamma^{\mu} \Psi}
 -m \bar{\Psi} \Psi - \frac{1}{4} F_{\mu \nu} F^{\mu \nu},
\end{equation}
where $\mathcal{D}_{\mu} = \br{\de{\mu} +\ii e A_{\mu}}$ and correspondingly $\mathcal{D}_{\mu}^{\dag} = \br{\overset{\leftharpoonup} {\de{\mu}} -\ii e A_{\mu}}$.
In order to describe the dynamics of the particles, we proceed by calculating the Dirac equation
\begin{equation}
\br{\ii \ga{\mu} \de{\mu} - e \ga{\mu} A_{\mu} \br{r} - m} \Psi \br{r} = 0 \label{Eq_Dirac1}
\end{equation}
and the adjoint Dirac equation
\begin{equation}
\bar{\Psi} \br{r} \br{\ii \ga{\mu} \overset{\leftharpoonup}{\de{\mu}} + e\ga{\mu} A_{\mu} \br{r} + m} = 0. \label{Eq_Dirac2}
\end{equation}

At his point we are able to define the density operator of the system 
\begin{equation}
 \hat{\mathcal C}_{\alpha \beta} \br{r,s} = \bar \Psi_{\beta} \br{r-s/2} \Psi_{\alpha} \br{r+s/2} - \Psi_{\alpha} \br{r+s/2} \bar \Psi_{\beta} \br{r-s/2},
\end{equation}
with the center-of-mass coordinate $r=\br{r_1 + r_2}/2$ and the relative coordinate $s = s_2 - s_1$. Depending on the dimension of the problem, the spinors
are written in either a four-dimensional representation or a two-dimensional representation. It is possible to perform the calculations in either formulation. In the Appendix, see \ref{App_Trans3}, \ref{App_Trans2} or \ref{App_Trans1}, detailed calculations of the
various possibilities are shown. However, we want to postpone this issue to the following section keeping the general notation introduced above.


As the Wigner operator is in principle formulated as the Fourier transform of the density operator, we go on
defining the covariant Wigner operator
\begin{equation}
\hat {\mathcal W}_{\alpha \beta} \br{r,p} = \frac{1}{2} \int d^4s \ \ee^{\ii p s} \ \mathcal{U} \br{A,r,s} \com{\bar{\Psi}_{\beta} \br{r-s/2}, \Psi_{\alpha} \br{r+s/2}}.
\end{equation}
However, as the density operator is not
gauge invariant under local $U \br{1}$ transformations we additionally have to introduce a Wilson line factor
\begin{equation}
 \mathcal{U} \br{A,r,s} = \exp \br{\ii e \int_{-1/2}^{1/2} d \xi \ A \br{r+ \xi s} \ s}.
\end{equation}

No path ordering is needed as we approximate the background fields being of Hartree type. By this means, we take the mean ensemble average of the applied field. Thus, we basically average over the quantum fluctuations of the fields.
An immediate consequence is that the gauge fields become ordinary c-numbers.
  
\section{Equations of motion}
\label{Kap_EoM}
We derive the equations of motion for the Wigner operator by taking into account the Dirac equation \eqref{Eq_Dirac1} 
and the adjoint Dirac equation \eqref{Eq_Dirac2}. This yields (see appendix \ref{App_EoM} for a detailed calculation)
\begin{alignat}{5}
&\br{\frac{1}{2} D_{\mu} \br{r,p} - \ii \Pi_{\mu} \br{r,p} } &&\ga{\mu} \hat{\mathcal W} \br{r,p} &&= -&&\ii \hat{\mathcal W} \br{r,p}, \\
&\br{\frac{1}{2} D_{\mu} \br{r,p} + \ii \Pi_{\mu} \br{r,p}} &&\hat{\mathcal W} \br{r,p} \ga{\mu} &&= &&\ii \hat{\mathcal W} \br{r,p},
\end{alignat}
with the pseudo-differential operators
\begin{alignat}{5}
 &D_{\mu} \br{r,p} &&= \de{\mu}^r &&- e &&\int_{-1/2}^{1/2} d \xi \ &&F_{\mu \nu} \br{r - \ii \xi \partial^p} \de{p}^{\nu} ,\\
 &\Pi_{\mu} \br{r,p} &&= p_{\mu} &&-\ii e &&\int_{-1/2}^{1/2} d \xi \ \xi \ &&F_{\mu \nu} \br{r - \ii \xi \partial^p} \de{p}^{\nu}.
\end{alignat}
Then we proceed by taking the vacuum expectation value (\ref{App_Avg}) of the Wigner operator yielding the covariant Wigner function
\begin{equation}
 \mathbbm{W} \br{r,p} = \langle \Phi \vert \hat{\mathcal W} \br{r,p} \vert \Phi \rangle.
\end{equation}

At this point, the consequences of a Hartree approximation become obvious. Taking the vacuum expectation value on the left hand side of the equations of motion we obtain terms of the form
\begin{alignat}{5}
&\langle \Phi \vert F_{\mu \nu} \br{r - \ii \xi \partial^p} \ \mathcal{U} \br{A,r,s} \com{\bar{\Psi}_{\beta} \br{r-s/2}, \psi_{\alpha} \br{r+s/2} } \vert \Phi \rangle.
\end{alignat}
As the electromagnetic fields are of Hartree type, we can replace the product of operators by a product of expectation values\cite{PhysRevD.44.1825} yielding
\begin{alignat}{5}
&\langle \Phi \vert F_{\mu \nu} \br{r - \ii \xi \partial^p} \vert \Phi \rangle \langle \Phi \vert \mathcal{U} \br{A,r,s} \com{\bar{\Psi}_{\beta} \br{r-s/2}, \psi_{\alpha} \br{r+s/2}} \vert \Phi \rangle.
\end{alignat}
Note, that we have just abandoned the quantum interactions between Dirac fields and the gauge fields. However, the vector potential is still present in
the second term. All in all, this leads to the equations of motion for the covariant Wigner function
\begin{alignat}{5}
 &\br{\frac{1}{2} D_{\mu} \br{r,p} -\ii \Pi_{\mu} \br{r,p} } &&\gamma^{\mu} \mathbbm W \br{r,p} &&= -&&\ii \mathbbm W \br{r,p}, \\
 &\br{\frac{1}{2} D_{\mu} \br{r,p} +\ii \Pi_{\mu} \br{r,p} } &&\mathbbm W \br{r,p} \gamma^{\mu} &&= &&\ii \mathbbm W \br{r,p}.
\end{alignat}

\section{Pair production in arbitrary dimensions}
Up to now, we have formulated all equations in a general covariant form without specifying the dimension of the underlying physics we want to describe.
The objective of this section is to demonstrate how one can obtain all transport equations by exploiting various symmetries of
the applied background fields. In this way, it is possible to show, that the equations of motion describing a lower dimensional system
can also be obtained when starting with the full formalism for $QED_{3+1}$. For the sake of completeness, a detailed derivation of the transport equations
in $n+1$ dimensions starting with a $QED_{n+1}$ Lagrangian is done in the Appendix, see \ref{App_Trans3}, \ref{App_Trans2} or \ref{App_Trans1}.

\vspace{5cm}

In order to properly describe matter dynamics in $3+1$ dimensions one finds a $4$-dimensional irreducible representation. Due to the fact, that the density function
is therefore $4 \times 4$ dimensional and the covariant Wigner function transforms as a Dirac gamma matrix
we can decompose it into $16$ covariant Wigner components
\begin{equation}
\mathbbm{W} = \frac{1}{4} \br{\mathbbm{1} \mathbbm{S} + \ii \gamma_5 \mathbbm{P} + \ga{\mu} \mathbbm{V}_{\mu} + \ga{\mu} \gamma_5 \mathbbm{A}_{\mu} + \sigma^{\mu \nu} \mathbbm{T}_{\mu \nu}}. 
\end{equation}

Due to the challenges stemming from working with the covariant Wigner function, we switch to an equal-time approach\cite{Ochs1998351}.
This is done by taking the energy average of the covariant Wigner function. The individual Wigner components 
are transformed as
\begin{align}
 \mathbbm{w} \br{\mathbf{x},\mathbf{p},t} = \int \frac{d p_0}{2 \pi} \mathbbm{W} \br{r,p},
\end{align}
where $\mathbf{p}$ denotes the particles kinetic momentum, while $\mathbf{x}$ describes the position of the particles. This yields
after some calculation, see appendix \ref{App_Trans3}, the transport equations in the equal-time approach 
  \begin{alignat}{4}
    & D_t \mathbbm{s}     && && -2 \boldsymbol{\Pi} \cdot \mathbbm{t_1} &&= 0, \label{eq_3_1} \\
    & D_t \mathbbm{p} && && +2 \boldsymbol{\Pi} \cdot \mathbbm{t_2} &&= -2\mathbbm{a}_0,  \\
    & D_t \mathbbm{v}_0 &&+ \mathbf{D} \cdot \mathbbm{v} && &&= 0,  \\
    & D_t \mathbbm{a}_0 &&+ \mathbf{D} \cdot \mathbbm{a} && &&= 2\mathbbm{p},  \\    
    & D_t \mathbbm{v} &&+ \mathbf{D} \ \mathbbm{v}_0 && +2 \boldsymbol{\Pi} \times \mathbbm{a} &&= -2\mathbbm{t_1},  \\    
    & D_t \mathbbm{a} &&+ \mathbf{D} \ \mathbbm{a}_0 && +2 \boldsymbol{\Pi} \times \mathbbm{v} &&= 0,  \\
    & D_t \mathbbm{t_1} &&+ \mathbf{D} \times \mathbbm{t_2} && +2 \boldsymbol{\Pi} \ \mathbbm{s} &&= 2\mathbbm{v},  \\    
    & D_t \mathbbm{t_2} &&- \mathbf{D} \times \mathbbm{t_1} && -2 \boldsymbol{\Pi} \ \mathbbm{p} &&= 0.  \label{eq_3_8}
  \end{alignat} 
The $16$ Wigner components are all connected and furthermore the fields are in general non-local as they appear
in the pseudo-differential operators in the following way: 
  \begin{alignat}{6}
     & D_t && = \de{t} &&+ e &&\int d\xi &&\mathbf{E} \br{\mathbf{x}+\ii \xi \boldsymbol{\nabla}_p,t} && ~ \cdot \boldsymbol{\nabla}_p,  \\
     & \mathbf{D} && = \boldsymbol{\nabla}_x &&+ e &&\int d \xi &&\mathbf{B} \br{\mathbf{x}+\ii \xi \boldsymbol{\nabla}_p,t} &&\times \boldsymbol{\nabla}_p,  \\
     & \boldsymbol{\Pi} && = \mathbf{p} &&- \ii e &&\int d \xi \xi &&\mathbf{B} \br{\mathbf{x}+\ii \xi \boldsymbol{\nabla}_p,t} &&\times \boldsymbol{\nabla}_p.
  \end{alignat}    
The initial conditions describing a vacuum state, see appendix \ref{Sec_Vac}, are such that all Wigner moments vanish except for the scalar and three-vector ones:
\begin{align}
  \mathbbm{s}_i = -\frac{2}{\omega},\ \mathbbm{v}_{i} = -\frac{2 \mathbf{p}}{\omega},  
\end{align}
with the one-particle energy $\omega = \sqrt{1 + \mathbf{p}^2}$. 
Besides, we are assuming a spatially unbounded physical problem, thus there are
no restrictions stemming from boundary conditions. In chapter four we will discuss the issue with
this kind of boundaries and how one can solve the corresponding system of PDEs.

With regard to our simulations we skip a detailed analysis of symmetries at this point. Instead, we will focus on discussing symmetries in lower dimensional systems. The only exception are cylindrically symmetric problems, which are discussed separately in section \ref{Kap_CylSym}.
  
\subsection{Pair production in the plane}    

The first reduction is done restricting the background field to a plane embedded in three-dimensional space. Without loss
of generality, we can choose a vector potential of the form
\begin{equation}
 \mathbf{A} \br{\mathbf{x},t} = A_x \br{x,z,t} \mathbf{e}_x + A_z \br{x,z,t} \mathbf{e}_z.
\end{equation}
Thus, we obtain for the electric and magnetic field 
\begin{align}
 &\mathbf{B} \br{\mathbf{x},t} = B(x,z,t) \mathbf{e}_y,\\
 &\mathbf{E} \br{\mathbf{x},t} = E_x \br{x,z,t} \mathbf{e}_x + E_z \br{x,z,t} \mathbf{e}_z.
\end{align}
The electric field can be seen as defined in the plane, while the magnetic field would then become a scalar quantity.
Such a choice of the fields eliminates all derivatives with respect to $p_y$ from the equations \eqref{eq_3_1} - \eqref{eq_3_8}. Hence, the 
momentum in direction of $y$ becomes an ordinary parameter. Fixing $p_y$ to zero limits the accessible
phase-space domain to a single plane. Consequently, the transport equations \eqref{eq_3_1} - \eqref{eq_3_8}
are modified yielding a less complex system of DEs  
  \begin{alignat}{6}
    & D_t \mathbbm{s}   &&  && && -2 \Pi_1 \mathbbm{t}_{1,1} &&-2 \Pi_3 \mathbbm{t}_{1,3} &&= 0, \label{eq2_1a} \\
    & D_t \mathbbm{v}_0 &&+ D_1 \mathbbm{v}_1 &&+ D_3 \mathbbm{v}_3 && && &&= 0,  \\   
    & D_t \mathbbm{v}_1 &&+ D_1 \mathbbm{v}_0 && && &&-2 \Pi_3 \mathbbm{a}_2 &&= -2\mathbbm{t}_{1,1},  \\    
    & D_t \mathbbm{v}_3 && &&+ D_3 \mathbbm{v}_0 && +2 \Pi_1 \mathbbm{a}_2 && &&= -2\mathbbm{t}_{1,3},  \\        
    & D_t \mathbbm{a}_2 && && &&-2 \Pi_1 \mathbbm{v}_3 &&+ 2 \Pi_3 \mathbbm{v}_1 &&= 0,  \\
    & D_t \mathbbm{t}_{1,1} && &&- D_3 \mathbbm{t}_{2,2} && +2 \Pi_1 \mathbbm{s} && &&= 2\mathbbm{v}_1,  \\    
    & D_t \mathbbm{t}_{1,3} &&+ D_1 \mathbbm{t}_{2,2} && && &&+2 \Pi_3 \mathbbm{s} &&= 2\mathbbm{v}_3,  \\    
    & D_t \mathbbm{t}_{2,2} &&+ D_1 \mathbbm{t}_{1,3} &&- D_3 \mathbbm{t}_{1,1} && && &&= 0. \label{eq2_8a}   
  \end{alignat}  
  The corresponding pseudo-differential operators are
  \begin{alignat}{6}
     & D_t &&= \de{t} &&+ e &&\int d\xi \left( \right. && E_x &&\br{x+\ii \xi \de{p_x},z+\ii \xi \de{p_z},t} \de{p_x}  \\
       & && && &&+ && E_z &&\br{x+\ii \xi \de{p_x},z+\ii \xi \de{p_z},t} \de{p_z} \left. \right), \notag \\
     & D_1 &&= \de{x} &&+e &&\int d \xi &&B &&\br{x+\ii \xi \de{p_x},z+\ii \xi \de{p_z},t} \de{p_z},  \\
     & D_3 &&= \de{z} &&- e &&\int d \xi &&B &&\br{x+\ii \xi \de{p_x},z+\ii \xi \de{p_z},t} \de{p_x},  \\
     & \Pi_1 &&= p_x &&- \ii e &&\int d \xi \xi &&B &&\br{x+\ii \xi \de{p_x},z+\ii \xi \de{p_z},t} \de{p_z}, \\
     & \Pi_3 &&= p_z &&+ \ii e &&\int d \xi \xi &&B &&\br{x+\ii \xi \de{p_x},z+\ii \xi \de{p_z},t} \de{p_x}.
  \end{alignat} 
  When performing the derivation of the transport equations using $4$-spinors
  and the Lagrangian in $QED_{2+1}$ we basically obtain the same system of DEs as in appendix \ref{App_Trans2}. Hence, we conclude, that the system \eqref{eq2_1a} - \eqref{eq2_8a}
  indeed describes pair production in a plane. 
  
  We can further reduce the problem by writing $A_z=0$ and $A_x \br{x,z,t} = A_x \br{z,t}$. Compared to the previous case, this additionally simplifies the situation. The electric field is still defined in the plane, but shows only one nonzero component. The corresponding magnetic field remains a scalar. Furthermore, the problem is now homogeneous in $x$.
  Therefore, the system
  of DEs \eqref{eq2_1a} - \eqref{eq2_8a} is further reduced taking the form  
  \begin{alignat}{6}
    & D_t \mathbbm{s}   &&  && && -2 \Pi_1 \mathbbm{t}_{1,1} &&-2 \Pi_3 \mathbbm{t}_{1,3} &&= 0, \label{eq_2_1b} \\
    & D_t \mathbbm{v}_0 &&+ D_1 \mathbbm{v}_1 &&+ D_3 \mathbbm{v}_3 && && &&= 0,  \\   
    & D_t \mathbbm{v}_1 &&+ D_1 \mathbbm{v}_0 && && &&-2 \Pi_3 \mathbbm{a}_2 &&= -2\mathbbm{t}_{1,1},  \\    
    & D_t \mathbbm{v}_3 && &&+ D_3 \mathbbm{v}_0 && +2 \Pi_1 \mathbbm{a}_2 && &&= -2\mathbbm{t}_{1,3},  \\        
    & D_t \mathbbm{a}_2 && && &&-2 \Pi_1 \mathbbm{v}_3 &&+ 2 \Pi_3 \mathbbm{v}_1 &&= 0,  \\
    & D_t \mathbbm{t}_{1,1} && &&- D_3 \mathbbm{t}_{2,2} && +2 \Pi_1 \mathbbm{s} && &&= 2\mathbbm{v}_1,  \\    
    & D_t \mathbbm{t}_{1,3} &&+ D_1 \mathbbm{t}_{2,2} && && &&+2 \Pi_3 \mathbbm{s} &&= 2\mathbbm{v}_3,  \\    
    & D_t \mathbbm{t}_{2,2} &&+ D_1 \mathbbm{t}_{1,3} &&- D_3 \mathbbm{t}_{1,1} && && &&= 0. \label{eq_2_8b}    
  \end{alignat}   
The corresponding initial conditions are
\begin{align}
  \mathbbm{s}_i = -\frac{2}{\omega},\ \mathbbm{v}_{1i} = -\frac{2 p_x}{\omega},\ \mathbbm{v}_{3i} = -\frac{2 p_z}{\omega},  
\end{align}
with $\omega = \sqrt{1 + p_x^2 + p_z^2}$.

  The pseudo-differential operators read
  \begin{alignat}{6}
      & D_t &&= \de{t} +&& e &&\int d\xi &&E \br{z+\ii \xi \de{p_z},t} \de{p_x},  \\
      & D_1 &&= &&e &&\int d \xi &&B \br{z+\ii \xi \de{p_z},t} \de{p_z},  \\
      & D_3 &&= \de{z} - &&e &&\int d \xi &&B \br{z+\ii \xi \de{p_z},t} \de{p_x},  \\
      & \Pi_1 &&= p_x - &&\ii e &&\int d \xi \xi &&B \br{z+\ii \xi \de{p_z},t} \de{p_z}, \\
      & \Pi_3 &&= p_z + &&\ii e &&\int d \xi \xi &&B \br{z+\ii \xi \de{p_z},t} \de{p_x}.
  \end{alignat}       

Hence, the decoupled equations are zero throughout the calculation and therefore neglected. 
Some remarks about the discrete symmetries of this system of DEs are in order. At first, without a rigorous proof we assume that the solution to \eqref{eq_2_1b}-\eqref{eq_2_8b} is unique. 
Secondly, we find that the equations \eqref{eq_2_1b}-\eqref{eq_2_8b} stay invariant when replacing $p_z \to -p_z$ as long
as the Wigner components transform as
\begin{alignat}{5}
 \ma{\mathbbm{s} &\br{z,p_x,p_z} \\ \mathbbm{v}_0 &\br{z,p_x,p_z} \\ \mathbbm{v}_1 &\br{z,p_x,p_z} \\ \mathbbm{v}_3 &\br{z,p_x,p_z} \\
 \mathbbm{a}_2 &\br{z,p_x,p_z} \\ \mathbbm{t}_{1,1} &\br{z,p_x,p_z} \\ \mathbbm{t}_{1,3} &\br{z,p_x,p_z} \\ \mathbbm{t}_{2,2} &\br{z,p_x,p_z} } \to
 \ma{&\mathbbm{s} &\br{z,p_x,-p_z} \\ -&\mathbbm{v}_0 &\br{z,p_x,-p_z} \\ &\mathbbm{v}_1 &\br{z,p_x,-p_z} \\ -&\mathbbm{v}_3 &\br{z,p_x,-p_z} \\
 -&\mathbbm{a}_2 &\br{z,p_x,-p_z} \\ &\mathbbm{t}_{1,1} &\br{z,p_x,-p_z} \\ &\mathbbm{t}_{1,3} &\br{z,p_x,-p_z} \\ &\mathbbm{t}_{2,2} &\br{z,p_x,-p_z} }. \label{eq_2_trafo}
\end{alignat}
This means that given a solution ``solution I'' consisting of the Wigner components $\mathbbm{w}_k \br{z,p_x,p_z}$, we can immediately construct another solution ``solution II'' $\tilde {\mathbbm{w}}_k \br{z,p_x,p_z}$. Due to the relations \eqref{eq_2_trafo} ``solution II'' can also be written in terms of $\mathbbm{w}_k \br{z,p_x,p_z}$. Due to the assumption, that only one solution exists both Wigner  vectors have to contain the same information. Hence, the following relations hold
\begin{alignat}{6}
 \mathbbm{w}_k \br{z,p_x,p_z} = \tilde {\mathbbm{w}}_k \br{z,p_x,p_z} \to \tilde {\mathbbm{w}}_k \br{z,p_x,p_z} = \pm \mathbbm{w}_k \br{z,p_x,-p_z},
\end{alignat}
where the sign depends upon the specific Wigner component, see \eqref{eq_2_trafo}. Hence, we conclude that the Wigner components $\mathbbm{w}_k \br{z,p_x,p_z}$ are either symmetric or antisymmetric in $p_z$.

A further symmetry is found in case the applied vector potential is either symmetric or antisymmetric in  $z$.
Assuming $A \br{z,t}$ is symmetric in $z$ the equations \eqref{eq_2_1b}-\eqref{eq_2_8b} are invariant under the replacement $z \to -z$ if the Wigner components
transform as
\begin{alignat}{5}
 \ma{\mathbbm{s} &\br{z,p_x,p_z} \\ \mathbbm{v}_0 &\br{z,p_x,p_z} \\ \mathbbm{v}_1 &\br{z,p_x,p_z} \\ \mathbbm{v}_3 &\br{z,p_x,p_z} \\
 \mathbbm{a}_2 &\br{z,p_x,p_z} \\ \mathbbm{t}_{1,1} &\br{z,p_x,p_z} \\ \mathbbm{t}_{1,3} &\br{z,p_x,p_z} \\ \mathbbm{t}_{2,2} &\br{z,p_x,p_z} } \to
 \ma{&\mathbbm{s} &\br{-z,p_x,p_z} \\ -&\mathbbm{v}_0 &\br{-z,p_x,p_z} \\ &\mathbbm{v}_1 &\br{-z,p_x,p_z} \\ &\mathbbm{v}_3 &\br{-z,p_x,p_z} \\
 &\mathbbm{a}_2 &\br{-z,p_x,p_z} \\ &\mathbbm{t}_{1,1} &\br{-z,p_x,p_z} \\ &\mathbbm{t}_{1,3} &\br{-z,p_x,p_z} \\ -&\mathbbm{t}_{2,2} &\br{-z,p_x,p_z} }. 
\end{alignat}
If $A \br{z,t}$, however, is antisymmetric in $z$ we observe a symmetry in $z$ and $p_x$. Under the replacement
\begin{align}
 z \to -z ,\qquad p_x \to -p_x,
\end{align}
the Wigner components obey
\begin{alignat}{5}
 \ma{\mathbbm{s} &\br{z,p_x,p_z} \\ \mathbbm{v}_0 &\br{z,p_x,p_z} \\ \mathbbm{v}_1 &\br{z,p_x,p_z} \\ \mathbbm{v}_3 &\br{z,p_x,p_z} \\
 \mathbbm{a}_2 &\br{z,p_x,p_z} \\ \mathbbm{t}_{1,1} &\br{z,p_x,p_z} \\ \mathbbm{t}_{1,3} &\br{z,p_x,p_z} \\ \mathbbm{t}_{2,2} &\br{z,p_x,p_z} } \to
 \ma{&\mathbbm{s} &\br{-z,-p_x,p_z} \\ -&\mathbbm{v}_0 &\br{-z,-p_x,p_z} \\ -&\mathbbm{v}_1 &\br{-z,-p_x,p_z} \\ &\mathbbm{v}_3 &\br{-z,-p_x,p_z} \\
 -&\mathbbm{a}_2 &\br{-z,-p_x,p_z} \\ -&\mathbbm{t}_{1,1} &\br{-z,-p_x,p_z} \\ &\mathbbm{t}_{1,3} &\br{-z,-p_x,p_z} \\ &\mathbbm{t}_{2,2} &\br{-z,-p_x,p_z} }
\end{alignat}
leaving \eqref{eq_2_1b}-\eqref{eq_2_8b} invariant.

At this point we want to suggest an additional transformation in order to reduce the system of DEs \eqref{eq_2_1b}-\eqref{eq_2_8b}. 
The idea is to mix the components via rotation and reflection matrices. Therefore, we introduce the mappings
\begin{alignat}{5}
 &\ma{\overline{\mathbbm{s}} \\ \underline{\mathbbm{s}} } &&= \ma{1 & 1 \\ 1 & -1} \ma{\mathbbm{s} \\ \mathbbm{a}_2 },\hspace{1.5cm} \label{eq_2_trafo1}
 &&\ma{\overline{\mathbbm{v}}_1 \\ \underline{\mathbbm{v}}_1 } &&= \ma{1 & -1 \\ 1 & 1} \ma{\mathbbm{v}_1 \\ \mathbbm{t}_{1,3} },\\ 
 &\ma{\overline{\mathbbm{v}}_3 \\ \underline{\mathbbm{v}}_3 } &&= \ma{1 & 1 \\ 1 & -1} \ma{\mathbbm{v}_3 \\ \mathbbm{t}_{1,1} },\ 
 &&\ma{\overline{\mathbbm{v}}_0 \\ \underline{\mathbbm{v}}_0 } &&= \ma{1 & -1 \\ 1 & 1} \ma{\mathbbm{v}_0 \\ \mathbbm{t}_{2,2} }. \label{eq_2_trafo4}
\end{alignat}
This results in a decoupling of the equations \eqref{eq_2_1b}-\eqref{eq_2_8b} into two separate non-trivial systems of DEs. Four transformed Wigner
components can be summarized in this first system of DEs
\begin{alignat}{6}
  & D_t \overline{\mathbbm{s}}   &&  && && -2 \Pi_1 \overline{\mathbbm{v}}_3 &&+ 2 \Pi_3 \overline{\mathbbm v}_1 &&= 0, \label{eq_2b_1}  \\
  & D_t \overline{\mathbbm{v}}_1 &&+D_1 \overline{\mathbbm{v}}_0 && && &&- 2 \Pi_3 \overline{ \mathbbm{s}} &&= -2\overline{\mathbbm{v}}_3,  \\    
  & D_t \overline{\mathbbm{v}}_3 && &&+D_3 \overline{\mathbbm{v}}_0 && +2 \Pi_1 \overline{\mathbbm{s}} && &&= 2 \overline{\mathbbm{v}}_1,  \\ 
  & D_t \overline{\mathbbm{v}}_0 &&+D_1 \overline{\mathbbm{v}}_1 &&+D_3 \overline{\mathbbm{v}}_3 && && &&= 0, \label{eq_2b_4} 
\end{alignat}
with initial conditions 
\begin{align}
  \overline{\mathbbm{s}}_i = -\frac{2}{\omega},\ \overline{\mathbbm{v}}_{1i} = -\frac{2 p_x}{\omega} ,\ \overline{\mathbbm{v}}_{3i} = -\frac{2 p_z }{\omega}.
\end{align}
The other components build up the second system of DEs
\begin{alignat}{6}
  & D_t \underline{\mathbbm{s}}   &&  && && +2 \Pi_1 \underline{\mathbbm{v}}_3 &&- 2 \Pi_3 \underline{\mathbbm v}_1 &&= 0, \label{eq_2c_1} \\
  & D_t \underline{\mathbbm{v}}_1 &&+D_1 \underline{\mathbbm{v}}_0 && && &&+ 2 \Pi_3 \underline{ \mathbbm{s}} &&= 2\underline{\mathbbm{v}}_3,  \\    
  & D_t \underline{\mathbbm{v}}_3 && &&+D_3 \underline{\mathbbm{v}}_0 && -2 \Pi_1 \underline{\mathbbm{s}} && &&= -2 \underline{\mathbbm{v}}_1,  \\ 
  & D_t \underline{\mathbbm{v}}_0 &&+D_1 \underline{\mathbbm{v}}_1 &&+D_3 \underline{\mathbbm{v}}_3 && && &&= 0, \label{eq_2c_4}
\end{alignat}
with initial conditions 
\begin{align}
  \underline{\mathbbm{s}}_i = -\frac{2}{\omega},\ \underline{\mathbbm{v}}_{1i} = -\frac{2 p_x}{\omega} ,\ \underline{\mathbbm{v}}_{3i} = -\frac{2 p_z }{\omega}.
\end{align}
In both cases the one-particle energy is given by $\omega = \sqrt{1 + p_x^2 + p_z^2}$. If we were to introduce a further transformation of the form
\begin{align}
 \acute {\overline{\mathbbm{w}}}_k = \overline{\mathbbm{w}}_k/2, \qquad \acute {\underline{\mathbbm{w}}}_k = \underline{\mathbbm{w}}_k/2,
\end{align}
we would immediately obtain the two $2$-spinor formulations for a $2+1$ dimensional problem calculated in appendix \ref{App_Trans2}.

Interestingly, we can transform the first system \eqref{eq_2b_1}-\eqref{eq_2b_4} into the second system \eqref{eq_2c_1}-\eqref{eq_2c_4}. In order to
do that, we have to introduce the transformation $p_z \to - p_z$ and find the relations
\begin{alignat}{6}
 &\overline{\mathbbm{s}} \br{- p_z} &&= &&\underline{\mathbbm{s}} \br{p_z} ,\ &&\overline{\mathbbm{v}}_1 \br{-p_z} = &&\underline{\mathbbm{v}}_1 \br{p_z} ,\\
 &\overline{\mathbbm{v}}_3 \br{- p_z} &&= -&&\underline{\mathbbm{v}}_3 \br{p_z} ,\ &&\overline{\mathbbm{v}}_0 \br{-p_z} = -&&\underline{\mathbbm{v}}_0 \br{p_z}. 
\end{alignat}
Following the discussion on symmetries above and assuming
\begin{align}
\overline{\mathbbm{w}} \br{z,p_x,p_z} = \ma{\overline{\mathbbm{s}} &\br{z,p_x,p_z} \\ \overline{\mathbbm{v}}_1 &\br{z,p_x,p_z} \\ \overline{\mathbbm{v}}_3 &\br{z,p_x,p_z} \\ \overline{\mathbbm{v}}_0 &\br{z,p_x,p_z}} 
\end{align}
is a solution to the system \eqref{eq_2b_1}-\eqref{eq_2b_4} then 
\begin{align}
\underline{\mathbbm{w}} \br{z,p_x,p_z} = \ma{&\underline{\mathbbm{s}} &\br{z,p_x,p_z} \\ &\underline{\mathbbm{v}}_1 &\br{z,p_x,p_z} \\ &\underline{\mathbbm{v}}_3 &\br{z,p_x,p_z} \\ &\underline{\mathbbm{v}}_0 &\br{z,p_x,p_z}} = \ma{&\overline{\mathbbm{s}} &\br{z,p_x,-p_z} \\ &\overline{\mathbbm{v}}_1 &\br{z,p_x,-p_z} \\ -&\overline{\mathbbm{v}}_3 &\br{z,p_x,-p_z} \\ -&\overline{\mathbbm{v}}_0 &\br{z,p_x,-p_z}} 
\end{align}
is a solution to the system \eqref{eq_2c_1}-\eqref{eq_2c_4}. 

Consequently, one has to solve only either of the two systems above as one can reconstruct the whole solution in the end.
Moreover, the reduction to $1+1$ dimensions is simple as both solutions have to coincide for $p_z=0$, thus making both systems of DEs equal.  
In section \ref{sec_observables} the implications of the transformation on the observables is discussed in more detail. 

We may check whether the symmetries found for the equations \eqref{eq_2_1b}-\eqref{eq_2_8b} are still valid. 
From the transformations \eqref{eq_2_trafo1} - \eqref{eq_2_trafo4} we immediately see, that $(\text{anti-})$symmetry in $z$ still holds for
a vector potential being symmetric in $z$ as long as
\begin{alignat}{6}
 \ma{\overline{\mathbbm{s}} &\br{z,p_x,p_z} \\ \overline{\mathbbm{v}}_1 &\br{z,p_x,p_z} \\ \overline{\mathbbm{v}}_3 &\br{z,p_x,p_z} \\ \overline{\mathbbm{v}}_0 &\br{z,p_x,p_z}} \to 
 \ma{&\overline{\mathbbm{s}} &\br{-z,p_x,p_z} \\ &\overline{\mathbbm{v}}_1 &\br{-z,p_x,p_z} \\ &\overline{\mathbbm{v}}_3 &\br{-z,p_x,p_z} \\ -&\overline{\mathbbm{v}}_0 &\br{-z,p_x,p_z}} ,
\end{alignat}
\begin{alignat}{6}
 \ma{\underline{\mathbbm{s}} &\br{z,p_x,p_z} \\ \underline{\mathbbm{v}}_1 &\br{z,p_x,p_z} \\ \underline{\mathbbm{v}}_3 &\br{z,p_x,p_z} \\ \underline{\mathbbm{v}}_0 &\br{z,p_x,p_z}} \to 
 \ma{&\underline{\mathbbm{s}} &\br{-z,p_x,p_z} \\ &\underline{\mathbbm{v}}_1 &\br{-z,p_x,p_z} \\ &\underline{\mathbbm{v}}_3 &\br{-z,p_x,p_z} \\ -&\underline{\mathbbm{v}}_0 &\br{-z,p_x,p_z}} .
\end{alignat}
However, in case of a vector potential fulfilling $A \br{-z,t} = - A\br{z,t}$
no further symmetry under discrete transformations is found. Moreover, neither of the systems \eqref{eq_2b_1}-\eqref{eq_2b_4}, \eqref{eq_2c_1}-\eqref{eq_2c_4} is (anti-)symmetric in $p_z$.
The decomposition shown above from $QED_{3+1}$ to $QED_{2+1}$ is of course not unique as there are multiple possibilities to embed a lower
dimensional physical system into a higher dimensional one. 

\vspace{5cm}

The derivation of the equations of motion in appendix \ref{App_Trans2} is done in the $xy$-plane, so we will also show the non-trivial transport equations 
in case of $\mathbf{A} \br{\mathbf{x},t} = A \br{y,t} \mathbf{e}_x$ and $p_z = 0$:
  \begin{alignat}{6}
    & D_t \mathbbm{s}   &&  && && -2 \Pi_1 \mathbbm{t}_{1,1} &&-2 \Pi_2 \mathbbm{t}_{1,2} &&= 0, \label{eq_2d_1} \\
    & D_t \mathbbm{v}_0 &&+ D_1 \mathbbm{v}_1 &&+ D_2 \mathbbm{v}_2 && && &&= 0,  \\   
    & D_t \mathbbm{v}_1 &&+ D_1 \mathbbm{v}_0 && && &&+2 \Pi_2 \mathbbm{a}_3 &&= -2\mathbbm{t}_{1,1},  \\    
    & D_t \mathbbm{v}_2 && &&+ D_2 \mathbbm{v}_0 && -2 \Pi_1 \mathbbm{a}_3 && &&= -2\mathbbm{t}_{1,2},  \\        
    & D_t \mathbbm{a}_3 && && &&+2 \Pi_1 \mathbbm{v}_2 &&- 2 \Pi_2 \mathbbm{v}_1 &&= 0,  \\
    & D_t \mathbbm{t}_{1,1} && &&+ D_2 \mathbbm{t}_{2,3} && +2 \Pi_1 \mathbbm{s} && &&= 2\mathbbm{v}_1,  \\    
    & D_t \mathbbm{t}_{1,2} &&- D_1 \mathbbm{t}_{2,3} && && &&+2 \Pi_2 \mathbbm{s} &&= 2\mathbbm{v}_2,  \\    
    & D_t \mathbbm{t}_{2,3} &&- D_1 \mathbbm{t}_{1,2} &&+ D_2 \mathbbm{t}_{1,1} && && &&= 0. \label{eq_2d_8}    
  \end{alignat}   
  Here, the pseudo-differential operators read
  \begin{alignat}{6}
     & D_t &&= \de{t} &&+ e &&\int d\xi &&E \br{y+\ii \xi \de{p_y},t} \de{p_x},  \\
     & D_1 &&= &&-e &&\int d \xi &&B \br{y+\ii \xi \de{p_y},t} \de{p_y},  \\
     & D_2 &&= \de{y} &&+ e &&\int d \xi &&B \br{y+\ii \xi \de{p_y},t} \de{p_x},  \\
     & \Pi_1 &&= p_x &&+ \ii e &&\int d \xi \xi &&B \br{y+\ii \xi \de{p_y},t} \de{p_y}, \\
     & \Pi_2 &&= p_y &&- \ii e &&\int d \xi \xi &&B \br{y+\ii \xi \de{p_y},t} \de{p_x}
  \end{alignat} 
and the initial conditions are
\begin{align}
  \mathbbm{s}_i = -\frac{2}{\omega},\ \mathbbm{v}_{1i} = -\frac{2 p_x}{\omega} ,\ \mathbbm{v}_{2i} = -\frac{2 p_y }{\omega}.
\end{align}
Symmetry analysis as well as the transformations provided in \eqref{eq_2_trafo1} - \eqref{eq_2_trafo4} hold in modified form also for
the system above.
  
\subsection{Pair production along a line}   
We proceed by restricting the problem to $1+1$ dimensions, see appendix \ref{App_Trans1} for an alternative derivation. In order to perform the reduction, we
choose(without loss of generality) the potential to be $\mathbf{A} \br{\mathbf{x},t} = A \br{x,t} \mathbf{e}_x$. 
Similarly to the previous case 
all derivatives with respect to $p_y$ and $p_z$ vanish. Hence, we are again free to choose values for the parameters $p_y$ and $p_z$. 
When taking $p_y=p_z=0$ and due to the homogeneity of the fields in $y$ and $z$ we obtain a $1+1$-dimensional problem. Consequently, the transport equations take the form
  \begin{alignat}{6}
    & D_t \mathbbm{s}   &&  && && -2 p_x \mathbbm{t}_{1,1} && &&= 0, \label{eq_1_1} \\
    & D_t \mathbbm{v}_0 &&+ \de{x} \mathbbm{v}_1 && && && &&= 0,  \\   
    & D_t \mathbbm{v}_1 &&+ \de{x} \mathbbm{v}_0 && && && &&= -2\mathbbm{t}_{1,1},  \\    
    & D_t \mathbbm{t}_{1,1} && && && +2 p_x \mathbbm{s} && &&= 2\mathbbm{v}_1, \label{eq_1_4} 
  \end{alignat} 
  with 
  \begin{align}
    D_t = \de{t} + e\int \ d\xi E_x \br{x+\ii \xi \de{p_x},t} \de{p_x}.      
  \end{align}
The only remaining initial conditions are
\begin{align}
  \mathbbm{s}_i = -\frac{2}{\omega},\ \mathbbm{v}_{1i} = -\frac{2 p_x}{\omega},
\end{align}
where $\omega = \sqrt{1 + p_x^2}$. 
Introducing the transformation
\begin{align}
 \acute{\mathbbm{w}}_k = \mathbbm{w}_k/2
\end{align}
we finally obtain the DHW equations for a $1+1$ dimensional problem.  

If the vector potential is symmetric in $x$, thus $A \br{x,t} = A \br{-x,t}$, we find
a symmetry in the transport equations if
\begin{alignat}{5}
 \ma{\mathbbm{s} &\br{x,p_x} \\ \mathbbm{v}_0 &\br{x,p_x} \\ \mathbbm{v}_1 &\br{x,p_x} \\ \mathbbm{t}_{1,1} &\br{x,p_x}} \to
 \ma{&\mathbbm{s} &\br{-x,p_x} \\ -&\mathbbm{v}_0 &\br{-x,p_x} \\ &\mathbbm{v}_1 &\br{-x,p_x} \\ &\mathbbm{t}_{1,1} &\br{-x,p_x}}. 
\end{alignat}
In case of $A \br{x,t} = -A \br{-x,t}$ we find another symmetry assuming
\begin{align}
 x \to -x ,\qquad p_x \to -p_x,
\end{align}
and
\begin{alignat}{5}
 \ma{\mathbbm{s} &\br{x,p_x} \\ \mathbbm{v}_0 &\br{x,p_x} \\ \mathbbm{v}_1 &\br{x,p_x} \\ \mathbbm{t}_{1,1} &\br{x,p_x}} \to
 \ma{&\mathbbm{s} &\br{-x,-p_x} \\ &\mathbbm{v}_0 &\br{-x,-p_x} \\ -&\mathbbm{v}_1 &\br{-x,-p_x} \\ -&\mathbbm{t}_{1,1} &\br{-x,-p_x}}. 
\end{alignat}

\section{Spatially homogeneous fields}
\label{Kap_SpatHom}
In this section we will investigate the consequences of taking the homogeneous limit for the vector potential. Then, the magnetic field vanishes and the electric field is of the form $E = E \br{t}$. Thereby, in total six Wigner components decouple from the system of PDEs \eqref{eq_3_1} - \eqref{eq_3_8} and 
we obtain the following system of equations
\begin{alignat}{4}
    & D_t \mathbbm{s}     && && -2 \mathbf{p} \cdot \mathbbm{t_1} &&= 0, \label{Spat1} \\    
    & D_t \mathbbm{v} && && +2\mathbf{p} \times \mathbbm{a} &&= -2\mathbbm{t_1},  \\    
    & D_t \mathbbm{a} && && +2 \mathbf{p} \times \mathbbm{v} &&= 0,  \\
    & D_t \mathbbm{t_1} && && +2 \mathbf{p} \ \mathbbm{s} &&= 2\mathbbm{v}, \label{Spat4}  
\end{alignat} 
  with the differential operator
  \begin{alignat}{6}
      & D_t &&= \de{t} \ +&& \ e && &&\mathbf{E} \br{t} \cdot \boldsymbol{\nabla}_p
  \end{alignat}   
and the initial conditions 
\begin{align}
  \mathbbm{s}_i = -\frac{2}{\omega},\ \mathbbm{v}_{i} = -\frac{2 \mathbf{p}}{\omega},
\end{align}  
where $\omega = \sqrt{1+ \mathbf{p}^2}$.
Introducing the canonical momentum via the transformation $\mathbf{q} = \mathbf{p} +e \mathbf{A} \br{t}$ 
the PDEs are reduced to a system of ODEs
\begin{alignat}{4}
    & \de{t} \mathbbm{s}     && && -2 \br{ \mathbf{q} -e \mathbf{A} \br{t}} \cdot \mathbbm{t_1} &&= 0, \\    
    & \de{t} \mathbbm{v} && && +2\br{ \mathbf{q} -e \mathbf{A} \br{t}} \times \mathbbm{a} &&= -2\mathbbm{t_1},  \\    
    & \de{t} \mathbbm{a} && && +2 \br{ \mathbf{q} -e \mathbf{A} \br{t}} \times \mathbbm{v} &&= 0,  \\
    & \de{t} \mathbbm{t_1} && && +2 \br{ \mathbf{q} -e \mathbf{A} \br{t}} \ \mathbbm{s} &&= 2\mathbbm{v}.   
\end{alignat}
A further reduction of this system is possible in case the applied vector potential is not three-dimensional. 
Using a vector potential of the type $\mathbf{A}  = A_x \br{t} \mathbf{e}_x + A_y \br{t} \mathbf{e}_y$ and fixing the transversal momentum $p_z$ to zero one obtains
 \begin{alignat}{4}
    & \de{t} \mathbbm{s}     && -2 \br{q_x -e A_x \brt}  \mathbbm{t}_{1,1} && -2 \br{q_y -e A_y \brt}  \mathbbm{t}_{1,2} &&= 0, \label{eq_hom_1} \\    
    & \de{t} \mathbbm{v}_1 && && + 2\br{q_y -e A_y \brt} \mathbbm{a}_3 &&= -2\mathbbm{t}_{1,1},  \\    
    & \de{t} \mathbbm{v}_2 && - 2\br{q_x -e A_x \brt} \mathbbm{a}_3 && &&= -2\mathbbm{t}_{1,2},  \\     
    & \de{t} \mathbbm{a}_3 && + 2\br{q_x -e A_x \brt} \mathbbm{v}_2 && - 2\br{q_y -e A_y \brt} \mathbbm{v}_1 &&= 0,  \\  
    & \de{t} \mathbbm{t}_{1,1} && +2 \br{q_x -e A_x \brt} \mathbbm{s} && &&= 2\mathbbm{v}_1 ,\\   
    & \de{t} \mathbbm{t}_{1,2} && && +2 \br{q_y -e A_y \brt} \mathbbm{s} &&= 2\mathbbm{v}_2.      \label{eq_hom_8}
\end{alignat}
We turn our attention to the discrete symmetries of the equations above. In case the vector potential transforms as
\begin{equation}
 \ma{A_x \brt \\ A_y \brt \\ 0} = \ma{-A_x \br{-t} \\ \ \ A_y \br{-t} \\ 0}
\end{equation}
the DEs are invariant under the replacement
\begin{align}
 t \to -t ,\qquad q_x \to -q_x,
\end{align}
in case of
\begin{alignat}{5}
 &\ma{&\mathbbm{s} &\br{t,q_x,q_y} \\ &\mathbbm{v}_1 &\br{t,q_x,q_y} \\ &\mathbbm{v}_2 &\br{t,q_x,q_y} \\ &\mathbbm{a}_3 &\br{t,q_x,q_y} \\ &\mathbbm{t}_{1,1} &\br{t,q_x,q_y} \\ &\mathbbm{t}_{1,2} &\br{t,q_x,q_y}} \to
 &\ma{&\mathbbm{s} &\br{-t,-q_x,q_y} \\ -&\mathbbm{v}_1 &\br{-t,-q_x,q_y} \\ &\mathbbm{v}_2 &\br{-t,-q_x,q_y} \\ &\mathbbm{a}_3 &\br{-t,-q_x,q_y} \\ &\mathbbm{t}_{1,1} &\br{-t,-q_x,q_y} \\ -&\mathbbm{t}_{1,2} &\br{-t,-q_x,q_y}}. 
\end{alignat}
If $A_x \brt$ is symmetric in $t$ and $A_y \brt$ is antisymmetric in $t$, we find a similar relation using a transformation of the form $t \to -t ,~ q_y \to -q_y$.
Moreover, we find a third symmetry for vector potentials obeying
\begin{equation}
 \ma{A_x \brt \\ A_y \brt \\ 0} = \ma{-A_x \br{-t} \\ -A_y \br{-t} \\ 0}.
\end{equation}
Then, invariance of the system of DEs \eqref{Spat1}-\eqref{Spat4} is ensured for the transformation
\begin{align}
 t \to -t ,\qquad q_x \to -q_x ,\qquad q_y \to -q_y ,
\end{align}
and
\begin{alignat}{5}
 &\ma{&\mathbbm{s} &\br{t,q_x,q_y} \\ &\mathbbm{v}_1 &\br{t,q_x,q_y} \\ &\mathbbm{v}_2 &\br{t,q_x,q_y} \\ &\mathbbm{a}_3 &\br{t,q_x,q_y} \\ &\mathbbm{t}_{1,1} &\br{t,q_x,q_y} \\ &\mathbbm{t}_{1,2} &\br{t,q_x,q_y}} \to
 &\ma{&\mathbbm{s} &\br{-t,-q_x,-q_y} \\ -&\mathbbm{v}_1 &\br{-t,-q_x,-q_y} \\ -&\mathbbm{v}_2 &\br{-t,-q_x,-q_y} \\ -&\mathbbm{a}_3 &\br{-t,-q_x,-q_y} \\ &\mathbbm{t}_{1,1} &\br{-t,-q_x,-q_y} \\ &\mathbbm{t}_{1,2} &\br{-t,-q_x,-q_y}}. 
\end{alignat}
Calculations for elliptically polarized fields can be found in the literature\cite{PhysRevD.89.085001,Li,Shen}. 

Equations \eqref{Spat1}-\eqref{Spat4} can be further decomposed in case one applies a vector potential of the form $\mathbf{A}  = A_x \br{t} \mathbf{e}_x$ or
$\mathbf{A}  = A_y \br{t} \mathbf{e}_y$ and is only interested in particles with vanishing transversal momenta. However, there is a more elegant
way in order to simplify the equations to study pair production in a homogeneous field exhibiting cylindrical symmetry. The corresponding transport equations are identical to the equations derived within QKT and therefore of special interest. This reduction will be examined in appendix \ref{App_Alter}.

\section{Cylindrically symmetric fields}
\label{Kap_CylSym}

In the following we will analyze a configuration where
$\mathbf{A} \br{\mathbf{x},t} = A \br{x,t} \mathbf{e}_x$.
Due to the special form we obtain a cylindrically symmetric problem, with $x$ the parallel direction and $y, \ z$ the  transversal directions. Moreover, the electric field also exhibits cylindrical symmetry and the magnetic field vanishes.
Exploiting the symmetry of the electric field, we introduce the coordinates $p_{\rho}$ and $\theta$, which transform as
\begin{alignat}{5}
 & p_x = p_x , \qquad p_y &&= p_{\rho} \Ct, \qquad p_z &&= p_{\rho} \St.
\end{alignat}
The one-particle quasi-energy is given by $\omega = \sqrt{1+p_x^2+p_{\rho}^2}$ and the initial conditions take the form
\begin{align}
 \mathbbm{s}_i = -\frac{2}{\omega},\ \mathbbm{v}_{1i} = -\frac{2 p_x}{\omega} ,\ \mathbbm{v}_{2i} = -\frac{2 p_{\rho} \Ct}{\omega},\ \mathbbm{v}_{3i} = -\frac{2 p_{\rho} \St}{\omega}.
\end{align}

The transport equations \eqref{eq_3_1}-\eqref{eq_3_8} are transformed accordingly yielding 
  \begin{alignat}{6}
    & D_t \mathbbm{s}     && && -2 p_x \mathbbm{t}_{1,1} &&- 2 p_{\rho} \Ct \mathbbm t_{1,2} &&- 2 p_{\rho} \St \mathbbm t_{1,3} &&= 0, \label{eq_Cy1} \\
    & D_t \mathbbm{p} && && +2 p_x \mathbbm{t}_{2,1} &&+ 2 p_{\rho} \Ct \mathbbm t_{2,2} &&- 2 p_{\rho} \St \mathbbm t_{2,3} &&= -2\mathbbm{a}_0,  \\
    & D_t \mathbbm{v}_0 &&+ \de{x} \mathbbm{v}_1 && && && &&= 0,  \\    
    & D_t \mathbbm{a}_0 &&+ \de{x} \mathbbm{a}_1 && && && &&= 2\mathbbm{p},  \\    
    & D_t \mathbbm{v}_1 &&+ \de{x} \mathbbm{v}_0 && &&+ 2 p_{\rho} \Ct \mathbbm{a}_3 &&- 2 p_{\rho} \St \mathbbm{a}_2 &&= -2\mathbbm{t}_{1,1},  \\    
    & D_t \mathbbm{v}_2 && && -2 p_x \mathbbm{a}_3 && &&+2 p_{\rho} \St \mathbbm{a}_1  &&= -2 \mathbbm{t}_{1,2},  \\        
    & D_t \mathbbm{v}_3 && && +2 p_x \mathbbm{a}_2 &&-2 p_{\rho} \Ct \mathbbm{a}_1 &&  &&= -2 \mathbbm{t}_{1,3},  \\ 
    & D_t \mathbbm{a}_1 &&+ \de{x} \mathbbm{a}_0 && &&+2 p_{\rho} \Ct \mathbbm{v}_3 &&+ 2 p_{\rho} \St \mathbbm{a}_2 &&= 0,  \\
    & D_t \mathbbm{a}_2 && && -2 p_x \mathbbm{v}_3 && &&+2 p_{\rho} \St \mathbbm{v}_1 &&= 0,  \\
    & D_t \mathbbm{a}_3 && &&+ 2 p_x \mathbbm{v}_2 &&-2 p_{\rho} \Ct \mathbbm{v}_1 && &&= 0,  \\      
    & D_t \mathbbm{t}_{1,1} && && +2 p_x \mathbbm{s} && && &&= 2\mathbbm{v}_1,  \\
    & D_t \mathbbm{t}_{1,2} &&-\de{x} \mathbbm{t}_{2,3} && &&+2 p_{\rho} \Ct \mathbbm{s} && &&= 2 \mathbbm{v}_2,  \\  
    & D_t \mathbbm{t}_{1,3} &&+\de{x} \mathbbm{t}_{2,2} && && &&+2 p_{\rho} \St \mathbbm{s} &&= 2 \mathbbm{v}_3,  \\      
    & D_t \mathbbm{t}_{2,1} && && -2 p_x \mathbbm{p} && && &&= 0, \\     
    & D_t \mathbbm{t}_{2,2} &&+ \de{x} \mathbbm{t}_{1,3} && &&-2 p_{\rho} \Ct \mathbbm{p} && &&= 0, \\  
    & D_t \mathbbm{t}_{2,3} &&- \de{x} \mathbbm{t}_{1,2} && && &&-2 p_{\rho} \St \mathbbm{p} &&= 0,  \label{eq_Cy16}
  \end{alignat}  
with 
\begin{align}
 D_t = \de{t} + e\int d\xi E \br{x+\ii \xi \de{p_x},t} \de{p_x}.
\end{align}

In the next step we introduce rotation and reflection matrices in order to transform half of the Wigner components. This mapping is written as
\begin{alignat}{6}
 &\ma{\acute{\mathbbm{v}} \\ \tilde{\mathbbm{v}} } &&= \ma{\Ct & \St \\ -\St & \Ct} \ma{\mathbbm{v}_2 \\ \mathbbm{v}_3 },\ 
 &&\ma{\tilde{\mathbbm{a}} \\ \acute{\mathbbm{a}} } &&= \ma{\Ct & \St \\ \St & -\Ct} \ma{\mathbbm{a}_2 \\ \mathbbm{a}_3 },\\ 
 &\ma{\acute{\mathbbm{t}}_1 \\ \tilde{\mathbbm{t}}_1 } &&= \ma{\Ct & \St \\ -\St & \Ct} \ma{\mathbbm{t}_{1,2} \\ \mathbbm{t}_{1,3} },\ 
 &&\ma{\tilde{\mathbbm{t}}_2 \\ \acute{\mathbbm{t}}_2 } &&= \ma{\Ct & \St \\ \St & -\Ct} \ma{\mathbbm{t}_{2,2} \\ \mathbbm{t}_{2,3} }.\ 
\end{alignat}
An immediate consequence of these
transformations is the decoupling of the system of DEs \eqref{eq_Cy1}-\eqref{eq_Cy16} into a trivial and a non-trivial subsystem. 
While we omit the trivial system of equations, the other subsystem reads
\begin{alignat}{5}
  & D_t \mathbbm{s}     && && -2 p_x \mathbbm{t}_{1,1} &&- 2 p_{\rho} \acute{\mathbbm t}_1 &&= 0,  \label{eq2_Cy1} \\
  & D_t \mathbbm{v}_1 &&+\de{x} \mathbbm{v}_0 && &&- 2 p_{\rho} \acute{ \mathbbm{a}} &&= -2\mathbbm{t}_{1,1},  \\    
  & D_t \acute{\mathbbm{v}} && && +2 p_x \acute{\mathbbm{a}} && &&= -2 \acute{\mathbbm{t}}_1,  \\ 
  & D_t \mathbbm{v}_0 &&+ \de{x} \mathbbm{v}_1 && && &&= 0,  \\ 
  & D_t \acute{\mathbbm{a}} && && - 2 p_x \acute{\mathbbm{v}} &&+2 p_{\rho} \mathbbm{v}_1 &&= 0,  \\
  & D_t \mathbbm{t}_{1,1} && && +2 p_x \mathbbm{s} && &&= 2\mathbbm{v}_1,  \\  
  & D_t \acute{\mathbbm{t}}_1 &&+ \de{x} \acute{\mathbbm{t}}_{2} && &&+2 p_{\rho} \mathbbm{s} &&= 2 \acute{\mathbbm{v}},  \\
  & D_t \acute{\mathbbm{t}}_2 &&+ \de{x} \acute{\mathbbm{t}}_{1} && && &&= 0,  \label{eq2_Cy8} 
\end{alignat}
with the corresponding initial conditions
\begin{align}
  \mathbbm{s}_i = -\frac{2}{\omega},\ \mathbbm{v}_{1i} = -\frac{2 p_x}{\omega} ,\ \acute{\mathbbm{v}}_i = -\frac{2 p_{\rho} }{\omega}.
\end{align}
We proceed by introducing further transformations 
\begin{alignat}{6}
 &\ma{\overline{\mathbbm{s}} \\ \underline{\mathbbm{s}} } &&= \ma{1 & 1 \\ 1 & -1} \ma{\mathbbm{s} \\ \acute{\mathbbm{a}} },\ 
 &&\ma{\overline{\mathbbm{v}} \\ \underline{\mathbbm{v}} } &&= \ma{1 & -1 \\ 1 & 1} \ma{\mathbbm{v}_1 \\ \acute{\mathbbm{t}}_1 },\\ 
 &\ma{\overline{\mathbbm{p}} \\ \underline{\mathbbm{p}} } &&= \ma{1 & 1 \\ -1 & 1} \ma{\acute{\mathbbm{v}} \\ \mathbbm{t}_{1,1} },\ 
 &&\ma{\overline{\mathbbm{v}}_0 \\ \underline{\mathbbm{v}}_0 } &&= \ma{1 & -1 \\ 1 & 1} \ma{\mathbbm{v}_0 \\ \acute{\mathbbm{t}}_2 }.\ 
\end{alignat}
As a result the system of DEs \eqref{eq2_Cy1}-\eqref{eq2_Cy8} decouples into two different, but non-trivial, systems of DEs.
In the following, the first system of equations yields
\begin{alignat}{5}
  & D_t \overline{\mathbbm{s}}     && && -2 p_x \overline{\mathbbm{p}} &&+ 2 p_{\rho} \overline{\mathbbm v} &&= 0, \label{eq3_Cy1} \\
  & D_t \overline{\mathbbm{v}} &&+\de{x} \overline{\mathbbm{v}}_0 && &&- 2 p_{\rho} \overline{ \mathbbm{s}} &&= -2\overline{\mathbbm{p}},  \\    
  & D_t \overline{\mathbbm{p}} && && +2 p_x \overline{\mathbbm{s}} && &&= 2 \overline{\mathbbm{v}},  \\ 
  & D_t \overline{\mathbbm{v}}_0 &&+ \de{x} \overline{\mathbbm{v}} && && &&= 0, \label{eq3_Cy4}  
\end{alignat}
with initial conditions 
\begin{align}
  \overline{\mathbbm{s}}_i = -\frac{2}{\omega},\ \overline{\mathbbm{v}}_{i} = -\frac{2 p_x}{\omega} ,\ \overline{\mathbbm{p}}_i = -\frac{2 p_{\rho} }{\omega}.
\end{align}
For the sake of completeness, the second system of DEs takes the form
\begin{alignat}{5}
  & D_t \underline{\mathbbm{s}}     && && -2 p_x \underline{\mathbbm{p}} &&- 2 p_{\rho} \underline{\mathbbm v} &&= 0, \label{eq3_Cy5} \\
  & D_t \underline{\mathbbm{v}} &&+\de{x} \underline{\mathbbm{v}}_0 && &&+ 2 p_{\rho} \underline{ \mathbbm{s}} &&= -2\underline{\mathbbm{p}},  \\    
  & D_t \underline{\mathbbm{p}} && && +2 p_x \underline{\mathbbm{s}} && &&= 2 \underline{\mathbbm{v}},  \\ 
  & D_t \underline{\mathbbm{v}}_0 &&+ \de{x} \underline{\mathbbm{v}} && && &&= 0, \label{eq3_Cy8}  
\end{alignat}
with initial conditions 
\begin{align}
  \underline{\mathbbm{s}}_i = -\frac{2}{\omega},\ \underline{\mathbbm{v}}_{i} = -\frac{2 p_x}{\omega} ,\ \underline{\mathbbm{p}}_i = \frac{2 p_{\rho} }{\omega}.
\end{align}
Both subsystems \eqref{eq3_Cy1}-\eqref{eq3_Cy4} and \eqref{eq3_Cy5}-\eqref{eq3_Cy8} can be reduced to the system of DEs describing pair production along a line, see appendix \ref{App_Trans1}, by fixing $p_{\rho}$ to zero.

At this point, it has to be mentioned that by solving \eqref{eq3_Cy1}-\eqref{eq3_Cy4} one can construct solutions for the system of equations \eqref{eq3_Cy5}-\eqref{eq3_Cy8}.
Moreover, it is absolutely sufficient to solve system \eqref{eq3_Cy1}-\eqref{eq3_Cy4} for $p_{\rho} \ge 0$ as the particle yield is equivalent for both subsystems.
This is best seen when both DEs are rewritten via introducing the mappings
 \begin{alignat}{6}
 &\ma{\overline{\mathbbm{u}} \\ \overline{\mathbbm{r}} } &&= \ma{p_{\rho} & -1 \\ 1 & p_{\rho}} \ma{\mathbbm{s} \\ \mathbbm{p} },\qquad 
 &\ma{\underline{\mathbbm{u}} \\ \underline{\mathbbm{r}} } &&= \ma{-p_{\rho} & -1 \\ 1 & -p_{\rho}} \ma{\mathbbm{s} \\ \mathbbm{p} }.  
\end{alignat}
The equations \eqref{eq3_Cy1}-\eqref{eq3_Cy4} can then be transformed into a system of PDEs reading
\begin{alignat}{5}
  & D_t \overline{\mathbbm{u}}     && && -2 p_x \overline{\mathbbm{r}} &&+ 2 \br{1+p_{\rho}^2} \overline{\mathbbm v} &&= 0, \\
  & D_t \overline{\mathbbm{v}} &&+\de{x} \overline{\mathbbm{v}}_0 && &&- 2 \overline{ \mathbbm{u}} &&= 0,  \\    
  & D_t \overline{\mathbbm{r}} && && +2 p_x \overline{\mathbbm{u}} && &&= 0,  \\ 
  & D_t \overline{\mathbbm{v}}_0 &&+ \de{x} \overline{\mathbbm{v}} && && &&= 0,  
\end{alignat}
and the second system \eqref{eq3_Cy5}-\eqref{eq3_Cy8} is accordingly transformed into 
\begin{alignat}{5}
  & D_t \underline{\mathbbm{u}}     && && -2 p_x \underline{\mathbbm{r}} &&+ 2 \br{1+p_{\rho}^2} \underline{\mathbbm v} &&= 0, \\
  & D_t \underline{\mathbbm{v}} &&+\de{x} \underline{\mathbbm{v}}_0 && &&- 2 \underline{ \mathbbm{u}} &&= 0,  \\    
  & D_t \underline{\mathbbm{r}} && && +2 p_x \underline{\mathbbm{u}} && &&= 0,  \\ 
  & D_t \underline{\mathbbm{v}}_0 &&+ \de{x} \underline{\mathbbm{v}} && && &&= 0,  
\end{alignat}
As the initial conditions are given by
\begin{align}
 \overline{\mathbbm{u}} = \underline{\mathbbm{u}} = 0 ,\qquad \overline{\mathbbm{r}} = \underline{\mathbbm{r}} = -\frac{2}{\omega} \br{1 + p_{\rho}^2},
\end{align}
the equivalence of these systems is evident. Hence, solving either of the systems of equations is enough in order to compute the observables also for the other system. 

Additionally, in case of $E \br{x,t} = E \br{-x,t}$ we observe a symmetry for $x \to -x$ and
\begin{alignat}{5}
 &\ma{&\overline{\mathbbm{s}} &\br{t,x,p_x} \\ &\overline{\mathbbm{v}} &\br{t,x,p_x} \\ &\overline{\mathbbm{p}} &\br{t,x,p_x} \\ &\overline{\mathbbm{v}}_0 &\br{t,x,p_x} } \to
 &\ma{&\overline{\mathbbm{s}} &\br{t,-x,p_x} \\ &\overline{\mathbbm{v}} &\br{t,-x,p_x} \\ -&\overline{\mathbbm{p}} &\br{t,-x,p_x} \\ &\overline{\mathbbm{v}}_0 &\br{t,-x,p_x} }. 
\end{alignat}
On the other hand, assuming an electric field that is antisymmetric in $x$ the equations \eqref{eq3_Cy1}-\eqref{eq3_Cy4} are invariant under a transformation
\begin{align}
 x \to -x ,\qquad p_x \to - p_x ,\qquad p_{\rho} \to - p_{\rho} ,
\end{align}
and
\begin{alignat}{5}
 &\ma{&\overline{\mathbbm{s}} &\br{t,x,p_x,p_{\rho}} \\ &\overline{\mathbbm{v}} &\br{t,x,p_x,p_{\rho}} \\ &\overline{\mathbbm{p}} &\br{t,x,p_x,p_{\rho}} \\ &\overline{\mathbbm{v}}_0 &\br{t,x,p_x,p_{\rho}} } \to
 &\ma{&\overline{\mathbbm{s}} &\br{t,-x,-p_x,-p_{\rho}} \\ -&\overline{\mathbbm{v}} &\br{t,-x,-p_x,-p_{\rho}} \\ -&\overline{\mathbbm{p}} &\br{t,-x,-p_x,-p_{\rho}} \\ &\overline{\mathbbm{v}}_0 &\br{t,-x,-p_x,-p_{\rho}} }. 
\end{alignat}

At this point, one remark regarding the historical QKT is in order\cite{Schmidt}. It is the limit of the DHW formalism in case of a spatially homogeneous, linearly polarized electric field\cite{PhysRevD.44.1825,PhysRevD.82.105026}. 
In cylindrical coordinates, in case the electric field is only time-dependent only the 
components($\overline{\mathbbm{s}},~ \overline{\mathbbm{v}} ,~ \overline{\mathbbm{p}}$) do not vanish. The detailed derivation can be found in appendix \ref{App_QKT}, while
we show the final equations of QKT here 
\begin{align}
  \ma{\dot{F} \\ \dot{G} \\ \dot{H}} = \ma{0 & W & 0 \\ -W & 0 & -2\omega \\ 0 & 2\omega & 0} \ma{F \\ G \\ H} + \ma{0 \\ W \\ 0}, \label{QKT} 
\end{align}
with initial conditions $F_i = G_i = H_i = 0$.
We have used abbreviations for the one-particle energy $\omega$, with the canonical momentum $q_x$ and $W \qt$:
\begin{align}
 \omega \qt = \sqrt{1 + p_{\rho}^2 + \br{q_x - eA \brt}^2} ,\quad W \qt = \frac{e E \brt \sqrt{1+p_{\rho}^2} }{\omega^2 \qt}.
\end{align}
The main advantage of this formulation is, that one directly works with the particle density. This is especially useful if one is not interested in other observables, because no final transformations of the components $F ,~ G ,~ H$ are necessary.

\section{Observables}
\label{sec_observables}
We have shown how the Wigner components in the DHW formalism for $QED_{3+1}$ are related to
their lower dimensional counterparts. Moreover, we have shown how the DHW equations simplify in case of homogeneous and/or cylindrically symmetric fields. However, one has to bear in mind, that these are only of theoretical use.
In order to describe a physical process, one has to provide observable quantities which could then be
verified/falsified in an experiment. The observables we formulate in the following are all determined by Noether's theorem.
The interested reader may have a look at the literature\cite{PhysRevD.44.1825,Hebenstreit}, where a more elaborated calculation can be found.
Nevertheless, we want to give an overview of useful observables and how they can be calculated in $n+1$ dimensions using the Wigner components.
At this point, we introduce the phase-space element
\begin{equation}
 d \Gamma = d^nx \ d^np.
\end{equation}
Note, that a factor $\br{2 \pi}^n$ is missing, thus we give all results per unit volume element.
Then, we choose, without loss of generalization, the $xy$-plane
in case of a $2+1$ dimensional problem and $x$ in case of a $1+1$ dimensional problem.

As gauge invariance of the QED action is described by symmetry under $U(1)$ transformations, the electric charge is given by
\begin{equation}
 \mathcal{Q}_{3+1} = e \int d \Gamma \ \mathbbm{v}_0 \brw.
\end{equation}
If one is working in $2+1$ dimensions one obtains different expressions for the electric charge depending on the basis set used
\begin{alignat}{6}
  \mathcal{Q}_{4-spinor} = e \int d \Gamma \ \mathbbm{v}_0 \brw, \\
  \mathcal{Q}_{1} = e \int d \Gamma \ \overline{\mathbbm{v}}_0 \brw, \\
  \mathcal{Q}_{2} = e \int d \Gamma \ \underline{\mathbbm{v}}_0 \brw. 
\end{alignat}
Due to the analysis of the DHW equations, one finds that these expressions are related 
\begin{equation}
 \mathcal{Q}_{4-spinor} = \frac{1}{2} \br{\mathcal{Q}_{1} + \mathcal{Q}_{2} }. \label{charge}
\end{equation}
In case of a $1+1$ dimensional formulation the expression for the electric charge reads
\begin{equation}
 \mathcal{Q}_{1+1} = e \int d \Gamma \ \mathbbm{v}_0 \br{x,p,t} = 2 \ e \int d \Gamma \ \acute{\mathbbm{v}}_0 \br{x,p,t}.
\end{equation}
Mind the additional factor $2$ in the expression above. This factor is not present in reference \cite{Hebenstreit} and cannot be computed within a $1+1$ dimensional derivation.
It shows up here, because the lower-dimensional observables stem from a $3+1$ dimensional calculation, where we have introduced spin-$1/2$ particles,
a concept that does not exist in $1+1$ dimensions.

Due to the fact, that the energy-momentum tensor is conserved, we obtain for the total energy 
\begin{equation}
 \mathcal{E}_{3+1} = \int d \Gamma \br{\mathbbm{s} \brw + \mathbf{p} \cdot \mathbbm{v} \brw} + \frac{1}{2} \int d^3x \br{\mathbf{E} \brx^2 + \mathbf{B}\brx^2}
\end{equation}
In $2+1$ dimensions we obtain for the gauge part
\begin{equation}
 \mathcal{E}_{gauge,2+1} = \frac{1}{2} \int d^2x \br{\mathbf{E} \brx^2 + B\brx^2}.
\end{equation}
Due to the fact, that magnetic fields do not exist in $1+1$ dimensions the gauge part of the energy yields
\begin{equation}
 \mathcal{E}_{gauge,1+1} = \frac{1}{2} \int dx \ E \br{x,t}^2.
\end{equation}
In addition, analysis of the matter part yields for $2+1$ dimensions the expressions
\begin{alignat}{5}
 \mathcal{E}_{4-spinor,matter} = \int d \Gamma \br{\mathbbm{s} \brw + \mathbf{p} \cdot \mathbbm{v} \brw} ,\\
 \mathcal{E}_{1,matter} = \int d \Gamma \br{\overline{\mathbbm{s}} \brw + \mathbf{p} \cdot \overline{\mathbbm{v}} \brw} ,\\
 \mathcal{E}_{2,matter} = \int d \Gamma \br{\underline{\mathbbm{s}} \brw + \mathbf{p} \cdot \underline{\mathbbm{v}} \brw}. 
\end{alignat}
They are related with each other similarly to \eqref{charge}.
In $1+1$ dimensions we obtain for the gauge part of the energy momentum tensor
\begin{equation}
 \mathcal{E}_{matter,1+1} = \ \int d \Gamma \br{\mathbbm{s} \br{x,p,t} + p \ \mathbbm{v} \br{x,p,t}}.
\end{equation}

The particle density and thus the particle yield basically make up the matter part of the total energy
\begin{equation}
 \mathcal{N} = \int d \Gamma \ \frac{\br{\mathbbm{s} \brw - \mathbbm{s}_i \br{\mathbf{p}}} + \mathbf{p} \cdot \br{ \mathbbm{v} \brw - \mathbbm{v}_i \br{\mathbf{p}}}}{\omega \br{\mathbf{p}}}.
\end{equation}
Hence, introducing the one-particle energy $\omega \br{\mathbf{p}} = \sqrt{1 + \mathbf{p}^2}$ we can assign an energy to a quasi-particle.
Additionally, one is usually interested in the total number of created particles. Thus, the vacuum offset is
removed by subtracting the vacuum part leading to $\mathcal N = 0$ for the vacuum state. As the distribution function gives the distribution of created matter/antimatter and the electric charge describes the distribution of charge we can easily obtain the distribution of the created electrons and positrons. This is simply done via calculating either $\mathcal{N}_{e^-} = \br{\mathcal{N} - \mathcal{Q}}/2$ for the electron distribution or $\mathcal{N}_{e^+} = \br{\mathcal{N} + \mathcal{Q}}/2$ for the positrons. In case of cylindrically symmetric problems the distribution function writes
\begin{align}
 \mathcal{N}_{Cyl} = 2& \ \int d \Gamma \ \frac{1}{\omega \br{p_x,p_{\rho}}} \left( \right. \br{\overline{\mathbbm{s}} \br{x,p_x,p_{\rho},t} - \overline{\mathbbm{s}}_i \br{p_x,p_{\rho}}} + \\
 &p_x \ \br{ \overline{\mathbbm{v}} \br{x,p_x,p_{\rho},t} - \overline{\mathbbm{v}}_i \br{p_x,p_{\rho}}} + p_{\rho} \ \br{ \overline{\mathbbm{p}} \br{x,p_x,p_{\rho},t} - \overline{\mathbbm{p}}_i \br{p_x,p_{\rho}}} \left. \right).
\end{align}
For a two dimensional problem formulated in a $4$-spinor representation one has to calculate
\begin{equation}
 \mathcal{N}_{4-spinor} = \int d \Gamma \ \frac{\br{\mathbbm{s} \brw - \mathbbm{s}_i \br{\mathbf{p}}} + \mathbf{p} \cdot \br{ \mathbbm{v} \brw - \mathbbm{v}_i \br{\mathbf{p}}}}{\omega \br{\mathbf{p}}}.
\end{equation}
In case of a two spinor formulation one basically has to solve two different systems of equations. Although these two solutions are related via a single variable transformation we state the full version of the distribution function here. In total, this reads
\begin{equation}
 \mathcal{N}_{4-spinor} = \frac{1}{2} \br{\mathcal{N}_1 + \mathcal{N}_2},
\end{equation}
where
\begin{align}
 \mathcal{N}_1 =& \int d \Gamma \ \frac{1}{\omega \br{\mathbf{p}}} \left( \right. \br{\overline{\mathbbm{s}} \br{\mathbf{x},\mathbf{p},t} - \overline{\mathbbm{s}}_i \br{\mathbf{p}}} + \\
 &p_x \ \br{ \overline{\mathbbm{v}} \br{\mathbf{x},\mathbf{p},t} - \overline{\mathbbm{v}}_i \br{\mathbf{p}}} + p_y \ \br{ \overline{\mathbbm{p}} \br{\mathbf{x},\mathbf{p},t} - \overline{\mathbbm{p}}_i \br{\mathbf{p}}} \left. \right)
\end{align}
and
\begin{align}
 \mathcal{N}_2 =& \int d \Gamma \ \frac{1}{\omega \br{\mathbf{p}}} \left( \right. \br{\underline{\mathbbm{s}} \br{\mathbf{x},\mathbf{p},t} - \underline{\mathbbm{s}}_i \br{\mathbf{p}}} + \\
 &p_x \ \br{ \underline{\mathbbm{v}} \br{\mathbf{x},\mathbf{p},t} - \underline{\mathbbm{v}}_i \br{\mathbf{p}}} + p_y \ \br{ \underline{\mathbbm{p}} \br{\mathbf{x},\mathbf{p},t} - \underline{\mathbbm{p}}_i \br{\mathbf{p}}} \left. \right).
\end{align}
For $1+1$ dimensions the expression for the distribution function takes the form
\begin{equation}
 \mathcal{N}_{1+1} = \int d \Gamma \ \frac{\br{\mathbbm{s} \br{x,p_x,t} - \mathbbm{s}_i \br{p_x}} + p_x \ \br{ \mathbbm{v} \br{x,p_x,t} - \mathbbm{v}_i \br{p_x}}}{\omega \br{p_x}}.
\end{equation}

Conservation of the energy-momentum tensor also leads to the definition of the total momentum
\begin{equation}
 \mathcal P_{3+1} = \int d \Gamma \br{\mathbf{p} \ \mathbbm{v}_0 \brw} + \int d^3x \ \mathbf{E} \brx \times \mathbf{B} \brx.
\end{equation}
In case the background field is defined in a plane the gauge part transforms as
\begin{equation}
 \mathcal P_{gauge,2+1} = \int d^2x \ma{\ \ E_y \brx B \brx \\ -E_x \brx B \brx}.
\end{equation}
In $1+1$ dimensions the gauge part vanishes entirely leading to
\begin{equation}
 \mathcal P_{gauge,1+1} = 0.
\end{equation}
The matter part follows the same rules as given for the charge density.

From the conservation of the angular momentum tensor one obtains for the Lorentz boost
\begin{equation}
 \mathcal K_{3+1} = t \mathcal P - \int d\Gamma \ \mathbf{x} \br{\mathbbm{s} \brw + \mathbf{p} \cdot \mathbbm{v} \brw} - \frac{1}{2} \int d^3x \ \mathbf{x} \br{\mathbf{E} \brx^2 + \mathbf{B} \brx^2}.
\end{equation}
We obtain in case of a $2+1$ dimensional physical system
\begin{equation}
 \mathcal K_{gauge,2+1} = - \frac{1}{2} \int d^2x \ \mathbf{x} \br{\mathbf{E} \brx^2 + B \brx^2}
\end{equation}
and in case of a $1+1$ dimensional system
\begin{equation}
 \mathcal{K}_{gauge,1+1} = - \frac{1}{2} \int dx \ x \ E \br{x,t}^2.
\end{equation}
Besides, the matter part obeys the same rules already given for the total energy.

In addition to the Lorentz boost, conservation of the angular momentum tensor also yields the total angular momentum
\begin{equation}
 \mathcal M_{3+1} = \int d \Gamma \br{\mathbf{x} \times \mathbf{p} \ \mathbbm{v}_0 \brw - \frac{1}{2} \mathbbm{a} \brw} + \int d^3x \ \mathbf{x} \times \mathbf{E} \brx \times \mathbf{B} \brx .
\end{equation}
When working in $2+1$ dimensions the total angular momentum is a scalar. Moreover, there seems to be an inconsistency as the angular momentum
looks different comparing $\mathcal{M}$ in the $4$-spinor representation with $\mathcal{M}$ in the $2$-spinor representation. Within the $4$-spinor formulation we obtain 
\begin{align}
 \mathcal M_{4-spinor} =& \int d\Gamma \br{ \br{x \ p_y - y \ p_x} \mathbbm{v}_0 \brw - \frac{1}{2} \mathbbm{a}_3 \brw} \\
  &- \int d^2x \br{x \ E_x \brx B \brx + y \ E_y \brx B \brx}, \notag
\end{align}
while the $2$-spinor formulation yields 
\begin{align}
 \mathcal M_{1} =& \int d\Gamma \br{ \br{x \ p_y - y \ p_x} \overline{\mathbbm{v}}_0 \brw - \frac{1}{2} \overline{\mathbbm{s}} \brw} \\
  &- \int d^2x \br{x \ E_x \brx B \brx + y \ E_y \brx B \brx} \notag
\end{align}
and
\begin{align}
 \mathcal M_{2} =& \int d\Gamma \br{ \br{x \ p_y - y \ p_x} \underline{\mathbbm{v}}_0 \brw + \frac{1}{2} \underline{\mathbbm{s}} \brw} \\
  &- \int d^2x \br{x \ E_x \brx B \brx + y \ E_y \brx B \brx}. \notag
\end{align}
It has to be noted, that one obtains different expressions depending on the basis set used.
This issue, however, clarifies when decomposing the DHW equations from a $3+1$ dimensional formulation to the various $2+1$ dimensional systems. 
From the transformation \eqref{eq_2_trafo1}-\eqref{eq_2_trafo4} one obtains
\begin{align}
 \mathbbm{a}_3 = \frac{1}{2} \br{\overline{\mathbbm{s}} - \underline{\mathbbm{s}}}, \qquad \mathbbm{v}_0 = \frac{1}{2} \br{\overline{\mathbbm{v}}_0 + \underline{\mathbbm{v}}_0}.
\end{align}
Hence, the following relation connects the three different expressions for the angular momentum
\begin{equation}
 \mathcal M_{4-spinor} = \frac{1}{2} \br{\mathcal M_{1} + \mathcal M_{2}}.
\end{equation}

The equations above contain all relevant information in order to calculate observable quantities. If one is using
one of the transformed systems derived within this chapter one has to translate the matter part of the equations correspondingly.


\section{Classical limit}
\label{Kap_Class}
The phase-space approach can also be used in order to describe e.g. a plasma. In plasma physics quantum effects play often only a minor role, therefore classical kinetic theory is a common approach.
In order to connect a quantum kinetic approach with its classical counterpart
 we describe how to obtain the Vlasov equation for the DHW formalism \cite{PhysRevA.48.1869}. However, we want
to approach the classical limit in a slightly different way compared to the literature. We start with the derivation of the Vlasov equation by analyzing the limit $\hbar \to 0$ of the transport equations already 
derived above. For the sake of convenience, we state the system of equations again:
  \begin{alignat}{4}
    & D_t \mathbbm{s}     && && -2 \boldsymbol{\Pi} \cdot \mathbbm{t_1} &&= 0, \label{eq_3_1d} \\
    & D_t \mathbbm{p} && && +2 \boldsymbol{\Pi} \cdot \mathbbm{t_2} &&= -2\mathbbm{a}_0,  \\
    & D_t \mathbbm{v}_0 &&+ \mathbf{D} \cdot \mathbbm{v} && &&= 0, \label{eq_3_3d} \\
    & D_t \mathbbm{a}_0 &&+ \mathbf{D} \cdot \mathbbm{a} && &&= 2\mathbbm{p},  \\    
    & D_t \mathbbm{v} &&+ \mathbf{D} \ \mathbbm{v}_0 && +2 \boldsymbol{\Pi} \times \mathbbm{a} &&= -2\mathbbm{t_1},  \\    
    & D_t \mathbbm{a} &&+ \mathbf{D} \ \mathbbm{a}_0 && +2 \boldsymbol{\Pi} \times \mathbbm{v} &&= 0,  \\
    & D_t \mathbbm{t_1} &&+ \mathbf{D} \times \mathbbm{t_2} && +2 \boldsymbol{\Pi} \ \mathbbm{s} &&= 2\mathbbm{v},  \\    
    & D_t \mathbbm{t_2} &&- \mathbf{D} \times \mathbbm{t_1} && -2 \boldsymbol{\Pi} \ \mathbbm{p} &&= 0, \label{eq_3_8d}
  \end{alignat} 
with the operators
  \begin{alignat}{6}
     & D_t && = \de{t} &&+ e &&\int d\xi &&\mathbf{E} \br{\mathbf{x}+\ii \xi \boldsymbol{\nabla}_p,t} && ~ \cdot \boldsymbol{\nabla}_p,  \\
     & \mathbf{D} && = \boldsymbol{\nabla}_x &&+ e &&\int d \xi &&\mathbf{B} \br{\mathbf{x}+\ii \xi \boldsymbol{\nabla}_p,t} &&\times \boldsymbol{\nabla}_p,  \\
     & \boldsymbol{\Pi} && = \mathbf{p} &&- \ii e &&\int d \xi \xi &&\mathbf{B} \br{\mathbf{x}+\ii \xi \boldsymbol{\nabla}_p,t} &&\times \boldsymbol{\nabla}_p.
  \end{alignat}  
Dimensional analysis of the system above shows, that
\begin{alignat}{6}
 &D_t &&= [\text{time}^{-1}] ,~ &&c \mathbf{D} &&= [\text{time}^{-1}] ,\\ 
 &\frac{c}{\hbar} \boldsymbol{\Pi} &&= [\text{time}^{-1}],~ &&\frac{c^2}{\hbar} &&= [\text{time}^{-1}].
\end{alignat}
Hence, in the limit $\hbar \to 0$ the system of PDEs simplifies to a system of algebraic equations.
These algebraic equations impose constraints on the classical Wigner components:
\begin{alignat}{6}
 &\mathbbm{t}_1 &&= - \mathbf{p} \times \mathbbm{a}, \label{DHW_constr1_1} \\
 &\mathbbm{a}_0 &&= - \mathbf{p} \cdot \mathbbm{t}_2, \\
 &\mathbbm{p} &&= 0, \\
 &\mathbbm{v} &&= \mathbf{p} \ \mathbbm{s}. \label{DHW_constr1_4}
\end{alignat}
Additionally, we may obtain two further relations between Wigner components implied from the constraint equations \eqref{App_constr3_1}-\eqref{App_constr3_8}.
As in classical physics the Wigner operator and correspondingly all its components are on shell, we can write
\begin{align}
 \mathbbm{w}_i \br{x,p} = \mathbbm{w}^{\pm}_i \br{\mathbf{x}, \mathbf{p}, t} \delta \br{p_0 \pm \omega},
\end{align}
with $\omega = \sqrt{1 + \mathbf{p}}$. This enables us to evaluate the integral over $d p_0$ in \eqref{App_constr3_1}-\eqref{App_constr3_8} analytically and subsequently
we obtain
\begin{align}
 \omega \mathbbm{s} = \pm \mathbbm{v}_0, \label{DHW_constr2_1} \\
 \mathbbm{a} = \pm \omega \mathbbm{t}_2. \label{DHW_constr2_2}
\end{align}
However, these findings could have been obtained also from the definition of the observables, see section \ref{sec_observables}. This is because the function $\mathbbm{s}$
yields the phase-space mass density and $\mathbbm{v}_0$ the phase-space charge density. In a similar fashion, one can conclude that $\mathbbm{a}$
gives the phase-space spin density, while $\mathbbm{t}_2$ determines the phase-space magnetic moment density. As a remark,
the function $\mathbbm{v}$ yields the phase-space current density.

In order to derive the Vlasov equation, we start with the equation(see \eqref{eq_3_3d})
\begin{equation}
 D_t \mathbbm{v}_0 + \mathbf{D} \cdot \mathbbm{v} = 0.
\end{equation}
Plugging in the constraints \eqref{DHW_constr1_1}-\eqref{DHW_constr1_4} as well as \eqref{DHW_constr2_1}-\eqref{DHW_constr2_2} yields
\begin{equation}
 \pm \br{\de{t} \mathbbm{s}^{\pm}} \pm \frac{1}{\omega^2} \br{\mathbf{E} \cdot \mathbf{p}} \mathbbm{s}^{\pm} \pm \br{\mathbf{E} \cdot \boldsymbol{\nabla}_p} \mathbbm{s}^{\pm}
 + \mathbf{v} \cdot \br{\boldsymbol{\nabla}_x \mathbbm{s}^{\pm}} + \frac{1}{\omega} \br{\mathbf{B} \times \boldsymbol{\nabla}_p} \cdot \mathbf{p} \mathbbm{s}^{\pm} = 0,
\end{equation}
where we have used the relation $\mathbf{v} = \mathbf{p}/ \omega$.
We proceed by writing 
\begin{equation}
 \mathbbm{s}^{\pm} \br{\mathbf{x}, \mathbf{p},t} = \frac{f^{\pm} \br{\mathbf{x}, \mathbf{p},t}}{\omega \br{\mathbf{p}}},
\end{equation}
which results in 
\begin{equation}
 \pm \br{\de{t} f^{\pm}} \pm \br{\mathbf{E} \cdot \boldsymbol{\nabla}_p} f^{\pm}
 + \mathbf{v} \cdot \br{\boldsymbol{\nabla}_x f^{\pm}} + \br{\mathbf{v} \times \mathbf{B}} \cdot \boldsymbol{\nabla}_p f^{\pm} = 0.
\end{equation}
When describing antiparticles we have to consider their negative momentum. Hence, we distinguish particles from antiparticles by introducing
\begin{alignat}{6}
 &f^{+} \br{\mathbf{x}, \mathbf{p},t} &&\to &&f^+ \br{\mathbf{x}, \mathbf{p},t},\\
 &f^{-} \br{\mathbf{x}, \mathbf{p},t} &&\to &&f^- \br{\mathbf{x}, -\mathbf{p},t},
\end{alignat}
which ultimately leads to 
\begin{alignat}{6}
\de{t} f^+ + \mathbf{v} \cdot \br{\boldsymbol{\nabla}_x f^+} + e \br{\mathbf{E} + \mathbf{v} \times \mathbf{B}} \cdot \boldsymbol{\nabla}_p f^+ = 0 , \label{vlasov_1} \\
\de{t} f^- + \mathbf{v} \cdot \br{\boldsymbol{\nabla}_x f^-} - e \br{\mathbf{E} + \mathbf{v} \times \mathbf{B}} \cdot \boldsymbol{\nabla}_p f^- = 0. \label{vlasov_2}
\end{alignat}
The equations above are the Vlasov equations governing the dynamics of distributions of charged particles and antiparticles in electromagnetic fields. Note, that the only difference is the sign of 
the electric charge. 
In addition, the total particle yield (without subtracting the particles initially present) takes the form
\begin{align}
 \mathcal{N} = \int d \Gamma \ \frac{\mathbbm{s} \brw + \mathbf{p} \cdot \mathbbm{v} \brw}{\omega \br{\mathbf{p}}}.
\end{align}
The particle rate in terms of $f^{\pm}$ instead of the Wigner functions yields
\begin{align}
 \mathcal{N} \br{\mathbf{x}, \mathbf{p},t} = \int d \Gamma f^{\pm} \br{\mathbf{x}, \mathbf{p},t},
\end{align}
showing that $f^{\pm} \br{\mathbf{x}, \mathbf{p},t}$ is indeed the particle distribution function.
Integration of \eqref{vlasov_1} and \eqref{vlasov_2} over the momentum space therefore leads to the continuity equations for particles/antiparticles
\begin{equation}
 \de{t} \rho^{\pm} + \boldsymbol{\nabla}_x \cdot \mathbf{j}^{\pm} = 0, 
\end{equation}
with
\begin{align}
 \rho^{\pm} = \int d^3p \ f^{\pm} ,\qquad \mathbf{j}^{\pm} = \int d^3 p \ \frac{\mathbf{p}}{\omega} \ f^{\pm}.
\end{align}
We conclude that pair production and annihilation is impossible as the total particle number is independent of
the total number of antiparticles. Due to the fact, that the total particle number is constant the following relation for particles holds
\begin{align}
 \frac{d}{dt} f \br{\mathbf{x}, \mathbf{p},t} = \de{t} f + \br{\de{t} \mathbf{x}} \cdot \boldsymbol{\nabla}_x f + \br{\de{t} \mathbf{p}} \cdot \boldsymbol{\nabla}_p f = 0. \label{vlas_lor}
\end{align}
Comparing the equations \eqref{vlas_lor} and \eqref{vlasov_1} we obtain the relativistic Lorentz force
\begin{equation}
 \frac{d \mathbf{p}}{dt} = \mathbf{E} + \mathbf{v} \times \mathbf{B}.
\end{equation}

\pagestyle{plain}
\addtocontents{toc}{\protect\clearpage}
\chapter{Solution strategies and models for the background field}
\pagestyle{fancy}
In order to solve the transport equations derived in chapter three we have to rely on numerical methods as only a few analytical results are available \cite{Hebenstreit}.
Our strategy is to discretize in the spatial and momentum domain leading to a system of ODEs with 
$N_x \times N_y \times N_z \times N_{p_x} \times N_{p_y} \times N_{p_z}$ components.

The number of possibilities in order to perform simulations involving partial differential equations are quite vast.
In principle one can distinguish three different classes: Finite Difference Methods(FDM), Finite Element Methods(FEM) and
Spectral Methods. Generally speaking, FDM can be seen as ``workhorses''. They are usually easy to code and provide reliable results. However, accuracy could be
a problem especially in case of higher dimensional problems as an error estimate yields a polynomial dependence on the grid spacing.

FEM on the other hand probably provide the most versatile solvers as there are many possibilities of how to optimize the convergence rate for almost arbitrary specifications.
The drawback is, that especially if one wants to benefit from all advantages the method offers, coding is likely to become cumbersome. 

Hence, we will focus on spectral methods(to be more precise pseudo-spectral methods) throughout this thesis. As the domain 
is simple and we do not expect discontinuous behaviour of the Wigner functions, pseudo-spectral methods hold as an ideal tool in order
to study pair production within the DHW approach. This is, because compared with other methods, one needs less grid points in order
to achieve a specific accuracy due to the high convergence rate. This has turned out to be crucial when working with field configurations including electric and magnetic fields.

For the time integration we have implemented a Dormand-Prince Runge-Kutta integrator of order 8(5,3), \cite{Press:2007:NRE:1403886}.
Depending on hardware resources as well as the system of DEs one wants to solve, an adaptive Runge-Kutta integrator of lower order or a multistep method
could be advantageous \cite{Press:2007:NRE:1403886,Hairer1,Hairer2}. As the results did not change when testing different integrators we refrain from a further analysis.

\section{Pseudo-spectral methods}
\label{Kap_PsMethods}

As pseudo-spectral methods, or spectral methods in general, are a very vast and rich topic, we will neither give an entire
description nor discuss their mathematical properties in detail. The interested reader is referred to Boyd\cite{Boyd} for a complete introduction into
this topic including pros and cons of spectral methods. In case one is primarily interested in applications, Trefethen\cite{Trefethen} is a good introductory book. 

In this thesis, however, we will keep the formulation simple and show only elementary context in order to
minimize the effort it takes to understand the technical background. 
The basic idea of spectral approaches is to approximate an unknown function $f \br{x}$ by
$N$ basis functions:
\begin{align}
 F \br{x} \approx F_N \br{x} = \sum_{i=0}^N \alpha_i \ f_i \br{x}.
\end{align}
Applying the operator of the differential equation $\mathcal{L}$ to the approximated function yields
\begin{equation}
 \mathcal R \br{x;\alpha_i} = \mathcal L F_N \br{x} - \mathcal F,
\end{equation}
where $\mathcal F$ is the full solution and $\mathcal R$ the residuum. The main task is then to minimize this residual function $\mathcal R$.
When working with pseudospectral methods one basically uses one high-order polynomial determined by all points in the domain to achieve this goal.  

The interesting aspect of pseudospectral methods is, that the order of the polynomial is not fixed. Hence, a higher number of grid points
increases the resolution leading to a much better accuracy. In Boyd\cite{Boyd} the corresponding truncation error is estimated with
\begin{align}
 \text{error} \approx \mathcal O \br{N^{-N}},
\end{align}
presupposing one is using ``optimal'' basis functions.
As can be seen, it is possible to obtain good results even for a modest number of grid points
making calculations even in a three-dimensional domain possible. 

As written in Boyd\cite{Boyd}, there are many possibilities in order to find appropriate basis functions $f_i \br{x}$ depending on the problem and
also the domain. In the following we will introduce the Fourier basis and discuss the most important implications.

\subsection*{Fourier basis}

We have chosen a Fourier basis mainly due to two reasons. First, a Fourier basis enables the possibility to perform derivatives via Fourier transforms\cite{Stegun}. Hence, we can use one of
the highly optimized libraries available for FFT\cite{FFT,Press:2007:NRE:1403886} and do not have to deal with derivative matrices. The second important reason is the way the grid points are sampled.
Using a Fourier basis the grid points are distributed equidistantly over the domain. This makes a difference in accuracy when compared to e.g. a Chebychev grid, where
the points are sampled closer near the boundary. As a consequence, one obtains also a high resolution in the central region of phase space.

In order to apply a Fourier basis, we have to sample the Wigner functions as well as the fields etc. on a periodic grid. In addition, we have to artificially introduce periodic boundary
conditions. However, this does not alter the problem as we will rewrite Wigner functions, such that they fall off sufficiently fast at the boundaries. One only has to insure, that
the function values at the grid points defined at the boundaries keep being zero throughout the time integration.

In the following we quickly go through the most important points when applying pseudospectral methods on a Fourier basis to a generic problem\cite{Trefethen}. The main idea is to
use the relation
\begin{align}
 \de{x} G \br{x} = \mathcal F_{x}^{-1} \Bigl( \mathcal F_{x} \br{ \de{x} G \br{x}} \Bigr)= \mathcal F_{x}^{-1} \br{\ii w_{x} \hat{G} \br{w_x}}.
\end{align}
Hence, we start by discretizing
the domain $\com{-\pi,\pi}$ such that we obtain $N$ grid points with a spacing of $h = 2 \pi / N$. We will denote the grid points with $x_i$, where $x_0 = x_N$, and
the function value on the grid points as $v_i = v \br{x_i}$. The discrete Fourier transform is then defined as
\begin{align}
 \hat v_k = h \sum_{j=1}^N \ee^{-\ii k x_j} \ v_j, \qquad k = -\frac{N}{2}+1,\ldots, \frac{N}{2}. \label{Fourier_k}
\end{align}
Accordingly, the inverse Fourier transform is defined as
\begin{align}
 v_j = \frac{1}{2 \pi} \sum_{k=-N/2+1}^{N/2} \ee^{\ii k x_j} \ \hat v_k, \qquad j = 1, \ldots,N. 
\end{align}
Note, that $k \in \com{-\pi/h, \pi/h}$, because one cannot distinguish wavenumbers differing by $2 \pi n/h$, with $n \in \mathbbm{N}$.

In order to perform a spectral derivative on $v \br{x}$ on proceeds as follows. At first, $\hat v_k$ is calculated via a Fourier transform of $v_j$.
Secondly, $\hat v_k$ is multiplied by $\br{\ii k}^n$, where $n$ denotes the $n$-th derivative. If one is interested in a differentiation of odd order
one has to set $\hat v_{N/2}$ to zero. Eventually, the derivative $V_j$ is obtained performing an inverse Fourier transform on $\br{\ii k}^n \hat v_k$.

\section{Pseudo-differential operators}
In order to solve any of the transport equations derived in chapter three we have to deal with the pseudo-differential operators. There are multiple ways
of how to treat these operators properly. In the following we are going to introduce two different schemes. At first, we describe an approximation working
for slowly varying fields. The second approach we are discussing is more sophisticated including the possibility to work also with background fields 
changing arbitrarily in space.

\subsection{Operator expansion} 
\label{Sol_Exp}
Whenever the background fields are only slowly varying in space, a Taylor expansion and subsequent truncation will be efficient\cite{Vasak1987462,PhysRevD.44.1825,Hebenstreit}. 
If truncated, e.g. at first order, the non-local operators
  \begin{alignat}{6}
     & D_t && = \de{t} &&+ e &&\int d\xi &&\mathbf{E} \br{\mathbf{x}+\ii \xi \boldsymbol{\nabla}_p,t} && ~ \cdot \boldsymbol{\nabla}_p,  \\
     & \mathbf{D} && = \boldsymbol{\nabla}_x &&+ e &&\int d \xi &&\mathbf{B} \br{\mathbf{x}+\ii \xi \boldsymbol{\nabla}_p,t} &&\times \boldsymbol{\nabla}_p,  \\
     & \boldsymbol{\Pi} && = \mathbf{p} &&- \ii e &&\int d \xi \xi &&\mathbf{B} \br{\mathbf{x}+\ii \xi \boldsymbol{\nabla}_p,t} &&\times \boldsymbol{\nabla}_p.
  \end{alignat}  
are transformed into the local operators
  \begin{alignat}{6}
     & D_t && = \de{t} &&+ e && \mathbf{E} \br{\mathbf{x},t} && \cdot \boldsymbol{\nabla}_p,  \\
     & \mathbf{D} && = \boldsymbol{\nabla}_x &&+ e && \mathbf{B} \br{\mathbf{x},t} &&\times \boldsymbol{\nabla}_p,
     \end{alignat}
     and
     \begin{alignat}{6}
     & \boldsymbol{\Pi} && = \mathbf{p} &&+ \frac{e}{12} \br{\tilde{\boldsymbol{\nabla}}_x \cdot \boldsymbol{\nabla}_p} &&\mathbf{B} \br{\mathbf{x},t} &&\times \boldsymbol{\nabla}_p,
  \end{alignat}  
where $\tilde{\boldsymbol{\nabla}}_x$ only works on $\mathbf{B} \br{\mathbf{x},t}$, resulting in an easier numerical treatment. In case the magnetic field is applied perpendicular to the electric field studying its influence on the pair production process results in a at least three-dimensional domain in phase space. If the vector potential obeys the form $\mathbf{A} = A \br{z,t} \mathbf{e}_x$ the non-trivial phase-space parameters are
$x ,~ p_x$ and $p_z$. Due to the fact, that RAM is limited, there are restrictions on the number of grid points we can use. Therefore, we have to
find a reasonable trade-off for the grid size per direction. 

The Taylor expanded differential operators are local in phase-space making it easier to define transformations in order to enhance numerical stability. The system \eqref{eq_2b_1}-\eqref{eq_2b_4}
\begin{alignat}{6}
  & D_t \overline{\mathbbm{s}}   &&  && && -2 \Pi_1 \overline{\mathbbm{v}}_3 &&+ 2 \Pi_3 \overline{\mathbbm v}_1 &&= 0, \\
  & D_t \overline{\mathbbm{v}}_1 &&+D_1 \overline{\mathbbm{v}}_0 && && &&- 2 \Pi_3 \overline{ \mathbbm{s}} &&= -2\overline{\mathbbm{v}}_3,  \\    
  & D_t \overline{\mathbbm{v}}_3 && &&+D_3 \overline{\mathbbm{v}}_0 && +2 \Pi_1 \overline{\mathbbm{s}} && &&= 2 \overline{\mathbbm{v}}_1,  \\ 
  & D_t \overline{\mathbbm{v}}_0 &&+D_1 \overline{\mathbbm{v}}_1 &&+D_3 \overline{\mathbbm{v}}_3 && && &&= 0, 
\end{alignat}
with the differential operators
\begin{alignat}{6}
     & D_t && = \de{t} &&+ e && E \br{z,t} && \de{p_x},  \\
     & D_1 && =  &&+ e && B \br{z,t} && \de{p_z}, \\
     & D_3 && = \de{z} &&- e && B \br{z,t} && \de{p_x}, \\
     & \Pi_1 && = p_x &&+ \frac{e}{12} && \br{\de{z} B \br{z,t}} && \de{p_z} \de{p_z}, \\
     & \Pi_3 && = p_z &&- \frac{e}{12} && \br{\de{z} B \br{z,t}} && \de{p_z} \de{p_x}.
\end{alignat}
will hold for demonstration purpose. At first we will switch from kinetic momentum to canonical momentum defining
\begin{equation}
 \mathbf{q} = \mathbf{p} + e \mathbf{A} \br{z,t}.
\end{equation}
The derivative operators transform accordingly
\begin{align}
 &\de{p_x} = \de{q_x},\ \ \de{p_z} = \de{q_z}, \ \ \de{t} = \de{\tilde{t}} - e E \br{\tilde z, \tilde t} \de{q_x},\ \ \de{z} = \de{\tilde{z}} + e B \br{\tilde z, \tilde t} \de{q_x}
\end{align}
resulting in the system of DEs
\begin{align}
  & \de{t} \overline{\mathbbm{s}}   &&  && && -2 \Pi_1 \overline{\mathbbm{v}}_3 &&+ 2 \Pi_3 \overline{\mathbbm v}_1 &&= 0, \label{sol_1_1} \\
  & \de{t} \overline{\mathbbm{v}}_1 &&+e B \br{z,t} \de{q_z} \overline{\mathbbm{v}}_0 && && &&- 2 \Pi_3 \overline{ \mathbbm{s}} &&= -2\overline{\mathbbm{v}}_3,  \\    
  & \de{t} \overline{\mathbbm{v}}_3 && &&+\de{z} \overline{\mathbbm{v}}_0 && +2 \Pi_1 \overline{\mathbbm{s}} && &&= 2 \overline{\mathbbm{v}}_1,  \\ 
  & \de{t} \overline{\mathbbm{v}}_0 &&+e B \br{z,t} \de{q_z} \overline{\mathbbm{v}}_1 &&+\de{z} \overline{\mathbbm{v}}_3 && && &&= 0, \label{sol_4_1}
\end{align}
where
\begin{alignat}{6}
     & \Pi_1 && = q_x - \frac{e}{m} A \br{z,t} &&+ \frac{e}{12} && \bigg( \de{z} B \br{z,t} \bigg) && \de{q_z} \de{q_z}, \\
     & \Pi_3 && = q_z &&- \frac{e}{12} && \bigg( \de{z} B \br{z,t} \bigg) && \de{q_z} \de{q_x}.     
\end{alignat}

Additionally, we introduce modified Wigner components
\begin{equation}
 \overline{\mathbbm{w}}_k^{v} \br{z, q_x, q_z} = \overline{\mathbbm{w}}_k \br{z, q_x, q_z} - \overline{\mathbbm{w}}_i \br{q_x, q_z},
\end{equation}
where $\mathbbm{w}_k \br{z, q_x, q_z}$ are the equal-time Wigner components and $\mathbbm{w}_i \br{z, q_x, q_z}$ the corresponding initial conditions.
Implementing such a transformation yields for the vacuum state $\mathbbm{w}_k^{v} \br{z, q_x, q_z} = 0$. Moreover, all Wigner 
functions become zero in the limit of taking $z ,~ q_x$ or $q_z$ to infinity. The drawback of using modified Wigner functions is, that one automatically
introduces inhomogeneous terms in the system of DEs \eqref{sol_1_1} - \eqref{sol_4_1}.
Calculating these terms, however, is straightforward within the operator expansion approach. Due to the Taylor expansion of the operators the inhomogeneities are reduced to
\begin{alignat}{5}
 &\de{t} \overline{\mathbbm{s}}_i &&= \de{t} \br{-\frac{2}{\omega}} &&= && \frac{2 e E \br{z,t} \bigl( q_x - e A \br{z,t} \bigr) }{\omega^3},\\
 &\de{t} \overline{\mathbbm{v}}_{1i} &&= \de{t} \br{-\frac{2 \br{q_x - e A \br{z,t}}}{\omega}} &&= -&&\frac{2 e E \br{z,t} \br{1+q_z^2}}{\omega^3},\\
 &\de{t} \overline{\mathbbm{v}}_{3i} &&= \de{t} \br{-\frac{2 q_z}{\omega}} &&= &&\frac{2 e E \br{z,t} q_z} {\omega^3},
\end{alignat}
\begin{alignat}{5}
 &2 \Pi_3 \overline{\mathbbm{v}}_{1i} - 2 \Pi_1 \mathbbm{v}_{3i} &&= &&\frac{q_z \ \br{\de{z} e B \br{z,t}} \ \br{3 + \omega^2}}{3 \omega^5}, \\
 &2 \overline{\mathbbm{v}}_{3i} -2 \Pi_3 \mathbbm{s}_i &&= -&&\frac{\br{\de{z} e B \br{z,t}} \bigl( q_x - e A \br{z,t} \bigr) q_z}{\omega^5} ,\\
 &2 \overline{\mathbbm{v}}_{1i} -2 \Pi_1 \mathbbm{s}_i &&= -&&\frac{\br{\de{z} e B \br{z,t}} \omega^2 -3 q_z^2}{3\omega^5},
\end{alignat}
with $\omega = \sqrt{1+\br{q_x - e A \br{z,t}}^2 + q_z^2}$. We have just shown the implications of truncation at first order. However, it could prove useful to take also the next-to-leading order into account. The calculations, that will be presented in chapter seven, could be improved in this way.

At last, we discuss possible further transformations of
the phase-space parameter set $\{z ,\ q_x ,\ q_z \}$ to a parameter set $\{\tilde z ,\ \tilde{q}_x ,\ \tilde{q}_z \}$:
\begin{alignat}{5}
 &\tilde z &&= a_1 \ \text{atanh} \br{z} + b_1 ,\\ 
   & && \hspace{5cm} z \in [-L_1, L_1] \to \tilde z \in [-\tilde L_1, \tilde L_1] ,\\
 &\tilde q_x &&= a_2 \ \text{atanh} \br{q_x} + b_2 ,\\
   & && \hspace{5cm} q_x \in [-L_2, L_2] \to \tilde q_x \in [-\tilde L_2, \tilde L_2] ,\\
 &\tilde q_z &&= \frac{2 L_3}{\pi} \text{atan} \br{\frac{1}{a_3} \tan \br{\frac{\pi q_z}{L_3}}} + b_3,\\ 
   & && \hspace{5cm} q_z \in [-L_3, L_3] \to \tilde q_z \in [-\tilde L_3, \tilde L_3],
\end{alignat}
where $a_i > 0$ and $b_i \in \mathbbm{R}$ for $i = 1 ,~ 2 , ~3$. 
The derivative operators transform accordingly
\begin{alignat}{6}
 &\de{\tilde{z}} &&= \frac{1}{a_1} \br{1 - \tilde{z}^2},\\
 &\de{\tilde{q}_x} &&= \frac{1}{a_2} \br{1 - \tilde{q}_x^2},\\
 &\de{\tilde{q}_z} &&= \frac{1}{2 a_3} \br{1 +a_3^2 + \br{a_3^2 - 1} \cos \br{\frac{\pi}{L} \tilde q_z } }.
\end{alignat}
The main advantage of introducing such a transformation lies in the possibility of covering a larger phase-space volume without increasing the number of
grid points. In case the parameters $a_i$ and $b_i$ are chosen properly one can even maintain the resolution at important parts in phase-space as the
``stretching'' mainly applies to the points closer to the boundary of the domain.

\subsection{Full solution}
For special field configurations, especially when they effectively result in a lower dimensional formalism, it is possible to treat the pseudo-differential operators exactly\cite{Hebenstreit}.
In case of a vector potential of the form $\mathbf{A} = A \br{x,t} \mathbf{e}_x$ we can work in a two-dimensional domain, because the only remaining derivatives are $\de{x}$ and $\de{p_x}$, see \eqref{eq3_Cy1}-\eqref{eq3_Cy4}. Presuppose one has access to
$\sim 100$ $MB$ RAM, one does not face any problems regarding memory. Hence, the grid size is no limiting factor in these calculations 
resulting in the possibility to choose a suitable domain without worrying about the resolution.

When performing calculations with a vector potential of the form $\mathbf{A} = A \br{x,t} \mathbf{e}_x$ only one pseudo-differential operator maintains its non-locality:
\begin{equation}
 D_t = \de{t} + e \int d\xi E \br{x+\ii \xi \de{p_x},t} \de{p_x} = \de{t} + \Delta.
\end{equation}
As all quantities are defined in terms of a Fourier basis we obtain
\begin{align}
 \Delta \ \mathbbm{w}_k \br{x, p_x, t} = \mathcal F_{p_x}^{-1} \br{ \mathcal F_{p_x} \br{e \int d\xi E \br{x+\ii \xi \de{p_x},t} \de{p_x} \mathbbm{w}_k \br{x, p_x, t}} }.
\end{align}
Similarly to the case described in section \ref{Sol_Exp} we Taylor expand the electric field, which yields
\begin{align}
 \Delta \ \mathbbm{w}_k \br{x, p_x, t} = \mathcal F_{p_x}^{-1} \br{ \mathcal F_{p_x} \br{e \int d\xi \sum_{n=0}^{\infty} \frac{1}{n!} E^{(n)} \br{x,t} \br{\ii \xi}^n \de{p_x}^{n+1} \mathbbm{w}_k \br{x, p_x, t}} }.
\end{align}
When taking the Fourier transform of the inner term, the derivatives with respect to $p_x$ are transformed into factors giving
\begin{align}
 \Delta \ \mathbbm{w}_k \br{x, p_x, t} = \mathcal F_{p_x}^{-1} \br{e \int d\xi \sum_{n=0}^{\infty} \frac{1}{n!} E^{(n)} \br{x,t} \br{\ii \xi}^n \br{\ii w_{p_x}}^{n+1} \hat{\mathbbm{w}}_k \br{x, w_{p_x}, t} }.
\end{align}
Collecting all terms in the sum we are left with a new expression for the derivative operator
\begin{equation}
 \Delta \ \mathbbm{w}_k \br{x, p_x, t} = \mathcal F_{p_x}^{-1} \br{\ii e w_{p_x} \int d\xi E \br{x-\xi w_{p_x},t} \hat{\mathbbm{w}}_k \br{x, w_{p_x}, t}}. \label{FF_1}
\end{equation}
Up to now, we have not posed any restrictions on the spatial dependence of the electric field. We assume, that all Wigner functions vanish for $x \to \pm \infty$. More precisely, restrict ourselves to field configurations that
fall off sufficiently fast in $x$. Additionally, we prefer to have an analytic expression for the integral given in \eqref{FF_1}. One possibility is to use a field of the form
\begin{equation}
 E \br{x,t} = E_0 \br{t} \exp \br{-\frac{x^2}{2 \lambda^2}}. 
\end{equation}
Performing the parameter integral in \eqref{FF_1} yields 
\begin{align}
 \int_{-1/2}^{1/2} d\xi \exp \br{-\frac{\br{x- \xi w_{p_x}}^2}{2 \lambda^2}} 
 = \sqrt{\frac{\pi}{2}} \frac{\lambda}{w_{p_x}} \br{\text{erf} \br{\frac{w_{p_x}+2x}{\sqrt{8} \lambda}}
 +\text{erf} \br{\frac{w_{p_x}-2x}{\sqrt{8} \lambda}} },
\end{align}
with $\text{erf} \br{x}$ being the error-function.
Defining 
\begin{equation}
 \mathcal S \br{x, w_{p_x}, t} = \sqrt{\frac{\pi}{2}} E_0 \brt \lambda \br{\text{erf} \br{\frac{w_{p_x}+2x}{\sqrt{8} \lambda}}
	      +\text{erf} \br{\frac{w_{p_x}-2x}{\sqrt{8} \lambda}} }
\end{equation}
equation \eqref{FF_1} takes the form
\begin{equation}
 \Delta \ \mathbbm{w}_k \br{x, p_x, t} = \mathcal F_{p_x}^{-1} \Bigl( \ii e \ \mathcal{S} \br{x, w_{p_x}, t} \ \hat{\mathbbm{w}}_k \br{x, w_{p_x}, t} \Bigr). 
\end{equation}
Then, we introduce modified Wigner functions in the same way as in reference\cite{Hebenstreit} leading to
\begin{equation}
 \mathbbm{w}_k^{v} \br{x, p_x, t} = \mathbbm{w}_k \br{x, p_x, t} - \mathbbm{w}_i \br{p_x}.
\end{equation}
Thereby, we obtain an inhomogeneous system of DEs.  
Applying the differential operator to $\mathbbm{s}_i \br{p_x}$ yields
\begin{equation}
D_t \mathbbm{s}_i \br{p_x} = -\mathcal F_{p_x}^{-1} \br{\ii e \sqrt{\frac{2}{\pi}} \ \mathcal S \br{x, w_{p_x}, -\infty} \ K_0 \br{\vert w_{p_x} \vert} }, 
\end{equation}
where $K_0 \br{\vert w_{p_x} \vert}$ is a modified Bessel function of the second kind.
When doing the same for $\mathbbm{v}_i \br{p_x}$ we are confronted with the fact, that 
\begin{align}
\mathcal F_{p_x} \mathbbm{v}_i \br{p_x} = \mathcal F_{p_x} \br{-\frac{p_x}{\sqrt{1 + p_x^2}}}
\end{align}
does only exist in terms of $\de{w_{p_x}} \text{sign} \br{w_{p_x}}$. Hence, we use the trick of performing $\de{p_x} \mathbbm{v}_i \br{p_x}$ first and 
only afterwards we apply $\mathcal F_{p_x}^{-1} \mathcal F_{p_x}$ to the inhomogeneous term. This yields
\begin{equation}
D_t \mathbbm{v}_i \br{p_x} = \mathcal F_{p_x}^{-1} \br{ e \sqrt{\frac{2}{\pi}} \ \text{sign} \br{w_{p_x}} \ \mathcal S \br{x,p_x,-\infty} \ K_1 \br{\vert w_{p_x} \vert } }, 
\end{equation}
where $K_1 \br{\vert w_{p_x} \vert}$ is again a modified Bessel function of the second kind.
N.B.: The calculation for $p_{\rho} \neq 0$, see \eqref{eq3_Cy1}-\eqref{eq3_Cy4}, is straightforward leading to $\overline{\mathbbm{p}}_i = p_{\rho} \ \overline{\mathbbm{s}}_i$.
  
\subsection{Filtering}
When discussing solution strategies for the DHW equations we also have to mention possible problems of spectral methods. Aliasing \cite{Boyd} is a prominent example. 
Aliasing basically means, that when working on a grid only
a finite number of wavenumbers are distinguishable. Frequencies outside the range $k \in \com{-\pi/h, \pi/h}$, see section \ref{Fourier_k},
appear as lower wavenumbers. This could then lead to instabilities during a calculation. In order to get rid of this unwanted effect so-called
dealiasing methods have been invented. For an overview on this subject the interested reader is referred to Boyd\cite{Boyd}.
  
One possibility in order to fix aliasing-related problems is to introduce a spectral filter. This means, that before applying the inverse Fourier transform
one artificially sets the highest spectral coefficients to zero thus filtering all higher wavenumbers. In chapter seven we have filtered one-third of the wavenumbers for the simulations regarding a nonzero magnetic field.
   
\section{Model for the background field}
\label{Kap_Field}
In chapter three we have derived the DHW equations and in section \ref{Kap_PsMethods} we have discussed their numerical treatment. The electric and magnetic field, however, have been seen as general input. In the following sections we will relate these fields to real experimental specifications. Hence, we will introduce various models for the time- and spatial dependence.

\subsection{Parameter scales}

In order to accurately describe pair production we begin identifying the relevant scales of the problem and thus discussing
the dependence of our fields on the various parameters. At this point the relation to atomic ionization processes helps in order
to come up with a numerically feasible model still capable describing all key aspects of the underlying physical process.

When modeling fields in atomic ionization processes one relies upon the so-called dipole approximation ignoring all spatial dependencies
of the produced Laser pulse. An analysis of the relevant scales involved in such a process explains the applicability of this approximation.
A Laser in a typical ionization experiment operates with an intensity of $I \sim 10^{14}$ $W/cm^2$ and a pulse length of $T_{pulse} \sim 100$ $fs$.
When using an optical Laser the photon energy is $\omega \sim 1$ $eV$ or in terms of wavelength $\lambda \sim 1000$ $nm$. The radius of an atom, however,
is $r_{Atom} \sim 10^{-10}$ $m$. Comparing the two scales one immediately finds out, that the Laser intensity basically does not change over
the size of an atom. A short check assuming a field propagating along $z$ with $|\mathbf{k}| = \omega$ yields
\begin{align}
 \exp \br{\ii \omega z} \approx \exp \br{\ii \ 10^{-10} \ 10^{6}} \approx 1 + \ii \ 10^{-4} \approx 1.
\end{align}
Neglecting the spatial dependence of the electric field is therefore a valid approximation. This implies also, that the model for
the vector potential is only time-dependent resulting in a vanishing magnetic field. 

We proceed by analyzing the relevant scales for pair production in a lab based experiment. High-power Lasers
are planned to achieve intensities of $I \sim 10^{26}$ $W/cm^2$ as well as photon energies of $\omega \sim 10$ $keV$. On the other hand,
it is possible to scatter a Laser pulse with a high-energy electron beam ($\omega_e \sim 10$ $GeV$). Hence, one can assume, that
the photon energies in a general experiment could exceed $10-100$ $keV$. We go on comparing these energies, which are especially important in order to
produce particles in the multiphoton regime, with the Compton wavelength of an electron, which is 
$\bar \lambda_C = 1 \ m^{-1} \approx 2 \cdot 10^{-6}$ $eV^{-1}$. Hence, in the multiphoton regime the spatial extent of the applied electric field
becomes relevant. In case of pure Schwinger pair production the photon energies are much lower and
the assumption of a purely time-dependent electric field should be valid.

\subsection{Model for the time-dependence}
The most important factor in this work is the time-dependence of the background fields. Therefore, there is a special emphasis finding
a suitable model in order to accurately describe the time-dependence of a Laser pulse. We will introduce the electric field as a function of frequency\cite{Moller} and subsequently derive the time-dependence via a Fourier transform. An electric pulse with a phase expanded up to second order takes the form
\begin{equation}
 \tilde E \br{\omega} = \varepsilon \ E_0 \ \tau \ \exp \br{-\frac{(\omega-\omega_0)^2 \tau^2}{2}} \exp \br{-\ii \br{ a+b \br{\omega-\omega_0} + c \br{\omega-\omega_0}^2 }}.
\end{equation}
We proceed by taking the Fourier transform of the electric field yielding
\begin{equation}
 E \br{t} =\frac{\varepsilon \ E_0 \ \tau}{\sqrt{\tau^2+2 \ii c}} \exp \br{-\ii \br{a-\omega_0 t} - \frac{ \br{b-t}^2}{2 \br{\tau^2+2 \ii c}}}.
\end{equation}
For the sake of convenience, we may express the electric field in real quantities, which leads to
\begin{equation}
E \brt = \frac{\varepsilon \ E_0 \ \tau}{\sqrt[4]{\tau^4+4 c^2}} \exp \br{-\frac{ \br{b-t}^2 \tau^2}{2 \br{\tau^4+4 c^2}}} 
  \cos \br{\omega_0 t - a -\frac{1}{2} \textrm{arctan} \br{\frac{2c}{\tau^2}} + \frac{c (b-t)^2}{4 c^2+\tau ^4}}.
\end{equation}
At this point, we may associate the variables $a ,\ b ,\ c$ and $\omega_0$ with physical quantities. The parameter $a$ directly controls the
phase of the subcycling field, while changing $b$ yields an offset in time. 
The parameter $c$ determines whether one has a so-called chirped pulse. This means that a value of $c \neq 0$ leads to a change in field frequency over one pulse period. Additionally, a non-vanishing $c$ leads to a stretching of the whole Laser pulse.
The parameter $\omega_0$ fixes the frequency of the pulsed field. In addition, when switching off the chirp, this parameter
can be associated with the energy of the photons. 
In the following sections, we will set $b = c = 0$ and obtain a field of the form
\begin{equation}
 E \brt = \varepsilon \ E_0 \exp \br{-\frac{t^2}{2 \tau^2}} \cos \br{\omega_0 t - a}.
\end{equation}

In addition, for specific calculations we have also replaced the Gaussian envelope function above by the numerically favorable function
\begin{equation}
 f \br{\frac{t}{\tau}} = \cos \br{\frac{t}{\tau}}^4.
\end{equation}

As it is shown above, we will use fields that fall off sufficiently fast for $t \to \pm \infty$. However,
one should remark, that analytical solutions for the Dirac equation in case of a constant vector potential and also a constant magnetic field exist.
A constant term in the vector potential leaves the distribution function $N \br{\mathbf{x},\mathbf{p}}$ unchanged. Applying a constant magnetic field, however, leads to the occurrence of Landau levels.
The interested reader is referred to \cite{Tarakanov,Hakim}.


\subsection{Spatially inhomogeneous background fields}
\label{Spat_Inh}
The simplest way of adding a spatial dependence to the electric field is to define a vector potential, which is directed along $\mathbf{e}_x$
and inhomogeneous in direction of $x$ \cite{Hebenstreit}. We clearly do not have wave propagation in this case. However, such a configuration
can be seen as an oversimplified model in order to account for focusing of a Laser beam. Its main advantage is, that we only have to deal with an electric field, because $\mathbf{B} = \boldsymbol{\nabla} \times \mathbf{A} \br{x,t} = 0$. 

In order to model the spatial inhomogeneity, we have chosen a model of the form \cite{Hebenstreit}:
\begin{align}
 g \br{\frac{x}{\lambda}} = \exp \br{-\frac{x^2}{2 \lambda^2}}.
\end{align}
Another simple inhomogeneity term can be modeled using a Lorentzian shaped function.
Such a focus can be achieved when a Gaussian beam, see section \ref{Kap_Exp}, is applied
\begin{equation}
 E \br{\mathbf{x}} = \varepsilon \ E_0 \frac{\omega_0}{\omega \br{x}} \exp \br{-\frac{y^2+z^2}{\omega \br{x}^2} - \ii k x - \ii k \frac{y^2+z^2}{2 R \br{x}} +\ii \zeta \br{x}}.
\end{equation}
In such a case, $R \br{x}$ is the radius of curvature and $\zeta \br{x}$ is the Gouy phase. In such a specific case the envelope function takes the form
\begin{align}
 g \br{\frac{x}{\lambda}} = \frac{\omega_0}{\omega \br{x}} = \frac{1}{\sqrt{1+ \br{\frac{x}{\lambda}}^2}}. 
\end{align}

\subsection{Model for the magnetic field}
Instead of simply proposing a vector potential, we want to motivate our choice for
a configuration where the magnetic field does not vanish.
In the simple case of two plane waves, polarized along $\mathbf{e}_z$ and propagating in $\pm x$-direction
\begin{alignat}{3}
 &\mathbf{E}_{\pm} = &&\cos \left( \omega \left( t \pm x \right) \right) \mathbf{e}_z ,\\ 
 &\mathbf{B}_{\pm} = \pm &&\cos \left( \omega \left( t \pm x \right) \right) \mathbf{e}_y,
\end{alignat}
we obtain a standing wave reading
\begin{alignat}{5}
 &\mathbf{E} = \mathbf{E}_{+} + \mathbf{E}_{-} &&= 2 \cos \left( \omega t \right) \cos \left( \omega x \right) \mathbf{e}_z \label{Comp_E},\\
 &\mathbf{B} = \mathbf{B}_{+} + \mathbf{B}_{-} &&= 2 \sin \left( \omega t \right) \sin \left( \omega x \right) \mathbf{e}_y. \label{Comp_B}     
\end{alignat}
Additionally, we have to take into account the finite size/length of the Laser pulse. Hence, we may write down
a vector potential of the form
\begin{equation}
 \mathbf{A} \br{x,t} = f \br{\frac{t}{\tau}} \ \sin \br{\omega t} \ g \br{\frac{x}{\lambda}} \mathbf{e}_z, 
\end{equation}
where $f \br{t/ \tau}$ and $g \br{x/ \lambda}$ define the temporal and spatial envelope of the field, respectively.  
Compared to \eqref{Comp_E}-\eqref{Comp_B}, the temporal dependency approximately fulfills the requirements of being a periodic field only if $\omega \tau \gg 1$. Regarding the spatial dependence, we focus on a single spatial peak of the vector potential. By this way, we can investigate the impact of magnetic fields in a simple way. Without any subcycle structure only the envelope function $g \br{x/ \lambda}$ determines the spatial pulse profile.

In chapter seven we discuss the results for a non-vanishing magnetic field. However, note that we had to perform computations on short time-scales due to numerical reasons.

\section{Final momentum distribution}

After finally determining the background fields we can investigate at which point in time pair production terminates. In other words, we want to know when the matter creation process reaches quasi-equilibrium. In our case, quasi-equilibrium means that the conversion
from energy to matter and vice-versa is completed and thus the particle momentum spectra converged. This is best shown mathematically by stating the
rate of change in the particle distribution. Hence, we compute the derivative of the particle rate with respect to time for vanishing background fields yielding
\begin{align}
  \mathcal{\dot N} \br{\mathbf{p}} = \int d^3x \ \frac{\mathbbm{\dot s} \brw + \mathbf{p} \cdot \mathbbm{\dot v} \brw}{\omega}. \label{equi_N}
\end{align}
From the equations \eqref{eq_3_1}-\eqref{eq_3_8} we obtain relations for a vanishing background field
\begin{alignat}{6}
 &\de{t} \mathbbm{s}     && && -2 \mathbf{p} \cdot \mathbbm{t_1} &&= 0 ,\\
 &\de{t} \mathbbm{v} &&+ \boldsymbol{\nabla}_x \cdot \mathbbm{v}_0 && +2 \mathbf{p} \times \mathbbm{a} &&= -2\mathbbm{t_1}. 
\end{alignat}
Plugging these relations into \eqref{equi_N} leads to
\begin{align}
  \mathcal{\dot N} \br{\mathbf{p}} = -\frac{1}{\omega} \int d^3x \ \br{\mathbf{p} \cdot \boldsymbol{\nabla}_x} \mathbbm{v}_0 \brw. 
\end{align}
Due to the fact, that all Wigner functions fall off for $\mathbf{x} \to \pm \infty$ the expression above vanishes. Hence, as soon as the background fields vanish
the particle momentum distribution becomes constant. 

\begin{figure}[tbh]
{\includegraphics[width=.495\textwidth]{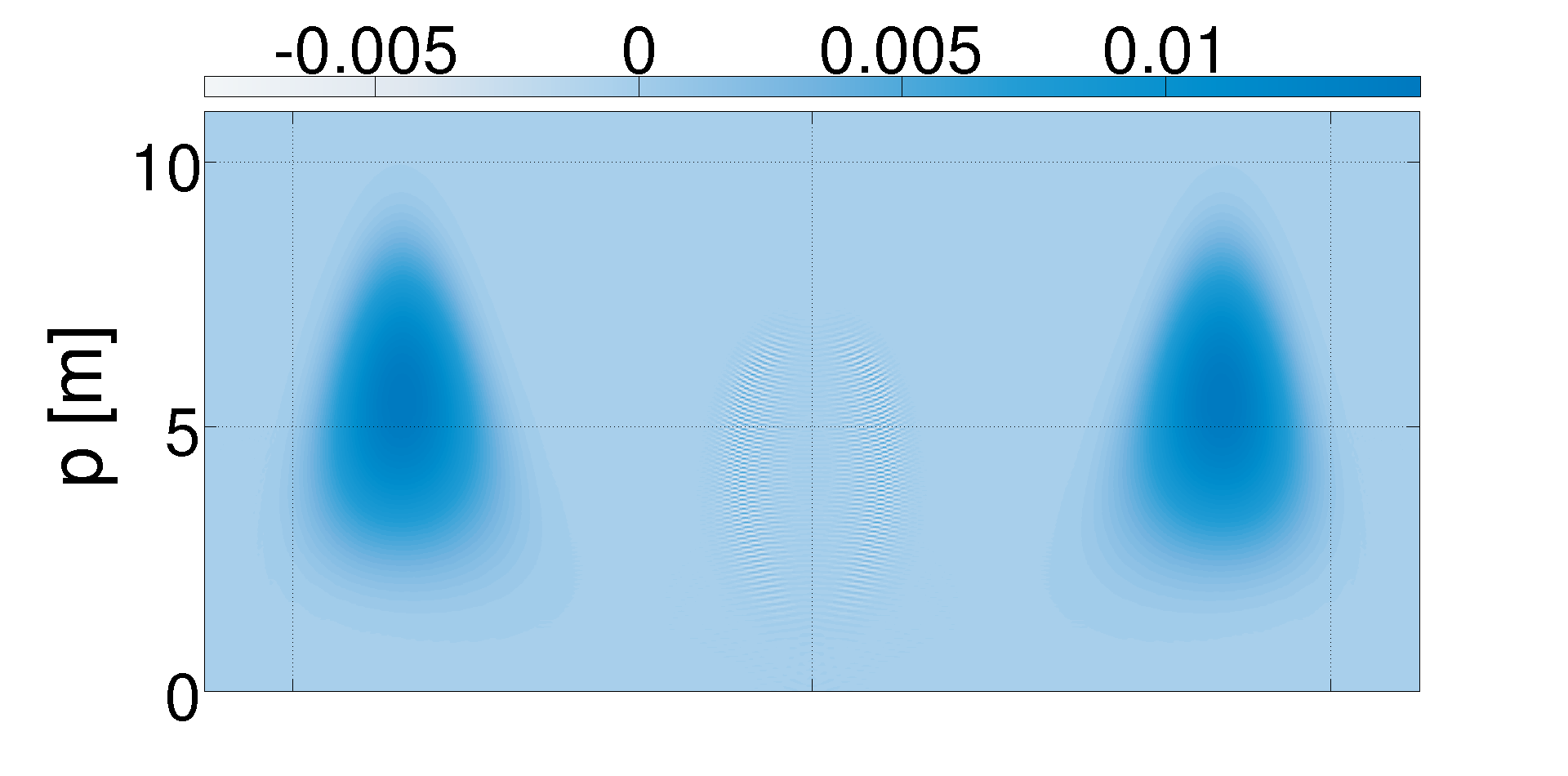}}   
{\includegraphics[width=.495\textwidth]{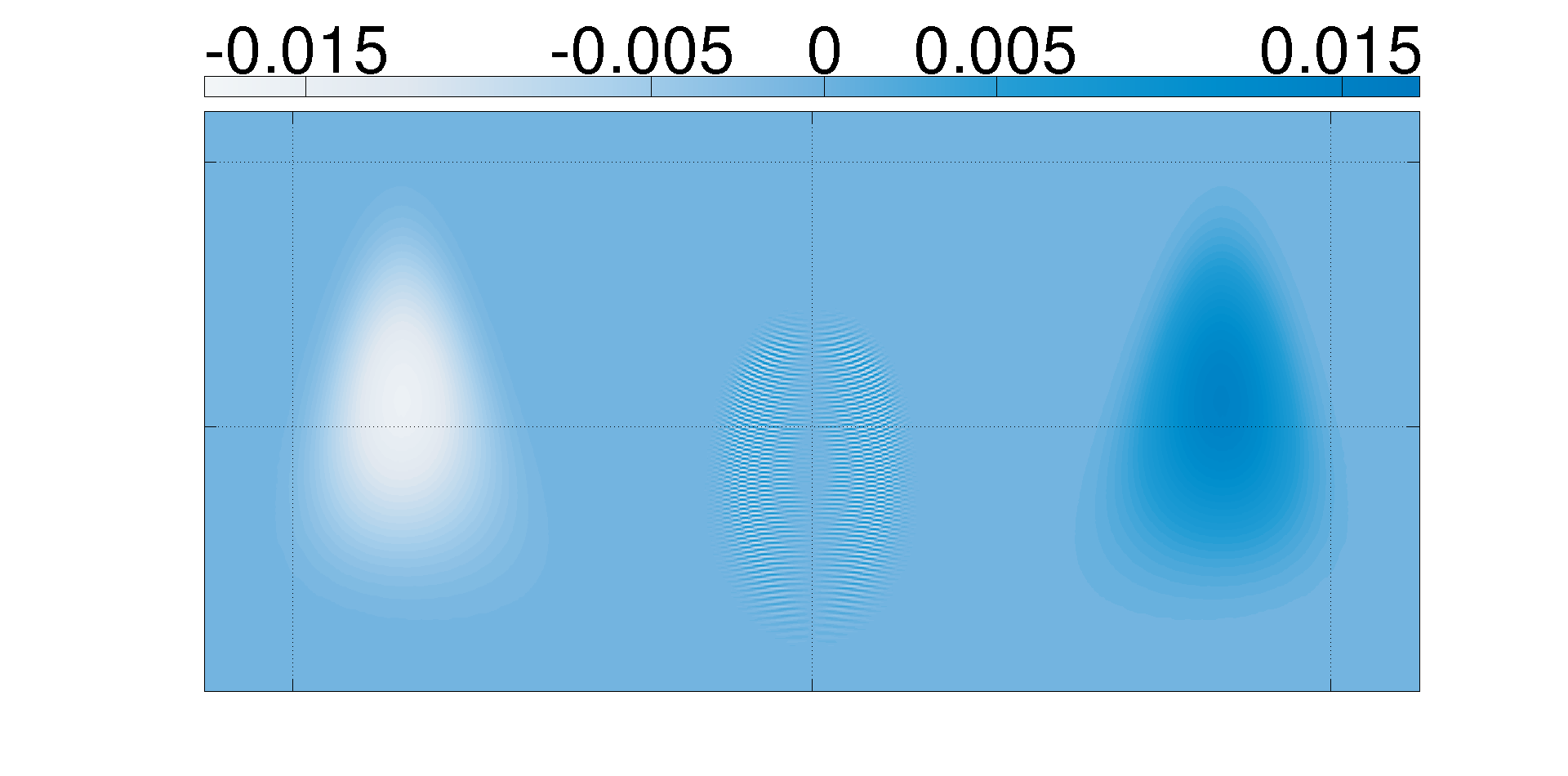}}   
{\includegraphics[width=.495\textwidth]{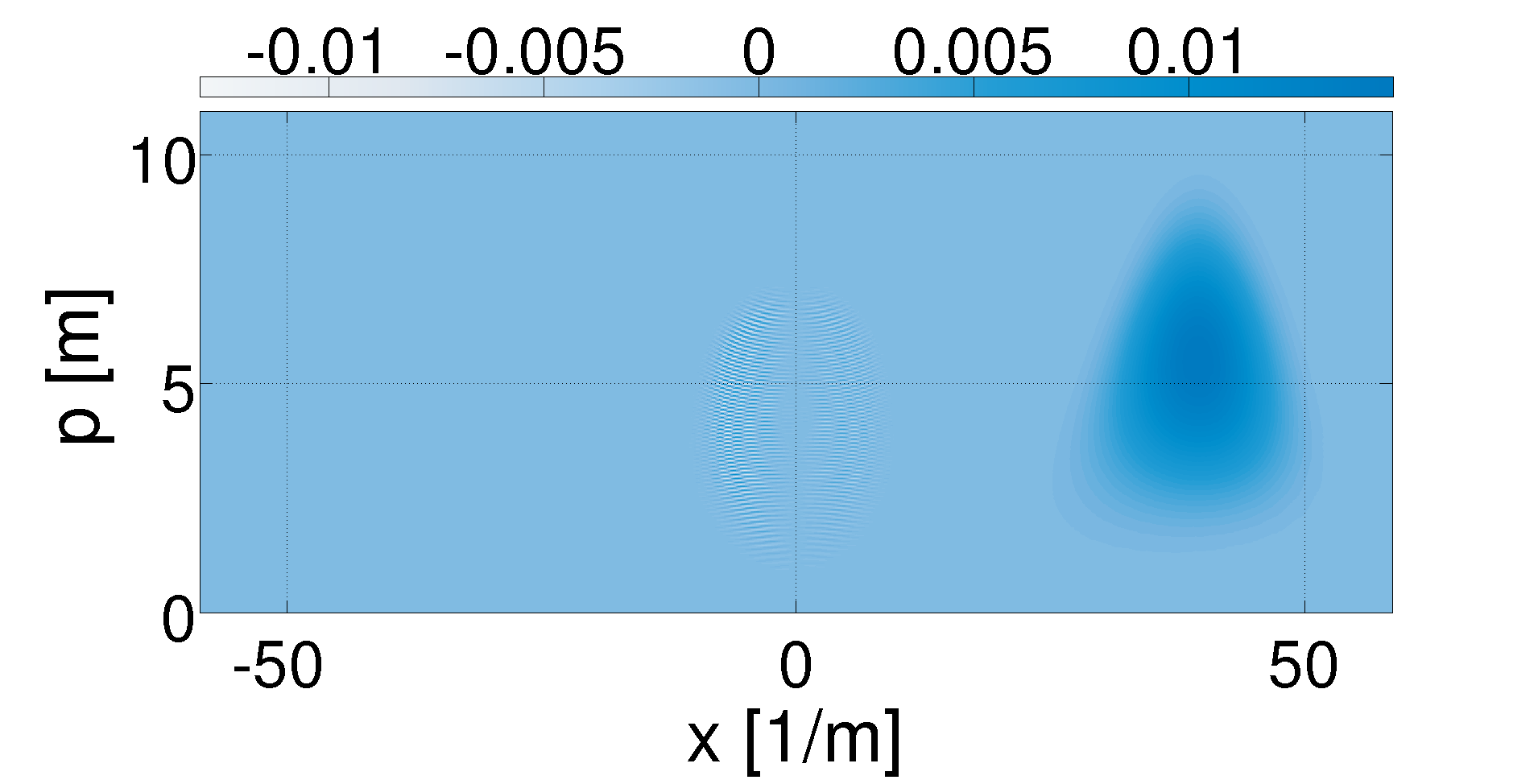}}  
{\includegraphics[width=.495\textwidth]{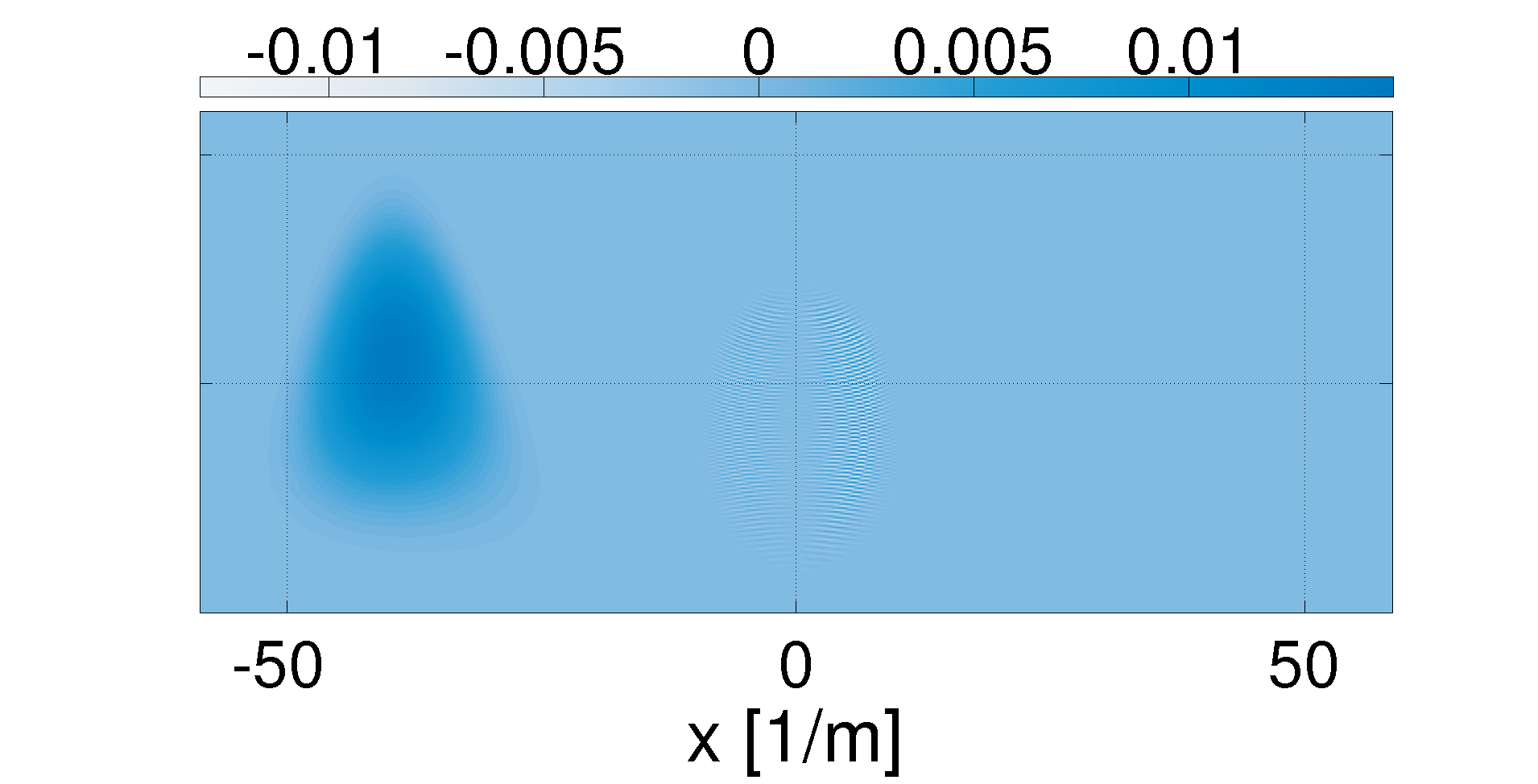}}      
\caption[Demonstration of particle-, charge-, positron- and electron distribution.]{Comparison of the particle distribution(upper left corner), charge distribution(upper right corner), positron distribution(lower left corner) and electron distribution(lower right corner) at final time $t_f$. Due to the form of the applied electric field a clear particle/antiparticle separation is visible. Parameters are taken from: \ref{TabApp_Semi0}.}
\label{Fig_FinalMom}
\end{figure}

The quasi-probability distribution function for quasi-equilibrium at times $t_f$ for an electric field exhibiting a single peak in the $xt$-plane is illustrated in Fig. \ref{Fig_FinalMom}. At $t_f$ the background field already vanished, thus the distribution function $N \br{p}$ has converged. However, the particles have obtained a non-vanishing momentum therefore they constantly move in $x$-space. Due to the fact, that electrons and positrons have opposite
charge they are accelerated in opposite directions, thus the charge separation. At this point, we want to highlight that the distribution function is not non-negative in phase space signaling quantum coherence. A fact, that is best seen in Fig. \ref{Fig_FinalMom} around the origin. Note that coarse graining \cite{PhysRevA.48.1869} or integrating out either the momentum or the spatial dependence will lead to non-negative probabilities.

\section{Semi-classical interpretation}
\label{Kap_SemiClass}
In order to support understanding the outcome of our calculations we want to introduce a semi-classical picture based upon analyzing single-particle trajectories. 
At this point we have to stress, that neither we claim to have control over the pair production process at intermediate times nor do we interpret the results at intermediate times. The fact, that the semi-classical picture takes input at all times has to be seen as a requirement in order to use this tool. Still, we do not give any interpretation or show results obtained from the underlying physics at intermediate times.

As already mentioned in section \ref{Kap_Dyn}, we focus on the relativistic Lorentz force in order to determine the particle trajectories. In this way, we obtain viable results for minimal numerical effort.
To start with, the Lorentz force describes the impact of an electromagnetic field on the trajectory of a charged particle. In order to study this influence we have to seed a particle in the background field at a specific position having a well defined momentum.

\begin{figure}[thb]
{\includegraphics[width=.495\textwidth]{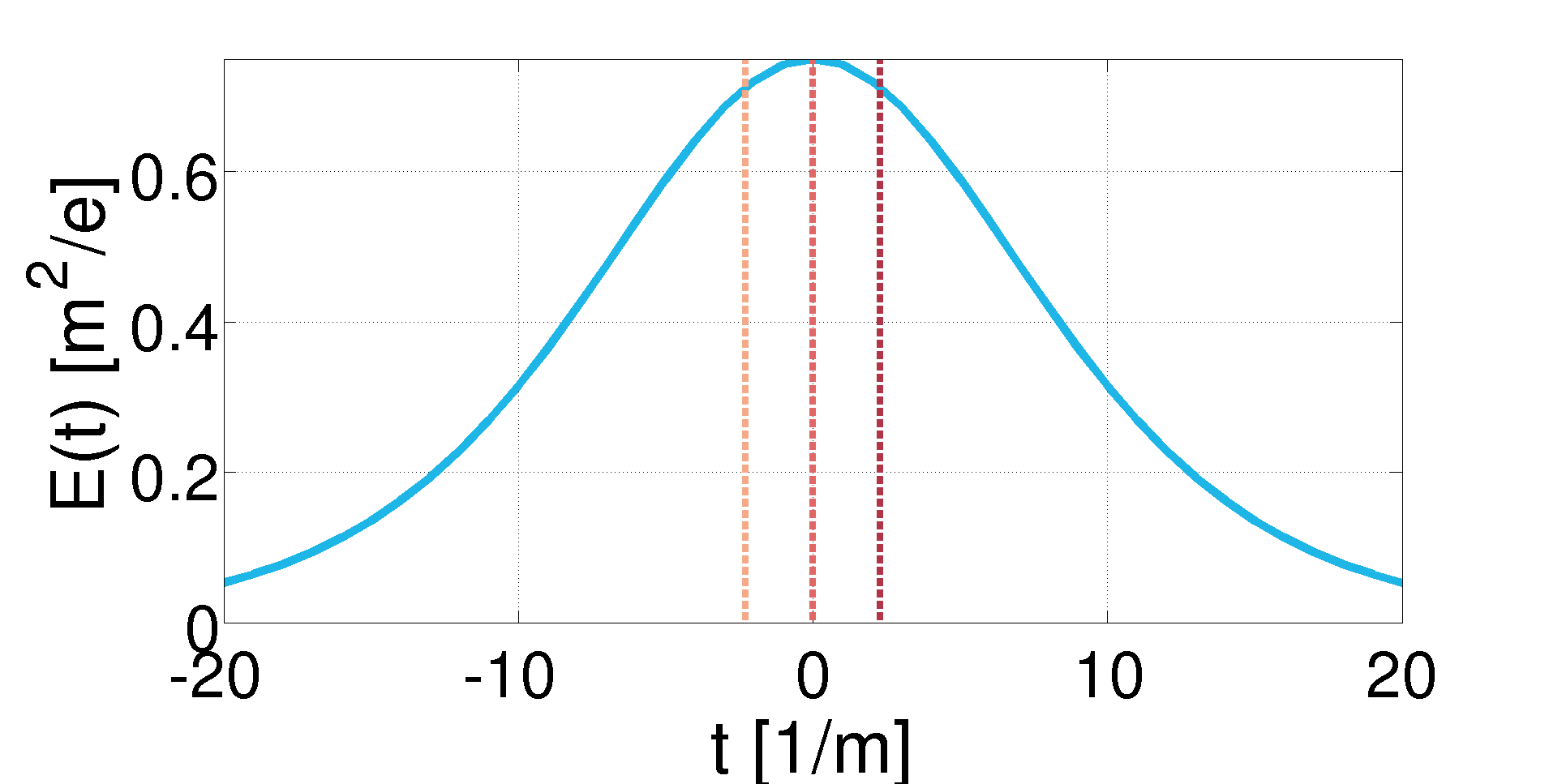}}      
{\includegraphics[width=.495\textwidth]{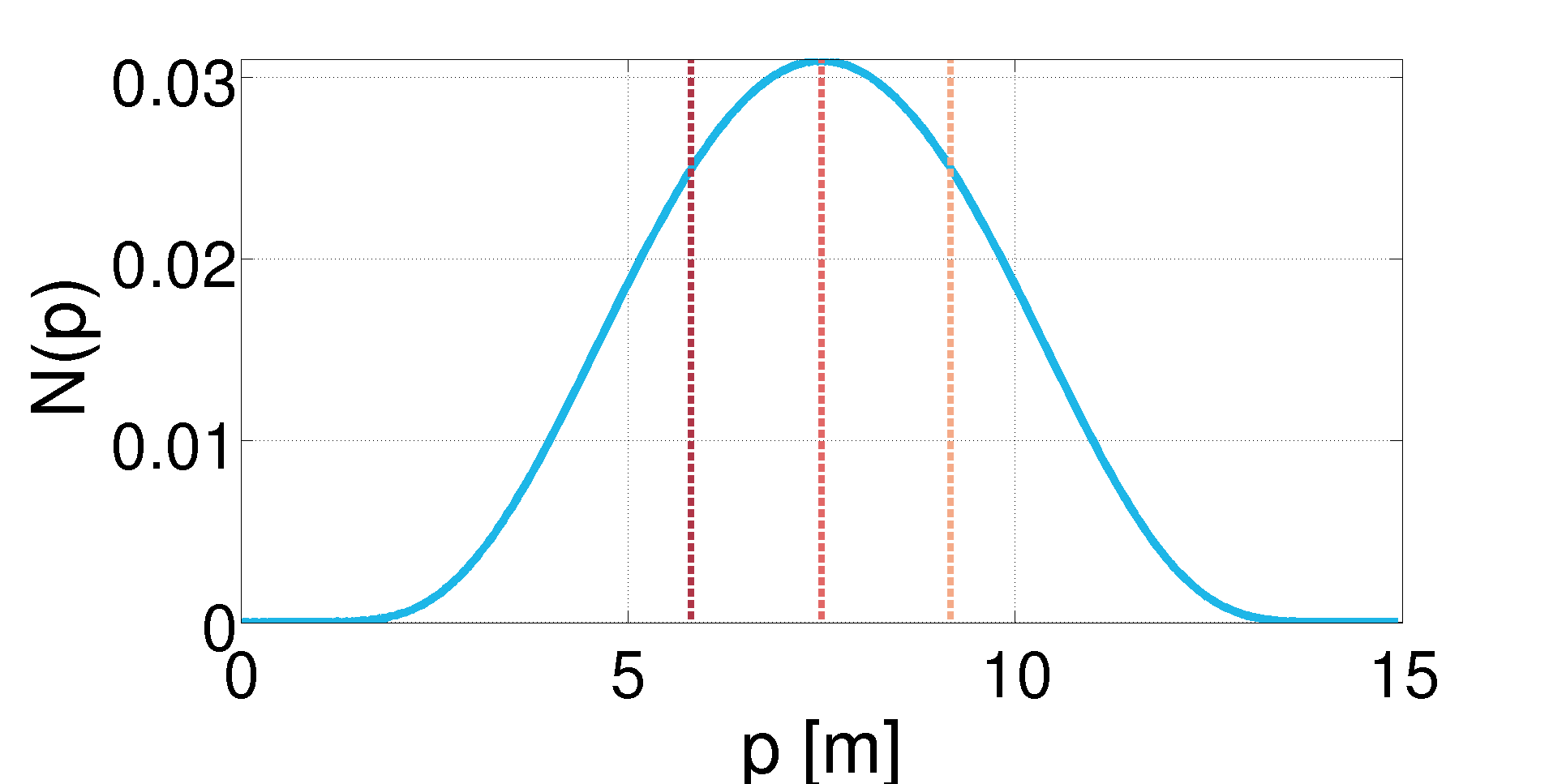}} 
\caption[Final particle distribution in case of a single-peaked, homogeneous electric field.]{Applying the homogeneous field given on the left-hand side(parameters: $\varepsilon = 0.75$, $\tau= 10$ $m^{-1}$, Tab. \ref{TabApp_Semi1}) we obtain the final particle distribution function on the right-hand side. The straight lines show a particle's final momentum when seeded at specific times $t_i$. The colors for seeding time and final momentum match.}
\label{Fig_Semi1}
\end{figure}

We begin introducing this method assuming a homogeneous electric field of the form
\begin{equation}
 E \brt = \varepsilon \ E_0 \ \text{sech}^2 \br{ \frac{t}{\tau}}. \label{Semi_E1}
\end{equation}
Evaluating the DHW equations we obtain a single-peaked particle distribution function being symmetric around $\tilde p = \varepsilon \tau$, see Fig. \ref{Fig_Semi1}. Within the semi-classical picture
we expect, that a higher field strength implies a larger particle production rate. Hence, we conclude that most particles are created at $t = 0$. The Lorentz force then predicts a final momentum of $p_f = \varepsilon \tau$ for particles created at $t=0$ with $p_i = 0$. As the applied electric field is symmetric in $t$, also the symmetry of the distribution function is obtained via seeding electrons at times $t_i = \pm \Delta t$. These findings are illustrated in Fig. \ref{Fig_Semi1}. 

\ctable[pos=hb,
caption = {Final particle momenta of electrons in an electric field ($\varepsilon = 0.5$, $\tau = 10$ $m^{-1}$) when
created at $x=0$ and $t = -10$ $m^{-1}$, $t=0$ or $t=10$ $m^{-1}$. The index of the momentum decodes the time of creation, with $p_0 = p(t_{i}=0)$. Particles become focused towards higher momenta as $\lambda$ decreases.},
cap={Semi-classical analysis of particle self-bunching.},
label = {Tab_Semi1},
mincapwidth = \textwidth,
]{lccccc}{}{
    \toprule
     $\lambda[1/m] - p[m]$ & $p_-$ & $p_0$ & $p_+$ & $\Delta p_-$ & $\Delta p_+$  \\
    \midrule
    $\infty$ & 8.80 & 5.00 & 1.19 & 3.80  & -3.80 \\
    \midrule
    $25$ & 7.98 & 4.78 & 1.17 & \hlight{3.2} & \hlight{-3.61} \\       
    \midrule    
    $10$ & 5.72 & 4.10 & 1.11 & \hlight{1.61} & \hlight{-2.98} \\   
    \midrule
    $5$ & 3.40 & 3.14 & 1.00 & \hlight{0.26} & \hlight{-2.13} \\         
    \bottomrule
}    

\begin{figure}[tbh]
{\includegraphics[width=.495\textwidth]{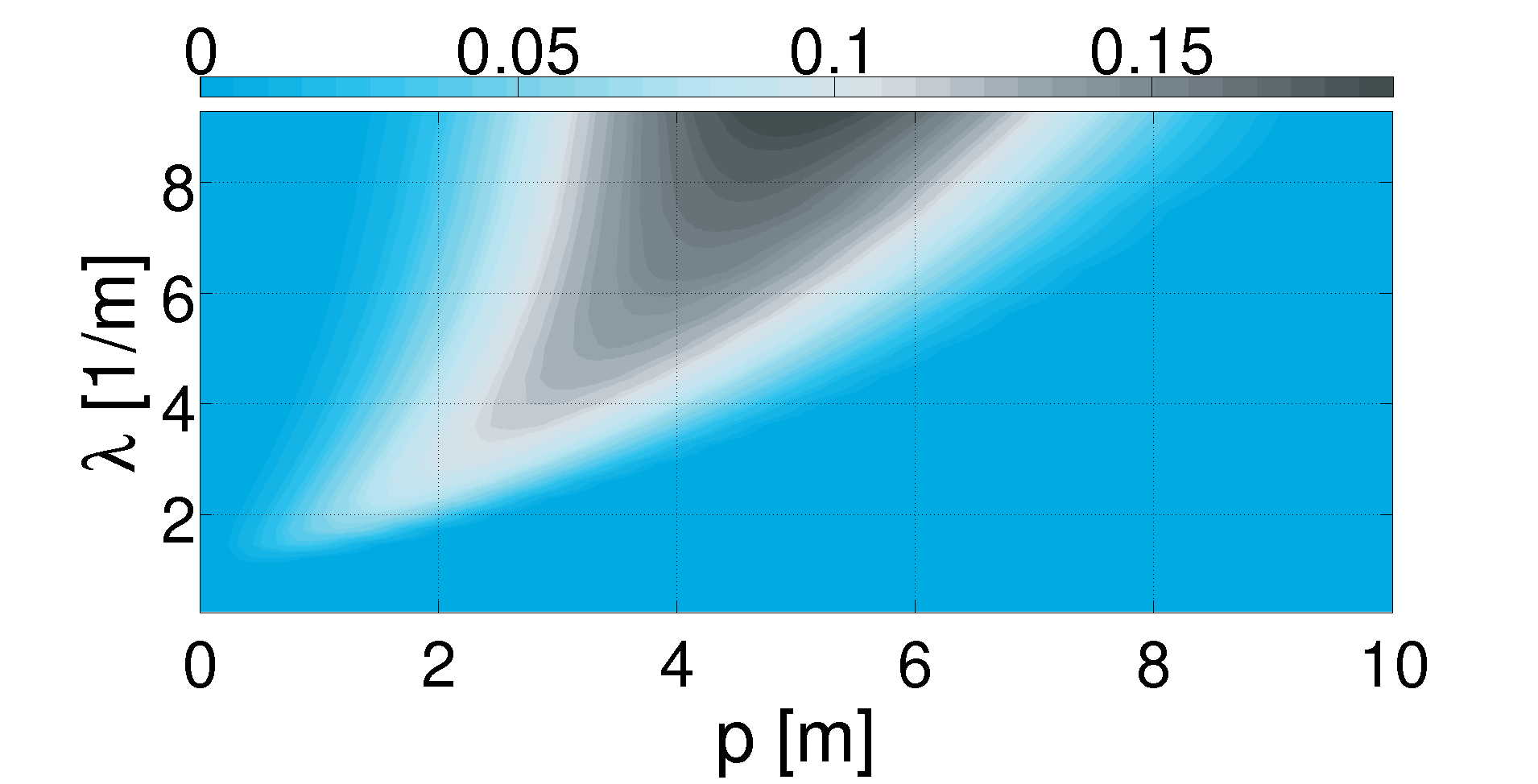}}      
{\includegraphics[width=.495\textwidth]{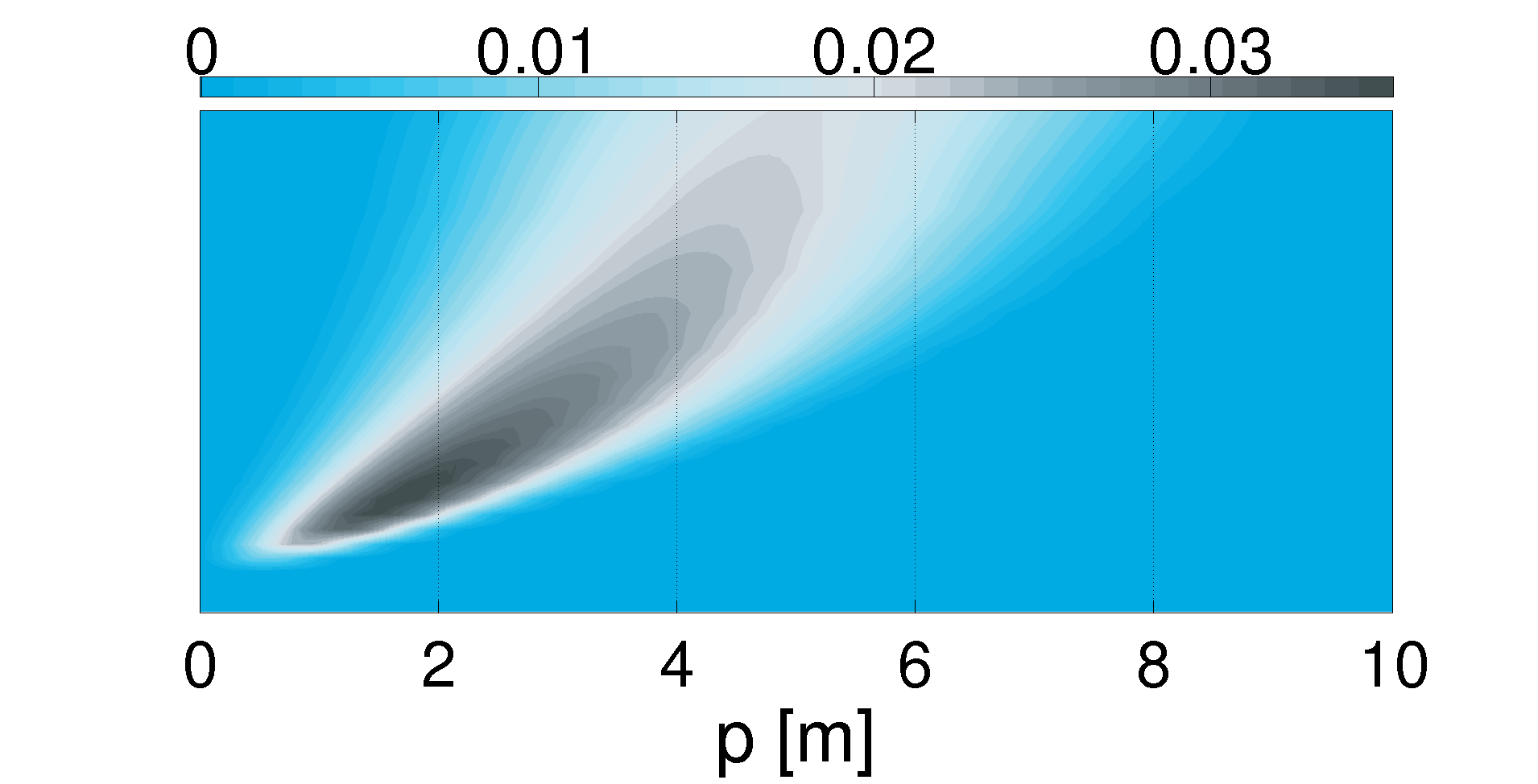}}    
\caption[2D picture of the distribution function for a single-peaked, inhomogeneous electric field.]{Distribution function(left-hand side) and normalized distribution function(right-hand side) for a single pulsed field with $\varepsilon = 0.75$, $\tau = 10$ $m^{-1}$. A shift towards lower momentum for
smaller spatial extent is observable. Additionally, the peak of the distribution function increases monotonously, while the normalized distribution function shows a maximum around $\lambda \approx 2$ $m^{-1}$. Parameters taken from: Tab. \ref{TabApp_Semi2}}
\label{Fig_Semi2}
\end{figure}

\begin{figure}[tbh]
\centering
{\includegraphics[width=.75\textwidth]{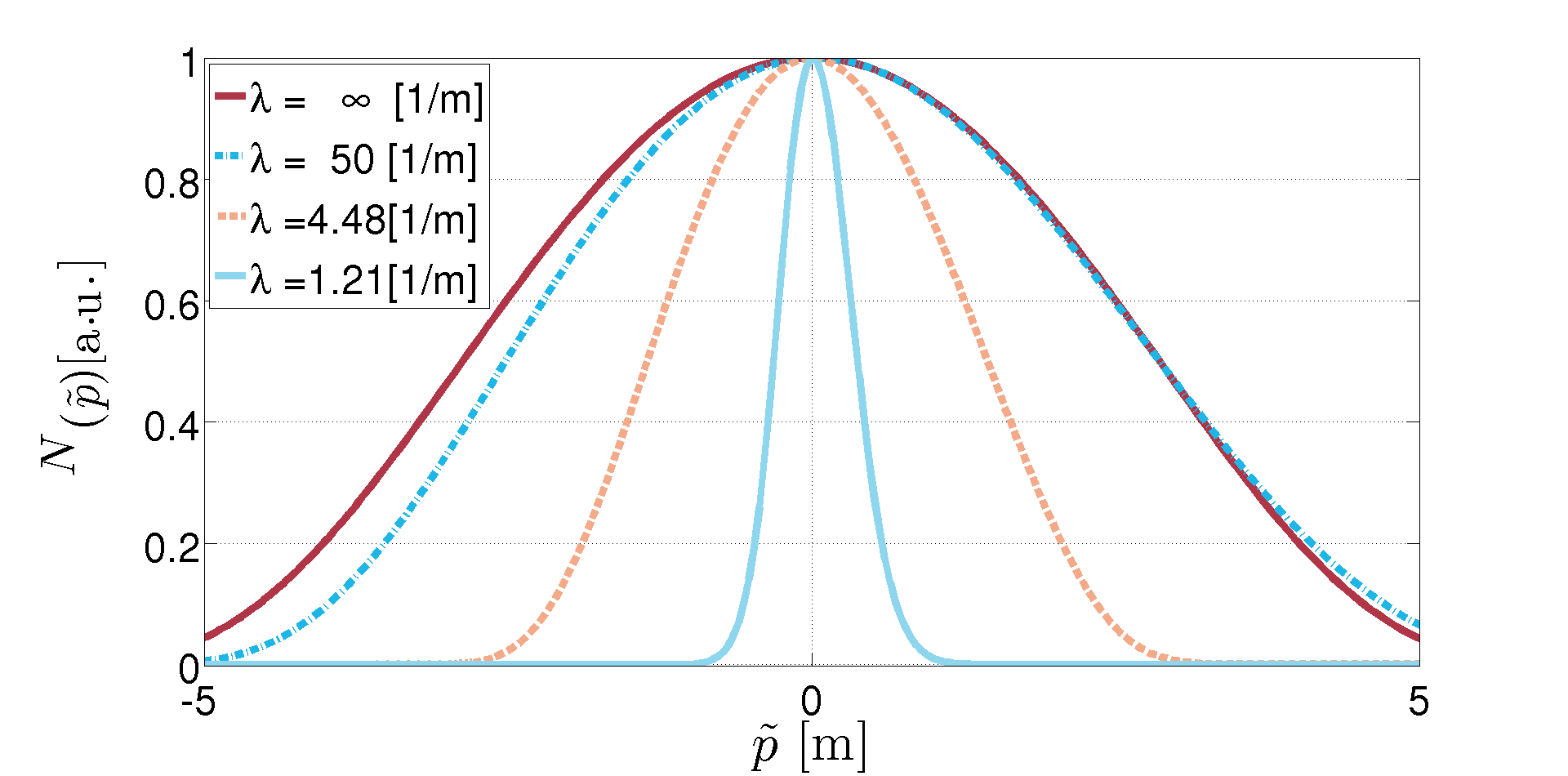}} 
\caption[Modified particle distribution function demonstrating particle self-bunching.]{Modified particle distribution function for a single pulsed field with $\varepsilon = 0.75$, $\tau = 10$ $m^{-1}$, Tab. \ref{TabApp_Semi3}. We have normalized all momentum distribution functions such that $\text{max} \br{N \br{\tilde p, \lambda}} = 1$. Additionally, we have shifted all functions, so that the peak value is always centered at $\tilde p=0$. In case of homogeneous fields ($\lambda \to \infty$ $m^{-1}$) the particle distribution function
is symmetric around the central peak position. All other lines do not display this symmetry and moreover the modified distribution functions become narrower the smaller $\lambda$ becomes.}
\label{Fig_Semi3}
\end{figure}

\newpage
We proceed investigating the impact of a spatial inhomogeneity on the particle distribution, thus we replace the electric field \eqref{Semi_E1} by
\begin{equation}
 E \br{x,t} = \varepsilon \ E_0 \ \text{sech}^2 \br{ \frac{t}{\tau}} \exp \br{-\frac{x^2}{2 \lambda^2}}. \label{Semi_E2}
\end{equation}
We can see in Fig. \ref{Fig_Semi2}, that the net momentum of the particle bunch is reduced the smaller $\lambda$ becomes. Careful analysis reveals, that the particle distribution is not symmetric around $\tilde p$ for inhomogeneous fields, see Fig. \ref{Fig_Semi3}. This self-bunching effect was first observed in reference \cite{PhysRevLett.107.180403}. We go on giving an explanation for this effect using semi-classical methods. Seeding electrons at different times within the field \eqref{Semi_E2} we indeed obtain an accumulation at higher final momenta, see Tab. \ref{Tab_Semi1}. Hence, the interpretation of a self-focusing of the particles within an electric field of the form \eqref{Semi_E2} is also supported by analyzing possible particle trajectories.

\vspace{5cm}

\begin{figure}[tbh]
{\includegraphics[width=.495\textwidth]{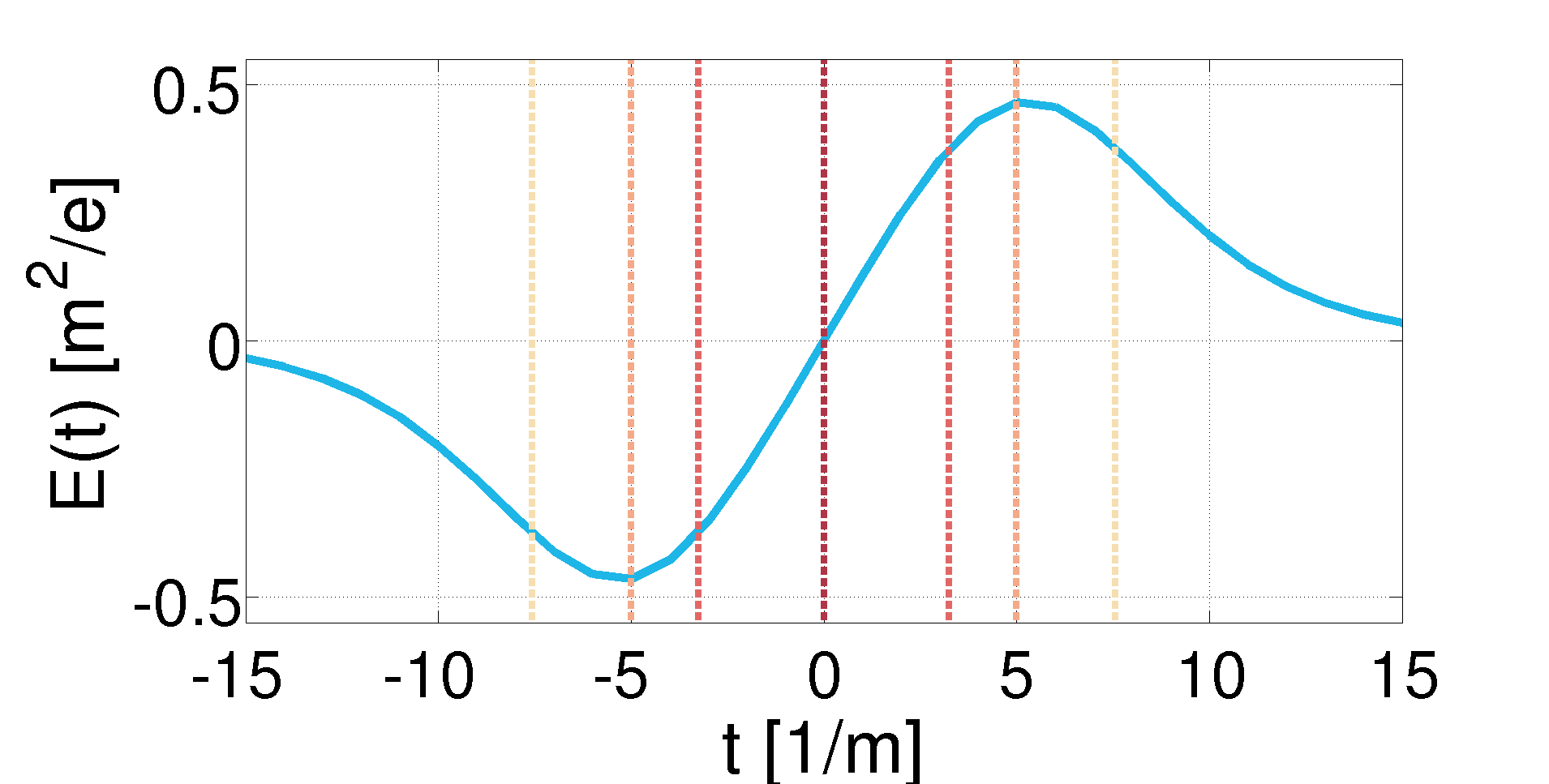}}      
{\includegraphics[width=.495\textwidth]{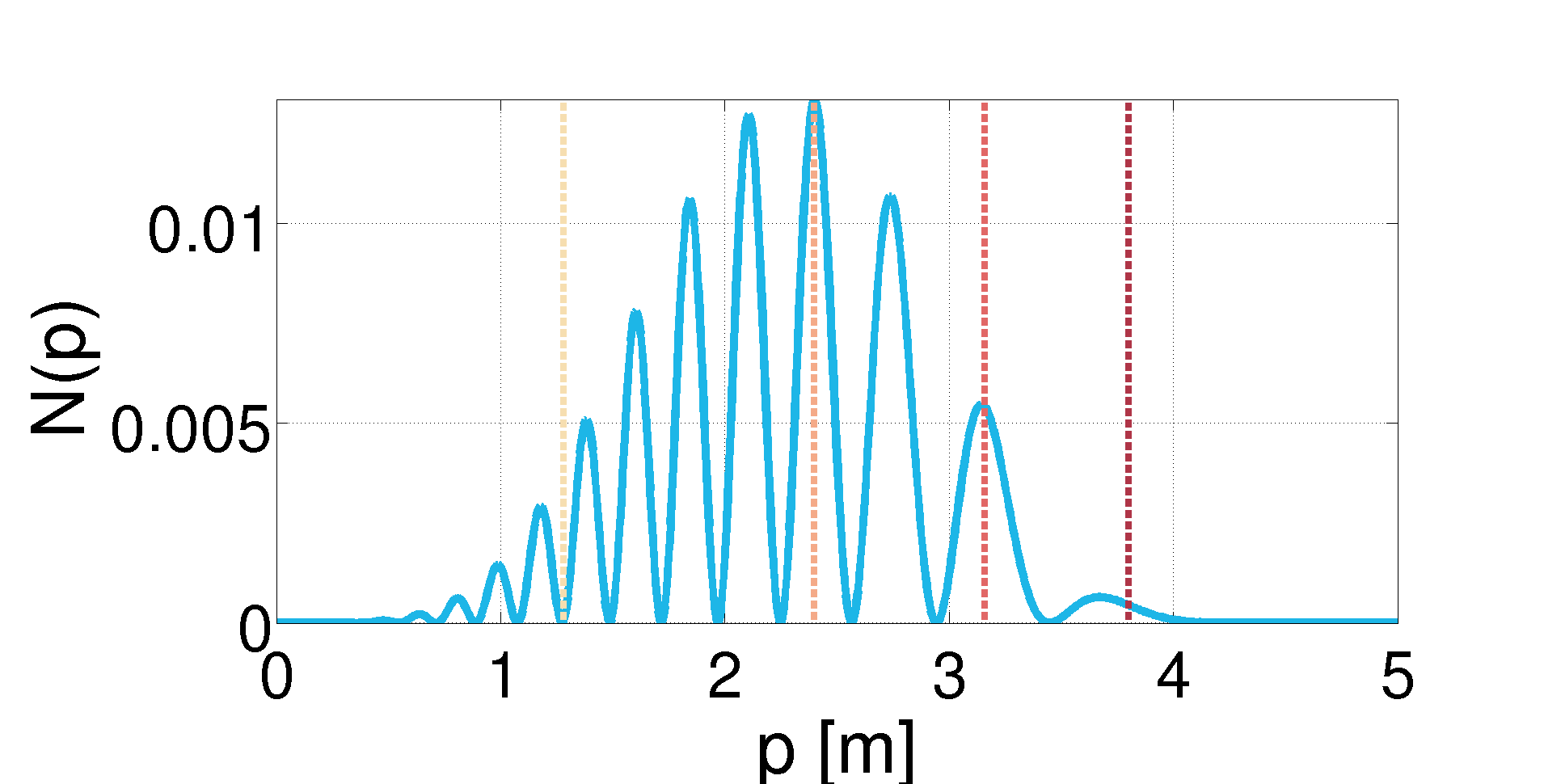}} 
\caption[Final particle distribution in case of a double-peaked, homogeneous background field.]{Particle distribution function for a homogeneous field of the form displayed on the left-hand side. The straight lines show a particle's final momentum when seeded at specific times $t_i$. The colors for seeding time and final momentum match.
Parameters: $\varepsilon=0.5$, $\tau=5$ $m^{-1}$. Additional information is given in: Tab. \ref{TabApp_Semi4}.}
\label{Fig_Semi4}
\end{figure}

However, it is neither possible to predict the distribution function peak position exactly nor it is surprising that this semi-classical interpretation runs into problems when $\lambda \sim 1 \ m^{-1}$.
To proceed, we want to study a field taking the form
\begin{equation}
 E \br{x,t} = \varepsilon \ E_0 \ \br{\text{sech}^2 \br{ \frac{t-t_0}{\tau}} - \text{sech}^2 \br{ \frac{t+t_0}{\tau}}} \exp \br{-\frac{x^2}{2 \lambda^2}}. \label{Semi_E3}
\end{equation}
In the beginning, we investigate the homogeneous limit $\lambda \to \infty$.
The electric field as well as the particle distribution function in case of $\lambda \to \infty$ is displayed in Fig. \ref{Fig_Semi4}. Oppositely to Fig. \ref{Fig_Semi1}, we can see an interference pattern to emerge. Again, we interpret the results applying a semi-classical picture. Due to the fact, that $|E \br{\mathbf{x},-t_0}|=|E \br{\mathbf{x},t_0}|$ we expect the same amount of particles is produced at $t = \pm t_0$. Calculating the trajectories of these particles assuming that they are created with vanishing momentum, we find that it is impossible to determine the particles origin when measured at position $\tilde x_f$ with final momentum $\tilde p_f$, because for every particle trajectory originating at $t = t_0$ and $x = 0$ we can identify an equivalent (for $t \ge t_0$) trajectory originating at $t = -t_0$. As the particle characteristics at a final time $t_f$ are indistinguishable we would therefore expect an interference pattern to occur, in agreement with a double slit experiment in time.

\begin{figure}[tbh]
{\includegraphics[width=.495\textwidth]{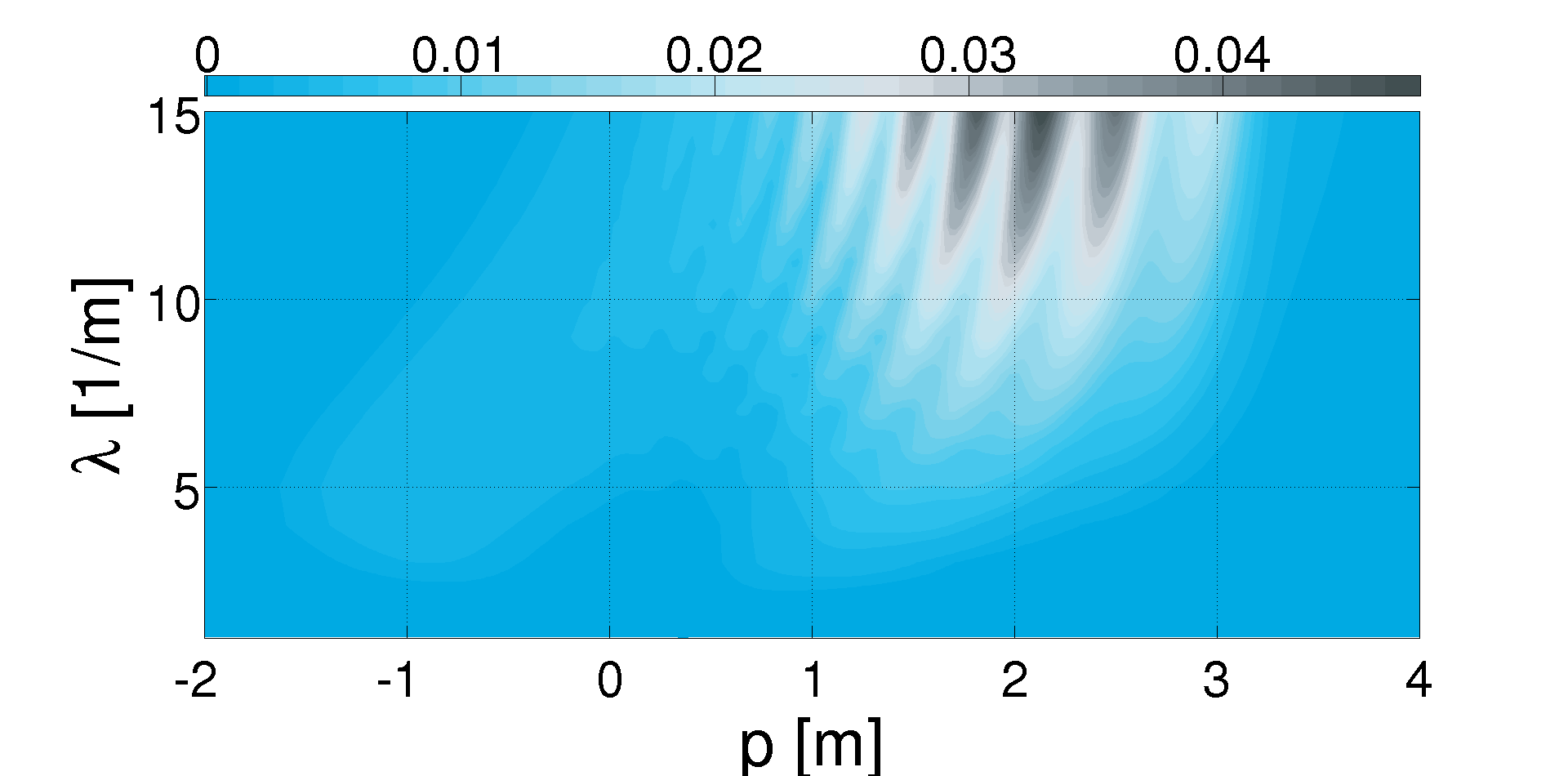}}      
{\includegraphics[width=.495\textwidth]{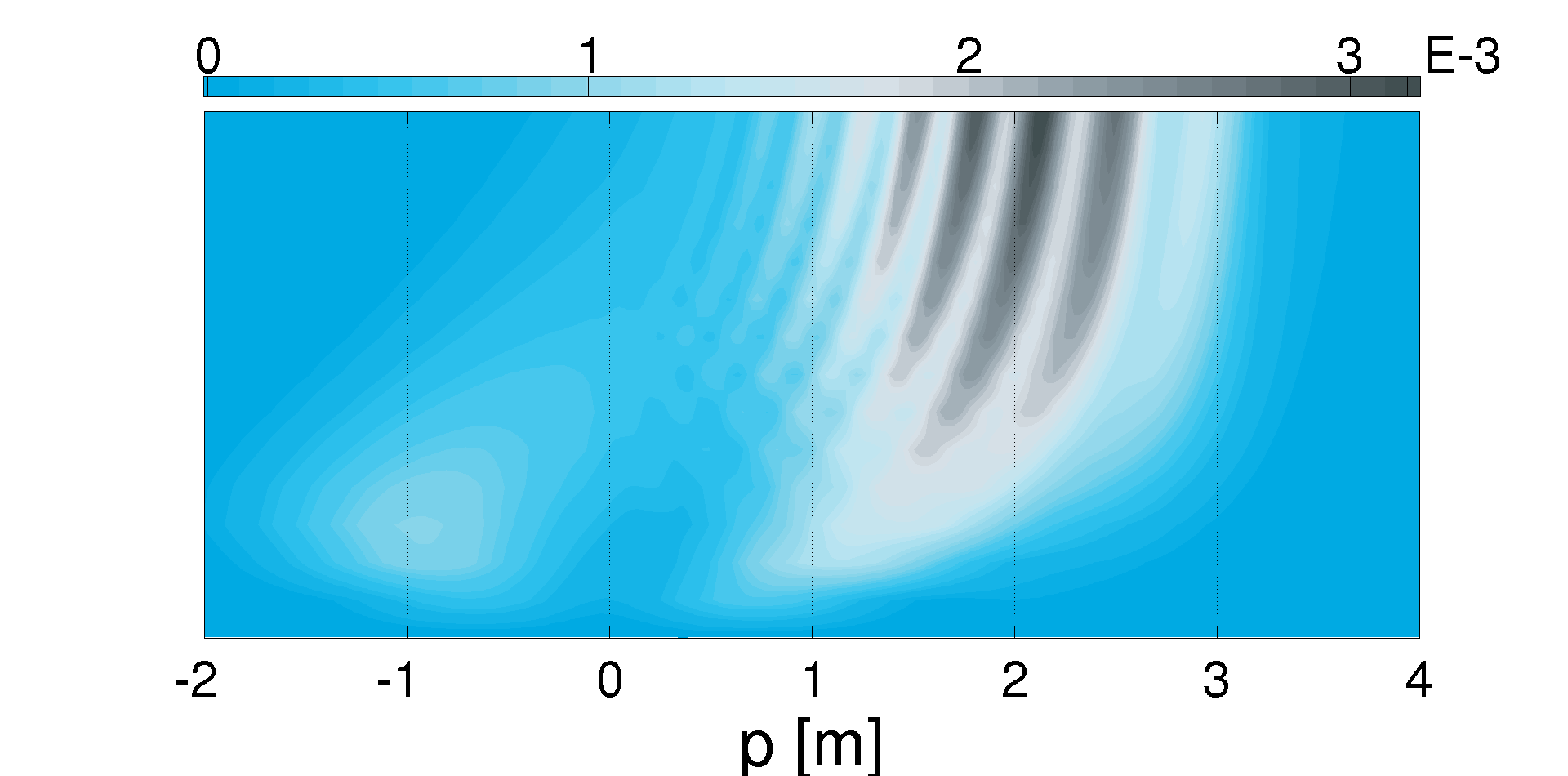}}    
{\includegraphics[width=.495\textwidth]{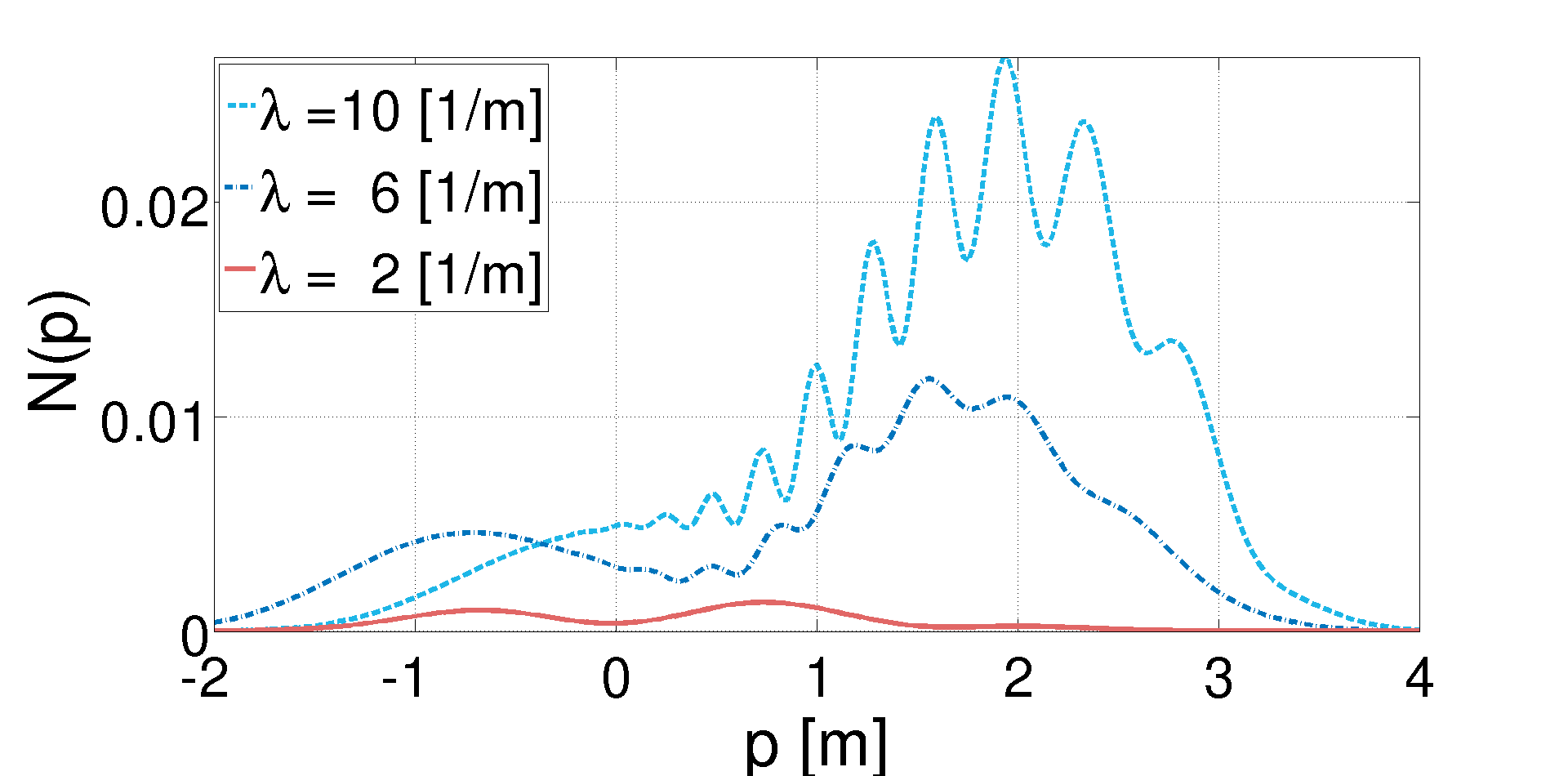}}      
{\includegraphics[width=.495\textwidth]{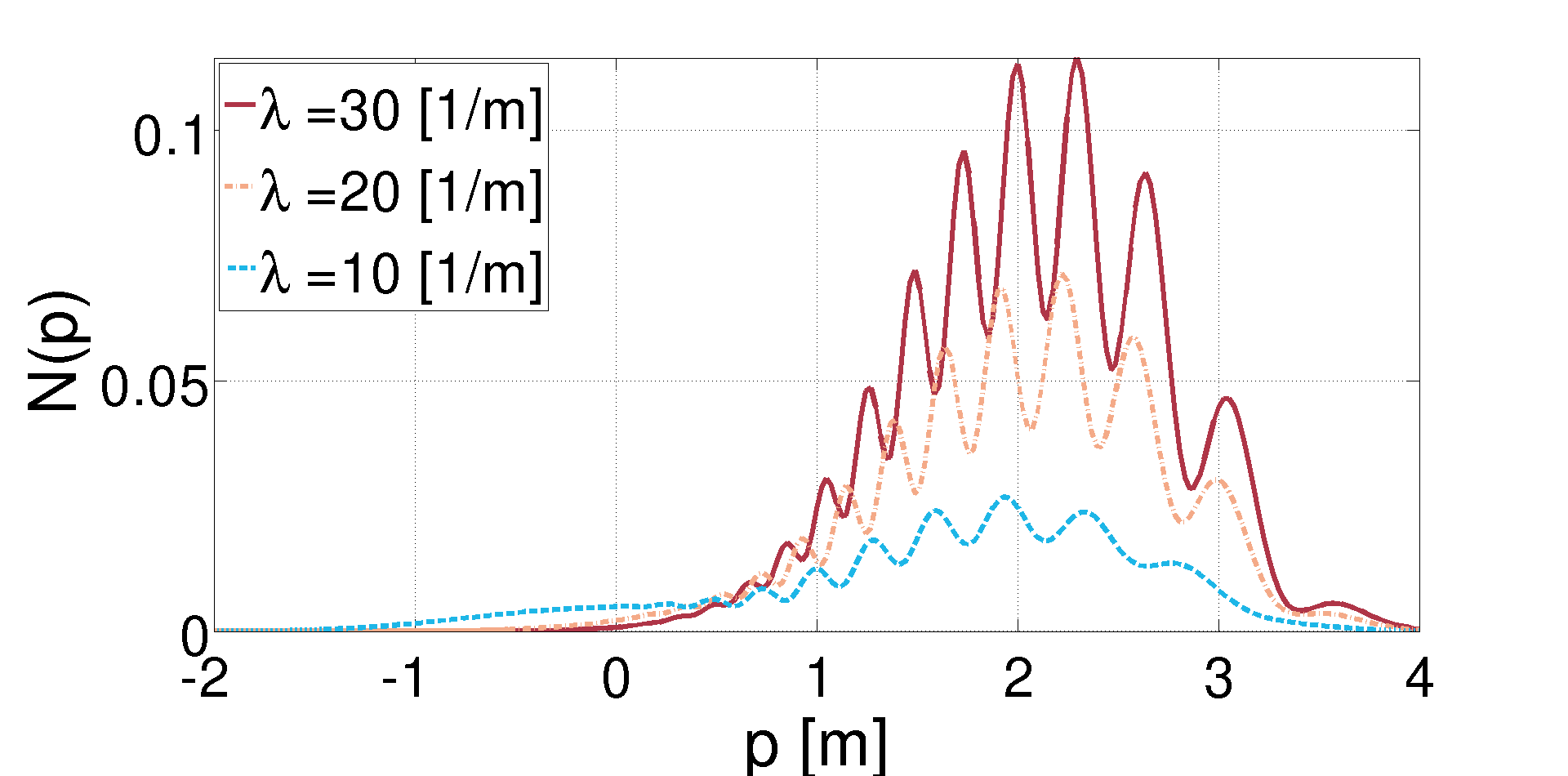}}  
\caption[2D picture of the distribution function for a double-peaked, inhomogeneous electric field.]{
Distribution function(left-hand side) and normalized distribution function(right-hand side) for double-pulsed field with field parameters $\varepsilon = 0.5$, $\tau=5$ $m^{-1}$. The interference pattern, which is clearly visible in the lower right plot vanishes as $\lambda$ approaches zero. Additionally, a shift of the particle bunch towards vanishing momentum is observable.  At $p \sim -1$ $m$ and $\lambda < 10$ $m^{-1}$ a second peak in the distribution function rises. Additional parameters: Tab. \ref{TabApp_Semi5}, \ref{TabApp_Semi6}, \ref{TabApp_Semi7}.}
\label{Fig_Semi5}
\end{figure}

We proceed examining a spatially inhomogeneous background field. Similarly to the case of a single pulsed field described above, we find a decrease in the total particle momenta. 
Additionally, if $\lambda$ becomes smaller particles created at $t = - t_0$ are substantially differently accelerated compared to particles produced at $t = t_0$. As this allows to distinguish the particles created at either of the peaks we would expect, that the interference pattern vanishes. In case $\lambda$ is chosen to be of the order of the Compton scale, we further find that particles created at $t = - t_0$ basically do not interfere with the second peak of the electric field. Hence, a clear double-peak structure in momentum space should emerge\cite{PhysRevLett.108.030401}. Comparison with Fig. \ref{Fig_Semi5} shows nice agreement with the considerations above evidencing the usefulness and correctness of the semi-classical analysis introduced herein.

\vspace{5cm}

At last, we want to examine an oscillating, spatially inhomogeneous electric field of the form
\begin{equation}
 E \br{x,t} = \varepsilon \ E_0 \ \exp \br{-\frac{t^2}{2 \tau^2}} \exp \br{-\frac{x^2}{2 \lambda^2}}. \label{Semi_E4}
\end{equation}
Performing a semi-classical investigation yields the following.
We assume particles are created at times when the electric field is maximal, so we start by seeding an electron at $t=0$. This electron subsequently oscillates due to the rapid change in the electric field strength. However, due to the spatial inhomogeneity the electron acquires a net momentum. Hence, we would expect, that particles produced within such a highly oscillating field obtain an additional boost in momentum space away from the origin. This effect is well known in plasma physics as the ponderomotive force
\begin{equation}
 \mathbf{F}_{pond} = -\frac{e^2 E_0^2 \ \varepsilon^2}{4 \ m \ \omega^2} \ \boldsymbol{\nabla}_x \br{\mathbf{E} \br{x}^2}.
\end{equation}

Indeed, we can find traces of this net force when studying the momentum map of the created particles.
However, we will postpone the analysis of this force and how it influences the particle momentum to chapter six. In this way, we are able to establish first the regime of multiphoton pair production in chapter five and are then able to expand the findings to spatially inhomogeneous fields.

\part{Results}
\pagestyle{plain}
\chapter{Multiphoton pair production}
\pagestyle{fancy}

\begin{figure}[hb]
\begin{center}
  \includegraphics[width=0.9\textwidth]{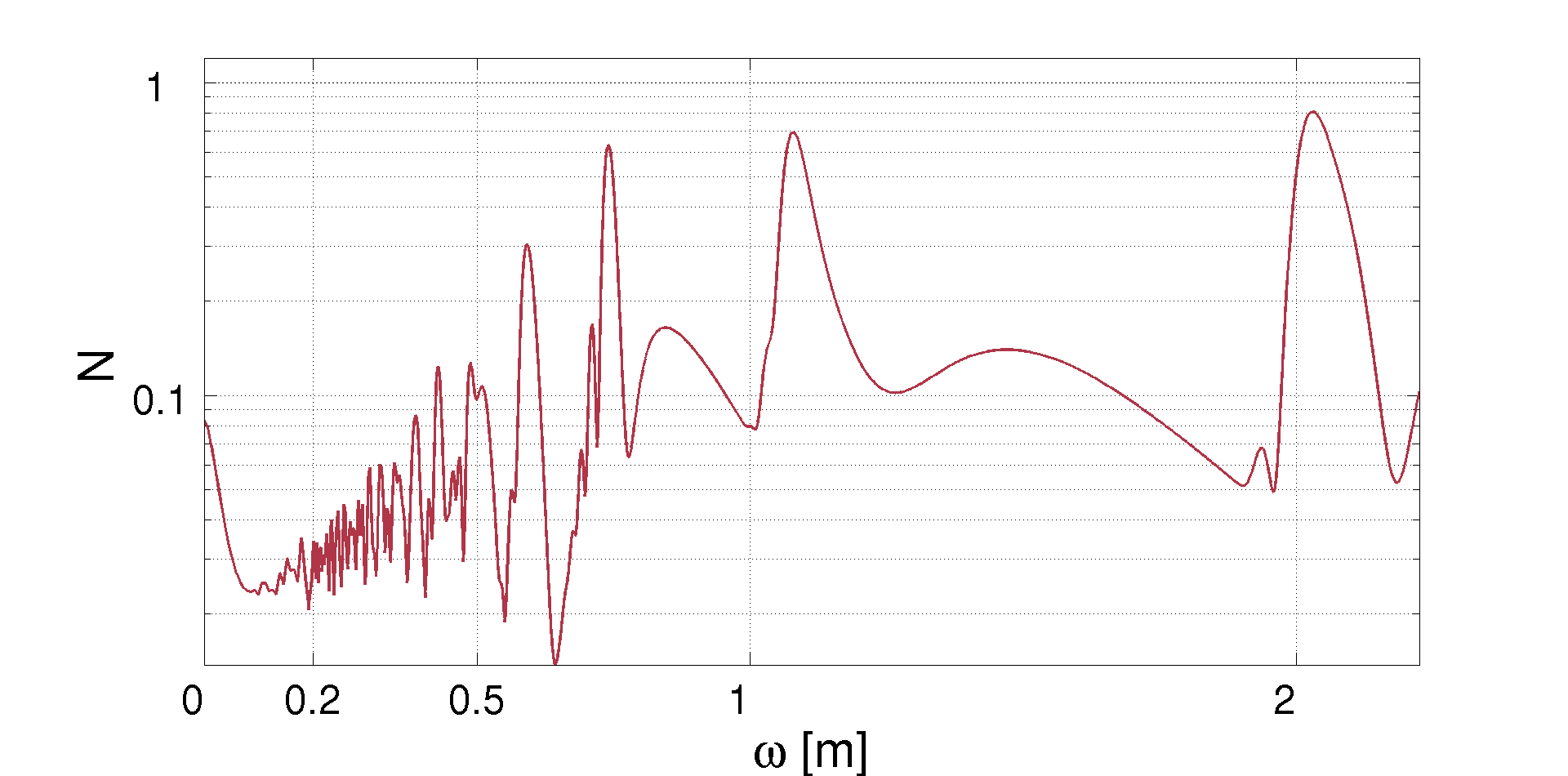}
\end{center}
\caption[Particle yield as a function of the field frequency.]{Logarithmic plot of the particle yield $N$ for various field frequencies $\omega$ for a pulse with finite temporal extent. The drop off at $\omega \sim 0.1$ $m$ signals a transition from a Schwinger dominated region($\omega \lesssim 0.1$ $m$) to an absorption dominated region($\omega \gtrsim 0.1$ $m$). The clear peaks in the region $\omega \ge 0.5$ $m$ are characteristically for multiphoton pair production in many-cycle pulses. Further parameters: Tab. \ref{Tab_EffM1_App}} 
\label{Fig_YieldFull}
\end{figure}

In Fig. \ref{Fig_YieldFull} we have illustrated the particle yield as a function of the field frequency. We employed a generic, pulsed electric field superimposed by a photon-energy dependent subcycle structure.

In total, we identify three different regimes. In case of low field frequencies $\omega$ Schwinger pair production dominates. The increase in the particle yield is due to the fact, that in the tunneling regime the pulse length is a major factor when determining the production rate. The lower $\omega$ the fewer oscillations in the field, thus the time interval the field is close to its peak value is extended. We also observe, that in Fig. \ref{Fig_YieldFull} the yield is noisy in the midrange. As the pulse length is finite, there is a regime where the employed electric field is basically a few-cycle pulse. Interpretation of the momentum output in the case of a few-cycle pulse is complicated and beyond the scope of this thesis. However, we will give example results in section \ref{App_CEP}. Of major importance in the chapters five and six is the high-frequency sector: the regime of multiphoton pair production. In this region the electric field is a many-cycle pulse exhibiting an oscillatory structure mimicking a monochromatic pulse.

\clearpage

In this chapter we will investigate particle creation in the regime of multiphoton pair production. This is done within a spatially homogeneous background field, which allows to employ QKT. In this way, we are able to precisely determine effective mass signatures, because QKT allows to perform calculations for electric fields lasting for up to $t \approx 10^4$ $m^{-1}$. 

\section{Effective mass signatures}

\footnote{This section is entirely based upon the paper Kohlf\"urst et al.\cite{PhysRevLett.112.050402}. The ideas originally developed in the paper are expanded and the arguments are supported by additional and more elaborate illustrations.} The main advantage of introducing the concept of an effective mass is the easier treatment of multiphoton pair production as displayed in the following. 

\subsection{Quantum kinetic theory and background field}

In order to study the implications of an effective mass we have to define an electric field first. In contrast to reference \cite{PhysRevLett.112.050402}, we decided to employ a homogeneous electric field pulse of the form
\begin{equation}
 E \brt = \varepsilon \ E_0 \cos^4 \br{\frac{t}{\tau}} \cos \br{\omega t}, \label{EffMass1}
\end{equation}
where $t \in \com{-\frac{\pi \tau}{2},\frac{\pi \tau}{2}}$. In this case, $\varepsilon$ defines the peak field strength, $\omega$ the field frequency and thus the ``photon'' energy, and $\tau$ determines the pulse duration. 

We will investigate pair production in a parameter region set by the Compton scale. 
Although these parameter values are not accessible in experiment at the moment, they could become approachable within future tailored X-Ray Laser technology\cite{Ringwald2001107}. We examine the implications of e.g. longer pulse lengths on the particle yield in section \ref{Sec_Xtr}. We are aware of the fact, that the model \eqref{EffMass1} does not resemble a real field configuration in an experiment. Rather, it can be seen as an approximation for a field occurring in an antinode of a standing-wave mode. In case one is interested in more realistic models one would have to take into account spatial inhomogeneities as well as magnetic components, see chapters six and seven as well as \cite{Hatsagortsyan,PhysRevLett.102.080402,PhysRevLett.107.180403}. 

As we are investigating pair production in a spatially homogeneous electric field, we employ QKT. Moreover, within this section we are only interested in the particles created parallel to the applied background field. Hence, throughout this chapter we basically investigate a $1+1$-dimensional problem. The corresponding transport equations can be formulated as in \eqref{QKT}:
\begin{align}
  \ma{\dot{F} \\ \dot{G} \\ \dot{H}} = \ma{0 & W & 0 \\ -W & 0 & -2\omega \\ 0 & 2\omega & 0} \ma{F \\ G \\ H} + \ma{0 \\ W \\ 0}, 
\end{align}
where
\begin{align}
 \omega \br{q_x,t} = \sqrt{m^2 + \br{q_x - eA \brt}^2} ,\quad W \br{q_x,t} = \frac{e E \brt m}{\omega^2 \br{q_x,t}}.
\end{align}
Additionally, we have applied vacuum initial conditions $F_i = G_i = H_i = 0$.

In order to discuss our findings we will show results for asymptotic times only. This includes the final particle momentum spectrum, given by $F \br{q_x,t \to \infty}$ as well as the particle yield:
\begin{equation}
 N = \int \ dq_x \ n \br{q_x} =  \int \ dq_x \ F \br{q_x,t \to \infty}.
\end{equation}
N.B.: The momentum $q_x$ is denoted as $qx$ in the following figures to improve readability of the axis labels.

\subsection{Concept of an effective mass}

The concept of an effective mass has been introduced in order to collect the particle-background interactions in a simplified manner. Instead of describing a particle together with all its field interactions, we treat the corresponding quasi-particle as it was free but has a different mass $m_{\ast}$. Naturally, one looses information when combining all possible effects to a single number. However, in case of an electron in a strong, many-cycle electromagnetic background field this information loss seems to be negligible small. Hence, examining observables with respect to effective mass signatures is a valid procedure as we will show in the following.

We have to emphasize, that we will show results where the effective mass appears as a directly accessible quantity. This is in contrast to previous investigations of the effective mass of electrons in a background field, where the photon emission spectrum has turned out to be the ``observable-to-go''.  This includes experiments regarding undulator fields in FEL\cite{PhysRevSTAB.10.034801,McNeil}, plane-waves\cite{Wolkow,Goldman,Nikishov,PhysRev.138.B740} as well as more general fields\cite{PhysRevE.77.036402} up to suggestions matching realistic experimental requirements\cite{PhysRev.150.1060,PhysRevD.3.1692,PhysRevA.80.053403,PhysRevA.81.022125,PhysRevA.83.022101,PhysRevA.83.032106,PhysRevA.85.046101,PhysRevA.85.046102,PhysRevLett.109.100402,PhysRevD.87.085040}. Moreover, the impact of the pulse shape has been taken into account and it has been found that the concept of an effective mass is by no means universal. Rather, it is an instrument introduced at a specific stage of the calculation with the goal to simplify complex interactions. In case of determining the motion of an electron in a periodic field, the effective mass is introduced via averaging over the (quasi-)four momentum. In lightcone gauge and in case of a monochromatic plane wave the effective mass takes the form\cite{Wolkow}:
\begin{equation}
m_{\ast} = m \sqrt{1+ \xi^2} ,\qquad \xi = \frac{e}{m} \sqrt{-\langle A^{\mu} A_{\mu} \rangle}. \label{Meff}
\end{equation}
Due to the fact, that we propose a vacuum initial state for all of our calculations, the applied field cannot be periodic in time. However, multiphoton pair production requires a many-cycle pulse. The applied electric field \eqref{EffMass1} is approximately monochromatic and therefore we will adopt the definition above. At this point it should be mentioned, that $m_{\ast}$ takes such a simple form, only because $\langle A_{\mu} \rangle^2$ is approximately zero in our case. The interested reader is referred to \cite{Kibble} for a more general treatment of the effective mass. 

We have to be clear on the fact, that the effective mass concept within a pair production process is only meaningful for long-pulsed, many-cycle fields. Otherwise, the time interval the particle spends within the background field is too short making it impossible to determine a single value summarizing all particle-field interactions. Furthermore, the concept of an effective mass is intimately connected with the concept of averaging over all interactions and thus averaging over the background field. Hence, we focus on field configurations which allow to easily extract an averaged value we can use in order to approximate the occurring interactions. 

In the following, we concentrate on directly accessible observables instead of examining the particle's photon emission. Hence, we come up with distinct features of an effective mass concept illustrating and discussing our results obtained for the particle momentum spectrum and the particle yield. As we are interested in particle creation from colliding Laser pulses we obtain a comparatively clean setting for investigating the threshold for pair production governed by an effective mass. Further information on this topic can be found in the literature\cite{Heinzl,PhysRevLett.109.100402,PhysRevLett.101.130404,PhysRevA.81.022122,PhysRevLett.108.240406,Nousch2012246,Smolyanski}.

The simulations are done employing QKT, which automatically includes non-Markovian memory effects\cite{Smolyanski,PhysRevD.58.125015,Schmidt}. Furthermore, we obtain the particles momentum spectrum, thus making it possible to compare with results obtained in atomic ionization. As already mentioned in the introductory chapters, there is a conceptual similarity between atomic ionization and pair production\cite{Delone,Popov2,Salieres,PhysRevD.73.065028,PhysRevD.72.105004,Hatsagortsyan}. We will make use of this fact in order to present and discuss our findings. 


\begin{figure}[htb]
 \includegraphics[width=0.5\textwidth]{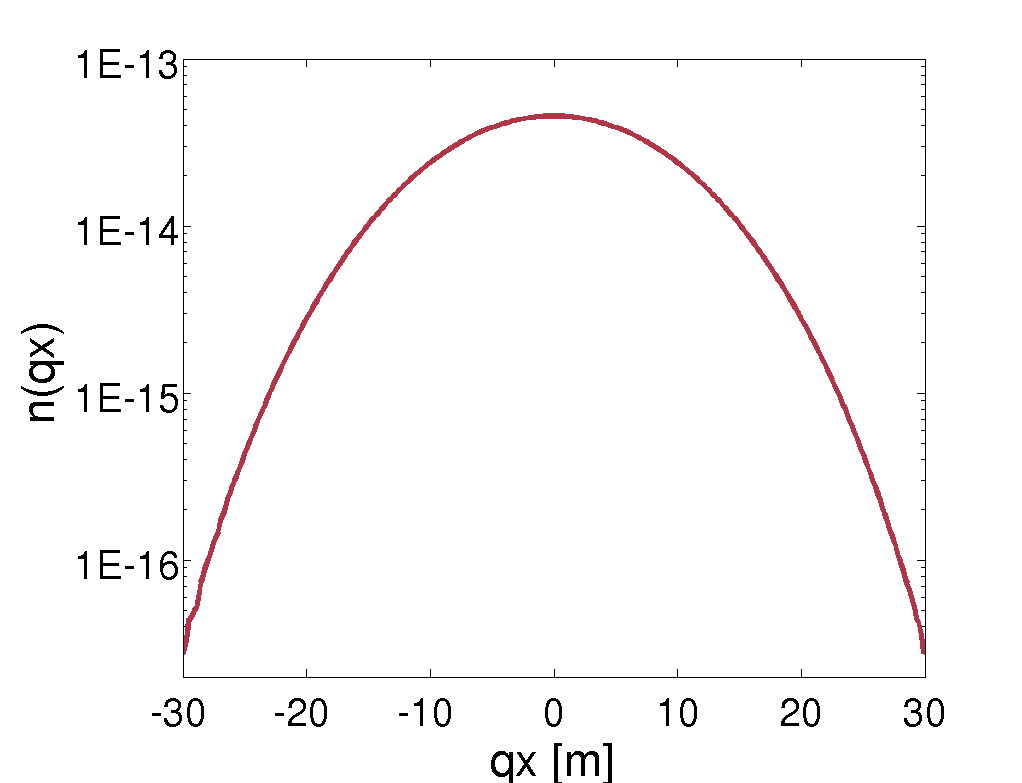}
 \includegraphics[width=0.5\textwidth]{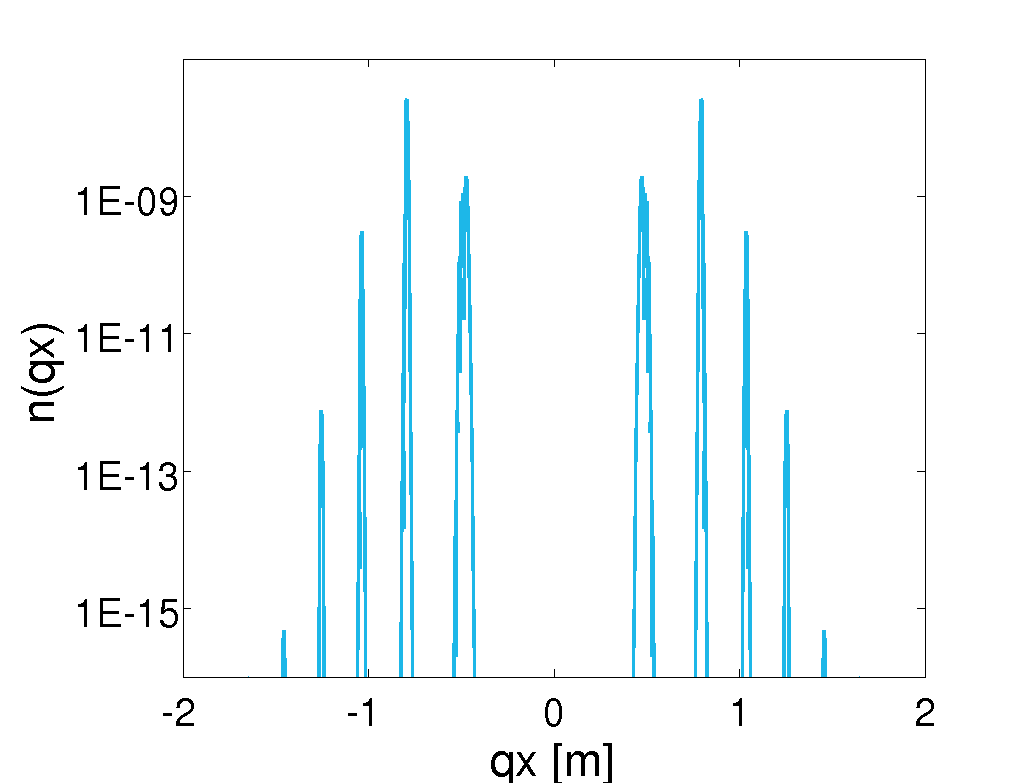}
 \caption[Momentum spectrum for Schwinger and multiphoton pair production.]{Logarithmic plot of the canonical momentum spectrum for Schwinger (left-hand side) and multiphoton pair production (right-hand side). The Schwinger spectrum shows a smooth distribution, while the multiphoton spectrum exhibits peaks due to the occurrence of $n+s$ absorption processes. Field parameters: $\varepsilon= 0.1,~ \tau= 1000$ $m^{-1}$, $\omega=0$ (left-hand side) and $\omega=0.322$ $m$(right-hand side). Further information in Tab. \ref{Tab_distr0}} 
\label{Fig_Distr} 
\end{figure}

At first we illustrate the different momentum distributions regarding tunneling and multiphoton pair production in Fig. \ref{Fig_Distr}. We observe a broad peak in momentum space in case of Schwinger pair production, because the background field is only slowly varying. The longer the total pulse length, the broader the electron distribution function. The picture changes dramatically in the multiphoton regime, where multiple small peaks of different height are resulting. This is a characteristic feature of multiphoton pair production. Besides particle creation via $n$ photons, the possibilities of $n+s$-photon absorption processes are non-zero. The similarities to atomic ionization and especially ATI spectra are striking and have been noticed in the literature\cite{Popov2,Salieres,PhysRevD.73.065028,PhysRevD.72.105004,Hatsagortsyan}.

\subsection{Particle yield}

\begin{figure}[ht]
\begin{center}
  \includegraphics[width=0.75\textwidth]{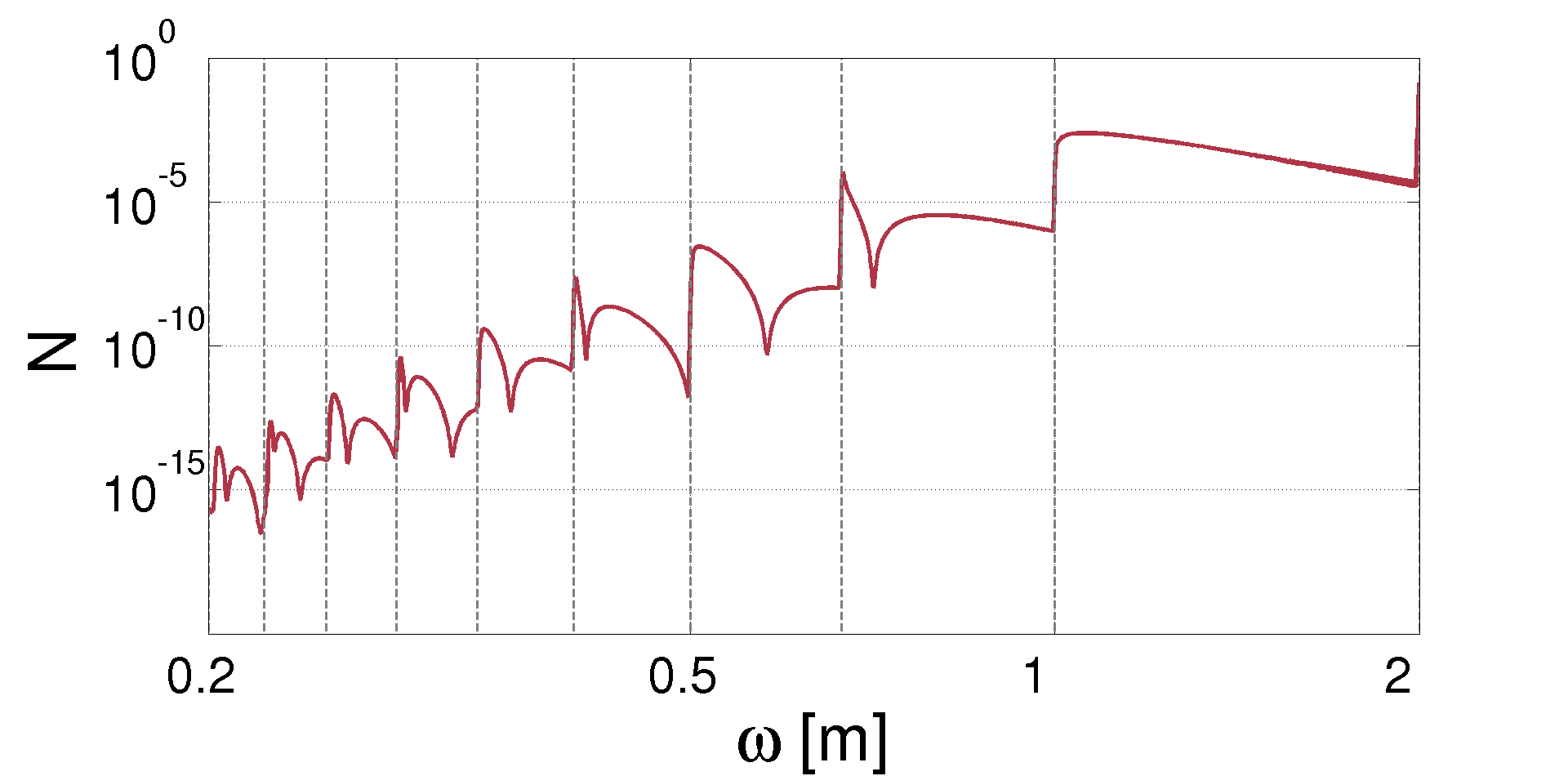}
\end{center}
\caption[Particle yield as a function of the field frequency in the multiphoton regime.]{Particle yield for various $\omega$ displayed in a log-log plot. The oscillations and thus the peaks can be seen as thresholds for $n$-photon processes. The dotted lines indicate a field-independent threshold $n \omega=2m$. These vertical lines are located slightly below the peaks in the particle yield. Field parameters: $\varepsilon= 0.05,~ \tau= 1000$ $m^{-1}$ and Tab. \ref{Tab_yield}}  
\label{Fig_Yield}
\end{figure}

We start investigating effective mass signatures by comparing the particle yield obtained from simulations with the naive assumption $n \omega = 2m$. 
In Fig. \ref{Fig_Yield}, the difference between bare mass threshold and simulation is displayed. The yield in Fig. \ref{Fig_Yield} shows an oscillatory structure. Moreover, the threshold peaks increase the higher the photon frequency becomes. This, however, is only true for rather clean configurations, where further effects like Pauli-blocking can be neglected, compare with Fig. \ref{Fig_YieldFull}. 


\begin{figure}[hbt]
 \includegraphics[width=0.5\textwidth]{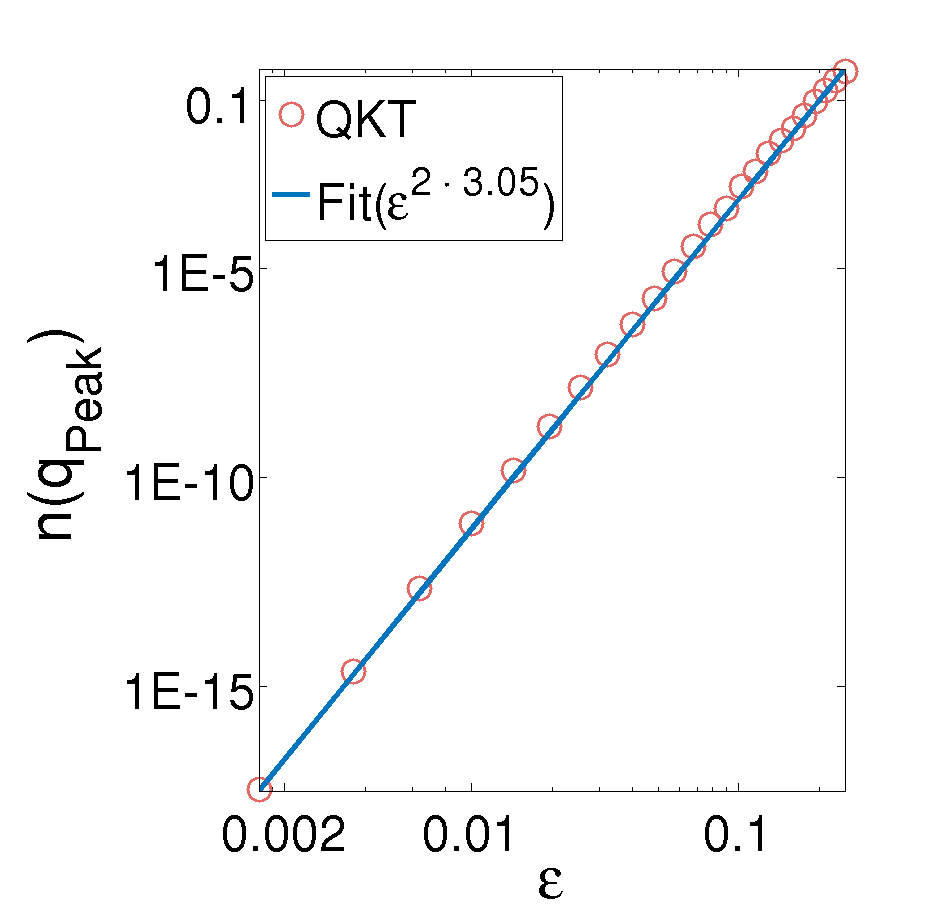}
 \includegraphics[width=0.5\textwidth]{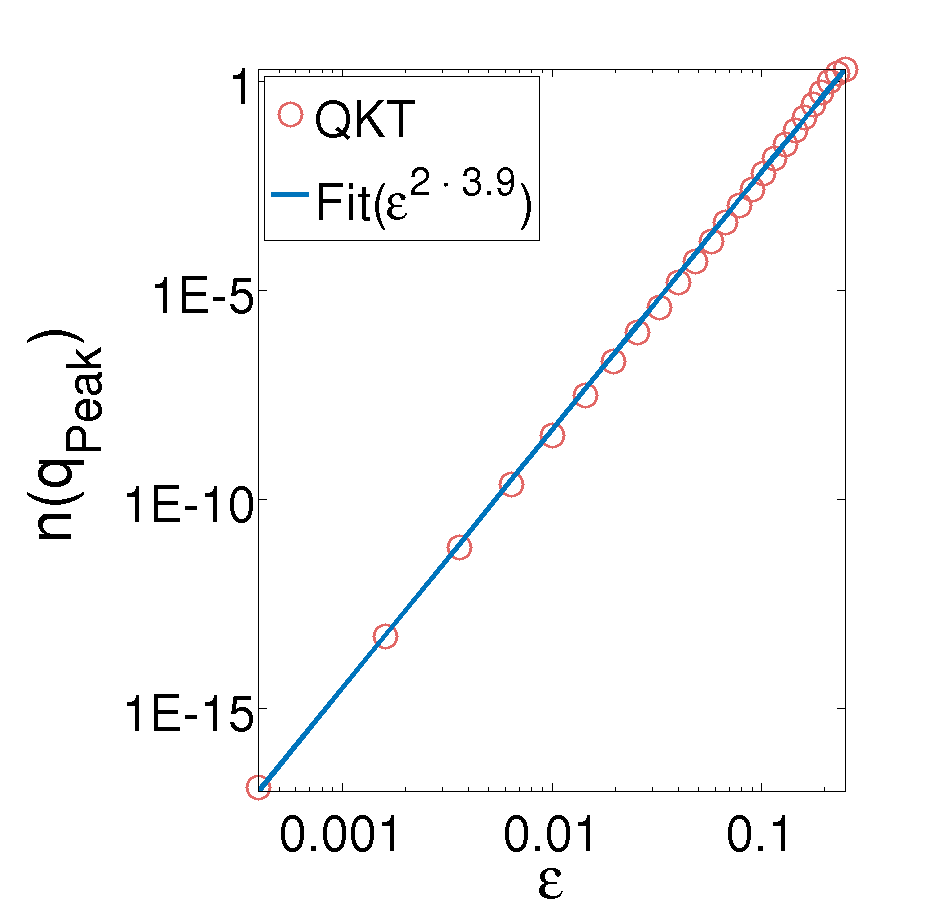}
 \caption[Monitoring the height of multiphoton peaks in momentum space as a function of the field strength.]{Double-log plot of two different peak heights in the momentum spectrum as a function of the field strength $\varepsilon$ for a pulse with $\tau=2000$ $m^{-1}$ and $\omega=0.7$ $m$. Monitoring the peak height and subsequent fitting reveals $3+s$ absorption processes. The fit on the left-hand side displays a $s=0$ absorption process and the right-hand side a $s=1$ process. Further parameters: Tab. \ref{Tab_peak}} 
 \label{Fig_Multiph}
\end{figure}

Additionally, the main peaks can be directly attributed to the $n$-photon thresholds in terms of multiphoton pair production. On the one hand, this interpretation is corroborated by similar findings in the literature\cite{PhysRevA.81.022122}. On the other hand, we can examine the particle spectrum around the corresponding peaks. By monitoring the height of the peaks in the spectrum and due to the fact, that the photon-field strength dependency rate is given by $\varepsilon^{2n}$ we obtain the number of photons involved by a fit as illustrated in Fig. \ref{Fig_Multiph}. 

\clearpage

A more direct way of determining the relevant process is given via comparison with the naive assumption $n \omega = 2 m$ (vertical lines in Fig. \ref{Fig_Yield}). Inspecting the underlying results for Fig. \ref{Fig_Yield}, we find a deviation between calculation and naive prediction. Interestingly, this deviation increases with $n$.  

This puzzle can be resolved using the effective mass model. Although the relation \eqref{Meff} has been derived for plane-waves we still suggest to employ it here. The averaging should then be restricted to one field oscillation around $t=0$. Ignoring the fact, that we work with finite pulses we propose
\begin{equation}
 m_{\ast} = m \sqrt{1+\xi^2} \approx m \sqrt{1 + \frac{e^2 E_0^2}{m^2} \frac{\varepsilon^2}{2 \omega^2}}. \label{Meff_Mod}
\end{equation}
A direct consequence of introducing this model is the fact, that the threshold energy for $n$-photon pair production has to be rewritten. The modified threshold condition then reads
\begin{equation}
 n \omega = 2 m_{\ast}. \label{Meff_Mod2}
\end{equation}

\begin{figure}[htb]
 \includegraphics[width=0.5\textwidth]{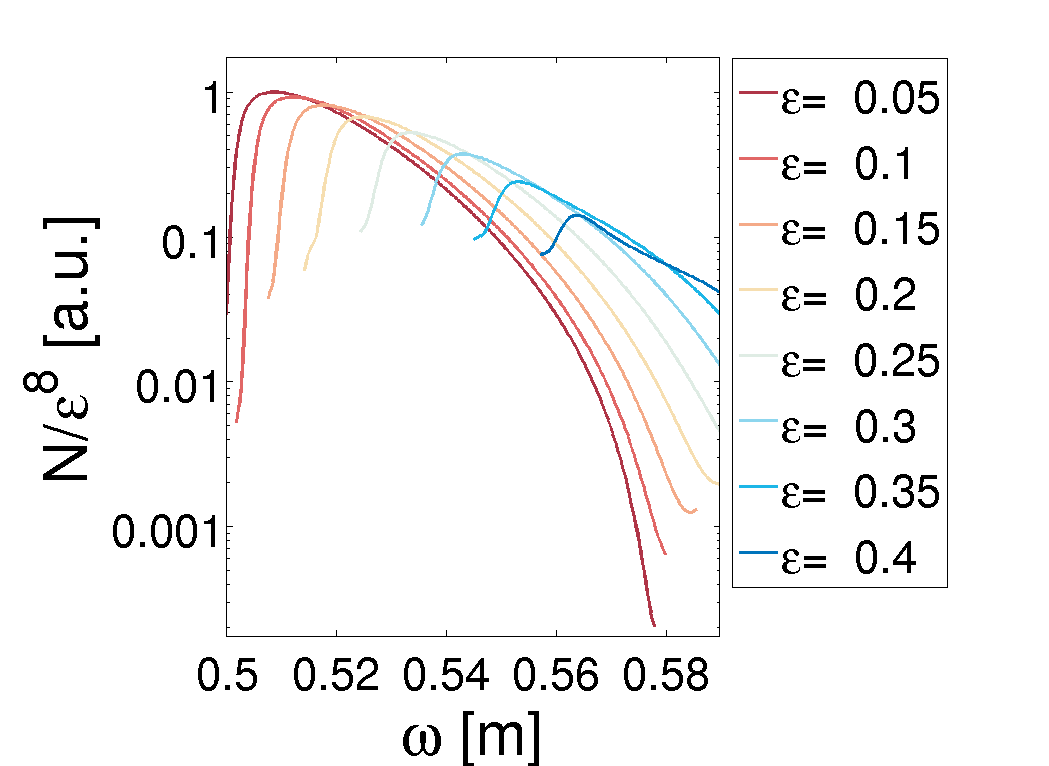}
 \includegraphics[width=0.5\textwidth]{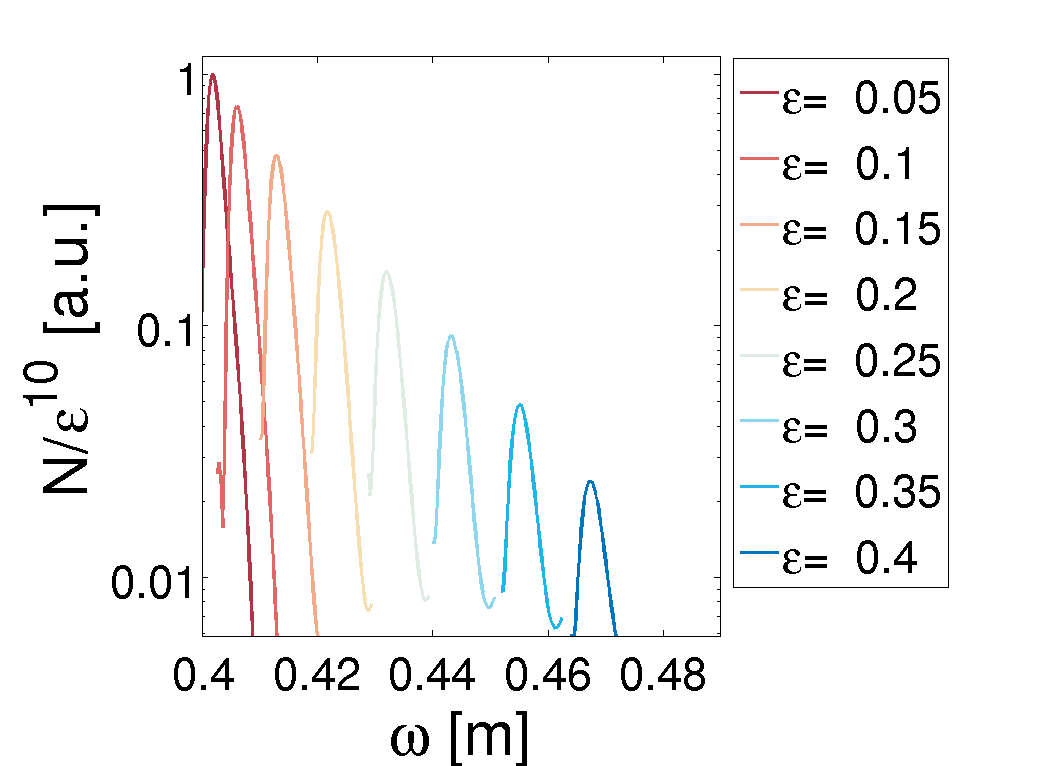}
 \includegraphics[width=0.5\textwidth]{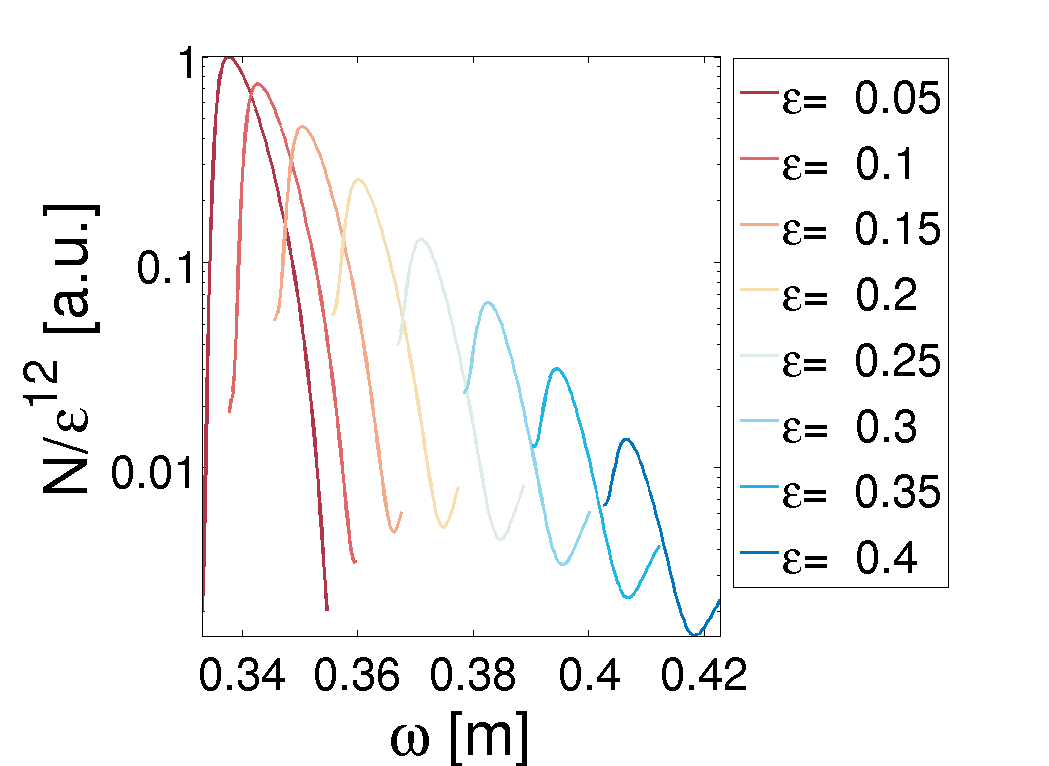}
 \includegraphics[width=0.5\textwidth]{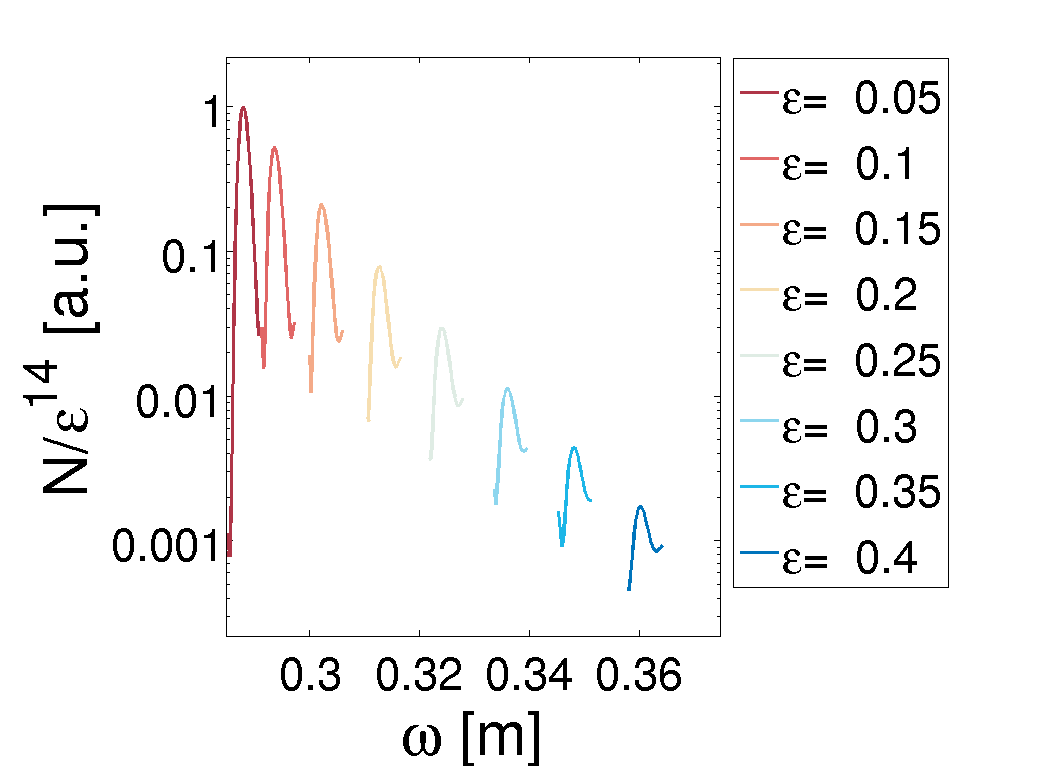} 
 \includegraphics[width=0.5\textwidth]{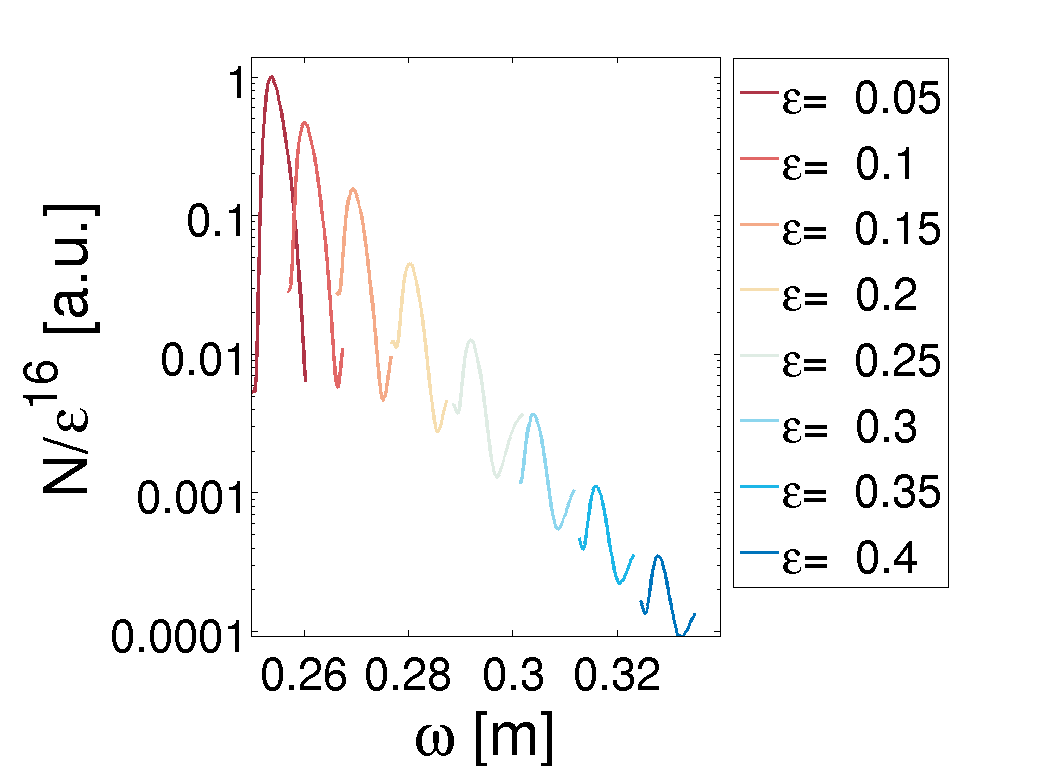}
 \includegraphics[width=0.5\textwidth]{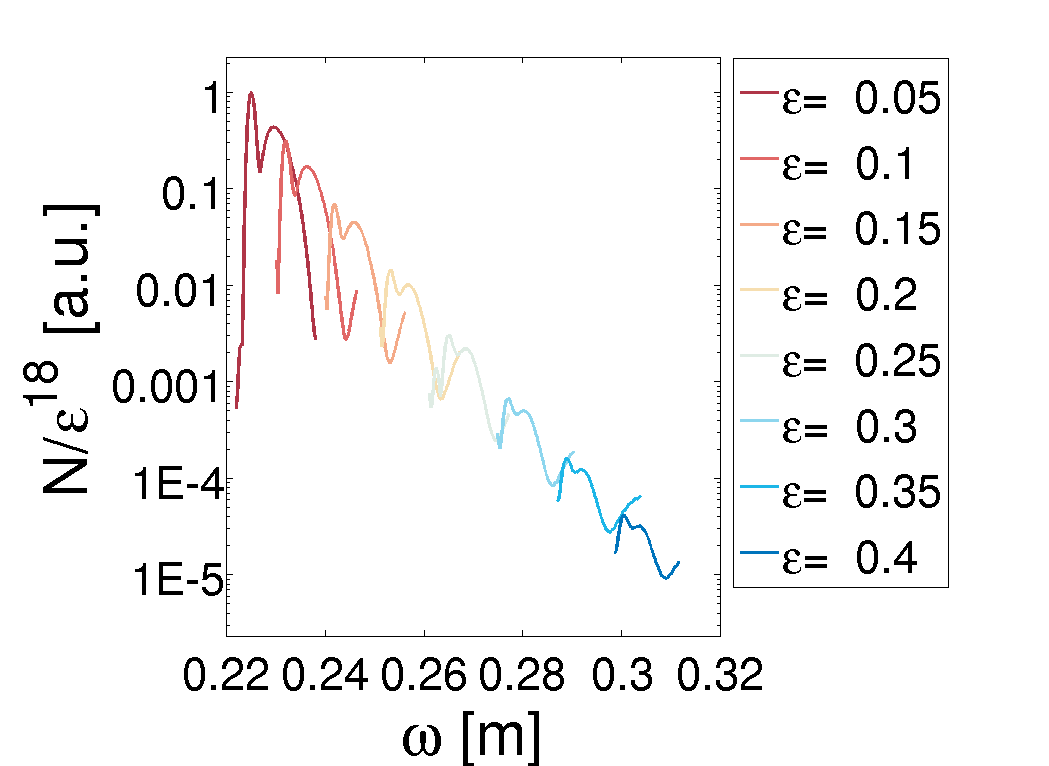} 
 \caption[Normalized particle yield for various $n$-photon absorption processes.]{Normalized particle yield for various $n$ photon absorption processes. The functions are displayed in a logarithmic plot for different photon energy intervals depending on the specified value for $n$. In all plots the main peak indicating a $n$-photon threshold is shifted to higher energies as the peak field strength is increased. For all plots we have used a pulse length of $\tau=800$ $m^{-1}$. The normalization constants are chosen according to the $\varepsilon^{2n}$ photon number relation. Additional information: Tab. \ref{Tab_EMass}} 
 \label{Fig_normYield}
\end{figure}

Combining equations \eqref{Meff_Mod} and \eqref{Meff_Mod2} we are able to resolve the threshold condition in terms of the photon energies
\begin{equation}
 \omega_n = \sqrt{\frac{2 m^2}{n^2} + \sqrt{\frac{4 m^4}{n^4} + \frac{2e^2 E_0^2 \ \varepsilon^2}{n^2}}} \label{Meff_w}
\end{equation}
as well as in terms of the effective mass at the $n$-photon threshold
\begin{equation}
 m_{\ast,n} = \sqrt{\frac{m^2}{2} + \frac{\sqrt{2 m^4 + n^2 e^2 E_0^2 \ \varepsilon^2}}{2 \sqrt{2}} }. \label{Meff_m}
\end{equation}

From the two relations above we obtain two distinctive predictions. According to equation \eqref{Meff_w} the threshold frequencies increase with the field strength. This characteristic feature of a field-dependent threshold is clearly visible in Fig. \ref{Fig_normYield}.  We have displayed all $n$-photon peaks starting with $n=4$ and ending with $n=9$. By this way, one can observe the subtle difference between a $n$-even and a $n$-odd photon process due to charge-conjugation invariance\cite{PhysRevA.81.022122,PhysRevLett.102.080402}. 

\begin{figure}[htb]
 \includegraphics[width=0.5\textwidth]{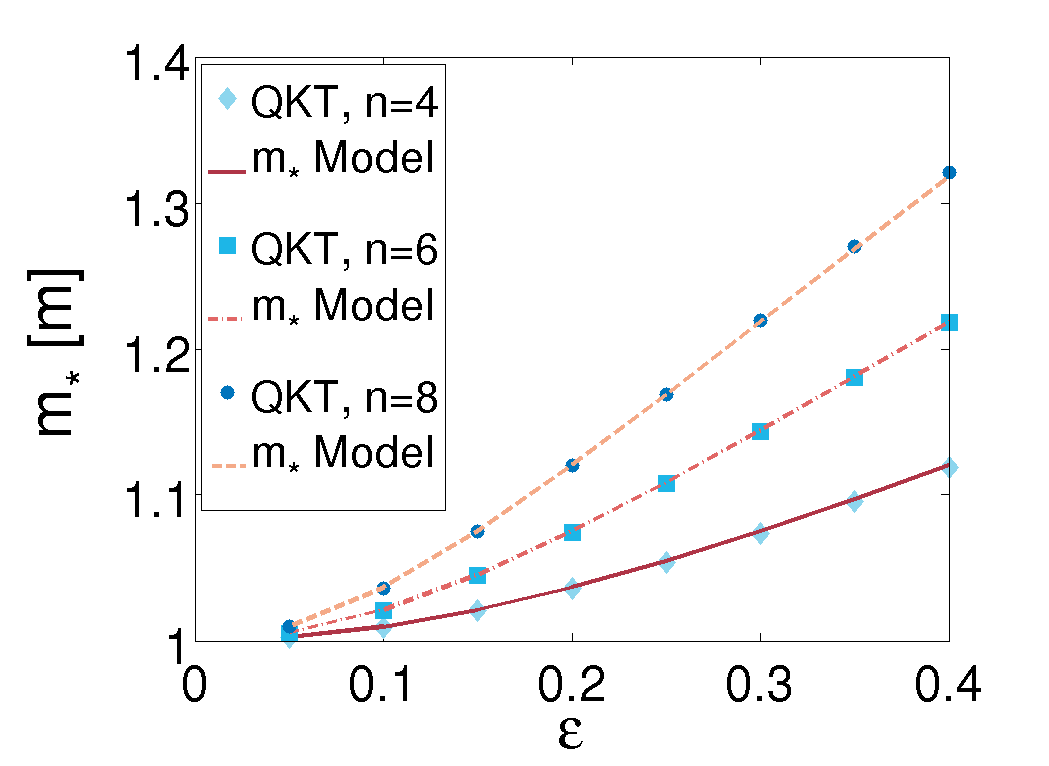}
 \includegraphics[width=0.5\textwidth]{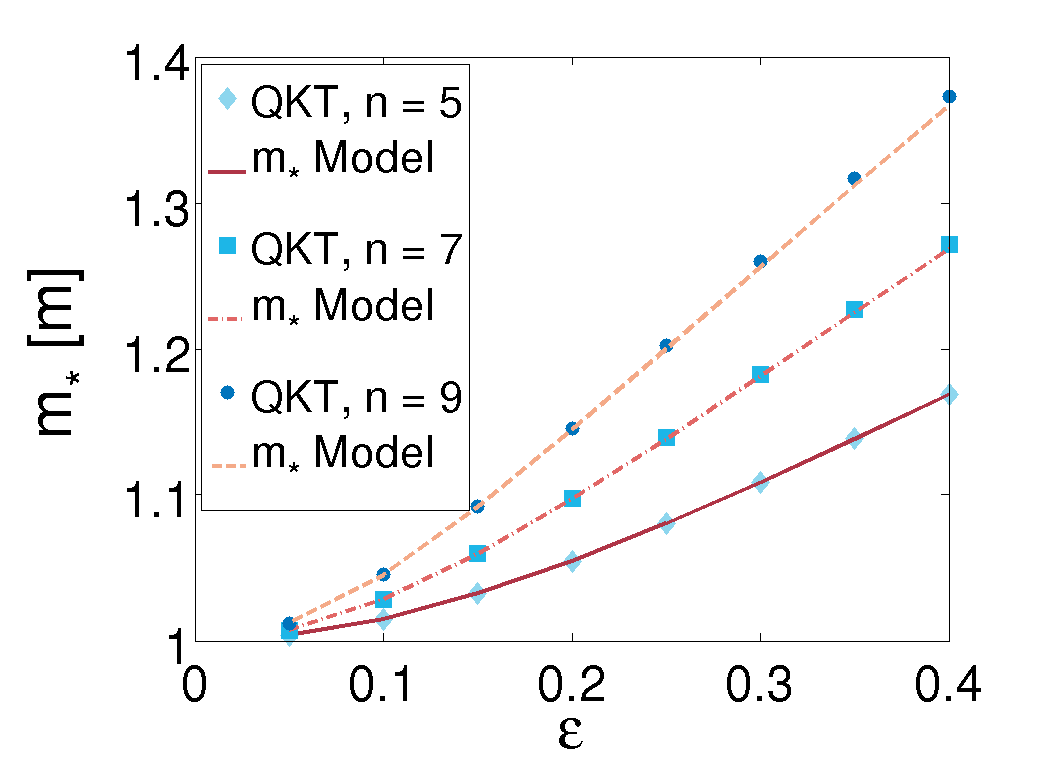}
 \caption[Comparison of the predictions of the effective mass model with the results obtained from a simulation regarding QKT.]{Comparison of the predictions of the effective mass model with the results obtained from simulation. The functions denoted with "QKT" are extracted from the particle yield, i.e.  from a numerical calculation ($\varepsilon_k$ and $m_{\ast}(\varepsilon_k)$). These values match within $1 \%$ tolerance the predictions of the effective mass model. In order to illustrate the results we used a pulse length of $\tau=800$ $m^{-1}$. Further information: Tab. \ref{Tab_EMass}} 
 \label{Fig_effMass}
\end{figure}

Secondly, \eqref{Meff_m} predicts an increase of the effective mass with the photon number $n$ as well as with the field strength. In Fig. \ref{Fig_effMass} we compare between simulation and simply evaluating \eqref{Meff_m}. One observes agreement of all corresponding lines up to high field strengths. Moreover, depending on the chosen parameter values one can see an increase of the effective mass of up to $40 \%$ despite the fact that we have not ``optimized'' the field in any way\cite{PhysRevLett.109.100402}. 

\clearpage

For the sake of completeness, we have to mention that we calculated the lines in Fig. \ref{Fig_normYield} for even photon numbers by simply evaluating the particle yield. For an entirely correct result one would have to compute the values for $m_{\ast}$ by considering the momentum spectrum. Due to charge-conjugation invariance there is a small deviation compared to the value obtained otherwise. This analysis is summarized in Tab. \ref{Tab_Ch6_2a}.

At this point another remark is in order. In Fig. \ref{Fig_Yield} it can be seen, that for photon frequencies slightly above the threshold an additional local maximum is formed. This seems to be related to the fact, that an additional push in order to separate particles from antiparticles is favorable for the total yield. A detailed analysis of this effect is beyond the scope of this thesis, but would be interesting as subject of further investigation.

\clearpage

\ctable[pos=hb,
caption = {Table displaying the results for the effective mass in case of even $n$ photon absorption obtained from particle yield ($m_{\ast,n}$) vs. computation using the particle distribution function ($m_{\ast,q_x}$). The peak in momentum space is located around $q_x \sim 0.15$ $m$.},
cap = {Results for the effective mass in case of even $n$-photon absorption.},
label =Tab_Ch6_2a, 
mincapwidth = \textwidth,
]{ l c c c c c c }{
}{
    \toprule
    $\varepsilon$ & $m_{\ast,4}$ $[m]$ & $m_{\ast,q_x}(n=4)$ $[m]$ & $m_{\ast,6}$ $[m]$ & $m_{\ast,q_x}(n=6)$ $[m]$ & $m_{\ast,8}$ $[m]$ & $m_{\ast,q_x}(n=8)$ $[m]$ \\
    \midrule
    0.05 & 1.0024 & 1.0024 & 1.0055 & 1.0054 & 1.0097 & 1.0095 \\
    \midrule    
    0.1 & 1.0095 & 1.0095 & 1.0211 & 1.0208 & 1.0363 & 1.0358 \\
    \midrule    
    0.15 & 1.0208 & 1.0207 & 1.0448 & 1.0443 & 1.0747 & 1.0738 \\
    \midrule    
    0.2 & 1.0357 & 1.0356 & 1.0744 & 1.0738 & 1.1201 & 1.1186 \\
    \midrule    
    0.25 & 1.0535 & 1.0533 & 1.1078 & 1.1070 & 1.1690 & 1.1670 \\
    \midrule    
    0.3 & 1.0736 & 1.0733 & 1.1435 & 1.1426 & 1.2194 & 1.2168 \\
    \midrule    
    0.35 & 1.0955 & 1.0950 & 1.1805 & 1.1795 & 1.2702 & 1.2669 \\
    \midrule    
    0.4 & 1.1189 & 1.1179 & 1.2183 & 1.2170 & 1.3210 & 1.3166 \\    
    \bottomrule
} 

\subsection{Particle momentum distribution}

In Fig. \ref{Fig_Distr} we have already seen, that the particle distribution function in the regime of multiphoton pair production exhibits characteristic peaks. Moreover, for Fig. \ref{Fig_Multiph} we were monitoring these peak heights in order to demonstrate the validity of the $n$-photon absorption picture. In the following we will focus on the peak positions and analyze their dependence on the field parameters. To this end, we deduce from the fact, that the energy of multiphoton pair production $\mathcal{E}_{(n+s)\gamma} = \mathcal{E}_{e^-} + \mathcal{E}_{e^+}$ is conserved, the following relation:
\begin{equation}
 \br{\frac{(n+s) \omega}{2}}^2 = m_{\ast}^2 + q_{n+s}^2. \label{Meff_En}
\end{equation}
In the relation above, $m_{\ast}$ holds as the particles mass and $q_{n+s}$ gives the canonical momentum of the outgoing particles. In addition $(n+s)$ denotes the number of photons involved, where $n$ represents the number of photons needed in order to reach the threshold for the electric field specified. Due to the fact, that $m_{\ast}$ is a function of the field parameters, see equation \eqref{Meff_m} as well as Fig. \ref{Fig_effMass}, the characteristic momentum $q_{n+s}$ is also field-dependent beyond the photon energy. 

\begin{figure}[htb]
\begin{center}
 \includegraphics[width=0.8\textwidth]{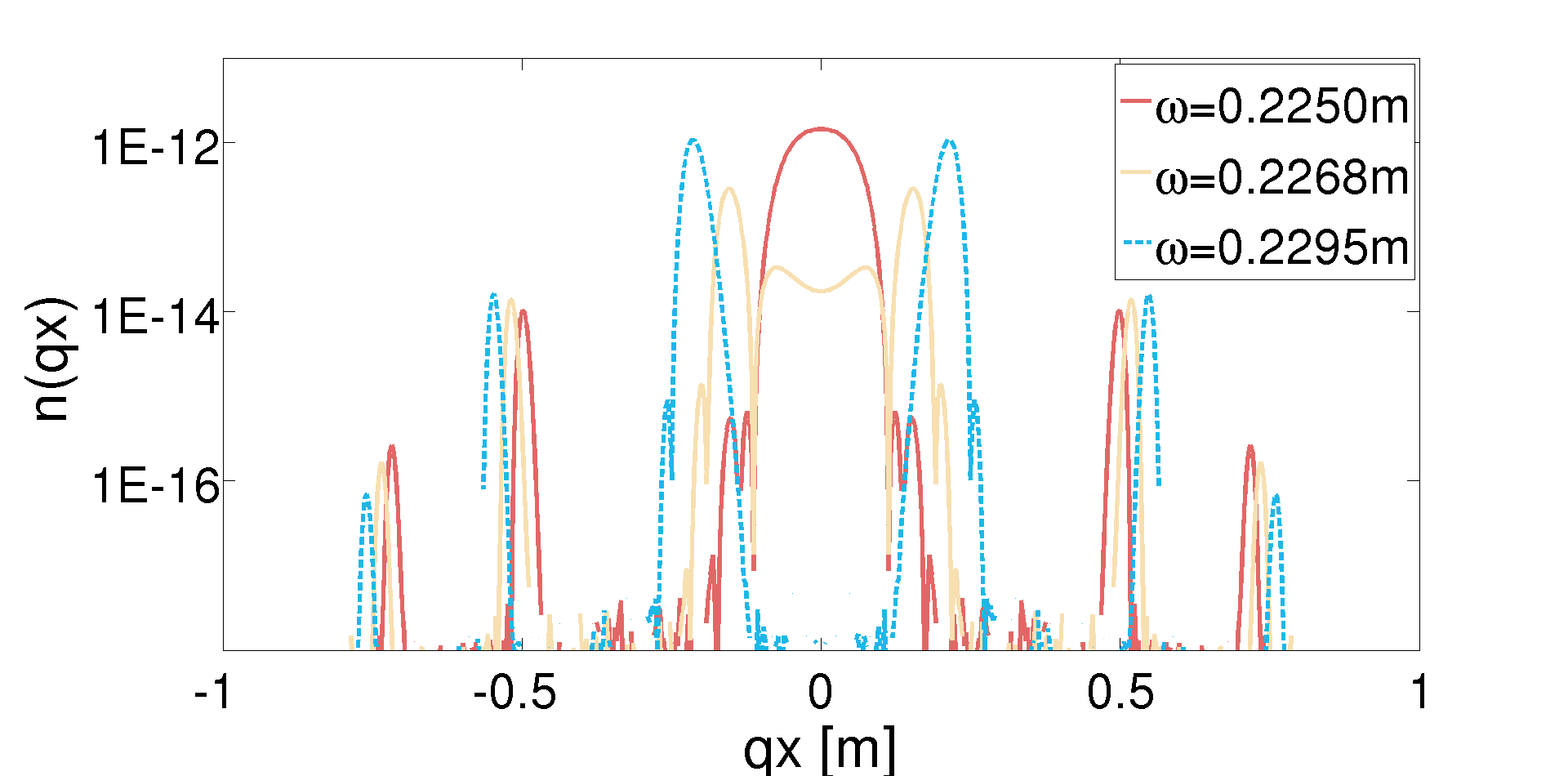}
\end{center}
\caption[Particle momentum spectrum in the regime of multiphoton pair production for various field frequencies.]{Particle momentum spectra in the regime of multiphoton pair production for pulses with $\varepsilon=0.05$ and $\tau=800$ $m^{-1}$. Every peak can be attributed to a $n+s$ absorption process, where $s=0,1,2, \ldots$. The higher the photon energy the higher the boost for the particles created. Further parameters: Tab. \ref{Tab_distr1}} 
\label{Fig_Distr1}
\end{figure}

From the energy conservation relation \eqref{Meff_En} we can read off the $n+s$ photon peaks in momentum space. At first we find, that there are contributions of $n+s$ photon processes to the $n$-photon peak in the yield. Moreover, in case of low field strengths the effective mass is barely distinguishable from the bare mass, see Fig. \ref{Fig_effMass}. As a result all peaks starting with the $n$-photon peak are visible in momentum space. Additionally, a $n+s$-photon peak is approximately separated from the $n+s+1$ peak by the photon energy $\omega$, see Fig. \ref{Fig_Distr1}.

\begin{figure}[htb]
 \begin{center}
   \includegraphics[width=0.6\textwidth]{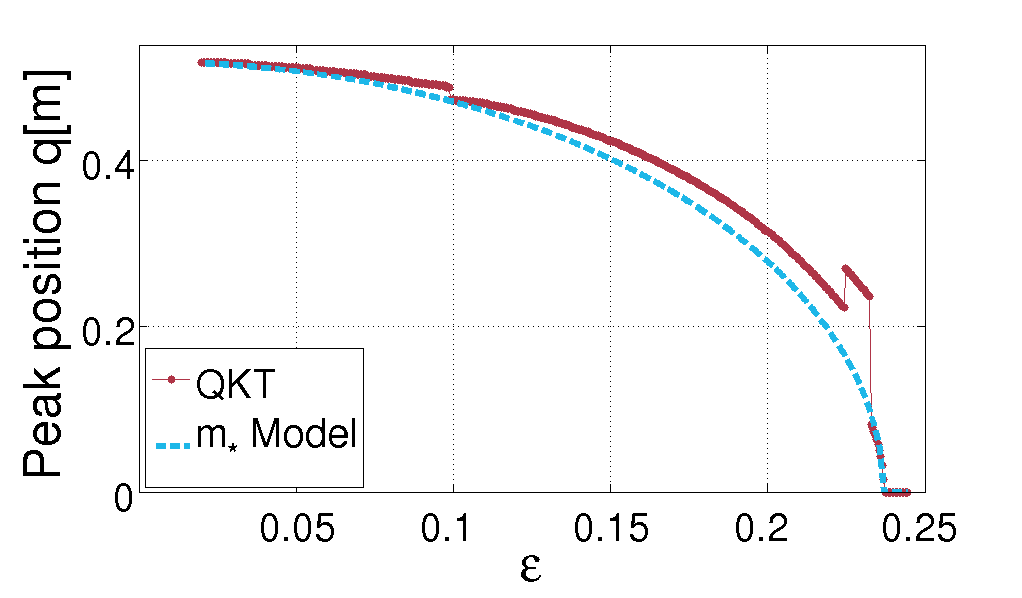}
 \end{center}   
\caption[Displaying channel closing by monitoring the $n+s=7$ photon peak position as a function of the field strength.]{Displaying channel closing by monitoring the $n+s=7$ photon peak position as a function of the field strength $\varepsilon$. The pulse length of the field is $\tau=1000$ $m^{-1}$ and the photon frequency $\omega=0.322$ $m$. Further parameters are provided in Tab. \ref{Tab_channel}}  
\label{Fig_Channel}
\end{figure}

However, this picture changes drastically in case of high field strengths and therefore high effective masses. This is due to the fact, that an increase of the effective mass $m_{\ast}$ for fixed photon energy $\omega$ inevitably results in a decrease of the characteristic momentum $q_{n+s}$. As a consequence, the characteristic photon peaks in momentum space move towards vanishing momentum as illustrated in Fig. \ref{Fig_Channel}. Close to the threshold of pair production we observe, that discontinuities appear. This happens, because the peak as a whole decreases and splits up resulting in multiple small peaks superimposing the main peak. As this renders a clear determination of the peak position impossible, we decided to plot only the position of the highest local maximum, thus the discontinuous behaviour.
N.B.: A similar effect is observable in Fig. \ref{Fig_Distr1} for $\omega = 0.2268$ $m$ around $q_x \approx 0$.

At one point $q_{n+s}$ becomes imaginary, because the photon energy from $n+s$ photons is not sufficient in order to produce particles with mass $m_{\ast}$. As a consequence the peak in the momentum distribution function vanishes entirely, an effect known under the name ``channel closing'' in atomic ionization\cite{Kopold}. This phenomenon can be accurately described via the effective mass model. The shift of the peaks in momentum space are nicely covered by \eqref{Meff_En}. Moreover, it makes the location of the threshold predictable making channel closing an interesting observable due to its sensitivity on the effective mass.

To summarize, we employed QKT in order to compute the effects of an effective mass in the regime of multiphoton pair production. Although this concept is not universal, we have identified various observables that show signatures of an effective particle mass.

\section{Pair production at extreme parameter regions}
\label{Sec_Xtr}
We have already investigated the concept of an effective mass in great detail. However, we limited the computations to parameter regions, that are numerically feasible but not experimentally accessible. In this section we want to put emphasize on ``extreme'' field configurations. In this way we are checking whether the investigations above are also valid for conditions, that could become within experimental reach in the near future.

\begin{figure}[htb]
 \includegraphics[width=0.5\textwidth]{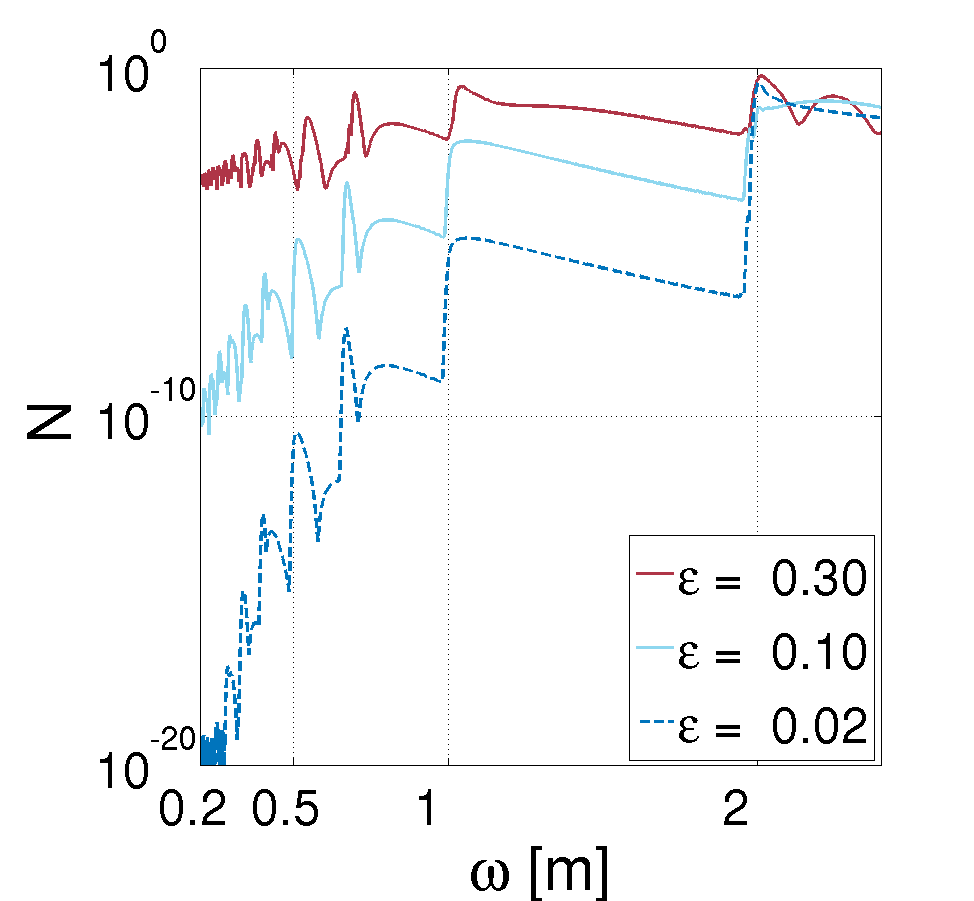}
 \includegraphics[width=0.5\textwidth]{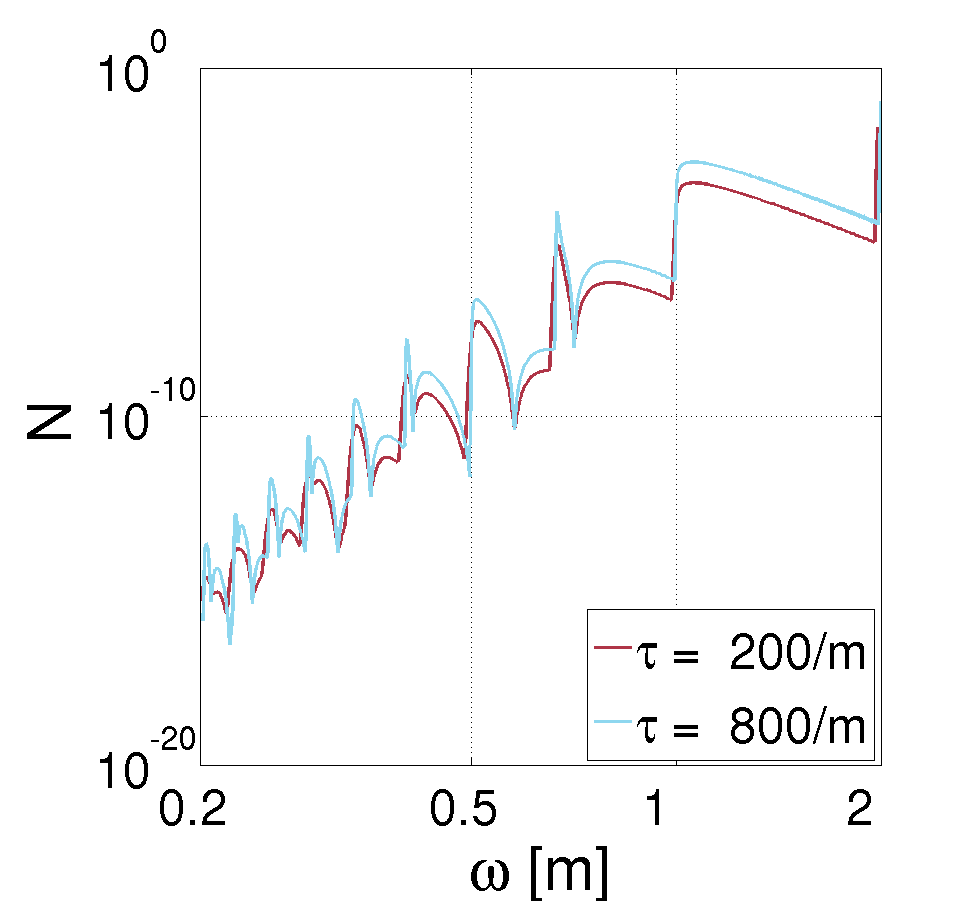} 
\caption[Plots of the particle yield for various field strengths and pulse lengths.]{Logarithmic plots of the particle yield for various field strengths (left-hand side) and pulse lengths (right-hand side). The characteristic structure of multiphoton pair production is conserved for the parameter sets given. Left: $\tau=200$ $m^{-1}$ and Tab. \ref{Tab_eps}. Right: $\varepsilon=0.05$ and Tab. \ref{Tab_yield_tau}.}  
\label{Fig_YieldX}
\end{figure}

We have illustrated in Fig. \ref{Fig_YieldX} the implications of various field strengths $\varepsilon$ and pulse lengths $\tau$ on the particle creation process. The most important indicator for the validity of the investigations in the previous section is given by the overall structure of the particle yield. 

From the definition of the Keldysh parameter, see section \ref{Sec_His}, $\gamma=\br{m \omega}/\br{e \varepsilon E_0}$ we can already deduce, that an increase of the field strength directly leads to a decrease of the dominance of multiphoton pair production. The Schwinger effect becomes more important, thus the multiphoton regime can be clearly identified for higher frequencies $\omega$ only. This is also reflected in Fig. \ref{Fig_YieldX} as the multiphoton peaks, easily determinable for $\varepsilon=0.02$, become noisy for higher field strengths. However, careful analysis shows, that the concept of an effective mass is still legitimate although complicated in the region of high $n$.     

Current Laser systems usually operate at the femtosecond scale, while the characteristic scale for pair production is the Compton wavelength. In order to extrapolate from the QED scale to the Laser scale we compare the particle yield for a pulse of $\tau=200$ $m^{-1}$ with a pulse of $\tau=800$ $m^{-1}$, see Fig. \ref{Fig_YieldX}. The Keldysh parameter proposed is independent of the pulse length, therefore we do not expect any further restrictions on the validity of the effective mass concept. 

\vspace{5cm}

However, due to the extended pulse length the overall number of field oscillations per pulse increases. Hence, for a given field frequency $\omega$, a few-cycle pulse can turn into a many-cycle pulse by simply increasing the pulse length. By this way, multiphoton signatures become pronounced. The corresponding consequences can be directly observed in  Fig. \ref{Fig_YieldX}. 

\begin{figure}[htb]
\begin{center}
 \includegraphics[width=0.5\textwidth]{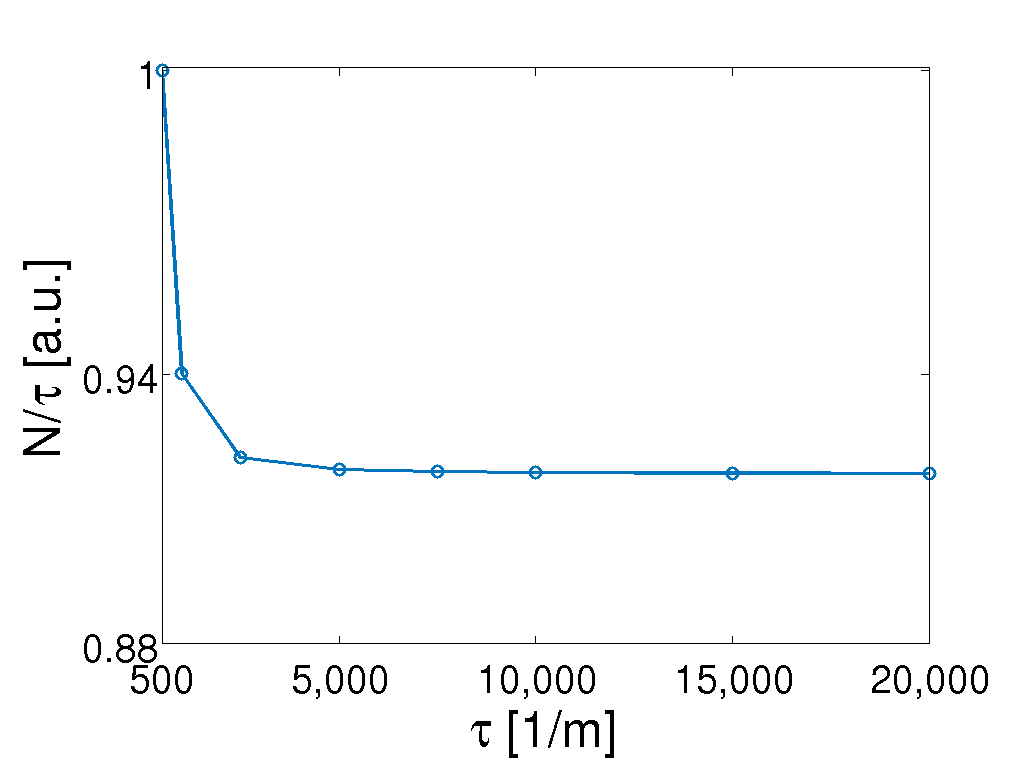}
\end{center}
\caption[Plot of the normalized particle yield for extremely long pulse lengths.]{Plot of the normalized particle yield as a function of the pulse length $\tau$ for $\varepsilon=0.01$ and $\omega=0.7$ $m$. For longer pulses the particle yield depends linearly on $\tau$. Additional parameters: Tab. \ref{Tab_tau}} 
\label{Fig_Tau}
\end{figure}

We have thus verified, that the concept of multiphoton pair production, effective mass etc. is still valid also for long-pulsed fields. The important question is now whether a quantitative extrapolation to femtosecond pulses is possible. To do this, we have performed simulations for various pulse lengths $\tau$, but fixed field strength $\varepsilon$ and field frequency $\omega$. The results are presented in Fig. \ref{Fig_Tau}, where we additionally normalized our findings by the pulse length. We find, that after passing a critical pulse length of roughly $\tau \approx 1000$ $m^{-1}$(for a specific parameter set) the normalized particle yield $N/ \tau$ becomes constant. Hence, the particle yield increases linearly with the pulse length for this parameter set. In order to answer the question proposed above, it should be feasible to extrapolate the findings to experimentally accessible regions as long as all of the further restrictions are met (low total particle number to avoid Pauli-blocking etc.). 

At this point, we have to discuss the implications of field frequencies exceeding the threshold for pair production. A field frequency of $\omega \approx 2$ $m$ plays a special role as one directly operates at the $1$-photon particle threshold. At least at this point backreaction\cite{PhysRevD.87.105006} cannot be neglected any more.
Additionally, QKT does not cover particle creation beyond one particle-antiparticle pair and also the Hartree approximation becomes questionable at this scale. 

\vspace{5cm}

\begin{figure}[htb]
\begin{center}
 \includegraphics[width=0.6\textwidth]{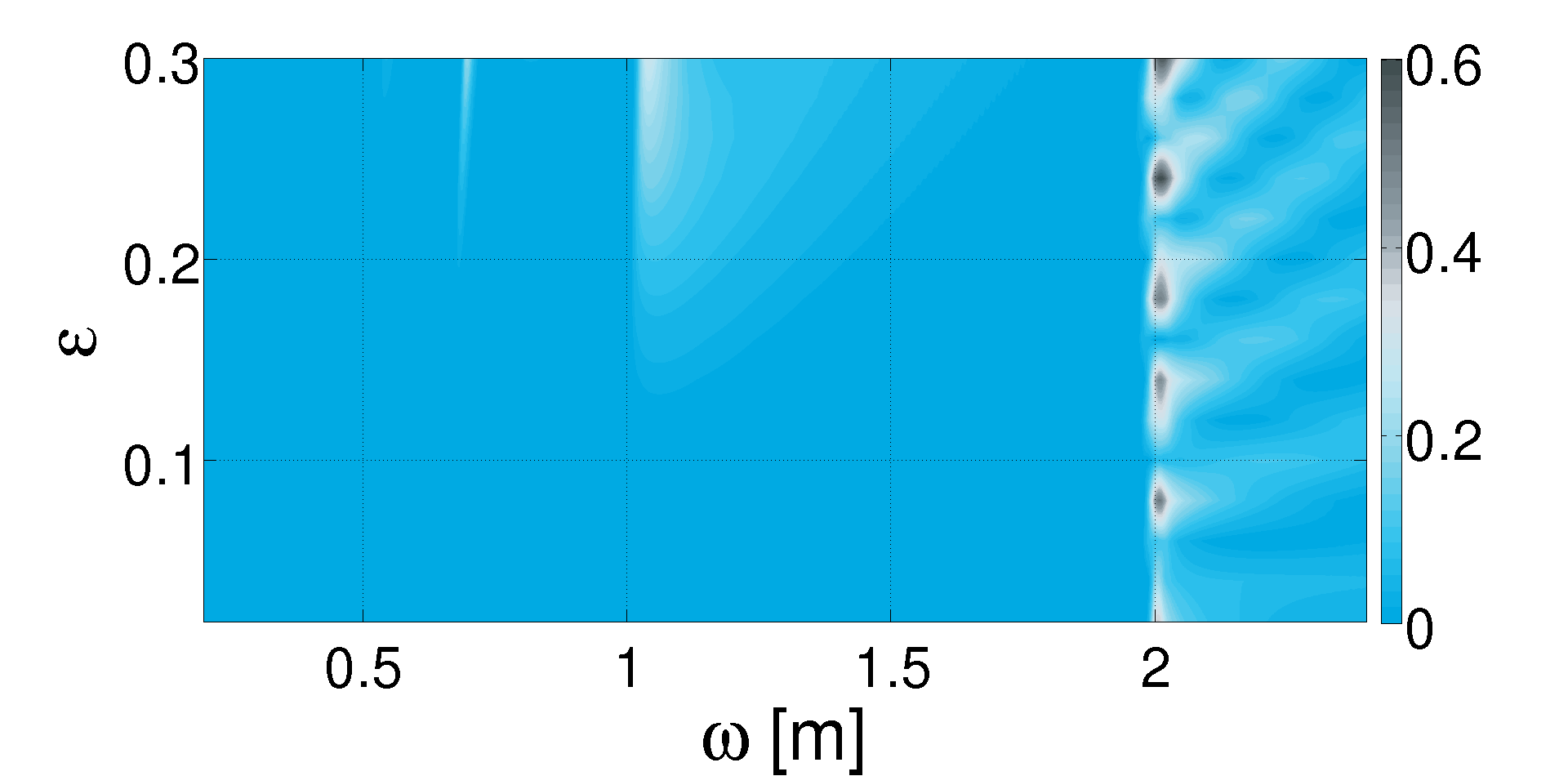} 
\end{center}
\caption[2D plot of the particle yield as a function of the field strength and the field frequency.]{2D plot of the particle yield as a function of the field strength $\varepsilon$ and the field frequency $\omega$, while $\tau=200$ $m^{-1}$. For increasing photon energy and high field strength opening up of the $n$ photon peaks is observable. Further parameters: Tab. \ref{Tab_eps}}  
\label{Fig_YieldX2}
\end{figure}

In Fig. \ref{Fig_YieldX2} one can nicely observe the onset of $n$-photon thresholds for increasing field strengths. However, the pattern at $\omega \gtrsim 2$ is not reliable as the assumptions underlying our calculations break down. Hence, we do not present any results with frequencies $\omega > 2m$ in this thesis.

\pagestyle{plain}
\chapter{Spatially inhomogeneous electric fields}
\pagestyle{fancy}
\label{Sec_InhField}
Throughout this chapter we consider spatially inhomogeneous electric fields still neglecting  magnetic fields. In this way, it is possible to focus on the consequences of a finite spatial extent. Exploiting cylindrical symmetry allows to employ the DHW formalism in a full pseudo-differential operator approach.

\section[Spatial focusing, full-operator DHW approach and background fields]{Spatial focusing}

The main point in the following studies is to investigate the impact of spatial focusing on the pair production process. In order to provide intensities capable of creating an experimentally verifiable amount of particles spatial focusing is of central importance. However, only little is known about ``optimal'' focusing. 

\subsection*{Full-operator DHW approach and background fields}

Investigations regarding spatially inhomogeneous electric fields require DHW transport equations beyond simple QKT. However, concentrating on field configurations where the magnetic field component vanishes limits the amount of possible models for the background field. We decided to use a pulse taking the form
\begin{equation}
 \mathbf{E} \br{x,t} = \varepsilon \ E_0 \ \exp \br{-\frac{x^2}{2 \lambda^2}} \ \cos^4 \br{\frac{t}{\tau}} \ \cos \br{\omega t} \mathbf{e}_x, \label{Cyl_E}
\end{equation}
where $t \in \com{-\frac{\pi \tau}{2},\frac{\pi \tau}{2}}$.
The variables $\varepsilon$, $\tau$, $\lambda$ and $\omega$ define the peak field strength, the total pulse length, the spatial extent and the photon frequency, respectively. By this way, we want to mimic the finite focus of a Laser pulse. Introducing a spatial extent of the form $\exp \br{-\frac{x^2}{2 \lambda^2}}$ offers the advantage of a full treatment of the pseudo-differential operators and enables us to compare with results in the literature \cite{PhysRevLett.107.180403}. A more realistic focus as presented in section \ref{Kap_Exp} as well as incorporation of transversal spatial extents are beyond the scope of this investigation.  

Due to the specific model \eqref{Cyl_E} we use, it is possible to employ the DHW formalism for cylindrically symmetric fields. The corresponding transport equations \eqref{eq3_Cy1}-\eqref{eq3_Cy4} read
\begin{alignat}{5}
  & D_t \ \br{\overline{\mathbbm{s}}^v-\frac{2m}{\omega}}     && && -2 p_x \overline{\mathbbm{p}}^v &&+ 2 p_{\rho} \overline{\mathbbm v}^v &&= 0, \label{Cyl_Tr1} \\
  & D_t \ \br{\overline{\mathbbm{v}}^v-\frac{2 p_x}{\omega}} &&+\de{x} \overline{\mathbbm{v}}^v_0 && &&- 2 p_{\rho} \overline{ \mathbbm{s}}^v &&= -2m\overline{\mathbbm{p}}^v,  \\    
  & D_t \ \br{\overline{\mathbbm{p}}^v-\frac{2 p_{\rho}}{\omega}} && && +2 p_x \overline{\mathbbm{s}}^v && &&= 2m \overline{\mathbbm{v}}^v,  \\ 
  & D_t \quad \ \overline{\mathbbm{v}}^v_0 &&+ \de{x} \overline{\mathbbm{v}}^v && && &&= 0, \label{Cyl_Tr4}
\end{alignat}
with the one-particle energy $\omega=\sqrt{m^2+p_x^2+ p_{\rho}^2}$ and the pseudo-differential operator
\begin{align}
 D_t = \de{t} + e\int d\xi E \br{x+\ii \xi \de{p_x},t} \de{p_x}.
\end{align}
Moreover, we have employed vacuum initial conditions
\begin{equation}
 \overline{\mathbbm{s}}^v = \overline{\mathbbm{v}}^v = \overline{\mathbbm{p}}^v = \overline{\mathbbm{v}}^v_0 = 0.
\end{equation}
The results presented in the following have been obtained through calculating the particle distribution function
\begin{alignat}{5}
  &\mathcal{N} &&= \ \int dx \ dp_x \ d p_{\rho} \ n \br{x,p_x,p_{\rho}} \\ 
  & &&= \int dx \ dp_x \ d p_{\rho} \ \frac{ m \overline{\mathbbm{s}}^v \br{x,p_x,p_{\rho}} + 
 p_x \ \overline{\mathbbm{v}}^v \br{x,p_x,p_{\rho}} + p_{\rho} \ \overline{\mathbbm{p}}^v \br{x,p_x,p_{\rho}}}{\omega \br{p_x,p_{\rho}}}.
\end{alignat}
Note the definition above holds for the particle distribution obtained for the system \eqref{Cyl_Tr1}-\eqref{Cyl_Tr4} only. If one is interested in the proper observable quantity one has to multiply with an additional factor of $2$, see section \ref{sec_observables}. Similarly to the previous chapter, we denote the parallel momentum with $px$ and the transversal momentum with $p \rho$ in the figures.

\section[Ponderomotive forces for pair production in $3+1$ dimensions]{Ponderomotive forces}

A spatially inhomogeneous background field enables us to study particle drifts. This is an important point, because we have to take plasma effects into account in order to describe real experimental situations. This is due to particle-field interactions after the creation process. However, within the DHW approach we do not have control over the pair production process at intermediate times. 
Employing a semi-classical picture is of great help shedding light on this issue. In this way, we can relate the findings obtained from investigating pair production with plasma physics in order to describe subsequent plasma dynamics. Naturally, this approach cannot cover all quantum effects, but it helps interpreting the results obtained.

The interaction of relativistic particles with high-intensity Laser pulses requires a relativistic treatment. An approach describing Laser-matter interactions and in particular expressions for the relativistic ponderomotive force can be found in the literature\cite{PhysRevLett.75.4622,Bauer,Manheimer}.
A concept proven useful in order to determine this force is again given by the effective mass.
Expanding the definition of the effective mass \eqref{Meff} by a spatial dependency of the electric field we obtain
\begin{equation}
 m_{\ast} \br{x} = m \sqrt{1+\tilde{\xi}\br{x}^2}, \quad \text{with} \quad \tilde{\xi}\br{x}^2 = \frac{e}{m} \sqrt{- \langle A_{\mu} \br{x,t} A^{\mu} \br{x,t} \rangle }.
\end{equation}
In case of the model \eqref{Cyl_E} this yields
 \begin{equation}
 m_{\ast} \br{x} \approx m \sqrt{1 + \frac{e^2 E_0^2}{m^2} \frac{\varepsilon^2}{2 \omega^2} \ \tilde E \br{x}^2 },
\end{equation}
where 
\begin{equation}
 \tilde E \br{x} = \exp \br{-\frac{x^2}{2 \lambda^2}}.
\end{equation}
The definition above allows us to easily motivate the introduction of forces related to the spatial inhomogeneity of the applied field. 
The ponderomotive four-force has been already obtained in Bauer et al.\cite{PhysRevLett.75.4622} reading:
\begin{equation}
 F_{pond} = -\br{\mathbf{v}_0 \cdot \boldsymbol{\nabla}_x m_{\ast},~ \boldsymbol{\nabla}_x m_{\ast} + \frac{\gamma_0 -1}{v_0^2} \ \br{\mathbf{v}_0 \cdot \boldsymbol{\nabla}_x m_{\ast} } \ \mathbf{v}_0}. \label{Pond_Rel}
\end{equation}
The quantity $\mathbf{v}_0$ denotes the oscillation center speed, which basically is the speed of a quasi-particle of mass $m_{\ast}$, and $\gamma_0$ is the corresponding Lorentz factor. One immediately observes, that the driving force is given by the spatial derivative of the effective mass. In case of the model \eqref{Cyl_E}, this yields
\begin{equation}
 \boldsymbol{\nabla}_x m_{\ast} = -\frac{e^2 E_0^2 \ \varepsilon^2}{4 \ m_{\ast} \ \omega^2} \ \de{x} \br{E \br{x}^2} \ \mathbf{e}_x = -\frac{e^2 E_0^2 \ \varepsilon^2}{2 \ m_{\ast} \ \omega^2} \ \frac{x}{\lambda^2} \exp \br{-\frac{x^2}{\lambda^2}} \ \mathbf{e}_x.
\end{equation}
N.B.: The general non-relativistic limit($\varepsilon/\omega \ll 1 ,~ \gamma_0 \to 1$), which was already given in chapter four, yields
\begin{equation}
 \mathbf{F}_{pond} = -\frac{e^2 E_0^2 \ \varepsilon^2}{4 \ m \ \omega^2} \ \boldsymbol{\nabla}_x \br{\mathbf{E} \br{x}^2}.
\end{equation}

\begin{figure}[htb]
\begin{center}
 \includegraphics[width=0.55\textwidth]{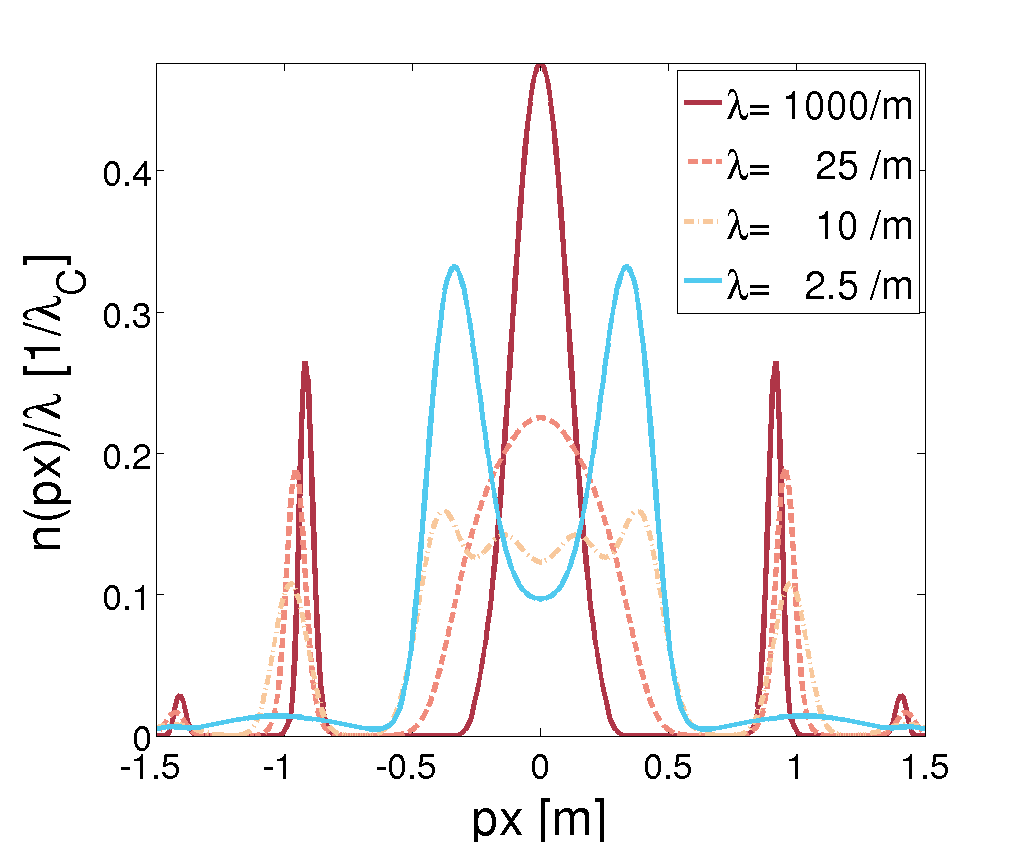}
 \end{center}
 \caption[Plot of the reduced particle distribution function in momentum space for various spatial extents.]{Plot of the reduced particle distribution function in $1D$-momentum space. The peaks at $\lambda=1000$ $m^{-1}$ are related to $n+s$-photon pair production. The lower $\lambda$ the stronger the ponderomotive force. An additional boost in momentum for lower $\lambda$ is observable. Pulse parameters: $\varepsilon=0.5$, $\tau=100$ $m^{-1}$, $\omega=0.7$ $m$. Additional information: Tab. \ref{Tab_distr2}} 
 \label{Fig_Pond1}
\end{figure}

One consequence of a spatially inhomogeneous background field is the additional particle acceleration due to ponderomotive forces. One expects, that the particles created are pushed away from the high-field to weak-field regions of the electric field. Additionally, the force \eqref{Pond_Rel} shows a maximum at $x=\lambda$ and its implications are observable for small spatial extent only, see Fig. \ref{Fig_Pond1} for an illustration. \\

As a remark, in equation \eqref{Pond_Rel} one obtains a second term depending on the speed of the oscillation center. Hence, particles with different transversal momentum interact differently with the applied background field. Signatures of this additional force term will be analyzed below, where we present results for $3+1$ dimensional problems.

%

In order to quantify our findings, we rely on a semi-classical interpretation. We assume, that particle pair production occurs at times when the absolute value of the applied electric field exhibits a local maximum. At first we have a look at a $n$-photon process, where particles are produced with momentum close to $p_x \approx 0$. If the field parameters have been chosen properly, this particle would stay within the strong field region until the background field vanishes. As long as the electric field lasts it forces the particle to follow its periodic oscillations. However, due to fact that the field is spatially inhomogeneous the force acting on the particle is different at different positions in space-time. Imagine the particle is created at $t_0=0$ and $x_0=0$ the electric field forces the particle in a certain direction. At the turning point the field strength has to be evaluated at $t_1=\pi/ \omega$ and $x=x(t_1)$, which is certainly lower than the field strength in the beginning. Next, one has to compare the fields at times $t_2=2\pi/ \omega$ vs. $t_3=3\pi/ \omega$.
This consideration holds for the whole particle trajectory, thus the particle is shifted towards a weak-field region resulting in a net force observable in the particle momentum. This effect is easily observable in Fig. \eqref{Fig_Pond1}.

Additionally, in Fig. \ref{Fig_Pond1} for $ \lambda=2.5$ $m^{-1}$ the first above-threshold peak at around $p_x \approx 0.9$ $m$ vanishes. This observation can be related to the fact, that in a semi-classical interpretation  particles could be kicked out of the high-field region within a few field cycles. We assume, that a particle with high initial momentum obtains an additional boost from the electric field. When the oscillating field changes sign this particle has already moved to a position $\tilde x$ where the electric field is comparatively weak. This weak field is not capable of reversing the direction of the particles movement. Hence, the particle will just move away from the strong field region effectively interfering only with a few-cycle pulse. Moreover, particles with high momentum created at different times experience the same effect, thus the possibilities of interfering particle trajectories are also limited. In this way, the high particle kinetic energy prevents a development of clear patterns in the distribution function, see Tab. \ref{Tab_Inh}. 

\ctable[pos=ht,
caption = {Evaluating the relativistic Lorentz force equation for particles seeded at $t_0(t=0)$ at position $x_0=0$ for a field of strength $\varepsilon=0.5$, length $\tau=100$ $m^{-1}$ and frequency $\omega=0.7$ $m$. The spatial extent $\lambda$ as well as the initial momentum $p_0$ have been varied. The final momenta $p_f$ are obtained at times $t_f(t=\pi/2 \tau)$ when the applied field already vanishes. Certain values are highlighted to show the impact of ponderomotive forces.},
cap = {Evaluating the relativistic Lorentz force equation in order to demonstrate the effect of ponderomotive forces in $1+1$ dimensions.},
label = Tab_Inh, 
mincapwidth = \textwidth,
]{ c c c}{
}{
    \toprule
    \hspace{1cm} $\lambda$ $[m^{-1}]$ \hspace{1cm} & \hspace{1cm} $p_0$ $[m]$ \hspace{1cm} & \hspace{1cm} $p_f$ $[m]$ \hspace{1cm} \\
    \midrule
    1000 & 0 & $10^{-7}$ \\  
    \midrule
    25 & 0 & 0.015 \\ 
    \midrule
    10 & 0 & 0.162 \\ 
    \midrule
    \hlight{2.5} & 0 & \hlight{0.444} \\    
    \midrule
    \midrule
    1000 & 0.92 & 0.92 \\  
    \midrule
    25 & 0.92 & 0.97 \\ 
    \midrule
    10 & 0.92 & 0.99 \\ 
    \midrule
    \hlight{2.5} & 0.92 & \hlight{1.12} \\    
    \midrule
    \midrule
    1000 & 1.4 & 1.4 \\  
    \midrule
    25 & 1.4 & 1.42 \\ 
    \midrule
    10 & 1.4 & 1.43 \\ 
    \midrule
    \hlight{2.5} & 1.4 & \hlight{1.65} \\     
    \bottomrule
} 

\begin{figure}[htb]
\begin{center}
 \includegraphics[width=0.8\textwidth]{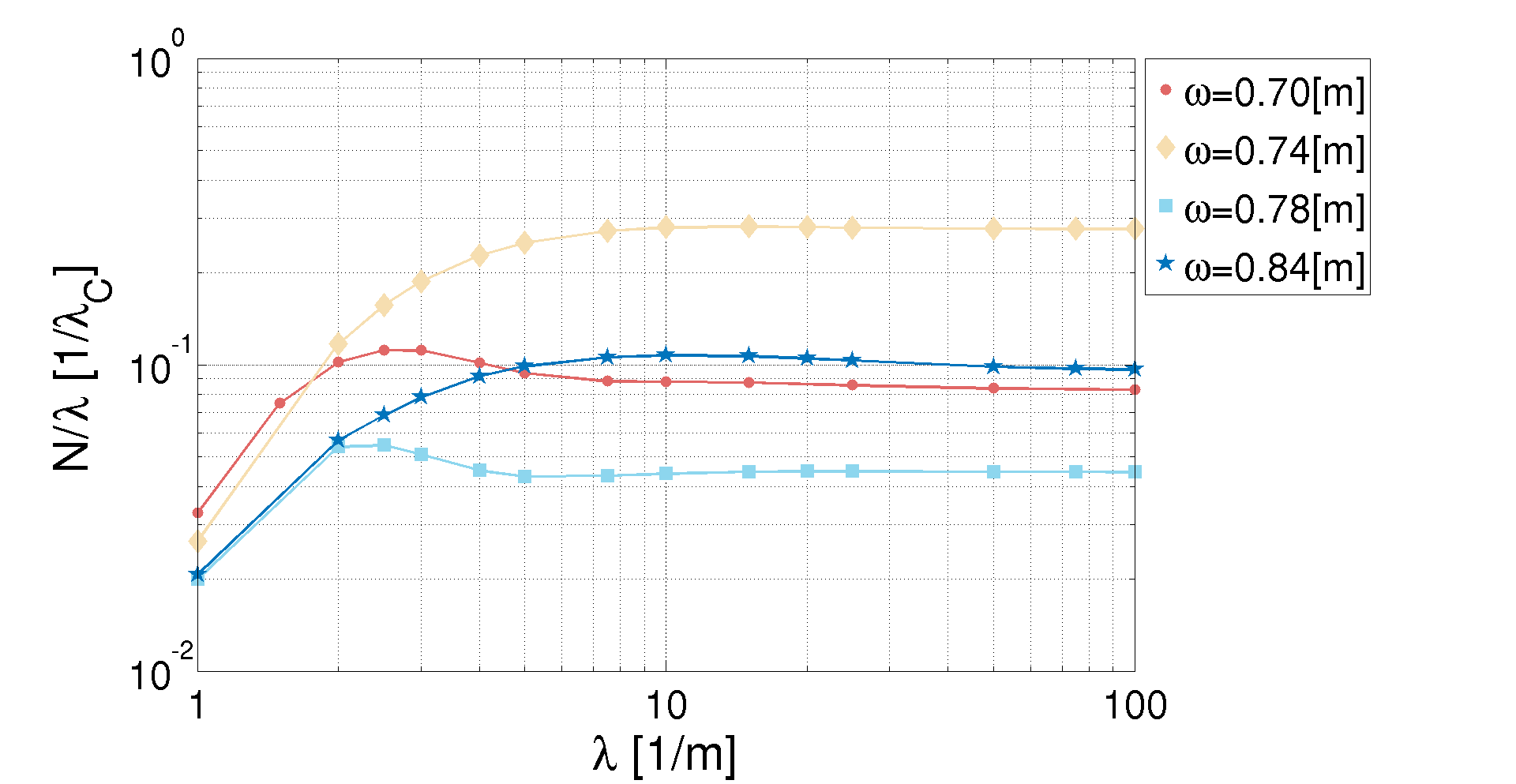} 
\end{center}
\caption[Plot of the reduced particle yield for various field frequencies as a function of the spatial extent.]{Log-log plot of the reduced particle yield $N/\lambda$ for various field frequencies $\omega$ as a function of the spatial extent $\lambda$. All lines converge to their  respective homogeneous limit for $\lambda \to \infty$. The increase of the particle yield for $\lambda < 10$ $m^{-1}$ can be related to the ponderomotive force. The reason for the drop-off at $\lambda \approx 1$ $m^{-1}$ can be traced back to a decrease in the total energy of the field. Further parameters: $\varepsilon=0.5$, $\tau=100$ $m^{-1}$ and Tab. \ref{Tab_rho1}} 
\label{Fig_Pond2}
\end{figure}

At this point we want to study how the changes in the particle distribution function translate into the particle yield, see Fig. \ref{Fig_Pond2}. However, in a realistic Laser setup, focusing goes hand-in-hand with an increase of the local field intensity. A fact, that has been completely neglected here, because we wanted to keep the model for the inhomogeneous electric field \eqref{Cyl_E} as simple as possible. 
When examining the particle yield for fixed spatial extent $\lambda$ one finds a non-monotonic dependence on the field frequency $\omega$. 
This is particularly interesting if one aims at optimizing the particle yield for a given range of Laser parameters. Due to the subtle interplay between photon energy and field focusing, tuning the experimental setup is a highly non-trivial task. The fact, that reducing the spatial extent does not directly lead to a decrease in the reduced particle yield is also noteworthy.
Moreover, we find a rapid drop-off in the particle yield for vanishing spatial extent. An effect, that has been observed previously \cite{PhysRevD.72.105004,Hebenstreit} and will be addressed in section \ref{Sec_Low} in more detail.

\subsection*{Pair production in $3+1$ dimensions}
\label{Sec_CylFig}
%

At this point we turn our attention to $3+1$ dimensional problems. This means, that we keep the model for the field \eqref{Cyl_E}, but take into account also the transversal particle momenta. Moreover, we want to focus on $3$-photon pair production in order to discuss various specialties of calculations for full momentum space.

\begin{figure}[htb]
\begin{center}
 \includegraphics[width=0.49\textwidth]{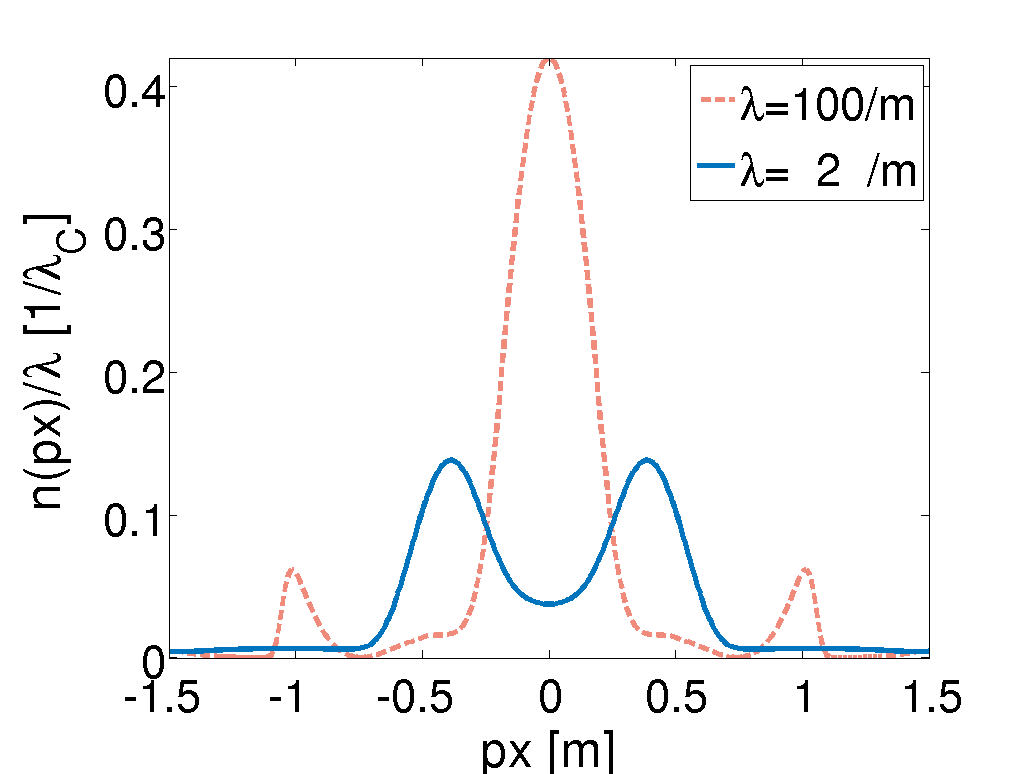}
 \includegraphics[width=0.49\textwidth]{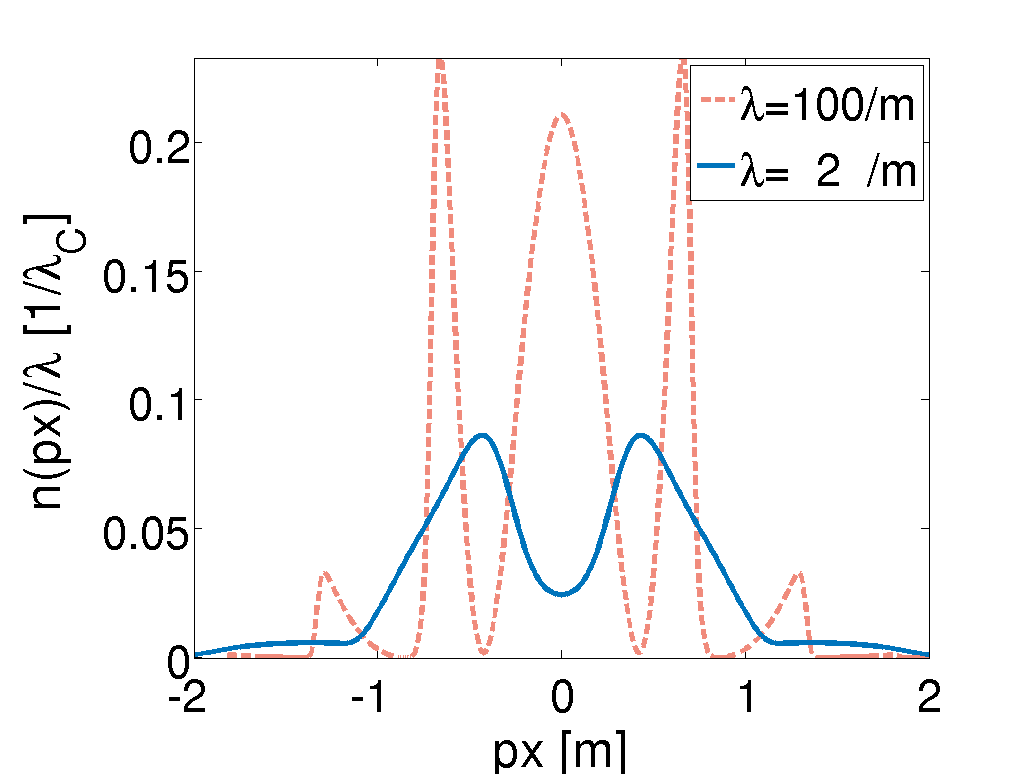}  
\end{center} 
\caption[Reduced particle distribution function for two different spatial extents as a function of the parallel momentum.]{Reduced particle distribution function for two different spatial extents $\lambda$ as a function of the parallel momentum $p_x$ for a field with $\varepsilon=0.5$, $\tau=100$ $m^{-1}$, $\omega=0.74$ $m$(left-hand side) an $\omega=0.84$ $m$(right-hand side). For $\lambda=100$ $m^{-1}$ various $n+s$ photon peaks are visible. The splitting of the main peak for $\lambda=2$ $m^{-1}$ can be related to the ponderomotive force. The vanishing of the above-threshold peaks can be interpreted in a semi-classical picture due to the short time span a particle stays within the strong field. Further data: Tab. \ref{Tab_rho2}} 
\label{Fig_Distr3D}
\end{figure}

\begin{figure}[htb]
 \includegraphics[width=0.5\textwidth]{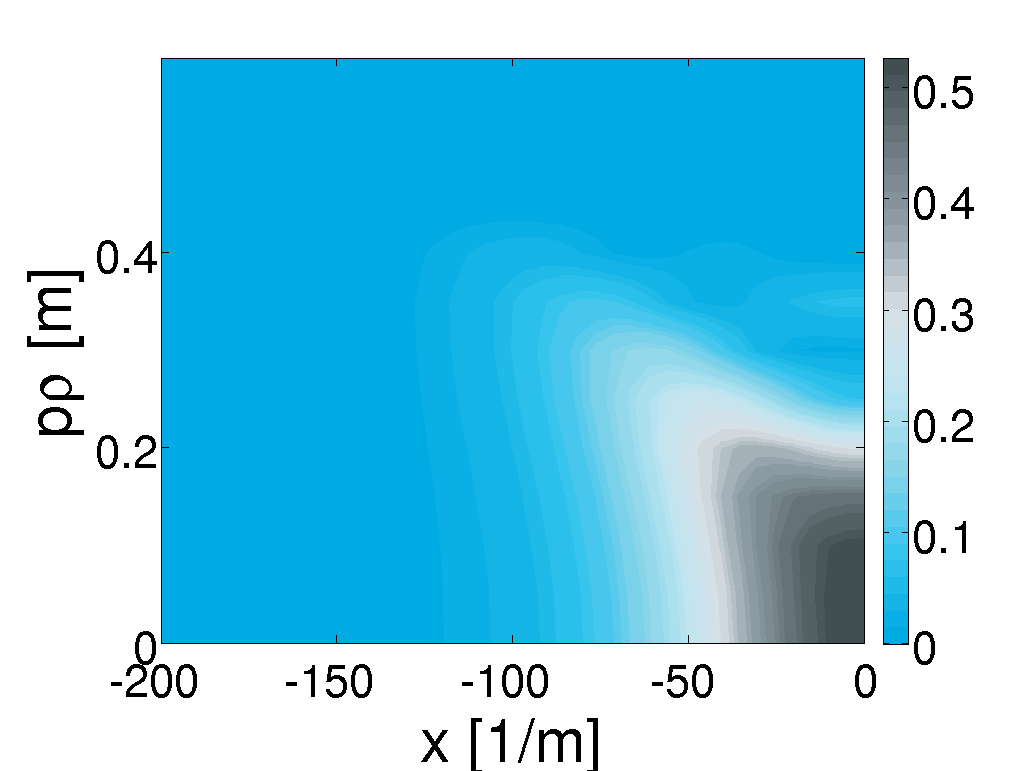}
 \includegraphics[width=0.5\textwidth]{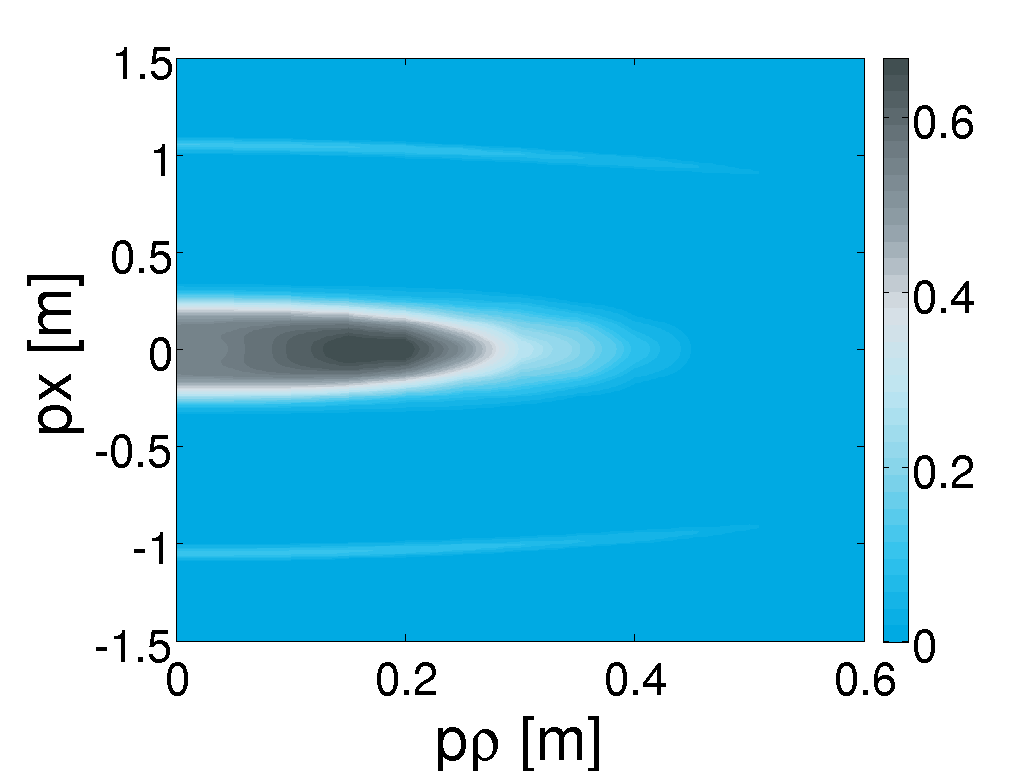}
 \includegraphics[width=0.5\textwidth]{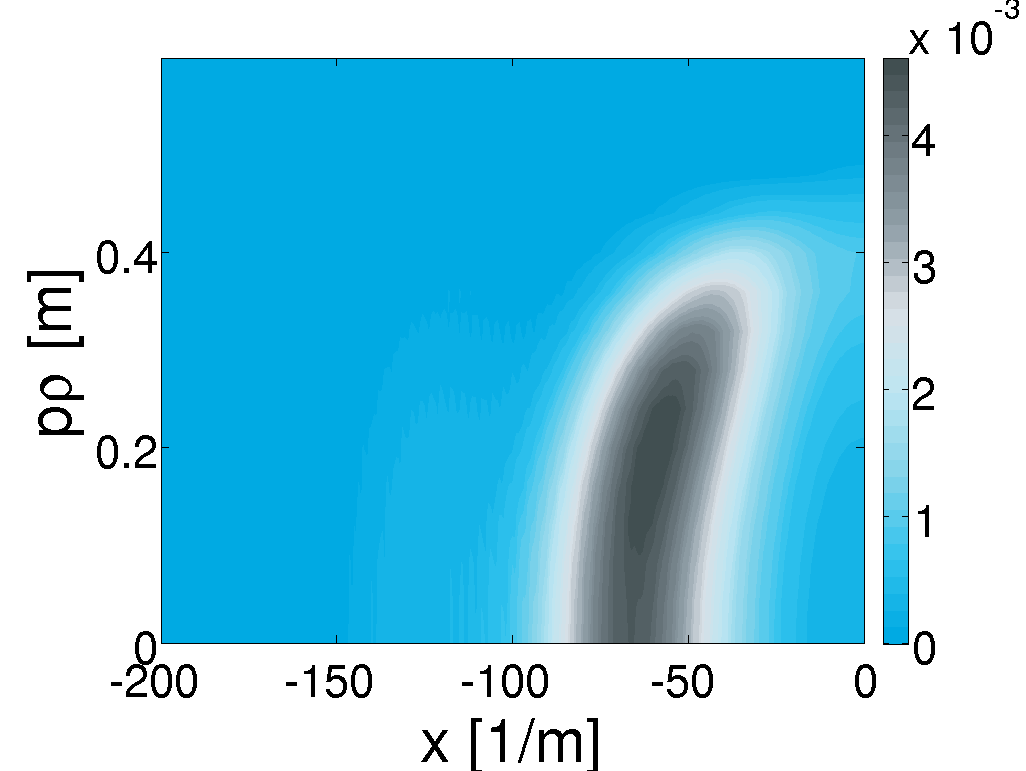}
 \includegraphics[width=0.5\textwidth]{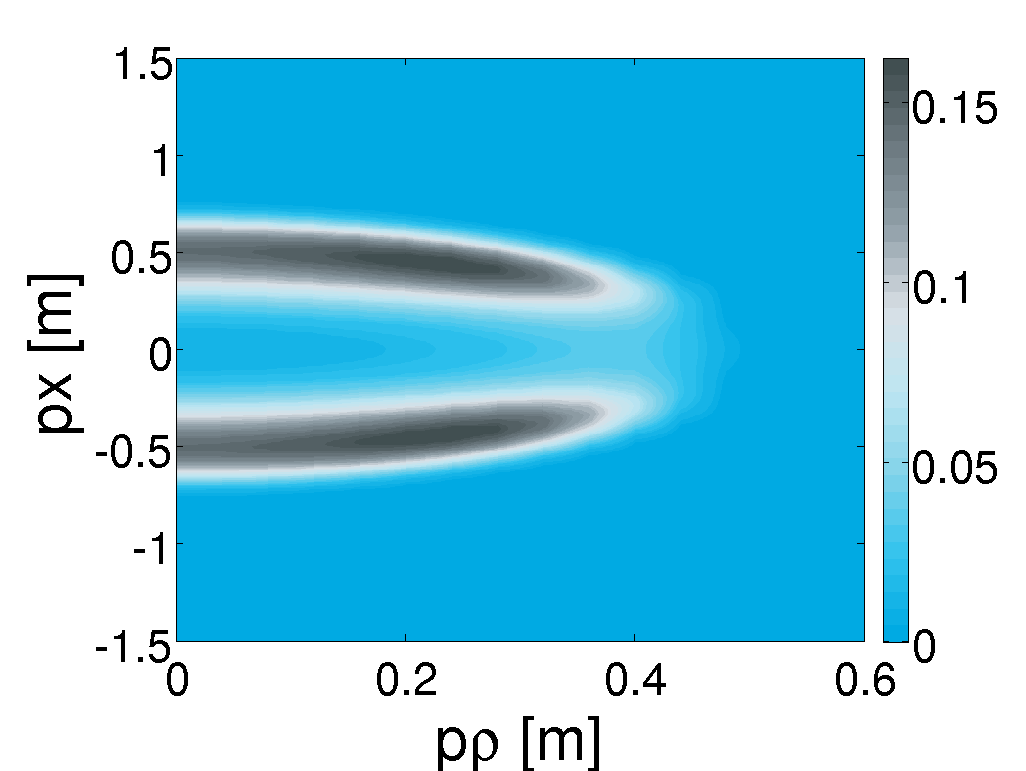}
 \includegraphics[width=0.5\textwidth]{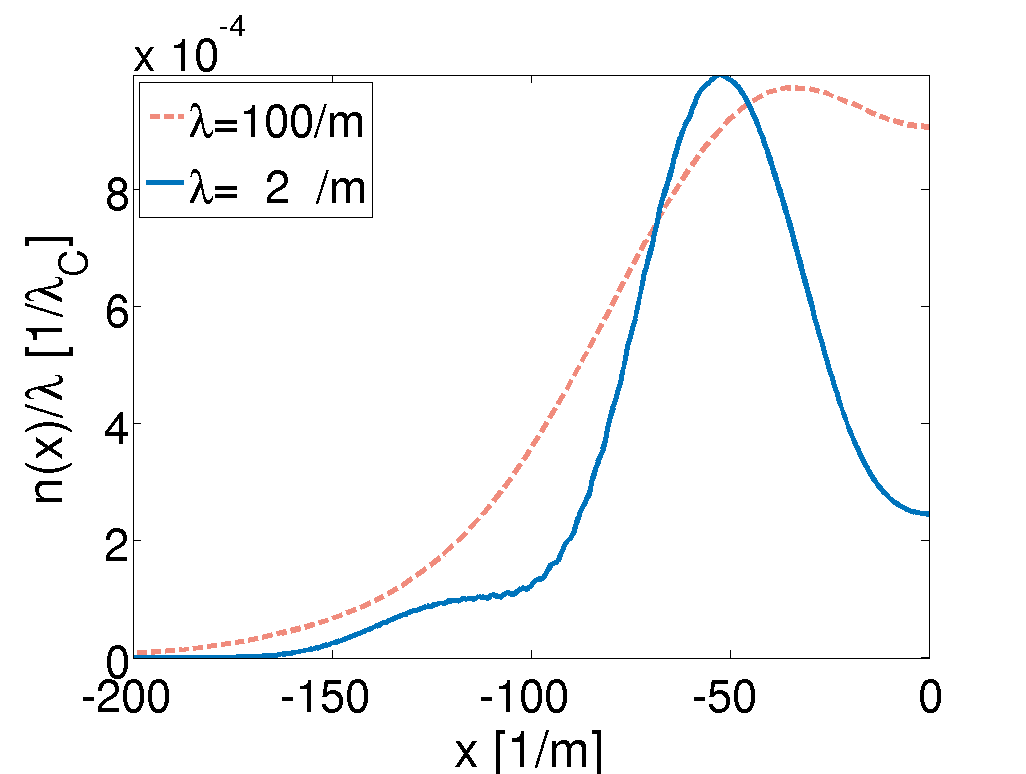}
 \includegraphics[width=0.5\textwidth]{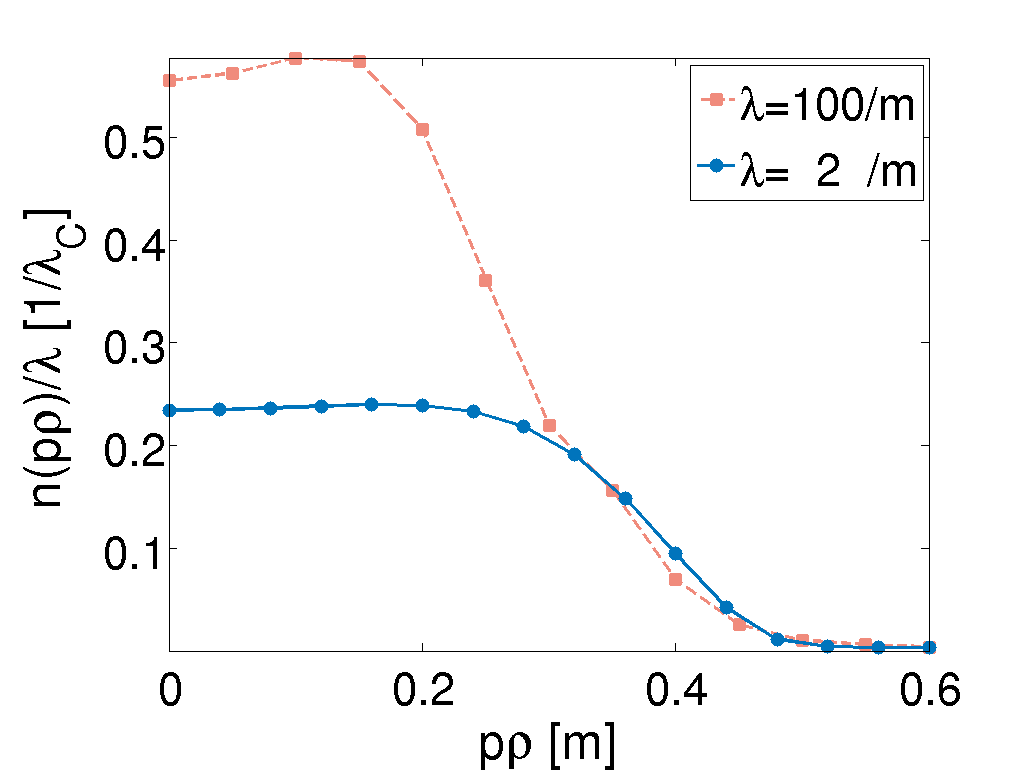}  
 \caption[Collection of figures and $2D$ plots displaying the particle distribution for different spatial extents and frequency $\omega=0.74$ $m$.]{Distribution functions for different spatial extents ($\lambda=100$ $m^{-1}$: topmost, $\lambda=2$ $m^{-1}$: middle) a field strength of $\varepsilon=0.5$, a pulse length of $\tau=100$ $m^{-1}$ and a field frequency of $\omega=0.74$ $m$. Particles are accelerated in $x$-direction for smaller spatial extent. The higher the transversal momentum $p_{\rho}$ the lower the effect. Additional data: Tab. \ref{Tab_rho2}} 
 \label{Fig_3D_1}
\end{figure}

\begin{figure}[htb]
 \includegraphics[width=0.5\textwidth]{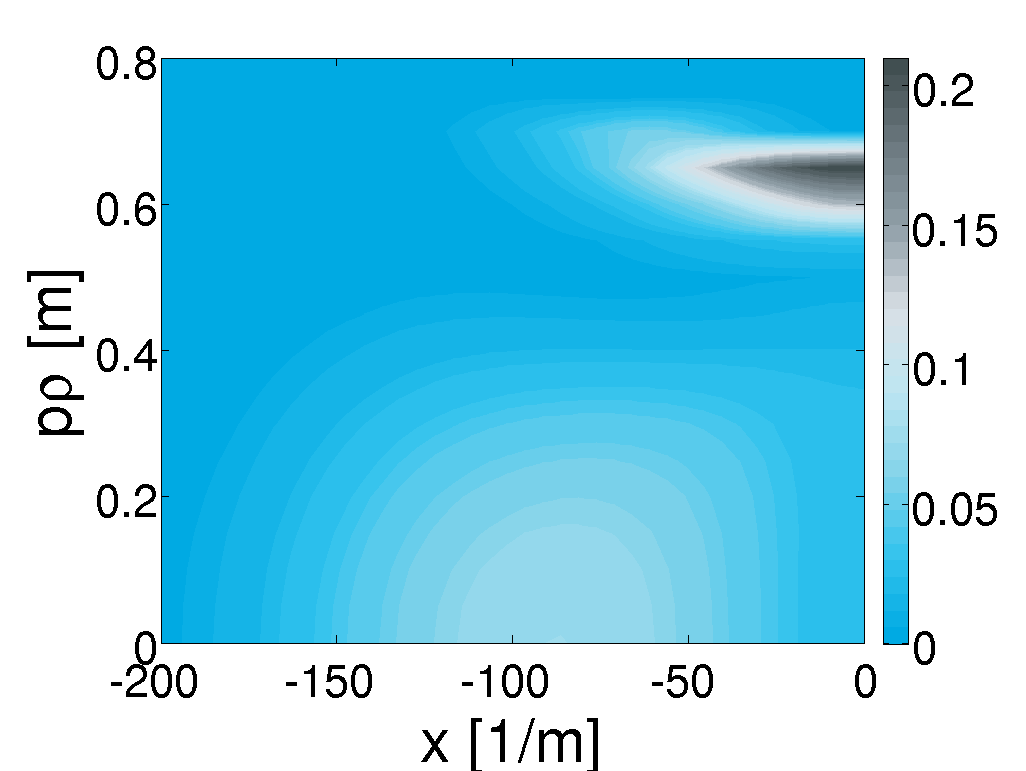}
 \includegraphics[width=0.5\textwidth]{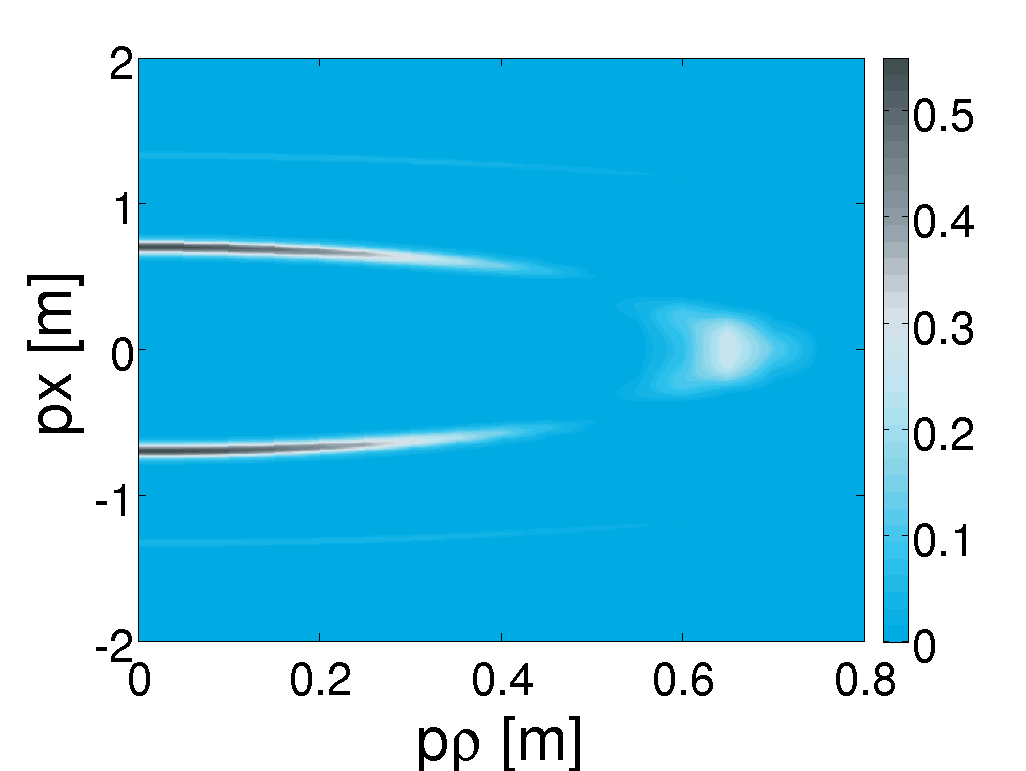}
 \includegraphics[width=0.5\textwidth]{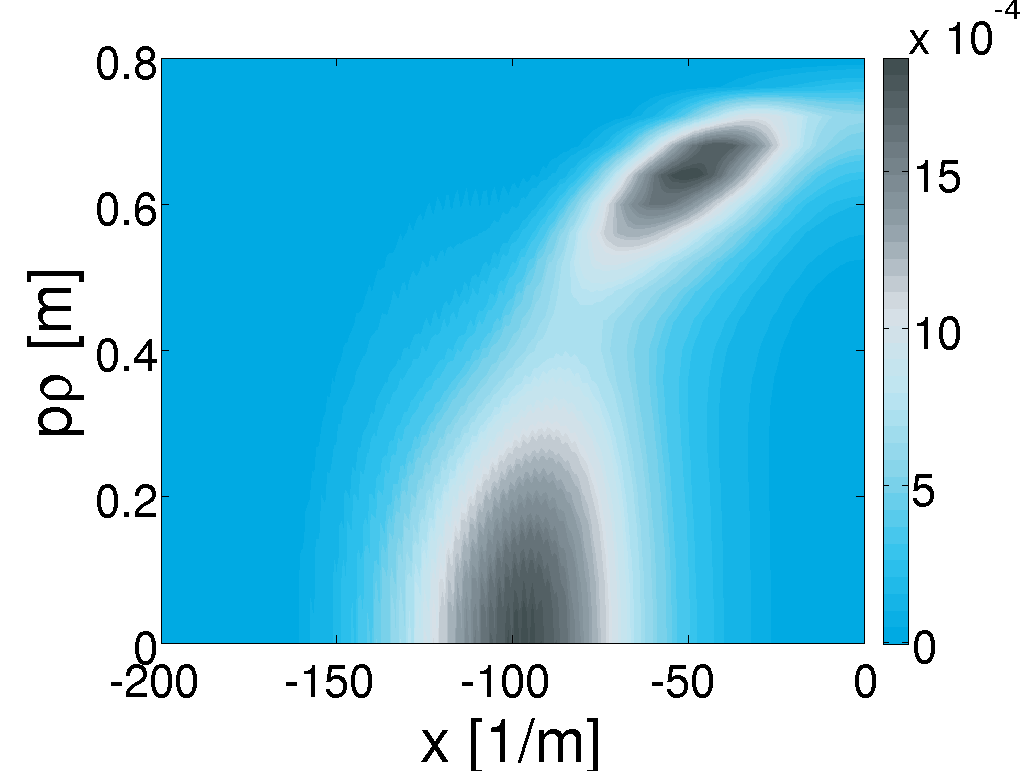}
 \includegraphics[width=0.5\textwidth]{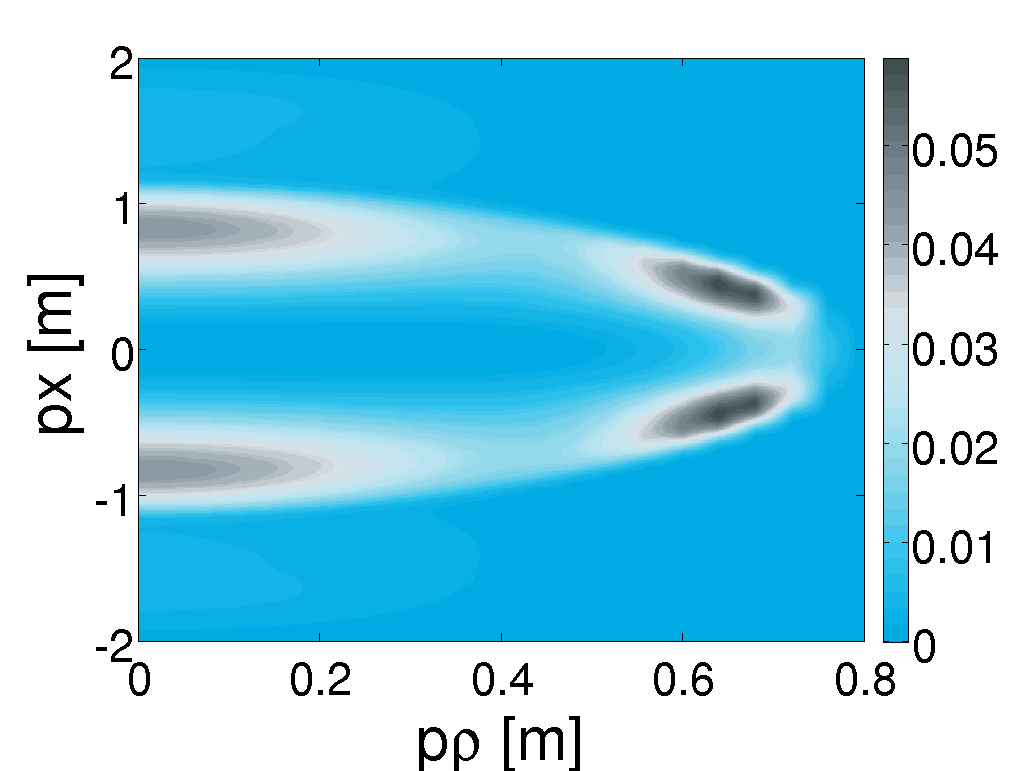}
 \includegraphics[width=0.5\textwidth]{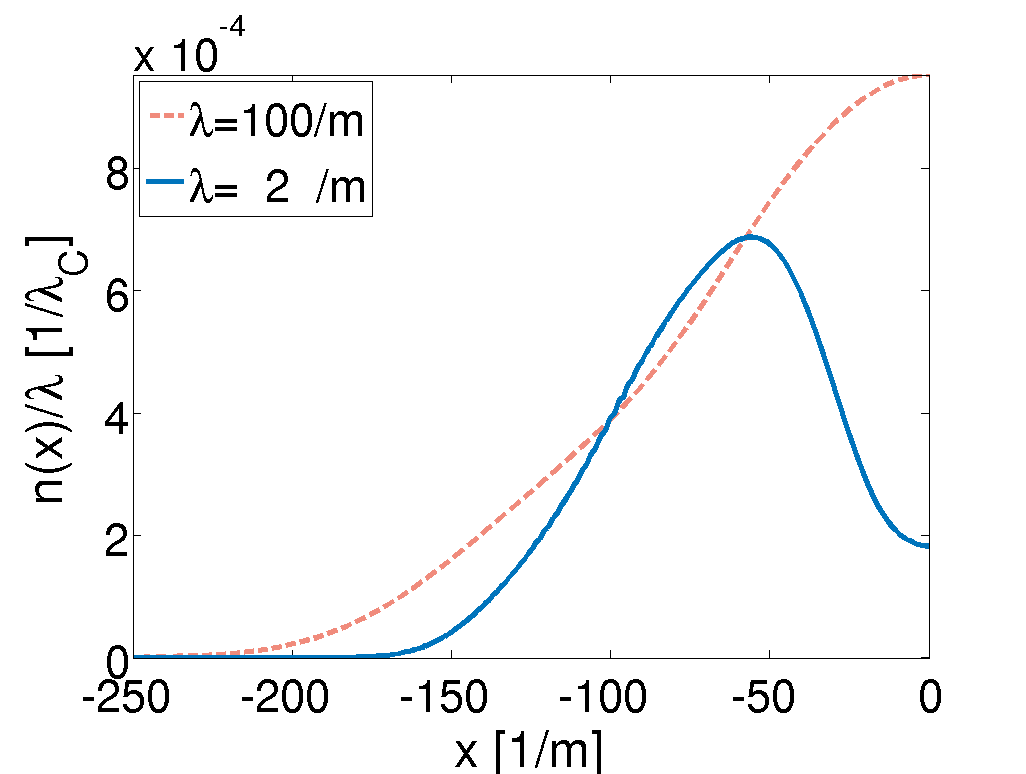}
 \includegraphics[width=0.5\textwidth]{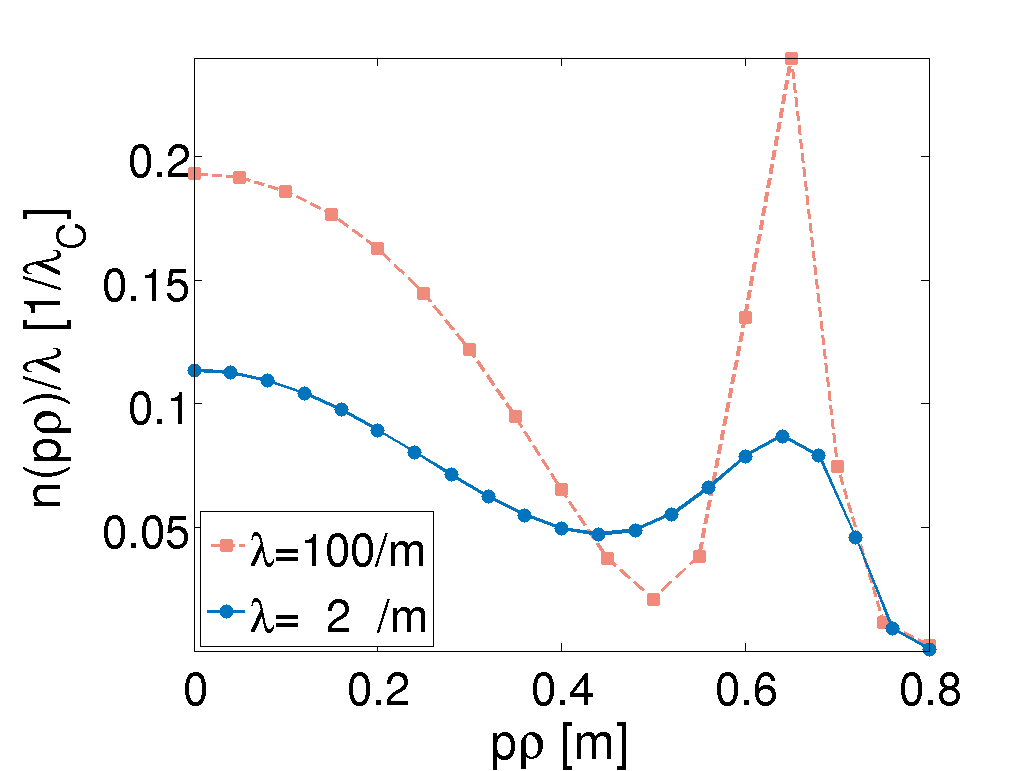} 
 \caption[Collection of figures and $2D$ plots displaying the particle distribution for different spatial extents and frequency $\omega=0.84$ $m$.]{Distribution functions for different spatial extents ($\lambda=100$ $m^{-1}$: topmost, $\lambda=2$ $m^{-1}$: middle), a field strength of $\varepsilon=0.5$, a pulse length of $\tau=100$ $m^{-1}$ and a field frequency of $\omega=0.84$ $m$. The splitting of the peak in the $p_{\rho} p_x$-plot at $\br{0.6,0}$ can be related to the ponderomotive force. The clear peaks for vanishing transversal momenta broaden when $\lambda$ becomes smaller. Additional data: Tab. \ref{Tab_rho2}}
 \label{Fig_3D_2}
\end{figure}

In Fig. \ref{Fig_Distr3D} two different particle distribution functions are shown. On the left-hand side, we see a main peak due to $3$-photon absorption and side peaks due to above threshold pair production. Ponderomotive forces then yield a splitting of the main peak and an overall boost similarly to Fig. \ref{Fig_Pond1}. The figure on the right-hand side needs further explanation.
We observe a peak around $p_x=0$ and side peaks at $p_x \approx \pm 0.66$ $m$. This picture, however, is highly misleading as one would expect that these peaks stem from a $n$ and a $n+1$ absorption process. This possible misinterpretation can be resolved by looking at the illustration displaying the result for the full momentum space (Fig. \ref{Fig_3D_2}). Multiphoton pair production leaves a characteristic ring-like structure in the momentum space superimposed by a substructure caused by quantum interference, see Fig. \ref{Fig_3D_1} and Fig. \ref{Fig_3D_2}. The occurrence of the peak around $p_x \approx 0$ in Fig. \ref{Fig_Distr3D} is caused by integration over the transversal direction only. Hence, the inner peaks at $|p_x| < 0.85$ $m$ can all be related to $n=3$-photon absorption. 
Additionally, due to the ponderomotive force the particle distribution in Fig. \ref{Fig_Distr3D} is distorted. Again, one can relate the main contribution to the peak formerly located at $p_x \approx 0$. The fact, that the additional peaks($p_x \approx \pm 1.3$ $m$) are also forced to higher momenta is hidden, compare Fig. \ref{Fig_Distr3D} with Fig. \ref{Fig_3D_2}.

In section \ref{Kap_SemiClass} we have already explained how to interpret the interference pattern in terms of a semi-classical analysis.
The explanation can also be applied in the case of a many-cycle pulse \eqref{Cyl_E} with multiple local maxima in $| E \br{x,t} |$. We assume again, that particles are created at each of these peaks. As we work in the multiphoton regime (Keldysh parameter $\gamma \gg 1$) we can relate the field frequency with the particles total energy. Hence, the kinetic energy of the particles is fixed, but the momentum direction is random. We would therefore expect a ring structure to form in momentum space when the applied electric field vanishes. Furthermore, in the idealized case of a periodic, spatially homogeneous field, we cannot measure a final particle momentum in order to trace back the particles trajectory to its origin. This is due to the fact, that there are multiple possibilities for the particle's trajectory with the momentum specified. This ambiguity can be related to the emergence of an interference pattern. An illustration of the results obtained for a $3+1$-dimensional problem is displayed in Fig. \eqref{Fig_3D_1} and Fig. \eqref{Fig_3D_2}.


\clearpage
We will continue using this semi-classical picture in order to interpret the difference in deflection for different transversal momenta, see Fig. \ref{Fig_3D_1} and Fig. \ref{Fig_3D_2}. The fact, that particles are accelerated differently due to their transversal momentum is already given by the relativistic ponderomotive force \eqref{Pond_Rel}. This is simply a consequence of relativistic mechanics.
The findings using semi-classical methods are summarized in Tab. \ref{Tab_Pond3D} for a special set of parameters. It is found, that the higher the transversal momentum is the lower their final parallel momentum becomes.

\ctable[pos=hb,
caption = {Particle acceleration within a spatially inhomogeneous electric field. The results have been obtained solving the relativistic Lorentz force equation for a field of strength $\varepsilon=0.5$, length $\tau=100$ $m^{-1}$, frequency $\omega=0.74$ $m$ and various spatial extent $\lambda$. The particle was seeded at $x=t=0$ with parallel momentum $p_{x,0}$ and transversal momentum $p_{\rho,0}$. The results obtained for the final parallel momentum $p_{x,f}$ can be related to the relativistic ponderomotive force.},
cap = {Evaluating the relativistic Lorentz force equation in order to demonstrate the effect of relativistic ponderomotive forces in $3+1$ dimensional cylindrically symmetric problems.},
label = Tab_Pond3D, 
mincapwidth = \textwidth,
]{ c c c}{
}{
    \toprule
    \hspace{1cm} $p_{\rho,0}$ $[m]$ \hspace{1cm} & \hspace{1cm} $\lambda$ $[m^{-1}]$ \hspace{1cm} & \hspace{1cm} $p_{x,f}$ $[m]$ \hspace{1cm} \\
    \midrule
    0 & 1000 & $7 \cdot 10^{-6}$ \\  
    \midrule
    0.5 & 1000 & $6 \cdot 10^{-6}$ \\  
    \midrule
    1 & 1000 & $4 \cdot 10^{-6}$ \\      
    \midrule
    \midrule
    0 & 20 & 0.021 \\  
    \midrule
    0.5 & 20 & 0.017 \\  
    \midrule
    1 & 20 & 0.011 \\      
    \midrule
    \midrule
    \hlight{0} & 10 & \hlight{0.13} \\  
    \midrule
    0.5 & 10 & 0.10 \\  
    \midrule
    \hlight{1} & 10 & \hlight{0.06} \\       
    \bottomrule
} 

\section{Schwinger pair production in strongly localized fields}
\label{Sec_Low}
We have already seen, that spatial focusing could enhance the reduced particle production rate. However, there are limits on the focusability. Here, we want to investigate at which point the process of particle creation terminates.
Using \eqref{Cyl_E} as a model for the field this translates into an investigation of $\lambda \sim \mathcal{O} \br{1 ~ m^{-1}}$. In addition we have fixed the field frequency $\omega$ to zero in order to concentrate on the regime of Schwinger pair production. 

In the literature \cite{Hebenstreit,PhysRevD.72.105004,PhysRevD.72.065001}, mainly fields resembling $1+1$ dimensional situations were considered. We want to extend these ideas to $3+1$ dimensions by taking into account also the particles produced with non-vanishing transversal momentum. Moreover, we want to identify the role the field strength $\varepsilon$ and the pulse duration $\tau$. 

\begin{figure}[htb]
\begin{center}
 \includegraphics[width=0.49\textwidth]{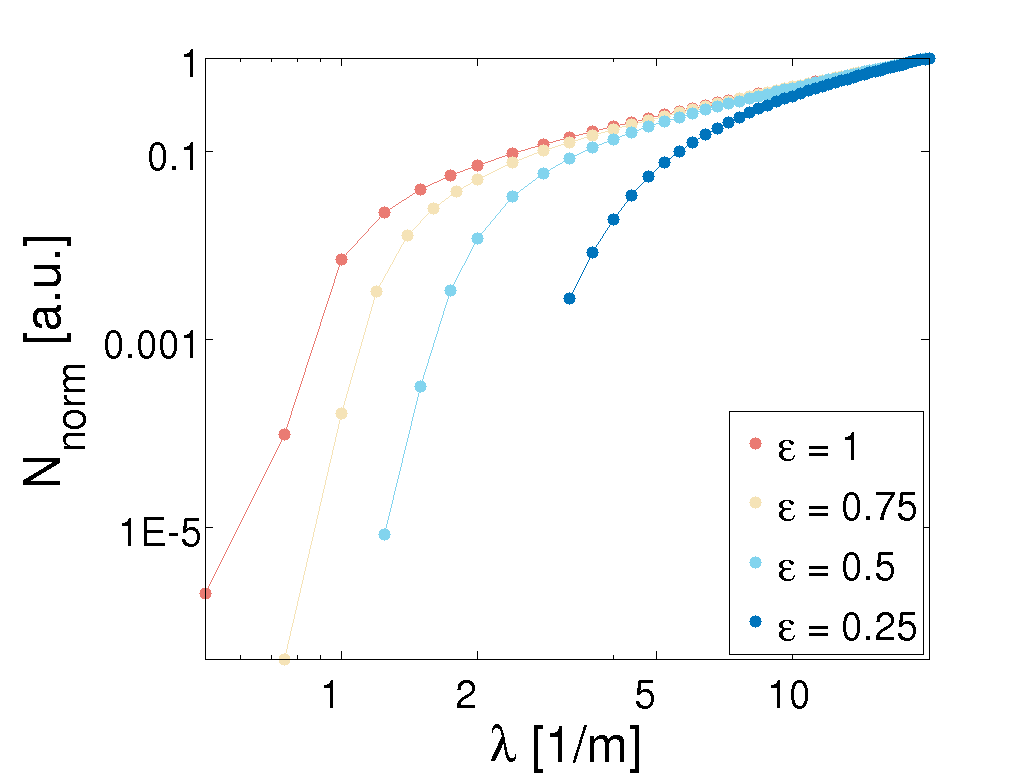}
 \includegraphics[width=0.49\textwidth]{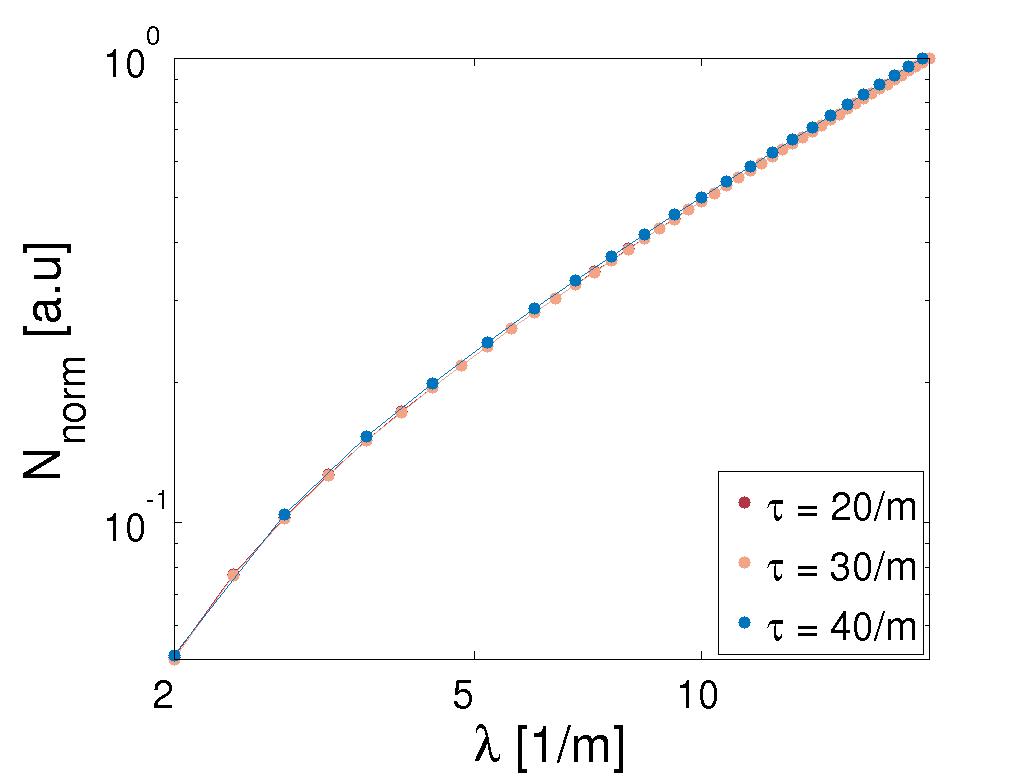}
\end{center}
\caption[Qualitative comparison of the particle yield for various field strengths and pulse lengths as a function of the spatial extent.]{Qualitative comparison of the particle yield for various field strengths $\varepsilon$ and pulse lengths $\tau$ as a function of the spatial extent $\lambda$. The data is normalized such, that $N_{norm} \br{\lambda=15} = 1$. The higher the field strength the later the drop-off in the particle yield sets in. On the right-hand side the impact of the pulse length is shown to be non-existent. Parameters for the left-hand side: $\tau=20$ $m^{-1}$, $\omega=0$, see also Tab. \ref{Tab_yield2b}. Parameters for the right-hand side: $\varepsilon=0.75$, $\omega=0$ and Tab. \ref{Tab_yield2}.} 
\label{Fig_Narrow}
\end{figure}

Both investigations are summarized in Fig. \ref{Fig_Narrow}. 

At first we have to note, that the Schwinger pair production process is in contrast to multiphoton particle creation, not driven by the field frequency but the work that the field can do on a charged particle. Hence, the relevant parameters are the fields strength and its extent in spatial and temporal directions. To put it simple, the larger the strong-field region of the applied electric field is the higher the Schwinger particle yield becomes. The question is then, what are the significant parameters when limiting this strong-field region in space? 

Following the arguments in the literature \cite{Hebenstreit,PhysRevD.72.105004,PhysRevD.72.065001}, we assume, that the background field has to do work in order to separate charges (particle/antiparticle). If this work done exceeds the mass of the particle-antiparticle pair we interpret the successful delocalization of the charges as a pair production process. Hence, pair production reduces to the relation
\begin{equation}
 e \int d \mathbf{x} \cdot \mathbf{E} \br{\mathbf{x}} \ge m_{e^-} + m_{e^+}. \label{E_Cond}
\end{equation}

We evaluate the electric field for fixed time $\tilde t$. This is due to the fact, that the electric field has to provide the energy at every instant of time. Increasing the total pulse length therefore does not effect our consideration on spatially localized fields, see Fig. \ref{Fig_Narrow}. Considering the model \eqref{Cyl_E}, we obtain for condition \eqref{E_Cond}:
\begin{equation}
 e \int dx \ \varepsilon \ E_0 \ \exp \br{-\frac{x^2}{2 \lambda^2}} \ge 2m.
\end{equation}
Integration on the left-hand side gives
\begin{equation}
 \int_{x_1}^{x_2} dx \exp \br{-\frac{x^2}{2 \lambda^2}} = \sqrt{\frac{\pi}{2}} \ \lambda \ \text{erf} \br{\frac{x}{\sqrt{2} \lambda}}.
\end{equation}
As for the limits of the integral: Here, we assume, that particles are created at peak field strength, i.e.$x=0$, and the work done by the field ranges to infinity. The limits can therefore be given as $x_1=0$ and $x_2=\infty$. This simplifies the expression above resulting in the following condition
\begin{equation}
 e \ \varepsilon \ E_0 \ \sqrt{\frac{\pi}{2}} \ \lambda \ge 2m.
\end{equation}
Eventually, this yields a condition on the spatial extent $\lambda$:
\begin{equation}
 \lambda \ge \sqrt{\frac{8}{\pi}} \frac{m}{\varepsilon \ e \ E_0}. \label{E_Cond2}
\end{equation}
Due to the uncertainties in defining appropriate integral limits the expression above is not unique, c.f. reference \cite{Hebenstreit}. Nevertheless, if condition \eqref{E_Cond2} is violated the applied electric field does not provide the necessary energy for Schwinger pair production.    
Investigating the left-hand side of Fig. \ref{Fig_Narrow} for fixed $N_{norm}$, we find indeed, that the drop-off for too small spatial extents can be related to the condition \eqref{E_Cond2}. Although this condition serves only as an estimate it gives the correct order for termination of the particle creation process. An analysis of the data obtained is given in Tab. \ref{Tab_Narrow}.

It should be noted, that we have integrated over full momentum space in order to obtain the particle yield. Despite the fact, that the electric field is homogeneous in transversal direction and thus the total energy is infinite, we observe a drop-off for small $\lambda$. The decisive quantity seems not to be the work the electric field could possibly provide in total, but the work the field can provide along the particle trajectory.


\ctable[pos=hb,
caption = {Comparing the values of the field strength $\varepsilon$ and the spatial extent $\lambda$ obtained from fixing $N_{norm}=0.01$ by interpolation. The yield is normalized such, that $N_{norm} \br{\lambda=15} = 1$ for all $\varepsilon$ individually. The product $\varepsilon$ times $\lambda$ should be constant according to the idealized model.},
cap = {Comparing the values of the field strength $\varepsilon$ and the spatial extent $\lambda$ when fixing the particle yield to a certain value.},
label = Tab_Narrow, 
mincapwidth = \textwidth,
]{ c c c}{
}{
    \toprule
    \hspace{1cm} $\varepsilon$ \hspace{1cm} & \hspace{1cm} $\lambda$ $[m^{-1}]$ \hspace{1cm} & \hspace{1cm} $\varepsilon \ \lambda$ $[m^{-1}]$ \hspace{1cm} \\
    \midrule
    1 & 1.056 & 1.056 \\  
    \midrule
    0.75 & 1.352 & 1.014\\ 
    \midrule
    0.5 & 1.955 & 0.977 \\  
    \midrule
    0.25 & 3.68 & 0.92 \\     
    \bottomrule
} 
\pagestyle{plain}
\chapter{Pair production in electric and magnetic fields}
\pagestyle{fancy}
The subject of investigation throughout chapter seven is the impact magnetic fields have on the pair creation process. Therefore, the time-dependent electric and magnetic fields are obtained by introducing a simple model for the underlying vector potential.
The key objective of this study is to pave the way for realistic investigations on the pair production process. Especially, when thinking about our preliminary considerations in chapter three, 
an elaborate investigation on pair production in electromagnetic fields is inevitable.

\section[Pair production in the plane: Operator-expanded DHW approach and model for the fields]{Pair production in the plane}

As magnetic fields cannot be defined in $1+1$ dimensions, we perform calculations in the plane. Hence, the electric field is generally given by a two-component vector and the magnetic field is a scalar quantity. Additionally, we are not free to choose any particular form for the fields, because they have to obey Maxwell equations.

\subsection*{Operator-expanded DHW approach and model for the fields}

Introducing a simple model for the vector potential pointing in $x$-direction and being inhomogeneous in $z$-direction we automatically obtain electromagnetic fields fulfilling the homogeneous Maxwell equations. Moreover, we propose a vector potential which falls off for asymptotic times. This greatly simplifies the calculations and also the interpretation of the results, because the background fields also vanish for $t \to \pm \infty$. In the following the first vector potential used is stated
\begin{align}
 \mathbf{A} \br{z,t} = \varepsilon \ E_0 \ \tau \left( \tanh \left( \frac{t+\tau}{\tau} \right) - \tanh \left( \frac{t-\tau}{\tau} \right) \right) \exp \br{-\frac{z^2}{2 \lambda^2}} \ \mathbf{e}_x. \label{Field1}
\end{align}
Correspondingly, the electric field takes the form
\begin{align}
 \mathbf{E} (z,t) = \quad \varepsilon \ E_0 \ \ \ \left( \text{sech}^2 \left( \frac{t-\tau}{\tau} \right) - \text{sech}^2 \left( \frac{t+\tau}{\tau} \right) \right)  
 \exp \br{-\frac{z^2}{2 \lambda^2}} \ \boldsymbol{e}_x
\end{align}
and the magnetic field yields
\begin{align}
 B \br{z,t} = -\varepsilon \ E_0 \ \tau \left( \tanh \left( \frac{t+\tau}{\tau} \right) - \tanh \left( \frac{t-\tau}{\tau} \right) \right) \exp \br{-\frac{z^2}{2 \lambda^2}} \ \frac{z}{\lambda^2}.
\end{align}
The fields are characterized in terms of the peak field strength $\varepsilon$, the temporal peak width $\tau$ and the spatial extent $\lambda$.
As the electric field exhibits a clear double-peak structure in time, at least the homogeneous limit should clearly show an interference pattern in $p_x$. Determining the correct resolution of this interference pattern therefore holds as a good test on the validity of our operator expansion. 

However, defining the electromagnetic fields in this way we were able to obtain results for short pulse lengths only. In the following we introduce an alternative   
field configuration with one dominant peak in time. By this way, the total pulse length could be greatly expanded due to the absence of interference patterns. The corresponding model for the vector field takes the form
\begin{alignat}{5}
 &\mathbf{A} \br{z,t} = &&- \varepsilon \ E_0 \ \tau \ \exp \br{-\frac{z^2}{2 \lambda^2}} \times \notag \\
 & &&\left( \tanh \left( \frac{t}{\tau} \right) - \frac{1}{2} \tanh \left( \frac{t-\tau}{\tau} \right) - \frac{1}{2} \tanh \left( \frac{t+\tau}{\tau} \right) \right)  \ \mathbf{e}_x. \label{Field2}
\end{alignat}
Again, the electric and magnetic field is obtained via taking the derivative of the vector potential yielding
\begin{align}
 &\mathbf{E} \br{z,t} = &&\varepsilon \ E_0 \ \exp \br{-\frac{z^2}{2 \lambda^2}} \times \notag \\ 
 & &&\left( \text{sech}^2 \left( \frac{t}{\tau} \right) - \frac{1}{2} \text{sech}^2 \left( \frac{t-\tau}{\tau} \right) - \frac{1}{2} \text{sech}^2 \left( \frac{t+\tau}{\tau} \right) \right)  \ \mathbf{e}_x, \\
 &B \br{z,t} =&& \varepsilon \ E_0 \ \tau \ \frac{z}{\lambda^2} \ \exp \br{-\frac{z^2}{2 \lambda^2}} \times \notag \\ 
 & &&\left( \tanh \left( \frac{t}{\tau} \right) - \frac{1}{2} \tanh \left( \frac{t-\tau}{\tau} \right) - \frac{1}{2} \tanh \left( \frac{t+\tau}{\tau} \right) \right). 
\end{align}
In the following, we will denote the two different configurations based upon the time-dependency of the electric field as ``double-peaked'' \eqref{Field1} and ``quasi single-peaked'' \eqref{Field2}.

The special form of the vector potentials \eqref{Field1} and \eqref{Field2} makes a calculation of the pair production process within the DHW approach feasible. We have decided to perform calculations with the following system(\eqref{eq_2b_1}-\eqref{eq_2b_4}):
\begin{alignat}{6}
  & D_t \overline{\mathbbm{s}}   &&  && && -2 \Pi_1 \overline{\mathbbm{v}}_3 &&+ 2 \Pi_3 \overline{\mathbbm v}_1 &&= 0, \label{eq_B_1}  \\
  & D_t \overline{\mathbbm{v}}_1 &&+D_1 \overline{\mathbbm{v}}_0 && && &&- 2 \Pi_3 \overline{ \mathbbm{s}} &&= -2m\overline{\mathbbm{v}}_3,  \\    
  & D_t \overline{\mathbbm{v}}_3 && &&+D_3 \overline{\mathbbm{v}}_0 && +2 \Pi_1 \overline{\mathbbm{s}} && &&= 2m \overline{\mathbbm{v}}_1,  \\ 
  & D_t \overline{\mathbbm{v}}_0 &&+D_1 \overline{\mathbbm{v}}_1 &&+D_3 \overline{\mathbbm{v}}_3 && && &&= 0, \label{eq_B_4}
\end{alignat}
where we have used abbreviations for the operators:
  \begin{alignat}{6}
      & D_t &&= \de{t} +&& e &&\int d\xi &&E \br{z+\ii \xi \de{p_z},t} \de{p_x},  \\
      & D_1 &&= &&e &&\int d \xi &&B \br{z+\ii \xi \de{p_z},t} \de{p_z},  \\
      & D_3 &&= \de{z} - &&e &&\int d \xi &&B \br{z+\ii \xi \de{p_z},t} \de{p_x},  \\
      & \Pi_1 &&= p_x - &&\ii e &&\int d \xi \xi &&B \br{z+\ii \xi \de{p_z},t} \de{p_z}, \\
      & \Pi_3 &&= p_z + &&\ii e &&\int d \xi \xi &&B \br{z+\ii \xi \de{p_z},t} \de{p_x}.
  \end{alignat} 
Vacuum initial conditions are given via
\begin{align}
  \overline{\mathbbm{s}}_i = -\frac{2}{\omega},\ \overline{\mathbbm{v}}_{1i} = -\frac{2 p_x}{\omega} ,\ \overline{\mathbbm{v}}_{3i} = -\frac{2 p_z }{\omega}.
\end{align}
In this way, we obtain physically meaningful results for least numerical costs. Moreover, the distribution function evaluated at asymptotic times reads
\begin{alignat}{5}
  &\mathcal{N} &&= \ \int dz \ dp_x \ dp_z \ n \br{z,p_x,p_z} \\ 
  & &&= \int dz \ dp_x \ dp_z \ \frac{ m \overline{\mathbbm{s}}^v \br{z,p_x,p_z} + 
 p_x \ \overline{\mathbbm{v}}^v \br{z,p_x,p_z} + p_z \ \overline{\mathbbm{p}}^v \br{z,p_x,p_z}}{\sqrt{m^2+p_x^2+p_z^2}}.
\end{alignat}
Further explanations on the modified Wigner functions 
\begin{equation}
 \overline{\mathbbm{w}}_k^{v} \br{z, p_x, p_z} = \overline{\mathbbm{w}}_k \br{z, p_x, p_z} - \overline{\mathbbm{w}}_i \br{p_x, p_z}
\end{equation} 
is given in section \ref{Sol_Exp}, see also \ref{sec_observables} for results regarding $QED_{2+1}$. 
For the sake of readability, the momenta in the following figures are again labelled with $px$ and $pz$.

\vspace{5cm}

\section{Effective field amplitude}

In the literature, pair production for $B \neq 0$ has been mainly investigated for constant fields\cite{Dunne1998322,PhysRevD.52.R3163,PhysRevA.90.032101,Kim2}. Only in a few calculations  spatially inhomogeneous or time-dependent magnetic fields have been taken into account \cite{DiPiazza2006520,PhysRevLett.102.080402}. However, already in the simple case of a time-dependent, purely electric background field diverse possibilities for pair production arise. Hence, it is of major importance to understand the impact of inhomogeneous magnetic fields on the particle creation process.

The Lorentz invariants in $QED_{3+1}$ take the form
\begin{equation}
 \mathcal F = -\frac{1}{4} F^{\mu \nu} F_{\mu \nu} = \frac{1}{2} \br{\mathbf{E}^2 - \mathbf{B}^2} ,\qquad \mathcal G = -\frac{1}{4} F^{\mu \nu} \tilde F_{\mu \nu} = \mathbf{E} \cdot \mathbf{B}.
\end{equation}
They play a crucial role in describing pair production. Following Dunne et al. \cite{DunneHEul} we will distinguish four different possibilities. In case of $\mathbf{E}^2 > 0$ and $\mathbf{B} = 0$ the situation resembles the Schwinger pair production process. Hence, the applied field creates particles by a rate of \begin{align}
\exp \br{-\frac{E_0 \ \pi}{e |\mathbf{E}|}}. \label{Schw_B}
\end{align}
In contrast, a background field where $\mathbf{B}^2 \ge 0$ and $\mathbf{E}=0$ is not capable of producing particles. If both, the electric as well as the magnetic field, are nonzero the Lorentz invariant $\mathcal G$ determines the outcome. In case of crossed fields, thus $\mathcal G = 0$, and $\mathcal F > 0$ the situation is the same as for the case of $\mathbf{E}^2 > 0$, $\mathbf{B}=0$. If  $\mathcal F \le 0$ particle creation terminates. The last possibility remaining is given by $\mathcal G \neq 0$. Then, the pair production rate is still described via an exponential factor \eqref{Schw_B}.
The prefactor, however, depends on the strength of the magnetic field. If $\mathbf{B} > 0$ the production rate is even enhanced. 

As we proposed to investigate pair production in the plane, $\mathcal G$ is always zero. The quantity $\mathcal F$, on the other hand, becomes a function of $z$ and $t$. Additionally, we have to translate the observations made in the literature to inhomogeneous background fields.
Hence, assuming that pair production is only possible where $E \br{z,t}^2 > B \br{z,t}^2$ we suggest a domain-dependent effective field amplitude
\begin{equation}
 \tilde E \br{z,t}^2 = E \br{z,t}^2 - B \br{z,t}^2. \label{Eq_FieldAmpl}
\end{equation}
Correspondingly, the modified field energy yields
\begin{equation}
 \mathcal{E} \br{E,B} = \int \ \tilde E \br{z,t}^2 \ \Theta \br{\tilde E \br{z,t}^2} \ dz \ dt,
\end{equation}
with the Heaviside function $\Theta \br{x}$.

\begin{figure}[htb]
\begin{center}
   \includegraphics[width=0.49\textwidth]{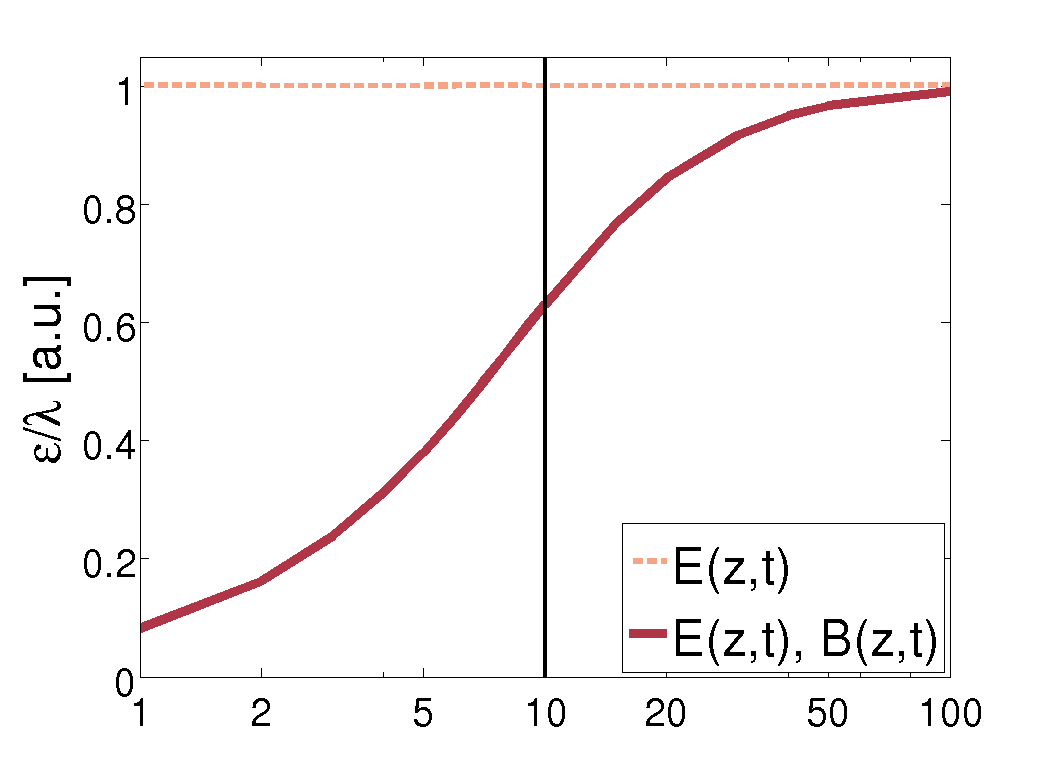}
   \includegraphics[width=0.49\textwidth]{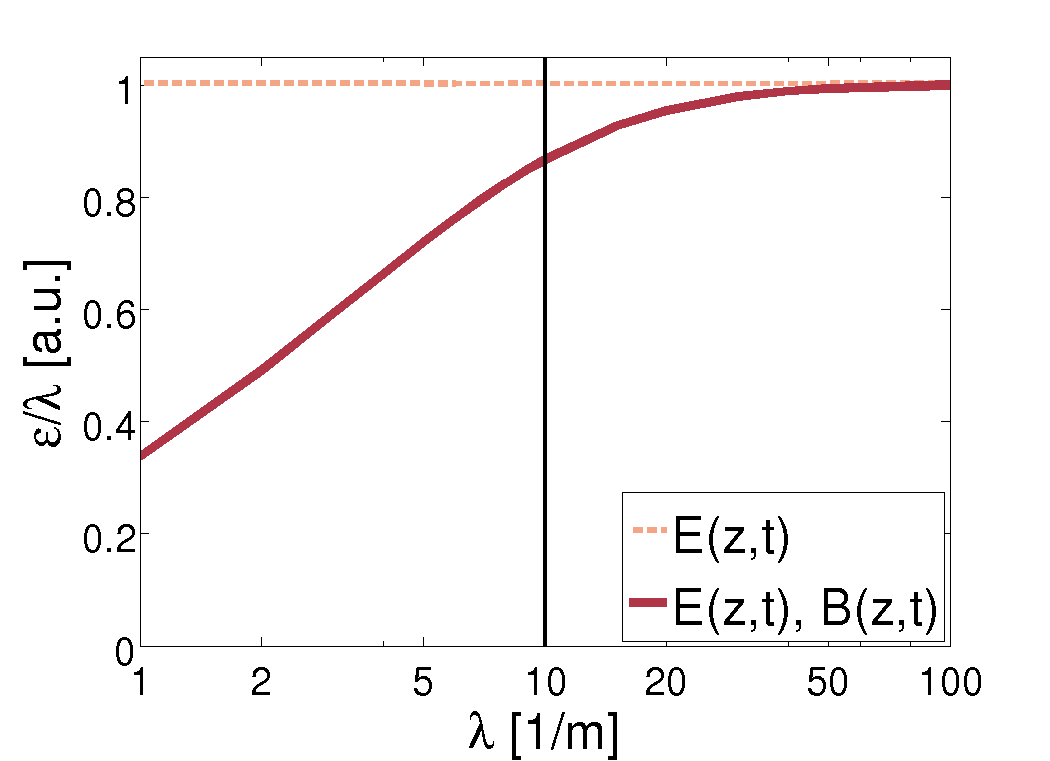}   
\end{center}
\caption[Plot of the modified effective field energy with and without a magnetic field.]{Log-lin plot of the modified effective field energy $\mathcal{E}$ normalized by the spatial extent $\lambda$ for a double-peaked field(left-hand) and for a single-peak dominated field(right-hand) with $\tau=10$ $m^{-1}$.  The unmodified electric field energy $\mathcal{E}\br{E,0}$ is a linear function of the spatial extent $\lambda$. How fast the modified energy $\mathcal{E}\br{E,B}/\lambda$ decreases depends on the structure of the background field.}
\label{Fig_BEn}
\end{figure}

The special form of the transport equations \eqref{eq_B_1}-\eqref{eq_B_4} enables us to estimate the contribution of the magnetic field for a given field configuration. The plots in Fig. \ref{Fig_BEn} hold as a reference where to expect a major difference in calculations with/without magnetic fields. As the modified energy $\mathcal{E}\br{E,B}$ shows a quantitative difference when comparing the different field configurations, we expect the onset of observable effects due to the magnetic field at different $\lambda$. 

\subsection{Particle distribution}

\begin{center}
\begin{figure}[htb]
 \includegraphics[width=0.5\textwidth]{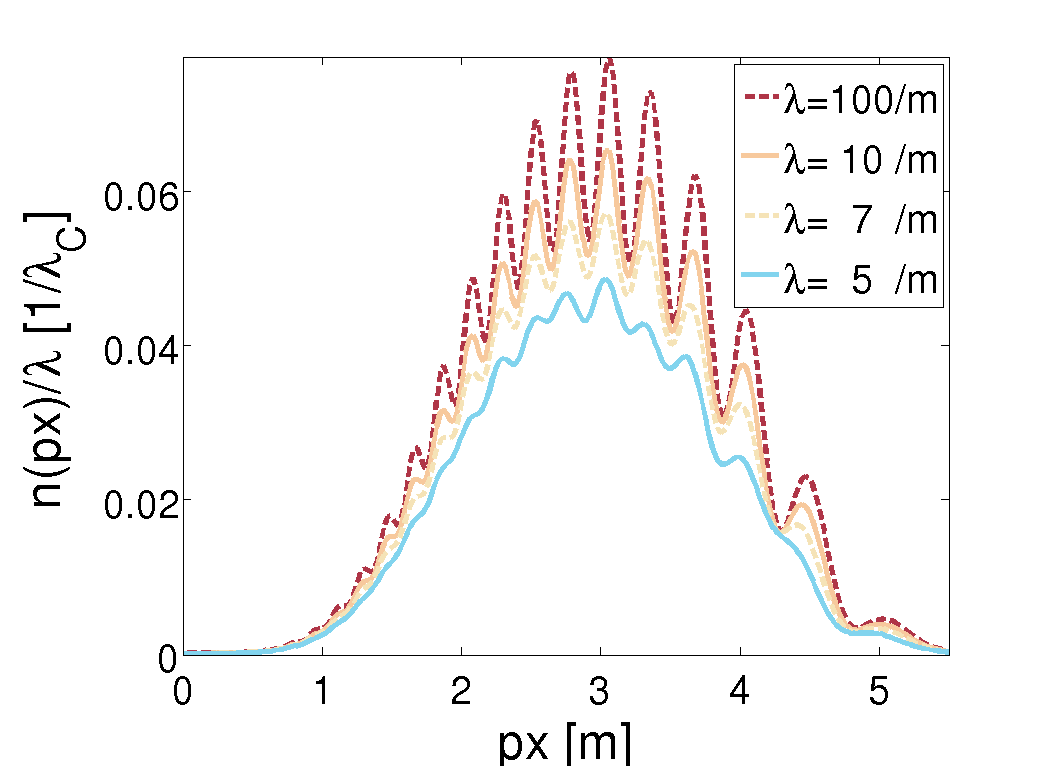}
 \includegraphics[width=0.5\textwidth]{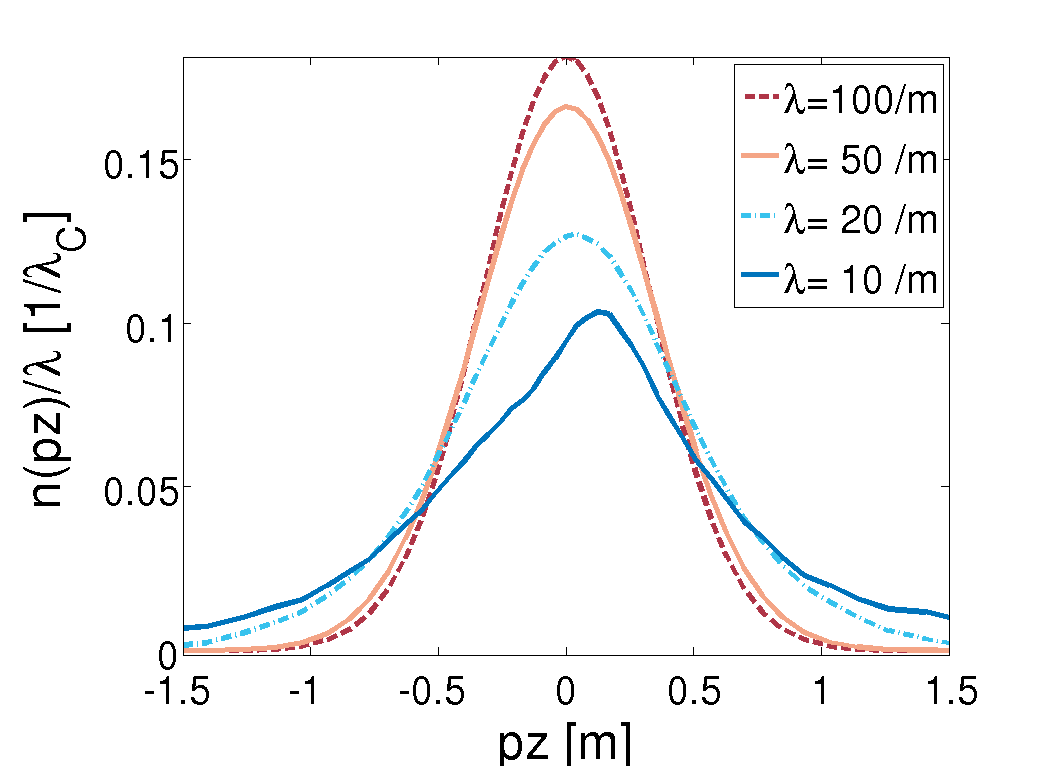} 
 \caption[Plots of the reduced particle density for the double-peak configuration as function of the different momenta.]{Reduced particle density for the double-peak configuration as a function of $p_x$(left) and $p_z$(right). Both distribution functions converge to the homogeneous limit. The deflection of the particle bunch for small $\lambda$ can be related to magnetic field effects. In addition, the interference pattern vanishes. Field parameters: $\varepsilon=0.707$ and $\tau=5$ $m^{-1}$. Further data: Tab. \ref{Tab_ShortInt_Th}}
 \label{Fig_BNp1}
\end{figure}
\end{center}

\begin{center}
\begin{figure}[h]
  \includegraphics[width=0.5\textwidth]{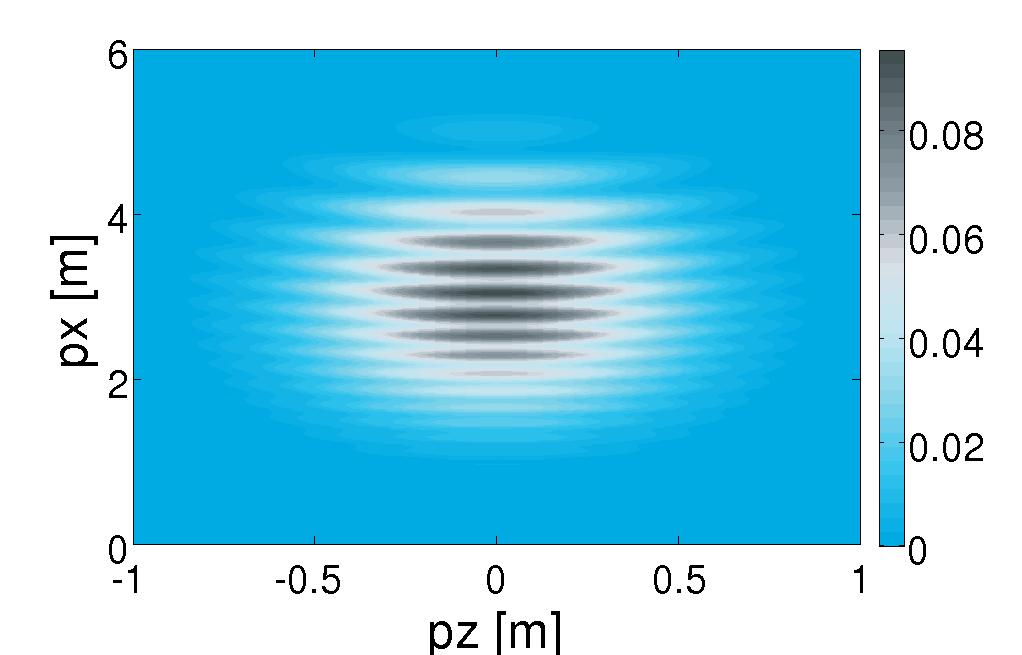}
  \includegraphics[width=0.5\textwidth]{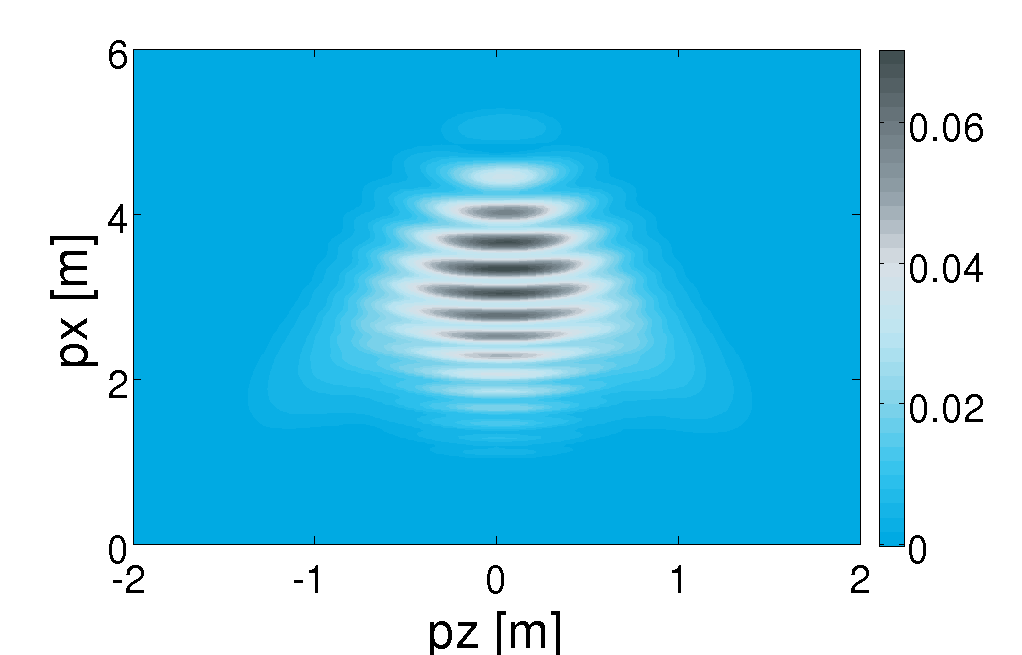}
  \includegraphics[width=0.5\textwidth]{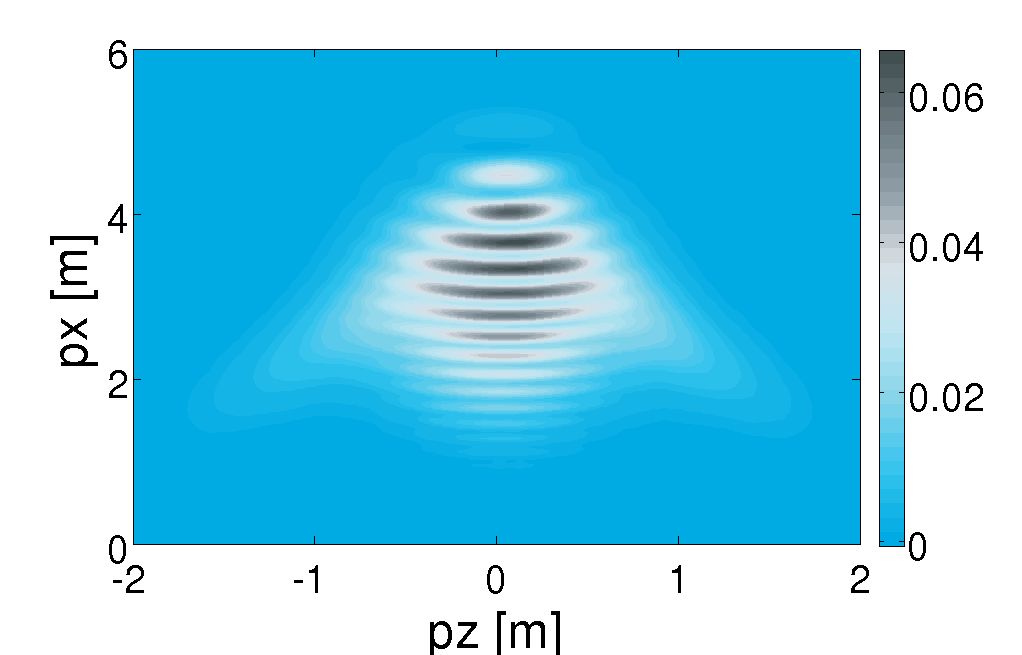}
  \includegraphics[width=0.5\textwidth]{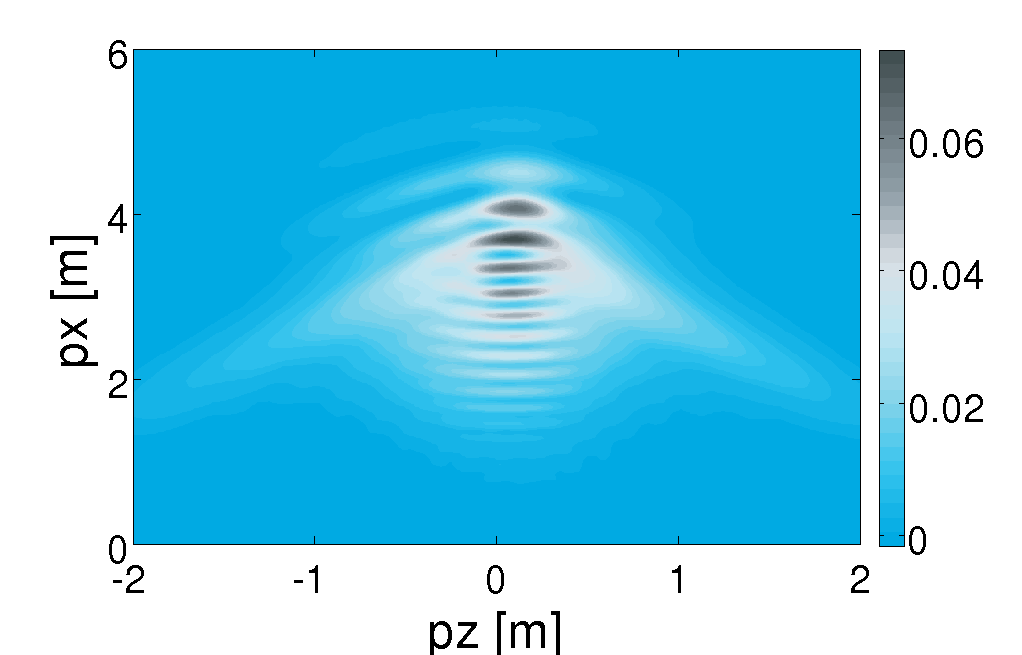}
  \includegraphics[width=0.5\textwidth]{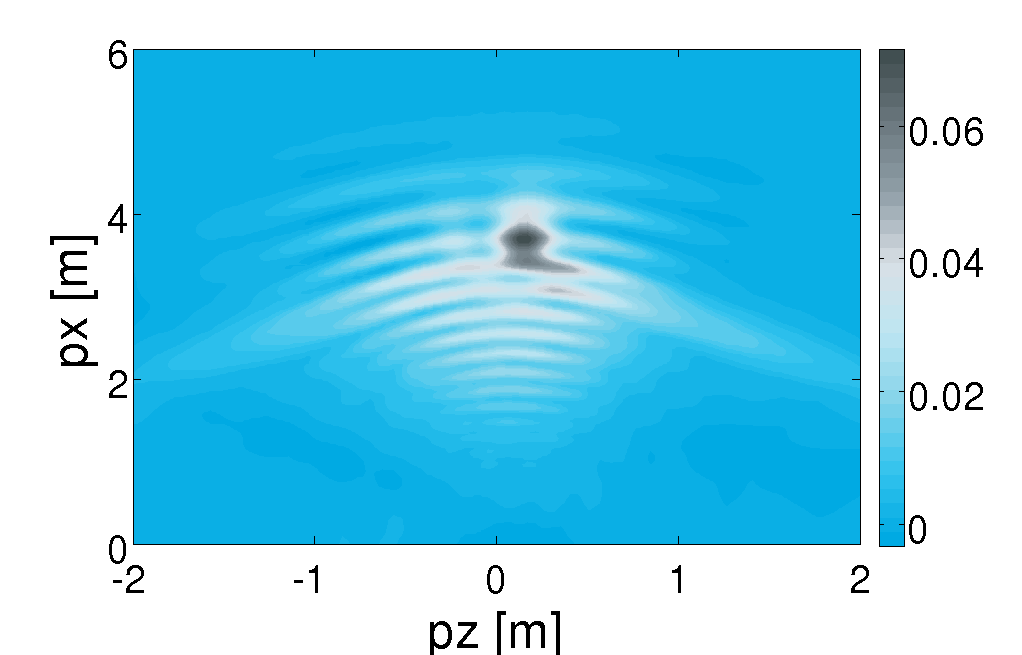}
  \includegraphics[width=0.5\textwidth]{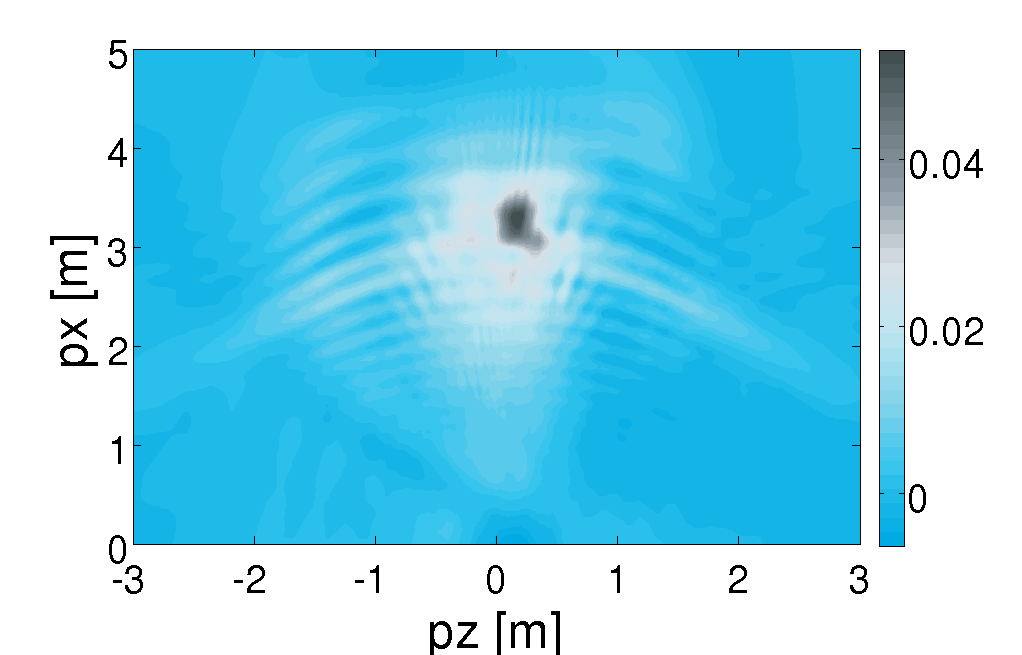}
  \caption[Momentum maps of the reduced particle distribution function for a double-peaked electric field and various spatial extents.]{Momentum maps of the reduced particle distribution function $n \br{p_x,p_z}/\lambda$ for the double-peaked configuration. Pulse parameters are: $\varepsilon=0.707$ and $\tau=5$ $m^{-1}$. The spatial extent $\lambda$ was varied from $\lambda=100$ $m^{-1}$(top left) to $\lambda=5$ $m^{-1}$(bottom right). Intermediate values are $\lambda=20,~ 15,~ 10,~ 7$ $m^{-1}$. The interference pattern can be related to the double-peak structure of the electric field. The stronger the magnetic field, the more the region with high particle density(dark grey) is concentrated. Further parameters: Tab. \ref{Tab_ShortB}}
  \label{Fig_BDistr1}
\end{figure}
\end{center}

We have performed calculations for the double-peaked as well as for the quasi single-peaked configuration. First, we concentrate on the double-peaked setup. We show the particle distribution function for various $\lambda$ in Fig. \ref{Fig_BNp1} and Fig. \ref{Fig_BDistr1}.

The particle bunch exhibiting an interference pattern for a homogeneous background becomes strongly localized and shifts to higher $p_z$ for smaller $\lambda$. Additionally, the oscillating function superimposing the distribution function vanishes.

In a semi-classical analysis the output can only be partially described. The fact, that the magnetic field strength significantly increases for smaller $\lambda$ renders approaches, which do not take quantum interactions into account, inconclusive. Nevertheless, we discuss the results obtained through evaluating the DHW equations in terms of a semi-classical approach.
Due to the double-peak structure of the electric field we expect pair production to happen either at $t=-\tau$ or $t=\tau$. Evaluation of the Lorentz force equation for the homogeneous field yields, that the electron bunch is located at $p_x \approx 3.4$ $m$, $p_z \approx 0$. Additionally, the distribution function should exhibit an interference pattern already discussed in section \ref{Kap_SemiClass}. A signal for the increasing importance of magnetic field interactions is the fact, that the $p_z$-symmetry in Fig. \ref{Fig_BNp1} is broken. Moreover, the disappearance of the oscillations in Fig. \ref{Fig_BNp1} for small $\lambda$ can be related to the strengthening of the magnetic field. Particles created at different times $t_0$ with different momentum in $z$-direction acquire a completely different final momentum. As only the magnetic field contributes to the $p_z$-momentum this result serves as another indicator for magnetic field contributions. We have summarized these findings in Tab. \ref{Tab_LorentzD}.

\ctable[pos=hb,
caption = {Evaluating the relativistic Lorentz force equation for the double-peaked configuration for particles seeded at $t_0=\pm \tau$ at position $x_0=0$ $m^{-1}$. The initial momenta $p_{x,0}$ and $p_{z,0}$ have been varied in order to obtain the corresponding final momenta $p_{x,f}$ and $p_{z,f}$ at asymptotic times. Parameters: Field strength $\varepsilon=0.707$, peak width $\tau=5$ $m^{-1}$ and spatial extent $\lambda=10$ $m^{-1}$.},
cap = {Evaluating the relativistic Lorentz force equation for a double-peaked electric field plus corresponding magnetic field.},
label = Tab_LorentzD, 
mincapwidth = \textwidth,
]{c c c c c}{
}{
    \toprule
    \hspace{0.2cm} $t_0$ $[m^{-1}]$ \hspace{0.2cm}  & \hspace{0.2cm} $p_{x,0}$ $[m]$ \hspace{0.2cm}  & \hspace{0.2cm} $p_{z,0}$ $[m]$ \hspace{0.2cm} & \hspace{0.2cm} $p_{x,f}$ $[m]$ \hspace{0.2cm} & \hspace{0.2cm} $p_{z,f}$ $[m]$ \hspace{0.2cm} \\
    \midrule
    -$\tau$ & 0 & 0 & 3.5 & 0 \\  
    \midrule
    $\tau$ & 0 & 0 & 3.5 & 0 \\
    \midrule
    -$\tau$ & 0.1 & 0 & 3.5 & 0 \\      
    \midrule
    $\tau$ & 0.1 & 0 & 3.5 & 0 \\        
    \midrule
    -$\tau$ & 0 & 0.1 & 3.39 & \hlight{-0.41} \\    
    \midrule
    $\tau$ & 0 & 0.1 & 3.4 & \hlight{0.25} \\   
    \bottomrule
} 

\clearpage


Next, we analyze the model \eqref{Field2}. In contrast to the previous configuration, the electric field exhibits a dominant peak in time. Due to this special feature, the contribution of the magnetic field for fixed spatial extent $\lambda$ is small compared to the double-peaked case. The fact, that the magnetic field plays only a minor role for $\lambda \gtrsim \tau$ is also reflected in the particle distribution. For $\lambda=100$ $m^{-1}$ we obtain a single peak in the distribution function without any interference pattern visible. Furthermore, the function $n \br{p_x}/\lambda$ remains nearly invariant for varying extent $\lambda$. The changes become visible, only if the distribution function is plotted as a function of $p_z$. 

\begin{center}
\begin{figure}[htb]
 \includegraphics[width=0.5\textwidth]{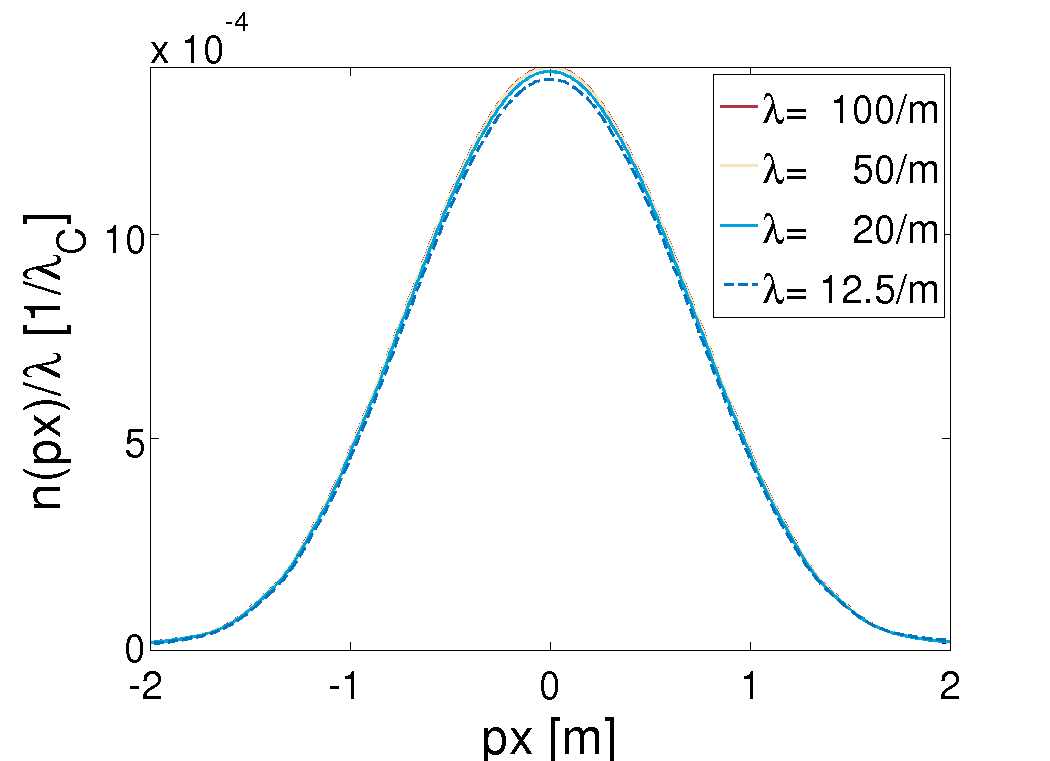}
 \includegraphics[width=0.5\textwidth]{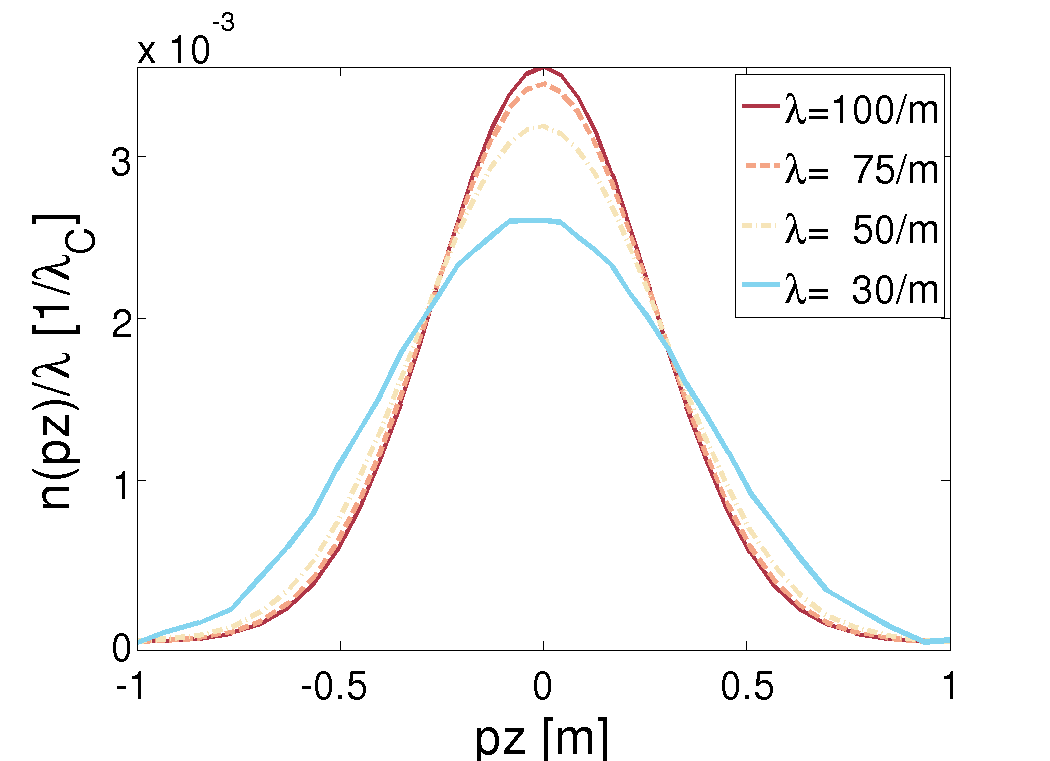} 
 \caption[Plots of the reduced particle density for a single-peak dominated electric field plus magnetic field as functions of the different momenta.]{Reduced particle density for the quasi single-peaked configuration as a function of $p_x$(left) and $p_z$(right). The broadening of the distribution function $n \br{p_z}/\lambda$ can be related to the presence of the magnetic field. Field parameters: $\varepsilon=0.707$ and $\tau=10$ $m^{-1}$. Further data: Tab. \ref{Tab_ShortInt_Th}}
 \label{Fig_BNp2}
\end{figure}
\end{center}

\begin{center}
\begin{figure}[H]
 \includegraphics[width=0.5\textwidth]{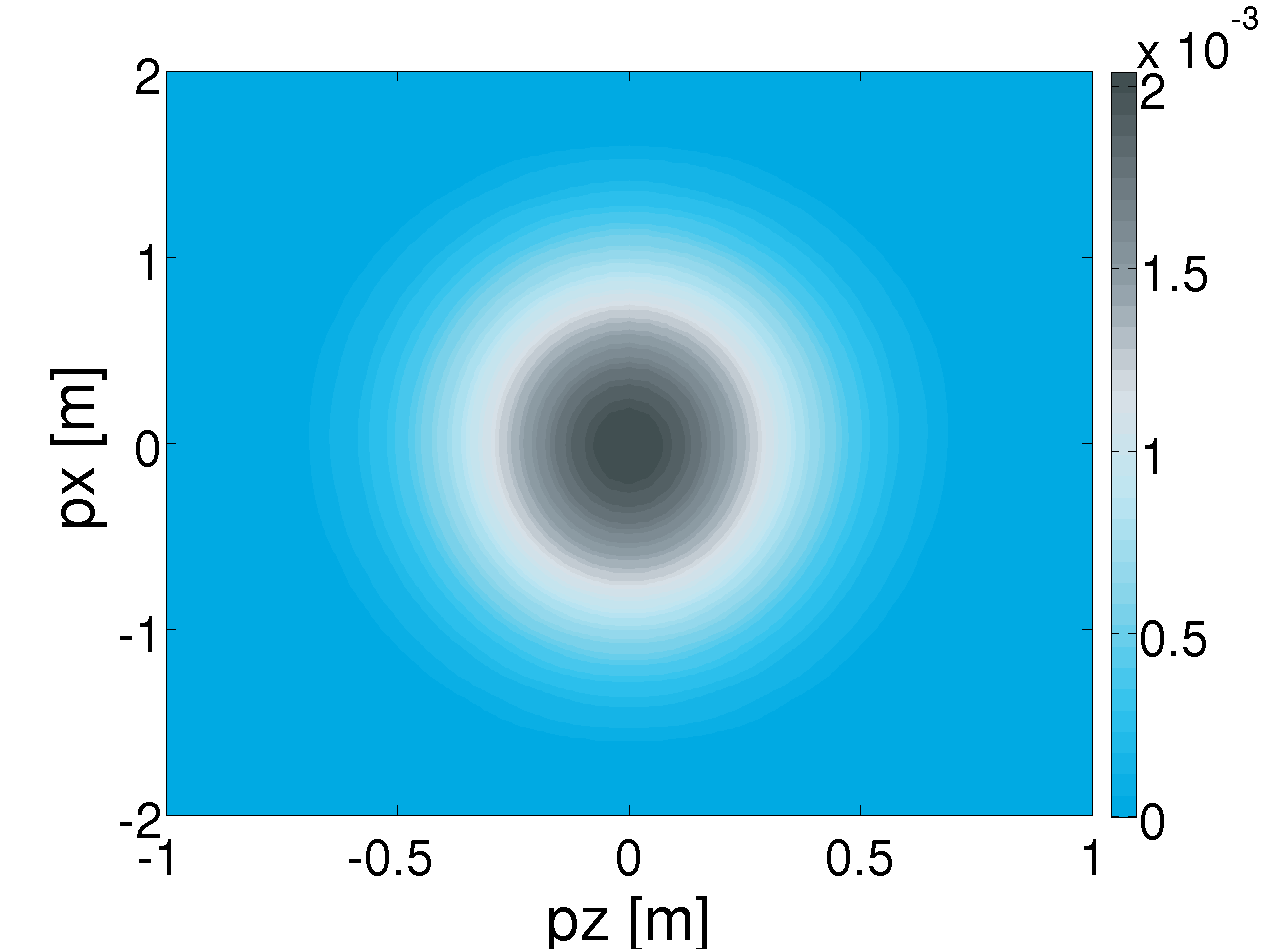}
 \includegraphics[width=0.5\textwidth]{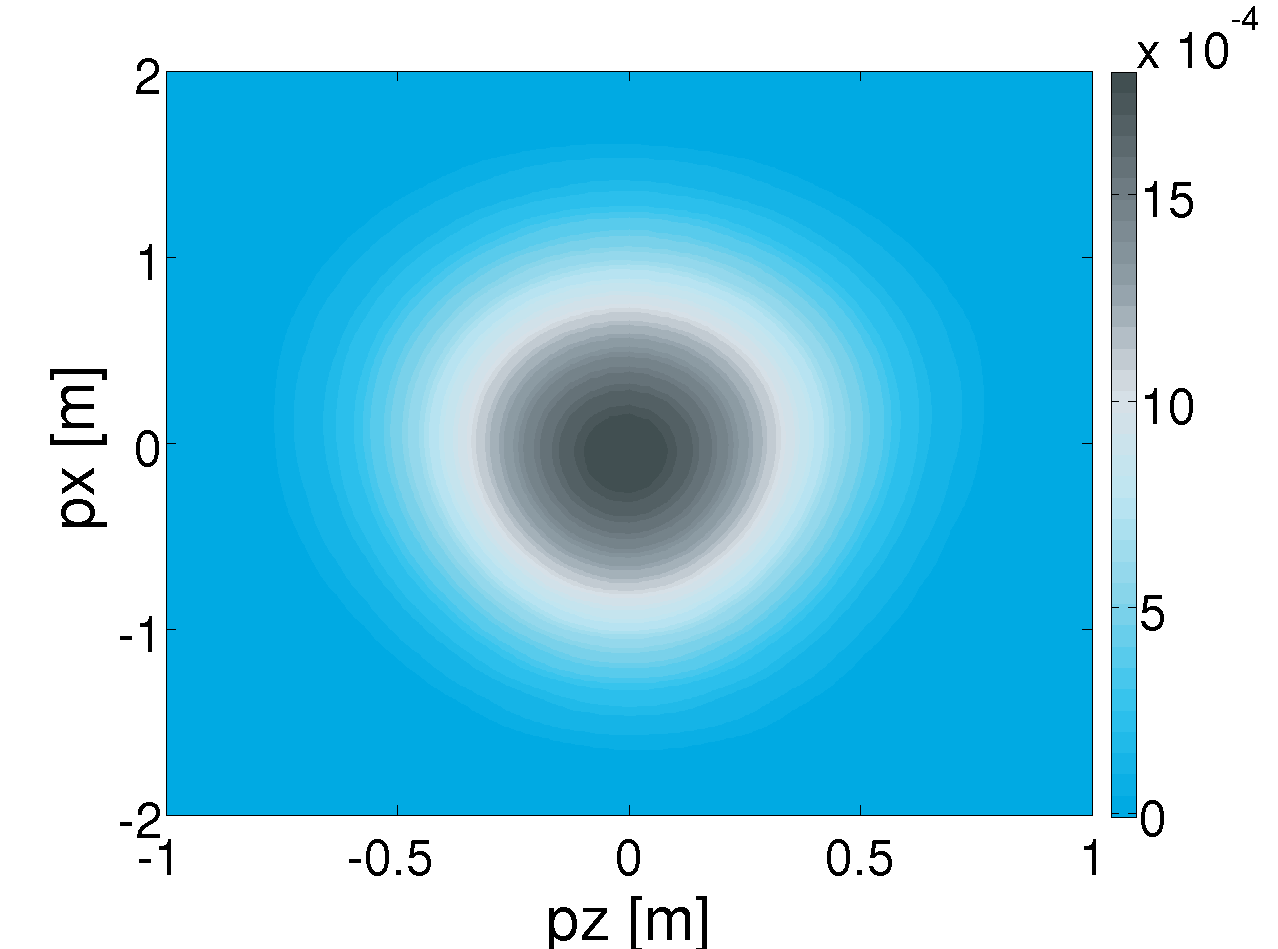}
 \includegraphics[width=0.5\textwidth]{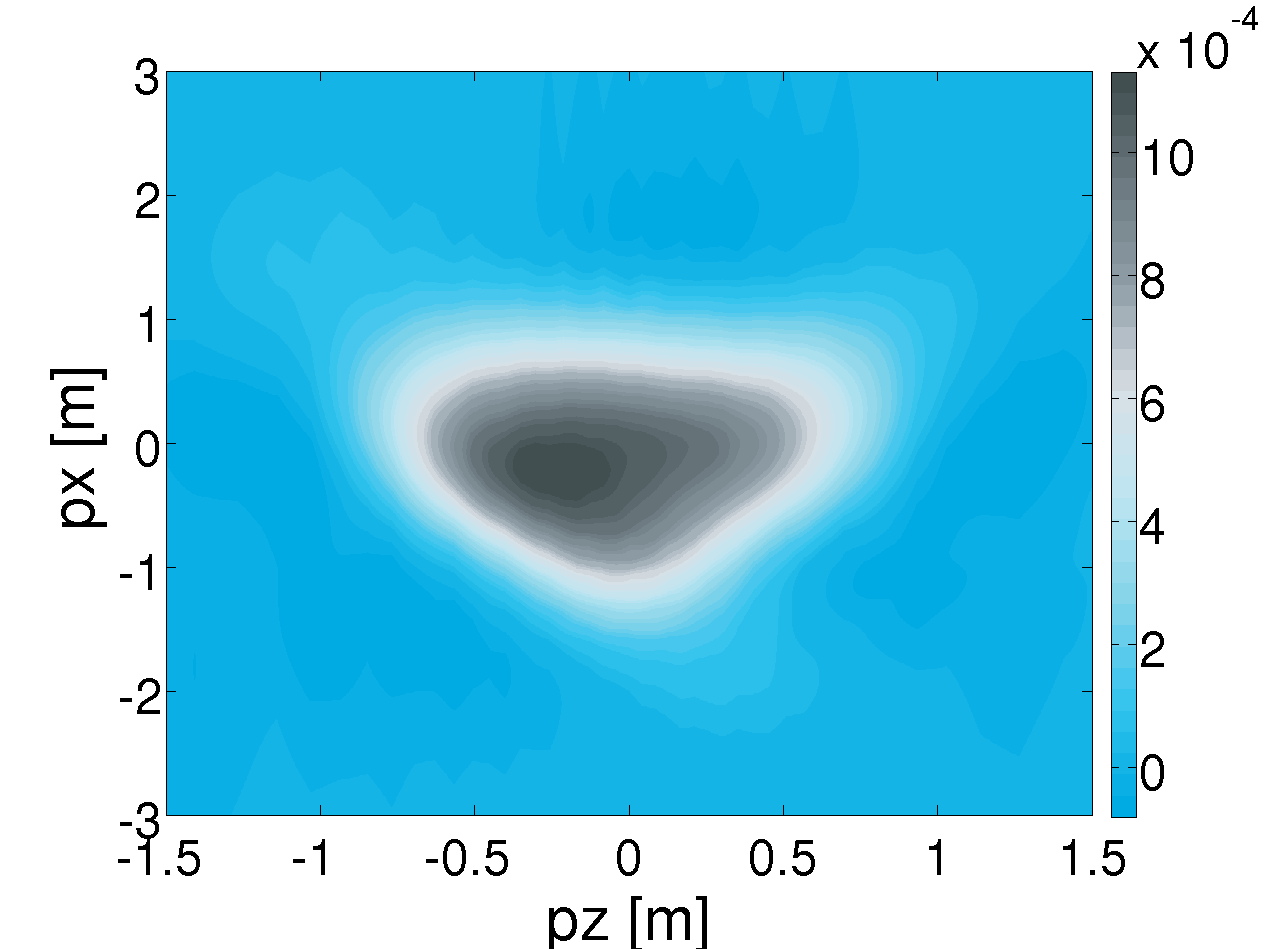}
 \includegraphics[width=0.5\textwidth]{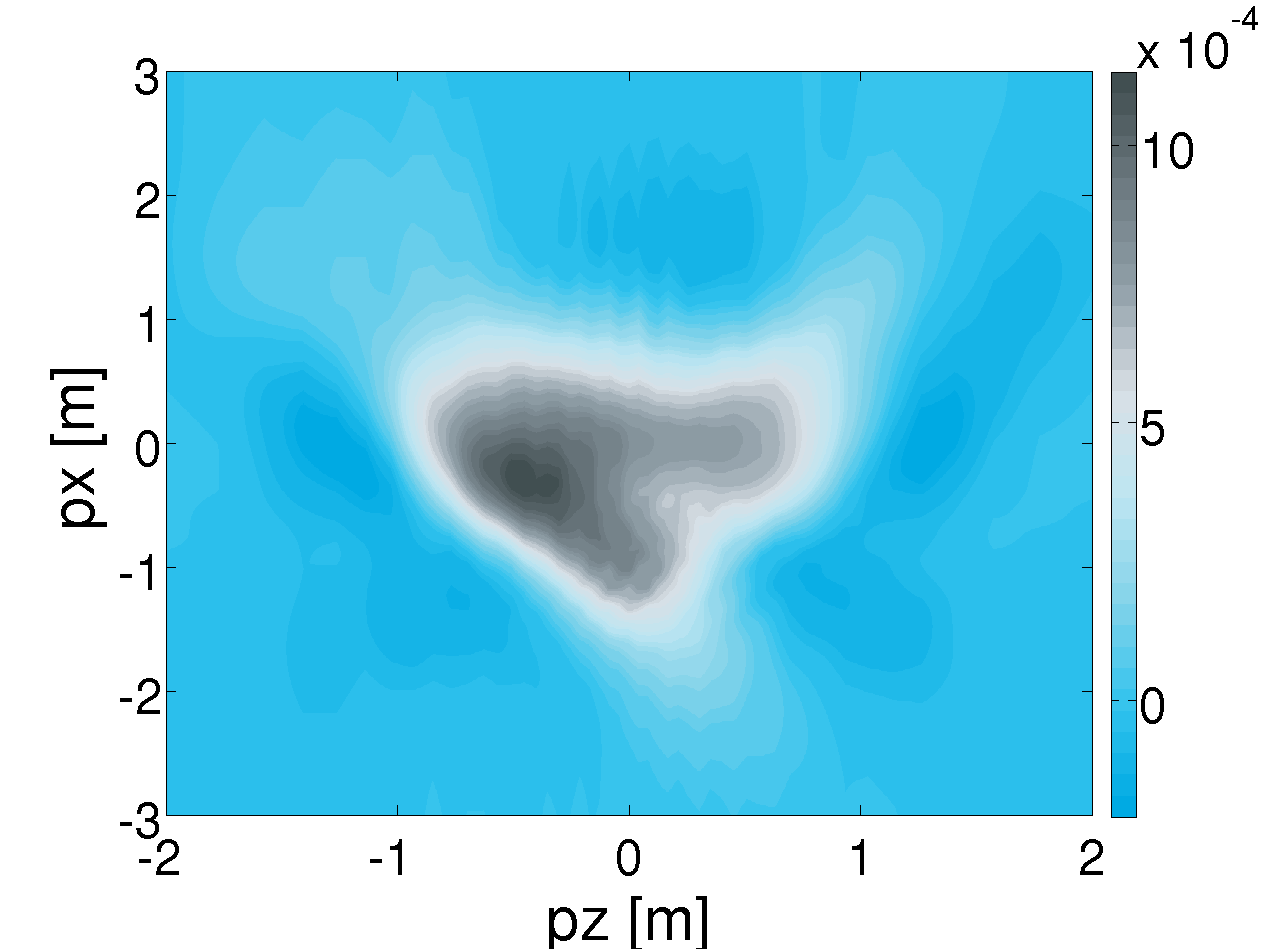}
 \includegraphics[width=0.5\textwidth]{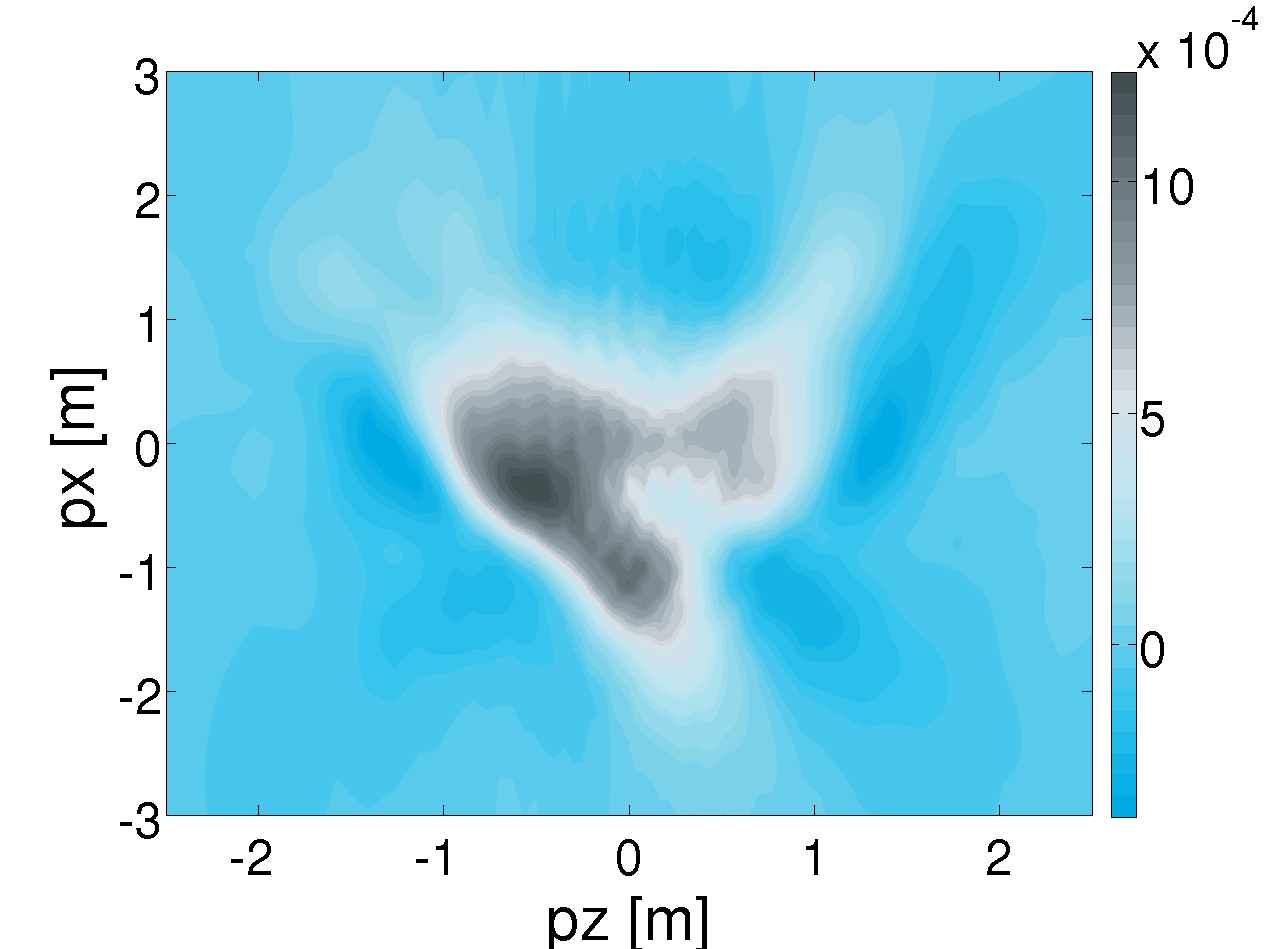}
 \includegraphics[width=0.5\textwidth]{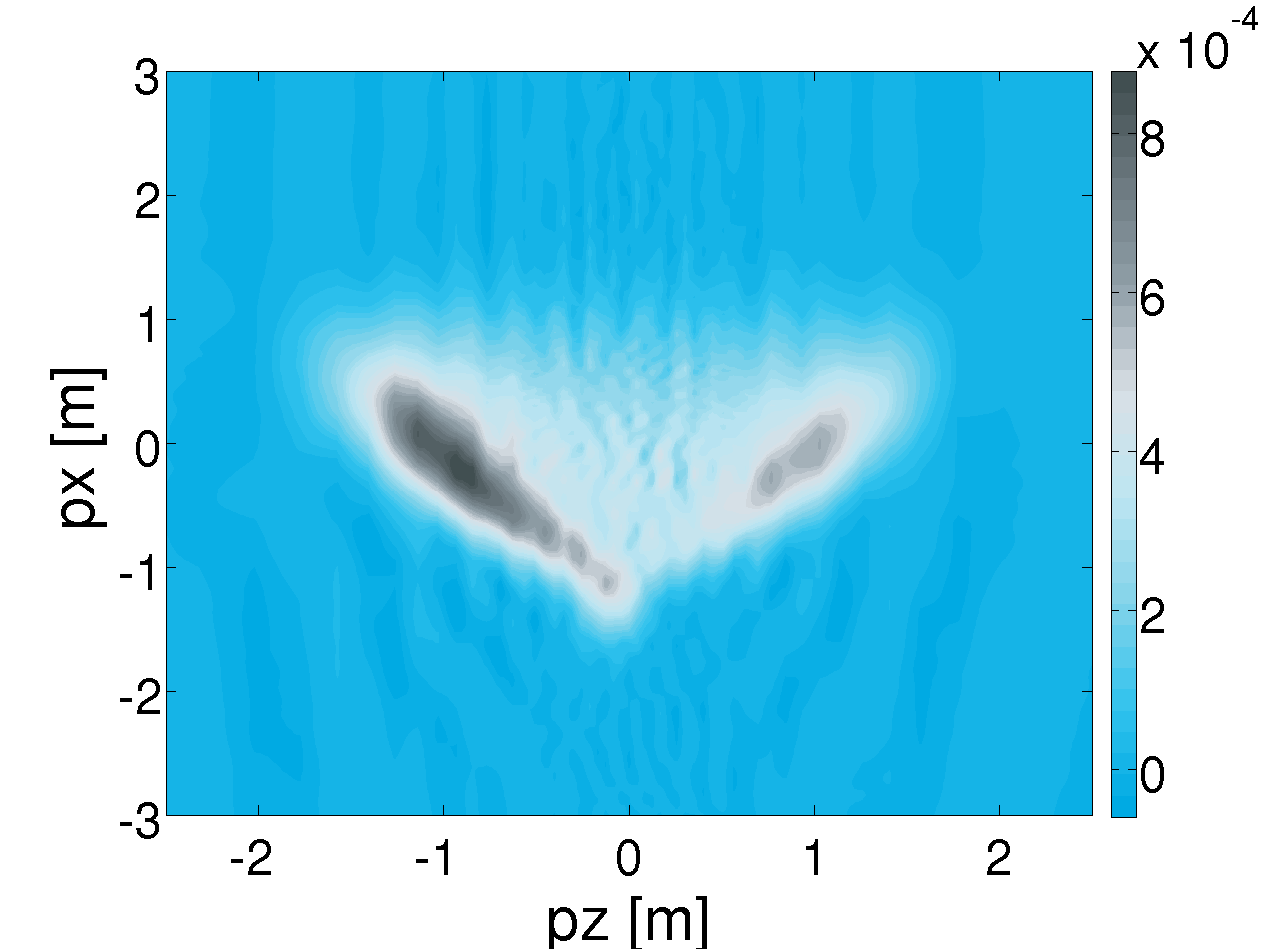}
\caption[Momentum maps of the reduced particle distribution function for an electric field exhibiting a single dominant peak plus corresponding magnetic field for various spatial extents.]{Reduced particle distribution function $n \br{p_x,p_z}/\lambda$ in full momentum space for the quasi single-peaked configuration. Corresponding parameters are: $\varepsilon=0.707$ and $\tau=10$ $m^{-1}$. The spatial extent $\lambda$ was varied from $\lambda=100$ $m^{-1}$(top left) to $\lambda=9$ $m^{-1}$(bottom right). Intermediate values are $\lambda=50,~ 20,~ 15,~ 12.5$ $m^{-1}$. The splitting of the particle bunch can be related to the finite spatial extent and thus a stronger magnetic field. Further parameters: Tab. \ref{Tab_Triple}}
\label{Fig_BDistr2}
\end{figure} 
\end{center}

The reason for this behaviour can be traced back to the finite spatial extent. Due to the increase of the magnetic field strength for smaller $\lambda$, the particles are mainly accelerated in $z$-direction, see Fig. \ref{Fig_BDistr2}. However, the magnetic field is not strong enough in order to particularly influence the specific process of pair production. Hence, one obtains the nearly constant reduced particle yield, shown in Fig. \ref{Fig_BNp2}.

\clearpage

\vspace{5cm}

In order to quantify our interpretation we rely on a semi-classical analysis.
In Tab. \ref{Tab_LorentzT} we present the data obtained via evaluating the Lorentz force equation for the model \eqref{Field2}. As the electric field configuration exhibits a single dominant peak in time, we assume that particle creation is most likely at $t \approx 0$, $x \approx 0$. Moreover, we can relate to model \eqref{Field2} approximately as a single-peak configuration and thus no interference pattern emerges.

\ctable[pos=hb,
caption = {Evaluating the relativistic Lorentz force equation for particles seeded at $t_0(t=0)$ at position $x_0=0$, because the background electric field exhibits a dominant peak in time. The corresponding parameters are given by the field strength $\varepsilon=0.707$, the peak width $\tau=10$ $m^{-1}$ and the spatial extent $\lambda$. The initial momenta $p_{x,0}$ and $p_{z,0}$ have been varied in addition to $\lambda$. The final momenta $p_{x,f}$ and $p_{z,f}$ are obtained at asymptotic times when the applied electric and magnetic fields already vanished.},
cap = {Evaluating the relativistic Lorentz force equation for an electric field exhibiting a dominant peak in time and an additional magnetic field.},
label = Tab_LorentzT, 
mincapwidth = \textwidth,
]{ c c c c c}{
}{
    \toprule
    \hspace{0.2cm} $\lambda$ $[m^{-1}]$ \hspace{0.2cm} & \hspace{0.2cm} $p_{x,0}$ $[m]$ \hspace{0.2cm}  & \hspace{0.2cm} $p_{z,0}$ $[m]$ \hspace{0.2cm} & \hspace{0.2cm} $p_{x,f}$ $[m]$ \hspace{0.2cm} & \hspace{0.2cm} $p_{z,f}$ $[m]$ \hspace{0.2cm} \\
    \midrule
    100 & 0 & 0 & 0.0 & 0 \\  
    \midrule
    100 & 0 & 0.5 & 0.0 & 0.52 \\
    \midrule
    100 & 0 & -0.5 & 0.0 & -0.52 \\    
    \midrule
    100 & 0.5 & 0 & 0.5 & 0 \\   
    \midrule
    100 & -0.5 & 0 & -0.5 & 0 \\      
    \midrule
    100 & 0.5 & 0.5 & 0.5 & 0.53 \\      
    \midrule
    100 & -0.5 & 0.5 & -0.5 & 0.51 \\ 
    \midrule
    \midrule
    10 & 0 & 0 & 0.0 & 0 \\  
    \midrule
    10 & 0 & 0.5 & 0.0 & 4.45 \\
    \midrule
    10 & 0 & -0.5 & 0.0 & -4.45 \\    
    \midrule
    10 & 0.5 & 0 & 0.5 & 0 \\   
    \midrule
    10 & -0.5 & 0 & -0.5 & 0 \\      
    \midrule
    10 & 0.5 & 0.5 & 0.5 & \hlight{8.53} \\      
    \midrule
    10 & -0.5 & 0.5 & -0.5 & \hlight{2.1} \\     
    \bottomrule
} 

\clearpage

Most striking results are obtained for $\lambda = 10$ $m^{-1}$ and $p_{z,0} \neq 0$. Compared with field configurations, where $\lambda = 100$ $m^{-1}$ the particles are strongly accelerated in $z$-direction. The last two lines of Tab. \ref{Tab_LorentzT} further reveal, that the initial value $p_{x,0}$ strongly affects the outcome. In terms of a semi-classical analysis this can be related to the fact, that first the particle acquires energy through the electric field resulting in an increase of $p_x$. As the background is independent of $x$, the third peak in $E \br{z,t}$ decelerates the particle effectively lowering $p_x$. However, simultaneously, the strength of the magnetic field increases thus boosting the particles momentum in $z$-direction. In this way it is possible to use the fields as a slingshot.
Furthermore, if the number of particles produced is independent of the sign of $p_{x,0}$ the results for $p_{x,f}$ and $p_{z,f}$ can be directly related to the V-shape of the particle distribution function, see Fig. \ref{Fig_BDistr2}. 

The symmetry-breaking in $p_z$ is probably related to field-spin interactions. However, at this point such an explanation would be speculative thus we retrain from an analysis and refer to future investigations.

\subsection{Particle yield}

\begin{figure}[htb]
\begin{center}
  \includegraphics[width=0.65\textwidth]{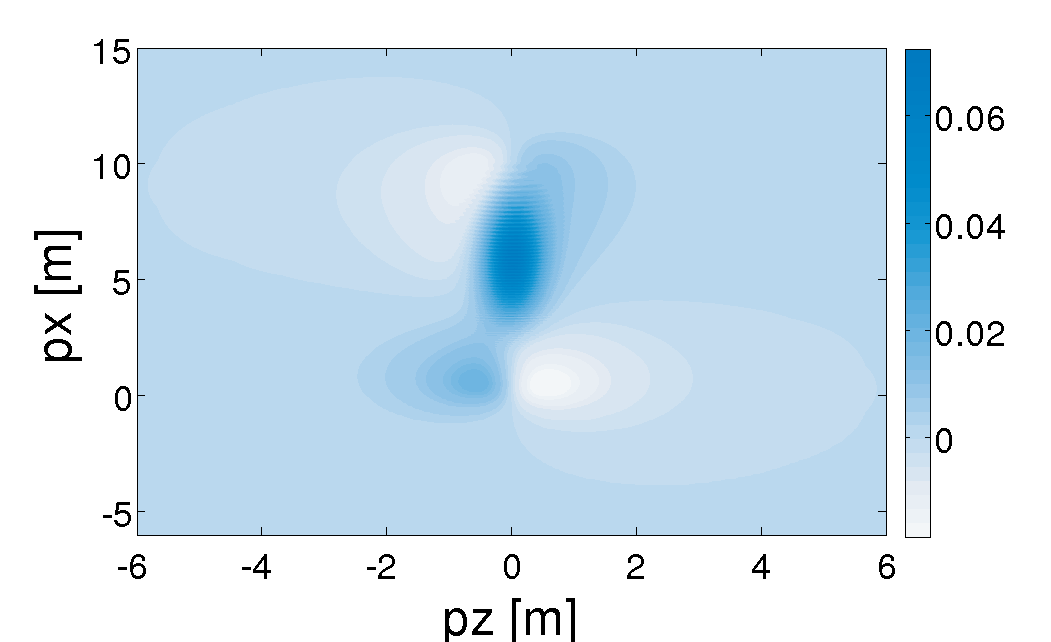}
\end{center}
\caption[Plot of the reduced quasi-distribution function in case the magnetic field is artificially set to zero.]{Showcase for a calculation with the magnetic field artificially fixed to zero. Extensive regions with negative ``particle distribution'' are identified. Compared to proper calculations ($B \br{z,t} \neq 0$) the reduced quasi-particle distribution is greatly broadened. Parameters used for the double-peaked electric field: $\varepsilon=0.707$, $\tau=10$ $m^{-1}$ and $\lambda=25$ $m^{-1}$. Further data: Tab. \ref{Tab_Long}}
\label{Fig_B0}
\end{figure}

Before we start examining the particle yield we show results of a simulation, where the magnetic field was artificially set to zero. In such a case, the transport equations(\eqref{eq_B_1}-\eqref{eq_B_4}) greatly simplify. However, due to violating the homogeneous Maxwell equations numerical artifacts show up in the particle distribution function. Fig. \ref{Fig_B0} holds as a showcase for such a calculation. We can still observe a peak in the distribution function at around $p_x= 7$ $m$, $p_z=0$  (which agrees with the semi-classical prediction). However, we spot large regions where the distribution becomes negative thus rendering any physical interpretation impossible. Still, we can compare the outcome with results obtained for different $\lambda$ in order to gather information about how much the magnetic field contributes. As this is better done integrating out all non-physical regions we focus on the particle yield only. In this way, we can also compare with the results for the modified field energy, Fig. \ref{Fig_BEn}.

\begin{center}
\begin{figure}[htb]
  \includegraphics[width=0.5\textwidth]{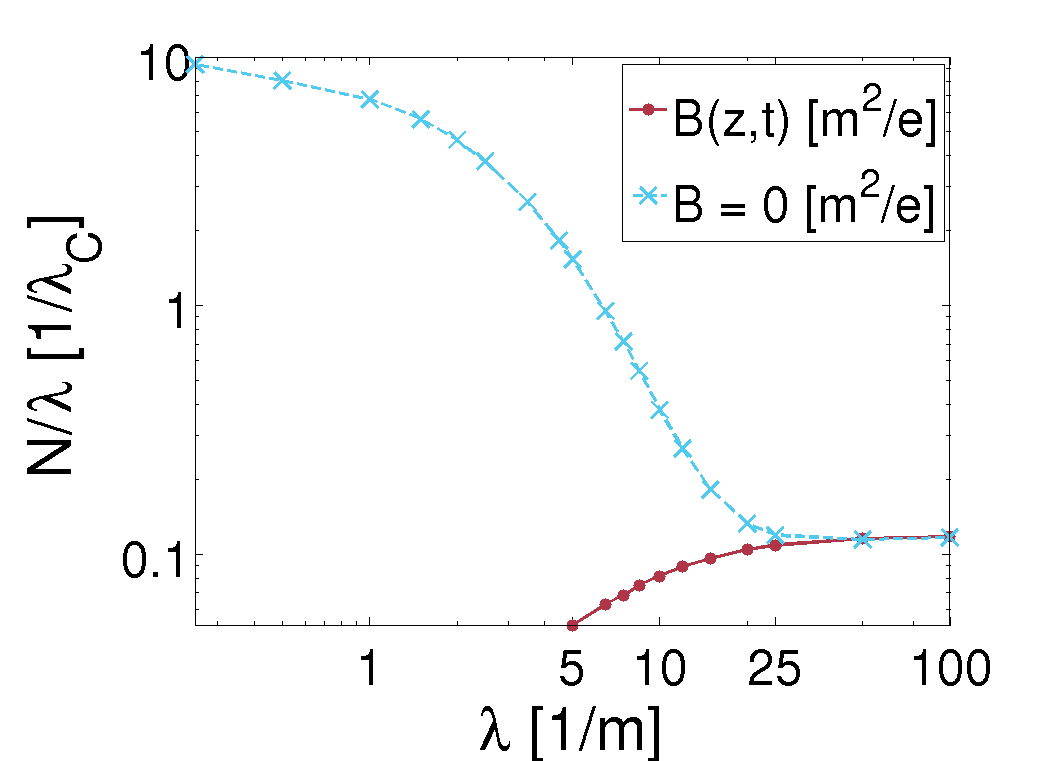}
  \includegraphics[width=0.5\textwidth]{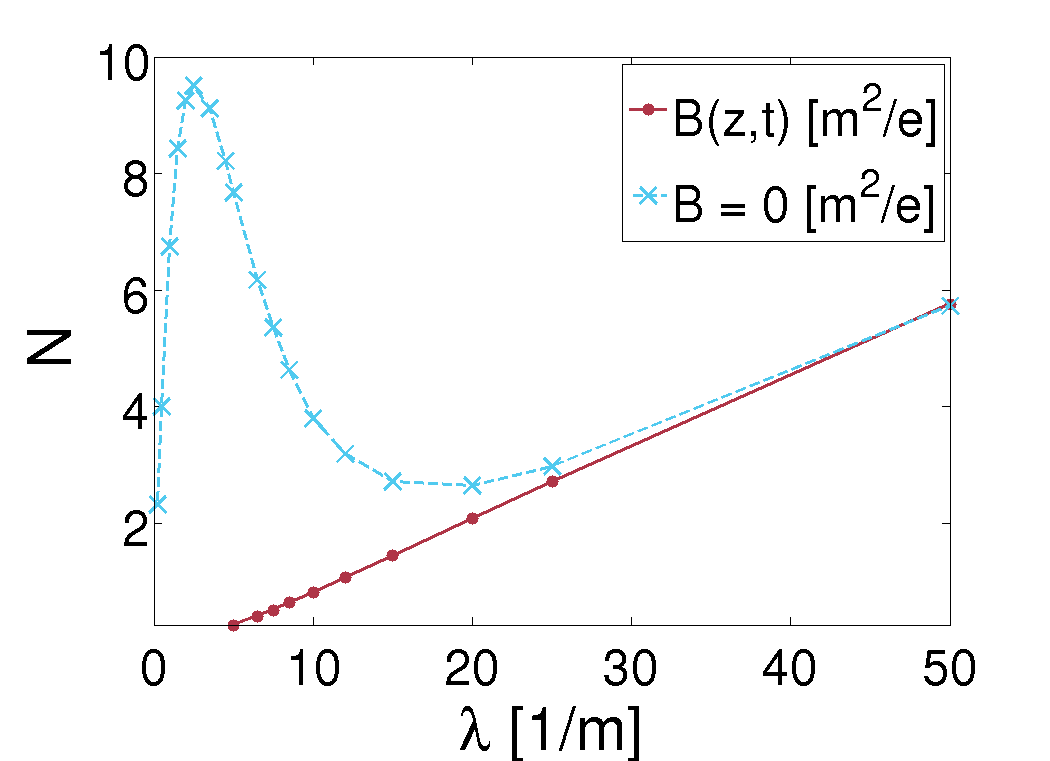}
  \caption[Particle yield as a function of the spatial extent.]{Reduced particle yield (left-hand) and particle yield (right-hand) drawn in a log-log and a lin-lin plot as a function of the spatial extent $\lambda$. The yield converges to the homogeneous result for $\lambda \gg 1$ $m^{-1}$. One clearly sees, that neglecting the magnetic fields leads to a sizable overestimation of the correct result. This behaviour can be related to the corresponding Lorentz invariants. We have used the double-peak model for the fields, where $\varepsilon=0.707$ and $\tau=10$ $m^{-1}$. Additional information can be found in the appendix: Tab. \ref{Tab_IntB}, Tab. \ref{Tab_IntB0}}
  \label{Fig_BYield}
\end{figure}
\end{center}

The field energy of the electric field is a linear function of its spatial extent $\lambda$. If no magnetic field existed, one would assume that the particle yield is also a linear function of $\lambda$. This holds in the limit of $\lambda \to \infty$, as displayed in Fig. \ref{Fig_BYield}, because the electric field is only weakly varying in $z$ and the magnetic field energy would have been comparatively small. 

We have already analyzed the impact of spatial limitations on the particle yield, see chapter six. Here, we also have to take the magnetic field into account. For this purpose, we introduced an effective field amplitude \eqref{Eq_FieldAmpl}. In this way, we may estimate the result for the yield by a simple model. The corresponding calculations are illustrated in Fig. \ref{Fig_BEn}, where for both field configurations a faster-than-linear decrease is shown. In Fig. \ref{Fig_BYield} we obtain a similar decrease in the yield for the double-peak configuration. At this point we have to remark, that numerical noise, which would render the distribution function uninterpretable, averages out in the particle yield. One issue is, that the drop-off in Fig. \ref{Fig_BYield} could also be related to a too low total energy(section \ref{Sec_Low}). Future investigation on this problem would be necessary to clarify this issue.  

\begin{figure}[htb]
\begin{center}
 \includegraphics[width=0.5\textwidth]{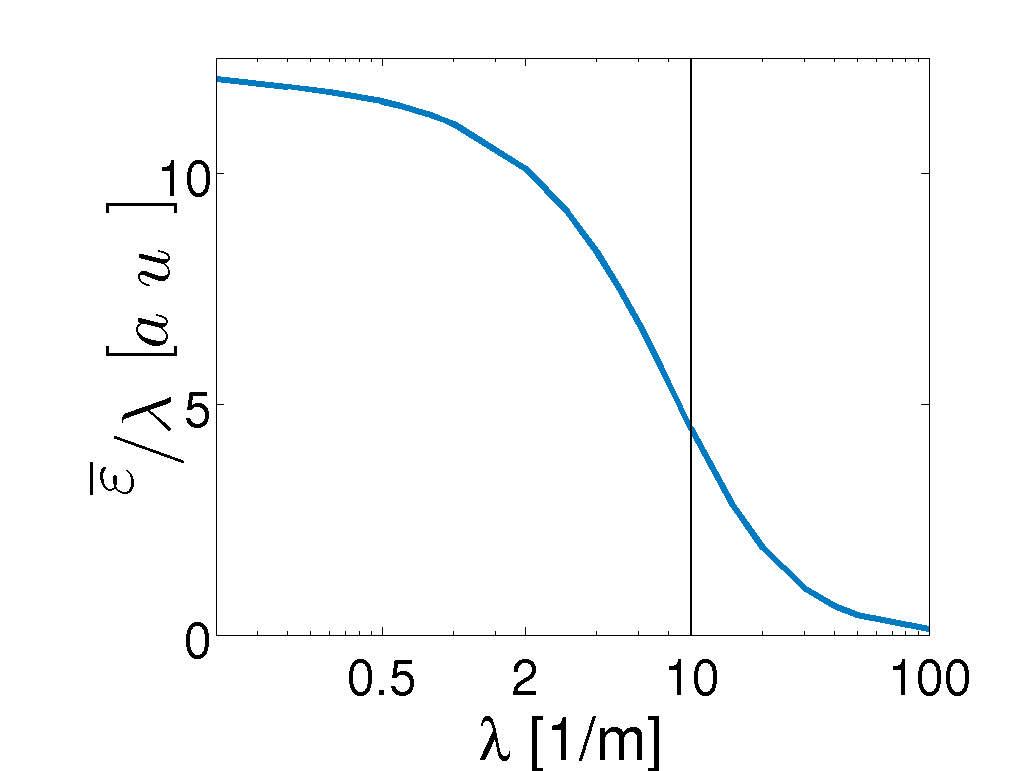} 
\end{center}
\caption[Plot of the overestimation of the effective field amplitude due to fixing $B$ to zero as a function of the spatial extent.]{Log-lin plot  of the overestimation of the effective field amplitude due to fixing $B$ to zero as a function of the spatial extent. The smaller $\lambda$ the more the magnetic field dominates the field energy. The model for the double-peaked electric field was used. The function is normalized such, that $\overline{\mathcal E}/\lambda=1$ for $\lambda=30$ $m^{-1}$.} 
\label{Fig_EB0}
\end{figure}

We concentrate on the differences between calculations, where the vector potential defines the fields and computations, where the magnetic field is artificially set to zero, thus not satisfying Maxwell equations. In the case of the double-peaked model \eqref{Field1} the actual field energy should be overestimated, because the magnetic field, working as a damping factor, is neglected. This overestimation, given in terms of the effective field energy, reads
\begin{equation}
 \overline{\mathcal E} = \mathcal E \br{E,0} - \mathcal E \br{E,B}. 
\end{equation}
An illustration for field strength $\varepsilon=0.707$ and peak width $\tau=10$ $m^{-1}$ is provided in Fig. \ref{Fig_EB0}. A comparison with DHW calculations shows, that the particle yield follows the same pattern, see Fig. \ref{Fig_BYield}. For $\lambda \gg \tau$ the contribution stemming from the magnetic field is negligible. The smaller $\lambda$ the stronger the magnetic field becomes. If $\lambda \approx \mathcal O \br{1 \ m^{-1}}$, the magnetic field is dominating exceeding the energy fraction of the electric field multiple times. In the limit $\lambda \to 0$ the work done by the fields is not sufficient in order to produce particles, see section \ref{Sec_Low}.

In conclusion, based upon numerical solutions of the DHW equations we have presented various possibilities allowing to determine the impact magnetic fields have on the pair production process.

\pagestyle{plain}
\addtocontents{toc}{\protect\clearpage}
\chapter{Summary and Outlook}
\pagestyle{fancy}

\section{Summary}

One of the main objectives of this thesis was to investigate electron-positron pair production in spatially inhomogeneous, time-dependent fields. A special focus was on multiphoton pair production and the effective mass concept. The results were obtained using the phase-space Wigner-formalism supporting the idea of treating phase-space approaches as an important tool in order to explain physics in strong fields and in particular electron-positron pair production. \\

In the introductory chapter we highlighted the possibilities Laser physics could offer in the future. Moreover, we introduced QED with a special emphasis on the future of strong-field QED. \\

The second chapter was devoted to introduce all concepts and ideas relevant for understanding pair production. We started with Maxwell equations, where we did not specify the charge density and the current. Nevertheless these equations restricted us to field configurations stemming from a predefined vector potential. Then, we went on discussing the Lorentz invariants, because they play a crucial role when describing particle creation in electromagnetic fields. 

As we presented experimental milestones in the past, we connected pair production with atomic physics. In this way, we defined a Keldysh parameter in order to easily differentiate between the different mechanisms of particle creation. Moreover, we showed how one can understand these mechanisms visualizing them in a Dirac sea picture. 

Furthermore, we mentioned the SLAC E-144 experiment and thus the first observations of pair production via light-light scattering. We discussed also Laser fields and in particular how to focus high-intensity Laser pulses. Additionally, we introduced the Dirac equation and subsequently the effective mass. \\

Last but not least, we presented the Wigner function approach and explained how to work with quantum kinetic methods. At that point, we already discussed the Hartree approximation in terms of QED. Then, we gave an overview over quantum kinetic methods in various dimensions and how they are related to each other. \\

In chapter three we entirely concentrated on a consistent derivation of the DHW formalism in arbitrary dimensions. Furthermore, we discussed various symmetries of the transport equations themselves. By that way, we also analyzed how symmetries in the background fields translate into the mathematical description of the pair production process. We even re-obtained the equations stemming from original quantum kinetic theory by employing cylindrical symmetry. In section \ref{sec_observables}, we derived various observables, amongst others the particle distribution and the charge distribution. Furthermore, we showed how the different formalisms are related to each other. In the last part of this chapter we demonstrated how to obtain the classical limit. Thereby, the relativistic Vlasov equation as well as a particle continuity equation were derived. \\

In chapter four we focused on computer methods and finding numerically feasible models for the background fields. As the DHW formalism leads to computationally challenging problems, we had to invest in sophisticated methods capable of overcoming these issues. Eventually, we found the advantages pseudo-spectral methods offer to be decisive for the project. Using these kind of solution techniques made it possible to perform computations with much lower restrictions on the background fields leading to a next step towards simulating pair production in a realistic experiment. Hence, we briefly introduced pseudo-spectral methods along with a discussion why the DHW equations are best formulated on a torus.

Furthermore, we discussed how to treat the pseudo-differential operators properly. We suggested to use an approximation for computationally demanding problems, but also showed how one can employ the full operators to the transport equations. 

Within this chapter we also analyzed various models for the background fields. This includes discussions on the time-dependence as well as on the spatial dependence. We further showed how one could model electric and magnetic fields in a standing wave.

We concluded this chapter by demonstrating the usefulness of a semi-classical approach. We showed how particle creation and subsequent field-particle interactions can be related. The examples given there held for demonstration purpose, such that we could refer to them at any point throughout the thesis. \\

In the second part of the thesis we presented the results obtained via solving the transport equations. For the sake of readability we divided the results into three different chapters. In chapter five we analyzed pair production in spatially homogeneous electric fields. In chapter six, the focus was on taking into account a finite spatial extent and in chapter seven we finally discussed calculations with non-vanishing magnetic field. \\

To be more specific, in chapter five we examined pair production in the regime of multiphoton absorption in terms of an effective mass concept. To this end, we showed under which circumstances this concept is appropriate in order to explain the results. Therefore, we chose an approximately monochromatic  many-cycle field, because in this way we obtained a clean environment for our investigations. Calculating the particle yield, we found the naive threshold assumption contradicting our results. Careful analysis of the yield's oscillatory structure revealed a remarkable agreement with the effective mass model. 

We proceeded searching for traces of the effective mass in the particle momentum distribution. Monitoring the multiphoton peak positions in momentum space we showed further agreement between DHW calculation and field-dependent predictions of the peak locations. This interesting feature, which is known as ``channel closing'' in atomic physics, holds as a promising candidate towards observation of effective mass signatures. Furthermore, we concluded that all investigations should also hold for theoretically accessible, experimental parameters. \\

Due to the improvements on the differential-equation solver, it was possible to reliably study pair production within the DHW formalism in an inhomogeneous background on longer time scales. Hence, it was possible to study multiphoton pair production for a background field mimicking spatial focusing. To this end, we expanded the effective mass concept to spatially inhomogeneous fields in chapter six. The special form of the background electric field further enabled us to test relativistic ponderomotive forces. Investigations of the particle spectrum indeed showed traces of these forces. Furthermore, we found a non-monotonic relation between focusing, field frequency and total particle yield. \\
 
In the end, we had a closer look at electric fields with field energy densities not exceeding the energy threshold for Schwinger pair production. It seemed, that the decisive quantity is the work that can be done at any instant of time. \\

At last, the influence of a time-dependent spatially inhomogeneous magnetic field on the pair production process was investigated. Evidence was found corroborating the idea, that a positive effective amplitude is crucial for the particle creation process. Furthermore, we showed that calculations for a weak magnetic field coincide with results obtained from extrapolation of the homogeneous case. At last, we demonstrated the significance of fulfilling homogeneous Maxwell equations. \\

\vspace{5cm}

In conclusion, we have shown how to formulate pair production in a phase-space approach taking advantage of symmetries. Furthermore, we have implemented and developed different algorithms in order to solve the transport equations. Finally, we have analyzed the outcome of our calculations by introducing the concept of an effective mass and adapting the background fields closer to experimental circumstances.

\section{Outlook}

Going through this thesis one finds several occasions, where continuation of our work seems to be promising. We will briefly comment on some selected topics.  \\

At several points throughout this thesis we just showed what was possible in first feasibility studies. 
This includes identifying observables connected with an effective mass as well as investigating spatial focusing. Additionally, further research on implications caused by a magnetic field seems to be worthwhile in order to obtain a better understanding of strong-field QED. \\ 

Although we have connected the DHW approach with its lower dimensional counterparts, only little is known about pair production in real $2+1$ dimensional structures, for example graphene. In principle, studying of such a problem should be possible within the DHW approach. However, this requires adapting the derivation to the graphene Lagrangian. In this way, we could possibly learn more about the matter creation process and depending on the outcome of the calculations this could open up new opportunities. \\

All of our results and concepts have been discussed for linearly polarized fields only. Investigation of rotating electric fields and therefore studies on arbitrary polarization are established in purely time-dependent electric fields only. An expansion to spatially inhomogeneous fields up to a complete simulation of $QED_{2+1}$-dimensional problems including a scalar magnetic field seems to be demanding but numerically feasible. This would mean a further step towards a realistic description of Laser fields in the interaction region. \\

Another possibly important aspect towards a complete description of the pair production process is given by the backreaction issue. 
Incorporating photon corrections in order to go beyond the mean-field description would be an additional piece of the complete puzzle.
Speaking in terms of related branches, such a beyond-mean-field approach could also help to better understand e.g. string-breaking in Quantum chromodynamics. \\

\part{Supplement}
\appendix
\captionsetup[figure]{list=no}
\captionsetup[table]{list=no}

\pagestyle{plain}
\chapter{Detailed calculations} 
\pagestyle{fancy} 
\section{Equations of motion}
\label{App_EoM}
We begin deriving the equations of motion defining the covariant density operator
\begin{equation}
 \hat {\mathcal C}_{\alpha \beta} \br{A,r,s} = \mathcal{U} \br{A,r,s} \com{\bar{\Psi}_{\beta} \br{r-s/2}, \psi_{\alpha} \br{r+s/2}},
\end{equation}
where
\begin{equation}
 \mathcal{U} \br{A,r,s} = \exp \br{\ii e \int_{-1/2}^{1/2} d \xi \ A \br{r+ \xi s} \ s}.
\end{equation}
From the definitions of the center-of-mass and the relative coordinates
\begin{align}
 r = \frac{r_1 + r_2}{2} ,\qquad s = r_2 - r_1,
\end{align}
we obtain the following derivative terms
\begin{align}
 \de{\mu}^{r_1} = \frac{1}{2} \de{\mu}^r - \de{\mu}^s ,\\
 \de{\mu}^{r_2} = \frac{1}{2} \de{\mu}^r + \de{\mu}^s.
\end{align}
We proceed by calculating 
\begin{equation}
 \de{\mu}^{r_i} \hat{\mathcal C} \br{A,r,s} = \br{\de{\mu}^{r_i} \mathcal{U} \br{A,r,s}} \com{\bar \Psi \br{r_1}, \Psi \br{r_2}} + \mathcal{U} \br{A,r,s} \br{\de{\mu}^{r_i} \com{\bar \Psi \br{r_1}, \Psi \br{r_2}}}, \label{App_Wigner1}
\end{equation}
where we have omitted the indices for readability. Calculation of the first term 
\begin{equation}
 \de{\mu}^{r_i} \mathcal{U} \br{A,r,s} = \br{\frac{1}{2} \de{\mu}^r \mp \de{\mu}^s} \mathcal{U} \br{A,r,s}, 
\end{equation}
yields 
\begin{alignat}{5}
 &\de{\mu}^s \mathcal{U} \br{A,r,s} &&= \mathcal{U} \br{A,r,s} \br{\ii e \int d \xi \br{A_{\mu} \br{r + \xi s} + \br{\de{\mu}^s A_{\nu} \br{r + \xi s}} s^{\nu} } },\\
 &\de{\mu}^r \mathcal{U} \br{A,r,s} &&= \mathcal{U} \br{A,r,s} \br{\ii e \int d \xi \br{\de{\mu}^r A_{\nu} \br{r + \xi s} } s^{\nu} }.
\end{alignat}
Introducing $\de{\mu}$ we can combine the results obtained in the equations above, including the derivatives, into
\begin{equation}
 \de{\mu}^{r_i} \mathcal{U} \br{A,r,s} = \mathcal{U} \br{A,r,s} \br{\ii e \int d \xi \br{\mp A_{\mu} \br{r + \xi s} + \br{\frac{1}{2} \mp \xi} \de{\mu} A_{\nu} \br{r + \xi s} s^{\nu}} }. \label{App_Eom1}
\end{equation}
We proceed using the definition of the field strength tensor $F_{\mu \nu} = \de{\mu} A_{\nu} - \de{\nu} A_{\mu}$ in order to write
\begin{equation}
 \de{\mu} A_{\nu} \br{r + \xi s} s^{\nu} = \Bigl( F_{\mu \nu} \br{r + \xi s} + \de{\nu} A_{\mu} \br{r + \xi s} \Bigr) s^{\nu}.
\end{equation}
Then we make use of the identity
\begin{equation}
  \de{\nu} A_{\mu} \br{r + \xi s} s^{\nu} = \de{\xi} A_{\mu} \br{r + \xi s} 
\end{equation}
and obtain
\begin{equation}
 \de{\mu} A_{\nu} \br{r + \xi s} s^{\nu} = F_{\mu \nu} \br{r + \xi s} s^{\nu} + \de{\xi} A_{\mu} \br{r + \xi s}.
\end{equation}
Hence, the integral kernel in \eqref{App_Eom1} can be rewritten
\begin{alignat}{5}
 &\ii e &&\int d \xi \br{\mp A_{\mu} \br{r + \xi s} + \br{\frac{1}{2} \mp \xi} \de{\mu} A_{\nu} \br{r + \xi s} s^{\nu}} \\
 =& \ \ii e &&\int d \xi \br{\mp A_{\mu} \br{r + \xi s} + \br{\frac{1}{2} \mp \xi} F_{\mu \nu} s^{\nu} + \br{\frac{1}{2} \mp \xi} \de{\xi} A_{\mu} \br{r + \xi s} } \\
 =& \ \ii e &&\int d \xi \br{\br{\frac{1}{2} \mp \xi} F_{\mu \nu} s^{\nu} + \de{\xi} \br{\br{\frac{1}{2} \mp \xi} A_{\mu} \br{r + \xi s} } }.
\end{alignat}
Eventually, we can integrate the second part in the equation above in order to obtain
\begin{equation}
 \de{\mu}^{r_i} \mathcal{U} \br{A,r,s} = \mathcal{U} \br{A,r,s} \br{\mp \ii e A_{\mu} \br{r \mp \frac{s}{2}} + \ii e \int d \xi \br{\frac{1}{2} \mp \xi} F_{\mu \nu} s^{\nu} }.
\end{equation}
If one performes the calculation with $\de{\mu}^{r_1}$ the second term in \eqref{App_Wigner1} multiplied by $\gamma^{\mu}$ is given by
\begin{equation}
 \de{\mu}^{r_1} \com{\bar \Psi \br{r_1}, \Psi \br{r_2}} \gamma^{\mu} = \br{ \de{\mu}^{r_1} \bar \Psi \br{r_1}} \Psi \br{r_2} \gamma^{\mu} - \Psi \br{r_2} \br{\de{\mu}^{r_1} \bar \Psi \br{r_1}} \gamma^{\mu}.
\end{equation}
Plugging in the Dirac equation \eqref{Eq_Dirac1} and introducing the commutator $\com{\bar \Psi \br{r_1}, \Psi \br{r_2}}$ yields
\begin{equation}
 \de{\mu}^{r_1} \com{\bar \Psi \br{r_1}, \Psi \br{r_2}} \gamma^{\mu} = \ii \com{\bar \Psi \br{r_1}, \Psi \br{r_2}} +\ii e \com{\bar \Psi \br{r_1}, \Psi \br{r_2}} \gamma^{\mu} A_{\mu} \br{r_1}.
\end{equation}
Combining all results we obtain the expression 
\begin{equation}
 \br{\br{\frac{1}{2} \de{\mu}^r - \de{\mu}^s} -\ii e \int d \xi \br{\frac{1}{2} - \xi} F_{\mu \nu} \br{r + \xi s} \ s^{\nu} } \hat{\mathcal C} \br{A,r,s} \gamma^{\mu} = \ii \hat{\mathcal C} \br{A,r,s}.
\end{equation}
On the other hand, performing the calculation using $\de{\mu}^{r_2}$ yields
\begin{equation}
 \br{\br{\frac{1}{2} \de{\mu}^r + \de{\mu}^s} -\ii e \int d \xi \br{\frac{1}{2} + \xi} F_{\mu \nu}\br{r + \xi s} \ s^{\nu} } \gamma^{\mu} \hat{\mathcal C} \br{A,r,s} = -\ii \hat{\mathcal C} \br{A,r,s}.
\end{equation}
At this point we may introduce the derivative operators
\begin{alignat}{5}
 &\tilde D_{\mu} \br{r,p} &&= \de{\mu}^r &&-\ii e &&\int d \xi \ &&F_{\mu \nu} \br{r + \xi s} \ s^{\nu} ,\\
 &\tilde \Pi_{\mu} \br{r,p} &&= \de{\mu}^s &&-\ii e &&\int d \xi \ \xi \ &&F_{\mu \nu} \br{r + \xi s} \ s^{\nu}, 
\end{alignat}
in order to write the equations of motion in a compact form
\begin{alignat}{5}
 &\br{\frac{1}{2} \tilde D_{\mu} \br{r,p} - \tilde{\Pi}_{\mu} \br{r,p}} &&\hat{\mathcal C} \br{A,r,s} \gamma^{\mu} &&= &&\ii \hat{\mathcal C} \br{A,r,s} ,\\
 &\br{\frac{1}{2} \tilde D_{\mu} \br{r,p} + \tilde{\Pi}_{\mu} \br{r,p}} &&\gamma^{\mu} \hat{\mathcal C} \br{A,r,s} &&= -&&\ii \hat{\mathcal C} \br{A,r,s}. 
\end{alignat}
In the last step we have to perform a Fourier transform from $s$-space to momentum space. Therefore, the derivatives $\de{\mu}^s$ as well as the terms
$s^{\nu}$ are transformed
\begin{align}
 \de{\mu}^s \to -\ii p_{\mu},\	s^{\nu} \to -\ii \de{p}^{\nu}
\end{align}
leading to the derivative operators
\begin{alignat}{5}
 &D_{\mu} \br{r,p} &&= \de{\mu}^r &&- e &&\int d \xi \ &&F_{\mu \nu} \br{r - \ii \xi \partial_p} \de{p}^{\nu} ,\\
 &\Pi_{\mu} \br{r,p} &&= p_{\mu} &&-\ii e &&\int d \xi \ \xi \ &&F_{\mu \nu} \br{r - \ii \xi \partial_p} \de{p}^{\nu}.
\end{alignat}
As the covariant Wigner operator $\hat{\mathcal W}$ is basically the Fourier transform of the covariant density operator we finally obtain the equations of motion
\begin{alignat}{5}
 &\br{\frac{1}{2} D_{\mu} \br{r,p} +\ii \Pi_{\mu} \br{r,p} } &&\hat{\mathcal W} \br{r,p} \gamma^{\mu} &&= &&\ii \hat{\mathcal W} \br{r,p}, \\
 &\br{\frac{1}{2} D_{\mu} \br{r,p} -\ii \Pi_{\mu} \br{r,p} } &&\gamma^{\mu} \hat{\mathcal W} \br{r,p} &&= -&&\ii \hat{\mathcal W} \br{r,p}.
\end{alignat}

\section{Energy averaging}
\label{App_Avg}
The derivative operators
\begin{alignat}{5}
 &D_{\mu} \br{r,p} &&= \de{\mu}^r &&- e &&\int d \xi \ &&F_{\mu \nu} \br{r - \ii \xi \partial_p} \de{p}^{\nu} ,\\
 &\Pi_{\mu} \br{r,p} &&= p_{\mu} &&-\ii e &&\int d \xi \ \xi \ &&F_{\mu \nu} \br{r - \ii \xi \partial_p} \de{p}^{\nu},
\end{alignat}
are split in a scalar and a vector part. This yields 
\begin{alignat}{5}
 &D_0 \br{\mathbf x,p,t} &&= \de{t} &&+ e &&\int d \xi \ &&\mathbf{E} \br{\mathbf x + \ii \xi \boldsymbol{\nabla}_p, t - \ii \xi \de{p_0}}  \cdot \boldsymbol{\nabla}_p ,\\
 &\mathbf D_k \br{\mathbf x,p,t} = \mathbf D \br{\mathbf x,p,t} &&= \boldsymbol{\nabla}_{x} &&+ e &&\int d \xi \ \ \left( \right. &&\mathbf E \br{\mathbf x + \ii \xi \boldsymbol{\nabla}_p, t - \ii \xi \de{p_0}}  \de{p_0} \\
  & && && && &&+ \mathbf B \br{\mathbf x + \ii \xi \boldsymbol{\nabla}_p, t - \ii \xi \de{p_0}}  \times \boldsymbol{\nabla}_p \left. \right) ,\\
 &\Pi_0 \br{\mathbf x,p,t} &&= p_0 &&+ \ii e &&\int d \xi \ \xi \ &&\mathbf{E} \br{\mathbf x + \ii \xi \boldsymbol{\nabla}_p, t - \ii \xi \de{p_0}}  \cdot \boldsymbol{\nabla}_p ,\\
 &\boldsymbol{\Pi}^k \br{\mathbf x,p,t} = \boldsymbol{\Pi} \br{\mathbf x,p,t} &&= \mathbf p &&+ \ii e &&\int d \xi \ \xi \left( \right. &&\mathbf{E} \br{\mathbf x + \ii \xi \boldsymbol{\nabla}_p, t - \ii \xi \de{p_0}}  \de{p_0} \\
  & && && && &&+ \mathbf B \br{\mathbf x + \ii \xi \boldsymbol{\nabla}_p, t - \ii \xi \de{p_0}}  \times \boldsymbol{\nabla}_p \left. \right.),  
\end{alignat}
where we have used $r = \br{\mathbf{x},t}$. 
N.B.: It has to be noted, that $\mathbf D_k \br{\mathbf x,p,t}$ and $\boldsymbol{\Pi}^k \br{\mathbf x,p,t}$ are given in covariant and contravariant quantities, respectively. However, for the sake of better readability we sacrifice the mathematically correct writing and only write vector components with lowered indices. 

Assuming, that we can Taylor expand the electromagnetic fields with respect to the temporal coordinate we obtain
\begin{align}
 &\mathbf E \br{\mathbf x + \ii \xi \boldsymbol{\nabla}_p, t - \ii \xi \de{p_0}} = \sum_{n = 0}^{\infty} \frac{1}{n!} \mathbf E^{(n)} \br{\mathbf x + \ii \xi \boldsymbol{\nabla}_p, t} \br{-\ii \xi \de{p_0}}^n ,\\
 &\mathbf B \br{\mathbf x + \ii \xi \boldsymbol{\nabla}_p, t - \ii \xi \de{p_0}} = \sum_{n = 0}^{\infty} \frac{1}{n!} \mathbf B^{(n)} \br{\mathbf x + \ii \xi \boldsymbol{\nabla}_p, t} \br{-\ii \xi \de{p_0}}^n,
\end{align}
with $\mathbf E^{(n)}$ and $\mathbf B^{(n)}$ giving the $n$-th derivative with respect to the temporal coordinate. We proceed applying these operators
to the Wigner function and averaging over the $p_0$ component. Additionally, we have to assume, that the Wigner function as well as 
all derivatives of the Wigner function with respect to $p_0$ vanish for $p_0 \to \pm \infty$. Furthermore, we define the equal-time Wigner components
\begin{align}
 \mathbbm{w} \br{\mathbf{x}, \mathbf{p}, t} = \int \frac{d p_0}{2 \pi} \mathbbm{W} \br{x,p}
\end{align}
and the first energy moments
\begin{align}
 \mathbbm{w}^{[1]} \br{\mathbf{x}, \mathbf{p}, t} = \int \frac{d p_0}{2 \pi} \ p_0 \ \mathbbm{W} \br{x,p}.
\end{align}
We find the relations
\begin{alignat}{5}
 &\int \frac{d p_0}{2 \pi} &&D_0 \br{\mathbf x,p,t} &&\mathbbm{W} \br{x, p} &&= D_t \br{\mathbf x, \mathbf p ,t} &&\mathbbm{w} \br{\mathbf x, \mathbf p ,t} ,\\
 &\int \frac{d p_0}{2 \pi} &&\mathbf D \br{\mathbf x,p,t} &&\mathbbm{W} \br{x,p} &&= \mathbf{D} \br{\mathbf x, \mathbf p ,t} &&\mathbbm{w} \br{\mathbf x, \mathbf p ,t}, \\
 &\int \frac{d p_0}{2 \pi} &&\Pi_0 \br{\mathbf x,p,t} &&\mathbbm{W} \br{x,p} &&= \Pi_t \br{\mathbf x, \mathbf p ,t} &&\mathbbm{w} \br{\mathbf x, \mathbf p ,t} + \mathbbm{w}^{[1]} \br{\mathbf x, \mathbf p ,t}, \\
 &\int \frac{d p_0}{2 \pi} &&\boldsymbol{\Pi} \br{\mathbf x,p,t} &&\mathbbm{W} \br{x,p} &&= \boldsymbol{\Pi} \br{\mathbf x, \mathbf p ,t} &&\mathbbm{w} \br{\mathbf x, \mathbf p ,t},
\end{alignat}
with the operators
  \begin{alignat}{6}
     & D_t \br{\mathbf x, \mathbf p ,t} && = \de{t} &&+ e &&\int d\xi &&\mathbf{E} \br{\mathbf{x}+\ii \xi \boldsymbol{\nabla}_p,t} && ~ \cdot \boldsymbol{\nabla}_p,  \\
     & \mathbf{D} \br{\mathbf x, \mathbf p ,t} && = \boldsymbol{\nabla}_x &&+ e &&\int d \xi &&\mathbf{B} \br{\mathbf{x}+\ii \xi \boldsymbol{\nabla}_p,t} &&\times \boldsymbol{\nabla}_p,  \\
     & \Pi_t \br{\mathbf x, \mathbf p ,t} &&= &&+\ii e &&\int d \xi \ \xi \ &&\mathbf{E} \br{\mathbf x +\ii \xi \boldsymbol{\nabla}_p,t} && ~ \cdot \boldsymbol{\nabla}_p, \\
     & \boldsymbol{\Pi} \br{\mathbf x, \mathbf p ,t} && = \mathbf{p} &&- \ii e &&\int d \xi \xi &&\mathbf{B} \br{\mathbf{x}+\ii \xi \boldsymbol{\nabla}_p,t} &&\times \boldsymbol{\nabla}_p.
  \end{alignat}

\section{Transport equations for $QED_{3+1}$}  
\label{App_Trans3}
First, we expand the covariant Wigner function using the Dirac bilinears \\
$\ga{A} = \{\mathbbm{1}, \gamma^{\mu}, \gamma_5, \gamma^{\mu} \gamma_5, \gamma^{\mu} \gamma^{\nu} - \gamma^{\nu} \gamma^{\mu}\}$
as basis functions. The covariant Wigner components hold as expansion coefficients, thus
we can write
\begin{equation}
\mathbbm{W} = \frac{1}{4} \br{\mathbbm{1} \mathbbm{S} + \ii \gamma_5 \mathbbm{P} + \ga{\mu} \mathbbm{V}_{\mu} + \ga{\mu} \gamma_5 \mathbbm{A}_{\mu} + \sigma^{\mu \nu} \mathbbm{T}_{\mu \nu}}, 
\end{equation}
with $\mu=0,1,2,3$ and $\sigma^{\mu \nu} = \frac{\ii}{2} \br{\gamma^{\mu} \gamma^{\nu} - \gamma^{\nu} \gamma^{\mu}}$. 
Then we insert this expansion into the equations of motion for the covariant Wigner function
\begin{alignat}{5}
 &\br{\frac{1}{2} D_{\mu} \br{r,p} +\ii \Pi_{\mu} \br{r,p} } &&\mathbbm W \br{r,p} \gamma^{\mu} &&= &&\ii \mathbbm W \br{r,p}, \\
 &\br{\frac{1}{2} D_{\mu} \br{r,p} -\ii \Pi_{\mu} \br{r,p} } &&\gamma^{\mu} \mathbbm W \br{r,p} &&= -&&\ii \mathbbm W \br{r,p}.
\end{alignat}
In order to progress in the derivation we have to introduce projection at this point.
As the Dirac bilinears obey the following relations
\begin{align}
 \ga{A} \gamma_A = \mathbbm{1} ,\qquad \text{tr} \br{\ga{A} \gamma_B} = 4 \delta_B^A ,\qquad \text{tr} \br{\ga{A}} = 0
\end{align}
also the following holds
\begin{align}
 \mathbbm{W} = \frac{1}{4} \sum_A f_A \ga{A}, \\
 f_A = \text{tr} \br{\gamma_A \mathbbm{W}}.
\end{align}

Using this together with the commutation and anti-commutation relations given in Tab. \ref{Tab_App_3}
we obtain a system of DEs for the covariant Wigner components.

\ctable[pos=t,
caption = {Commutator and anti-commutator relations of all Dirac bilinears in $QED_{3+1}$ with $\ga{\mu}$.},
label =Tab_App_3, 
mincapwidth = \textwidth,
]{ l c c }{
}{
    \toprule
    \hspace{2cm} & $\acom{\ga{\mu}, .}$ & $\com{\ga{\mu}, .}$ \\
    \midrule
    $\unit$ & 2$\ga{\mu}$ & 0 \\
    \midrule
    $\ga{5}$ & 0 & 2$\ga{\mu}\ga{5}$ \\
    \midrule        
    $\ga{\nu}$ & 2 $\eta^{\mu \nu}$ & -2$\ii \sigma^{\mu \nu}$ \\
    \midrule
    $\ga{\nu}\ga{5}$ & $\epsilon^{\mu \nu \rho \kappa} \sigma_{\rho \kappa}$ & 2$\eta^{\mu \nu}\ga{5}$ \\
    \midrule     
    $\sigma^{\nu \rho}$ & -2 $\epsilon^{\mu \nu \rho \kappa} \gamma_{\kappa} \gamma_5$ &\hspace{0.25cm} 2$\ii \br{\eta^{\mu \nu}\ga{\rho} - \eta^{\mu \rho}\ga{\nu}}$ \hspace{0.25cm} \\
    \bottomrule
}

The equations take the form
  \begin{alignat}{5}
    & D^{\mu} \mathbbm{V}_{\mu} && && && &&= 0,  \\   
    & D^{\mu} \mathbbm{A}_{\mu} && && && &&= 2 \mathbbm{P},  \\      
    &  && && && \Pi^{\mu} \mathbbm{V}_{\mu} &&= \mathbbm{S},  \\    
    & && && && \Pi^{\mu} \mathbbm{A}_{\mu} &&= 0,       
  \end{alignat}
  \begin{alignat}{5}
    & D^{\mu} \mathbbm{S} && && +4&& \Pi_{\nu} \mathbbm{T}^{\nu \mu} &&= 0,  \\ 
    & \Pi^{\mu} \mathbbm{S} && && -&& D_{\nu} \mathbbm{T}^{\nu \mu} &&= \mathbbm{V}^{\mu},  
  \end{alignat}
  \begin{alignat}{6}
    & D_{\mu} \mathbbm{P} && && - 2 &&\epsilon^{\mu \nu \rho \kappa} \Pi_{\nu} \mathbbm{T}_{\rho \kappa} && &&= -2 \mathbbm{A}_{\mu},  \\        
    & \Pi^{\mu} \mathbbm{P} && && +\frac{1}{2} &&\epsilon^{\mu \nu \rho \kappa} D_{\nu} \mathbbm{T}_{\rho \kappa} && &&= 0,             
  \end{alignat}    
\begin{alignat}{6}
    2& \epsilon^{\mu \nu \rho \kappa} \Pi_{\rho} \mathbbm{A}_{\kappa} && && && + \left( \right. D^{\mu} \mathbbm{V}^{\nu} &&- D^{\nu} \mathbbm{V}^{\mu} \left. \right) &&= 4 \mathbbm{T}^{\mu \nu},  \\  
    -\frac{1}{2} & \epsilon^{\mu \nu \rho \kappa} D_{\rho} \mathbbm{A}_{\kappa} && && &&+ \left( \right. \Pi^{\mu} \mathbbm{V}^{\nu} &&- \Pi^{\nu} \mathbbm{V}^{\mu} \left. \right) &&= 0. 
\end{alignat}   
When integrating over $p_0$ these equations become time-evolution equations for the single-time Wigner components
  \begin{alignat}{4}
    & D_t \mathbbm{s}     && && -2 \boldsymbol{\Pi} \cdot \mathbbm{t_1} &&= 0,  \\
    & D_t \mathbbm{p} && && +2 \boldsymbol{\Pi} \cdot \mathbbm{t_2} &&= -2\mathbbm{a}_0,  \\
    & D_t \mathbbm{v}_0 &&+ \mathbf{D} \cdot \mathbbm{v} && &&= 0,  \\
    & D_t \mathbbm{a}_0 &&+ \mathbf{D} \cdot \mathbbm{a} && &&= 2\mathbbm{p},  \\    
    & D_t \mathbbm{v} &&+ \mathbf{D} \ \mathbbm{v}_0 && +2 \boldsymbol{\Pi} \times \mathbbm{a} &&= -2\mathbbm{t_1},  \\    
    & D_t \mathbbm{a} &&+ \mathbf{D} \ \mathbbm{a}_0 && +2 \boldsymbol{\Pi} \times \mathbbm{v} &&= 0,  \\
    & D_t \mathbbm{t_1} &&+ \mathbf{D} \times \mathbbm{t_2} && +2 \boldsymbol{\Pi} \ \mathbbm{s} &&= 2\mathbbm{v},  \\    
    & D_t \mathbbm{t_2} &&- \mathbf{D} \times \mathbbm{t_1} && -2 \boldsymbol{\Pi} \ \mathbbm{p} &&= 0. 
  \end{alignat} 
In the system of DEs above we have decomposed the anti-symmetric tensor $\mathbbm{t}_{\mu \nu}$ into the two vectors
\begin{equation}
\mathbbm{t}_1 = 2 \mathbbm{t}^{i0} \mathbf{e}_i ,\ \mathbbm{t}_2 = \epsilon_{ijk} \mathbbm{t}^{jk} \mathbf{e}_i.
\end{equation}
Additionally, the pseudo-differential operators are given
  \begin{alignat}{6}
     & D_t && = \de{t} &&+ e &&\int d\xi &&\mathbf{E} \br{\mathbf{x}+\ii \xi \boldsymbol{\nabla}_p,t} && ~ \cdot \boldsymbol{\nabla}_p,  \\
     & \mathbf{D} && = \boldsymbol{\nabla}_x &&+ e &&\int d \xi &&\mathbf{B} \br{\mathbf{x}+\ii \xi \boldsymbol{\nabla}_p,t} &&\times \boldsymbol{\nabla}_p,  \\
     & \boldsymbol{\Pi} && = \mathbf{p} &&- \ii e &&\int d \xi \xi &&\mathbf{B} \br{\mathbf{x}+\ii \xi \boldsymbol{\nabla}_p,t} &&\times \boldsymbol{\nabla}_p.
  \end{alignat}   
Moreover, we obtain $16$ additional equations describing the relation between the single-time Wigner components and the first energy moments.

They read
  \begin{alignat}{7}
    & \mathbbm{s}^{(1)} && + \Pi_t \mathbbm{s}  && - \frac{1}{2} \mathbf{D} \mathbbm{t}_1 && &&  &&  &&= \mathbbm{v}_0, \label{App_constr3_1} \\
    & \mathbbm{p}^{(1)} && + \Pi_t \mathbbm{p}  && + \frac{1}{2} \mathbf{D} \mathbbm{t}_2 && &&  &&  &&= 0, \\    
    & \mathbbm{v}_0^{(1)} && + \Pi_t \mathbbm{v}_0  && && && - \boldsymbol{\Pi} \cdot \mathbbm{v} && &&= \mathbbm{s}, \\    
    & \mathbbm{a}_0^{(1)} && + \Pi_t \mathbbm{a}_0  && && && - \boldsymbol{\Pi} \cdot \mathbbm{a} && &&= 0, \\        
    & \mathbbm{v}^{(1)}  && + \Pi_t \mathbbm{v}   && &&  + \frac{1}{2} \mathbf{D} \times \mathbbm{a} && - \boldsymbol{\Pi} \ \mathbbm{v}_0 && &&= 0, \\           
    & \mathbbm{a}^{(1)}  && + \Pi_t \mathbbm{a}   && &&  + \frac{1}{2} \mathbf{D} \times \mathbbm{v} && - \boldsymbol{\Pi} \ \mathbbm{a}_0 && &&= -\mathbbm{t}_2, \\     
    & \mathbbm{t}_1^{(1)}  && + \Pi_t \mathbbm{t}_1   && &&  + \frac{1}{2} \mathbf{D} \ \mathbbm{s} && - \boldsymbol{\Pi} \times \mathbbm{t}_2 && &&= 0, \\       
    & \mathbbm{t}_2^{(1)}  && + \Pi_t \mathbbm{t}_2   && &&  - \frac{1}{2} \mathbf{D} \ \mathbbm{p} && +\boldsymbol{\Pi} \times \mathbbm{t}_1 && &&= -\mathbbm{a}, \label{App_constr3_8}         
  \end{alignat}    
where we have used the operator
\begin{align}
 \Pi_t \brw = \ii e \int d \xi \ \xi \ \mathbf{E} \br{\mathbf{x} + \ii \xi \boldsymbol{\nabla}_p,t} \cdot \boldsymbol{\nabla}_p.
\end{align}
As it was shown in Ochs et al.\cite{Ochs1998351}, the constraint equations are always fulfilled in Hartree approximation.

\section{Transport equations for $QED_{2+1}$}  
\label{App_Trans2}
There are in total three different ways of defining the basis matrices, if one wants to describe 
QED in $2+1$ dimensions\cite{Bashir,Raya}. Either one works with $4$-spinors and therefore in a reducible representation or one chooses an irreducible two-dimensional basis. 
However, there are two different irreducible basis sets. Hence, we write down the basis matrices we used in order to 
develop the system of transport equations describing pair production in a plane. 

\subsection{Four-Spinor formulation}
At first, we expand the covariant Wigner function using the Dirac bilinears as basis functions. As $\gamma^3$ and $\gamma^5$ play a special
role in $QED_{2+1}$ the expansion can be written in the form
\begin{align}
\mathbbm{W} &= \frac{1}{4} \sum_A \mathbbm{W}_A \ga{A} \\
 &=\frac{1}{4} \br{\mathbbm{1} \mathbbm{S} + \ga{\mu} \mathbbm{V}_{\mu} + \sigma^{\mu \nu} \mathbbm{T}_{\mu \nu} + \ga{\mu} \ga{3} \mathbbm{G}_{\mu} 
  + \ga{\mu} \ga{5} \mathbbm{A}_{\mu} + \ii \ga{5} \mathbbm{P} + \ii \ga{3} \mathbbm{H} + \ga{3}\ga{5} \mathbbm{J}}, \label{App_Wign4}
\end{align}
with $\mu=0,1,2$ and $\sigma^{\mu \nu} = \frac{\ii}{2} \com{\ga{\mu}, \ga{\nu}} $. We have chosen the following basis in order to perform the calculation
\begin{alignat}{6}
  &\ga{0} && = \ma{
  1 & 0 & 0 & 0 \\
  0 & -1 & 0 & 0 \\
  0 & 0 & -1 & 0 \\
  0 & 0 & 0 & 1},\  
  &&\ga{1} && = \ma{
  0 & \ii & 0 & 0 \\
  \ii & 0 & 0 & 0 \\
  0 & 0 & 0 & -\ii \\
  0 & 0 & -\ii & 0},\
  &&\ga{2} && = \ma{
  0 & 1 & 0 & 0 \\
  -1 & 0 & 0 & 0 \\
  0 & 0 & 0 & -1 \\
  0 & 0 & 1 & 0}, \\
  &\ga{3} && = \ma{
  0 & 0 & \ii & 0 \\
  0 & 0 & 0 & \ii \\
  -\ii & 0 & 0 & 0 \\
  0 & -\ii & 0 & 0}, \  
  &&\ga{5} && = \ma{
  0 & 0 & \ii & 0 \\
  0 & 0 & 0 & \ii \\
  \ii & 0 & 0 & 0 \\
  0 & \ii & 0 & 0}, \
  &&\ga{3} \ga{5} && = \ma{
  -1 & 0 & 0 & 0 \\
  0 & -1 & 0 & 0 \\
  0 & 0 & 1 & 0 \\
  0 & 0 & 0 & 1}, 
\end{alignat}  
\begin{alignat}{6}
  &\sigma^{01} && = \ma{
  0 & -1 & 0 & 0 \\
  1 & 0 & 0 & 0 \\
  0 & 0 & 0 & -1 \\
  0 & 0 & 1 & 0},\  
  &&\sigma^{02} && = \ma{
  0 & \ii & 0 & 0 \\
  \ii & 0 & 0 & 0 \\
  0 & 0 & 0 & \ii \\
  0 & 0 & \ii & 0}, \
  &&\sigma^{12} && = \ma{
  1 & 0 & 0 & 0 \\
  0 & -1 & 0 & 0 \\
  0 & 0 & 1 & 0 \\
  0 & 0 & 0 & -1}, \\
  &\ga{0} \ga{3} && = \ma{
  0 & 0 & \ii & 0 \\
  0 & 0 & 0 & -\ii \\
  \ii & 0 & 0 & 0 \\
  0 & -\ii & 0 & 0}, \  
  &&\ga{1} \ga{3} && = -\ma{
  0 & 0 & 0 & 1 \\
  0 & 0 & 1 & 0 \\
  0 & 1 & 0 & 0 \\
  1 & 0 & 0 & 0}, \       
  &&\ga{2} \ga{3} && = -\ma{
  0 & 0 & 0 & \ii \\
  0 & 0 & -\ii & 0 \\
  0 & \ii & 0 & 0 \\
  -\ii & 0 & 0 & 0}, 
\end{alignat}  
\begin{alignat}{6}  
  &\ga{0} \ga{5} && = \ma{
  0 & 0 & \ii & 0 \\
  0 & 0 & 0 & -\ii \\
  -\ii & 0 & 0 & 0 \\
  0 & \ii & 0 & 0}, \  
  &&\ga{1} \ga{5} && = -\ma{
  0 & 0 & 0 & -1 \\
  0 & 0 & -1 & 0 \\
  0 & 1 & 0 & 0 \\
  1 & 0 & 0 & 0}, \       
  &&\ga{2} \ga{5} && = -\ma{
  0 & 0 & 0 & \ii \\
  0 & 0 & -\ii & 0 \\
  0 & -\ii & 0 & 0 \\
  \ii & 0 & 0 & 0}.     
\end{alignat}
We proceed, replacing the covariant Wigner function in the equations of motion
\begin{alignat}{5}
 &\br{\frac{1}{2} D_{\mu} \br{r,p} +\ii \Pi_{\mu} \br{r,p} } &&\mathbbm W \br{r,p} \gamma^{\mu} &&= &&\ii \mathbbm W \br{r,p}, \\
 &\br{\frac{1}{2} D_{\mu} \br{r,p} -\ii \Pi_{\mu} \br{r,p} } &&\gamma^{\mu} \mathbbm W \br{r,p} &&= -&&\ii \mathbbm W \br{r,p},
\end{alignat}
with the right hand side of \eqref{App_Wign4}.
Via projection methods
\begin{align}
 \mathbbm{W}_A = \text{tr} \br{\gamma_A \mathbbm{W}}
\end{align}
and with the aid of the relations given in Tab. \ref{Tab_App_4er} we obtain a system of DEs
for the covariant Wigner components.
\ctable[
caption = {Commutator and anti-commutator relations of the gamma matrices with all Dirac bilinears in $QED_{2+1}$ in $4$-spinor representation.},
label =Tab_App_4er, 
mincapwidth = \textwidth,
]{ l c c }{
}{
    \toprule
    \hspace{2cm} & $\acom{\ga{\mu}, .}$ & $\com{\ga{\mu}, .}$ \\
    \midrule
    $\unit$ & 2$\ga{\mu}$ & 0 \\
    \midrule
    $\ga{\nu}$ & 2 $\eta^{\mu \nu}$ & -2$\ii \sigma^{\mu \nu}$ \\
    \midrule
    $\sigma^{\nu \rho}$ & -2 $\epsilon^{\mu \nu \rho} \ga{3} \ga{5}$ &\hspace{0.25cm} 2$\ii \br{\eta^{\mu \nu}\ga{\rho} - \eta^{\mu \rho}\ga{\nu}}$ \hspace{0.25cm} \\
    \midrule
    $\ga{\nu}\ga{3}$ & -2$\ii \epsilon^{\mu \nu \rho} \gamma_{\rho} \ga{5}$ & 2$\eta^{\mu \nu}\ga{3}$ \\
    \midrule
    $\ga{\nu}\ga{5}$ & -2$\ii \epsilon^{\mu \nu \rho} \gamma_{\rho} \ga{3}$ & 2$\eta^{\mu \nu}\ga{5}$ \\
    \midrule 
    $\ga{5}$ & 0 & 2$\ga{\mu}\ga{5}$ \\
    \midrule
    $\ga{3}$ & 0 & 2$\ga{\mu}\ga{3}$ \\
    \midrule
    $\ga{3} \ga{5}$ & -$\epsilon^{\mu \nu \rho} \sigma_{\nu \rho}$ & 0 \\
    \bottomrule
}

They take the form
  \begin{alignat}{5}
    & D^{\mu} \mathbbm{V}_{\mu} && && && &&= 0,  \\   
    & D^{\mu} \mathbbm{A}_{\mu} && && && &&= 2 \mathbbm{P},  \\      
    &  && && && \Pi^{\mu} \mathbbm{V}_{\mu} &&= \mathbbm{S},  \\    
    & && && && \Pi^{\mu} \mathbbm{A}_{\mu} &&= 0,       
  \end{alignat}
  \begin{alignat}{5}
    & D^{\mu} \mathbbm{S} && && +&& 4 \Pi_{\nu} \mathbbm{T}^{\nu \mu} &&= 0,  \\ 
    & D^{\mu} \mathbbm{G}_{\mu} && && && &&= 2 \mathbbm{H},  \\  
    & \Pi^{\mu} \mathbbm{S} && && -&& D_{\nu} \mathbbm{T}^{\nu \mu} &&= \mathbbm{V}^{\mu},  \\  
    & \Pi^{\mu} \mathbbm{G}_{\mu} && && && &&= 0,      
  \end{alignat}
  \begin{alignat}{5}
    & D_{\mu} \mathbbm{P} && && - 2\ii \ma{\Pi_2 \mathbbm{G}_1 - \Pi_1 \mathbbm{G}_2 \\ \Pi_2 \mathbbm{G}_0 - \Pi_0 \mathbbm{G}_2 \\ \Pi_0 \mathbbm{G}_1 - \Pi_1 \mathbbm{G}_0} && &&= -2\ii \mathbbm{A}_{\mu},  \\         
    & D_{\mu} \mathbbm{H} && && - 2\ii \ma{\Pi_2 \mathbbm{A}_1 - \Pi_1 \mathbbm{A}_2 \\ \Pi_2 \mathbbm{A}_0 - \Pi_0 \mathbbm{A}_2 \\ \Pi_0 \mathbbm{A}_1 - \Pi_1 \mathbbm{A}_0} && &&= -2\ii \mathbbm{G}_{\mu},  
  \end{alignat}  
  \begin{alignat}{5}
    & && && && \epsilon^{\mu \nu \rho} \Pi_{\mu} \mathbbm{T}_{\nu \rho} &&= -\mathbbm{J},  \\  
    & \br{D^{\mu} \mathbbm{V}^{\nu} - D^{\nu} \mathbbm{V}^{\mu}} && && -2&& \epsilon^{\mu \nu \rho} \Pi_{\rho} \mathbbm{J} &&= 4 \mathbbm{T}^{\mu \nu},   
  \end{alignat}   
  \begin{alignat}{5}
    & && && \frac{\ii}{2} \ma{D_2 \mathbbm{G}_1 - D_1 \mathbbm{G}_2 \\ D_2 \mathbbm{G}_0 - D_0 \mathbbm{G}_2 \\ D_0 \mathbbm{G}_1 - D_1 \mathbbm{G}_0} &&+ \Pi_{\mu} \mathbbm{P} &&= 0,  \\         
    & && && \frac{\ii}{2} \ma{D_2 \mathbbm{A}_1 - D_1 \mathbbm{A}_2 \\ D_2 \mathbbm{A}_0 - D_0 \mathbbm{A}_2 \\ D_0 \mathbbm{A}_1 - D_1 \mathbbm{A}_0} &&+ \Pi_{\mu} \mathbbm{H} &&= 0,   
  \end{alignat}    
  \begin{alignat}{5}
    & \epsilon^{\mu \nu \rho} D_{\mu} \mathbbm{T}_{\nu \rho} && && && &&= 0,  \\  
    \frac{1}{2} & \epsilon^{\mu \nu \rho} D_{\rho} \mathbbm{J} && && &&+ \br{\Pi^{\mu} \mathbbm{V}^{\nu} - \Pi^{\nu} \mathbbm{V}^{\mu}} &&= 4 \mathbbm{T}^{\mu \nu}. 
  \end{alignat}   
In order to obtain the equal-time formalism we continue by integrating over $p_0$. This yields the time-evolution equations
  \begin{alignat}{6}
    & D_t \mathbbm{s}   &&  && && -2 \Pi_1 \mathbbm{t}_{1,1} && -2 \Pi_2 \mathbbm{t}_{1,2} &&= 0, \\
    & D_t \mathbbm{v}_0 &&+ D_1 \mathbbm{v}_1 &&+ D_2 \mathbbm{v}_2 && && &&= 0,  \\   
    & D_t \mathbbm{v}_1 &&+ D_1 \mathbbm{v}_0 && && &&+2 \Pi_2 \mathbbm{j} &&= -2\mathbbm{t}_{1,1},  \\    
    & D_t \mathbbm{v}_2 && &&+ D_2 \mathbbm{v}_0 && -2 \Pi_1 \mathbbm{j} && &&= -2\mathbbm{t}_{1,2},  \\        
    & D_t \mathbbm{j} && && &&+2 \Pi_1 \mathbbm{v}_2 &&- 2 \Pi_2 \mathbbm{v}_1 &&= 0,  \\
    & D_t \mathbbm{t}_{1,1} && &&+ D_2 \mathbbm{t}_2 && +2 \Pi_1 \mathbbm{s} && &&= 2\mathbbm{v}_1,  \\    
    & D_t \mathbbm{t}_{1,2} &&- D_1 \mathbbm{t}_2 && && &&+2 \Pi_2 \mathbbm{s} &&= 2\mathbbm{v}_2,  \\    
    & D_t \mathbbm{t}_2 &&- D_1 \mathbbm{t}_{1,2} &&+ D_2 \mathbbm{t}_{1,1} && && &&= 0,    
  \end{alignat}   
where we have used the notation
\begin{align}
 \mathbbm{t}_1 = 2 \mathbbm{t}^{i0} \mathbf{e}_i,\ \mathbbm{t}_2 = 2 \mathbbm{t}^{12}.
\end{align}
The operators are defined via
  \begin{alignat}{6}
     & D_t &&= \de{t} &&+ e &&\int d\xi \left( \right. && E_x &&\br{x+\ii \xi \de{p_x},y+\ii \xi \de{p_y},t} \de{p_x}  \\
       & && && &&+ && E_y &&\br{x+\ii \xi \de{p_x},y+\ii \xi \de{p_y},t} \de{p_y} \left. \right), \notag \\
     & D_1 &&= \de{x} &&-e &&\int d \xi &&B &&\br{x+\ii \xi \de{p_x},y+\ii \xi \de{p_y},t} \de{p_y},  \\
     & D_2 &&= \de{y} &&+ e &&\int d \xi &&B &&\br{x+\ii \xi \de{p_x},y+\ii \xi \de{p_y},t} \de{p_x},  \\
     & \Pi_1 &&= p_x &&+ \ii e &&\int d \xi \xi &&B &&\br{x+\ii \xi \de{p_x},y+\ii \xi \de{p_y},t} \de{p_y}, \\
     & \Pi_2 &&= p_y &&- \ii e &&\int d \xi \xi &&B &&\br{x+\ii \xi \de{p_x},y+\ii \xi \de{p_y},t} \de{p_x}.
  \end{alignat} 
For the sake of completeness, we also mention the trivial second set of equations
  \begin{alignat}{6}
    & D_t \mathbbm{a}_0   &&+D_1 \mathbbm{a}_1  &&+D_2 \mathbbm{a}_2 && && &&= 2 \mathbbm{p}, \\
    & D_t \mathbbm{g}_0   &&+D_1 \mathbbm{g}_1  &&+D_2 \mathbbm{g}_2 && && &&= 2 \mathbbm{h}, \\    
    & D_t \mathbbm{p} && && && +2 \ii \Pi_1 \mathbbm{g}_2 &&-2 \ii \Pi_2 \mathbbm{g}_1 &&= -2 \ii \mathbbm{a}_0, \\        
    & D_t \mathbbm{h} && && && +2 \ii \Pi_1 \mathbbm{a}_2 &&-2 \ii \Pi_2 \mathbbm{a}_1 &&= -2 \ii \mathbbm{g}_0, \\        
    & D_t \mathbbm{g}_1 &&- D_1 \mathbbm{g}_0 && && &&+2 \ii \Pi_2 \mathbbm{p} &&= 0,  \\    
    & D_t \mathbbm{g}_2 && &&- D_2 \mathbbm{g}_0 && -2 \ii \Pi_1 \mathbbm{p} && &&= 0,  \\
    & D_t \mathbbm{a}_1 &&- D_1 \mathbbm{a}_0 && && &&+2 \ii \Pi_2 \mathbbm{h} &&= 0,  \\    
    & D_t \mathbbm{a}_2 && &&- D_2 \mathbbm{a}_0 && -2 \ii \Pi_1 \mathbbm{h} && &&= 0,      
  \end{alignat} 
with all Wigner components being constantly zero due to the initial conditions used.  
In addition one obtains constraint equations 
  \begin{alignat}{7}
    & \mathbbm{s}^{(1)} && + \Pi_t \mathbbm{s}  && - \frac{1}{2} D_1 \mathbbm{t}_{1,1} && - \frac{1}{2} D_2 \mathbbm{t}_{1,2} &&  && &&= \mathbbm{v}_0, \\  
    & \mathbbm{v}_0^{(1)} && + \Pi_t \mathbbm{v}_0  && && && - \Pi_1 \mathbbm{v}_1 && -\Pi_2 \mathbbm{v}_2 &&= \mathbbm{s}, \\
    & \mathbbm{v}_1^{(1)} && + \Pi_t \mathbbm{v}_1 && + \frac{1}{2} D_2 \mathbbm{j} && &&- \Pi_1 \mathbbm{v}_0 && &&= 0, \\     
    & \mathbbm{v}_2^{(1)} && + \Pi_t \mathbbm{v}_2 && - \frac{1}{2} D_1 \mathbbm{j} && && &&- \Pi_2 \mathbbm{v}_0 &&= 0, \\   
    & \mathbbm{j}^{(1)}  && + \Pi_t \mathbbm{j} && + \frac{1}{2} D_1 \mathbbm{v}_2 && - \frac{1}{2} D_2 \mathbbm{v}_1 && && &&= - \mathbbm{t}_{2} \\
    & \mathbbm{t}_{1,1}^{(1)} && + \Pi_t \mathbbm{t}_{1,1}  && + \frac{1}{2} D_1 \mathbbm{s} && && && - \Pi_2 \mathbbm{t}_{2} &&= 0, \\ 
    & \mathbbm{t}_{1,2}^{(1)} && + \Pi_t \mathbbm{t}_{1,2}  && && + \frac{1}{2} D_2 \mathbbm{s} && + \Pi_1 \mathbbm{t}_{2} && &&= 0, \\        
    & \mathbbm{t}_{2}^{(1)} && + \Pi_t \mathbbm{t}_{2}  && && && - \Pi_1 \mathbbm{t}_{1,2} && + \Pi_2 \mathbbm{t}_{1,1} &&= - \mathbbm{j},  
\end{alignat}
\begin{alignat}{7}
    & \mathbbm{a}_0^{(1)} && + \Pi_t \mathbbm{a}_0  && && && - \Pi_1 \mathbbm{a}_1 && -\Pi_2 \mathbbm{a}_2 &&= 0, \\ 
    & \mathbbm{g}_0^{(1)} &&  + \Pi_t \mathbbm{g}_0  && && && - \Pi_1 \mathbbm{g}_1 && -\Pi_2 \mathbbm{g}_2 &&= 0, \\    
    & \mathbbm{p}^{(1)} &&  + \Pi_t \mathbbm{p}  && + \frac{\ii}{2} D_1 \mathbbm{g}_2 && - \frac{\ii}{2} D_2 \mathbbm{g}_1 && && &&= 0, \\      
    & \mathbbm{h}^{(1)} && + \Pi_t \mathbbm{h}  && + \frac{\ii}{2} D_1 \mathbbm{a}_2 && - \frac{\ii}{2} D_2 \mathbbm{a}_1 && && &&= 0, \\    
    & \mathbbm{g}_{1}^{(1)} && + \Pi_t \mathbbm{g}_1   && &&- \frac{\ii}{2} D_2 \mathbbm{p} && - \Pi_1 \mathbbm{g}_0 && &&= \mathbbm{a}_2, \\
    & \mathbbm{g}_{2}^{(1)} && + \Pi_t \mathbbm{g}_2  &&  + \frac{\ii}{2} D_1 \mathbbm{p}  && && &&- \Pi_2 \mathbbm{g}_0 &&= -\mathbbm{a}_1, \\   
    & \mathbbm{a}_{1}^{(1)}  && + \Pi_t \mathbbm{a}_1   && &&  - \frac{\ii}{2} D_2 \mathbbm{h} && - \Pi_1 \mathbbm{a}_0 && &&= \mathbbm{g}_2, \\     
    & \mathbbm{a}_{2}^{(1)} && + \Pi_t \mathbbm{a}_2  && + \frac{\ii}{2} D_1 \mathbbm{h} && && &&- \Pi_2 \mathbbm{a}_0 &&= -\mathbbm{g}_1, 
  \end{alignat}    
  with the operator
\begin{align}
 \Pi_t \brw = \ii e \int d \xi \ \xi \ \mathbf{E} \br{x + \ii \xi \de{p_x},y + \ii \xi \de{p_y},t} \cdot \ma{\de{p_x} \\ \de{p_y}}
\end{align}  
implying a relation between first energy moments and equal-time Wigner components.
As was shown in Ochs et al.\cite{Ochs1998351}, the constraint equations are automatically fulfilled in Hartree approximation. 

\subsection{Two-Spinor formulation}
We expand the covariant Wigner function into the basis functions $\gamma^A=\{ \mathbbm{1}, \gamma^{\mu}\}$ yielding
\begin{equation}
\mathbbm{W} = \frac{1}{2} \sum_A \mathbbm{W}_A \gamma^A = \frac{1}{2} \br{\mathbbm{1} \mathbbm{S} + \ga{\mu} \mathbbm{V}_{\mu}}.
\end{equation}
As mentioned above there are two different ways of how to define gamma matrices. 
The first set of matrices(we will refer to this set as set $I$) reads
\begin{align}
  &\ga{0} && = \sigma_3 = \ma{
  1 & 0 \\
  0 & -1},\  
  &&\ga{1} && = \ii \sigma_1 = \ma{
  0 & \ii \\
  \ii & 0},\
  &&\ga{2} && = \ii \sigma_2 = \ma{
  0 & 1 \\
  -1 & 0},
\end{align}
while set $II$ takes the form
\begin{align}
  &\ga{0} && = \sigma_3 = \ma{
  1 & 0 \\
  0 & -1},\  
  &&\ga{1} && = \ii \sigma_1 = \ma{
  0 & \ii \\
  \ii & 0},\
  &&\ga{2} && = -\ii \sigma_2 = \ma{
  0 & -1 \\
  1 & 0}.
\end{align}
Then, we plug in the expanded Wigner function into the equations of motion
\begin{alignat}{5}
 &\br{\frac{1}{2} D_{\mu} \br{r,p} +\ii \Pi_{\mu} \br{r,p} } &&\mathbbm W \br{r,p} \gamma^{\mu} &&= &&\ii \mathbbm W \br{r,p}, \\
 &\br{\frac{1}{2} D_{\mu} \br{r,p} -\ii \Pi_{\mu} \br{r,p} } &&\gamma^{\mu} \mathbbm W \br{r,p} &&= -&&\ii \mathbbm W \br{r,p}.
\end{alignat}

\ctable[pos=t,
caption = {Commutator and anti-commutator relations of all Dirac bilinears in $QED_{2+1}$ with $\ga{\mu}$ using $2$-spinors. The multiple signs
correspond to the two different basis sets.},
label =Tab_App_2a, 
mincapwidth = \textwidth,
]{ l c c }{
}{
    \toprule
    \hspace{2cm} & $\acom{\ga{\mu}, .}$ & $\com{\ga{\mu}, .}$ \\
    \midrule
    $\unit$ & 2$\ga{\mu}$ & 0 \\
    \midrule    
    $\ga{\nu}$ & 2 $\eta^{\mu \nu}$ & $\mp 2 \ii \epsilon^{\mu \nu \kappa} \gamma_{\kappa}$ \\
    \bottomrule
}

Using projection methods 
\begin{align}
 \mathbbm{W}_A = \text{tr} \br{\gamma_A \mathbbm{W}}
\end{align}
and the (anti-)commutator relations given in Tab. \ref{Tab_App_2a} 
we obtain a system of DEs for the components of the covariant Wigner function
\begin{alignat}{5}
&\Pi_{\mu} \mathbbm{V}^{\mu} &&= \mathbbm{S}, \\
&D_{\mu} V^{\mu} &&= 0,
\end{alignat}
\begin{alignat}{5}
&\Pi^{\mu} \mathbbm{S} &&\pm \frac{1}{2} &&\epsilon^{\mu \nu \rho} D_{\nu} \mathbbm{V}_{\rho} &&= \mathbbm{V}^{\mu} , \label{App_Sp2_1} \\
&D^{\mu} \mathbbm{S} &&\mp 2 &&\epsilon^{\mu \nu \rho} \Pi_{\nu} \mathbbm{V}_{\rho} &&= 0. \label{App_Sp2_2}
\end{alignat}
In the equations above the difference in the basis sets is reflected in the different signs in \eqref{App_Sp2_1} - \eqref{App_Sp2_2}.


We proceed by performing the integration over $p_0$ in order to obtain a system of equations for the single-time Wigner coefficients 
\begin{alignat}{6}
  & D_t \mathbbm{s}   &&  && && \mp 2 \Pi_1 \mathbbm{v}_2 && \pm 2 \Pi_2 \mathbbm v_1 &&= 0, \\
  & D_t \mathbbm{v}_1 &&+D_1 \mathbbm{v}_0 && && && \mp 2 \Pi_2 \mathbbm{s} &&= \mp 2\mathbbm{v}_2,  \\    
  & D_t \mathbbm{v}_2 && &&+D_2 \mathbbm{v}_0 && \pm 2 \Pi_1 \mathbbm{s} && &&= \pm 2 \mathbbm{v}_1,  \\ 
  & D_t \mathbbm{v}_0 &&+D_1 \mathbbm{v} &&+D_2 \mathbbm{v}_2 && && &&= 0,
\end{alignat}
where the operators are defined via
  \begin{alignat}{6}
     & D_t &&= \de{t} &&+ e &&\int d\xi \left( \right. && E_x &&\br{x+\ii \xi \de{p_x},y+\ii \xi \de{p_y},t} \de{p_x}  \\
       & && && &&+ && E_y &&\br{x+\ii \xi \de{p_x},y+\ii \xi \de{p_y},t} \de{p_y} \left. \right), \notag \\
     & D_1 &&= \de{x} &&-e &&\int d \xi &&B &&\br{x+\ii \xi \de{p_x},y+\ii \xi \de{p_y},t} \de{p_y},  \\
     & D_2 &&= \de{y} &&+ e &&\int d \xi &&B &&\br{x+\ii \xi \de{p_x},y+\ii \xi \de{p_y},t} \de{p_x},  \\
     & \Pi_1 &&= p_x &&+ \ii e &&\int d \xi \xi &&B &&\br{x+\ii \xi \de{p_x},y+\ii \xi \de{p_y},t} \de{p_y}, \\
     & \Pi_2 &&= p_y &&- \ii e &&\int d \xi \xi &&B &&\br{x+\ii \xi \de{p_x},y+\ii \xi \de{p_y},t} \de{p_x}.
  \end{alignat} 
Furthermore, we find the constraint equations
\begin{alignat}{7}
&\mathbbm{v}_0^{(1)} &&+ \Pi_t \mathbbm{v}_0 && && &&- \Pi_1 \mathbbm{v}_1 &&- \Pi_2 \mathbbm{v}_2 &&= \mathbbm{s} ,\\
&\mathbbm{s}^{(1)} &&+ \Pi_t \mathbbm{s} &&\mp \frac{1}{2} D_1 \mathbbm{v}_2 && \pm \frac{1}{2} D_2 \mathbbm{v}_1 && && &&= \mathbbm{v}_0 ,\\
&\mathbbm{v}_{2}^{(1)} &&+ \Pi_t \mathbbm{v}^2 &&\pm \frac{1}{2} D_1 \mathbbm{s} && && &&- \Pi_2 \mathbbm{v}_0 &&= 0 ,\\
&\mathbbm{v}_{1}^{(1)} &&+ \Pi_t \mathbbm{v}^1 && &&\mp \frac{1}{2} D_2 \mathbbm{s} &&- \Pi_1 \mathbbm{v}_0 && &&= 0,
\end{alignat}
with
\begin{align}
 \Pi_t \brw = \ii e \int d \xi \ \xi \ \mathbf{E} \br{x + \ii \xi \de{p_x},y + \ii \xi \de{p_y},t} \cdot \ma{\de{p_x} \\ \de{p_y}}
\end{align}  
Again, these constraint equations are automatically fulfilled in a Hartree approximation.

\section{Transport equations for $QED_{1+1}$}
\label{App_Trans1}
When investigating the DHW formalism in $QED_{1+1}$ one finds only four distinct Dirac bilinears. This is due to the fact,  
that one works in a two-dimensional basis. Hence, the Wigner operator can be expanded in terms of $\gamma^A=\{\mathbbm{1}, \gamma^5, \gamma^1, \gamma^2\}$ leading to
\begin{equation}
\mathbbm{W} = \frac{1}{2} \sum_A \mathbbm{W}_A \gamma^A = \frac{1}{2} \br{\mathbbm{1} \mathbbm{S} + \ii \gamma^5 \mathbbm{P} + \ga{\mu} \mathbbm{V}_{\mu}}, 
\end{equation}

\ctable[
caption = {Commutator and anti-commutator relations of all Dirac bilinears in $QED_{1+1}$ with $\ga{\mu}$.},
label =Tab_App_1, 
mincapwidth = \textwidth,
]{ l c c }{
}{
    \toprule
    \hspace{2cm} & $\acom{\ga{\mu}, .}$ & $\com{\ga{\mu}, .}$ \\
    \midrule
    $\unit$ & 2$\ga{\mu}$ & 0 \\
    \midrule
    $\ga{5}$ & 0 & -2$\epsilon^{\mu \rho}\gamma_{\rho}$ \\
    \midrule        
    $\ga{\nu}$ & 2 $\eta^{\mu \nu}$ & 2$\epsilon^{\mu \nu} \gamma_{5}$ \\
    \bottomrule
}

with $\mu=0,1$. Plugging in this expansion into the equations of motion
\begin{alignat}{5}
 &\br{\frac{1}{2} D_{\mu} \br{r,p} +\ii \Pi_{\mu} \br{r,p} } &&\mathbbm W \br{r,p} \gamma^{\mu} &&= &&\ii \mathbbm W \br{r,p}, \\
 &\br{\frac{1}{2} D_{\mu} \br{r,p} -\ii \Pi_{\mu} \br{r,p} } &&\gamma^{\mu} \mathbbm W \br{r,p} &&= -&&\ii \mathbbm W \br{r,p},
\end{alignat}
and applying the relations summarized in Tab. \ref{Tab_App_1} together with 
\begin{align}
 \mathbbm{W}_A = \text{tr} \br{\gamma_A \mathbbm{W}}
\end{align}
we are able to determine
a system of DEs containing all covariant Wigner components.
The equations can be written as 
\begin{alignat}{5}
    & D^{\mu} \mathbbm{V}_{\mu} &&= 0,  \\        
    &  \Pi^{\mu} \mathbbm{V}_{\mu} &&= \mathbbm{S},    
  \end{alignat}
  \begin{alignat}{5}
    & D^{\mu} \mathbbm{S} +&& 2 &&\epsilon^{\mu \nu} \Pi_{\nu} \mathbbm{P} &&= 0,  \\ 
    & \Pi^{\mu} \mathbbm{S} -&&\frac{1}{2} &&\epsilon^{\mu \nu} D_{\nu} \mathbbm{P} &&= \mathbbm{V}^{\mu},  
  \end{alignat}  
\begin{alignat}{5}
    &  D^{\mu} \mathbbm{V}^{\nu} &&- D^{\nu} \mathbbm{V}^{\mu} &&= -2 \epsilon^{\mu \nu} \mathbbm{P},  \\  
    &  \Pi^{\mu} \mathbbm{V}^{\nu} &&- \Pi^{\nu} \mathbbm{V}^{\mu} &&= 0. 
\end{alignat}   

Integrating over $p_0$ results in a system of DEs for the equal-time Wigner functions
  \begin{alignat}{6}
    & D_t \mathbbm{s}   &&  && && -2 p_x \mathbbm{p} && &&= 0, \\
    & D_t \mathbbm{v}_0 &&+ \de{x} \mathbbm{v}_1 && && && &&= 0,  \\   
    & D_t \mathbbm{v}_1 &&+ \de{x} \mathbbm{v}_0 && && && &&= -2\mathbbm{p},  \\    
    & D_t \mathbbm{p} && && && +2 p_x \mathbbm{s} && &&= 2\mathbbm{v}_1,     
  \end{alignat} 
  with 
\begin{align}
 D_t = \de{t} + e\int d\xi E_x \br{x+\ii \xi \de{p_x},t} \de{p_x}. 
\end{align}
Additionally, we obtain the constraint equations
  \begin{alignat}{7}
    & \mathbbm{s}^{(1)} && + \Pi_t \mathbbm{s}  && - \frac{1}{2} \de{x} \mathbbm{p} &&  &&= \mathbbm{v}_0, \\   
    & \mathbbm{v}_0^{(1)} && + \Pi_t \mathbbm{v}_0  && &&- p_x \mathbbm{v} &&= \mathbbm{s}, \\    
    & \mathbbm{v}^{(1)}  && + \Pi_t \mathbbm{v}   && &&-p_x \mathbbm{v}_0 &&= 0, \\           
    & \mathbbm{p}^{(1)}  && + \Pi_t \mathbbm{p}   &&  + \frac{1}{2} \de{x} \mathbbm{s} && &&= 0,   
  \end{alignat} 
where 
\begin{align}
 \Pi_t \br{x,p_x,t} = \ii e \int d \xi \ \xi \ E \br{x + \ii \xi \de{p_x},t} \de{p_x}.
\end{align}
  
\section{Alternative formulations}
\label{App_Alter}
Especially in case of spatially homogeneous, elliptically polarized electric fields, it is convenient to use a modified system of DEs. Instead of working with
Wigner components and calculating observables afterwards, the emphasis is on including the particle distribution function directly into the DE.
For the sake of completeness, we will show how one can derive such a modified system using ``Method of characteristics''. The reader is encouraged to
apply this method to the equations derived in chapter three, which are most convenient to work with.

We begin stating the equations in case of a spatially homogeneous field:
\begin{alignat}{4}
    & D_t \mathbbm{s}     && && -2 \mathbf{p} \cdot \mathbbm{t_1} &&= 0, \\    
    & D_t \mathbbm{v} && && +2\mathbf{p} \times \mathbbm{a} &&= -2\mathbbm{t_1},  \\    
    & D_t \mathbbm{a} && && +2 \mathbf{p} \times \mathbbm{v} &&= 0,  \\
    & D_t \mathbbm{t_1} && && +2 \mathbf{p} \ \mathbbm{s} &&= 2\mathbbm{v},   
\end{alignat} 
  with the differential operator
  \begin{alignat}{6}
      & D_t &&= \de{t} +&& e && &&\mathbf{E} \br{t} \cdot \boldsymbol{\nabla}_p.  
  \end{alignat}  
This equation can be brought into the form
\begin{equation}
 D_t \mathbbm{w}_k = \overline{\mathbf{M}} \mathbbm{w}_k.
\end{equation}
The idea is to change the basis in order to include the distribution function directly into the DEs. Hence,
we write down a general basis expansion
\begin{align}
 \de{t} \mathbbm{w}_k = \overline{\mathbf{M}} \mathbbm{w}_k
 = -2 \ \overline{\mathbf{M}} \sum_{i=1}^{10} \alpha_i \mathbf{e}_i,
\end{align}
where $\alpha_i$ denote the expansion coefficients and $\mathbf{e}_i$ the basis vectors.
We want to have only one remaining non-vanishing coefficient for the vacuum case. Hence, we choose the first basis vector to be
\begin{align}
 \mathbf{e}_1 = \frac{1}{\omega} \ma{1 , q_x - e A_x \brt , q_y - e A_y \brt , q_z - e A_z \brt , 0 , 0 , 0 , 0 , 0 , 0}^T.
\end{align}
For convenience we define $\mathbf{p} \brt = \mathbf{q}-e \mathbf{A} \brt$.
We find the relations
\begin{equation}
\overline{\mathbf{M}} \mathbf{e}_1 = 0,\ \de{t} \mathbf{e}_1 = \frac{e E_x}{\omega} \mathbf{e}_2 + \frac{e E_y}{\omega} \mathbf{e}_3 + \frac{e E_z}{\omega} \mathbf{e}_4,       
\end{equation}
where we have defined the vectors
\begin{alignat}{5}
 \mathbf{e}_2 = -\frac{1}{\omega^2} \ma{p_x , p_x^2 - \omega^2, p_x p_y, p_x p_z, 0 ,0, 0, 0, 0, 0}^T, \\
 \mathbf{e}_3 = -\frac{1}{\omega^2} \ma{p_y , p_x p_y, p_y^2 - \omega^2 , p_y p_z, 0 ,0, 0, 0, 0, 0}^T, \\
 \mathbf{e}_4 = -\frac{1}{\omega^2} \ma{p_z , p_x p_z , p_y p_z , p_z^2 - \omega^2, 0 ,0, 0, 0, 0, 0}^T. \\ 
\end{alignat}
Calculating the derivative with respect to $t$ yields
\begin{alignat}{5}
 \de{t} \mathbbm{g}_2 = \frac{-e E_x \omega^2 + e E_x p_x^2 + e E_y p_x p_y + e E_z p_x p_z}{\omega^3} \mathbf{e}_1 - \frac{e E_x p_x}{\omega^2} \mathbf{e}_2 
 - \frac{e E_y p_x}{\omega^2} \mathbf{e}_3 - \frac{e E_z p_x}{\omega^2} \mathbf{e}_4 ,\\
 \de{t} \mathbbm{g}_3 = \frac{-e E_y \omega^2 + e E_x p_x p_y + e E_y p_y^2 + e E_z p_y p_z}{\omega^3} \mathbf{e}_1 - \frac{e E_x p_y}{\omega^2} \mathbf{e}_2 
 - \frac{e E_y p_y}{\omega^2} \mathbf{e}_3 - \frac{e E_z p_y}{\omega^2} \mathbf{e}_4 ,\\
 \de{t} \mathbbm{g}_4 = \frac{-e E_z \omega^2 + e E_x p_x p_z + e E_y p_y p_z + e E_z p_z^2}{\omega^3} \mathbf{e}_1 - \frac{e E_x p_z}{\omega^2} \mathbf{e}_2 
 - \frac{e E_y p_z}{\omega^2} \mathbf{e}_3 - \frac{e E_z p_z}{\omega^2} \mathbf{e}_4. 
\end{alignat}
Applying the matrix $\overline{\mathbf{M}}$ to these vectors gives
\begin{alignat}{8}
 &\overline{\mathbf{M}} \mathbf{e}_2 &&= && &&- 2p_z \mathbf{e}_6 &&+ 2 p_y \mathbf{e}_7 &&+ 2 \mathbf{e}_8, && &&\\
 &\overline{\mathbf{M}} \mathbf{e}_3 &&= &&2p_z \mathbf{e}_5 && &&- 2 p_x \mathbf{e}_7 && &&+ 2 \mathbf{e}_9 &&,\\
 &\overline{\mathbf{M}} \mathbf{e}_4 &&= -&&2p_y \mathbf{e}_5 && + 2 p_x \mathbf{e}_6 && && && &&+ 2 \mathbf{e}_{10}. 
\end{alignat}
The vectors $\mathbf{e}_i ,\ i \in \com{5,10}$  are all unit vectors with the only non-vanishing term 
placed in their $i^{th}$ row. Hence, all derivatives with respect to $t$ give zero. However, we still have
to apply the matrix $\overline{\mathbf{M}}$ to all vectors. This yields
\begin{alignat}{6}
 &\overline{\mathbf{M}} \mathbf{e}_5 &&= && && &&-2 p_z \mathbf{e}_3 &&+2 p_y \mathbf{e}_4   ,\\
 &\overline{\mathbf{M}} \mathbf{e}_6 &&= && &&2 p_z \mathbf{e}_2 && &&- 2 p_x \mathbf{e}_4 ,\\
 &\overline{\mathbf{M}} \mathbf{e}_7 &&= &&- &&2 p_y \mathbf{e}_2 &&+2 p_x \mathbf{e}_3 &&, 
\end{alignat}
\begin{alignat}{8}
 &\overline{\mathbf{M}} \mathbf{e}_8 &&= -2 \br{1 + p_x^2} &&\mathbf{e}_2 &&- 2p_x p_y &&\mathbf{e}_3 &&- 2 p_x p_z &&\mathbf{e}_4 ,\\
 &\overline{\mathbf{M}} \mathbf{e}_9 &&= -2 p_x p_y &&\mathbf{e}_2 &&- 2 \br{1 + p_y^2} &&\mathbf{e}_3 &&- 2 p_y p_z &&\mathbf{e}_4 ,\\
 &\overline{\mathbf{M}} \mathbf{e}_{10} &&= -2 p_x p_z &&\mathbf{e}_2 &&- 2 p_y p_z &&\mathbf{e}_3 &&- 2 \br{1 + p_z^2} &&\mathbf{e}_4. 
\end{alignat}
Plugging in all of the relations above into the equation $\de{t} \mathbbm{w} = \overline{\mathbf{M}} \mathbbm{w}$ and subsequent projection
of the basis vectors yields the rewritten system of DEs. Additionally, we want to identify the vacuum with a vanishing particle distribution function, thus
we may introduce it via $F = 1- \alpha_1$. Hence, we obtain the fully rewritten system with the distribution function included
\begin{alignat}{5}
 \de{t} F &= \frac{ \br{-e E_x \omega^2 + p_x e\mathbf{ E} \cdot \mathbf{p}}}{\omega^3} \alpha_2 \\
 &+ \frac{\br{-e E_y \omega^2 + p_y e\mathbf{E} \cdot \mathbf{p}}}{\omega^3} \alpha_3 \notag \\
 &+ \frac{\br{-e E_z \omega^2 + p_z e\mathbf{E} \cdot \mathbf{p}}}{\omega^3} \alpha_4 \notag,
\end{alignat}
\begin{alignat}{5}
 \de{t} \alpha_2 &= \frac{e E_x}{\omega} \br{F-1} + \frac{e E_x p_x}{\omega^2} \alpha_2 + \frac{e E_x p_y}{\omega^2} \alpha_3 + \frac{e E_x p_z }{\omega^2} \alpha_4 \\
  &+2 p_z \alpha_6 - 2 p_y \alpha_7 - 2 \br{1 + p_x^2} \alpha_8 - 2 p_x p_y \alpha_9 - 2 p_x p_z \alpha_{10}, \notag \\ 
 \de{t} \alpha_3 &= \frac{e E_y}{\omega} \br{F-1} + \frac{e E_y p_x}{\omega^2} \alpha_2 + \frac{e E_y p_y}{\omega^2} \alpha_3 + \frac{e E_y p_z}{\omega^2} \alpha_4 \\
  &-2 p_z \alpha_5 + 2 p_x \alpha_7 - 2 p_x p_y \alpha_8 - 2 \br{1 + p_y^2} \alpha_9 - 2 p_y p_z \alpha_{10}, \notag \\ 
 \de{t} \alpha_4 &= \frac{e E_z}{\omega} \br{F-1} + \frac{e E_z p_x}{\omega^2} \alpha_2 + \frac{e E_z p_y}{\omega^2} \alpha_3 + \frac{e E_z p_z}{\omega^2} \alpha_4 \\
  &+2 p_y \alpha_5 - 2 p_x \alpha_6 - 2 p_x p_z \alpha_8 - 2 p_y p_z \alpha_9 - 2 \br{1 + p_z^2} \alpha_{10}, \notag
\end{alignat}
\begin{alignat}{7}
 &\de{t} \alpha_5 &&= && &&2 p_z \alpha_3 && - &&2p_y \alpha_4 &&,\\
 &\de{t} \alpha_6 &&= -&&2 p_z \alpha_2 && && +&& 2p_x \alpha_4 &&,\\ 
 &\de{t} \alpha_7 &&= &&2 p_y \alpha_2 -&& 2p_x \alpha_3, && && &&
\end{alignat}
\begin{alignat}{5}
 &\de{t} \alpha_8 &&= 2 \alpha_2 ,\\
 &\de{t} \alpha_9 &&= 2 \alpha_3 ,\\
 &\de{t} \alpha_{10} &&= 2 \alpha_4. 
\end{alignat}
If one is interested in a background field exhibiting special characteristics a further reduction is of course possible.
As we have only changed the basis it is still possible to obtain, for example, the equations resulting
from QKT or the system of equations used in reference \cite{PhysRevD.89.085001}.

\subsection*{Quantum Kinetic Theory}
\label{App_QKT}
In order to reduce the system of DEs \eqref{eq3_Cy1}-\eqref{eq3_Cy4} to the equations obtained via QKT, we have to
take the homogeneous limit in the electric field $E \br{x,t} \to E \br{t}$. The derivatives with respect to $x$
vanish automatically and thus we can write
\begin{alignat}{5}
  & D_t \overline{\mathbbm{s}}     && && -2 p_x \overline{\mathbbm{p}} &&+ 2 p_{\rho} \overline{\mathbbm v} &&= 0, \\
  & D_t \overline{\mathbbm{v}} && && &&- 2 p_{\rho} \overline{ \mathbbm{s}} &&= -2\overline{\mathbbm{p}},  \\    
  & D_t \overline{\mathbbm{p}} && && +2 p_x \overline{\mathbbm{s}} && &&= 2 \overline{\mathbbm{v}},  
\end{alignat}  
with $D_t = \de{t} + e E \br{t} \de{p_x}$.
Next, we transform the kinetic momentum into the canonical momentum
\begin{alignat}{6}
 &p_x &&= q_x -e A \br{t} ,\hspace{1.5cm} &&t &&= \tilde{t}, \\
 &\de{p_x} &&= \de{q_x},\ &&\de{t} &&= \de{\tilde{t}} -e E \br{\tilde t} \de{q_x}.
\end{alignat}
Due to this mapping, we have successfully transformed the system of PDEs into a system of ODEs
\begin{alignat}{5}
  & \de{t} \overline{\mathbbm{s}}     && && -2 \br{q_x-e A} \overline{\mathbbm{p}} &&+ 2 p_{\rho} \overline{\mathbbm v} &&= 0,  \\
  & \de{t} \overline{\mathbbm{v}} && && &&- 2 p_{\rho} \overline{ \mathbbm{s}} &&= -2\overline{\mathbbm{p}},  \\    
  & D_t \overline{\mathbbm{p}} && && +2 \br{q_x -e A } \overline{\mathbbm{s}} && &&= 2 \overline{\mathbbm{v}}.   
\end{alignat} 
In the following we will demonstrate, that the system above is indeed the same obtained from QKT. We will rewrite the equations using a more convenient
vector-matrix notation. This gives
\begin{align}
 \de{t} \ma{\overline{\mathbbm{s}} \\ \overline{\mathbbm{p}} \\ \overline{\mathbbm{v}}} = \overline{\mathbf{M}} \ma{\overline{\mathbbm{s}} \\ \overline{\mathbbm{p}} \\ \overline{\mathbbm{v}}}
 = \overline{\mathbf{M}} \sum_{i=1}^3 \alpha_i \mathbf{e}_i, \label{eq_QKT_Cy1}
\end{align}
where $\alpha_i$ denote the expansion coefficients, $\mathbf{e}_i$ the basis vectors and $\overline{\mathbf{M}}$ is a $3 \times 3$ matrix.
Choosing the first basis vector to be 
\begin{align}
 \mathbf{e}_1 = -\frac{1}{\omega} \ma{1 \\ p_{\rho} \\ q_x -e A },
\end{align}
where $\omega = \sqrt{1 + \br{q_x -e A}^2 + p_{\rho}^2}$
we immediately see that we obtain the initial conditions via $\alpha_1 = 1$ and $\alpha_2 = \alpha_3 = 0$. 
Next, we find 
\begin{align}
 \overline{\mathbf{M}} \mathbf{e}_1 = \ma{0 \\ 0 \\ 0},\ \de{t} \mathbf{e}_1 = \frac{e E \sqrt{1 + p_{\rho}^2}}{\omega^2} \mathbf{e}_2,
\end{align}
with
\begin{align}
 \mathbf{e}_2 = \frac{1}{\omega \sqrt{1 + p_{\rho}^2}} \ma{\ \ q_x -e A \\ p_{\rho} \br{q_x -e A } \\ - \br{1 + p_{\rho}^2}}.
\end{align}
We proceed by calculating
\begin{align}
 \overline{\mathbf{M}} \mathbf{e}_2 = -2 \omega \mathbf{e}_3,\ \de{t} \mathbf{e}_2 = -\frac{e E \sqrt{1 + p_{\rho}^2}}{\omega^2} \mathbf{e}_1, 
\end{align}
where
\begin{align}
 \mathbf{e}_3 = \frac{1}{\sqrt{1 + p_{\rho}^2}}\ma{-p_{\rho} \\ 1 \\ 0}.
\end{align}
Doing the same for the last vector we obtain
\begin{align}
 \overline{\mathbf{M}} \mathbf{e}_3 = 2 \omega \mathbf{e}_2,\ \de{t} \mathbf{e}_3 = 0.
\end{align}
Inserting all of the above quantities into equation \eqref{eq_QKT_Cy1} we obtain one equation for every basis vector
\begin{alignat}{4}
 \mathbf{e}_1:& \qquad &\dot{\alpha}_1 &-& \frac{e E \sqrt{1+ p_{\rho}^2}}{\omega^2} \alpha_2 &=& 0, \\
 \mathbf{e}_2:& \qquad &\dot{\alpha}_2 &+& \frac{e E \sqrt{1+ p_{\rho}^2}}{\omega^2} \alpha_1 &=& 2 \omega \alpha_3, \\
 \mathbf{e}_3:& \qquad &\dot{\alpha}_3 & & &=& -2 \omega \alpha_2. 
\end{alignat}
In the end, we introduce the familiar quantities 
\begin{align}
 \alpha_1 = F - 1,\ \alpha_2 = G,\ \alpha_3 = H,
\end{align}
which yields the equations for QKT with appropriate initial conditions
\begin{align}
  \ma{\dot{F} \\ \dot{G} \\ \dot{H}} = \ma{0 & W & 0 \\ -W & 0 & -2\omega \\ 0 & 2\omega & 0} \ma{F \\ G \\ H} + \ma{0 \\ W \\ 0}.
\end{align}

\section{Vacuum initial conditions}
\label{Sec_Vac}
In order to obtain proper initial conditions describing for example a vacuum state we
have to solve the Dirac equation. The solution for the free
Dirac equation can be found in many textbooks. Nevertheless, we
explicitly calculate the vacuum initial conditions for a $2+1$ dimensional system for the
sake of a complete description.

We begin defining the basis matrices. We choose them to be
\begin{alignat}{6}
  &\ga{0} && = \sigma_3 = \ma{
  1 & 0 \\
  0 & -1},\  
  &&\ga{1} && = \ii \sigma_1 = \ma{
  0 & \ii \\
  \ii & 0},\
  &\ga{2} && = \ii \sigma_2 = \ma{
  0 & 1 \\
  -1 & 0}
\end{alignat} 
plus the unit matrix $\mathbbm{1}$.  
We proceed writing up the Dirac equation and, more important, the Dirac equation squared:
\begin{align}
 \br{\ii \gamma^{\mu} \de{\mu} -e \gamma^{\mu} A_{\mu} \br{r,p} - \mathbbm{1}m} \psi \br{r,p} = 0, \\
 \br{D^{\mu} D_{\mu} + \frac{\ii e}{2} \gamma^{\mu} \gamma^{\nu} F_{\mu \nu} \br{r,p} + \mathbbm{1} m}  \Phi \br{r,p} = 0,
\end{align}
where $D_{\mu} = \de{\mu} + \ii e A_{\mu} \br{r,p}$. As we are only interested in the vacuum solutions we have
$A_{\mu} = 0$ and thus $F_{\mu \nu} = 0$. Hence, we obtain
one equation each for every component of $\Phi \br{r,p} =g_1 \br{r,p} \ma{1 \\ 0} + g_2 \br{r,p} \ma{0 \\ 1}$:
\begin{align}
 \br{\de{t}^2 - \de{x}^2 - \de{y}^2 + m^2} g_1 \br{r,p} &= 0 ,\\
 \br{\de{t}^2 - \de{x}^2 - \de{y}^2 + m^2} g_2 \br{r,p} &= 0. 
\end{align}
We find two solutions for these equations
\begin{equation}
 g_k^{\pm} \br{r,p} = \ee^{\mp\ii \omega t} \ee^{\ii p_x x} \ee^{\ii p_y y},
\end{equation}
where $k = 1,2$ and $\omega^2 = m^2 + p_x^2 + p_y^2$. Consequently, we can write the spinor $\psi$ as a linear combination of all solutions
\begin{align}
 \psi \br{r,p} &= \ma{1 \\ \frac{p_y -\ii p_x}{\omega + m}} g_1^+ \br{r,p} + \ma{1 \\ \frac{p_y - \ii p_x}{-\omega + m}} g_1^- \br{r,p} \notag \\
  &+ \ma{\frac{-\ii p_x - p_y}{-\omega + m} \\ 1} g_2^+ \br{r,p} + \ma{\frac{-\ii p_x - p_y}{\omega+m} \\ 1} g_2^- \br{r,p}.
\end{align}
However, these are the solutions for the quadratic Dirac equation. In order to get rid of the redundant solutions we have to pick either $g^+$ or $g^-$.
When choosing $g^+$ the solutions are given by the two spinors
\begin{alignat}{5}
 &\psi_1 \br{r,p} &&= \ma{1 \\ \frac{p_y- \ii p_x}{\omega + m}} &&\ee^{-\ii p x} = u \br{p} &&\ee^{-\ii p x}, \\
 &\psi_2 \br{r,p} &&= \ma{\frac{p_y+ \ii p_x}{\omega - m} \\ 1} &&\ee^{-\ii p x} = v \br{p} &&\ee^{-\ii p x}.
\end{alignat}
We will associate the first spinor $\psi_1$ with particles and the second spinor $\psi_2$ with antiparticles leading to
\begin{equation}
 v \br{-p} = \ma{\frac{p_y+ \ii p_x}{\omega + m} \\ 1}.
\end{equation}
At first we have to properly normalize the spinors:
\begin{alignat}{6}
 &u^{\dag} \br{p} \cdot u \br{p} &&= 1+ \frac{p_y+\ii p_x}{\omega + m} \frac{p_y-\ii p_x}{\omega + m} &&= 1 + \frac{\omega-m}{\omega+m} &&= \frac{2 \omega}{\omega + m} &&= N_1^2 \\
\text{and}& \notag \\
 &v^{\dag} \br{-p} \cdot v \br{-p} &&= \frac{p_y-\ii p_x}{\omega + m} \frac{p_y+\ii p_x}{\omega + m} + 1 &&= \frac{\omega-m}{\omega+m} + 1 &&= \frac{2 \omega}{\omega + m} &&= N_2^2.
\end{alignat}
Then, we proceed by calculating the completeness relations
\begin{align}
 u \br{p} \cdot \bar u \br{p} &= \frac{\omega + m}{2 \omega} \ma{1 \\ \frac{p_y - \ii p_x}{\omega +m}} \ma{1 & \frac{p_y+\ii p_x}{\omega+m}} \ma{1 & 0 \\ 0 & -1}\\
 &= \frac{1}{2 \omega} \ma{\omega + m & -p_y-\ii p_x\\ p_y-\ii p_x & -\br{\omega - m}} = \frac{1}{2 \omega} \br{\omega \gamma_0 - p_x \gamma_1 - p_y \gamma_2 + \mathbbm{1} m} \\
 v \br{-p} \cdot \bar v \br{-p} &= \frac{1}{2 \omega} \br{\omega \gamma_0 - p_x \gamma_1 - p_y \gamma_2 - \mathbbm{1} m}.
\end{align}
We combine the expressions above with the definition for the free equal-time Wigner function
\begin{equation}
 \mathbbm{w}_{\alpha \beta} \br{A=0,\mathbf{x},\mathbf{p}} = \frac{1}{2} \int d^3s \ \ee^{-\ii \mathbf{p} \cdot \mathbf{s}} \com{\bar \psi_{\beta} \br{\mathbf{x}-\frac{\mathbf{s}}{2}}, \psi_{\alpha} \br{\mathbf{x} + \frac{\mathbf{s}}{2}} }.
\end{equation}
Writing the spinors in terms of creation and annihilation operators
\begin{align}
 \psi_{\alpha} \br{r,p} = u_{\alpha} \hat a + v_{\alpha} \hat{b^+}
\end{align}
and obeying commutator and anti-commutator relations yields
\begin{equation}
 \mathbbm{w}_i = v \br{-p} \cdot \bar v \br{-p} - u \br{p} \cdot \bar u \br{p} = \frac{1}{\omega} \br{-\gamma_1 p_x - \gamma_2 p_y - \mathbbm{1} m}.
\end{equation}
Hence, we obtain the following initial conditions
 \begin{equation}
  \mathbbm{s}_i = -\frac{1}{\omega},\ \mathbbm{v}_{1i} = -\frac{p_x}{\omega},\ \mathbbm{v}_{2i} = -\frac{p_x}{\omega}.
 \end{equation}
A calculation using $4$-spinors and therefore $16$ basis matrices leads to a similar result:
\begin{equation}
 \mathbbm{s}_i = -\frac{2}{\omega},\ \mathbbm{v}_{1i} = -\frac{2p_x}{\omega},\ \mathbbm{v}_{2i} = -\frac{2p_y}{\omega} ,\ \mathbbm{v}_{3i} = -\frac{2p_z}{\omega},
\end{equation}
with $\omega = \sqrt{1 + \mathbf{p}^2}$.

\section{Carrier Envelope Phase in the multiphoton regime}
\label{Sec_CEP}
Analyzing the transport equations, one finds, that for multiphoton pair production and therefore many-cycle pulses the carrier envelope phase(CEP) is irrelevant. This can be shown mathematically assuming, that the time-dependence of the electric field takes, e.g., the form
\begin{equation}
 E \br{t} = \cos \br{\frac{t}{\tau}}^4 \ \cos \br{\omega t + \phi}. 
\end{equation}
Then we introduce the variable $\tilde t = t + \phi / \omega = t + t_0$ with $t_0 \in \com{-\frac{\pi}{2},\frac{\pi}{2}}/ \omega$. This yields
\begin{equation}
 E \br{\tilde t} = \cos \br{\frac{\tilde t - t_0}{\tau}}^4 \ \cos \br{\omega \tilde t} 
\end{equation}
and the initial conditions transform to $\tilde t_i = -\frac{\pi}{2} \tau + t_0$. In case $\omega \tau$ is sufficiently large the initial condition gives approximately  $\tilde t_i \approx -\frac{\pi}{2} \tau$. 

Furthermore, the electric field yields
\begin{align}
 E \br{\tilde t} = \cos \br{\frac{\tilde t - t_0}{\tau}}^4 \ \cos \br{\omega \tilde t} \approx  \cos \br{\frac{\tilde t}{\tau}}^4 \ \cos \br{\omega \tilde t}.
\end{align}
Hence, the phase $\phi$ is irrelevant in the regime of multiphoton pair production.
In case of $\tau \to \infty$ and thus an infinitely extended pulse the approximation above is even exact.
An immediate consequence of this results is that all considerations concerning symmetries are true in the multiphoton regime independent whether $A \brt$ is symmetric or antisymmetric in $t$.

\section{Relativistic Lorentz force}  

A nice tool in order to estimate the influence of the applied fields on the trajectory of the created particles is the relativistic Lorentz force.
The equations of motion stemming from taking into account the Lorentz force are easier to solve and interpret than computations via the Dirac equation.
Although the Lorentz force does not take quantum effects into account, it turned out to be a useful
tool in order to understand, for example, particle deflection, self-bunching and also interference effects.
Despite the fact, that the Lorentz force is prominent in many textbooks and lecture notes we will provide a short summary here.
Formulated in an invariant manner the Lorentz force reads
\begin{equation}
 \frac{dP^{\mu}}{d \tau} = e F^{\mu \nu} U_{\nu},
\end{equation}
where $F^{\mu \nu}$ is the electromagnetic field strength tensor, $P^{\mu}$ is the four-momentum and $U_{\nu}$ the four-velocity.
We proceed by choosing a coordinate frame. Then the field strength tensor is 
\begin{align}
 F^{\mu \nu} = \ma{0 & -E_x &-E_y & -E_z \\ E_x & 0 & -B_z & B_y \\ E_y & B_z & 0 & -B_x \\ E_z & -B_y & B_x & 0}.
\end{align}
Moreover we can write
\begin{align}
 \frac{dP^{\mu}}{d \tau} = \gamma \frac{dP^{\mu}}{d t},\ \text{with} \ \gamma = \frac{1}{\sqrt{1-\mathbf{v}^2}}
\end{align}
and using the fact, that
\begin{align}
 U_{\mu} = \br{\gamma, -\gamma v_x, -\gamma v_y, -\gamma v_z}
\end{align}
we obtain
\begin{equation}
 \frac{d P ^{m}}{dt} = e \br{\mathbf{E} + \mathbf{v} \times \mathbf{B}}_m.
\end{equation}
As
\begin{align}
 v_m = \frac{d x_m}{dt}
\end{align}
and by rewriting the left hand side of the equation
\begin{equation}
 P^m = U^m = \gamma v_m = \gamma \frac{dx_m}{dt},
\end{equation}
we obtain the relativistic Lorentz force in a coordinate frame
\begin{equation}
 \frac{d}{dt} \br{\gamma \dot x_m} = e \br{\mathbf{E} \br{\mathbf{x},t} + \dot{\mathbf{x}} \times \mathbf{B} \br{\mathbf{x},t}}_m.
\end{equation}

\pagestyle{plain}
\chapter{Additional tables and figures}

\section{Figures}
\subsection{Particle yield for homogeneous fields}

\begin{figure}[tbh]
\begin{center}
  \includegraphics[width=0.7\textwidth]{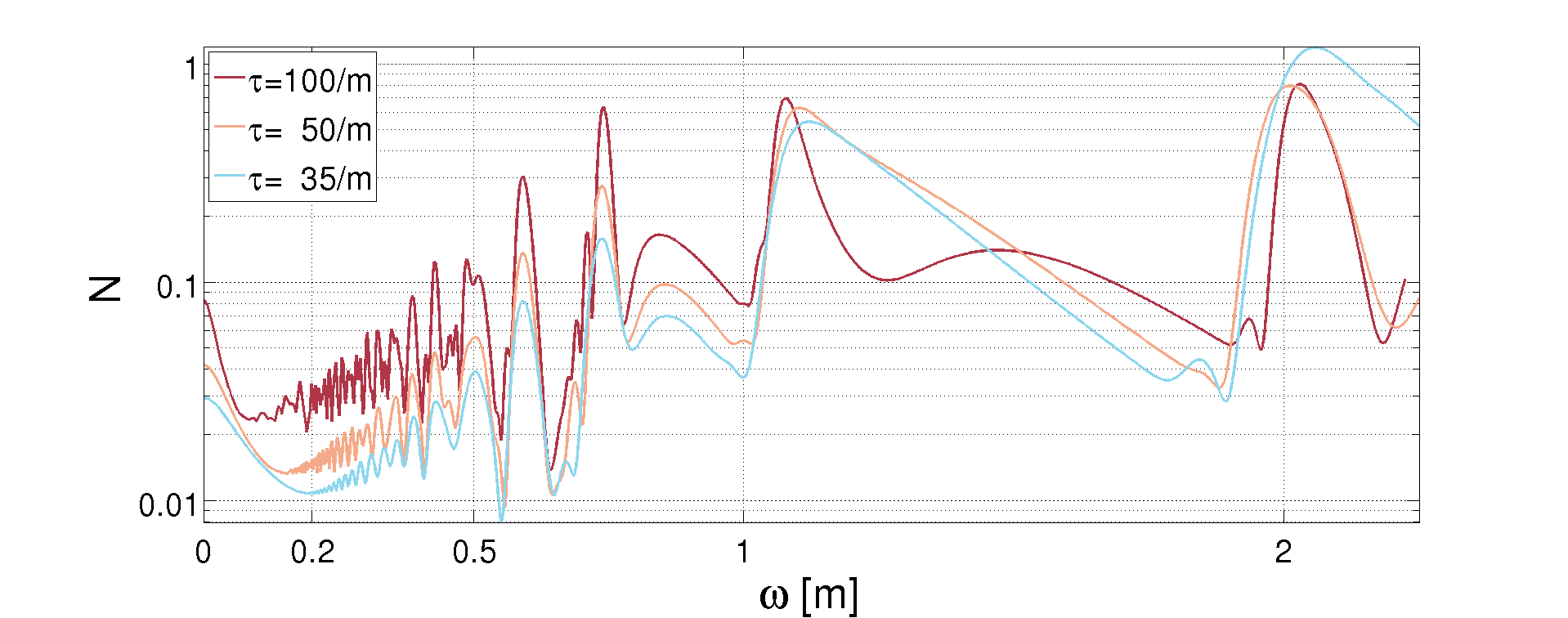}
\end{center}  
\caption{Electric field: $E \brt = \varepsilon \ E_0 \cos^4 \br{\frac{t}{\tau}} \cos \br{\omega t}$. Parameters: Tab. \ref{Tab_EffM1_App}}  
\label{FigApp_1}
\end{figure}

\begin{figure}[tbh]
\begin{center}
  \includegraphics[width=0.65\textwidth]{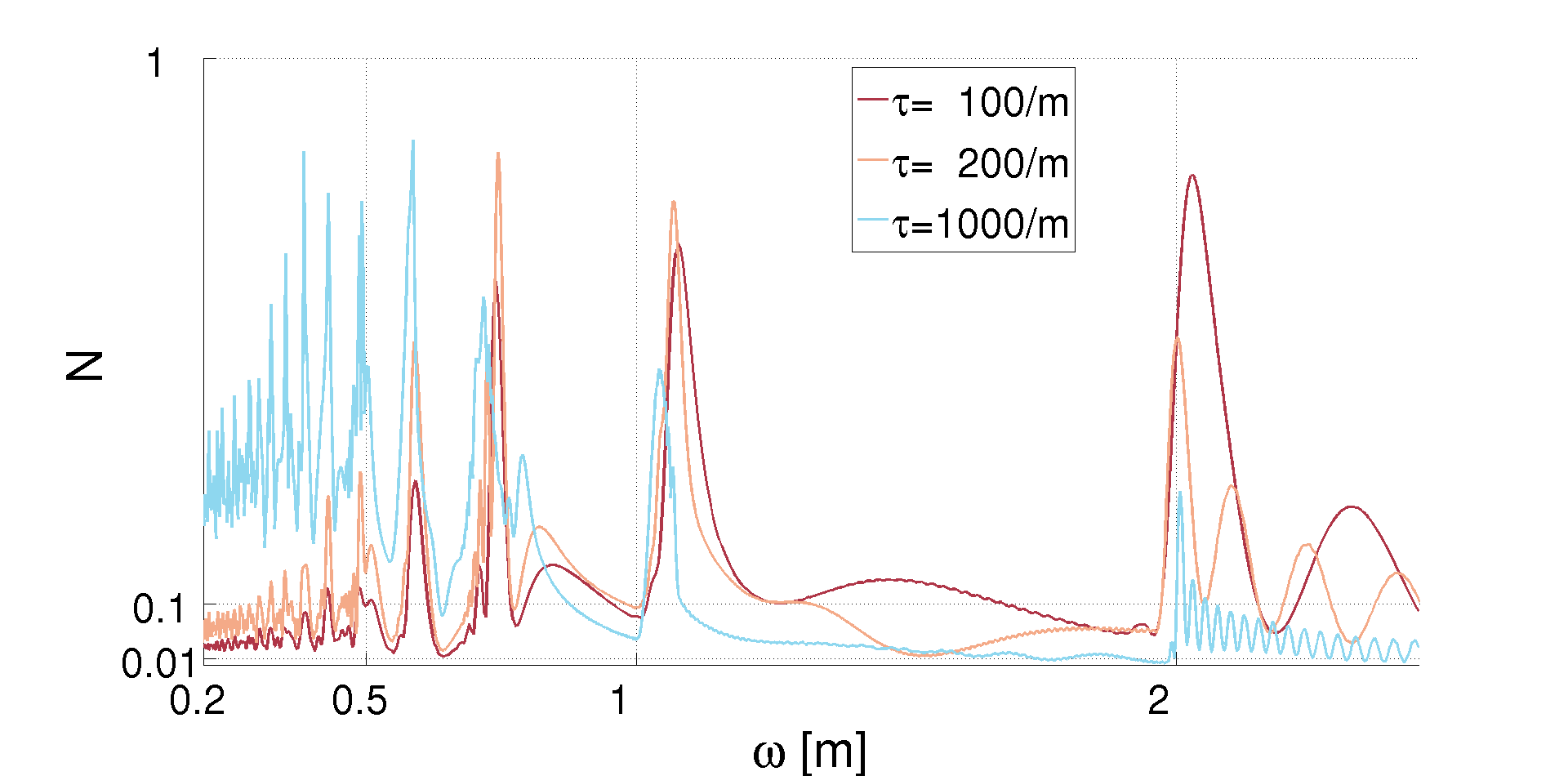}
\end{center}
\caption{Electric field: $E \brt = \varepsilon \ E_0 \cos^4 \br{\frac{t}{\tau}} \cos \br{\omega t}$. Parameters: Tab. \ref{Tab_EffM2_App}}
\label{FigApp_2}
\end{figure}

\subsection{Carrier Envelope Phase regime}
\label{App_CEP}

\begin{figure}[htb]
 \includegraphics[width=0.5\textwidth]{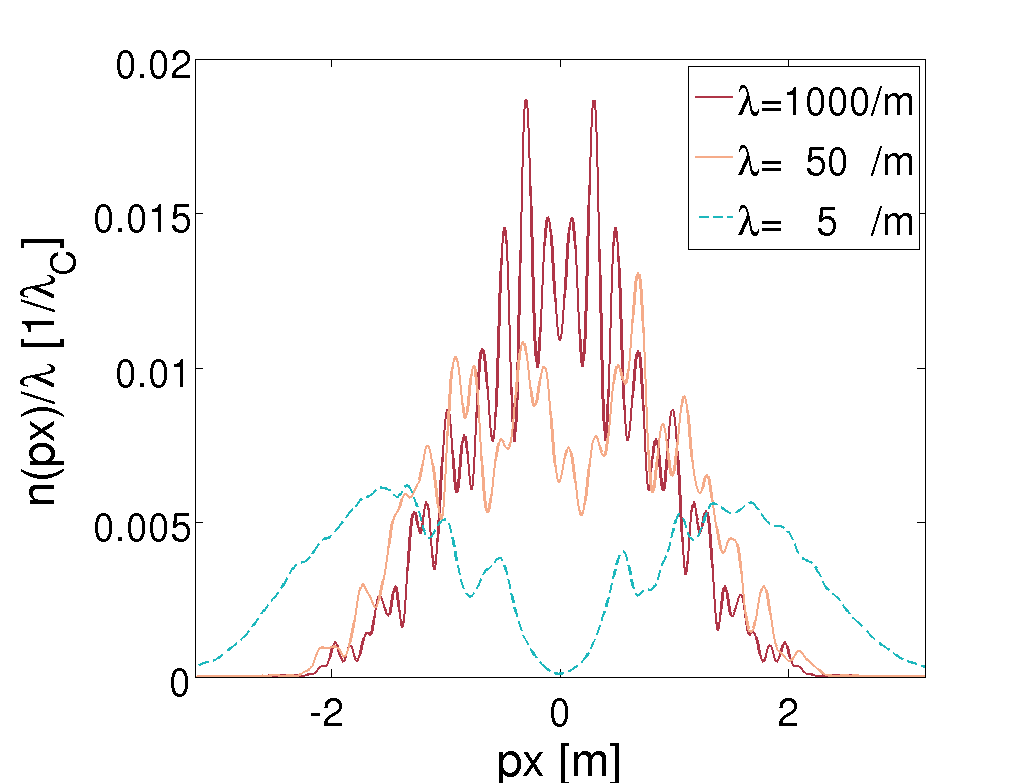}
 \includegraphics[width=0.5\textwidth]{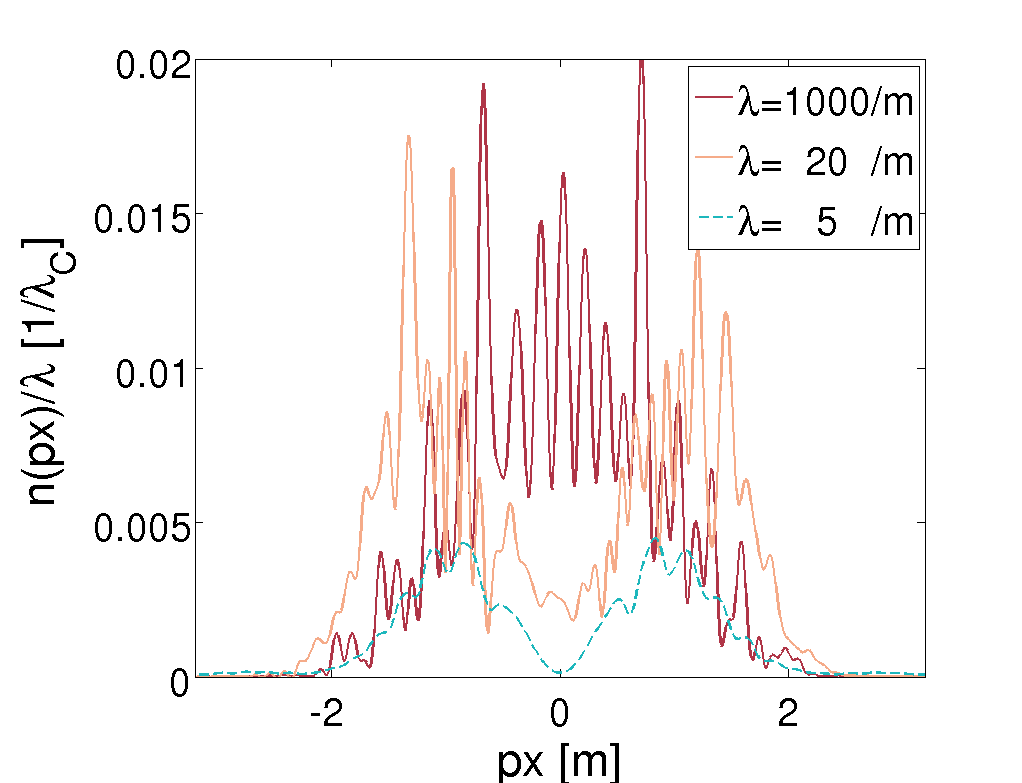}
 \caption{Electric field: $E \br{x,t} = \varepsilon \ E_0 \ \exp \br{-\frac{x^2}{2 \lambda^2} } \cos^4 \br{\frac{t}{\tau}} \cos \br{\omega t + \phi}$. $\phi=0$ on the left-hand side and $\phi=\pi/2$ on the right-hand side. Additional parameters: Tab. \ref{Tab_distr3}} 
 \label{FigApp_CEP}
\end{figure}

\subsection{Short pulsed, time-symmetric magnetic field}

\begin{center} 
\begin{figure}[H]
 \includegraphics[width=0.5\textwidth]{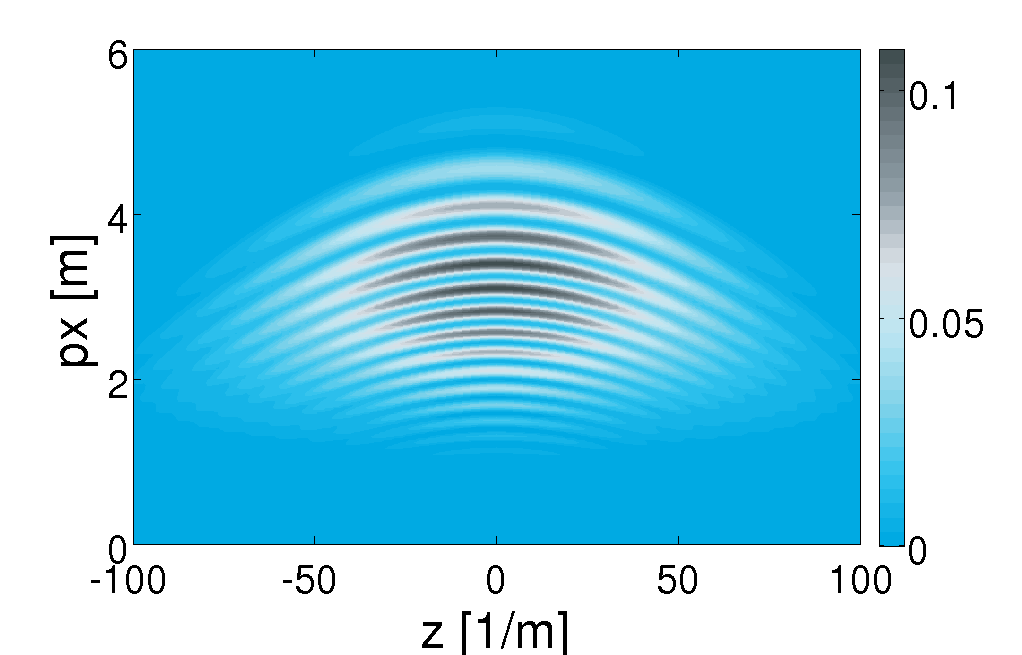}
 \includegraphics[width=0.5\textwidth]{./Fig/206/4.png}
 \includegraphics[width=0.5\textwidth]{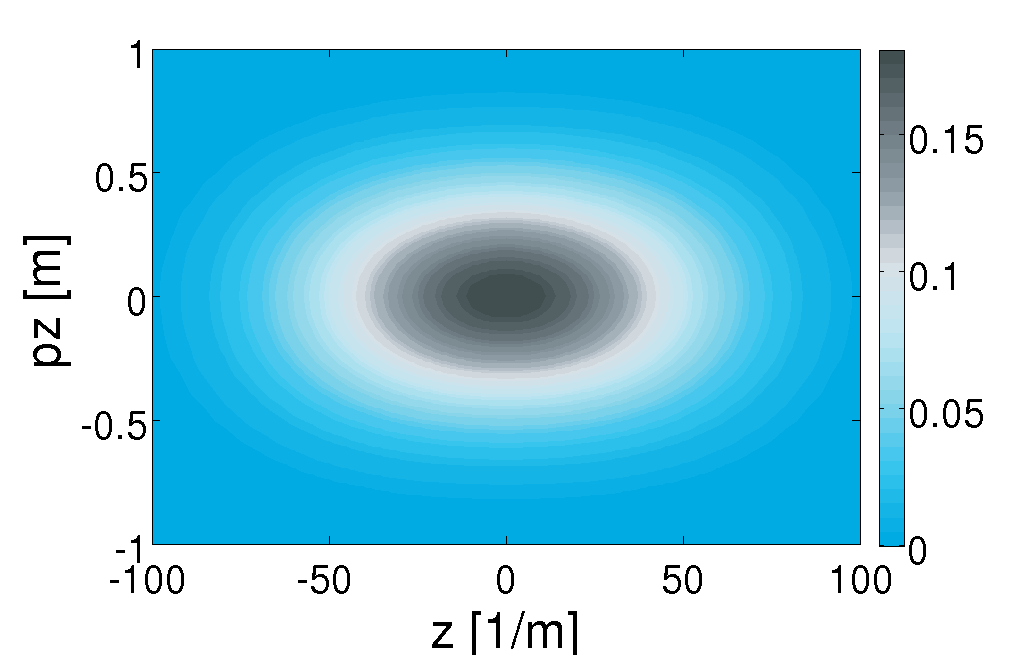} 
 \caption{Electric field: $\mathbf{E} (z,t) = \varepsilon \ E_0 \ \left( \text{sech}^2 \left( \frac{t-\tau}{\tau} \right) - \text{sech}^2 \left( \frac{t+\tau}{\tau} \right) \right) \exp \br{-\frac{z^2}{2 \lambda^2}} \ \boldsymbol{e}_x$. Magnetic field: $B \br{z,t} = -\varepsilon \ E_0 \ \tau \left( \tanh \left( \frac{t+\tau}{\tau} \right) - \tanh \left( \frac{t-\tau}{\tau} \right) \right) \exp \br{-\frac{z^2}{2 \lambda^2}} \ \frac{z}{\lambda^2}$. Parameters: Tab. \ref{Tab_ShortB}} 
 \label{FigApp_3}
\end{figure} 
\end{center}

\begin{center}
\begin{figure}[H]
 \includegraphics[width=0.5\textwidth]{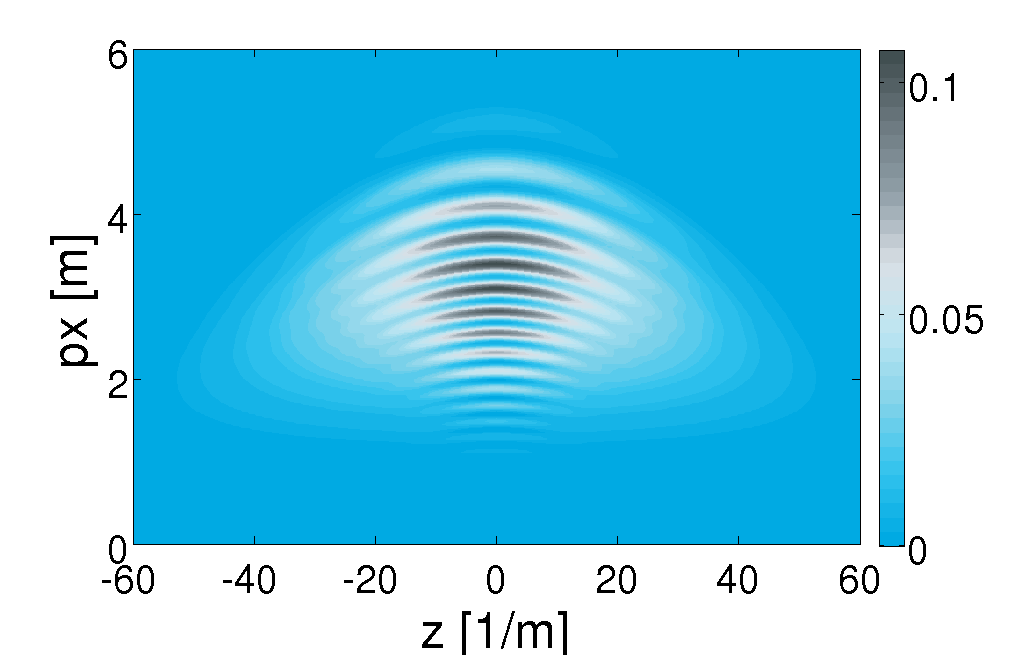}
 \includegraphics[width=0.5\textwidth]{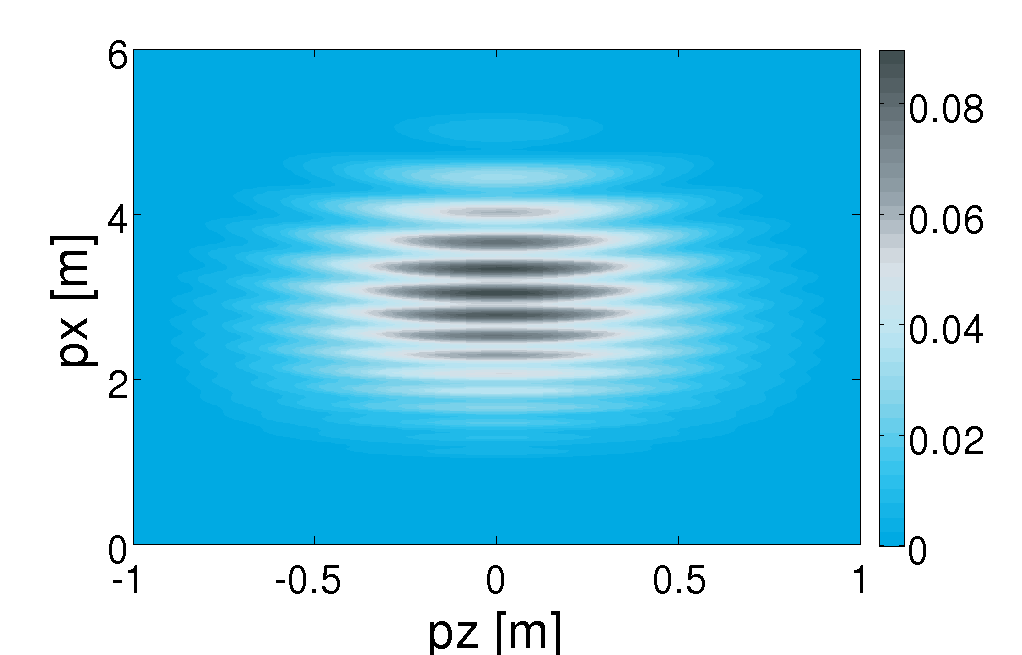}
 \includegraphics[width=0.5\textwidth]{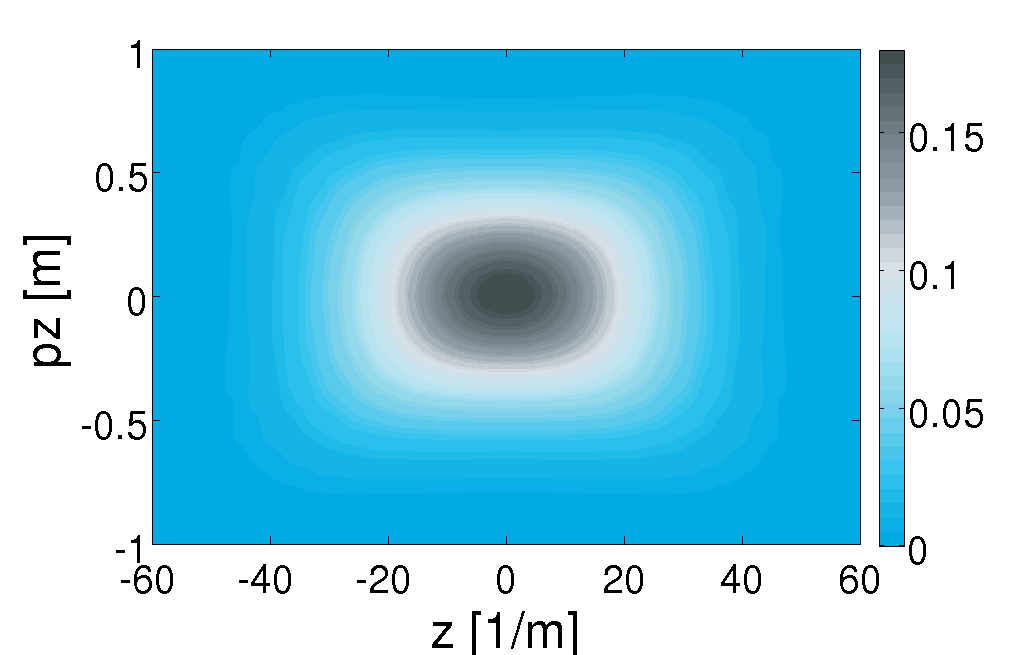} \\
 \includegraphics[width=0.5\textwidth]{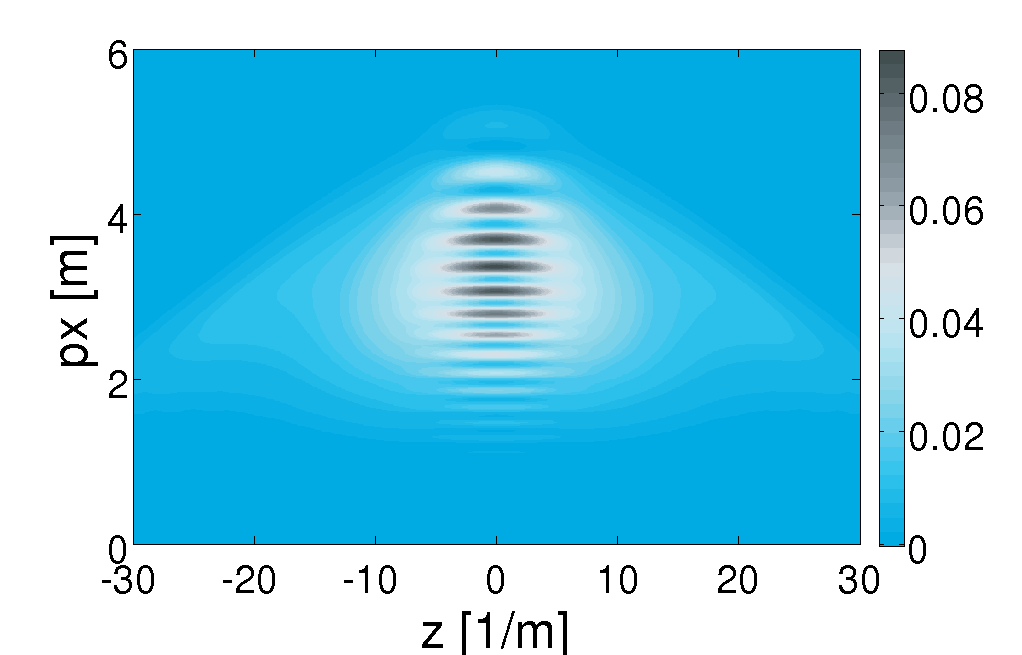}
 \includegraphics[width=0.5\textwidth]{./Fig/206/10.png}
 \includegraphics[width=0.5\textwidth]{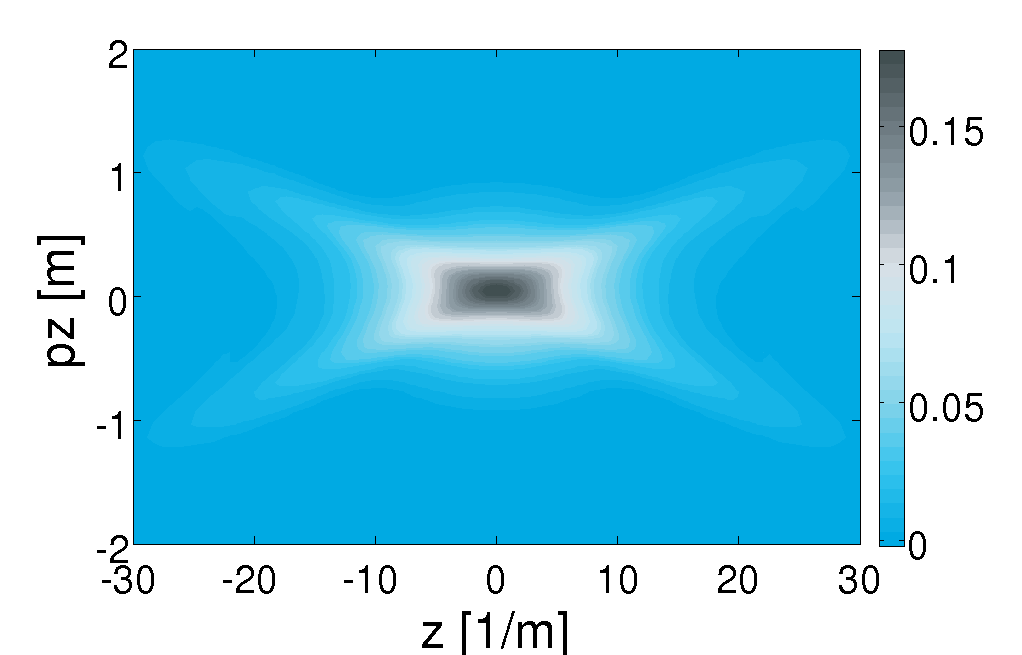} 
 \caption{Electric field: $\mathbf{E} (z,t) = \varepsilon \ E_0 \ \left( \text{sech}^2 \left( \frac{t-\tau}{\tau} \right) - \text{sech}^2 \left( \frac{t+\tau}{\tau} \right) \right) \exp \br{-\frac{z^2}{2 \lambda^2}} \ \boldsymbol{e}_x$. Magnetic field: $B \br{z,t} = -\varepsilon \ E_0 \ \tau \left( \tanh \left( \frac{t+\tau}{\tau} \right) - \tanh \left( \frac{t-\tau}{\tau} \right) \right) \exp \br{-\frac{z^2}{2 \lambda^2}} \ \frac{z}{\lambda^2}$. Parameters: Tab. \ref{Tab_ShortB}}  
 \label{FigApp_4}
\end{figure} 
\end{center}

\begin{center}
\begin{figure}[H]
 \includegraphics[width=0.5\textwidth]{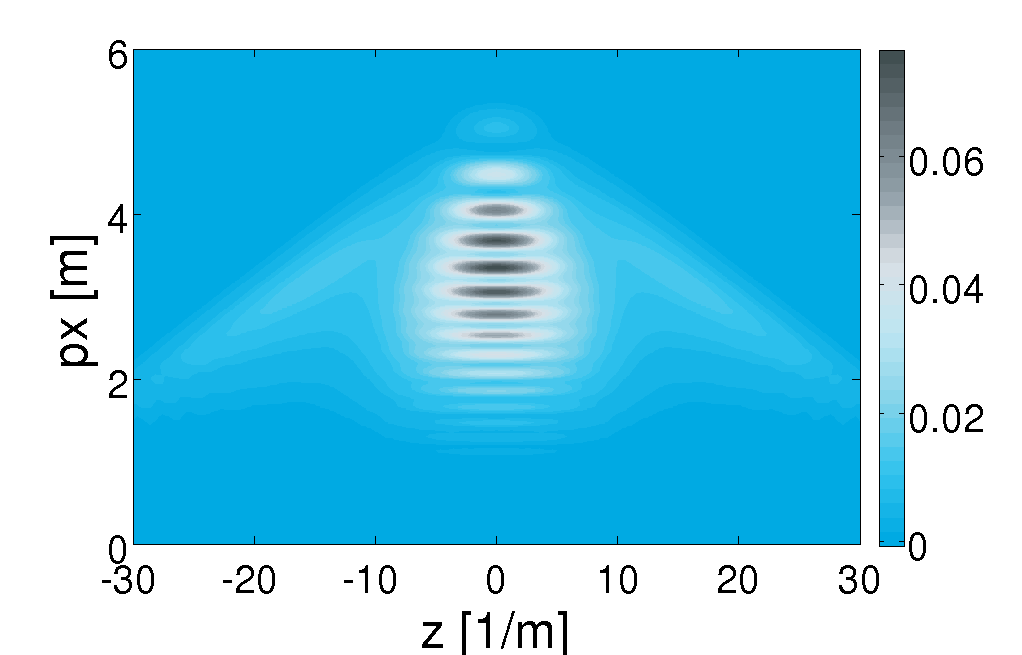}
 \includegraphics[width=0.5\textwidth]{./Fig/206/13.png}
 \includegraphics[width=0.5\textwidth]{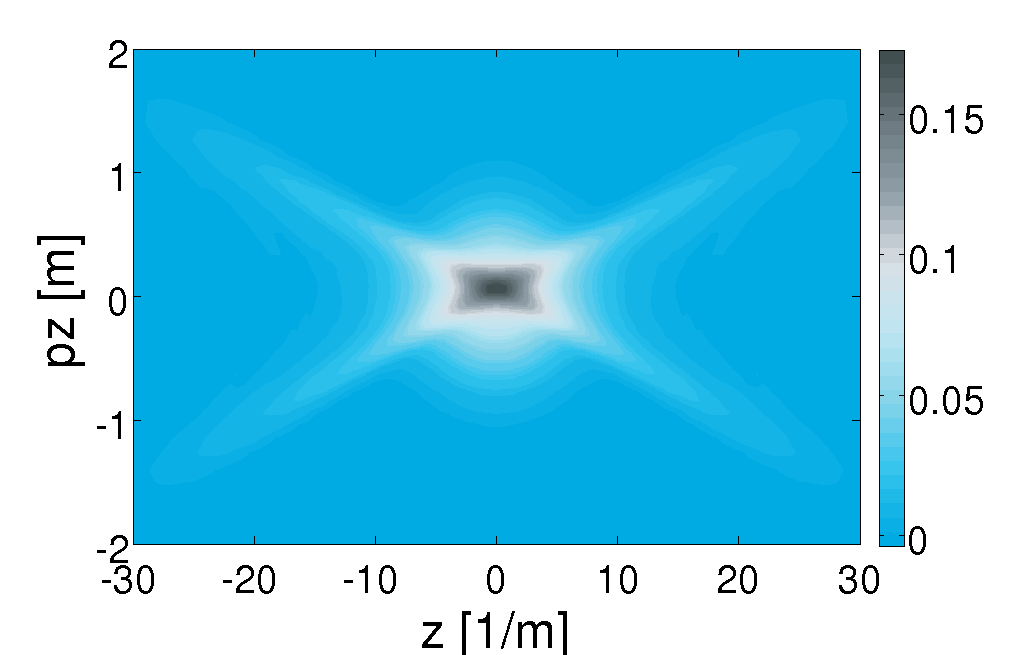} \\
 \includegraphics[width=0.5\textwidth]{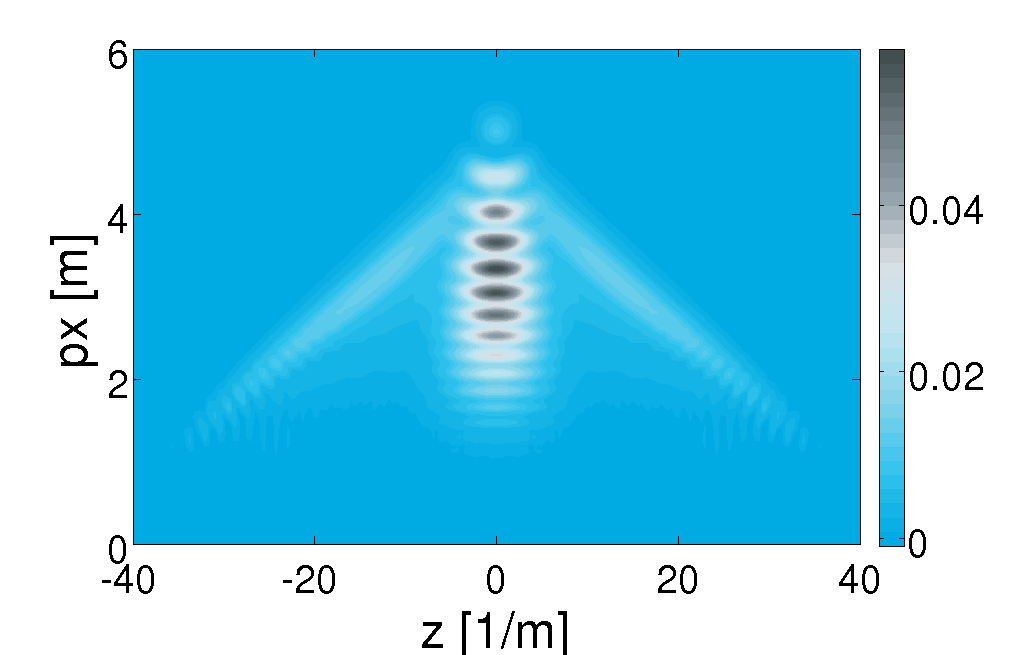}
 \includegraphics[width=0.5\textwidth]{./Fig/206/25.png}
 \includegraphics[width=0.5\textwidth]{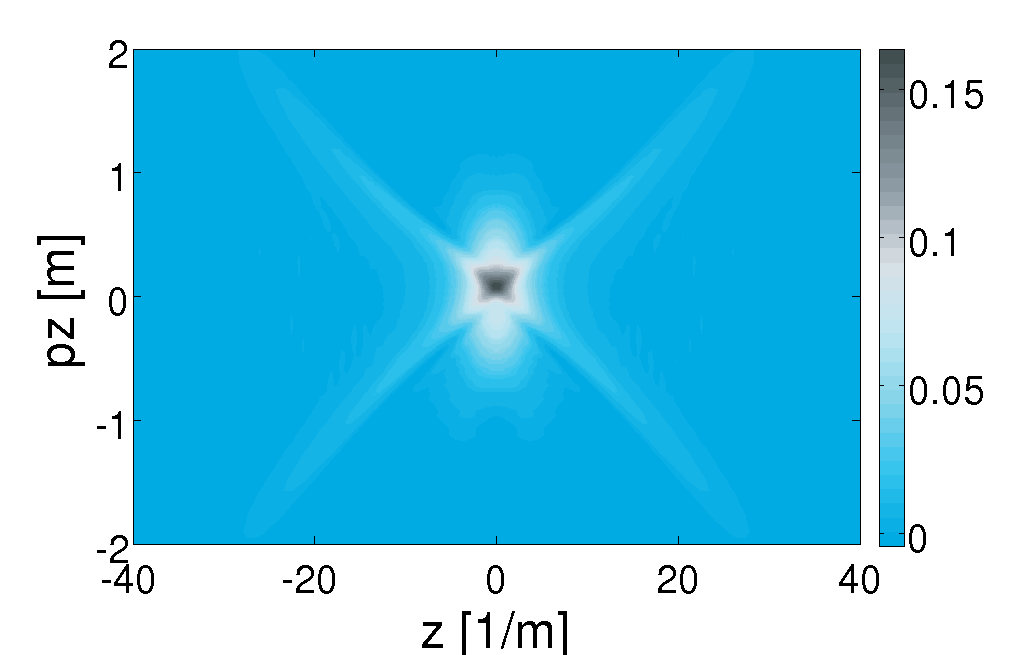} 
\caption{Electric field: $\mathbf{E} (z,t) = \varepsilon \ E_0 \ \left( \text{sech}^2 \left( \frac{t-\tau}{\tau} \right) - \text{sech}^2 \left( \frac{t+\tau}{\tau} \right) \right) \exp \br{-\frac{z^2}{2 \lambda^2}} \ \boldsymbol{e}_x$. Magnetic field: $B \br{z,t} = -\varepsilon \ E_0 \ \tau \left( \tanh \left( \frac{t+\tau}{\tau} \right) - \tanh \left( \frac{t-\tau}{\tau} \right) \right) \exp \br{-\frac{z^2}{2 \lambda^2}} \ \frac{z}{\lambda^2}$. Parameters: Tab. \ref{Tab_ShortB}}
\label{FigApp_5}
\end{figure} 
\end{center}

\begin{center}
\begin{figure}[H]
 \includegraphics[width=0.5\textwidth]{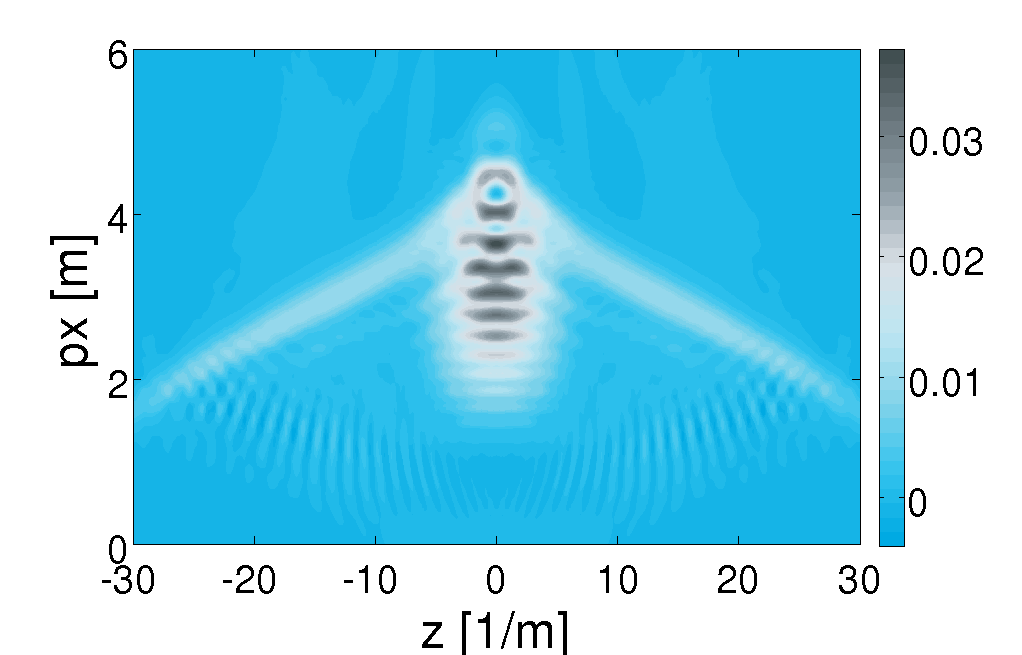}
 \includegraphics[width=0.5\textwidth]{./Fig/206/19.png}
 \includegraphics[width=0.5\textwidth]{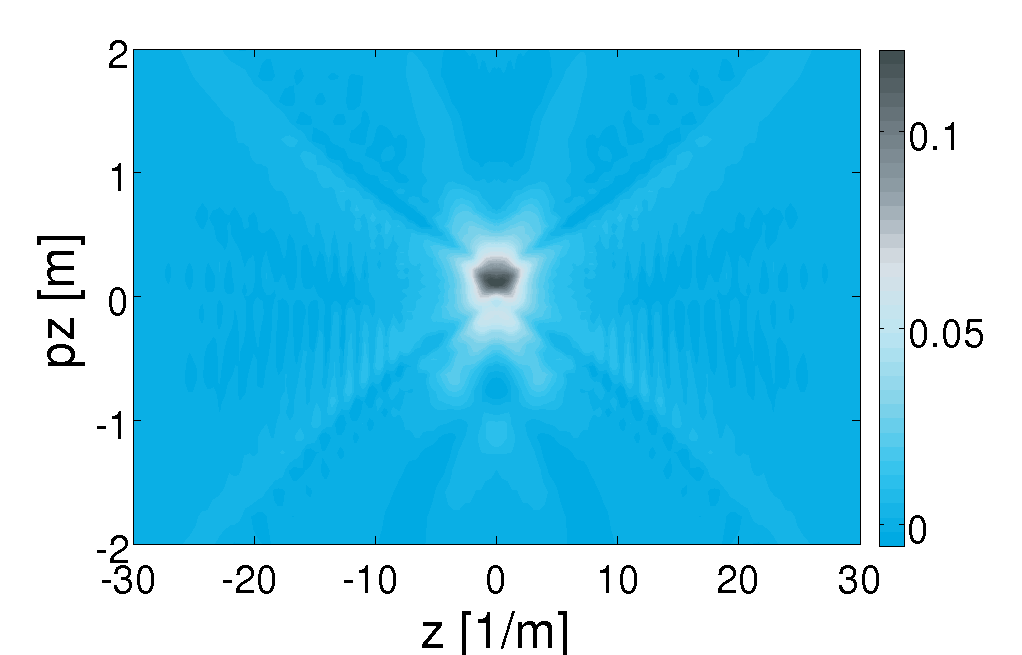} \\
 \includegraphics[width=0.5\textwidth]{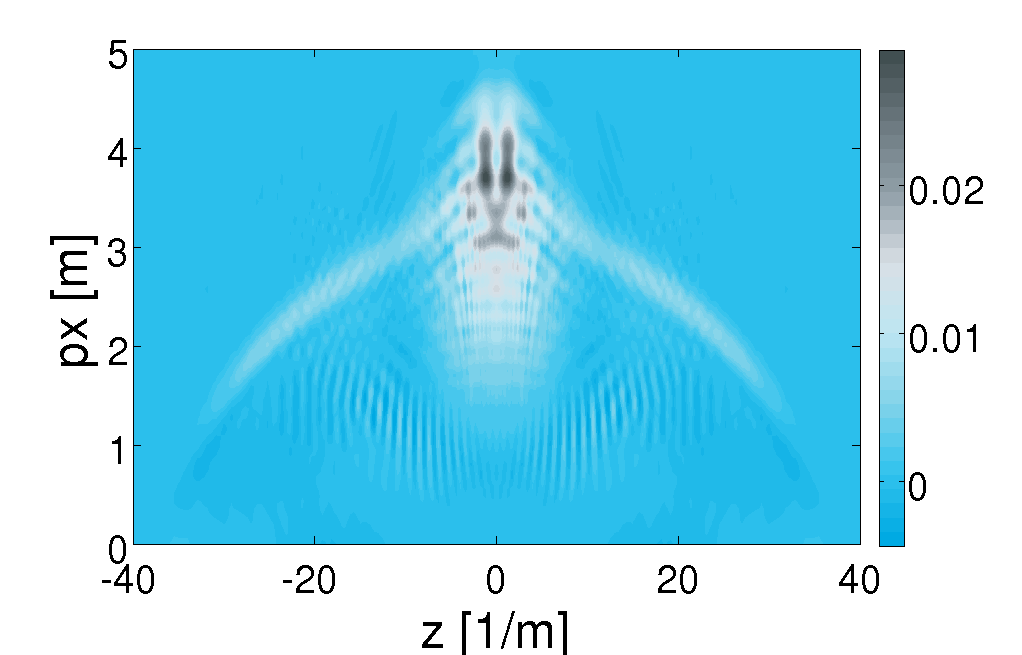}
 \includegraphics[width=0.5\textwidth]{./Fig/206/28.png}
 \includegraphics[width=0.5\textwidth]{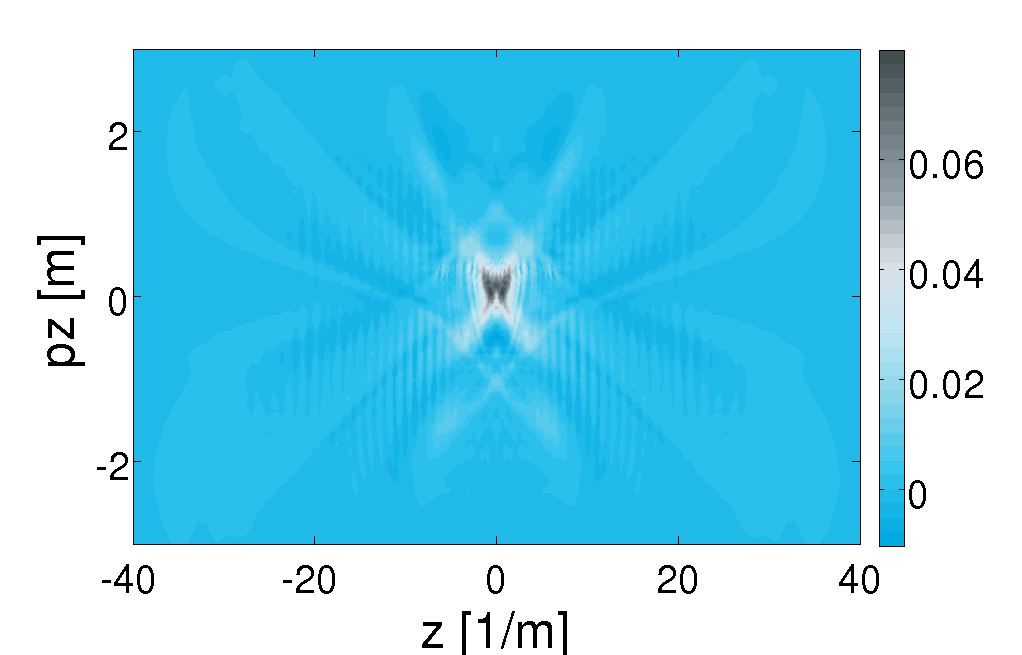} 
\caption{Electric field: $\mathbf{E} (z,t) = \varepsilon \ E_0 \ \left( \text{sech}^2 \left( \frac{t-\tau}{\tau} \right) - \text{sech}^2 \left( \frac{t+\tau}{\tau} \right) \right) \exp \br{-\frac{z^2}{2 \lambda^2}} \ \boldsymbol{e}_x$. Magnetic field: $B \br{z,t} = -\varepsilon \ E_0 \ \tau \left( \tanh \left( \frac{t+\tau}{\tau} \right) - \tanh \left( \frac{t-\tau}{\tau} \right) \right) \exp \br{-\frac{z^2}{2 \lambda^2}} \ \frac{z}{\lambda^2}$. Parameters: Tab. \ref{Tab_ShortB}}
\label{FigApp_6}
\end{figure} 
\end{center}


\subsection{Long pulsed, time-symmetric magnetic field}
 
\begin{center}
\begin{figure}[tbh]
 \includegraphics[width=0.5\textwidth]{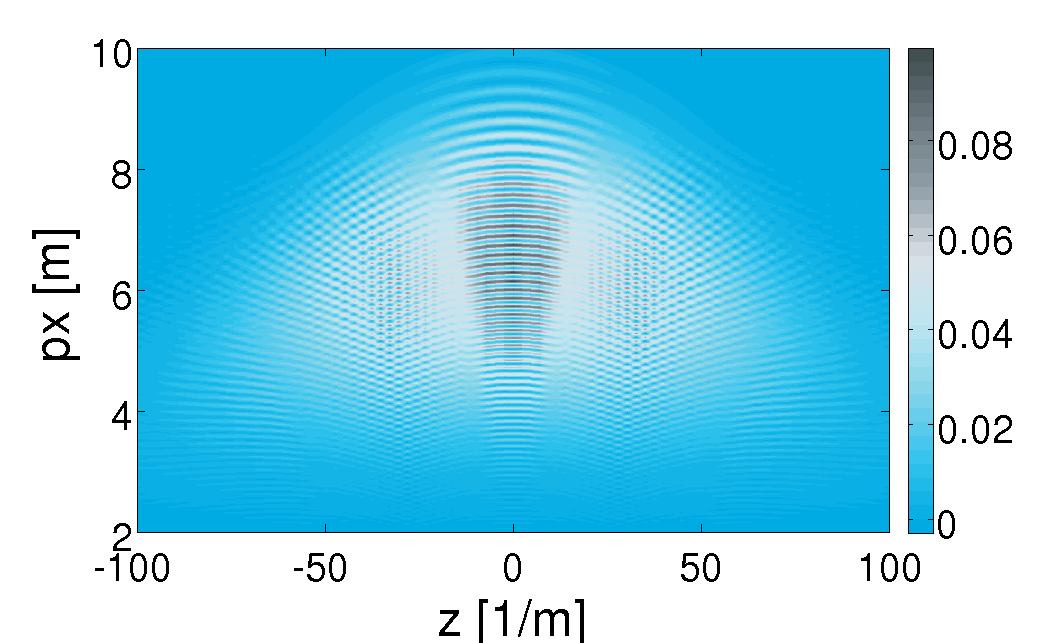}
 \includegraphics[width=0.5\textwidth]{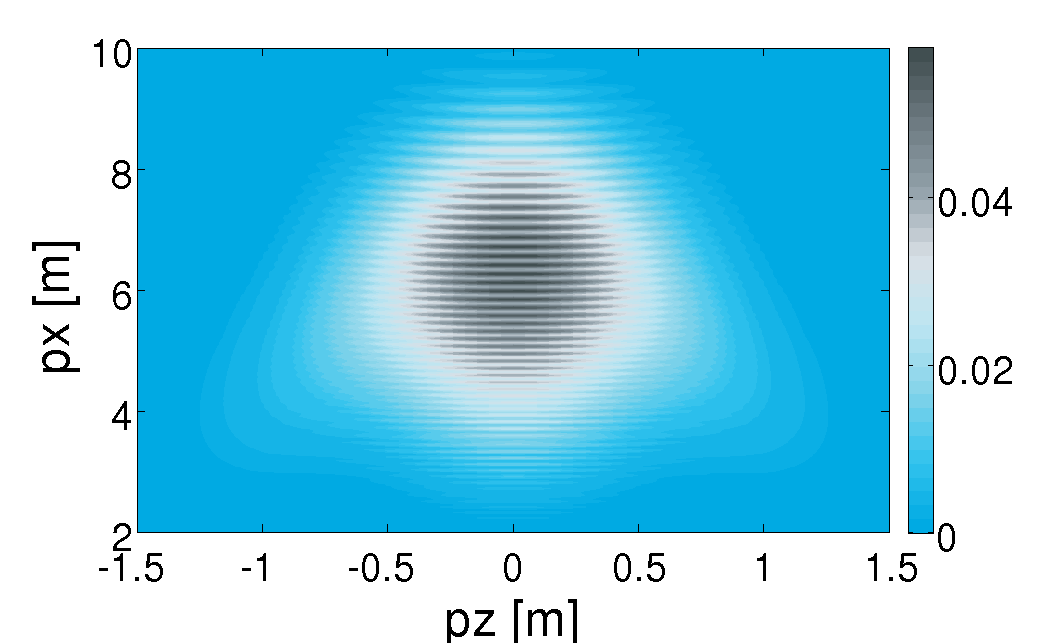}
 \includegraphics[width=0.5\textwidth]{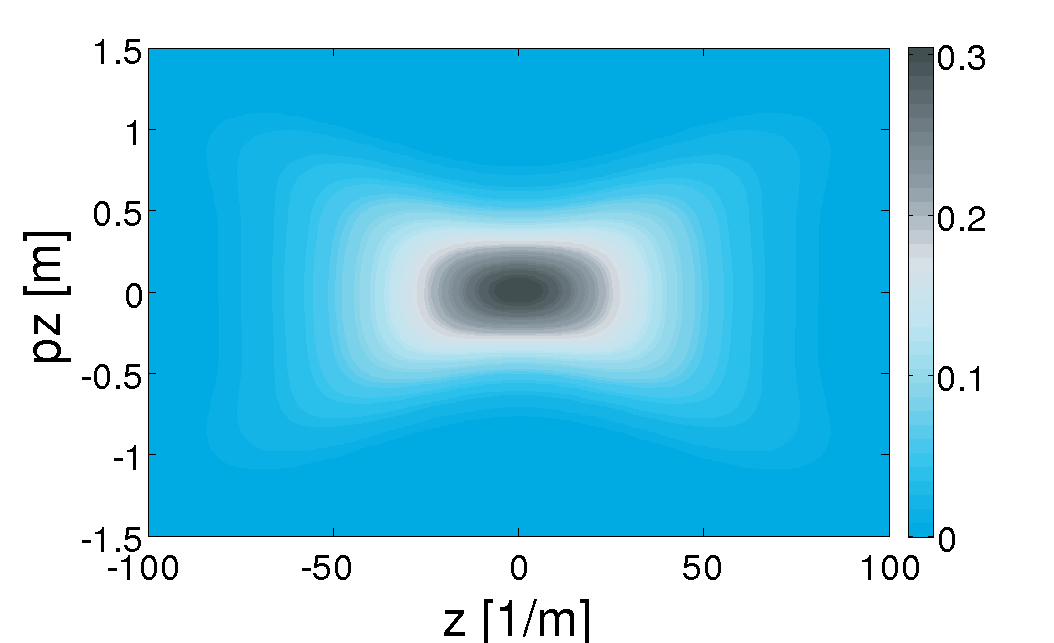}
\caption{Electric field: $\mathbf{E} (z,t) = \varepsilon \ E_0 \ \left( \text{sech}^2 \left( \frac{t-\tau}{\tau} \right) - \text{sech}^2 \left( \frac{t+\tau}{\tau} \right) \right) \exp \br{-\frac{z^2}{2 \lambda^2}} \ \boldsymbol{e}_x$. Magnetic field: $B \br{z,t} = -\varepsilon \ E_0 \ \tau \left( \tanh \left( \frac{t+\tau}{\tau} \right) - \tanh \left( \frac{t-\tau}{\tau} \right) \right) \exp \br{-\frac{z^2}{2 \lambda^2}} \ \frac{z}{\lambda^2}$. Parameters: Tab. \ref{Tab_LongB}}
\label{FigApp_7}
\end{figure} 
\end{center}

\begin{center}
\begin{figure}[tbh]
 \includegraphics[width=0.5\textwidth]{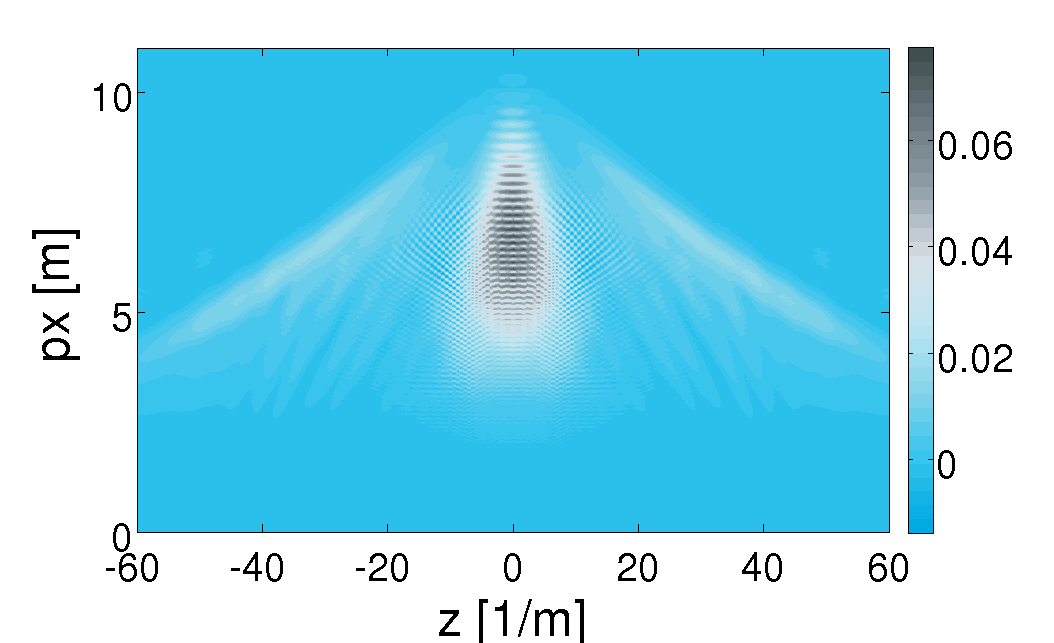}
 \includegraphics[width=0.5\textwidth]{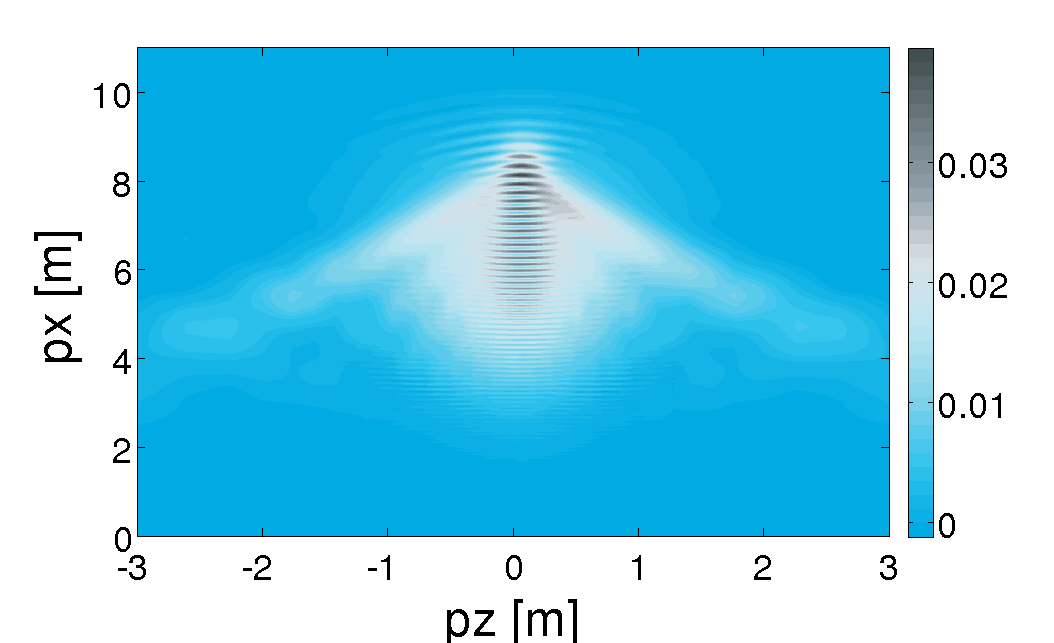}
 \includegraphics[width=0.5\textwidth]{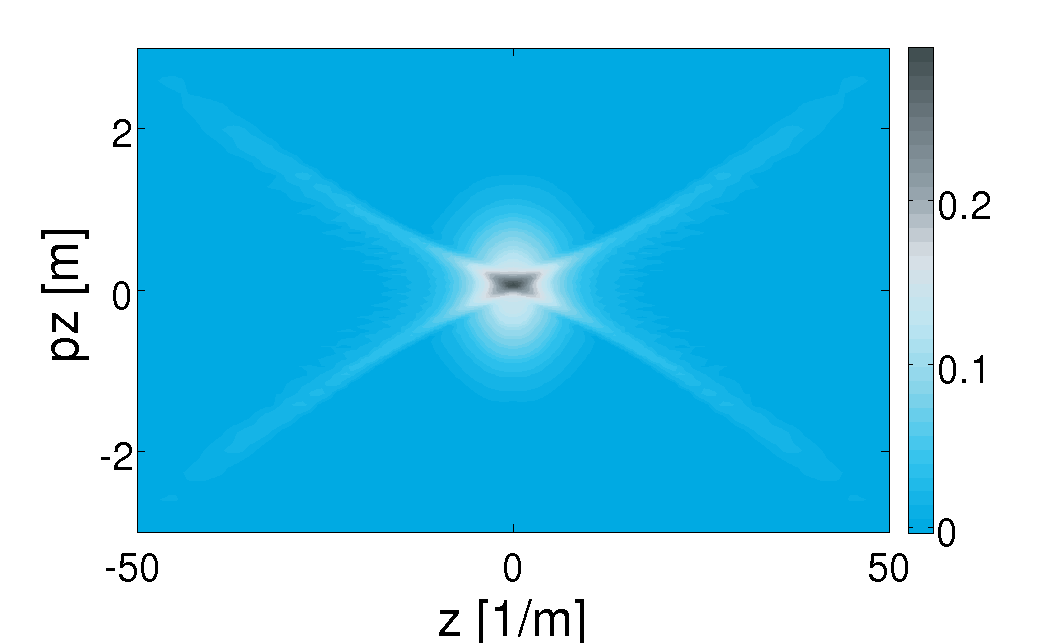} \\
 \includegraphics[width=0.5\textwidth]{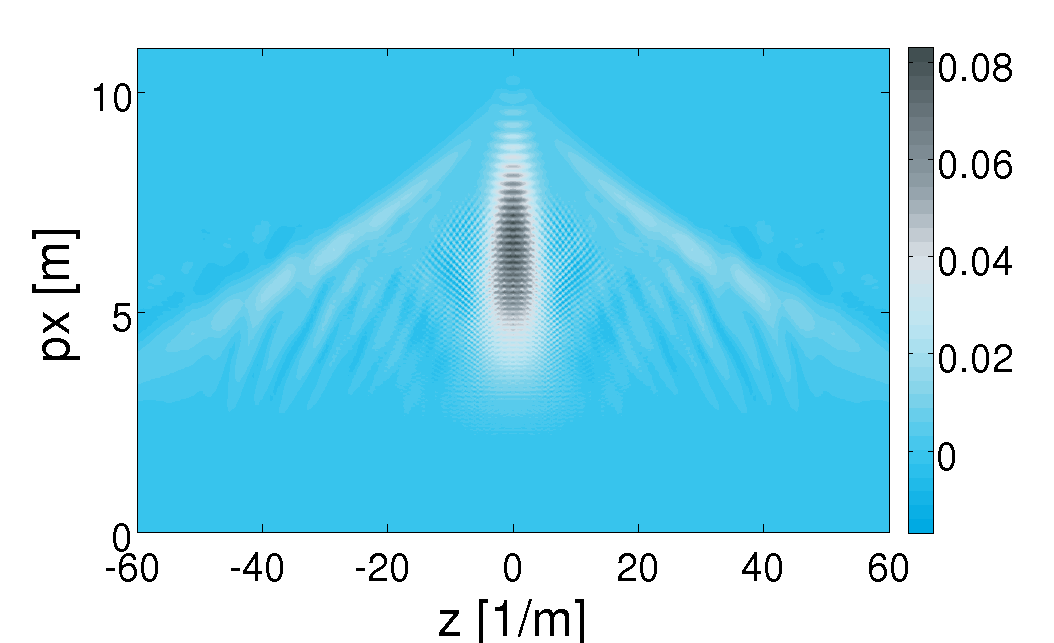}
 \includegraphics[width=0.5\textwidth]{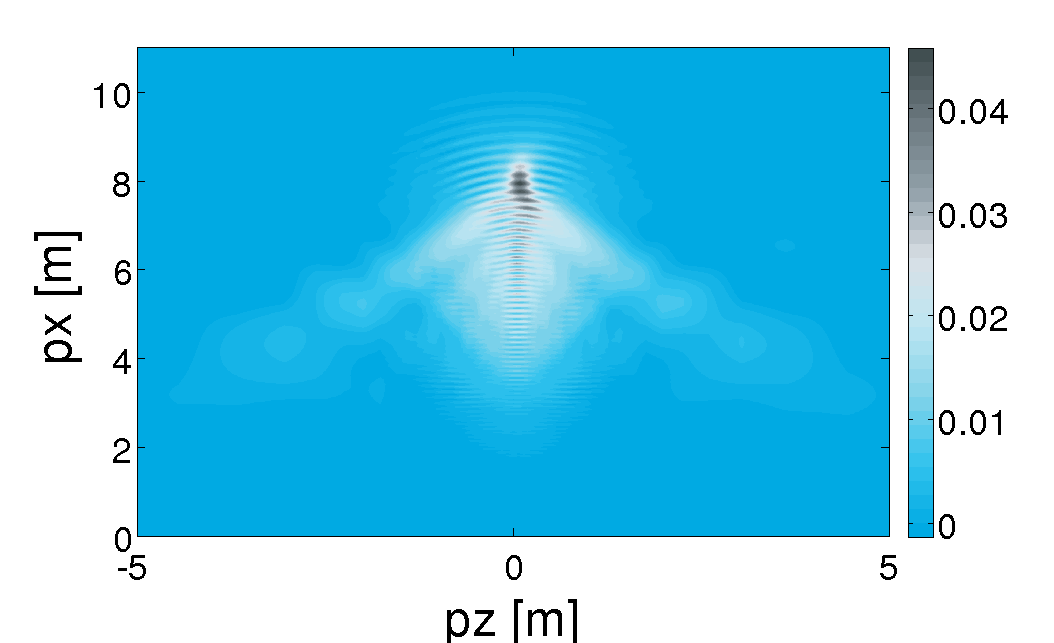}
 \includegraphics[width=0.5\textwidth]{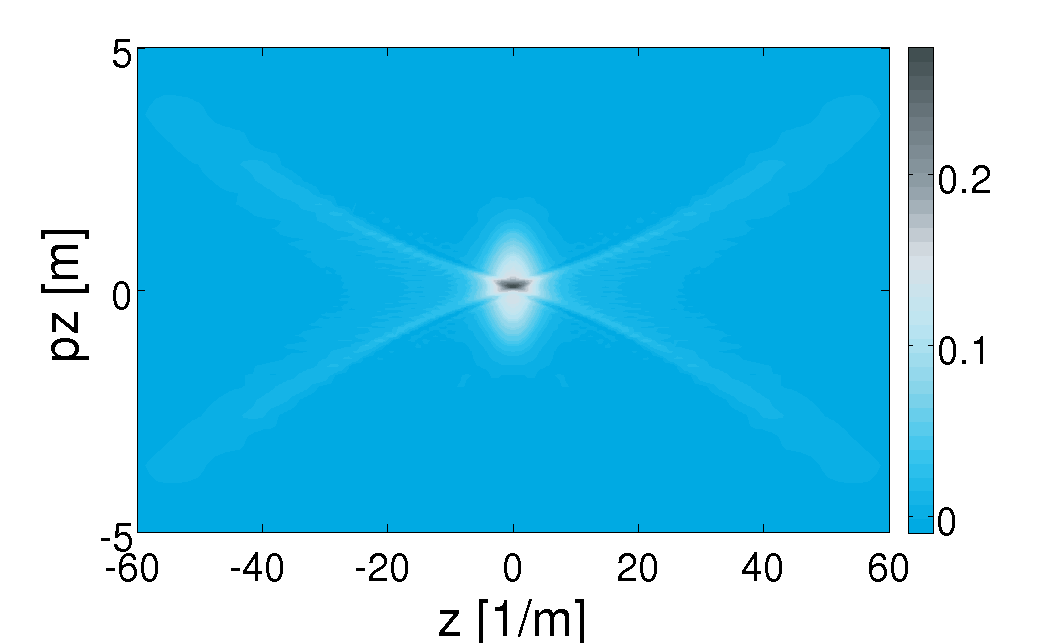} 
\caption{Electric field: $\mathbf{E} (z,t) = \varepsilon \ E_0 \ \left( \text{sech}^2 \left( \frac{t-\tau}{\tau} \right) - \text{sech}^2 \left( \frac{t+\tau}{\tau} \right) \right) \exp \br{-\frac{z^2}{2 \lambda^2}} \ \boldsymbol{e}_x$. Magnetic field: $B \br{z,t} = -\varepsilon \ E_0 \ \tau \left( \tanh \left( \frac{t+\tau}{\tau} \right) - \tanh \left( \frac{t-\tau}{\tau} \right) \right) \exp \br{-\frac{z^2}{2 \lambda^2}} \ \frac{z}{\lambda^2}$. Parameters: Tab. \ref{Tab_LongB}}
\label{FigApp_8}
\end{figure} 
\end{center}

\begin{center}
\begin{figure}[tbh]
 \includegraphics[width=0.5\textwidth]{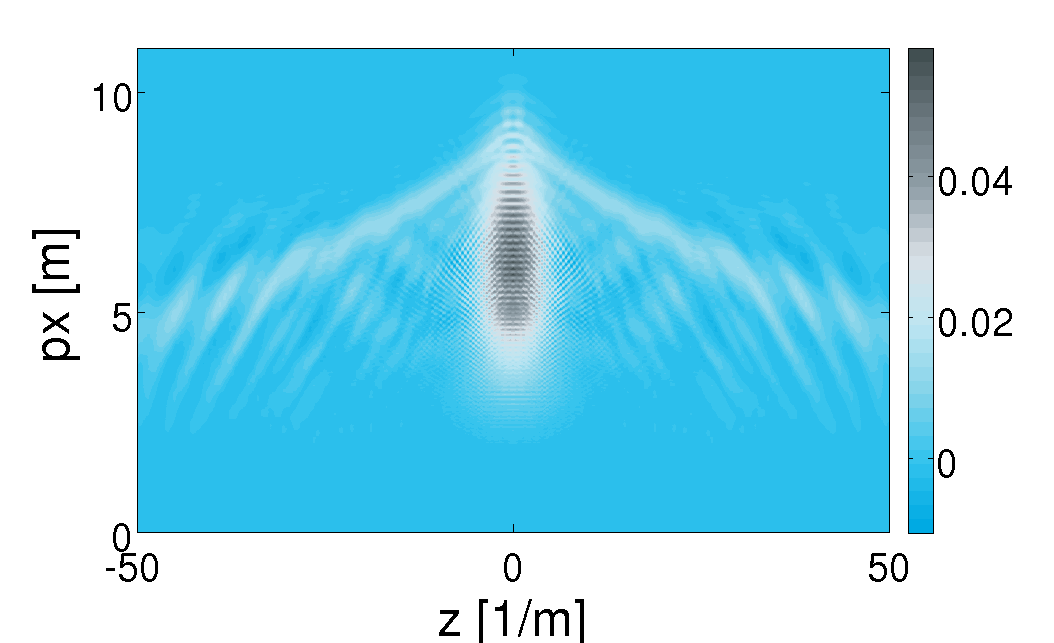}
 \includegraphics[width=0.5\textwidth]{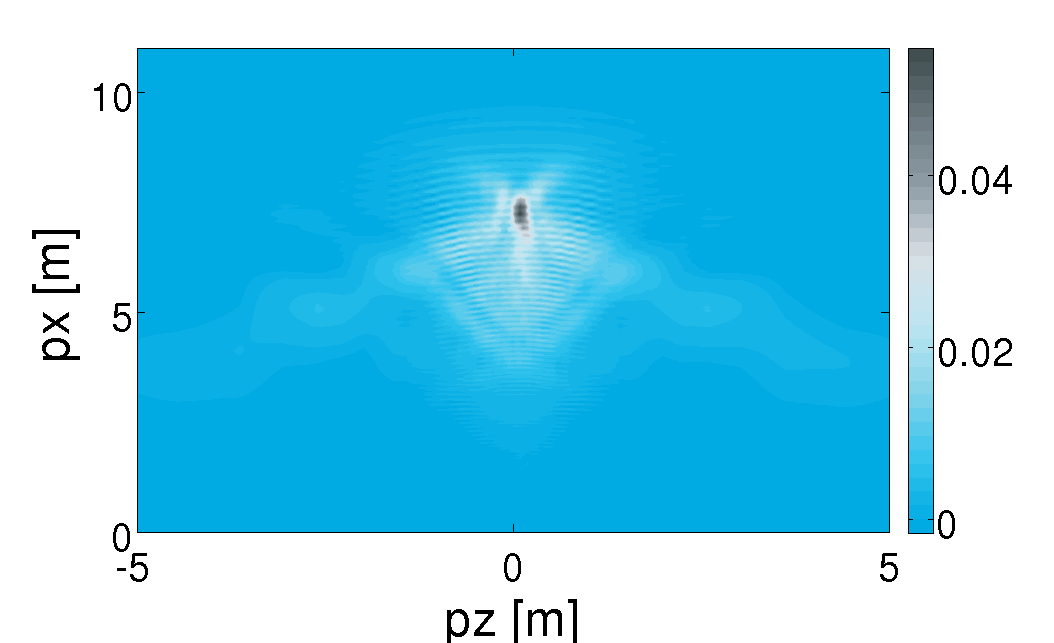}
 \includegraphics[width=0.5\textwidth]{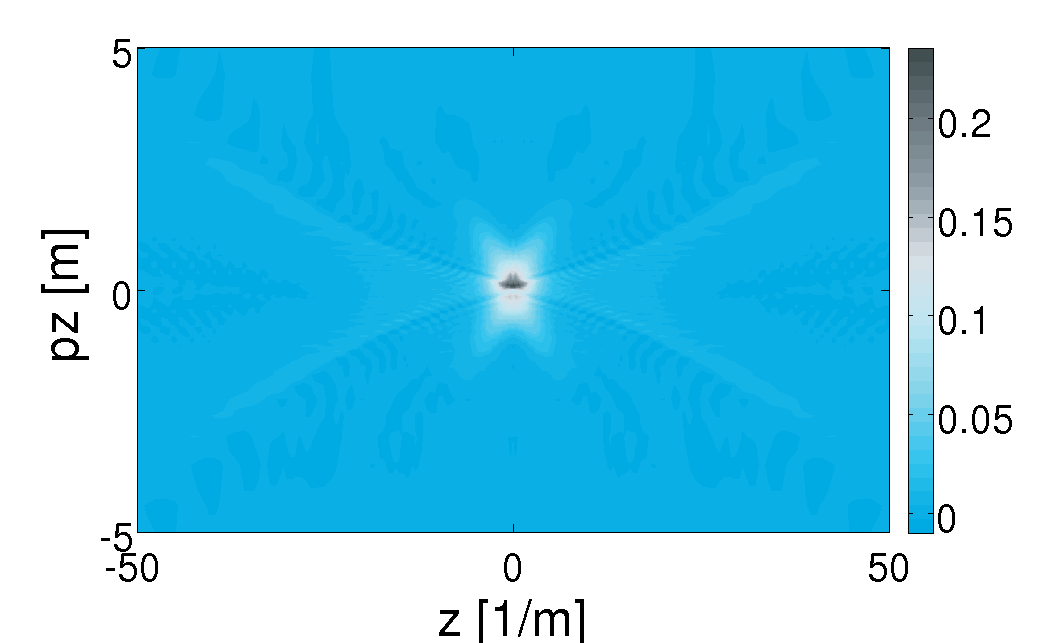} 
\caption{Electric field: $\mathbf{E} (z,t) = \varepsilon \ E_0 \ \left( \text{sech}^2 \left( \frac{t-\tau}{\tau} \right) - \text{sech}^2 \left( \frac{t+\tau}{\tau} \right) \right) \exp \br{-\frac{z^2}{2 \lambda^2}} \ \boldsymbol{e}_x$. Magnetic field: $B \br{z,t} = -\varepsilon \ E_0 \ \tau \left( \tanh \left( \frac{t+\tau}{\tau} \right) - \tanh \left( \frac{t-\tau}{\tau} \right) \right) \exp \br{-\frac{z^2}{2 \lambda^2}} \ \frac{z}{\lambda^2}$. Tab. \ref{Tab_LongB}}
\label{FigApp_9}
\end{figure} 
\end{center}

\begin{center}
\begin{figure}[tbh]
 \includegraphics[width=0.5\textwidth]{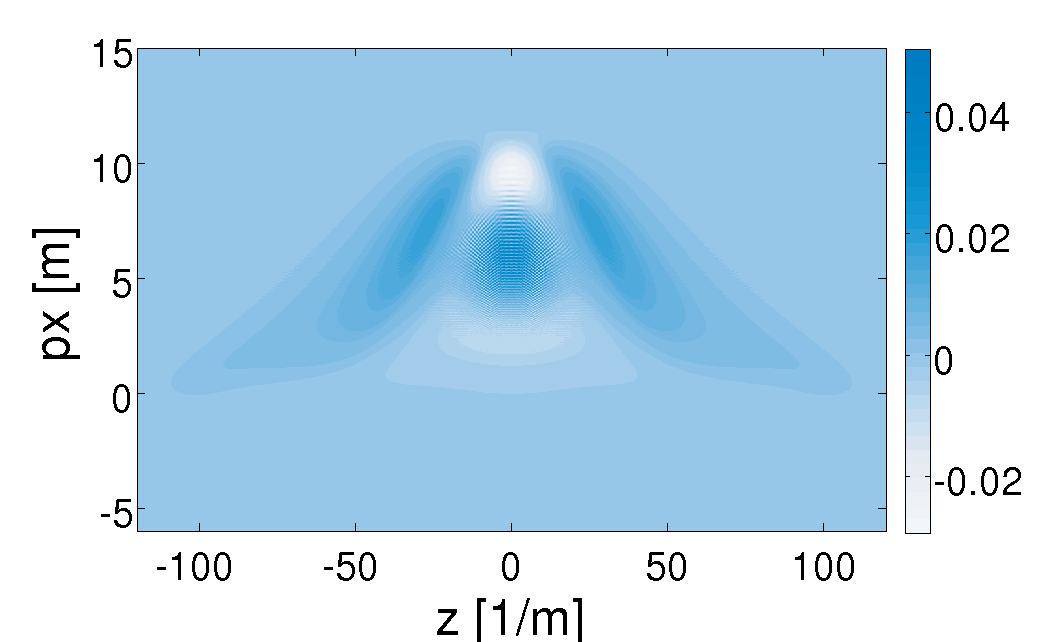}
 \includegraphics[width=0.5\textwidth]{./Fig/151/1b.png}
 \includegraphics[width=0.5\textwidth]{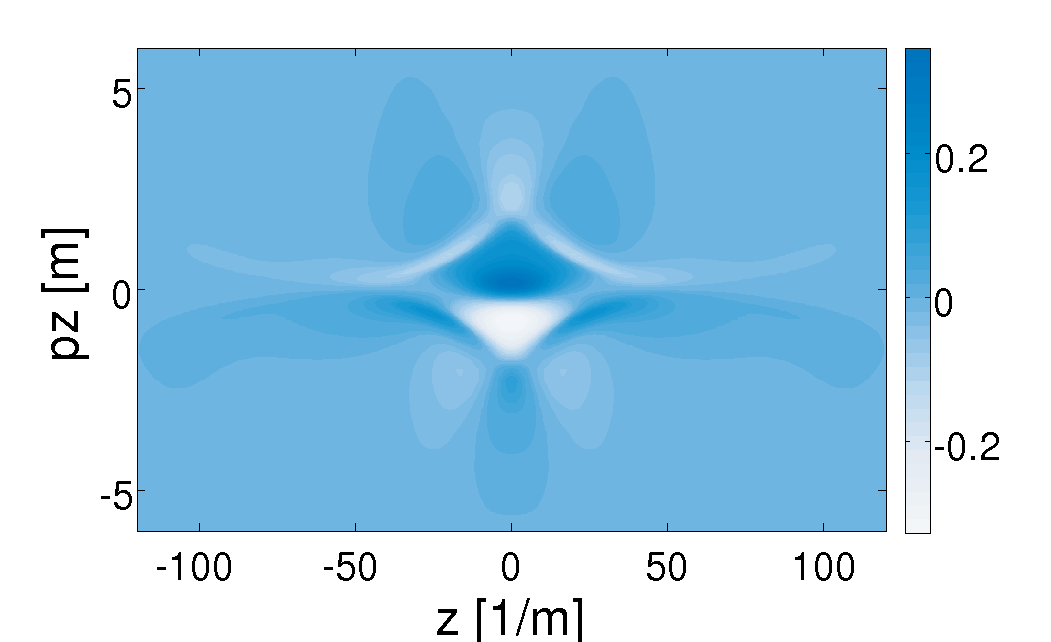}
\caption{Electric field: $\mathbf{E} (z,t) = \varepsilon \ E_0 \ \left( \text{sech}^2 \left( \frac{t-\tau}{\tau} \right) - \text{sech}^2 \left( \frac{t+\tau}{\tau} \right) \right) \exp \br{-\frac{z^2}{2 \lambda^2}} \ \boldsymbol{e}_x$. Magnetic field: $B = 0$. Parameters: Tab. \ref{Tab_Long}}
\label{FigApp_10}
\end{figure} 
\end{center}

\clearpage
\subsection{Time-antisymmetric magnetic field}

\begin{center}
\begin{figure}[tbh]
 \includegraphics[width=0.43\textwidth]{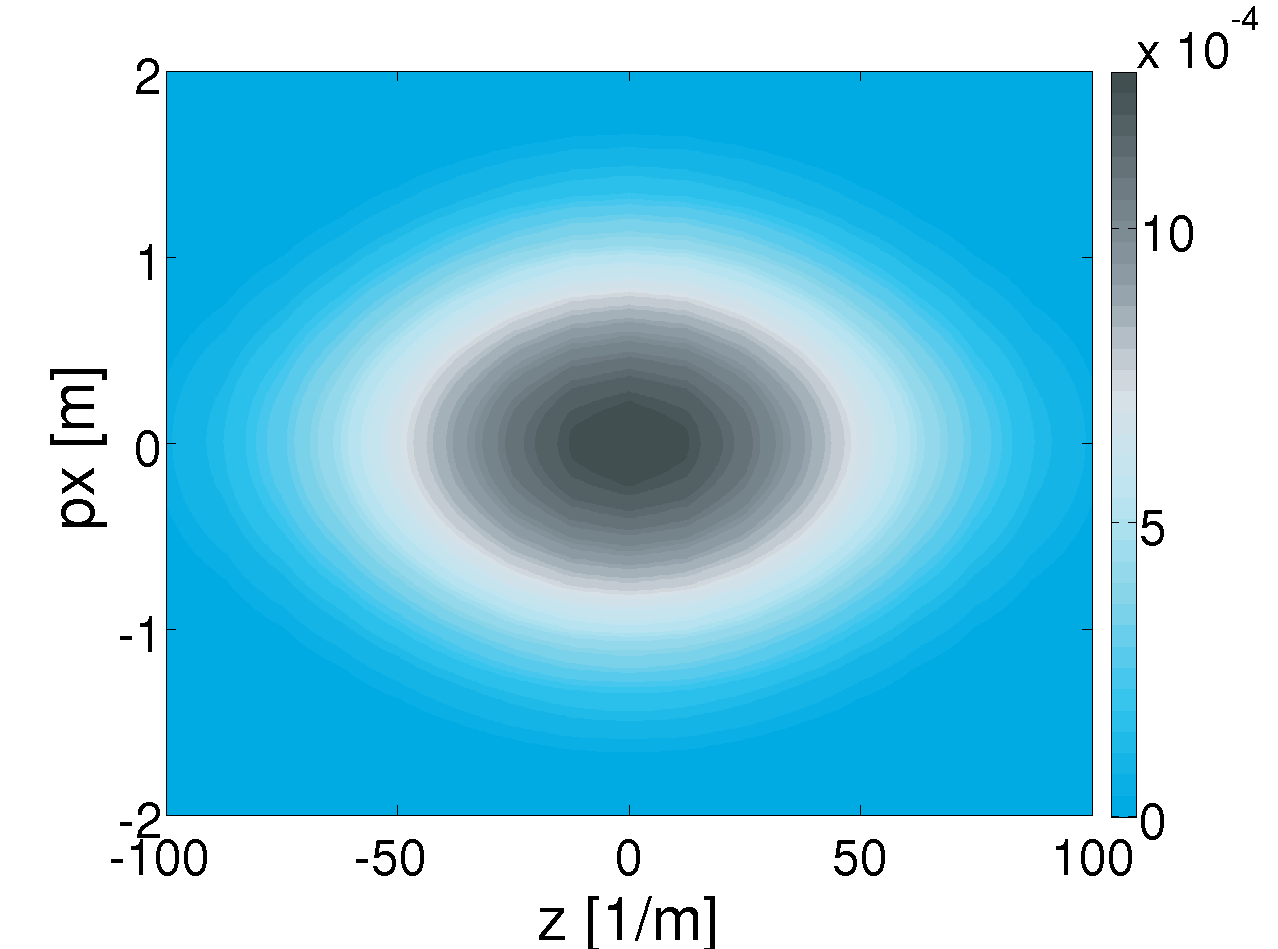}
 \includegraphics[width=0.43\textwidth]{./Fig/141/1.png}
 \includegraphics[width=0.43\textwidth]{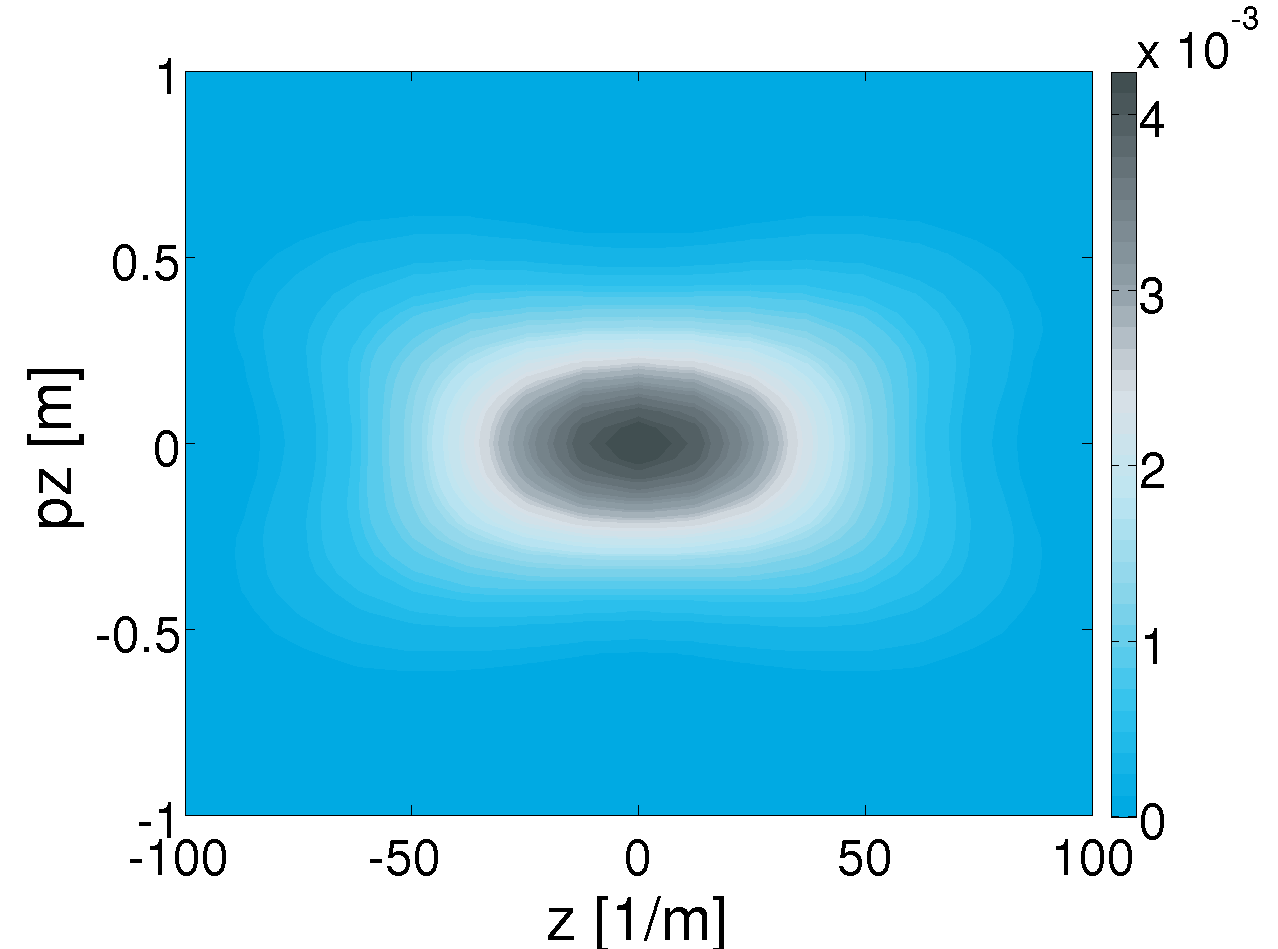}
\caption{Electric field: $\mathbf{E} \br{z,t} = \varepsilon E_0 \exp \br{-\frac{z^2}{2 \lambda^2}}
 \left( \text{sech}^2 \left( \frac{t}{\tau} \right) - \frac{1}{2} \text{sech}^2 \left( \frac{t-\tau}{\tau} \right) - \frac{1}{2} \text{sech}^2 \left( \frac{t+\tau}{\tau} \right) \right) \mathbf{e}_x$. Magnetic field: $B \br{z,t} = \varepsilon E_0 \tau \frac{z}{\lambda^2} \exp \br{-\frac{z^2}{2 \lambda^2}} \times \notag \left( \tanh \left( \frac{t}{\tau} \right) - \frac{1}{2} \tanh \left( \frac{t-\tau}{\tau} \right) - \frac{1}{2} \tanh \left( \frac{t+\tau}{\tau} \right) \right)$. Parameters: Tab. \ref{Tab_Triple}}
 \label{FigApp_11}
\end{figure}
\end{center}

\begin{center}
\begin{figure}[tbh]
 \includegraphics[width=0.43\textwidth]{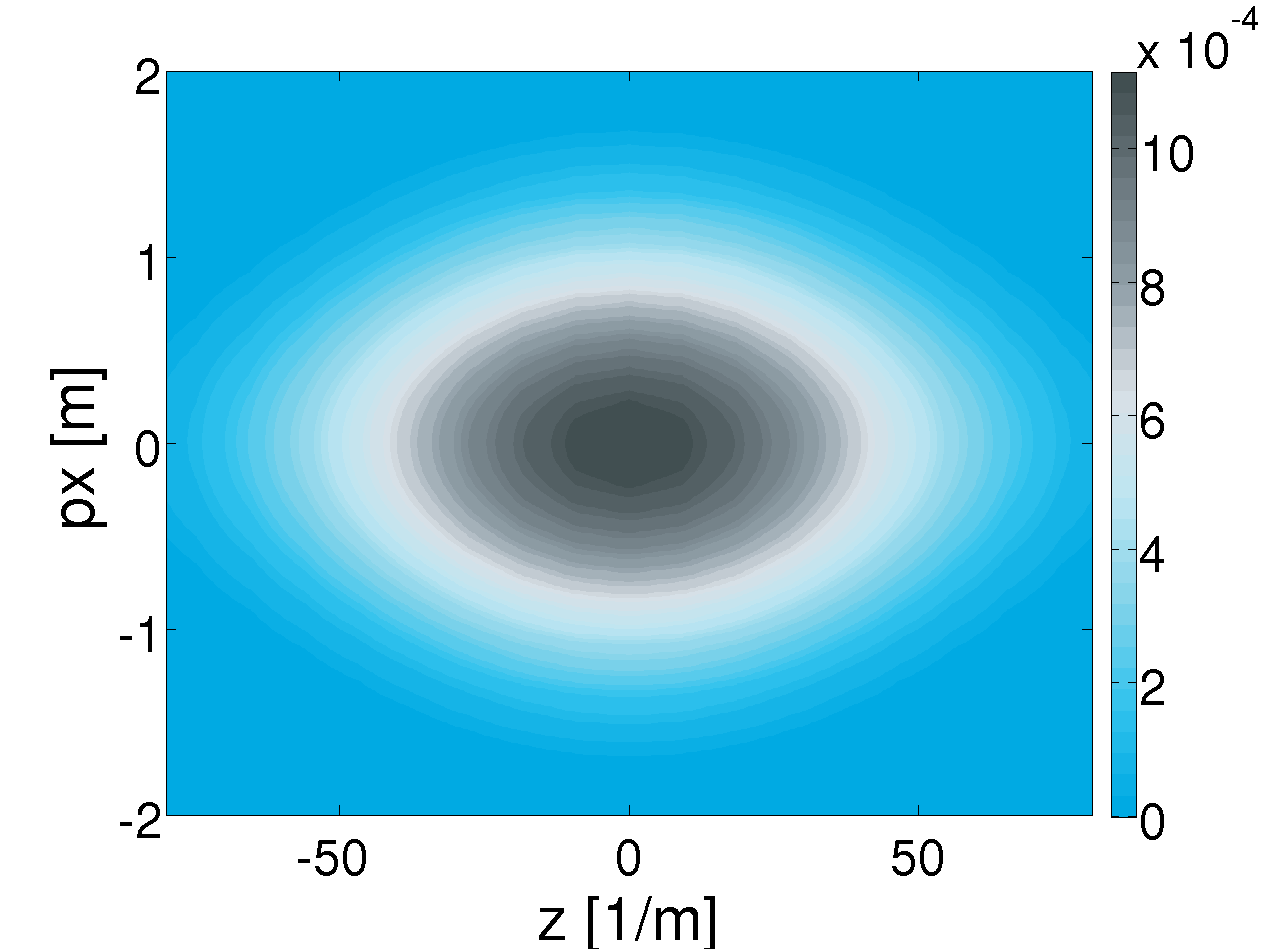}
 \includegraphics[width=0.43\textwidth]{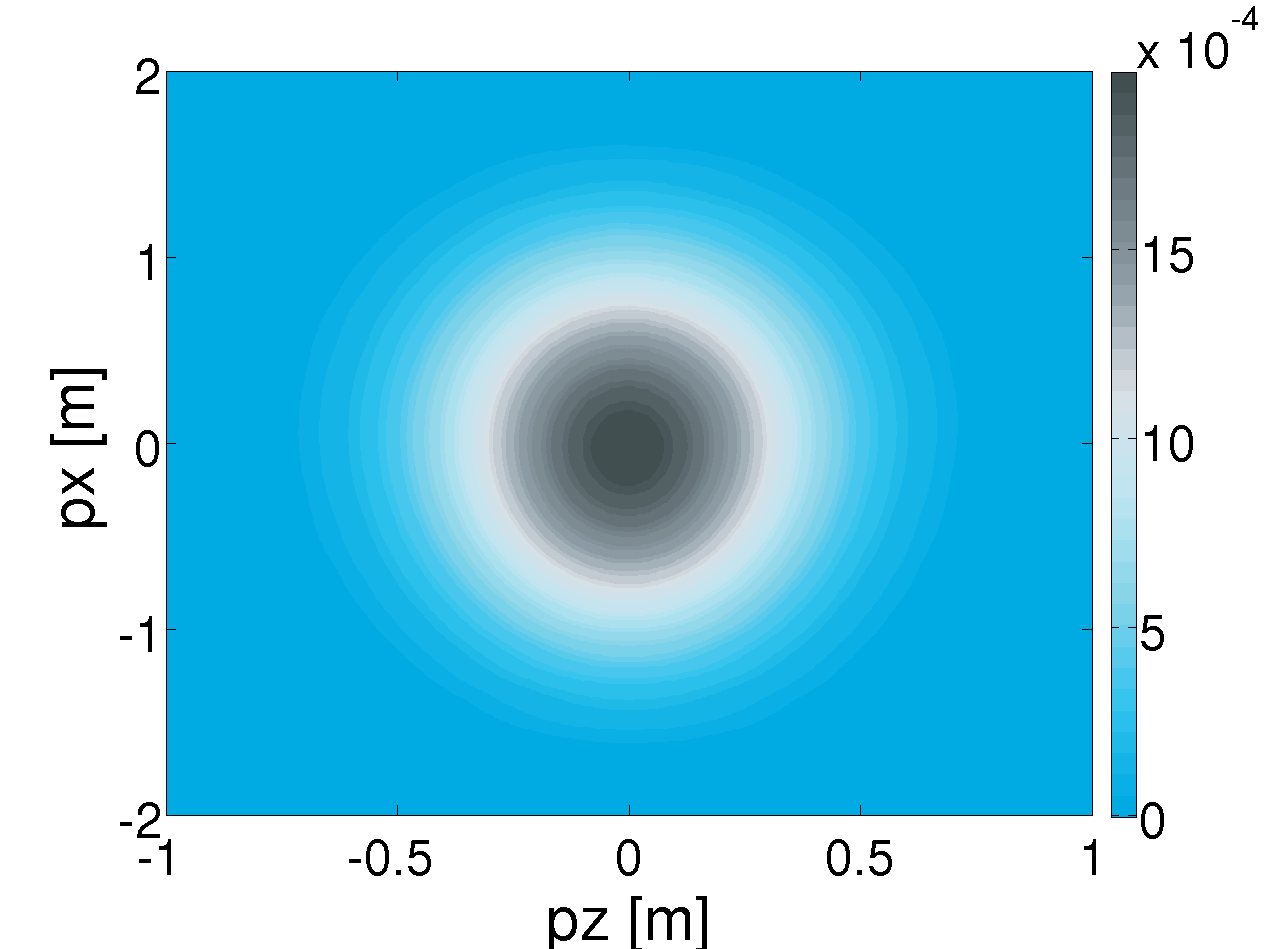}
 \includegraphics[width=0.43\textwidth]{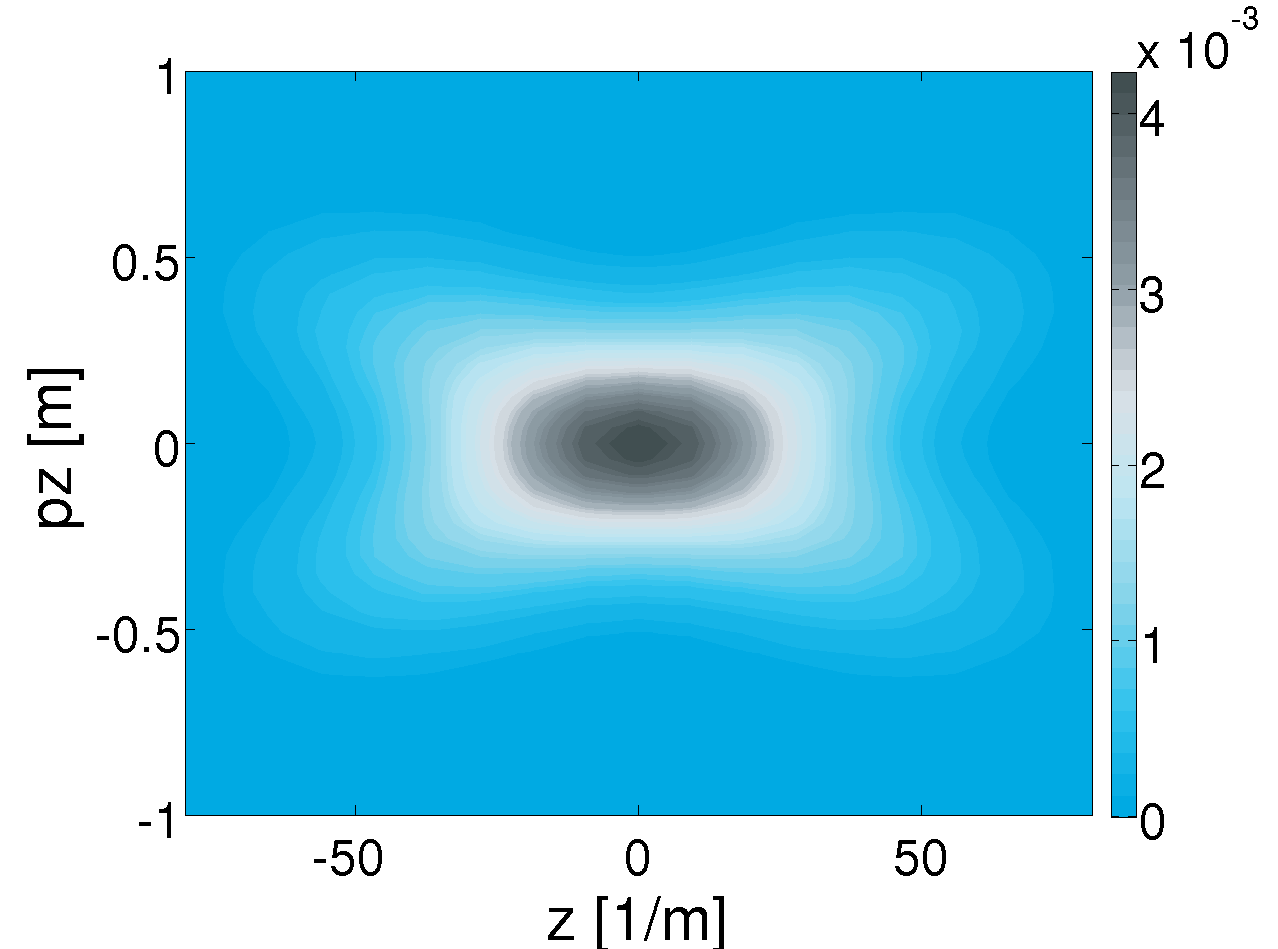} \\
 \includegraphics[width=0.43\textwidth]{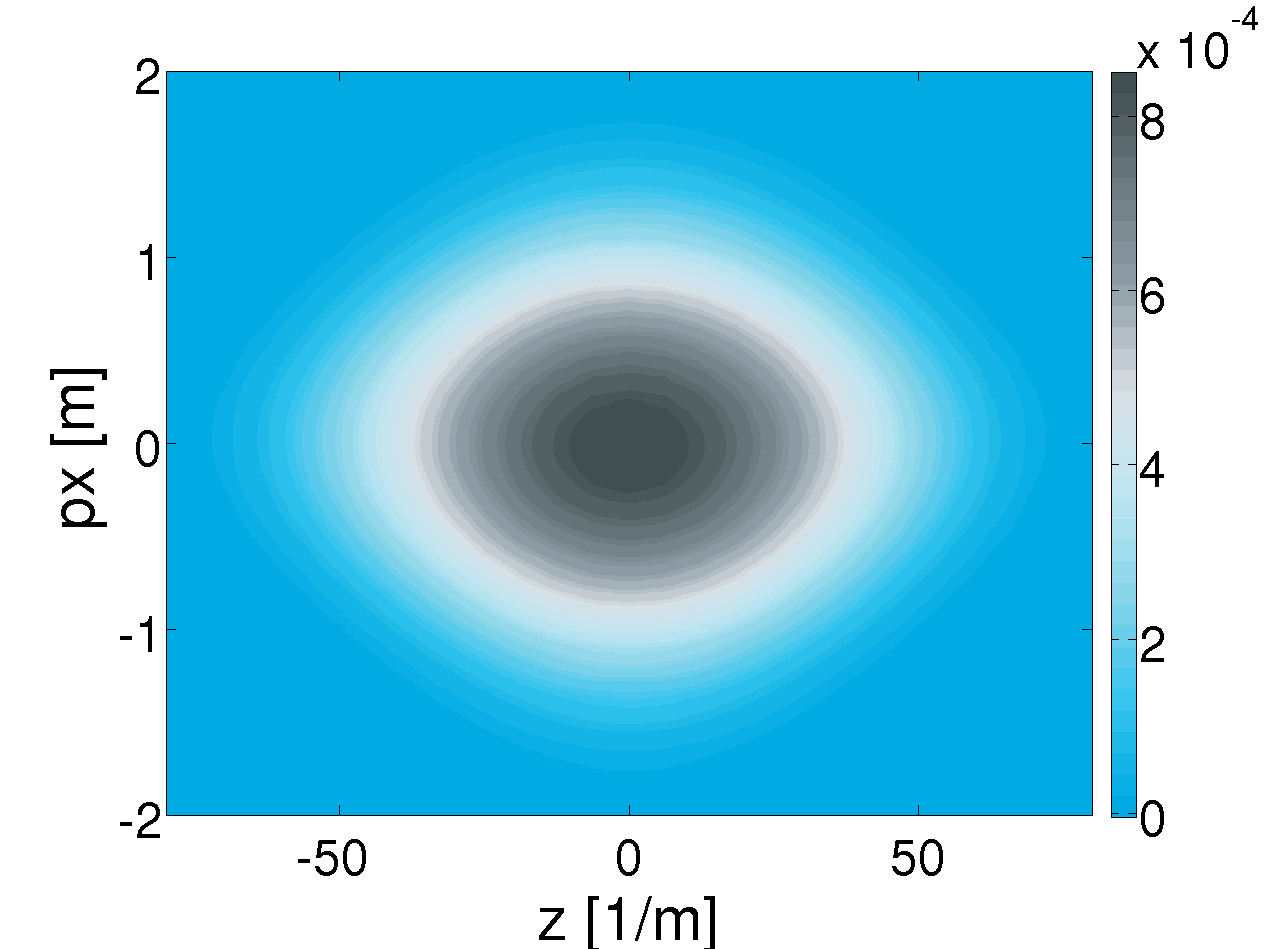}
 \includegraphics[width=0.43\textwidth]{./Fig/141/7.png}
 \includegraphics[width=0.43\textwidth]{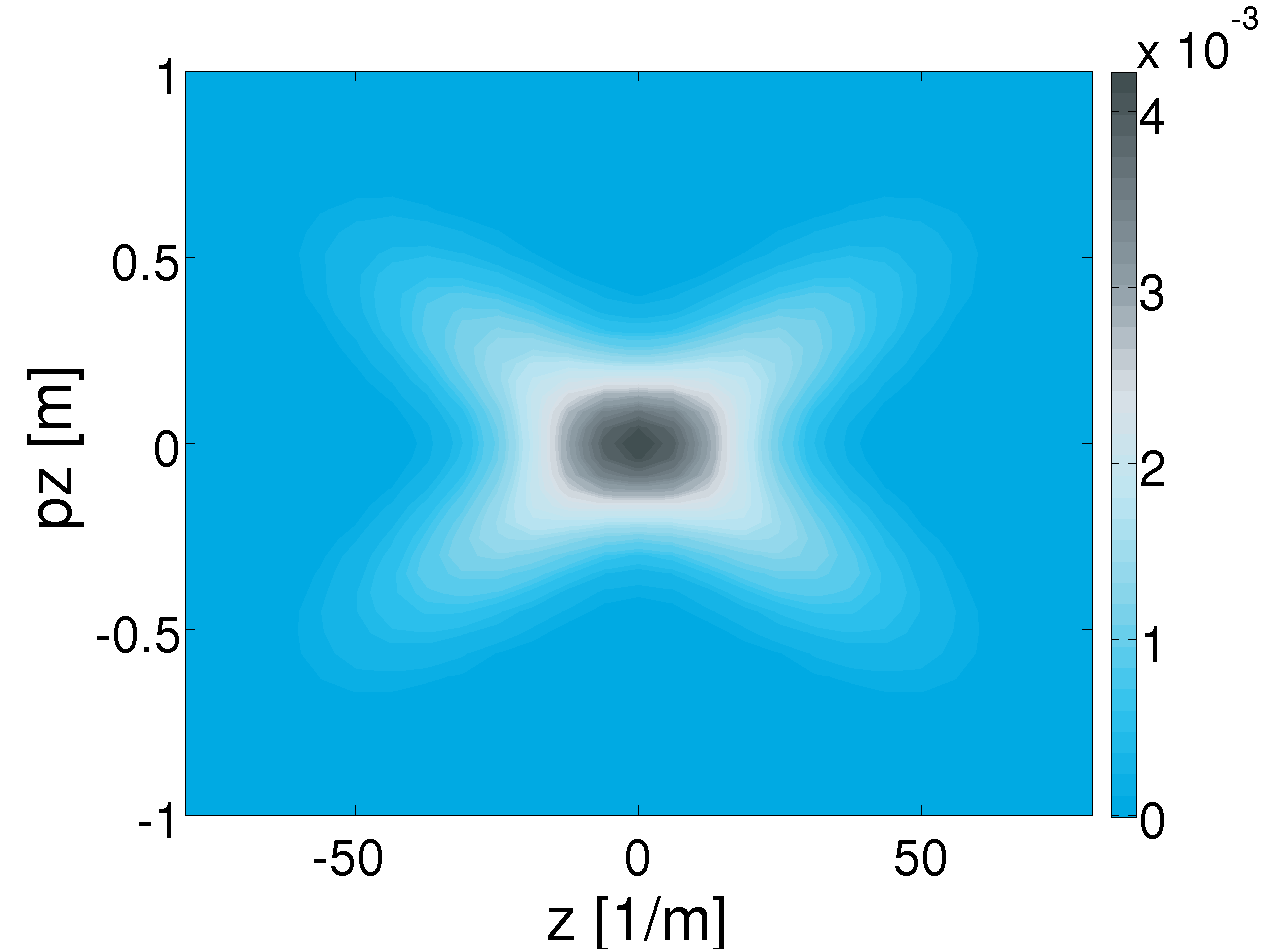} 
\caption{Electric field: $\mathbf{E} \br{z,t} = \varepsilon E_0 \exp \br{-\frac{z^2}{2 \lambda^2}}
 \left( \text{sech}^2 \left( \frac{t}{\tau} \right) - \frac{1}{2} \text{sech}^2 \left( \frac{t-\tau}{\tau} \right) - \frac{1}{2} \text{sech}^2 \left( \frac{t+\tau}{\tau} \right) \right) \mathbf{e}_x$. Magnetic field: $B \br{z,t} = \varepsilon E_0 \tau \frac{z}{\lambda^2} \exp \br{-\frac{z^2}{2 \lambda^2}} \times \notag \left( \tanh \left( \frac{t}{\tau} \right) - \frac{1}{2} \tanh \left( \frac{t-\tau}{\tau} \right) - \frac{1}{2} \tanh \left( \frac{t+\tau}{\tau} \right) \right)$.  Parameters: Tab. \ref{Tab_Triple}}
 \label{FigApp_12}
\end{figure}
\end{center}
 
\begin{center}
\begin{figure}[tbh]
 \includegraphics[width=0.43\textwidth]{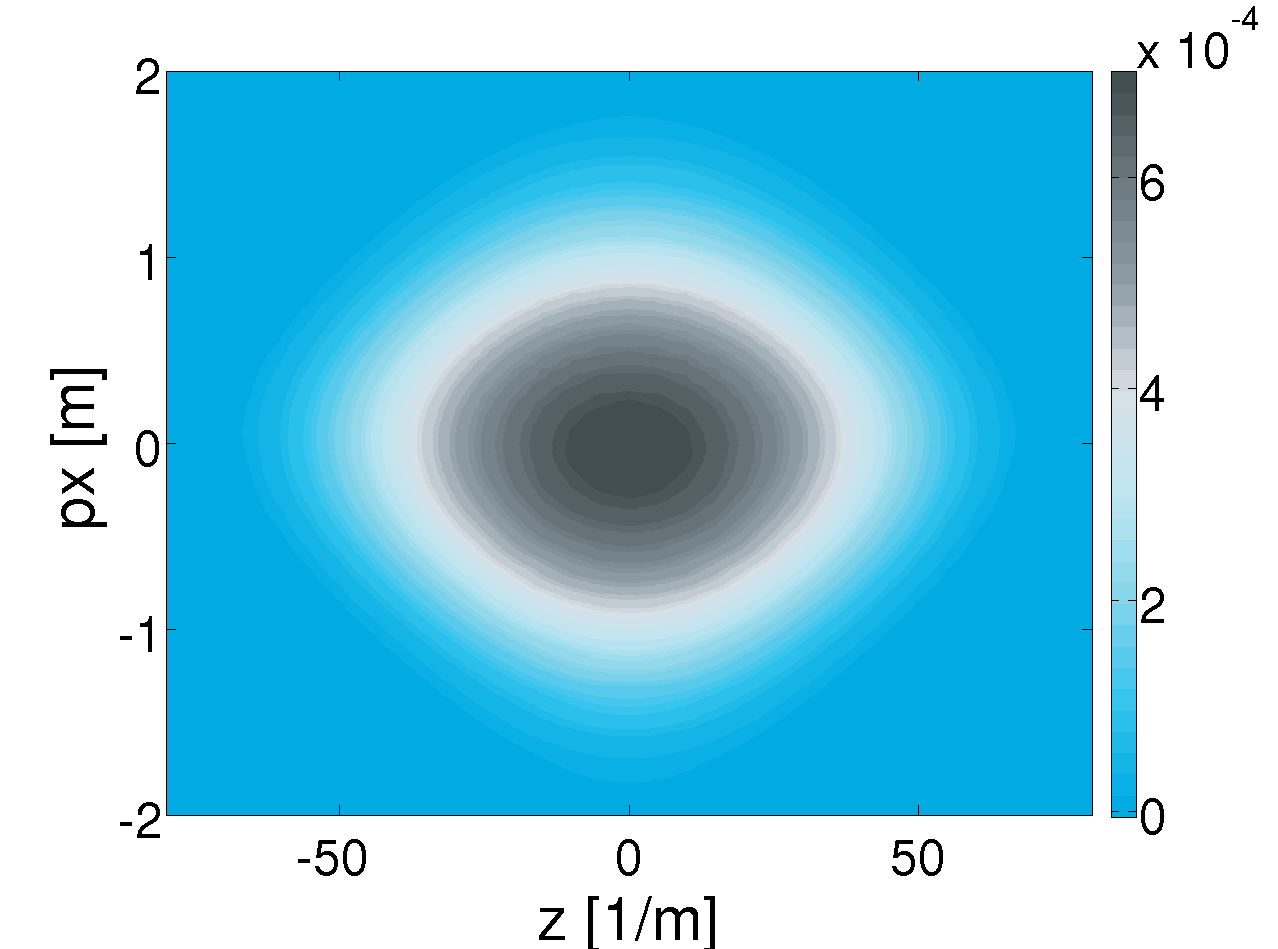}
 \includegraphics[width=0.43\textwidth]{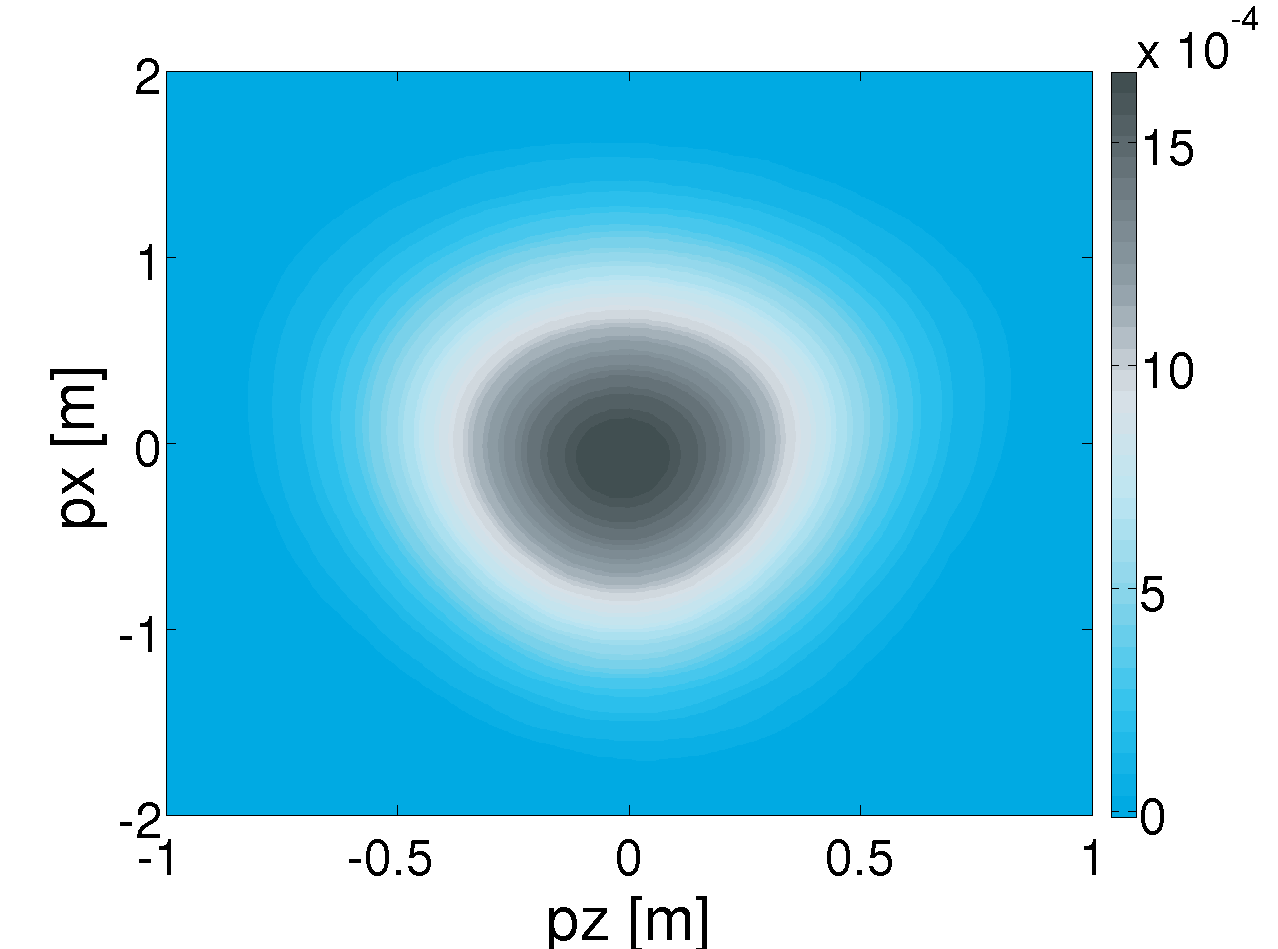}
 \includegraphics[width=0.43\textwidth]{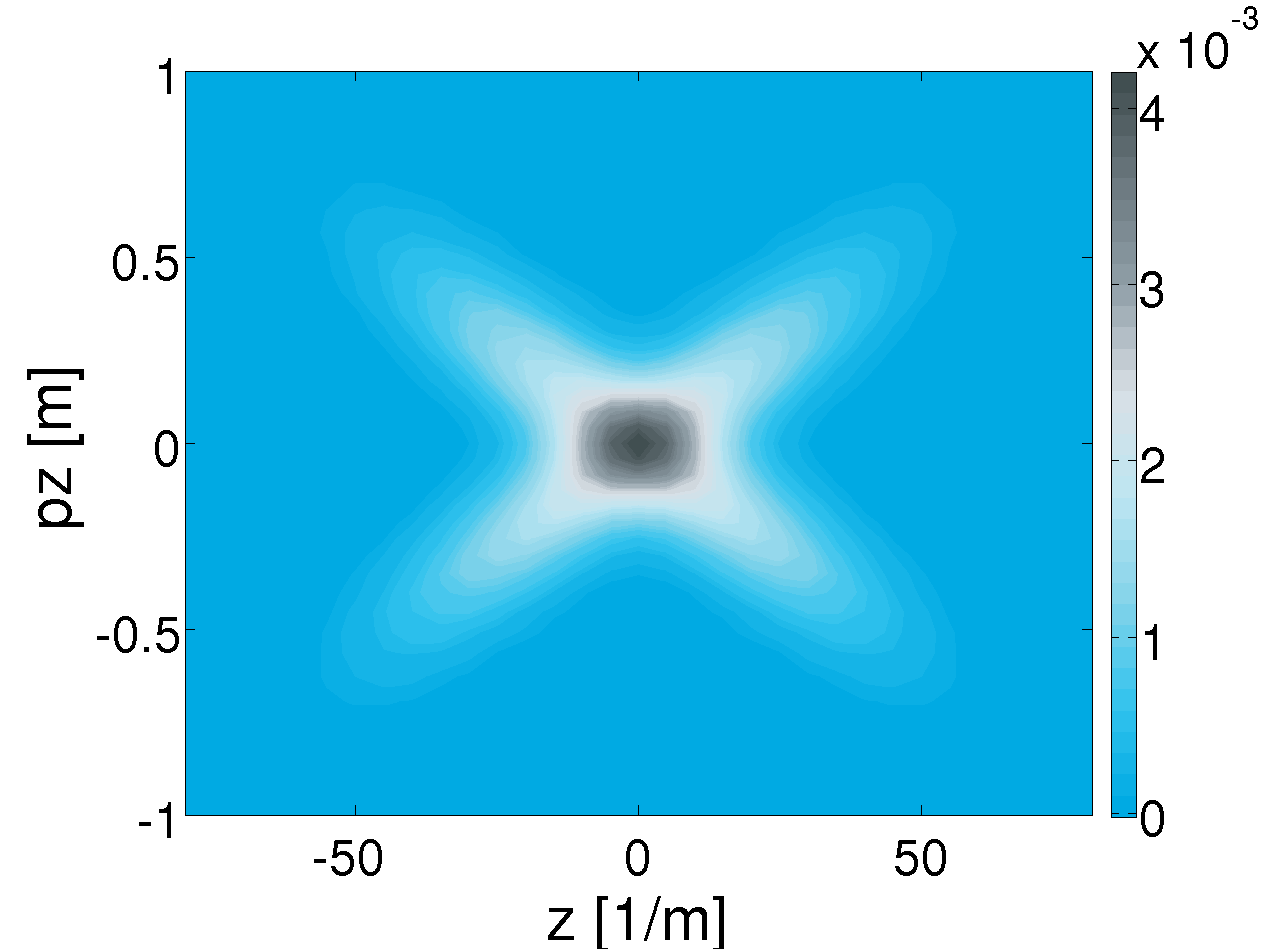} \\
 \includegraphics[width=0.43\textwidth]{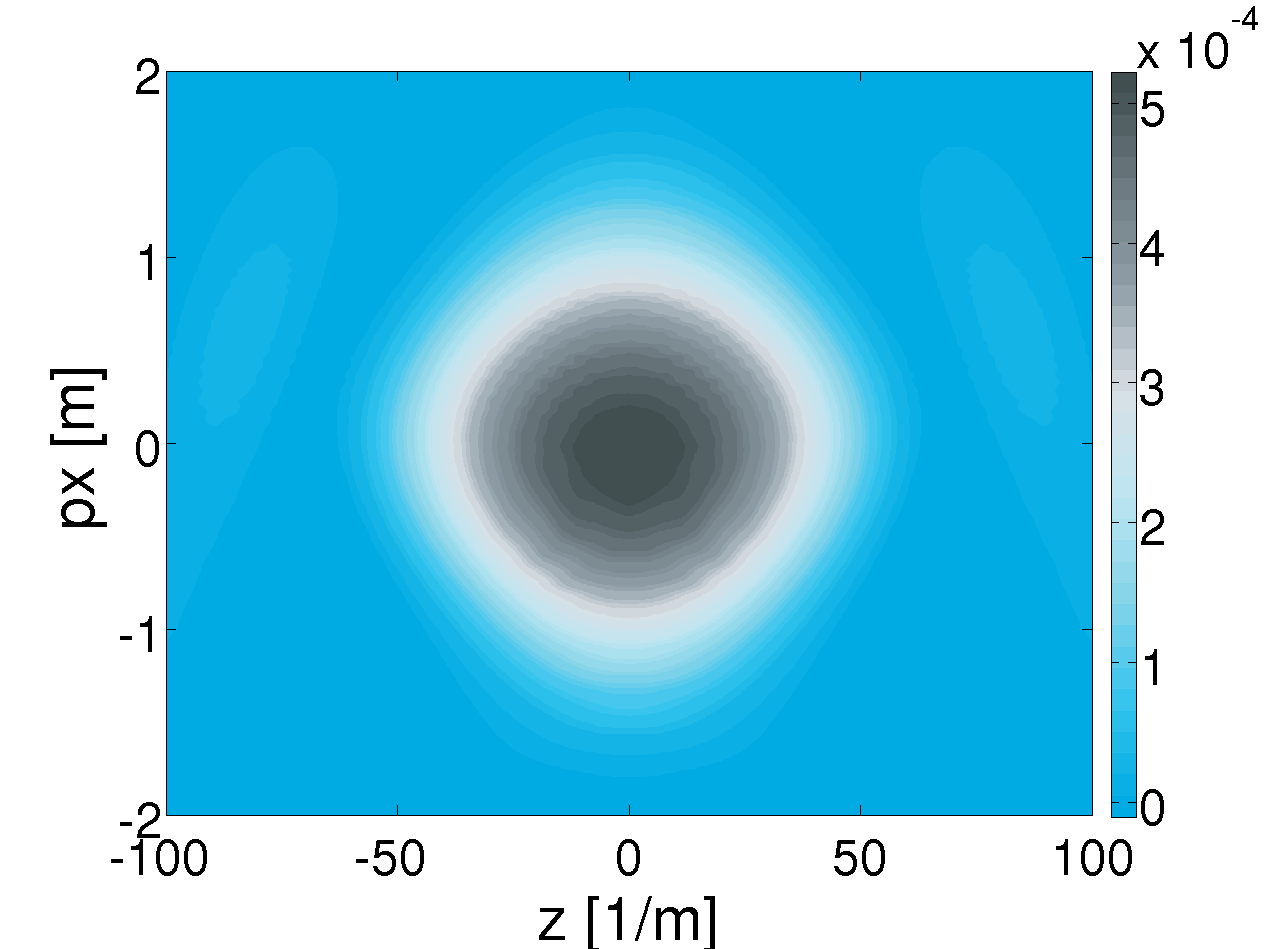}
 \includegraphics[width=0.43\textwidth]{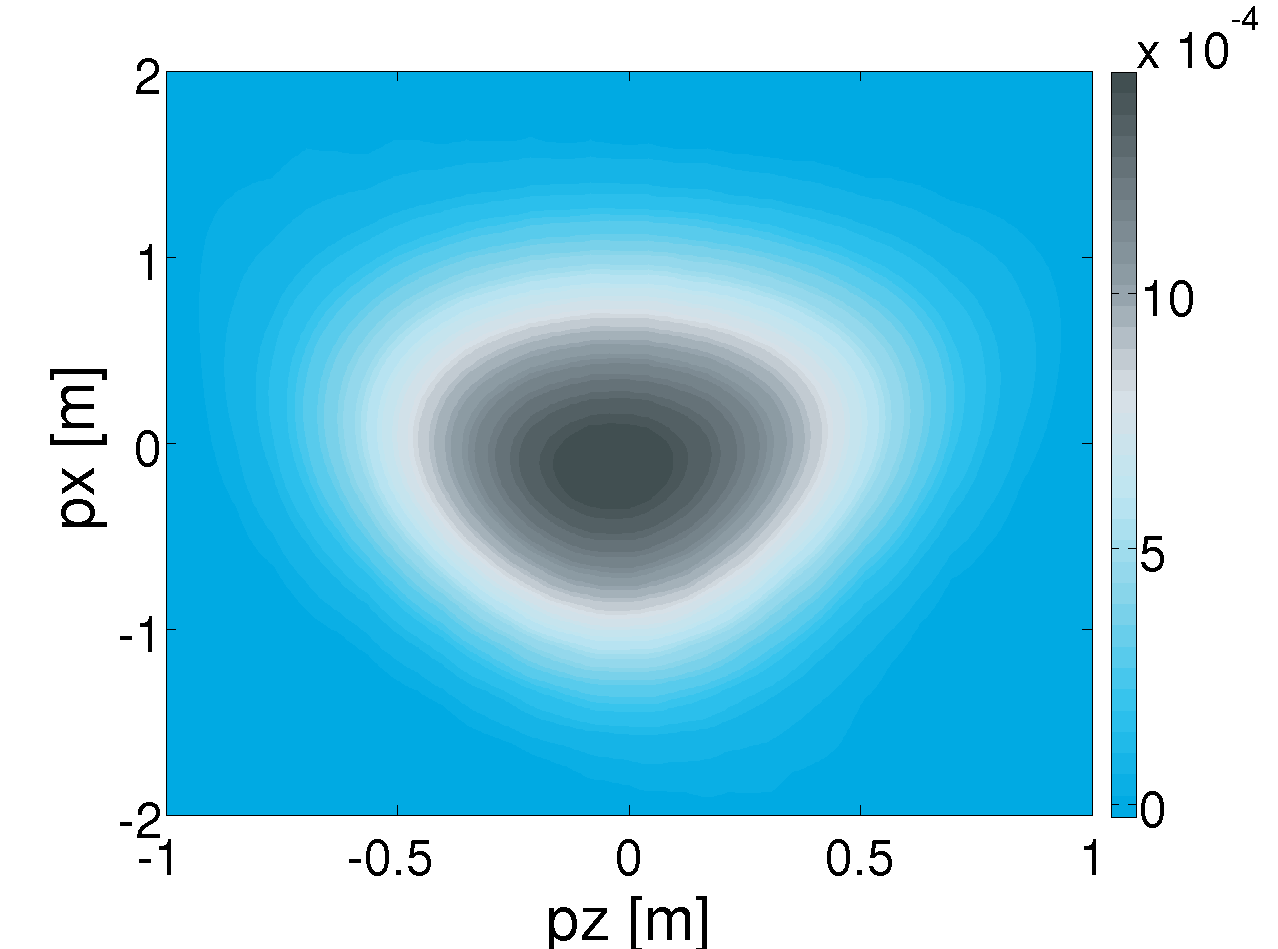}
 \includegraphics[width=0.43\textwidth]{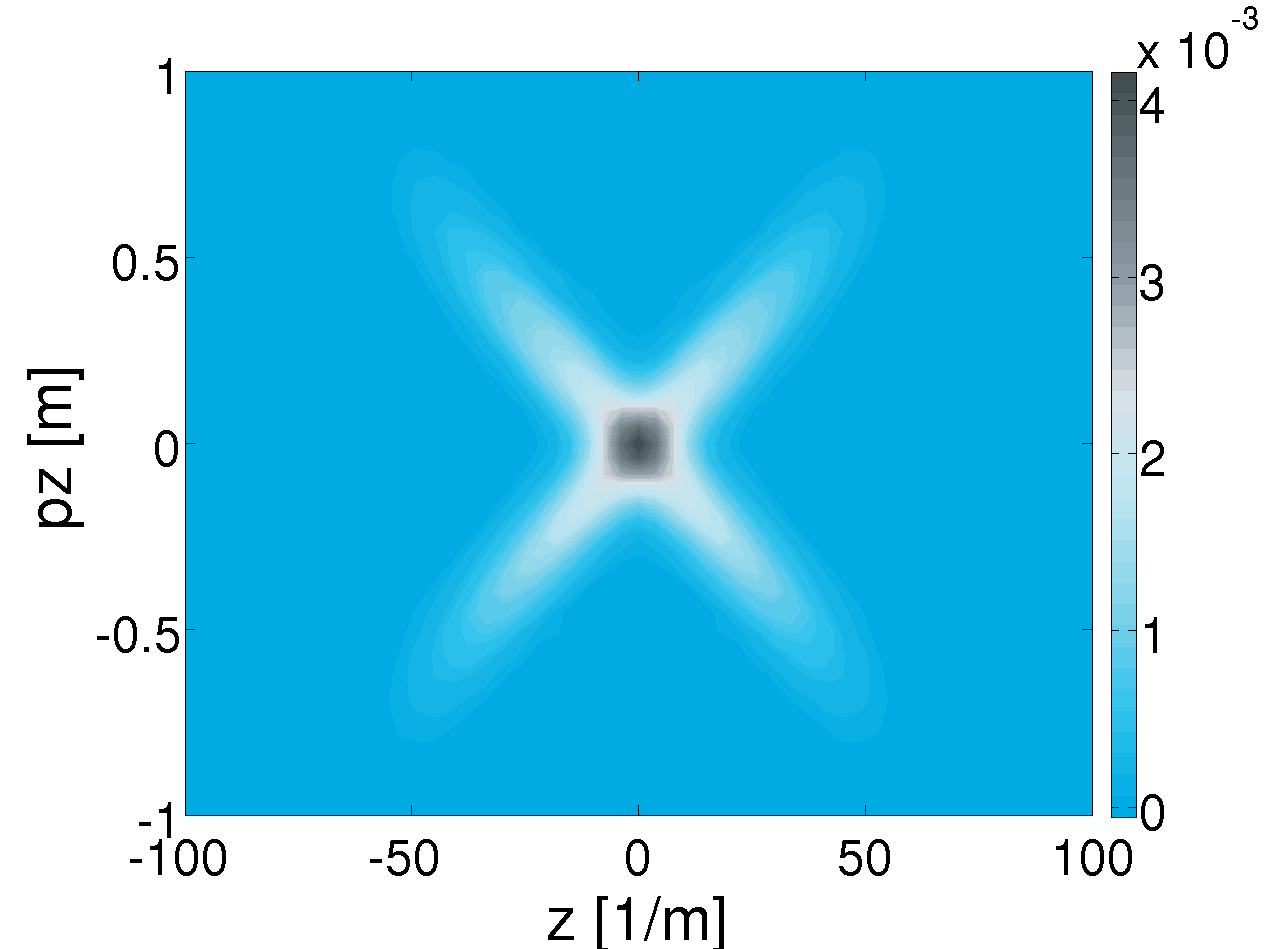} 
\caption{Electric field: $\mathbf{E} \br{z,t} = \varepsilon E_0 \exp \br{-\frac{z^2}{2 \lambda^2}}
 \left( \text{sech}^2 \left( \frac{t}{\tau} \right) - \frac{1}{2} \text{sech}^2 \left( \frac{t-\tau}{\tau} \right) - \frac{1}{2} \text{sech}^2 \left( \frac{t+\tau}{\tau} \right) \right) \mathbf{e}_x$. Magnetic field: $B \br{z,t} = \varepsilon E_0 \tau \frac{z}{\lambda^2} \exp \br{-\frac{z^2}{2 \lambda^2}} \times \notag \left( \tanh \left( \frac{t}{\tau} \right) - \frac{1}{2} \tanh \left( \frac{t-\tau}{\tau} \right) - \frac{1}{2} \tanh \left( \frac{t+\tau}{\tau} \right) \right)$.  Parameters: Tab. \ref{Tab_Triple}}
 \label{FigApp_13}
\end{figure} 
\end{center}

\begin{center}
\begin{figure}[tbh]
 \includegraphics[width=0.43\textwidth]{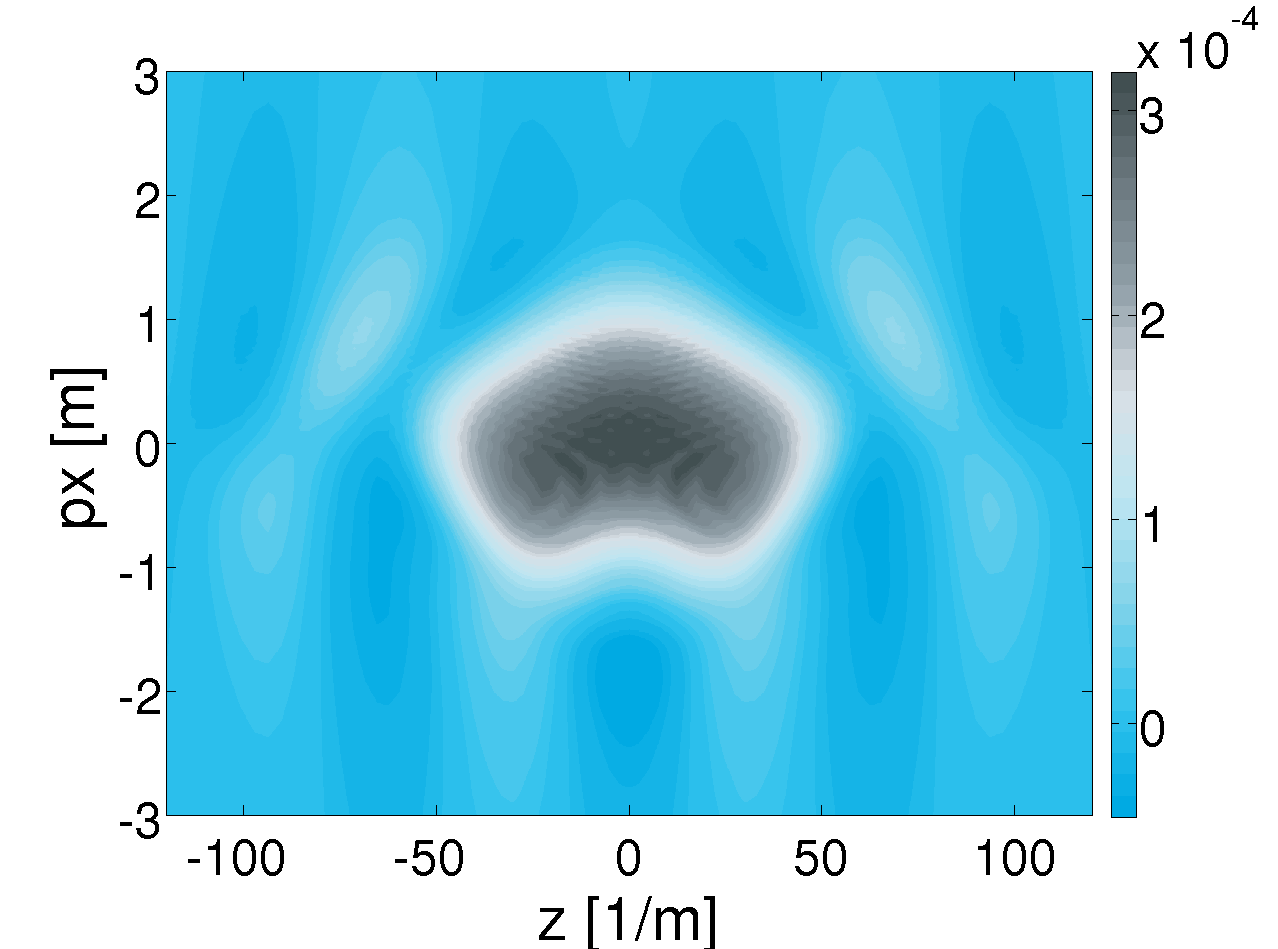}
 \includegraphics[width=0.43\textwidth]{./Fig/141/16.png}
 \includegraphics[width=0.43\textwidth]{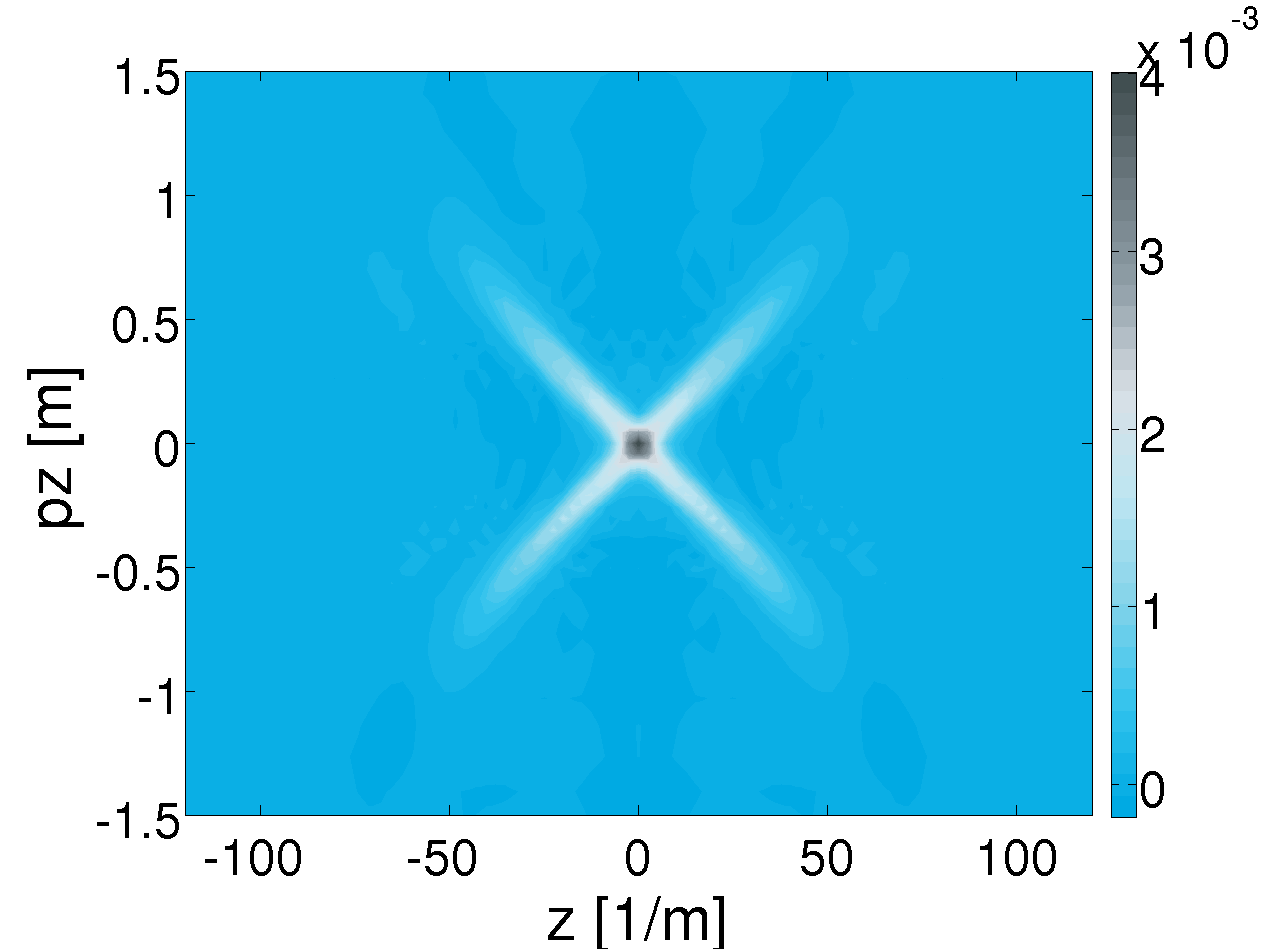} \\
 \includegraphics[width=0.43\textwidth]{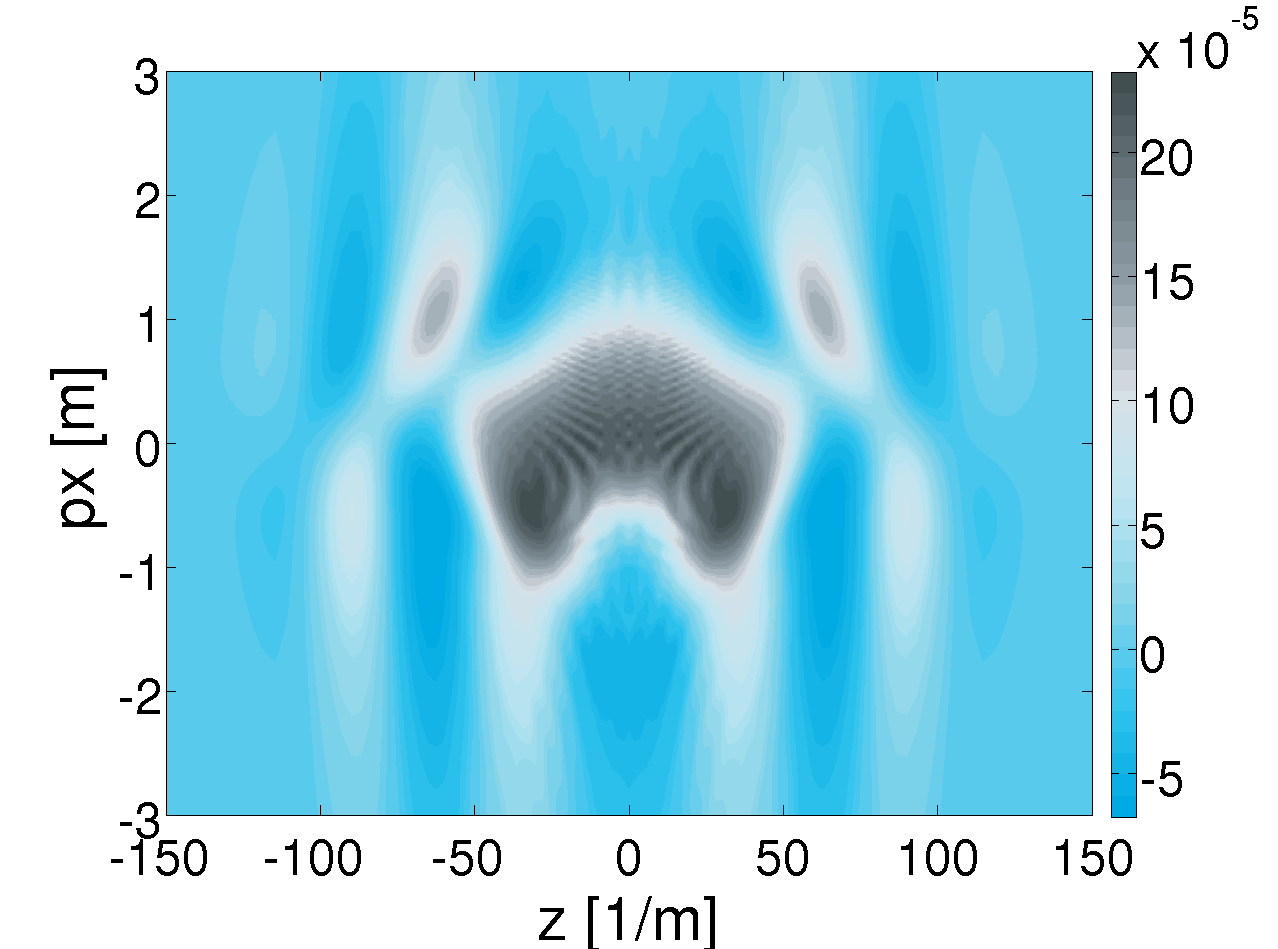}
 \includegraphics[width=0.43\textwidth]{./Fig/141/19.png}
 \includegraphics[width=0.43\textwidth]{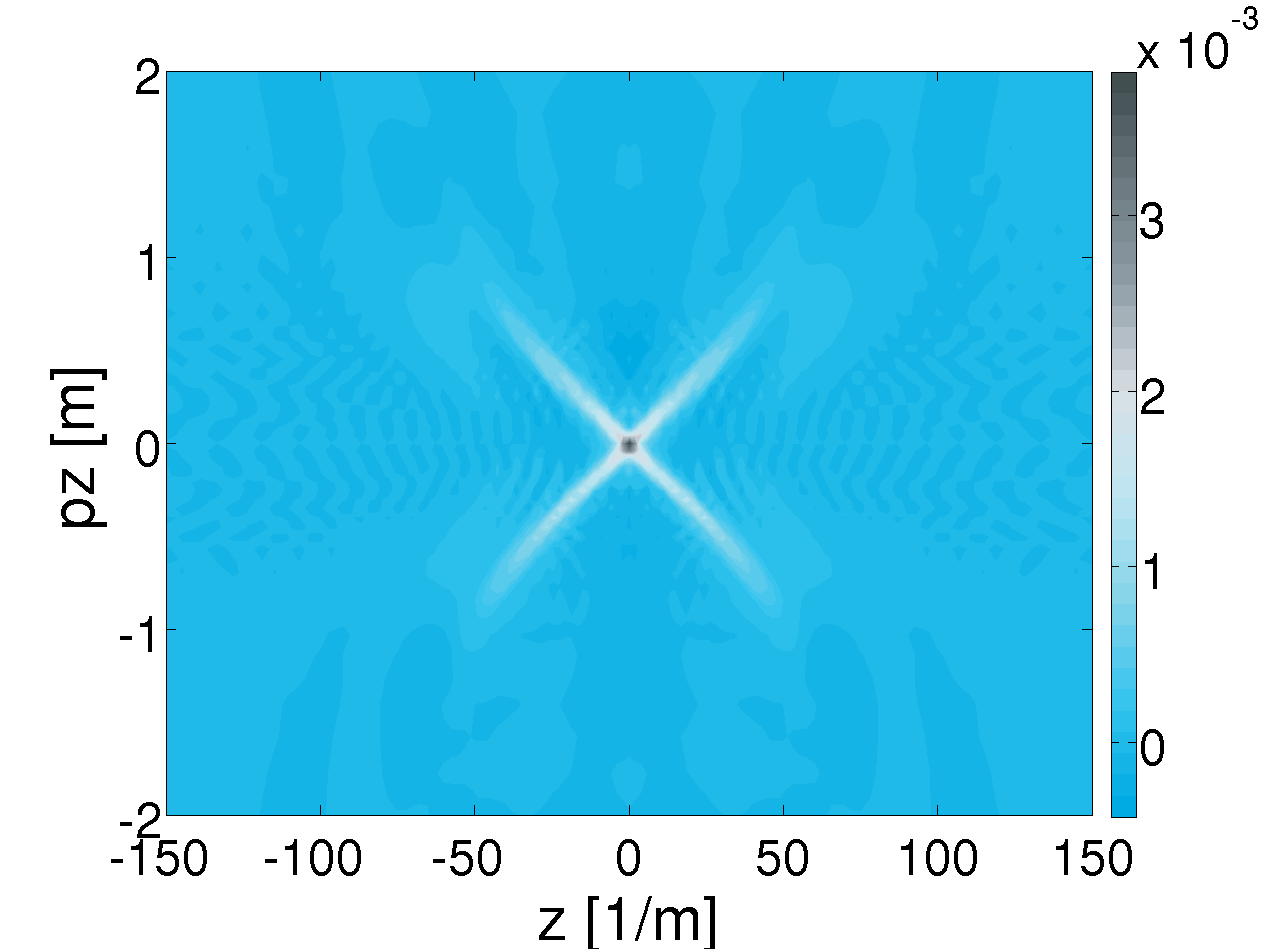} 
\caption{Electric field: $\mathbf{E} \br{z,t} = \varepsilon E_0 \exp \br{-\frac{z^2}{2 \lambda^2}}
 \left( \text{sech}^2 \left( \frac{t}{\tau} \right) - \frac{1}{2} \text{sech}^2 \left( \frac{t-\tau}{\tau} \right) - \frac{1}{2} \text{sech}^2 \left( \frac{t+\tau}{\tau} \right) \right) \mathbf{e}_x$. Magnetic field: $B \br{z,t} = \varepsilon E_0 \tau \frac{z}{\lambda^2} \exp \br{-\frac{z^2}{2 \lambda^2}} \times \notag \left( \tanh \left( \frac{t}{\tau} \right) - \frac{1}{2} \tanh \left( \frac{t-\tau}{\tau} \right) - \frac{1}{2} \tanh \left( \frac{t+\tau}{\tau} \right) \right)$.  Parameters: Tab. \ref{Tab_Triple}}
 \label{FigApp_14}
\end{figure}
\end{center}

\begin{center}
\begin{figure}[tbh]
 \includegraphics[width=0.43\textwidth]{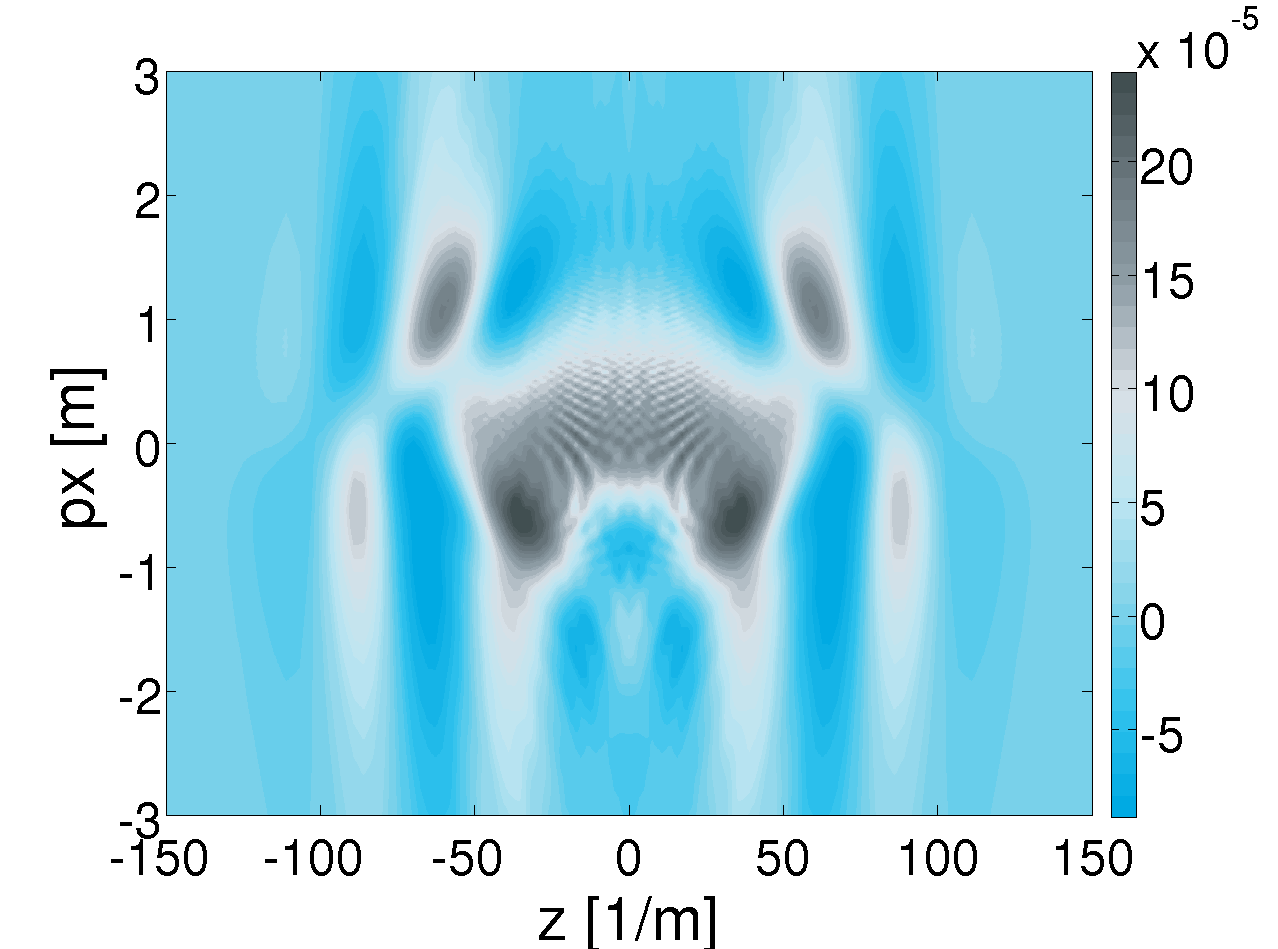}
 \includegraphics[width=0.43\textwidth]{./Fig/141/22.png}
 \includegraphics[width=0.43\textwidth]{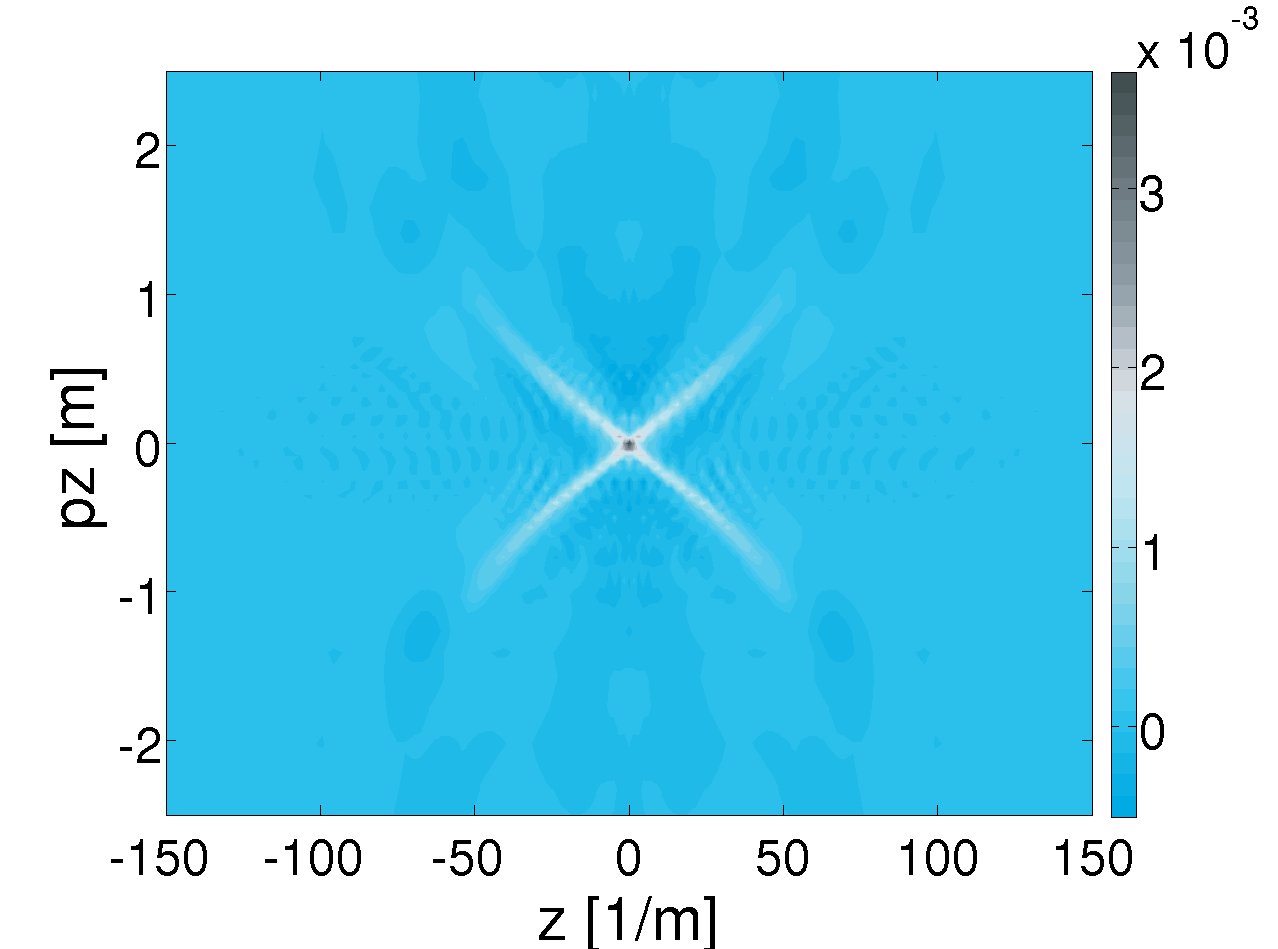} \\
 \includegraphics[width=0.43\textwidth]{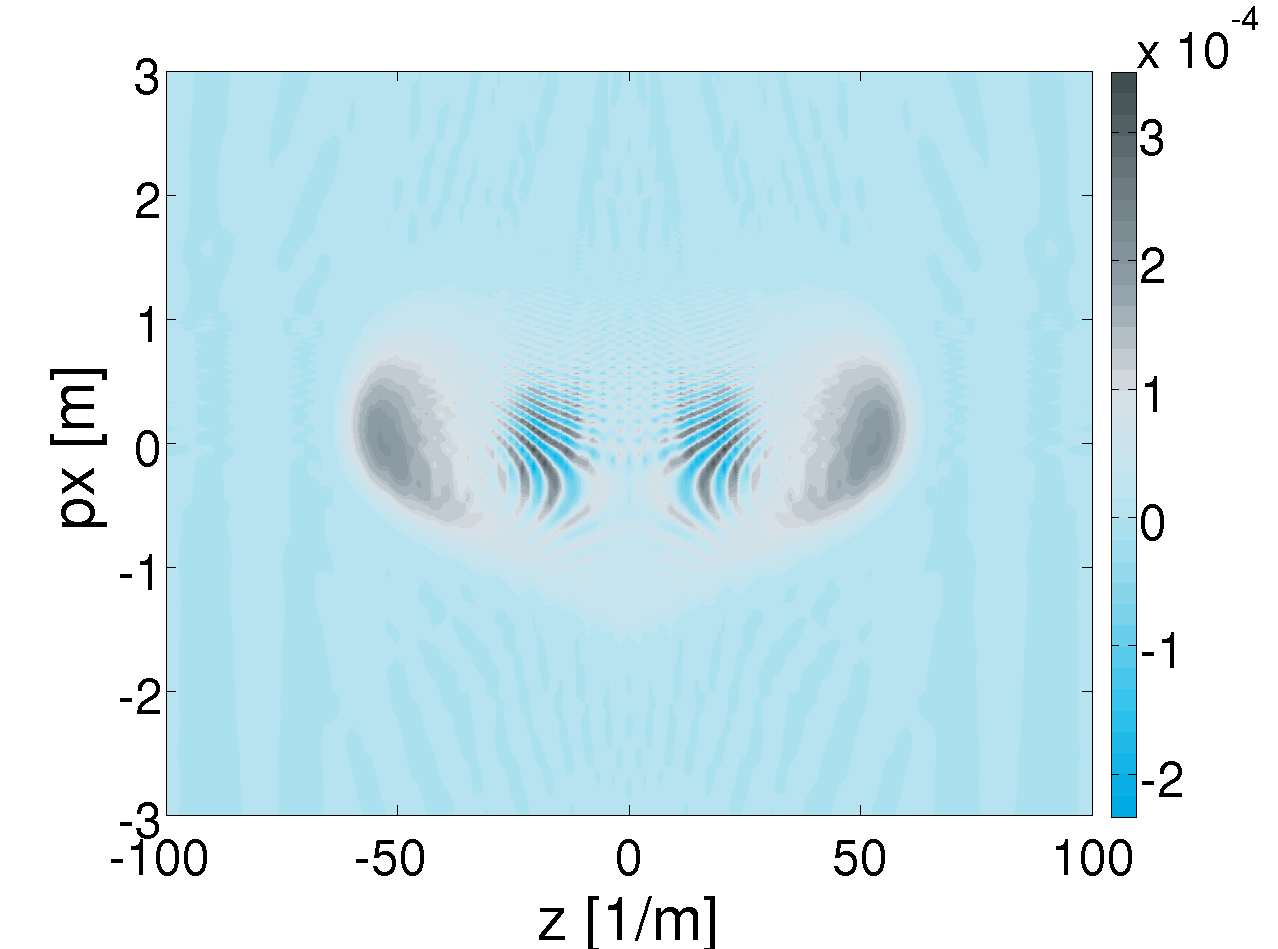}
 \includegraphics[width=0.43\textwidth]{./Fig/141/25.png}
 \includegraphics[width=0.43\textwidth]{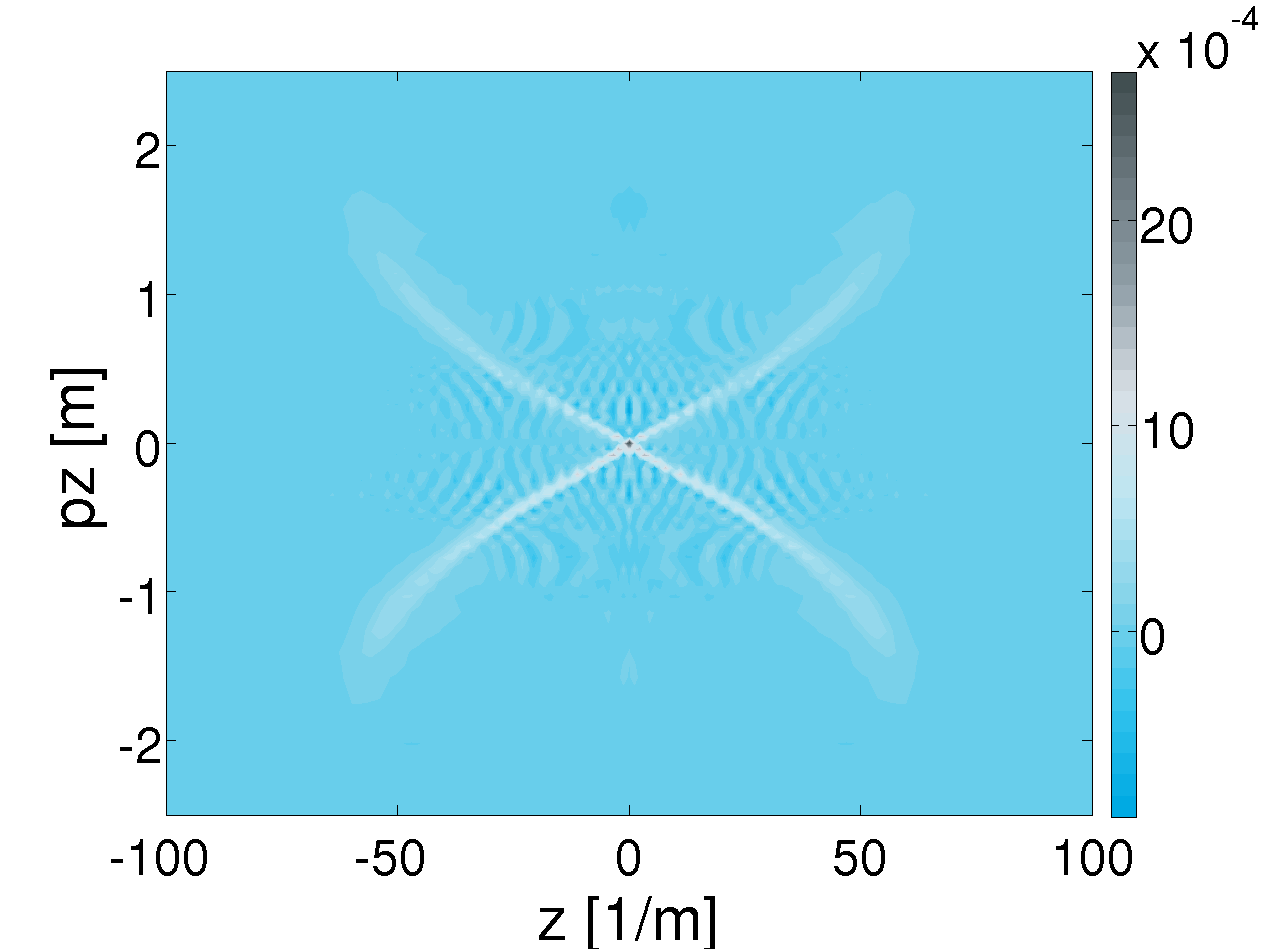} 
\caption{Electric field: $\mathbf{E} \br{z,t} = \varepsilon E_0 \exp \br{-\frac{z^2}{2 \lambda^2}}
 \left( \text{sech}^2 \left( \frac{t}{\tau} \right) - \frac{1}{2} \text{sech}^2 \left( \frac{t-\tau}{\tau} \right) - \frac{1}{2} \text{sech}^2 \left( \frac{t+\tau}{\tau} \right) \right) \mathbf{e}_x$. Magnetic field: $B \br{z,t} = \varepsilon E_0 \tau \frac{z}{\lambda^2} \exp \br{-\frac{z^2}{2 \lambda^2}} \times \notag \left( \tanh \left( \frac{t}{\tau} \right) - \frac{1}{2} \tanh \left( \frac{t-\tau}{\tau} \right) - \frac{1}{2} \tanh \left( \frac{t+\tau}{\tau} \right) \right)$.  Parameters: Tab. \ref{Tab_Triple}}
 \label{FigApp_15}
\end{figure}
\end{center}

\clearpage
\begin{landscape}
\section{Tables}

\ctable[pos=h,
caption = {The computation has been performed using MATLAB code. The results are illustrated in Fig. \ref{Fig_FinalMom}.},
label = TabApp_Semi0, 
mincapwidth = \textwidth,
]{ c c c c c c c }{
}{
    \toprule
    $\tau$ $[m^{-1}]$ & $\varepsilon$ & $\lambda$ $[m^{-1}]$ & $L_{p_x}$ $[m]$ & $N_{p_x}$ & $L_{x}$ $[m]$ & $N_{x}$ \\
    \midrule
    10 & 0.75 & 9.28 & [-11,11] & 512 & [-58.56,58.56] & 512  \\               
    \bottomrule
}

\ctable[pos=h,
caption = {The computation has been performed using MATLAB code. The results are illustrated in Fig. \ref{Fig_Semi1}.},
label = TabApp_Semi1, 
mincapwidth = \textwidth,
]{ c c c c }{
}{
    \toprule
    $\tau$ $[m^{-1}]$ & $\varepsilon$ & $L_{p_x}$ $[m]$ & $N_{p_x}$ \\
    \midrule
    10 & 0.75 & [-15,15] & 512  \\          
    \bottomrule
}

\ctable[pos=h,
caption = {The computation has been performed using MATLAB code. The second line yields the specification for generating the intermediate results. The results are illustrated in Fig. \ref{Fig_Semi2}.},
label = TabApp_Semi2, 
mincapwidth = \textwidth,
]{ c c c c c c c }{
}{
    \toprule
    $\tau$ $[m^{-1}]$ & $\varepsilon$ & $\lambda$ $[m^{-1}]$ & $L_{p_x}$ $[m]$ & $N_{p_x}$ & $L_{x}$ $[m]$ & $N_{x}$ \\
    \midrule
    10 & 0.75 & 0.238 & [-11,11] & 512 & [-40.47,40.47] & 512  \\      
    \midrule
    10 & 0.75 & $\lambda_i$ & [-11,11] & 512 & [-40-2$\lambda_i$,40+2$\lambda_i$] & 512  \\        
    \midrule
    10 & 0.75 & 9.28 & [-11,11] & 512 & [-58.56,58.56] & 512  \\         
    \bottomrule
}

\ctable[pos=h,
caption = {The computation has been performed using MATLAB code. Due to the fact, that we wanted to work with the same code for all configurations the case $\lambda \to \infty$ is approximated by the setup given in the first line. The results are illustrated in Fig. \ref{Fig_Semi3}.},
label = TabApp_Semi3, 
mincapwidth = \textwidth,
]{ c c c c c c c }{
}{
    \toprule
    $\tau$ $[m^{-1}]$ & $\varepsilon$ & $\lambda$ $[m^{-1}]$ & $L_{p_x}$ $[m]$ & $N_{p_x}$ & $L_{x}$ $[m]$ & $N_{x}$ \\
    \midrule
    10 & 0.75 & 10000 & [-15,15] & 512 & [-1,1] & 4  \\      
    \midrule
    10 & 0.75 & 50 & [-15,15] & 512 & [-70,70] & 512  \\
    \midrule
    10 & 0.75 & 4.479 & [-11,11] & 512 & [-48.95,48.95] & 512  \\
    \midrule
    10 & 0.75 & 1.21 & [-11,11] & 512 & [-42.42,42.42] & 512  \\      
    \bottomrule
}

\ctable[pos=h,
caption = {The computation has been performed using MATLAB code. The results are illustrated in Fig. \ref{Fig_Semi4}.},
label = TabApp_Semi4, 
mincapwidth = \textwidth,
]{ c c c c }{
}{
    \toprule
    $\tau$ $[m^{-1}]$ & $\varepsilon$ & $L_{p_x}$ $[m]$ & $N_{p_x}$ \\
    \midrule
    5 & 0.5 & [-12.5,12.5] & 4096  \\          
    \bottomrule
}

\ctable[pos=h,
caption = {The computation has been performed using MATLAB code. Line $2$ provides the necessary information in order to obtain the intermediate parameter values. The results are illustrated in Fig. \ref{Fig_Semi5}.},
label = TabApp_Semi5, 
mincapwidth = \textwidth,
]{ c c c c c c c }{
}{
    \toprule
    $\tau$ $[m^{-1}]$ & $\varepsilon$ & $\lambda$ $[m^{-1}]$ & $L_{p_x}$ $[m]$ & $N_{p_x}$ & $L_{x}$ $[m]$ & $N_{x}$ \\
    \midrule
    5 & 0.5 & 1 & [-12.5,12.5] & 1024 & [-42,42] & 512  \\ 
    \midrule
    5 & 0.5 & 1+$j$ & [-12.5,12.5] & 1024 & [-40-2$\lambda$,40+2$\lambda$] & 512  \\       
    \midrule
    5 & 0.5 & 15 & [-12.5,12.5] & 1024 & [-70,70] & 512  \\       
    \bottomrule
}

\ctable[pos=h,
caption = {The computation has been performed using MATLAB code. Line $2$ provides the necessary information in order to obtain the intermediate parameter values. The results are illustrated in Fig. \ref{Fig_Semi5}.},
label = TabApp_Semi6, 
mincapwidth = \textwidth,
]{ c c c c c c c }{
}{
    \toprule
    $\tau$ $[m^{-1}]$ & $\varepsilon$ & $\lambda$ $[m^{-1}]$ & $L_{p_x}$ $[m]$ & $N_{p_x}$ & $L_{x}$ $[m]$ & $N_{x}$ \\
    \midrule
    5 & 0.5 & 2 & [-12.5,12.5] & 1024 & [-44,44] & 512  \\ 
    \midrule
    5 & 0.5 & 6 & [-12.5,12.5] & 1024 & [-52,52] & 512  \\     
    \midrule
    5 & 0.5 & 10 & [-12.5,12.5] & 1024 & [-60,60] & 512  \\     
    \bottomrule
}

\ctable[pos=h,
caption = {The computation has been performed using MATLAB code. Line $2$ provides the necessary information in order to obtain the intermediate parameter values. The results are illustrated in Fig. \ref{Fig_Semi5}.},
label = TabApp_Semi7, 
mincapwidth = \textwidth,
]{ c c c c c c c }{
}{
    \toprule
    $\tau$ $[m^{-1}]$ & $\varepsilon$ & $\lambda$ $[m^{-1}]$ & $L_{p_x}$ $[m]$ & $N_{p_x}$ & $L_{x}$ $[m]$ & $N_{x}$ \\
    \midrule
    5 & 0.5 & 10 & [-12.5,12.5] & 1024 & [-60,60] & 512  \\ 
    \midrule
    5 & 0.5 & 20 & [-12.5,12.5] & 1024 & [-80,80] & 512  \\     
    \midrule
    5 & 0.5 & 30 & [-12.5,12.5] & 1024 & [-110,110] & 512  \\     
    \bottomrule
}

\ctable[pos=h,
caption = {The computation has been performed applying QKT. The results are illustrated in Fig. \ref{Fig_YieldFull} and Fig. \ref{FigApp_1}.},
label =Tab_EffM1_App, 
mincapwidth = \textwidth,
]{ c c c c }{
}{
    \toprule
    $\tau$ $[m^{-1}]$ & $\varepsilon$ & $L_{q_x}$ $[m]$ & $N_{q_x}$ \\
    \midrule
    35 & 0.5 & [-20,20] & 4096   \\
    \midrule
    50 & 0.5 & [-50,50] & 8192   \\    
    \midrule
    100 & 0.5 & [-100,100] & 16384   \\    
    \bottomrule
}

\end{landscape}
\begin{landscape}

\ctable[pos=h,
caption = {The computation has been performed applying QKT. The results are illustrated in Fig. \ref{Fig_Distr}.},
label =Tab_distr0, 
mincapwidth = \textwidth,
]{ c c c c c }{
}{
    \toprule
    $\tau$ $[m^{-1}]$ & $\varepsilon$ & $\omega$ $[m]$ & $L_{q_x}$ $[m]$ & $N_{q_x}$ \\
    \midrule
    1000 & 0.1 & 0 & [-50,50] & 512   \\
    \midrule
    1000 & 0.1 & 0.322 & [-5,5] & 4096   \\    
    \bottomrule
}

\ctable[pos=h,
caption = {The computation has been performed applying QKT. The results are illustrated in Fig. \ref{Fig_Yield}.},
label =Tab_yield, 
mincapwidth = \textwidth,
]{ c c c c c c }{
}{
    \toprule
    $\tau$ $[m^{-1}]$ & $\varepsilon$ & $\omega$ $[m]$ & $N_{\omega}$ & $L_{q_x}$ $[m]$ & $N_{q_x}$ \\
    \midrule
    1000 & 0.05 & [0.2,1.4] & 4001 & [-4,4] & 2048 \\  
    \midrule
    1000 & 0.05 & [1.401,2.0] & 600 & [-4,4] & 2048 \\      
    \bottomrule
}

\ctable[pos=h,
caption = {The computation has been performed applying QKT. The results are illustrated in Fig. \ref{Fig_Multiph}.},
label =Tab_peak, 
mincapwidth = \textwidth,
]{ c c c c c c }{
}{
    \toprule
    $\tau$ $[m^{-1}]$ & $\varepsilon$ & $N_{\varepsilon}$ & $\omega$ $[m]$ & $L_{q_x}$ $[m]$ & $N_{q_x}$ \\
    \midrule
    2000 & $[1,25]^2/50^2$ & 25 & 0.7 & [-3,3] & 4096 \\       
    \bottomrule
}

\ctable[pos=h,
caption = {The computation has been performed applying QKT. The results are illustrated in Fig. \ref{Fig_YieldX}.},
label =Tab_eps, 
mincapwidth = \textwidth,
]{ c c c c c c c }{
}{
    \toprule
    $\tau$ $[m^{-1}]$ & $\varepsilon$ & $N_{\epsilon}$ & $\omega$ $[m]$ & $N_{\omega}$ & $L_{q_x}$ $[m]$ & $N_{q_x}$ \\
    \midrule
    200 & [0.02,0.3] & 15 & [0.2,1.4] & 4001 & [-3.5,3.5] & 512 \\  
    \midrule
    200 & [0.22,0.3] & 15 & [1.401,2.4] & 1000 & [-3.5,3.5] & 512 \\      
    \bottomrule
}

\ctable[pos=h,
caption = {The computation has been performed applying QKT. The results are illustrated in Fig. \ref{Fig_YieldX}.},
label =Tab_yield_tau, 
mincapwidth = \textwidth,
]{ c c c c c c }{
}{
    \toprule
    $\tau$ $[m^{-1}]$ & $\varepsilon$ & $\omega$ $[m]$ & $N_{\omega}$ & $L_{q_x}$ $[m]$ & $N_{q_x}$ \\
    \midrule
    200 & 0.05 & [0.2,1.4] & 4001 & [-4,4] & 2048 \\  
    \midrule
    200 & 0.05 & [1.401,2.0] & 600 & [-4,4] & 2048 \\      
    \bottomrule
}

\ctable[pos=h,
caption = {The computation has been performed applying QKT. The results are illustrated in Fig. \ref{Fig_Tau}.},
label =Tab_tau, 
mincapwidth = \textwidth,
]{ c c c c c }{
}{
    \toprule
    $\tau$ $[m^{-1}]$ & $\varepsilon$ & $\omega$ $[m]$ & $L_{q_x}$ $[m]$ & $N_{q_x}$ \\
    \midrule
    500 & 0.01 & 0.7 & [-1.2,1.2] & 8192 \\
    \midrule
    1000 & 0.01 & 0.7 & [-1.2,1.2] & 8192 \\
    \midrule
    2500 & 0.01 & 0.7 & [-1.2,1.2] & 8192 \\
    \midrule
    5000 & 0.01 & 0.7 & [-1.2,1.2] & 8192 \\
    \midrule
    7500 & 0.01 & 0.7 & [-1.2,1.2] & 8192 \\
    \midrule
    10000 & 0.01 & 0.7 & [-1.2,1.2] & 8192 \\    
    \midrule
    15000 & 0.01 & 0.7 & [-1.2,1.2] & 8192 \\    
    \midrule
    20000 & 0.01 & 0.7 & [-1.2,1.2] & 8192 \\    
    \bottomrule
}

\ctable[pos=h,
caption = {The computation has been performed applying QKT. The results are illustrated in Fig. \ref{Fig_Channel}.},
label =Tab_channel, 
mincapwidth = \textwidth,
]{ c c c c c c }{
}{
    \toprule
    $\tau$ $[m^{-1}]$ & $\varepsilon$ & $N_{\varepsilon}$ & $\omega$ $[m]$ & $L_{q_x}$ $[m]$ & $N_{q_x}$ \\
    \midrule
    1000 & [0.02,0.24455] & 500 & 0.322 & [-0.75,0.75] & 1024   \\  
    \bottomrule
}

\ctable[pos=h,
caption = {The computation has been performed applying QKT. The results are illustrated in Fig. \ref{Fig_normYield}.},
label =Tab_EMass, 
mincapwidth = \textwidth,
]{ c c c c c c c }{
}{
    \toprule
    $\tau$ $[m^{-1}]$ & $\varepsilon$ & $N_{\varepsilon}$ & $\omega$ $[m]$ & $N_{\omega}$ & $L_{q_x}$ $[m]$ & $N_{q_x}$ \\
    \midrule
    800 & [0.05,0.4] & 8 & [0.222,0.3117] & 300 & [-3.5,3.5] & 2048 \\    
    \midrule
    800 & [0.05,0.4] & 8 & [0.25,0.3397] & 300 & [-3.5,3.5] & 2048 \\  
   \midrule    
    800 & [0.05,0.4] & 8 & [0.285,0.3747] & 300 & [-3.5,3.5] & 2048 \\ 
    \midrule    
    800 & [0.05,0.4] & 8 & [0.333,0.4227] & 300 & [-3.5,3.5] & 2048 \\
   \midrule
    800 & [0.05,0.4] & 8 & [0.4,0.4897] & 300 & [-3.5,3.5] & 2048 \\
    \midrule
    800 & [0.05,0.4] & 8 & [0.5,0.5897] & 300 & [-3.5,3.5] & 2048 \\ 
    \bottomrule
}

\ctable[pos=h,
caption = {The computation has been performed applying QKT. The results are illustrated in Fig. \ref{Fig_Distr1}.},
label =Tab_distr1, 
mincapwidth = \textwidth,
]{ c c c c c }{
}{
    \toprule
    $\tau$ $[m^{-1}]$ & $\varepsilon$ & $\omega$ $[m]$ & $L_{q_x}$ $[m]$ & $N_{q_x}$ \\
    \midrule
    800 & 0.05 & 0.225 & [-3.5,3.5] & 2048 \\
    \midrule
    800 & 0.05 & 0.2268 & [-3.5,3.5] & 2048 \\
   \midrule
    800 & 0.05 & 0.2295 & [-3.5,3.5] & 2048 \\     
    \midrule
    800 & 0.05 & 0.2536 & [-3.5,3.5] & 2048 \\    
    \bottomrule
}

\ctable[pos=h,
caption = {The computation has been performed applying DHW formalism in cylindrical coordinates. The information below holds for both calculations, $\phi=0$ and $\phi = \pi/2$. The results are illustrated in Fig. \ref{FigApp_CEP}.},
label =Tab_distr3, 
mincapwidth = \textwidth,
]{ c c c c c c c c }{
}{
    \toprule
    $\tau$ $[m^{-1}]$ & $\lambda$ $[m^{-1}]$ & $\varepsilon$ & $\omega$ $[m]$ & $L_{p_x}$ $[m]$ & $N_{p_x}$ & $L_{x}$ $[m]$ & $N_{x}$ \\
    \midrule
    100 & 1000 & 0.5 & 0.2 & [-10,10] & 2048 & [-6000,6000] & 512 \\
    \midrule
    100 & 50 & 0.5 & 0.2 & [-10,10] & 2048 & [-320,320] & 512 \\     
    \midrule
    100 & 20 & 0.5 & 0.2 & [-10,10] & 2048 & [-320,320] & 512 \\ 
   \midrule
    100 & 5 & 0.5 & 0.2 & [-10,10] & 2048 & [-320,320] & 1024 \\       
    \bottomrule
}

\ctable[pos=h,
caption = {The computation has been performed applying DHW formalism in cylindrical coordinates. The results are illustrated in Fig. \ref{Fig_Pond1}.},
label =Tab_distr2, 
mincapwidth = \textwidth,
]{ c c c c c c c c }{
}{
    \toprule
    $\tau$ $[m^{-1}]$ & $\lambda$ $[m^{-1}]$ & $\varepsilon$ & $\omega$ $[m]$ & $L_{p_x}$ $[m]$ & $N_{p_x}$ & $L_{x}$ $[m]$ & $N_{x}$ \\
    \midrule
    100 & 1000 & 0.5 & 0.7 & [-20,20] & 4096 & [-6000,6000] & 256 \\
    \midrule
    100 & 25 & 0.5 & 0.7 & [-20,20] & 4096 & [-320,320] & 256 \\ 
   \midrule
    100 & 10 & 0.5 & 0.7 & [-20,20] & 4096 & [-320,320] & 256 \\
    \midrule
    100 & 2.5 & 0.5 & 0.7 & [-20,20] & 4096 & [-320,320] & 256 \\           
    \bottomrule
}

\ctable[pos=h,
caption = {The computation has been performed applying DHW formalism in cylindrical coordinates. The results are illustrated in Fig. \ref{Fig_Pond2}.},
label =Tab_rho1, 
mincapwidth = \textwidth,
]{ c c c c c c c c }{
}{
    \toprule
    $\tau$ $[m^{-1}]$ & $\lambda$ $[m^{-1}]$ & $\varepsilon$ & $\omega$ $[m]$ & $L_{p_x}$ $[m]$ & $N_{p_x}$ & $L_{x}$ $[m]$ & $N_{x}$ \\
    \midrule
    100 & 1000 & 0.5 & 0.7 & [-20,20] & 4096 & [-6000,6000] & 256 \\
    \midrule
    100 & 100 & 0.5 & 0.7 & [-20,20] & 4096 & [-650,650] & 512 \\    
    \midrule
    100 & 50-1 & 0.5 & 0.7 & [-20,20] & 4096 & [-320,320] & 512 \\    
   \midrule
    100 & 1000 & 0.5 & 0.74 & [-20,20] & 4096 & [-6000,6000] & 512 \\
   \midrule
    100 & 100 & 0.5 & 0.74 & [-20,20] & 4096 & [-650,650] & 512 \\
   \midrule
    100 & 75 & 0.5 & 0.74 & [-20,20] & 4096 & [-450,450] & 512 \\
   \midrule
    100 & 50-10 & 0.5 & 0.74 & [-20,20] & 4096 & [-320,320] & 512 \\    
   \midrule
    100 & 7.5-1 & 0.5 & 0.74 & [-20,20] & 4096 & [-320,320] & 1024 \\   
   \midrule
    100 & 1000 & 0.5 & 0.78 & [-20,20] & 4096 & [-6000,6000] & 512 \\
   \midrule
    100 & 100 & 0.5 & 0.78 & [-20,20] & 4096 & [-650,650] & 512 \\
   \midrule
    100 & 75 & 0.5 & 0.78 & [-20,20] & 4096 & [-450,450] & 512 \\
   \midrule
    100 & 50-10 & 0.5 & 0.78 & [-20,20] & 4096 & [-320,320] & 512 \\    
   \midrule
    100 & 7.5-1 & 0.5 & 0.78 & [-20,20] & 4096 & [-320,320] & 1024 \\      
   \midrule
    100 & 1000 & 0.5 & 0.84 & [-20,20] & 4096 & [-6000,6000] & 512 \\
   \midrule
    100 & 100 & 0.5 & 0.84 & [-20,20] & 4096 & [-650,650] & 512 \\
   \midrule
    100 & 75 & 0.5 & 0.84 & [-20,20] & 4096 & [-450,450] & 512 \\
   \midrule
    100 & 50-10 & 0.5 & 0.84 & [-20,20] & 4096 & [-320,320] & 512 \\    
   \midrule
    100 & 7.5-1 & 0.5 & 0.84 & [-20,20] & 4096 & [-320,320] & 1024 \\        
    \bottomrule
}

\ctable[pos=h,
caption = {The computation has been performed applying DHW formalism in cylindrical coordinates. The results are illustrated in Fig. \ref{Fig_Distr3D} and Fig. \ref{Fig_3D_1}.},
label =Tab_rho2, 
mincapwidth = \textwidth,
]{ c c c c c c c c c c }{
}{
    \toprule
    $\tau$ $[m^{-1}]$ & $\lambda$ $[m^{-1}]$ & $\varepsilon$ & $\omega$ $[m]$ & $L_{p_x}$ $[m]$ & $N_{p_x}$ & $L_{p_{\perp}}$ $[m]$ & $N_{p_{\perp}}$ & $L_{x}$ $[m]$ & $N_{x}$ \\
    \midrule
    100 & 100 & 0.5 & 0.74 & [-5,5] & 1024 & [-1.5,0] & 31 & [-650,650] & 256 \\
   \midrule
    100 & 100 & 0.5 & 0.84 & [-5,5] & 1024 & [-1.5,0] & 31 & [-650,650] & 256 \\
   \midrule
    100 & 2 & 0.5 & 0.74 & [-10,10] & 2048 & [-1.76,0] & 45 & [-320,320] & 1024 \\    
   \midrule
    100 & 2 & 0.5 & 0.84 & [-10,10] & 2048 & [-0.92,0] & 24 & [-320,320] & 1024 \\    
    \bottomrule
}

\ctable[pos=h,
caption = {The computation has been performed applying DHW formalism in cylindrical coordinates.},
label =Tab_yield1, 
mincapwidth = \textwidth,
]{ c c c c c c c c c c }{
}{
    \toprule
    $\tau$ $[m^{-1}]$ & $\lambda$ $[m^{-1}]$ & $\varepsilon$ & $\omega$ $[m]$ & $L_{p_x}$ $[m]$ & $N_{p_x}$ & $L_{p_{\perp}}$ $[m]$ & $N_{p_{\perp}}$ & $L_{x}$ $[m]$ & $N_{x}$ \\
    \midrule
    20 & 1 & 0.75 & 0 & [-12,12] & 1024 & [0,1.2] & 25 & [-60,60] & 512 \\
    \midrule    
    20 & 2 & 0.75 & 0 & [-12.5,12.5] & 1024 & [0,1.2] & 25 & [-60,60] & 512 \\    
    \midrule
    20 & 3.5 & 0.75 & 0 & [-12.5,12.5] & 1024 & [0,1.2] & 25 & [-60,60] & 512 \\    
    \midrule
    20 & 5 & 0.75 & 0 & [-12.5,12.5] & 1024 & [0,1.2] & 25 & [-60,60] & 512 \\      
    \midrule
    20 & 7.5 & 0.75 & 0 & [-14,14] & 1024 & [0,1.2] & 25 & [-65,65] & 512 \\       
    \midrule    
    20 & 10 & 0.75 & 0 & [-16,16] & 1024 & [0,1.2] & 25 & [-70,70] & 512 \\       
    \midrule
    20 & 15 & 0.75 & 0 & [-18,18] & 1024 & [0,1.7] & 35 & [-75,75] & 512 \\ 
    \midrule    
    20 & 20 & 0.75 & 0 & [-20,20] & 1024 & [0,1.7] & 35 & [-82.5,82.5] & 256 \\         
    \bottomrule
}

\clearpage

\ctable[pos=h,
caption = {The computation has been performed applying DHW formalism in cylindrical coordinates. The intermediate points have been obtained on irregular distances. The results are illustrated in Fig. \ref{Fig_Narrow}.},
label =Tab_yield2b, 
mincapwidth = \textwidth,
]{ c c c c c c c c }{
}{
    \toprule
    $\tau$ $[m^{-1}]$ & $\lambda$ $[m^{-1}]$ & $\varepsilon$ & $\omega$ $[m]$ & $L_{p_x}$ $[m]$ & $N_{p_x}$ & $L_{x}$ $[m]$ & $N_{x}$ \\
    \midrule
    20 & 0.5 & 1.0 & 0 & [-12,12] & 512 & [-60,60] & 256 \\
    \midrule    
    20 & 20 & 1.0 & 0 & [-27,27] & 512 & [-82.5,82.5] & 256 \\       
    \midrule
    20 & 0.5 & 0.75 & 0 & [-12.5,12.5] & 1024 & [-60,60] & 512 \\
    \midrule    
    20 & 2 & 0.75 & 0 & [-12.5,12.5] & 512 & [-60,60] & 256 \\
    \midrule
    20 & 20 & 0.75 & 0 & [-20,20] & 512 & [-82.5,82.5] & 256 \\
    \midrule
    20 & 0.5 & 0.5 & 0 & [-12.5,12.5] & 512 & [-60,60] & 256 \\
    \midrule    
    20 & 20 & 0.5 & 0 & [-20,20] & 512 & [-82.5,82.5] & 256 \\        
    \midrule
    20 & 2 & 0.25 & 0 & [-12.5,12.5] & 512 & [-60,60] & 256 \\
    \midrule    
    20 & 20 & 0.25 & 0 & [-20,20] & 512 & [-82.5,82.5] & 256 \\       
    \bottomrule
}

\ctable[pos=h,
caption = {The computation has been performed applying DHW formalism in cylindrical coordinates. The intermediate points have been obtained on irregular distances. The results are illustrated in Fig. \ref{Fig_Narrow}.},
label =Tab_yield2, 
mincapwidth = \textwidth,
]{ c c c c c c c c }{
}{
    \toprule
    $\tau$ $[m^{-1}]$ & $\lambda$ $[m^{-1}]$ & $\varepsilon$ & $\omega$ $[m]$ & $L_{p_x}$ $[m]$ & $N_{p_x}$ & $L_{x}$ $[m]$ & $N_{x}$ \\
    \midrule
    20 & 2 & 0.75 & 0 & [-12.5,12.5] & 512 & [-60,60] & 256 \\
    \midrule
    20 & 20 & 0.75 & 0 & [-20,20] & 512 & [-82.5,82.5] & 256 \\    
   \midrule
    30 & 2 & 0.75 & 0 & [-12.5,12.5] & 512 & [-60,60] & 256 \\
    \midrule
    30 & 20 & 0.75 & 0 & [-20,20] & 512 & [-82.5,82.5] & 256 \\       
   \midrule
    40 & 2 & 0.75 & 0 & [-15,15] & 1024 & [-100,100] & 256 \\
    \midrule
    40 & 19.6 & 0.75 & 0 & [-26,26] & 1024 & [-100,100] & 256 \\          
    \bottomrule
}

\ctable[pos=h,
caption = {The computation has been performed applying DHW formalism in $2+1$ dimensions formulated with $2$-spinors(system $I$). 
Pseudo-differential operators are truncated at $n=2$. The results are illustrated in Fig. \ref{Fig_BDistr1}, Fig. \ref{FigApp_3}, Fig. \ref{FigApp_4}, Fig. \ref{FigApp_5} and Fig. \ref{FigApp_6}.},
label =Tab_ShortB, 
mincapwidth = \textwidth,
]{ c c c c c c c c c c c c c }{
}{
    \toprule
    $\tau$ $[m^{-1}]$ & $\lambda$ $[m^{-1}]$ & $\varepsilon$ & $L_{p_x}$ $[m]$ & $N_{p_x}$ & $L_{p_z}$ $[m]$ & $N_{p_z}$ & $L_{x}$ $[m]$ & $N_{x}$ & $b_2$ & $a_2$ & $a_3$ & $a_1$ \\
    \midrule
    5 & 100 & 0.707 & [-0.99,0.99] & 512 & [-16,16] & 64 & [-0.9995,0.9995] & 256 & 2.5 & 6 & 12 & 300 \\   
    \midrule
    5 & 50 & 0.707 & [-0.99,0.99] & 512 & [-16,16] & 64 & [-0.9995,0.9995] & 256 & 2.5 & 6 & 12 & 150 \\   
    \midrule    
    5 & 20 & 0.707 & [-0.99,0.99] & 512 & [-16,16] & 64 & [-0.9995,0.9995] & 256 & 2.5 & 6 & 12 & 60 \\   
    \midrule
    5 & 15 & 0.707 & [-0.99,0.99] & 512 & [-16,16] & 64 & [-0.9995,0.9995] & 256 & 2.5 & 6 & 12 & 45 \\      
    \midrule
    5 & 10 & 0.707 & [-0.99,0.99] & 512 & [-15,15] & 128 & [-0.99,0.99] & 256 & 2.5 & 6 & 12 & 40 \\
    \midrule
    5 & 7 & 0.707 & [-0.99,0.99] & 512 & [-16,16] & 64 & [-0.9995,0.9995] & 256 & 2.5 & 6 & 12 & 21 \\
    \midrule
    5 & 5 & 0.707 & [-0.99,0.99] & 512 & [-15,15] & 128 & [-0.99,0.99] & 256 & 2.5 & 6 & 12 & 20 \\      
    \bottomrule
}

\ctable[pos=h,
caption = {The computation has been performed applying DHW formalism in $2+1$ dimensions formulated with $2$-spinors(system $I$). 
Pseudo-differential operators are truncated at $n=2$. The results are illustrated in Fig. \ref{Fig_BNp2} and Fig. \ref{Fig_BNp1}.},
label =Tab_ShortInt_Th, 
mincapwidth = \textwidth,
]{ c c c c c c c c c c c c c }{
}{
    \toprule
    $\tau$ $[m^{-1}]$ & $\lambda$ $[m^{-1}]$ & $\varepsilon$ & $L_{p_x}$ $[m]$ & $N_{p_x}$ & $L_{p_z}$ $[m]$ & $N_{p_z}$ & $L_{x}$ $[m]$ & $N_{x}$ & $b_2$ & $a_2$ & $a_3$ & $a_1$ \\
    \midrule
    5 & 100 & 0.707 & [-0.99,0.99] & 512 & [-16,16] & 64 & [-0.9995,0.9995] & 256 & 2.5 & 6 & 12 & 300 \\    
    \midrule
    5 & 50 & 0.707 & [-0.99,0.99] & 512 & [-16,16] & 64 & [-0.9995,0.9995] & 256 & 2.5 & 6 & 12 & 150 \\
    \midrule
    5 & 20 & 0.707 & [-0.99,0.99] & 512 & [-16,16] & 64 & [-0.9995,0.9995] & 256 & 2.5 & 6 & 12 & 60 \\
    \midrule
    5 & 10 & 0.707 & [-0.99,0.99] & 512 & [-16,16] & 64 & [-0.9995,0.9995] & 256 & 2.5 & 6 & 12 & 30 \\     
    \midrule
    5 & 7 & 0.707 & [-0.99,0.99] & 512 & [-16,16] & 64 & [-0.9995,0.9995] & 256 & 2.5 & 6 & 12 & 21 \\
    \midrule
    5 & 5 & 0.707 & [-0.99,0.99] & 512 & [-15,15] & 128 & [-0.99,0.99] & 256 & 2.5 & 6 & 12 & 20 \\     
    \bottomrule
}

\ctable[pos=h,
caption = {The computation has been performed applying DHW formalism in $2+1$ dimensions formulated with $2$-spinors(system $I$). 
Pseudo-differential operators are truncated at $n=2$. The results are illustrated in Fig. \ref{FigApp_7}, Fig. \ref{FigApp_8} and Fig. \ref{FigApp_9}.},
label =Tab_LongB, 
mincapwidth = \textwidth,
]{ c c c c c c c c c c c c c }{
}{
    \toprule
    $\tau$ $[m^{-1}]$ & $\lambda$ $[m^{-1}]$ & $\varepsilon$ & $L_{p_x}$ $[m]$ & $N_{p_x}$ & $L_{p_z}$ $[m]$ & $N_{p_z}$ & $L_{x}$ $[m]$ & $N_{x}$ & $b_2$ & $a_2$ & $a_3$ & $a_1$ \\
    \midrule
    10 & 100 & 0.707 & [-0.99,0.99] & 1024 & [-15,15] & 64 & [-0.995,0.995] & 256 & 6 & 6 & 10 & 300 \\
    \midrule
    10 & 25 & 0.707 & [-0.99,0.99] & 1024 & [-15,15] & 64 & [-0.995,0.995] & 256 & 6 & 6 & 10 & 75 \\  
    \midrule
    10 & 20 & 0.707 & [-0.99,0.99] & 1024 & [-15,15] & 64 & [-0.995,0.995] & 256 & 6 & 6 & 10 & 60 \\
    \midrule
    10 & 15 & 0.707 & [-0.99,0.99] & 1024 & [-15,15] & 64 & [-0.995,0.995] & 256 & 6 & 6 & 10 & 45 \\    
    \bottomrule
} 

\ctable[pos=h,
caption = {The computation has been performed applying DHW formalism in $2+1$ dimensions formulated with $2$-spinors(system $I$). 
Pseudo-differential operators are truncated at $n=2$. Additionally, $B$ is fixed to zero throughout the calculation. The results are illustrated in Fig. \ref{Fig_B0} and Fig. \ref{FigApp_10}.},
label =Tab_Long, 
mincapwidth = \textwidth,
]{ c c c c c c c c c c c c c }{
}{
    \toprule
    $\tau$ $[m^{-1}]$ & $\lambda$ $[m^{-1}]$ & $\varepsilon$ & $L_{p_x}$ $[m]$ & $N_{p_x}$ & $L_{p_z}$ $[m]$ & $N_{p_z}$ & $L_{x}$ $[m]$ & $N_{x}$ & $b_2$ & $a_2$ & $a_3$ & $a_1$ \\
    \midrule
    10 & 25 & 0.707 & [-0.99,0.99] & 1024 & [-15,15] & 64 & [-0.995,0.995] & 256 & 6 & 6 & 10 & 75 \\
    \bottomrule
}

\ctable[pos=h,
caption = {The computation has been performed applying DHW formalism in $2+1$ dimensions formulated with $2$-spinors(system $I$). 
Pseudo-differential operators are truncated at $n=2$ or $n=4$ depending on $\lambda$. The results are illustrated in Fig. \ref{Fig_BYield}.},
label =Tab_IntB, 
mincapwidth = \textwidth,
]{ c c c c c c c c c c c c c }{
}{
    \toprule
    $\tau$ $[m^{-1}]$ & $\lambda$ $[m^{-1}]$ & $\varepsilon$ & $L_{p_x}$ $[m]$ & $N_{p_x}$ & $L_{p_{z}}$ $[m]$ & $N_{p_{z}}$ & $L_{z}$ $[m]$ & $N_{z}$ & $b_2$ & $a_2$ & $a_3$ & $a_1$ \\
    \midrule
    10 & 100 & 0.707 & [-0.99,0.99] & 1024 & [-15,15] & 64 & [-0.995,0.995] & 256 & 6 & 6 & 10 & 300 \\  
    \midrule    
    10 & 50 & 0.707 & [-0.99,0.99] & 1024 & [-15,15] & 64 & [-0.995,0.995] & 256 & 6 & 6 & 10 & 150 \\   
    \midrule    
    10 & 25 & 0.707 & [-0.99,0.99] & 1024 & [-15,15] & 64 & [-0.995,0.995] & 256 & 6 & 6 & 10 & 75 \\     
    \midrule
    10 & 20 & 0.707 & [-0.99,0.99] & 1024 & [-15,15] & 64 & [-0.995,0.995] & 256 & 6 & 6 & 10 & 80 \\       
    \midrule
    10 & 15 & 0.707 & [-0.99,0.99] & 1024 & [-15,15] & 64 & [-0.995,0.995] & 256 & 6 & 6 & 10 & 45 \\       
    \midrule
    10 & 12 & 0.707 & [-0.99,0.99] & 1024 & [-20,20] & 64 & [-0.9995,0.9995] & 256 & 6 & 6 & 12 & 30 \\
    \midrule    
    10 & 10 & 0.707 & [-0.99,0.99] & 1024 & [-15,15] & 64 & [-0.995,0.995] & 256 & 6 & 6 & 10 & 30 \\  
    \midrule
    10 & 8.5 & 0.707 & [-0.99,0.99] & 1024 & [-20,20] & 64 & [-0.9995,0.9995] & 256 & 6 & 6 & 12 & 21.25 \\
    \midrule
    10 & 7.5 & 0.707 & [-0.999,0.999] & 512 & [-22,22] & 128 & [-0.999,0.999] & 256 & 6 & 3 & 14 & 26.25 \\    
    \midrule
    10 & 6.5 & 0.707 & [-0.99,0.99] & 1024 & [-20,20] & 64 & [-0.9995,0.9995] & 256 & 6 & 6 & 12 & 19.5 \\    
    \midrule
    10 & 5 & 0.707 & [-0.99,0.99] & 512 & [-20,20] & 64 & [-0.9995,0.9995] & 256 & 6 & 6 & 12 & 15 \\       
    \bottomrule
}

\ctable[pos=h,
caption = {The computation has been performed applying DHW formalism in $2+1$ dimensions formulated with $2$-spinors(system $I$). 
Pseudo-differential operators are truncated at $n=2$. Additionally, $B$ is fixed to zero throughout the calculation. The results are illustrated in Fig. \ref{Fig_BYield}.},
label =Tab_IntB0, 
mincapwidth = \textwidth,
]{ c c c c c c c c c c c c c }{
}{
    \toprule
    $\tau$ $[m^{-1}]$ & $\lambda$ $[m^{-1}]$ & $\varepsilon$ & $L_{p_x}$ $[m]$ & $N_{p_x}$ & $L_{p_{z}}$ $[m]$ & $N_{p_{z}}$ & $L_{z}$ $[m]$ & $N_{z}$ & $b_2$ & $a_2$ & $a_3$ & $a_1$ \\
    \midrule
    10 & 100 & 0.707 & [-0.99,0.99] & 1024 & [-15,15] & 64 & [-0.995,0.995] & 256 & 6 & 6 & 10 & 300 \\   
    \midrule    
    10 & 50 & 0.707 & [-0.99,0.99] & 1024 & [-15,15] & 64 & [-0.995,0.995] & 256 & 6 & 6 & 10 & 150 \\   
    \midrule
    10 & 25 & 0.707 & [-0.99,0.99] & 1024 & [-15,15] & 64 & [-0.995,0.995] & 256 & 6 & 6 & 10 & 75 \\ 
    \midrule
    10 & 20 & 0.707 & [-0.99,0.99] & 1024 & [-20,20] & 64 & [-0.9995,0.9995] & 256 & 6 & 6 & 12 & 50 \\
    \midrule
    10 & 15 & 0.707 & [-0.99,0.99] & 1024 & [-20,20] & 64 & [-0.9995,0.9995] & 256 & 6 & 6 & 12 & 37.5 \\
    \midrule
    10 & 12 & 0.707 & [-0.99,0.99] & 1024 & [-20,20] & 64 & [-0.9995,0.9995] & 256 & 6 & 6 & 12 & 30 \\ 
    \midrule
    10 & 10 & 0.707 & [-0.99,0.99] & 1024 & [-20,20] & 64 & [-0.9995,0.9995] & 256 & 6 & 6 & 12 & 25 \\
    \midrule
    10 & 8.5 & 0.707 & [-0.99,0.99] & 1024 & [-20,20] & 64 & [-0.9995,0.9995] & 256 & 6 & 6 & 12 & 21.25 \\
    \midrule
    10 & 7.5 & 0.707 & [-0.99,0.99] & 1024 & [-20,20] & 64 & [-0.9995,0.9995] & 256 & 6 & 6 & 12 & 18.75 \\   
    \midrule
    10 & 6.5 & 0.707 & [-0.99,0.99] & 1024 & [-16,16] & 64 & [-0.9995,0.9995] & 256 & 6 & 6 & 12 & 26 \\   
    \midrule    
    10 & 5 & 0.707 & [-0.99,0.99] & 1024 & [-20,20] & 64 & [-0.9995,0.9995] & 256 & 6 & 6 & 12 & 20 \\
    \midrule
    10 & 4.5 & 0.707 & [-0.99,0.99] & 1024 & [-20,20] & 64 & [-0.9995,0.9995] & 256 & 6 & 6 & 12 & 18 \\
    \midrule
    10 & 3.5 & 0.707 & [-0.99,0.99] & 1024 & [-20,20] & 64 & [-0.9995,0.9995] & 256 & 6 & 6 & 12 & 14 \\          
    \midrule    
    10 & 2.5 & 0.707 & [-0.99,0.99] & 1024 & [-20,20] & 64 & [-0.9995,0.9995] & 256 & 6 & 6 & 12 & 10 \\          
    \midrule
    10 & 2 & 0.707 & [-0.99,0.99] & 1024 & [-20,20] & 64 & [-0.9995,0.9995] & 256 & 6 & 6 & 14 & 8 \\
    \midrule    
    10 & 1.5 & 0.707 & [-0.99,0.99] & 1024 & [-15,15] & 64 & [-0.9995,0.9995] & 256 & 6 & 6 & 12 & 6 \\
    \midrule
    10 & 1 & 0.707 & [-0.99,0.99] & 1024 & [-20,20] & 64 & [-0.9995,0.9995] & 256 & 6 & 6 & 14 & 4 \\        
    \midrule    
    10 & 0.5 & 0.707 & [-0.99,0.99] & 1024 & [-15,15] & 64 & [-0.9995,0.9995] & 256 & 6 & 6 & 12 & 2 \\  
    \midrule
    10 & 0.25 & 0.707 & [-0.99,0.99] & 1024 & [-12,12] & 64 & [-0.9995,0.9995] & 256 & 6 & 6 & 14 & 1 \\     
    \bottomrule
}

\ctable[pos=h,
caption = {The computation has been performed applying DHW formalism in $2+1$ dimensions formulated with $2$-spinors(system $I$). 
Pseudo-differential operators are truncated at $n=2$. The results are illustrated in Fig. \ref{Fig_BDistr2}, Fig. \ref{FigApp_11}, Fig. \ref{FigApp_12}, Fig. \ref{FigApp_13}, Fig. \ref{FigApp_14} and Fig. \ref{FigApp_15}.},
label =Tab_Triple, 
mincapwidth = \textwidth,
]{ c c c c c c c c c c c c }{
}{
    \toprule
    $\tau$ $[m^{-1}]$ & $\lambda$ $[m^{-1}]$ & $\varepsilon$ & $L_{p_x}$ $[m]$ & $N_{p_x}$ & $L_{p_z}$ $[m]$ & $N_{p_z}$ & $L_{x}$ $[m]$ & $N_{x}$ & $a_2$ & $a_3$ & $a_1$ \\
    \midrule
    10 & 100 & 0.707 & [-0.99,0.99] & 1024 & [-16,16] & 64 & [-0.99,0.99] & 128 & 6 & 12 & 800 \\
    \midrule
    10 & 75 & 0.707 & [-0.99,0.99] & 1024 & [-16,16] & 64 & [-0.99,0.99] & 128 & 6 & 12 & 600 \\
    \midrule    
    10 & 50 & 0.707 & [-0.99,0.99] & 1024 & [-16,16] & 64 & [-0.99,0.99] & 128 & 6 & 12 & 400 \\    
    \midrule
    10 & 40 & 0.707 & [-0.99,0.99] & 1024 & [-16,16] & 64 & [-0.99,0.99] & 128 & 6 & 12 & 320 \\    
    \midrule    
    10 & 30 & 0.707 & [-0.99,0.99] & 1024 & [-16,16] & 64 & [-0.99,0.99] & 128 & 6 & 12 & 240 \\    
    \midrule    
    10 & 20 & 0.707 & [-0.99,0.99] & 1024 & [-16,16] & 64 & [-0.99,0.99] & 128 & 6 & 12 & 160 \\     
    \midrule
    10 & 15 & 0.707 & [-0.99,0.99] & 1024 & [-16,16] & 64 & [-0.99,0.99] & 128 & 6 & 12 & 120 \\
    \midrule
    10 & 12.5 & 0.707 & [-0.99,0.99] & 1024 & [-16,16] & 64 & [-0.99,0.99] & 128 & 6 & 12 & 100 \\
    \midrule
    10 & 10 & 0.707 & [-0.99,0.99] & 1024 & [-16,16] & 64 & [-0.99,0.99] & 128 & 6 & 12 & 80 \\      
    \bottomrule
}

\ctable[pos=h,
caption = {The computation has been performed applying QKT. The results are illustrated in Fig. \ref{FigApp_2}.},
label =Tab_EffM2_App, 
mincapwidth = \textwidth,
]{ c c c c }{
}{
    \toprule
    $\tau$ $[m^{-1}]$ & $\varepsilon$ & $L_{q_x}$ $[m]$ & $N_{q_x}$ \\
    \midrule
    100 & 0.5 & [-3.5,3.5] & 256   \\
    \midrule
    200 & 0.5 & [-3.5,3.5] & 512   \\    
    \midrule
    100 & 0.5 & [-3.5,3.5] & 2048   \\    
    \bottomrule
}

\end{landscape}

\chapter{Matlab solver}
\label{App_Matlab}

In this chapter we provide a solver in MATLAB instead of stating pseudocode. The advantage is, that presupposing MATLAB is available one can immediately start to perform calculations regarding pair production within the DHW approach. In addition we have listed additional figures to demonstrate the efficiency of the code.

\section{Code}
\subsection{main.m}
\begin{lstlisting}
%Christian Kohlfürst
%July 26, 2015

%Fourier-Basis
%Full solution
%Symmetrie in x

function main								%First File called main.m

clc; clear all; 
%cd('/hosts/nashome/Matlab')				%Choose right directory

for id = 9999:9999,							%Number of runs	 

%Grid and initial data
Nx = 2; Np = 128; Lp = 25;
%Parameters for the electric field
eps = 0.75; tau = 1; ll = 100000; 		




%Fix the parameters that depend on the applied field
Lx = 75+2.5*ll; hx = Lx/Nx; Nx = Nx+1; 
hp = Lp/Np;
x = hx*(0:Nx-1); p = hp*(0:Np-1)'; 
[xx,pp] = meshgrid(x,p);

t = -5*tau; Tf = 5*tau; 					%Initial and final time, 

%Initial conditions
k1 = (pi/Lx)*[0:(Nx-2) 0 (-Nx+2):-1];
w = sqrt(1+(pp-Lp/2).^2);
w0 = 0*w;
w1 = 0*w;
w2 = 0*w;
w3 = 0*w;

%First inhomogeneity
pfac = 8; nn = 2^12*pfac; range = Lp*pfac; vv = nn/Np/pfac;
deltap = range/nn; deltay = 2*pi/(nn*deltap);

%Grid
gridp = (linspace(0,nn-1,nn)-nn/2) * deltap;
gridy(1,1:nn/2+1) = linspace(0,nn/2,nn/2+1) * deltay;
gridy(1,nn/2+2:nn) = -gridy(1,nn/2:-1:2);
gridy(1,nn/2+1) = 0; 					%Odd derivative, hence fix this point to zero

%Sampling
samplesp(1,1:nn/2) = 1./sqrt(1+gridp(nn/2+1:nn).^2);
samplesp(1,nn/2+1:nn) = 1./sqrt(1+gridp(1:nn/2).^2);

%Performing the derivative
in = zeros(nn,Nx);
in0 = zeros(Np,Nx);
for j = 1:Nx,
  samples = fft(samplesp)/sqrt(nn) * (1i*deltap*sqrt(nn/2/pi)) ...
    *ll*sqrt(pi/2).*(erf((gridy+2*x(j))/sqrt(8)/ll) + erf((gridy-2*x(j))/sqrt(8)/ll));
  samples = ifft(samples)*sqrt(nn);
  in(1:nn/2,j) = real(samples(nn/2+1:nn)) * deltay*sqrt(nn/2/pi);
  in(nn/2+1:nn,j) = real(samples(1:nn/2)) * deltay*sqrt(nn/2/pi);
  in0(:,j) = in(nn*(2*pfac-2)/4/pfac+1:vv:nn*(2*pfac+2)/4/pfac,j);
end

%Second inhomogeneity
pfac = 8; nn = 2^12*pfac; range = Lp*pfac; vv = nn/Np/pfac;
deltap = range/nn; deltay = 2*pi/(nn*deltap);




%Grid
gridp = (linspace(0,nn-1,nn)-nn/2) * deltap;
gridy(1,1:nn/2+1) = linspace(0,nn/2,nn/2+1) * deltay;
gridy(1,nn/2+2:nn) = -gridy(1,nn/2:-1:2);

%Sampling
samplesp(1,1:nn/2) = 1./sqrt(1+gridp(nn/2+1:nn).^2).^3;
samplesp(1,nn/2+1:nn) = 1./sqrt(1+gridp(1:nn/2).^2).^3;

%Derivative
in = zeros(nn,Nx);
in1 = zeros(Np,Nx);
for j = 1:Nx,
  samples = fft(samplesp) * (deltap*sqrt(nn/2/pi));
  samples(2:nn) = samples(2:nn) * ll*sqrt(pi/2)./(gridy(2:nn)+10^(-15)) ...
    .*(erf((gridy(2:nn)+10^(-15)+2*x(j))/sqrt(8)/ll) ...
    + erf((gridy(2:nn)+10^(-15)-2*x(j))/sqrt(8)/ll));
  samples(1) = samples(1) * exp(-x(j)^2/2/ll^2);
  samples = ifft(samples);
  in(1:nn/2,j) = real(samples(nn/2+1:nn)) * deltay*sqrt(nn/2/pi);
  in(nn/2+1:nn,j) = real(samples(1:nn/2)) * deltay*sqrt(nn/2/pi);
  in1(:,j) = in(nn*(2*pfac-2)/4/pfac+1:vv:nn*(2*pfac+2)/4/pfac,j);
end


%Grid in xs-space
hs = 2*pi/(Np*hp);
s(1,1:Np/2+1) = linspace(0,Np/2,Np/2+1) * hs;
s(1,Np/2+2:Np) = -s(1,Np/2:-1:2);
[xx,ss] = meshgrid(x,s);

%ODE Solver
options=odeset('RelTol',1e-10,'AbsTol',1e-10,'Stats','on'); 
[T,WW] = ode113(@Dgl,[t (t+Tf)/2 Tf],[w0(:); w1(:); w2(:); w3(:)],...
  options,eps,tau,ll,Nx,Np,Lp,k1,in0,in1,xx,pp,ss,hp,hs);
%Plot the distribution function
for kk = length(T):length(T),
    %Reshape
    w0 = reshape(WW(kk,0*Nx*Np+1:1*Nx*Np),Np,Nx);
    w1 = reshape(WW(kk,1*Nx*Np+1:2*Nx*Np),Np,Nx);
    w2 = reshape(WW(kk,2*Nx*Np+1:3*Nx*Np),Np,Nx);
    w3 = reshape(WW(kk,3*Nx*Np+1:4*Nx*Np),Np,Nx);
   %Teilchendichte ausrechnen
   np = (w0 + (pp-Lp/2).*w2)./w;
   figure(1)
   contourf(np,50); colorbar; shading flat;colormap('jet'); title(num2str(T(kk)));
end
    
end

\end{lstlisting}

\subsection{Dgl.m}
\begin{lstlisting}
%Christian Kohlfürst
%July 26, 2015

%Second file, specifying the ODE

function dW = Dgl(t,w,eps,tau,ll,Nx,Np,Lp,k1,in0,in1,xx,pp,ss,deltap,deltas)
  %Reshape
  w0 = reshape(w(1:Nx*Np),Np,Nx);
  w1 = reshape(w(Nx*Np+1:2*Nx*Np),Np,Nx);
  w2 = reshape(w(2*Nx*Np+1:3*Nx*Np),Np,Nx);
  w3 = reshape(w(3*Nx*Np+1:4*Nx*Np),Np,Nx);
  %Compute the electric field at time t
  Er = 1i*eps*sqrt(pi/2)*ll * (erf((ss+2*xx)/sqrt(8)/ll) + erf((ss-2*xx)/sqrt(8)/ll)) *...
    (1/cosh(t/tau)^2 - 0.5/cosh(t/tau-1)^2 - 0.5/cosh(t/tau+1)^2);  
  %Calculate time step
  w0new = 2*(pp-Lp/2).*w3 + eps*in0/cosh(t/tau)^2 - derivs(w0,Nx,Np,deltap,deltas,Er);
  w1new = - derivs(w1,Nx,Np,deltap,deltas,Er) - derivxc(w2,k1,Nx,Np);
  w2new = -2*w3 + eps*in1/cosh(t/tau)^2 - derivs(w2,Nx,Np,deltap,deltas,Er) ...
                 - derivxs(w1,k1,Nx,Np);
  w3new = -2*(pp-Lp/2).*w0 + 2*w2 - derivs(w3,Nx,Np,deltap,deltas,Er);
  %Reshape
  dW = [w0new(:); w1new(:); w2new(:); w3new(:)]; 
end


%Symmetric functions with respect to x
function ux = derivxc(v,k1,Nx,Np)
    ux = zeros(Np,Nx);
    for i = 1:Np
        v_full = [fliplr(v(i,:)) v(i,2:Nx-1)];
        v_full = fft(v_full);
        v_full = ifft(1i*k1.*v_full);
        ux(i,:) = [v_full(Nx:end) v_full(1)];
    end    
end

%Antiymmetric functions with respect to x
function ux = derivxs(v,k1,Nx,Np)
    ux = zeros(Np,Nx);
    for i = 1:Np
        v_full = [0 -fliplr(v(i,2:Nx-1)) 0 v(i,2:Nx-1)];
        v_full = fft(v_full);
        v_full = ifft(1i*k1.*v_full);
        ux(i,:) = [v_full(Nx:end) v_full(1)];
    end    
end

%Derivative with respect to s
function us = derivs(v,Nx,Np,deltap,deltas,Er)
    us = zeros(Np,Nx);
    samples = zeros(1,Np);
    for j = 1:Nx
        samples(1:Np/2) = v(Np/2+1:Np,j);
        samples(Np/2+1:Np) = v(1:Np/2,j);
        samples = fft(samples) .* transpose(Er(:,j));
        samples = ifft(samples);
        us(1:Np/2,j) = deltas*deltap*Np/2/pi * real(samples(Np/2+1:Np));
        us(Np/2+1:Np,j) = deltas*deltap*Np/2/pi * real(samples(1:Np/2));
    end
end

\end{lstlisting}

\newpage
\section{Example calculations}

\begin{figure}[tbh]
{\includegraphics[width=.495\textwidth]{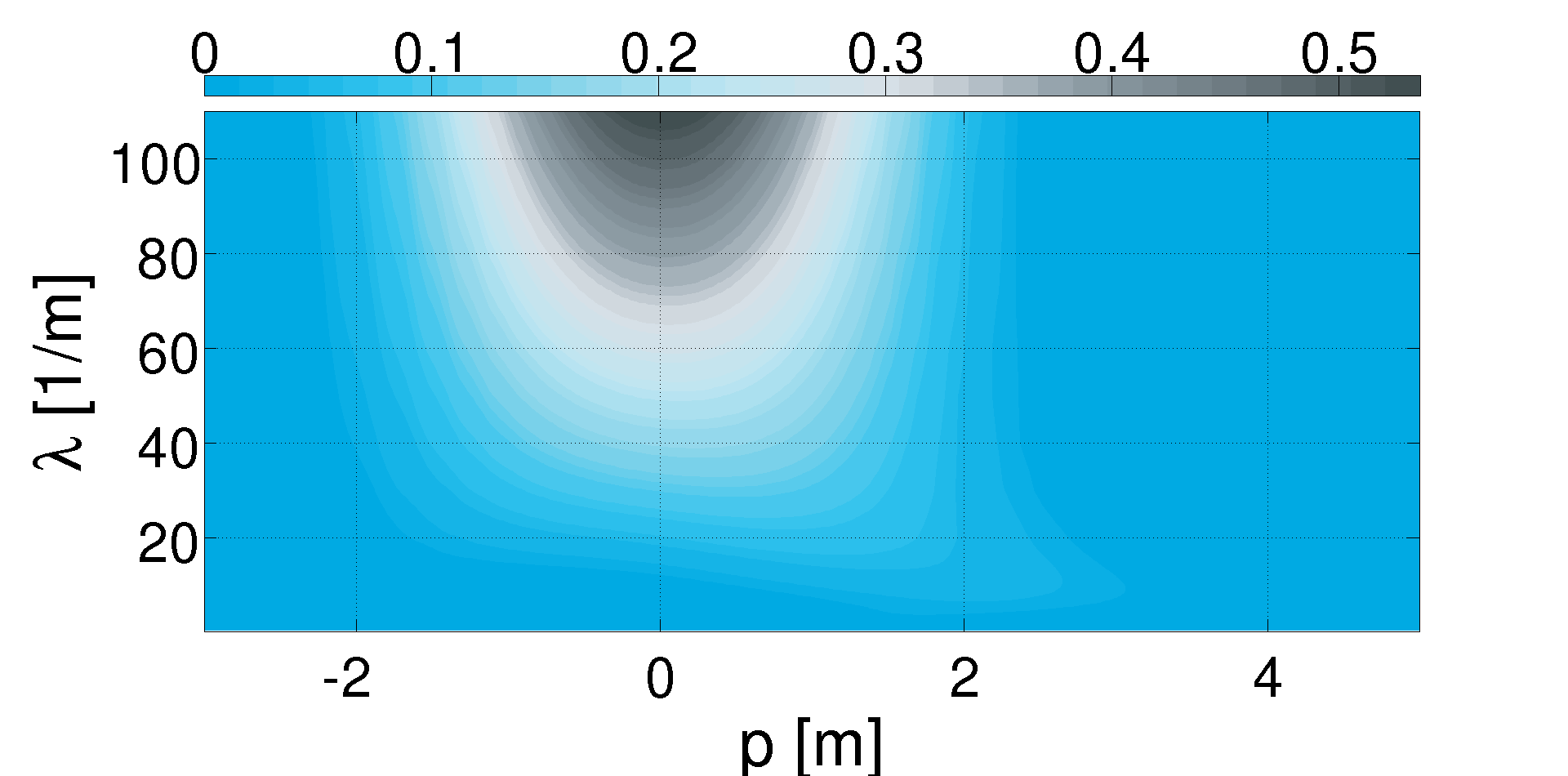}}      
{\includegraphics[width=.495\textwidth]{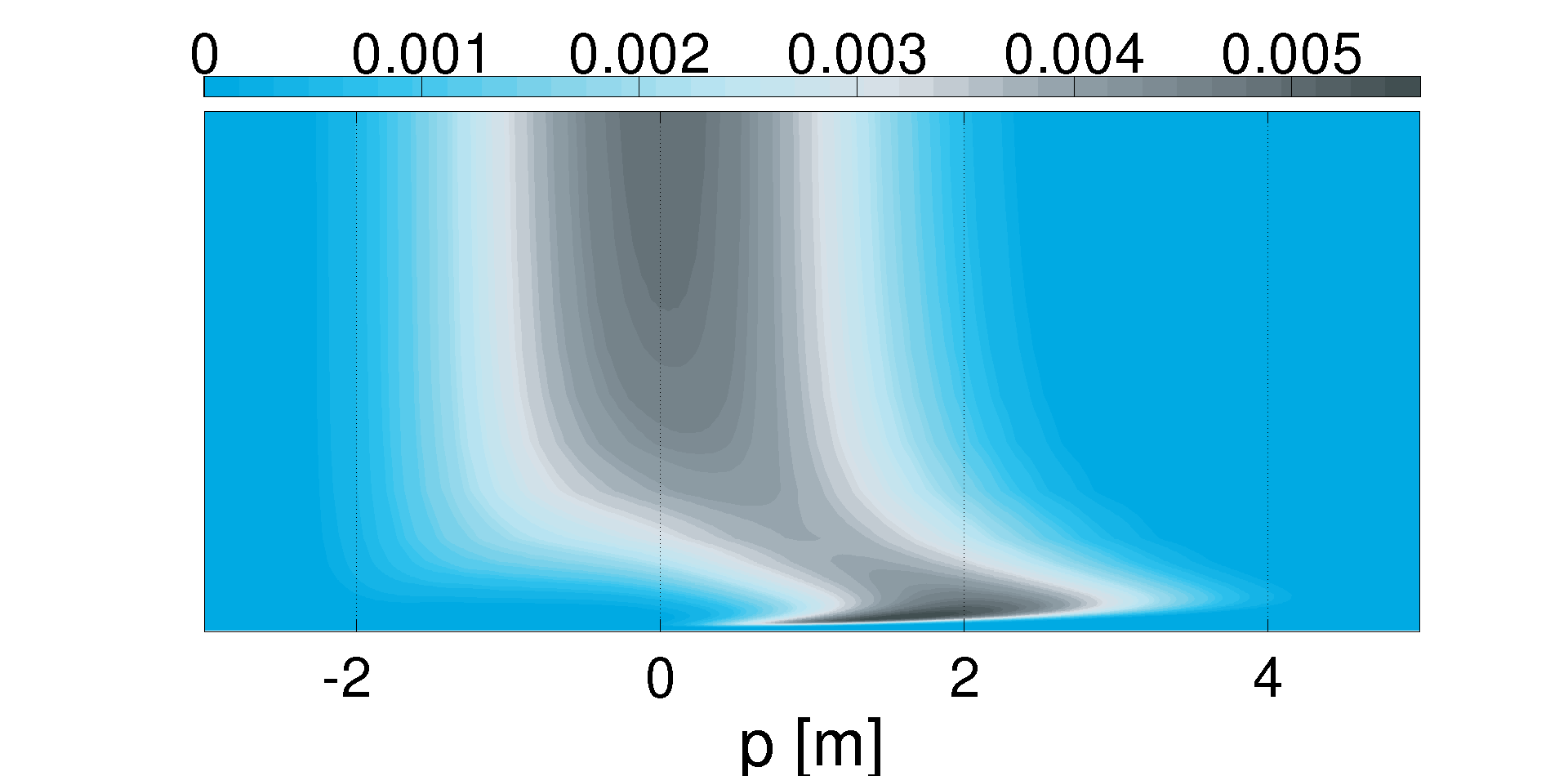}}    
\caption{Distribution function(left-hand side) and normalized distribution function(right-hand side) for a symmetric few-cycle pulse
focusing on the region $\lambda < 15$ $m^{-1}$. The bunch of particles is shifted towards positive momentum, but has a turning point at $\lambda \approx 5$ $m^{-1}$. Self-bunching effects becomes visible as the normalized particle distribution is maximal at $p \approx 2$ $m$ and $\lambda \approx 2$ $m^{-1}$. Parameters: $\varepsilon = 1.0$, $\tau=10$ $m^{-1}$ and Tab. \ref{Tab_Matlab1}.}
\label{FigApp1}
\end{figure}

\begin{figure}[tbh]
{\includegraphics[width=.495\textwidth]{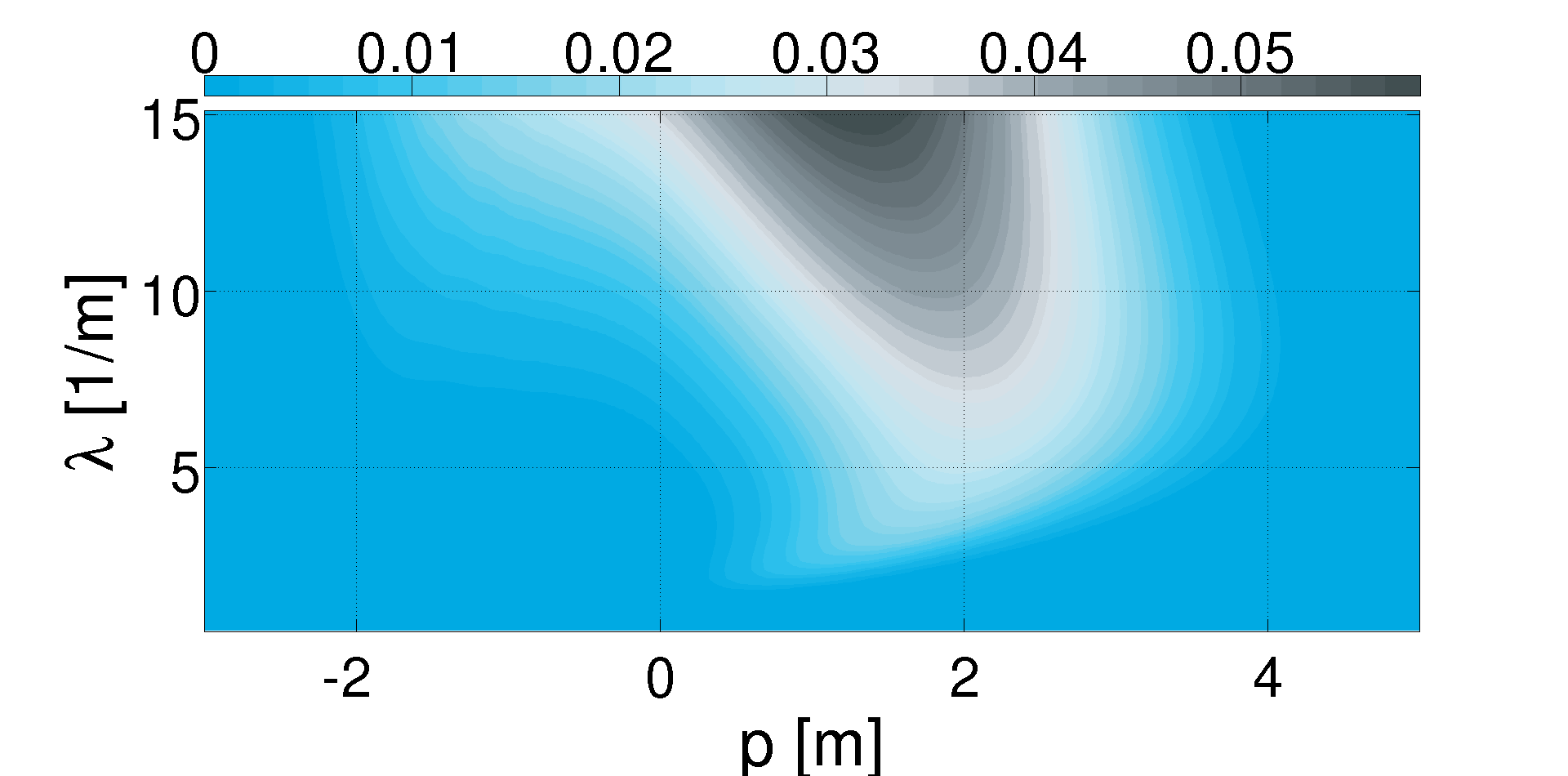}}      
{\includegraphics[width=.495\textwidth]{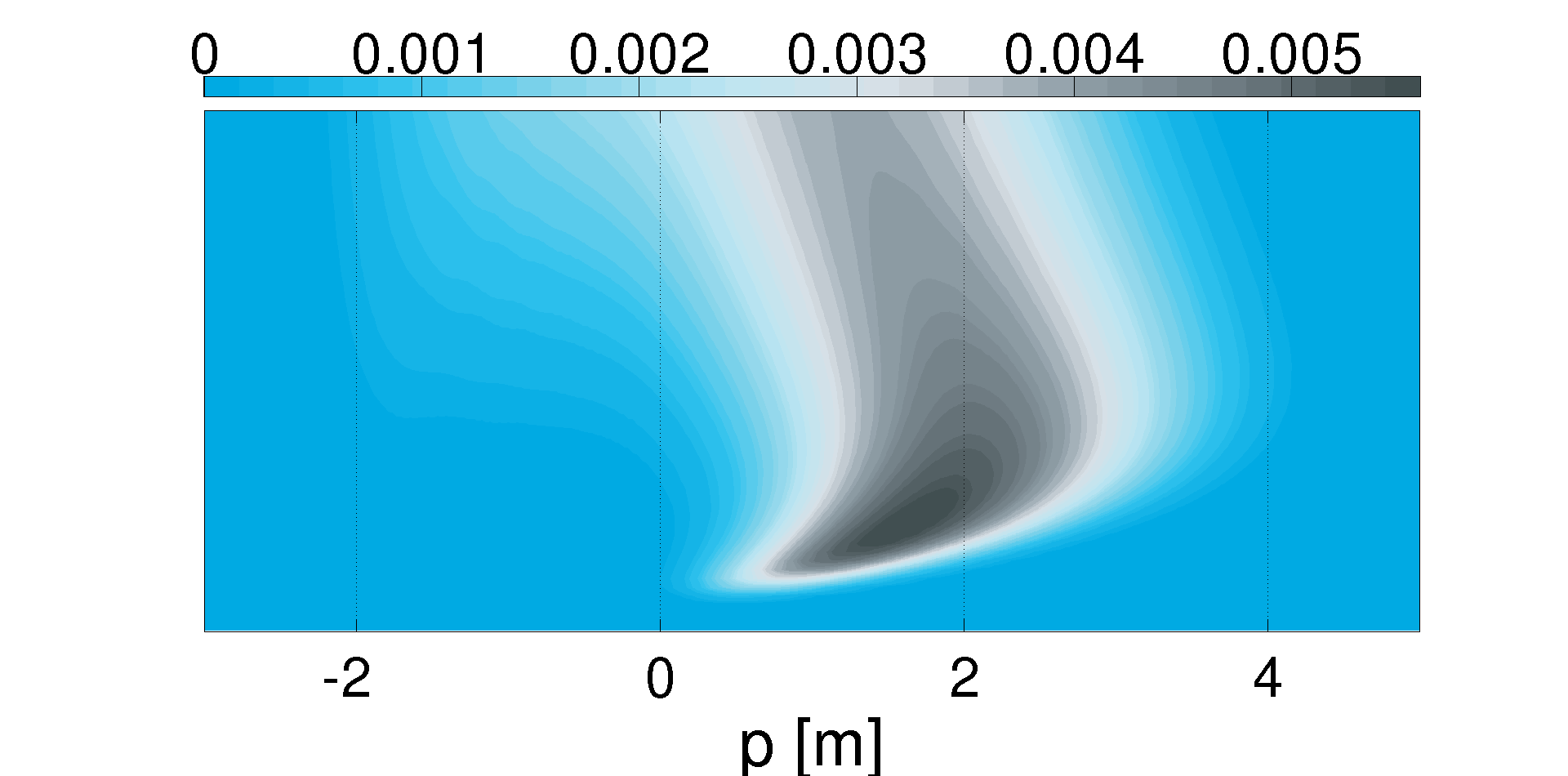}}    
\caption{Distribution function(left-hand side) and normalized distribution function(right-hand side) for a few-cycle pulse, symmetric in $t$. One can distinguish two regions separated around $\lambda \approx 10$ $m^{-1}$. For $\lambda \gg 10$ $m^{-1}$ the distribution function increases monotonously. For small spatial extent particle deflection and self-bunching become important.
Parameters: $\varepsilon = 1.0$, $\tau=10$ $m^{-1}$ and Tab. \ref{Tab_Matlab1}.}
\label{FigApp2}
\end{figure}

\begin{figure}[tbh]
\centering
{\includegraphics[width=.75\textwidth]{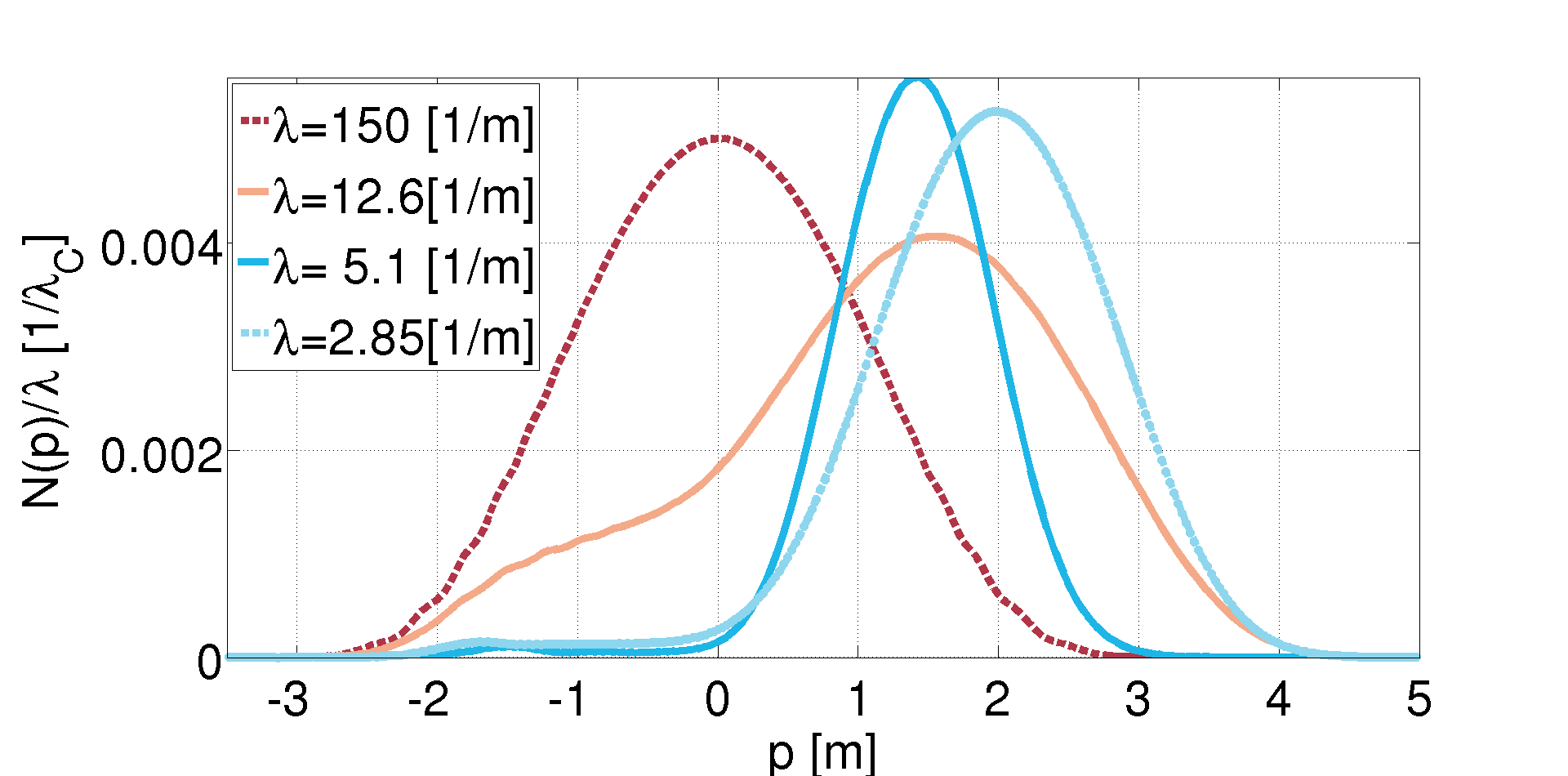}} 
\caption{Normalized particle distribution function $N(p)/ \lambda$ for a few-cycle pulse, symmetric in $t$. For $\lambda = 150$ $m^{-1}$ the particle distribution is symmetric around $p = 0$. For smaller $\lambda$ this symmetry vanishes as the particles net acceleration is greater than zero and in addition self-bunching effects become visible. Parameters: $\varepsilon = 1.0$, $\tau=10$ $m^{-1}$ and Tab. \ref{Tab_Matlab2}.}
\label{FigApp3}
\end{figure}

\clearpage
\section{Parameter tables}

\ctable[pos=hb,
caption = {The computation has been performed using the MATLAB code above. The lines $2$ and $5$ provide the necessary information in order to obtain the intermediate parameter values. The results are illustrated in Fig. \ref{FigApp1} and Fig. \ref{FigApp2}.},
label = Tab_Matlab1, 
mincapwidth = \textwidth,
]{ c c c c c c c }{
}{
    \toprule
    $\tau$ $[m^{-1}]$ & $\varepsilon$ & $\lambda$ $[m^{-1}]$ & $L_{p_x}$ $[m]$ & $N_{p_x}$ & $L_{x}$ $[m]$ & $N_{x}$ \\
    \midrule
    10 & 1.0 & 0.35 & [-15,15] & 1024 & [-75.7,75.7] & 1024  \\       
    \midrule
    10 & 1.0 & 0.35+0.25 $i$ & [-15,15] & 1024 & [-75-2$\lambda$,75+2$\lambda$] & 1024  \\   
    \midrule
    10 & 1.0 & 15.1 & [-15,15] & 1024 & [-105.2,105.2] & 1024  \\       
    \midrule
    10 & 1.0 & 20 & [-15,15] & 1024 & [-115,115] & 1024  \\   
    \midrule
    10 & 1.0 & 20+10 $j$ & [-15,15] & 1024 & [-75-2$\lambda$,75+2$\lambda$] & 1024  \\   
    \midrule
    10 & 1.0 & 110 & [-15,15] & 1024 & [-295,295] & 1024  \\      
    \bottomrule
}

\ctable[pos=hb,
caption = {The computation has been performed using the MATLAB code above. The results are illustrated in Fig. \ref{FigApp3}.},
label = Tab_Matlab2, 
mincapwidth = \textwidth,
]{ c c c c c c c }{
}{
    \toprule
    $\tau$ $[m^{-1}]$ & $\varepsilon$ & $\lambda$ $[m^{-1}]$ & $L_{p_x}$ $[m]$ & $N_{p_x}$ & $L_{x}$ $[m]$ & $N_{x}$ \\
    \midrule
    10 & 1.0 & 150 & [-15,15] & 1024 & [-345,345] & 1024  \\       
    \midrule
    10 & 1.0 & 12.6 & [-15,15] & 1024 & [-100.2,100.2] & 1024  \\       
    \midrule
    10 & 1.0 & 5.1 & [-15,15] & 1024 & [-85.2,85.2] & 1024  \\     
    \midrule
    10 & 1.0 & 2.85 & [-15,15] & 1024 & [-80.7,80.7] & 1024  \\      
    \bottomrule
}

\addcontentsline{toc}{chapter}{Bibliography}
\bibliography{books,article,articleB,articlePond,articleMass}
\bibliographystyle{unsrt}
\end{document}